\documentstyle[12pt,fleqn]{article}
\begin{document}
\begin{titlepage}
\raggedleft{\rm IASSNS-HEP-97/100\\ hep-th/9710046\\ October 1997}
\bigskip
\begin{center} \Large\bf
Black Holes and Solitons in String Theory
\end{center}
\bigskip
\begin{center}\rm
Donam Youm
\footnote{e-mail: youm@sns.ias.edu}\\
\smallskip\normalsize\it
School of Natural Sciences, Institute for Advanced
Study\\ Olden Lane, Princeton, NJ 08540
\end{center}
\rm\normalsize
\bigskip
\begin{abstract}
We review various aspects of classical solutions in string theories.  
Emphasis is placed on their supersymmetry properties, their special roles 
in string dualities and microscopic interpretations.  
Topics include black hole solutions in string theories on tori and $N=2$ 
supergravity theories; $p$-branes; microscopic interpretation of black hole 
entropy.  We also review aspects of dualities and BPS states.  
\end{abstract}
\end{titlepage}
\tableofcontents
\newpage
\begin{sloppypar}

\section{Outline of the Review}\label{intro}

It is a purpose of this review to discuss recent development in 
black hole and soliton physics in string theories.  
Recent rapid and exciting development in string dualities 
over the last couple of years changed our view on string theories.  
Namely, branes and other types of classical solutions that 
were previously regarded as irrelevant to string theories 
are now understood as playing important roles in  
non-perturbative aspects of string theories; these solutions 
are required to exist within string spectrum by recently 
conjectured string dualities.  Particularly, $D$-branes which are 
identified as non-perturbative string states that carry charges 
in R-R sector have classical $p$-brane solutions in string effective 
field theories as their long-distance limit description.  $p$-branes, 
other types of classical stringy solutions and fundamental strings 
are interrelated via web of recently conjectured string dualities.   
Much of progress has been made in constructing various $p$-brane and 
other classical solutions in string theories in an attempt to understand 
conjectured (non-perturbative) string dualities.  We review such progress 
in this paper.  
    
In particular, we discuss black hole solutions in string effective field 
theories in details.  Recent years have been active period 
for constructing black hole solutions in string theories.  
Construction of black hole solutions in heterotic string on tori with 
the most general charge configurations is close to completion.  
(As for rotating black holes in heterotic string on $T^6$, one charge 
degree of freedom is missing for describing the most general charge 
configuration.) 
Also, significant work has been done on a special class of black holes 
in $N=2$ supergravity theories.  These solutions, called {\it double 
extreme} black holes, are characterized by constant scalars and 
correspond to the minimum energy configurations among extreme solutions.  

Among other things, study of black holes and other classical solutions 
in string theories is of particular interest since these allow to address 
long-standing problems in quantum gravity such as microscopic interpretation 
of black hole thermodynamics within the framework of 
superstring theory.  In this review, we concentrate on recent remarkable 
progress in understanding microscopic origin of black hole entropy.  
Such exciting developments were prompted by construction of general class 
of solutions in string theories and realization that non-perturbative R-R 
charges are carried by $D$-branes.  Within subset of solutions with 
restricted range of parameters, the Bekenstein-Hawking entropy has been 
successfully reproduced by stringy microscopic calculations.  

Since the subject reviewed in this paper is broad and rapidly 
developing,  it would be a difficult task to survey every aspects 
given limited time and space.  The author made an effort 
to cover as many aspects as possible, especially emphasizing aspects of 
supergravity solutions, but there are still many issues missing in this 
paper such as stringy microscopic interpretation of black hole radiation, 
M(atrix) theory description of black holes and the most recent developments 
in $N=2$ black holes and $p$-branes.  
The author hopes that some of missing issues will be covered by other 
forthcoming review paper by Maldacena \cite{MALrep}.  
The review is organized as follows.  
Chapters \ref{bps} and \ref{dual} are introductory chapters where we 
discuss basic facts on solitons and string dualities which are necessary 
for understanding the remaining chapters.  In these two chapters, 
we especially illuminate relations between BPS solutions 
and string dualities.  In chapter \ref{n4bh}, we summarize recently 
constructed general class of black hole solutions in heterotic string 
on tori.  We show explicit generating solutions in each spacetime 
dimensions and discuss their properties.  
In chapter \ref{n2bh}, we review aspects of black holes in $N=2$ 
supergravity theories.  We discuss principle of a minimal central 
charge, double extreme solutions and quantum corrections.   
In chapter \ref{pbr}, we summarize recent development in $p$-branes.  
Here, we show how $p$-branes and other related solutions fit into 
string spectrum and discuss their symmetry properties under string 
dualities.  We systematically study various single-charged $p$-branes and 
multi-charged $p$-branes (dyonic $p$-branes and intersecting $p$-branes) 
in different spacetime dimensions. 
We also discuss black holes in type-II string on tori as 
special cases and their embedding to $p$-branes in higher-dimensions.  
In chapters \ref{ent} and \ref{dbr}, we summarize the recent exciting 
development in microscopic interpretation of black hole entropy within 
the framework of string theories.  
We discuss Sen's calculation of statistical entropy of 
electrically charged black holes, Tseytlin's work on statistical 
entropy of dyonic black holes within chiral null model and 
$D$-brane interpretation of black hole entropy.  

\section{Soliton and BPS State}\label{bps}
 
Solitons are defined as time-independent, non-singular, localized 
solutions of classical equations of motion with finite energy (density) 
in a field theory \cite{RAJ82,CALhs030}.  Such solutions in 
$D$ spacetime dimensions are alternatively called {\it $p$-branes} 
\cite{DUFkl,DUF203} if they are localized in $D-1-p$ spatial coordinates 
and independent of the other $p$ spatial coordinates, where $p< D-1$.  
For example, the $p=0$ case (0-brane) has a characteristic of point 
particles and is also called a black hole; 
$p=1$ case is called a string; $p=2$ case is a membrane.   
The main concern of this paper is on the $p=0$ case, but we discuss 
the extended objects ($p\geq 1$) in higher dimensions as embeddings of 
black holes and in relations to string dualities.  

As non-perturbative solutions of field theories, solitons have 
properties different from perturbative solutions in field theories.  
First, the mass of solitons is inversely proportional to some powers of 
dimensionless coupling constants in field theories.  So, in the 
regime where the perturbative approximations are valid (i.e. 
weak-coupling limit), the mass of solitons is arbitrarily large and the 
soliton states decouple from the low energy effective theories.  
So, their contributions to quantum effects are negligible.  Their contribution 
to full dynamics becomes significant in the strong coupling regime.   
Second, solitons are characterized by ``topological charges'', rather 
than by ``Noether charges''.  Whereas the Noether charges are associated with 
the conservation laws associated with continuous symmetry of the theory, 
the topological conservation laws are consequence of topological properties 
of the space of non-singular finite-energy solutions.  The space of 
nonsingular finite energy solutions is divided into several disconnected 
parts.  It takes infinite amount of energy to make a transition from one 
sector to another, i.e. it is not possible to make a transition 
to the other sector through continuous deformation.   
Third, the solitons with fixed topological charges are additionally  
parameterized by a finite set of numbers called ``moduli''.  
Moduli or alternatively called collective coordinates are parameters labeling 
different degenerate solutions with the same energy.  The space of solutions 
of fixed energy is called moduli space.  
The moduli of solitons are associated with symmetries of the 
solutions.  For example, due to the translational invariance of the 
Yang-Mills-Higgs Lagrangian, the monopole solution sitting at the origin 
has the same energy as the one at an arbitrary point in ${\bf R}^3$; 
the associated collective coordinates are the center of mass coordinates  
of the monopole.  In addition, there are collective coordinates 
associated with the gauge invariance of the theory.  Note, monopole 
carries charge of the $U(1)$ gauge group which is broken from 
the Non-Abelian ($SU(2)$) one at infinity (where the Higgs 
field takes its value at the gauge symmetry breaking vacuum).  
Thus, only relevant gauge transformations of Non-Abelian gauge 
group that relate different points in moduli space are those that 
do not approach identity at infinity, i.e. those that reduce to non-trivial 
$U(1)$ gauge transformations at infinity.   

Another important characteristic of solitons is that they are 
the minimum energy configurations for given topological charges, i.e. 
the energy of solitons saturates the Bogomol'nyi bound \cite{BOG,PRAs35}.  
The lower bound is determined by the topological charges, e.g. 
the winding number for strings and $U(1)$ gauge charge carried by black 
holes.  The original calculation \cite{BOG} of the energy 
bound for a soliton in flat spacetime involves taking complete 
square of the energy density $T_{tt}$; the minimum energy is saturated 
if the complete square terms are zero. 
Solitons therefore satisfy the first order differential equations
\footnote{Note, the stress-energy tensor is second 
order in derivatives of spacetime coordinates.} 
(``complete square terms'' $=0$), the so-called Bogomol'nyi or 
self-dual equations.  An example is the (anti) self-dual condition 
$F_{\mu\nu}=\pm\star F_{\mu\nu}$ for Yang-Mills instantons \cite{BELpst}.  
Another example is the magnetic monopoles \cite{HOO79,POL20} 
in an $SU(2)$ Yang-Mills theory, which satisfy the first order differential 
equation $B^i=\pm D^i\Phi$ relating the magnetic field $B^i$ to the Higgs 
field $\Phi$.  Here, the Higgs field takes its values at the minimum of 
the potential $V(\Phi)$, where the non-Abelian gauge group $SU(2)$ is 
spontaneously broken down to the Abelian $U(1)$ gauge group.  

The energy of solitons in asymptotically flat curved spacetime 
is given by the ADM mass \cite{ARNdm122,ABBd,HAWh13}, 
i.e. a Poincar\'e invariant conserved energy of gravitating systems.  
The ADM mass is defined in terms of a surface integral of the conserved 
current $J^{\mu} = T^{\mu\nu}K_{\nu}$ over a space-like hypersurface at 
spatial infinity.  Here, $T^{\mu\nu}$ is the energy-momentum tensor 
density and $K_{\nu}$ is a time-like Killing vector of the asymptotic 
spacetime.  The so-called positive-energy theorem 
\cite{CHOm,SHOy42,SHOy79,SHOy81,SHOy48,WIT80,NES,ISRn85,GIBhhp,HUL90} 
proves that the ADM mass of gravitating systems is always positive.  
In such proofs, one calculates the energy associated with a small 
deviation around the background spacetime and finds it always positive, 
implying that the background spacetime (Minkowski or anti-De Sitter 
space-time) is a stable vacuum configuration.  
The proof of the positive energy theorem, first given in \cite{WIT80} 
and refined covariantly in \cite{NES}, involves the volume and the 
surface integrals (related through the Stokes theorem) of Nester's 2-form, 
which is defined in terms of a spinor and its gravitational covariant 
derivative.  Such proofs have an advantage of being easily generalized to 
supergravity theories.  The positive energy theorem proves that the ADM 
mass of gravitating systems is always positive, provided the spinor 
satisfies the Witten's condition and the matter stress energy tensor, 
if any, satisfies the dominant energy condition. 

One way of proving positivity of the energy of solitons 
in curved spacetime is by embedding the solutions 
into (extended) supergravity theories \cite{GIBh82,GIBhhp} as solutions 
to equations of motion.  In this case, the Nester's form is defined in 
terms of the supersymmetry parameters and their supercovariant derivatives.  
Then, the surface integral yields the supercharge anticommutation relations 
of extended supersymmetry, i.e. the 4-momentum term plus the central 
charge term.  The 4-momentum in the surface integral is the ADM 
4-momentum \cite{ARNdm122,ABBd} of the soliton and the central charge 
corresponds to the topological charge carried by the soliton 
\cite{OLIw78,OSB83,DEAgit}; the soliton behaves as if a particle 
carrying the corresponding 4-momentum and quantum numbers.  
This is a reminiscent of BPS states in extended supergravities.  
One can think of solitons as realizations of states in supermultiplet 
carrying central charges of extended supersymmetry \cite{GIBp84,KAL29}.  
In fact, for each Killing spinor, defined as a spinor field which is 
covariant with respect to the supercovariant derivative, one can 
define a conserved anticommuting supercharge, whose anticommutation 
relation is just the surface integral of the Nester's 2-form. 

The integrand of the volume integral of the Nester's 2-form yields sum 
of terms bilinear in supersymmetry variations of the fermionic fields in the 
supergravity theory.  Since such terms are positive semidefinite operators, 
provided the (generalized) Witten's condition \cite{WIT80} and the dominant 
energy condition for the matter stress-energy tensor are satisfied, the 
terms in the surface integral have to be non-negative, leading to the 
inequality ``(ADM mass) $\geq$ (the maximum eigenvalue of the topological 
charge term)''.  Again a reminiscent of the mass bound for the states in 
the BPS supermultiplet.  This bound is saturated iff the supersymmetry 
variations of fermions are all zero.  The equations obtained by setting 
the supersymmetry variations of fermions equal to zero are called the 
Killing spinor equations. 
These are a system of first order differential equations satisfied by the 
minimum energy configuration among solutions with the same 
topological charge.  Such a configuration is a bosonic configuration which 
is invariant under supersymmetry transformations and therefore is  
called supersymmetric.

The necessary and sufficient condition for the existence of supersymmetric 
solution is the existence of ``non-zero'' superconvariantly constant 
spinors, i.e. Killing spinors.  Note, such Killing spinors 
define supercharges, which act on the lowest spin state to build up 
supermultiplets of superalgebra.  Killing spinors are Goldstone modes 
of broken supersymmetries;  for each supersymmetry preserved, the 
corresponding supercharge is projected onto zero norm states, and the 
rest of supercharges are associated with Goldstone spinor degrees of 
freedom originated from broken supersymmetries.  
The number of supercharges which are projected onto
the zero norm states is determined by the number of distinct eigenvalues of 
the central charge matrix.  In the language of 
solitons, such central charge matrix is determined by the charge 
configurations of solitons.  Alternatively, one can determine 
the number of supersymmetries preserved by the solitons from the 
spinor constraints, which are byproducts of the Killing spinor equations 
along with self-dual or the first order differential equations.  The number 
of constraints on the Killing spinors are again determined by the charge 
configuration of the solitons.  These constraints determine the 
number of independent spinor degrees of freedom, i.e. the number 
of supersymmetries preserved by the soliton.  
Thus, the number of supersymmetries preserved by solitons 
is intrinsically related to the topological charge configurations of 
solitons through either the number of eigenvalues of central matrix or 
the number of constraints on the Killing spinors.  

In the following, we elaborate on ideas discussed in the above 
in a more precise and concrete way, by quantifying ideas and giving 
some examples.  First, we discuss how the physical parameters (mass, angular 
momenta, etc) are defined from solitons.  
Then, we discuss the BPS multiplets of extended supersymmetry theories.    
Finally, we discuss positive energy theorem of general relativity 
and extended supergravity theories.

\subsection{Physical Parameters of Solitons}\label{bpspar}

We discuss how to define physical parameters 
(e.g. the ADM mass, angular momenta, $U(1)$ charges) of  
gravitating systems.  This serves to fix our conventions for 
defining parameters of solitons.  
The classical solutions near the space-like infinity can be regarded 
as the ``imprints'' of the ADM mass, angular momenta and 
electric/magnetic charges of the source.  

First, we discuss the parameters of spacetime metric.  
The physical parameters are defined with reference to the background 
(asymptotic) spacetime.
We assume that the spacetime is asymptotically Minkowski at space-like 
infinity, since the solitons under consideration in this review satisfy 
this condition.  

We consider the following general form of action in $D$ spacetime 
dimensions:
\begin{equation}
S=\int\sqrt{-g}d^{D}x\left({1\over{16\pi G^D_N}}{\cal R}+
{\cal L}_{mat}\right),
\label{genact}
\end{equation}
where $G^D_N$ is the $D$-dimensional Newton's constant (related to 
the Plank constant $\kappa_D$ as $\kappa^2_D=8\pi G^D_N$) and 
${\cal L}_{mat}$ is the matter Lagrangian density.  For the signature 
of the metric $g_{\mu\nu}$, we take the mostly positive convention 
$(-+\cdots +)$.  From (\ref{genact}), one obtains the following Einstein 
field equations for gravitation:
\begin{equation}
G_{\mu\nu}={\cal R}_{\mu\nu}-{1\over 2}g_{\mu\nu}{\cal R}=
8\pi G^N_DT^{mat}_{\mu\nu}, 
\label{genein}
\end{equation}
where the matter stress-energy tensor $T^{mat}_{\mu\nu}$ is defined as
\begin{equation}
T^{mat}_{\mu\nu}\equiv {2\over{\sqrt{-g}}}{{\partial(\sqrt{-g}
{\cal L}_{mat})}\over{\partial g^{\mu\nu}}}, 
\label{matstres}
\end{equation}
where $T^{mat}_{ij}$ are stresses, $T^{mat}_{0i}$ are momentum densities 
and $T^{mat}_{00}$ is the mass-energy density ($i,j=1,...,D-1$). 

In order to measure the mass, the momenta and the angular momenta 
of gravitating systems, one usually goes to the external spacetime 
far away from the source.  In this region, the gravitational field 
is weak and, therefore, the Einstein's field equations 
(\ref{genein}) take the form linear in the deviation 
$h_{\mu\nu}$ of the metric $g_{\mu\nu}$ from the flat one 
$\eta_{\mu\nu}$ ($g_{\mu\mu}=\eta_{\mu\nu}+h_{\mu\nu}$, $|h_{\mu\nu}| 
\ll 1$).  This linearize field equations
\footnote{In the linearized field theory, the spacetime vector indices 
are raised and lowered by the Minkowski metric $\eta_{\mu\nu}$.} 
have the invariance under the infinitesimal coordinate transformations 
($x^{\mu}\to x^{\mu}+\xi^{\mu}$) $h_{\mu\nu}\to h_{\mu\nu}-
\partial_{\nu}\xi_{\mu}-\partial_{\mu}\xi_{\nu}$, which resembles the 
gauge transformation of $U(1)$ gauge fields.  
(The linearized Riemann tensor, Einstein tensor, etc 
are examples of invariants under this transformation.)  
By using this gauge-invariance, one can fix the gauge by imposing the 
``Lorentz gauge'' condition $\partial_{\nu}(h^{\mu\nu}-{1\over 2}
\eta^{\mu\nu}h^{\alpha}_{\ \alpha})=0$.  
This gauge condition is left invariant under the gauge transformations 
satisfying $\xi^{\alpha,\beta}_{\ \ \,\beta}=0$.  In this gauge, the 
Einstein's equations take the form, which resembles the 
Maxwell's equations:
\begin{equation}
\nabla^2h_{\mu\nu}=-16\pi G^D_N\left(T^{mat}_{\mu\nu}-{1\over{D-2}}
\eta_{\mu\nu}T^{mat}\right)\equiv-16\pi G^D_N\bar{T}^{mat}_{\mu\nu},
\label{lineineqn}
\end{equation}
where $\nabla^2=\partial_{\alpha}\partial^{\alpha}$ is the  
flat $(D-1)$-dimensional space Laplacian and $T\equiv T^{\mu}_{\ \mu}$.  

The linearized Einstein's equations (\ref{lineineqn}) have the following 
general solution that resembles the retarded wave solution of 
the Maxwell's equations: 
\begin{eqnarray}
h_{\mu\nu}(x^i)&=&{{16\pi G^D_N}\over{(D-3)\Omega_{D-2}}}
\int{{\bar{T}_{\mu\nu}(t-|\vec{x}-\vec{y}|,\vec{y})}\over
{|\vec{x}-\vec{y}|^{D-3}}}d^{D-1}y
\cr
&=&{{16\pi G^D_N}\over{(D-3)\Omega_{D-2}}}{1\over{r^{D-3}}}\int
\bar{T}_{\mu\nu}d^{D-1}y
\cr
& &+{{16\pi G^D_N}\over{\Omega_{D-2}}}
{{x^k}\over{r^{D-1}}}\int y^k\bar{T}_{\mu\nu}d^{D-1}y+\cdots,
\label{lineinsol}
\end{eqnarray}
where $\Omega_{n}={{2\pi^{{n+1}\over 2}}\over{\Gamma({{n+1}\over 2})}}$ 
is the area of $S^n$ and $i,k=1,...,D-1$ are spatial indices.  
Note, the ADM $D$-momentum vector $P^{\mu}$ and angular momentum tensor 
$J^{\mu\nu}$ of the gravitating system are defined as
\begin{equation}
P^{\mu}=\int T^{\mu 0}d^{D-1}x,\ \ \ \ \ \ 
J^{\mu\nu}=\int(x^{\mu}T^{\nu 0}-x^{\nu}T^{\mu 0})d^{D-1}x.
\label{gravpara}
\end{equation}
In particular, the time component of $P^{\mu}$ is the ADM mass $M$, 
i.e. $M=P^0$.  

With a suitable choice of coordinate basis, one can put the spatial 
components $J^{ij}$ ($i,j=1,...,D-1$) of $J^{\mu\nu}$ in the following 
form expressed in terms of the angular momenta $J_k$ 
($k=1,...,[{{D-1}\over 2}]$) in each rotational plane:
\begin{equation}
[J^{ij}]=\left(\matrix{0&J_1& & & \cr -J_1&0& & & \cr 
& &0&J_2& \cr & &-J_2&0& \cr & & & &\ddots}\right),
\label{angmom}
\end{equation}
where for the even $D$ the last row and column have zero entries.  

In obtaining the general leading order expression for the metric, 
one chooses the rest frame ($P^i=0$) with the origin of coordinates 
at the center of mass of the system ($\int x^iT^{00}d^{D-1}x=0$).  
In this frame, $J^{0i}=0$, $J^{ij}=2\int x^iT^{j0}d^{D-1}x$ and 
$g_{\mu\nu}$ takes the form:
\begin{eqnarray}
ds^2&=&-\left[1-{{16\pi G^D_N}\over{(D-2)\Omega_{D-2}}}{M\over{r^{D-3}}}
+{\cal O}\left({1\over{r^{D-1}}}\right)\right]dt^2
\cr
& &-\left[{{16\pi G^D_N}\over{\Omega_{D-2}}}{{x^k}\over{r^{D-1}}}J^{ki}
+{\cal O}\left({1\over{r^{D-1}}}\right)\right]dtdx^i
\cr
& &+\left[\left(1+{{16\pi G^D_N}\over{(D-2)(D-3)\Omega_{D-2}}}
{M\over{r^{D-3}}}\right)\delta_{ij}\right.
\cr
& &+\left.({\rm gravitational\ radiation\ terms})\right]dx^idx^j.
\label{asymmet}
\end{eqnarray}
Note, the leading order terms of the asymptotically Minkowski 
metric is time-independent and is determined uniquely by the ADM 
mass $M$ and the intrinsic angular momenta $J^{ij}$ of the source.  

The general action (\ref{genact}) contains the following kinetic term
\footnote{We omit the dilaton factor in the kinetic term for 
the sake of the argument.} 
for a $d$-form potential $A_d={1\over{d!}}A_{\mu_1\cdots\mu_d}dx^{\mu_1}
\wedge\cdots\wedge dx^{\mu_d}$ with field strength $F_{d+1}=dA_d$:
\begin{equation}
S_{d-{\rm form}}={1\over{16\pi G^D_N}}\int d^Dx\sqrt{-g}
\left({1\over{2(d+1)!}}F^2_{d+1}\right). 
\label{dformact}
\end{equation}
Note, in this kinetic term, $G^D_N$ is absorbed in the action in contrast 
to the form of the matter term in (\ref{genact}).   The field equations 
and Bianchi identities of $A_d$ are
\begin{equation}
d\star F_{d+1}=2\kappa^2_D(-)^{d^2}\star J_d,\ \ \ \ \ \ 
dF_{d+1}=0,
\label{dformeqs}
\end{equation}
where $J_d$ is the rank $d$ source current and $\star$ denotes the Hodge-dual 
transformation in $D$ spacetime dimensions, i.e. $(\star A_d)^{\mu_1\cdots
\mu_{D-d}}\equiv{1\over{d!}}\epsilon^{\mu_1\cdots\mu_D}(A_d)_{\mu_{D-d+1}
\cdots\mu_D}$ with $\epsilon^{01\cdots D-1}=1$.   

The soliton that carries the ``Noether'' electric 
charge $Q_d$ under $A_d$ is an elementary extended object with 
$d$-dimensional worldvolume, called $(d-1)$-brane, and has the electric 
source $J$ coming from the $\sigma$-model action of the $(d-1)$-brane.  
The ``topological'' magnetic charge $P_{\tilde{d}}$ of $A_d$ is carried 
by a solitonic (i.e. singularity and source free) object with 
$\tilde{d}$-dimensional worldvolume, called $(\tilde{d}-1)$-brane, where 
$\tilde{d}\equiv D-d-2$.  
The ``Noether'' electric and the ``topological'' magnetic charges 
of $A_d$ are defined as
\begin{eqnarray}
Q_d&\equiv& \sqrt{2}\kappa_D\int_{{\cal M}_{D-d}}\,(-)^{d^2}\star J_d =
{1\over{\sqrt{2}\kappa_D}}\int_{S^{D-d-1}}\star F_{d+1}, 
\cr
P_{\tilde{d}}&\equiv&{1\over{\sqrt{2}\kappa_D}}\int_{S^{d+1}}\,F_{d+1}.
\label{dformagelec}
\end{eqnarray}
These charges obey the Dirac quantization condition \cite{NEP31,TEI167}:
\begin{equation}
{{Q_dP_{\tilde{d}}}\over{4\pi}}={n\over 2},\ \ \ n\in {\bf Z}.
\label{dformdirac}
\end{equation}
The electric and magnetic charges of $A_d$ have dimensions 
$[Q_d]=L^{-{1\over 2}(D-2d-2)}$ and $[P_{\tilde{d}}]=
L^{{1\over 2}(D-2d-2)}$, respectively.  Electric/magnetic 
charges are dimensionless when $D=2(d+1)$.  Examples are point-like 
particles ($d=1$) in $D=4$, strings ($d=2$) in $D=6$ \cite{SCH171,GRElpt384} 
and membranes ($d=3$) in $D=8$ \cite{GRElpt384,IZQlpt}.  

From (\ref{dformagelec}), one sees that the Ans\"atze for $F_{d+1}$ 
for the soliton that carries electric or magnetic charge of $A_d$ are
respectively given by:
\begin{equation}
\star F_{d+1}=\sqrt{2}\kappa_DQ_d\varepsilon_{\tilde{d}+1}/
\Omega_{\tilde{d}+1}, \ \ \ \ \ 
F_{d+1}=\sqrt{2}\kappa_DP_{\tilde{d}}\varepsilon_{d+1}/\Omega_{d+1},
\label{dformansat}
\end{equation}
where $\varepsilon_n$ denotes the volume form on $S^n$, and the electric 
and magnetic charges of $A_d$ are defined from the asymptotic behaviors:
\begin{equation}
A_d\sim {{\omega_d}\over{\sqrt{2}\kappa_D}}{{Q_d}\over{r^{D-d-2}}}, 
\ \ \ \ 
F_{d+1}\sim {{\Omega_{d+1}}\over{\sqrt{2}\kappa_D}}
{{P_{\tilde{d}}}\over{r^{d+1}}},
\label{dformelmgasym}
\end{equation}  
where $r$ is the transverse distance from the $(d-1)$-brane, 
$\omega_{d}$ is the volume form for the $(d-1)$-brane worldvolume 
and $\Omega_{d+1}$ is the volume form of $S^{d+1}$ surrounding the brane.  

From the elementary $(d-1)$-brane, one finds that 
the electric charge $Q_d$ is related to the tension $T_d$ of the 
$(d-1)$-brane in the following way:
\begin{equation}
Q_d=\sqrt{2}\kappa_DT_d(-)^{(D-d)(d+1)}.
\label{dformelec}
\end{equation}
Here, $T_d$ has dimensions $[T_d]=ML^{d-1}$ in the unit $c=1$ and 
therefore is interpreted as mass per unit $(d-1)$-brane volume.  
In particular, for a 0-brane ($d=1$) the tension $T_1$ is the mass.   
The Dirac quantization condition (\ref{dformdirac}), together 
with (\ref{dformelec}), yields the following form of the magnetic 
charge $P_{\tilde{d}}$ of $A_d$:
\begin{equation}
P_{\tilde{d}}={{2\pi n}\over{\sqrt{2}\kappa_DT_d}}(-)^{(D-d)(d+1)}, 
\ \ \ \ n\in{\bf Z}.
\label{dformag}
\end{equation}

We comment on the ADM mass of $(d-1)$-branes.
Note, in deriving (\ref{asymmet}) we assumed that the 
metric $g_{\mu\nu}$ depends on all the spatial coordinates.  
So, (\ref{asymmet}) applies only to the 0-brane type 
soliton (or black holes).  The $(d-1)$-branes do not depend on 
the $(d-1)$ longitudinal coordinates internal to the $(d-1)$-brane 
and therefore the Laplacian in (\ref{lineineqn}) is replaced by the flat 
$(D-d-1)$-dimensional one.   As a consequence, in particular, the 
$(t,t)$-component of the metric has the asymptotic behavior:
\begin{equation}
g_{tt}\sim -\left(1-{{16\pi G^D_N}\over{(D-2)\Omega_{D-d-1}}}
{{M_d}\over{V_{d-1}}}{1\over{r^{D-d-2}}}\right).
\label{ttpbranasym}
\end{equation}
Here, the ADM mass $M_d$ of the $(d-1)$-brane is defined as 
$M_d\equiv\int T^{00}d^{D-1}x=V_{d-1}\int T^{00}d^{D-d}x$, where 
$V_{d-1}$ is the volume of the $(d-1)$-dimensional space internal 
to the $(d-1)$-brane. So, for $(d-1)$-branes it is the ADM mass 
``density'' $\rho_d\equiv{{M_d}\over{V_{d-1}}}=
\int T^{00}d^{D-d}x$ that has the well-defined meaning.  
As an example, we consider the elementary BPS $(d-1)$-brane in 
$D$ dimensions.  The leading order asymptotic behavior of the 
$(t,t)$-component of the metric of $(d-1)$-brane carrying 
one unit of the $d$-form electric charge is
\footnote{When $\tilde{d}=0$, e.g. a string in $D=4$, 
the metric is asymptotically logarithmically divergent.  In this case, 
the ADM mass density is determined from volume integral of 
the $(t,t)$-component of the gravitational energy-momentum 
pseudo-tensor \cite{LANl}.} 
\begin{equation}
g_{tt}\sim -\left(1-{{D-d-2}\over{D-2}}{{c^{(D)}_d}\over{r^{D-d-2}}}
\right),\ \ \ 
\tilde{d}>0,
\label{leadttmet}
\end{equation}
where $c^{(D)}_d\equiv {{2\kappa^2_DT_d}\over{\tilde{d}
\Omega_{\tilde{d}+1}}}$ is the unit $(d-1)$-brane electric 
charge and $r\equiv(x^2_1+\cdots+x^2_{D-d-2})^{1/2}$ is the radial 
coordinate of the transverse space.  For $(d-1)$-branes carrying $m$ 
units of the basic electric charge, $c^{(D)}_d$ in (\ref{leadttmet}) is 
replaced by $mc^{(D)}_d$.  
From (\ref{ttpbranasym}) and (\ref{leadttmet}), one obtains the 
following ADM mass density of the $(d-1)$-brane carrying one 
unit of electric charge:
\begin{equation}
\rho_{d}=T_d={1\over{\sqrt{2}\kappa_D}}|Q_d|.
\label{unitelecadm}
\end{equation}

\subsection{Supermultiplets of Extended Supersymmetry}\label{bpssusy}

\subsubsection{Spinors in Various Dimensions}

Before we discuss the BPS states in extended supersymmetry theories, 
we summarize the basic properties of spinors for each spacetime 
dimensions $D$.  More details can be found, for example, in 
\cite{WET211,WET222,STR2,KUGt221}.  We assume that there is only one 
time-like coordinate.  The types of super-Poincar\'e algebra satisfied 
by supercharges depend on $D$.  The superalgebra is classified according 
to the fundamental spinor representations of the homogeneous group 
$SO(1,D-1)$ and the vector representation of the automorphism group 
that supercharges belong to.  The pattern of superalgebra repeats with 
$D$ mod 8.  

In even $D$, one can define $\gamma^5$-like matrix 
$\bar{\gamma}\equiv\eta\gamma^0\gamma^1\cdots\gamma^{D-1}$ which 
anticommutes with $\gamma^{\mu}$ and has the property $\bar{\gamma}^2=1$ 
(implying $\eta^2=(-1)^{(D-2)/2}$), required for constructing a 
projection operator.   So, the $2^{[D/2]}$ complex component 
{\it Dirac spinor} $\psi$, which is defined to transform as 
$\delta\psi=-{1\over 2}\varepsilon_{\mu\nu}\Sigma^{\mu\nu}\psi$ 
($\Sigma^{\mu\nu}\equiv-{1\over 4}[\gamma^{\mu},\gamma^{\nu}]$) 
under the infinitesimal Lorentz transformation, in even 
$D$ is decomposed into 2 inequivalent {\it Weyl spinors} 
$\psi_+={1\over 2}(1+\bar{\gamma})\psi$ and $\psi_-={1\over 2}(1-
\bar{\gamma})\psi$ with $2^{D/2-1}$ complex components each.  

We discuss the reality properties of spinors.  One can always find a matrix 
$B$ satisfying $\Sigma^{\mu\nu\,*}=B\Sigma^{\mu\nu}B^{-1}$.    
$B$ defines the {\it charge conjugation} operation:
\begin{equation}
\psi \to \psi^c={\cal C}\psi\equiv B^{-1}\psi^*.
\label{chconjop}
\end{equation}
By definition, the charge conjugation operator $\cal C$ commutes 
with the Lorentz generators $\Sigma^{\mu\nu}$, implying that 
$\psi$ and $\psi^c$ have the same Lorentz transformation properties. 
If ${\cal C}^2=1$, or equivalently $BB^*=1$, the Dirac spinor 
$\psi$ can be reduced to a pair of {\it Majorana spinors} (i.e. eigenstates 
of ${\cal C}$) $\psi_A={1\over 2}(1+{\cal C})\psi$ and $\psi_B={1\over 2}
(1-{\cal C})\psi$.  This is possible in $D=2,3,4,8,9$ mod 8.  First, 
in odd $D$, Majorana spinors are necessarily self-conjugate 
under $\cal C$ and are always {\it real}.  So, 
the Dirac spinors in odd $D$ are real [pseudoreal] in $D=1,3$ mod 8 
[$D=5,7$ mod 8].  In even $D$, Majorana spinors can be either 
complex or real depending on whether $\psi$ and $\psi^c$ 
have the same or opposite helicity.  Namely, since ${\cal C}\bar{\gamma}=
(-1)^{(D-2)/2}\bar{\gamma}{\cal C}$, $\psi$ and $\psi^c$ 
have the same [opposite] helicity for even [odd] $(D-2)/2$, i.e. $D=2$ mod 8 
[$D=4,8$ mod 8].  So, in even $D$, the Dirac spinors are real 
[complex] or pseudoreal for $D=2$ mod 8 [$D=4,8$ mod 8] or $D=6$ mod 8, 
respectively.  In particular, in $D=2$ mod 8, both the Weyl and the 
Majorana conditions are satisfied, and therefore 
in this case the Dirac spinor $\psi$ is called {\it Majorana-Weyl}.  

We saw that supercharges $Q_i$ ($i=1,...,N$), 
transforming as spinors under $SO(1,D-1)$, 
have different chirality and reality properties depending on $D$.  
The set $\{Q_i\}$, furthermore, transforms as a vector under an 
{\it automorphism group}, with $i$ acting as a vector index.  
The automorphism group depends on the reality properties of $\{Q_i\}$.  
The automorphism group is $SO(N)$, $USp(N)$ or $SU(N)\times U(1)$ for 
real, pseudoreal or complex case, respectively.  In $D=2$ mod 8 and 
$D=6$ mod 8, the pair of Weyl spinors with opposite chiralities are 
not related via ${\cal C}$ and therefore are independent:  
the automorphism groups are $SO(N_+)\times SO(N_+)$ and $USp(N_+)\times 
USp(N_+)$ in $D=2$ mod 8 and $D=6$ mod 8, respectively, where 
$N_{+}+N_{-}=N$.  The central charge $Z^{IJ}$ transforms as a 
rank 2 tensor under the automorphism group with $(I,J)$ 
acting as tensor indices.   In $D=0,1,7$ mod 8 [$D=3,4,5$ mod 
8], the central charge has the symmetry property $Z^{IJ}=Z^{JI}$ 
[$Z^{IJ}=-Z^{JI}$].

The number $N$ of supercharges $Q^I$ in each $D$ is restricted 
by the physical requirement that particle helicities should not exceed 2 
when compactified to $D=4$ \cite{NAH135,ARAd86,CUR,BERhwn,FRE}. 
This limits the maximum $D$ with 1 time-like coordinate and consistent 
supersymmetric theory to be 11 with $N=1$ supersymmetry, i.e. 32 
supercharge degrees of freedom.  This corresponds to $N=8$ supersymmetry 
in $D=4$ when compactified on $T^7$.  In $D<11$, the number of 
spinor degrees of freedom cannot exceed that of $N=1$, $D=11$ 
theory.  For the pseudoreal cases, i.e. $D=5,6,7$ mod 8, 
only even $N$ are possible.  

\subsubsection{Central Charges and Super-Poincar\'e Algebra}

We discuss types of central charge $Z^{IJ}$ one would expect in 
the super Poincar\'e algebra.  According to a theorem by Haag et. al. 
\cite{HAAls88}, within a unitary theory of point-like particle interactions 
in $D=4$ the central charge can be only Lorentz scalar.  
However, in the presence of $p$-branes ($p\geq 1$), 
central charges $Z^{IJ}_{\mu_1\cdots\mu_p}$ transforming as Lorentz 
tensors can be present in the superalgebra without violating 
unitarity of interactions \cite{DEAgit}.   In fact, as will be shown, 
it is the Lorentz tensor type central charges in higher dimensions 
that are responsible for the missing central charge degrees of freedom 
in lower dimensions when the higher-dimensional superalgebra is compactified 
with an assumption that no Lorentz tensor type central charges are present 
\cite{TOW048}.  

The Lorentz tensor type central charges appear in the supersymmetry algebra 
schematically in the form:
\begin{equation}
\{Q^I_{\alpha},Q^J_{\beta}\}=\delta^{IJ}({\cal C}\gamma^{\mu})_{\alpha\beta}
P_{\mu}+\sum_{p=0,1,...}({\cal C}\gamma^{\mu_1\cdots\mu_{p}})_{\alpha\beta}
Z^{IJ}_{\mu_1\cdots\mu_{p}},
\label{tenssusyalg}
\end{equation}
where $P_{\mu}$ is $D$-dimensional momentum, 
$I,J=1,...,N$ label supersymmetries and $\alpha,\beta$ 
are spinor indices in $D$ dimensions.  
Here, $({\cal C}\gamma^{\mu})_{\alpha\beta}$ in (\ref{tenssusyalg}) is 
replaced by $({\cal C}\gamma^{\mu}{\cal P}^{\pm})_{\alpha\beta}$ for 
positive or negative chiral Majorana spinors 
\footnote{A Majorana spinor $Q$ is defined as $\bar{Q}=Q^T{\cal C}$, where 
the bar denotes the Dirac conjugate.  
The positive or negative chiral spinor $Q_{\pm}$ is defined as 
$\bar{\gamma}Q_{\pm}=\pm Q_{\pm}$.} 
$Q^I_{\pm}$ (e.g. type-IIB theory), where ${\cal P}_{\pm}$ 
projects on the positive or negative chirality subspace, and also similarly 
for $({\cal C}\gamma^{\mu_1\cdots\mu_{p}})_{\alpha\beta}$.  
Note, $Z^{IJ}_{\mu_1\cdots\mu_{p}}$ commute with $Q^I_{\alpha}$ and 
$P_{\mu}$, but transform as second rank tensors under the Lorentz 
transformation, and therefore are central with respect to supertranslation 
algebra, only. 

The number of central charge degrees of freedom is determined by the 
number of all the possible $(I,J)$ in (\ref{tenssusyalg}) 
\cite{BAR54,BAR200,BAR55,BAR061}.  
In the sum term in (\ref{tenssusyalg}), one has to take into account the 
overcounting due to the Hodge-duality between $p$ and $D-p$ forms 
($Z^{IJ}_{\mu_1\cdots\mu_p}\sim Z^{IJ}_{\mu_1\cdots\mu_{D-p}}$).  
When $p=D-p$, $Z^{IJ}_{\mu_1\cdots\mu_p}$ are self-dual or anti-self dual.  
(For this case, the degrees of freedom are halved.)  
$(I,J)$ on $Z^{IJ}_{\mu_1\cdots\mu_p}$ are defined to have 
the same permutation symmetry as $(\alpha,\beta)$ in  
$\gamma^{\mu_1\cdots\mu_p}_{\alpha\beta}$ so that 
$\gamma^{\mu_1\cdots\mu_p}_{\alpha\beta}Z^{IJ}_{\mu_1\cdots\mu_p}$ 
is symmetric under the simultaneous exchanges of indices in the pairs 
$(I,J)$ and $(\alpha,\beta)$ so that they have the same symmetry property 
(under the exchange of the indices) as the left hand side of 
(\ref{tenssusyalg}).  Namely, only terms associated with  
$\gamma^{\mu}_{\alpha\beta}$ or $\gamma^{\mu_1\cdots\mu_p}_{\alpha\beta}$ 
that are either symmetric or antisymmetric under the exchange of 
$\alpha$ and $\beta$ can be present on the right hand side of 
(\ref{tenssusyalg}).

\subsubsection{Central Charges and $\kappa$-Symmetry}\label{bpssusykp}

The $p$-form central charge $Z^{IJ}_{\mu_1\cdots\mu_{p}}$ 
in (\ref{tenssusyalg}) arises from the surface term of the 
Wess-Zumino (WZ) term in the $p$-brane worldvolume action \cite{DEAgit}.  

Before we discuss this point, we summarize how WZ term emerges in 
the $p$-brane worldvolume action \cite{TOW732,TOW39}.  In the 
Green-Schwarz (GS) formalism \cite{GREs136,GREs243,HENm152,BERst189} 
of the supersymmetric $p$-brane worldvolume action, one achieves 
manifest spacetime supersymmetry by generalizing spacetime with bosonic 
coordinates $X^{\mu}$ ($\mu=0,1,...,D-1$) and global Lorentz symmetry to 
superspace $\Sigma$ with coordinates $Z^M=(X^{\mu},\theta^{\alpha})$ and 
super-Poincar\'e invariance.  
Here, $\alpha$ is a $D$-dimensional spacetime spinor index and the 
spacetime spinor $\theta^{\alpha}$ takes an additional index $I$ 
($I=1,..,N$) for $N$-extended supersymmetry theories, i.e. 
$\theta^{I\,\alpha}$.  
Fields in the GS action are regarded as maps from the worldvolume $W$ 
to $\Sigma$.  The worldvolume $W$ of a $p$-brane has coordinates 
$\xi^i=(\tau,\sigma_1,...,\sigma_p)$ with worldvolume vector index 
$i=0,1,...,p$.  We denote an immersion from $W$ to $\Sigma$ as 
$\phi:W\to\Sigma$.  The pullback $\phi^*$ of a form in $\Sigma$ by 
$\phi$ induces a form in $W$.  To generalize the bosonic $p$-brane 
worldvolume Lagrangian density 
${\cal L}_{bos}=T_p[-{\rm det}\left(\partial_iX^{\mu}(\xi)\partial_j
X^{\nu}(\xi)\eta_{\mu\nu}\right)]^{1/2}$ to be invariant 
under the supersymmetry transformation as well as local reparameterization 
and global Poincar\'e transformations, one introduces a 
supertranslation invariant $D$-vector-valued 1-form 
$\Pi^{\mu}\equiv dX^{\mu}-i\bar{\theta}\gamma^{\mu}d\theta$.  
This corresponds to the spacetime component of the left-invariant 1-form 
$\Pi^A=(\Pi^{\mu},\Pi^{\alpha}=d\theta^{\alpha})$ on $\Sigma$.  
The simplest and straightforward supersymmetric generalization of the bosonic 
worldvolume $\sigma$-model action for a $p$-brane is
\cite{BERst189,ACHetw198,GREs136,GREs243,HENm152,HUGlp180,BERst185,BERstt340,DEAgit}
\begin{equation}
S_1=T_p\int_W d^{p+1}\xi\sqrt{-{\rm det}\left(\Pi^{\mu}_i(\xi)\Pi^{\nu}_j
(\xi)\eta_{\mu\nu}\right)},
\label{gspbract}
\end{equation}
where $T_p$ is the $p$-brane tension and $\Pi^{\mu}_i$ is the 
$\xi^i$-component of the pullback of the 1-form $\Pi^{\mu}$ in $\Sigma$
by $\phi$, i.e. $(\phi^*\,\Pi^A)(\xi)=\Pi^A_i(\xi)d\xi^i$ with 
$\Pi^{\mu}_i=\partial_iX^{\mu}-i\bar{\theta}\gamma^{\mu}\partial_i\theta$ 
and $\Pi^{\alpha}_i=\partial_i\theta^{\alpha}$ ($\partial_i\equiv\partial/
\partial\xi^i$).  (\ref{gspbract}) is manifestly invariant under 
the global super-Poincar\'e and local reparameterization transformations, 
but is not invariant under a local fermionic symmetry, called 
`$\kappa$-symmetry' \cite{GREs136,HUGlp180,CEDvnsw159,BERt173,AGAps080}, 
which is essential for equivalence of the GS and NSR 
formalisms of the worldvolume action.  To make (\ref{gspbract})
invariant under the $\kappa$-symmetry, one introduces an 
additional term $S_{WZ}$, called `Wess-Zumino (WZ) action', into 
(\ref{gspbract}).  
To construct the Wess-Zumino (WZ) action for a $p$-brane, one introduces 
the super-Poincar\'e invariant closed $(p+2)$-form $h_{(p+2)}$ on 
$\Sigma$.  Such closed $(p+2)$-forms exist only for restricted 
values of $D$ and $p$.  The complete listing of the values of $(D,p)$ 
are found in \cite{ACHetw198}.  The maximum values of allowed $D$ and $p$ 
are $D_{max}=11$ and $p_{max}=5$, which can also be determined by the 
worldvolume bose-fermionic degrees of freedom matching condition 
discussed in the next paragraph.  
The super-Poincar\'e invariant closed $(p+2)$-form, in general, has the form 
$h_{(p+2)}=\Pi^{\mu_1}\cdots\Pi^{\mu_{p+2}}d\bar{\theta}\gamma_{\mu_1\cdots
\mu_{p+2}}d\theta$.  Since $h_{(p+2)}$ is closed, one can locally write 
$h_{(p+2)}$ in terms of a $(p+1)$-form $b_{(p+1)}$ (on $\Sigma$) 
as $h_{(p+2)}=db_{(p+1)}$.  Then, a super-Poincar\'e invariant WZ action 
for a $p$-brane is obtained by integrating $b_{(p+1)}$ over $W$ 
\cite{BERst189,ACHetw198,GREs136,GREs243,HENm152,HUGlp180,BERst185,BERstt340,DEAgit}:
\begin{equation}
S_{WZ}=T_p\int d^{p+1}\xi \phi^*\,b_{(p+1)}.
\label{pbrwzterm}
\end{equation}
Note, whereas $S_1$ and $S_{WZ}$ are individually invariant under 
the local reparameterization and global super-Poincar\'e 
transformations, the $\kappa$-symmetry is preserved only in the complete 
action $S_p=S_1+S_{WZ}$.  The $\kappa$-symmetry gauges way the half of the 
degrees of freedom of the spinor $\theta$, thereby only $1/2$ of  
spacetime supersymmetry is linearly realized as worldvolume supersymmetry 
\cite{HUGp278}.  To summarize, the invariance under the $\kappa$-symmetry 
necessitates the introduction of $b_{(p+1)}$ (on $\Sigma$) via the WZ term;   
$b_{(p+1)}$ couples to the worldvolume of the $p$-branes and becomes the 
origin of the central charge term in the supersymmetry algebra.  

We comment on the allowed values of $p$ and the number $N$ of spacetime 
supersymmetry for each $D$.  This is determined \cite{DUFl390} by 
matching the worldvolume bosonic degrees $N_B$ and 
fermionic degrees $N_F$ of freedom.  First, we consider the case where 
the worldvolume theory corresponds to {\it scalar} supermultiplet 
(with components given by scalars and spinors).  By choosing the static 
gauge (defined by $X^{\mu}(\xi)=(X^i(\xi),Y^m(\xi))=(\xi^i,Y^m(\xi))$, with 
$i=0,1,...,p$ and $m=p+1,...,D-1$), one finds that the number of on-shell 
bosonic degrees of freedom is $N_B=D-p-1$.  We denote the number of 
supersymmetries and the number of real components of the 
minimal spinor in $D$-dimensional spacetime [$(p+1)$-dimensional worldvolume] 
as $N$ and $M$ [$n$ and $m$], respectively.  Then, since the 
$\kappa$-symmetry and the on-shell condition each halves the number of 
fermionic degrees of freedom, the number of on-shell fermionic degrees of 
freedom is $N_F={1\over 2}mn={1\over 4}MN$.  The allowed values of $N$ and 
$p$ for each $D$ is determined by the worldvolume supersymmetry condition 
$N_B=N_F$, i.e. $D-p-1={1\over 2}mn={1\over 4}MN$.  
The complete listing of values of $N$ and $p$ are found in  
\cite{DUFl390,DUFkl}.  The maximum number of $D$ in which this condition 
can be satisfied is $D_{max}=11$ ($p=2$) with $M=32$ and $N=1$.  
So, for other cases ($D<11$), $MN\leq 32$.  Similarly, the 
maximum value of $p$ for which this condition can be satisfied is 
$p_{max}=5$.  The ``fundamental'' super $p$-branes \cite{ACHetw198} that 
satisfy this condition are $(D,p)=(11,2),(10,5),(6,3),(4,2)$.  
The 4 sets of $p$-branes obtained from these ``fundamental'' 
super $p$-branes through double-dimensional reduction are named the 
octonionic, quaternionic, complex and real sequences.  
Note, in addition to scalars and spinors, there are also higher spin 
fields on the worldvolume \cite{DUFl390}: vectors or antisymmetric tensors.  
First, we consider {\it vector} supermultiplets.  Since a worldvolume vector 
has $(p-1)$ degrees of freedom, the worldvolume supersymmetry condition 
$N_B=N_F$ becomes $D-2={1\over 2}mn={1\over 4}MN$.  This 
condition introduces additional points in the brane-scan.  Vector 
supermultiplets exist only for $3\leq p\leq 9$ and the bose-fermi 
matching condition can be satisfied in $D=4,6,10$, only.  
Second, we consider {\it tensor} worldvolume supermultiplets.  
In $p+1=6$ worldvolume dimensions, there exists a chiral 
$(n_+,n_-)=(2,0)$ tensor supermultiplet $(B^-_{\mu\nu},\lambda^I,
\phi^{[IJ]})$, $I,J=1,...,4$, with a self-dual 3-form field strength, 
corresponding to the $D=11$ 5-brane.  The decomposition of this 
$(2,0)$ supermultiplet under $(1,0)$ into a tensor multiplet with 1 
scalar and a hypermultiplet with 4 scalars, followed by truncation to 
just the tensor multiplet, leads to worldvolume theory of  
5-brane in $D=7$.

\subsubsection{Central Charges and Topological Charges}

We illustrate how Lorentz tensor type central charges 
(associated with $p$-branes) arise in the supersymmetry algebra 
\cite{DEAgit}.  Since the action $S_p$ has the manifest super-Poincar\'e 
invariance, one can construct supercharges $Q^i_{\alpha}$ from the 
conserved Noether currents $j_{\alpha}$ associated with super-Poincar\'e 
symmetry.  
Whereas (\ref{gspbract}) is invariant under the super-Poincar\'e 
variation, i.e. $\delta_{S}S_1=0$, the integrand of the 
WZ action (\ref{pbrwzterm}) is only quasi-invariant.  Namely, since 
$\delta_Sb_{(p+1)}=d\Delta_{(p)}$ for some $p$-form $\Delta_{(p)}$, the 
integrand of $S_{WZ}$ transforms by total spatial derivative: $T_p\delta_S
(\phi^*\,b_{(p+1)})=d(\phi^*\,\Delta_{(p)})\equiv\partial_i\Delta^i_{(p)} 
d\tau\wedge d\sigma^1\wedge\cdots\wedge d\sigma^p$.  It is $\Delta_{(p)}$ 
that induces `topological charge' which becomes central charge in 
the super-Poincar\'e algebra.  
Generally, when a Lagrangian density $\cal L$ is quasi-invariant under 
some transformation, i.e. $\delta_A{\cal L}=\partial_i\Delta^i_A$, 
the associated Noether current $j^i_A$ contains an ``anomalous'' term 
$\Delta^i_A$: $j^i_A=\delta_AZ^M{{\partial{\cal L}}\over{\partial(\partial_i
Z^M)}}-\Delta^i_A$.  Such an anomalous term modifies the 
algebra of the conserved charges $Q^A=\int d^p\sigma j^{\tau}_A$ 
to include a topological (or central) terms $A_{AB}$.  

For a $p$-brane, the WZ action (\ref{pbrwzterm}) gives rise to the central 
term in the supersymmetry algebra of the form:
\begin{equation}
A_{\alpha\beta}=T_p({\cal C}\gamma^{\mu_1\cdots\mu_p})_{\alpha\beta}
\int d^p\sigma j^{\tau\,\mu_1\cdots\mu_p}_T,
\label{pbrcentsusy}
\end{equation}
where $j^{\tau\,\mu_1\cdots\mu_p}_T$ is the (worldvolume) time-component 
of the topological current density $j^{i\,\mu_1\cdots\mu_p}_T=
\epsilon^{ij_1\cdots j_p}\partial_{j_1}X^{\mu_1}(\xi)\cdots
\partial_{j_p}X^{\mu_p}(\xi)$.   
So, $p$-form central charges in supersymmetry algebra
\footnote{In the supersymmetry algebra (\ref{tenssusyalg}), the $p$-brane 
tension $T_p$ is set equal to 1.}  
(\ref{tenssusyalg}) has the following general form \cite{BAR55} given 
by the surface integral of a $(p+1)$-form local current 
$J^{IJ}_{\mu_0\mu_1\cdots\mu_p}(x)$ over a space-like surface 
embedded in $D$-dimensional spacetime:
\begin{equation}
Z^{IJ}_{\mu_1\cdots\mu_p}=\int d^{D-1}\Sigma^{\mu_0}\,
J^{IJ}_{\mu_0\mu_1\cdots\mu_p}(x).  
\label{pformcent}
\end{equation}
Here, the $(p+1)$-form local current $J^{IJ}_{\mu_0\mu_1\cdots\mu_p}(x)$ 
has contributions from individual $p$-branes with the coordinates 
$X^a_{\mu}(\tau,\sigma_1,...,\sigma_p)$ and charges $z^{IJ}_a$ 
(index $a$ labeling each $p$-brane)
\footnote{So, a $p$-form central charge is related to boundaries of the 
$p$-brane.  For example for a string $(p=1)$, 
$Z_{\mu_1}\sim X_{\mu_1}(0)-X_{\mu_1}(\pi)$.}:
\begin{eqnarray}
J^{IJ}_{\mu_0\mu_1\cdots\mu_p}(x)&=&
\int d\tau d\sigma_1\cdots d\sigma_p\sum_a z^{IJ}_a\delta^{(D)}
\left(x-X^a(\tau,\sigma_1,...,\sigma_p)\right)
\cr
& &\times\partial_{\tau}X^a_{[\mu_0}\cdots\partial_{\sigma_p}
X^a_{\mu_p]}(\tau,\sigma_1,...,\sigma_p).
\label{formcurren}
\end{eqnarray}
This $(p+1)$-form current is coupled to a $(p+1)$-form 
gauge potential $A^{IJ}_{\mu_0\mu_1\cdots\mu_p}(x)$ of the low energy 
effective supergravity in the following way:
\begin{eqnarray}
S&\sim&\int d^Dx\,A^{\mu_0\mu_1\cdots\mu_p}_{IJ}(x)
J^{IJ}_{\mu_0\mu_1\cdots\mu_p}(x)
\cr
&=&\sum_a\int d\tau d\sigma_1\cdots d\sigma_p\,
A^{\mu_0\mu_1\cdots\mu_p}_{IJ}(X^a)\partial_{\tau}X^a_{[\mu_0}
\cdots\partial_{\sigma_p}X^a_{\mu_p]}z^{IJ}_a,
\label{ppotcurcoup}
\end{eqnarray}
where $z^{IJ}_a$ are the charges of $A^{IJ}_{\mu_0\mu_1\cdots\mu_p}(x)$ 
carried by the $a$-th $p$-brane with the coordinates $X^a_{\mu}$.  
The field equation for $A^{IJ}_{\mu_0\mu_1\cdots\mu_p}(x)$ is:
\begin{equation}
\partial^{\lambda}\partial_{[\lambda}A^{IJ}_{\mu_0\mu_1\cdots\mu_p]}(x)=
J^{IJ}_{\mu_0\mu_1\cdots\mu_p}(x).
\label{pfomreleq}
\end{equation}
So, one can think of $Z^{IJ}_{\mu_1\cdots\mu_p}$ as being related 
to charges of $A^{IJ}_{\mu_0\mu_1\cdots\mu_p}(x)$ with the charge 
source given by $p$-branes with their worldvolumes coupled to 
$A^{IJ}_{\mu_0\mu_1\cdots\mu_p}(x)$.  There is a one-to-one correspondence 
between $A^{IJ}_{\mu_0\mu_1\cdots\mu_p}(x)$ in the effective supergravity 
theory and $Z^{IJ}_{\mu_0\mu_1\cdots\mu_p}$ in the superalgebra, i.e. 
there are as many central extensions as form fields in the effective 
supergravity.  

\subsubsection{$S$-Theory}

The maximally extended superalgebra has 32 {\it real} degrees of freedom in 
the set $\{Q^I_{\alpha}\}$ of supercharges, i.e. $N=1$ supersymmetry 
in $D=11$ or $N=8$ supersymmetry in $D=4$.  So, the right 
hand side of (\ref{tenssusyalg}) has at most ${{32\times 33}\over 2}=528$ 
degrees of freedom; the sum of $D$ degrees of freedom of the momentum 
operator $P^{\mu}$ and the degrees of freedom of central charges 
$Z^{IJ}_{\mu_1\cdots\mu_p}$ in (\ref{tenssusyalg}) has to be 528.  
This is the main reason for the necessity of existence of $p$-branes in 
higher dimensions \cite{TOW048}; $N=1$ supersymmetry in $D=11$ without 
central charge has only 11 degrees of freedom on the right hand side of 
(\ref{tenssusyalg}).   

The extended superalgebra (\ref{tenssusyalg}) can be 
derived by compactifying either a type-A superalgebra in $D=2+10$ or 
a type-B superalgebra in $D=3+10$ (the so-called ``$S$-theories'') 
\cite{BAR54,BAR200,BAR55,BAR061}, with the Lorentz symmetries $SO(10,2)$ 
and $SO(9,1)\otimes SO(2,1)$, respectively.  The superalgebra of 
the type-IIA(B) theory is obtained by compactifying the type-A(B) algebra.  
First, the type-A algebra has the form:
\begin{equation}
\{Q_{\alpha},Q_{\beta}\}=\gamma^{M_1M_2}_{\alpha\beta}Z_{M_1M_2}
+\gamma^{M_1\cdots M_6}_{\alpha\beta}Z^+_{M_1\cdots M_6},
\label{asuper}
\end{equation}
where $M_i=0^{\prime},0,1,...,10$ is a $D=12$ vector index with 2 
time-like indices $0^{\prime},0$.   Note, in $D=12$ with 2 time-like 
coordinates, only gamma matrices which are (anti)symmetric are 
$\gamma^{M_1M_2}$ and $\gamma^{M_1\cdots M_6}$, with $\gamma^{M_1\cdots 
M_6}$ being self-dual.  So, one has terms involving 2-form central 
charge $Z_{M_1M_2}$ and self-dual 6-form central charge 
$Z^+_{M_1\cdots M_6}$, without momentum operator $P^M$ term, 
in (\ref{asuper}).  Generally, in the $D<12$ supersymmetry algebra 
compactified from the type-A algebra, the spinor indices $a$ and $\alpha$ 
of 32 spinors $Q^a_{\alpha}$ are regarded as those of $SO(c+1,1)$ and 
$SO(D-1,1)$ Lorentz groups, respectively, where $c$ is  
the number of compactified dimensions from the point of view of $D=10$ 
string theory.  Second, the type-B algebra has the form:
\begin{equation}
\{Q^{\bar{a}}_{\bar{\alpha}},Q^{\bar{b}}_{\bar{\beta}}\}=
\gamma^{\mu}_{\bar{\alpha}\bar{\beta}}
(c\tau_i)^{\bar{a}\bar{b}}P^i_{\mu}+
\gamma^{\mu_1\mu_2\mu_3}_{\bar{\alpha}\bar{\beta}}
c^{\bar{a}\bar{b}}Y_{\mu_1\mu_2\mu_3}+
\gamma^{\mu_1\cdots\mu_5}_{\bar{\alpha}\bar{\beta}}
(c\tau_i)^{\bar{a}\bar{b}}X^i_{\mu_1\cdots\mu_5},
\label{bsuper}
\end{equation}
where the indices are divided into the $D=10$ ones $\bar{\alpha},\bar{\beta}=
1,2,...,16$ and $\mu=0,1,...,9$ with the Lorentz group $SO(1,9)$, and 
the $D=3$ ones $\bar{a},\bar{b}=1,2$ and $i=0^{\prime},1^{\prime},
2^{\prime}$ with the Lorentz group $SO(1,2)$.  (The barred [unbarred] 
indices are spinor [spacetime vector] indices.)  Here, $\gamma^{\mu}$ 
[$\tau_i$] are gamma matrices of the $SO(1,9)$ [$SO(1,2)$] Clifford 
algebra and $c^{\bar{a}\bar{b}}=i\sigma^{\bar{a}\bar{b}}_2=
\epsilon^{\bar{a}\bar{b}}$.  

We discuss the maximal extended superalgebras that 
follow from the type-A algebra.  First, the following  
$N=1$, $D=11$ superalgebra is obtained from (\ref{asuper}) by 
compactifying the $0^{\prime}$-coordinate:
\begin{equation}
\{Q_{\alpha},Q_{\beta}\} = (\Gamma^{\mu}{\cal C})_{\alpha\beta}P_{\mu}
+(\Gamma^{\mu_1\mu_2}{\cal C})_{\alpha\beta}Z_{\mu_1\mu_2}+ 
(\Gamma^{\mu_1\cdots\mu_5}{\cal C})_{\alpha\beta}X_{\mu_1\cdots\mu_5}, 
\label{11dsuper}
\end{equation}
where each term on the right hand side emerges from the terms in 
(\ref{asuper}) as
\begin{eqnarray}
Z_{M_1M_2}&\to&P_{\mu}\oplus Z_{\mu_1\mu_2}\ \ \ \ \ 66=11+55
\cr
Z^+_{M_1\cdots M_6}&\to&X_{\mu_1\cdots\mu_5}\ \ \ \ \ \ \ \ \ \ 462=462.
\label{twelelev}
\end{eqnarray}
The central charges $Z_{\mu_1\mu_2}$ and $Z_{\mu_1\cdots\mu_5}$ are 
associated with the $M\,2$ and $M\,5$ branes, respectively.   
The maximal extended superalgebras in $D<11$ are obtained by 
compactifying the $D=11$ supertranslation algebra (\ref{11dsuper}) on  
tori.   The central charge degrees of freedom in lower dimensions are 
counted by adding the contribution from the internal momentum $P_m$ 
($m=1,...,11-D$) and the number of ways of wrapping $M\,2$ and $M\,5$ 
branes on cycles of $T^{11-D}$ in obtaining various $p$-branes in lower 
dimensions.  Schematically, decompositions of the terms on the 
right hand side of (\ref{11dsuper}) are:
\begin{eqnarray}
P_{\mu}&=&P_{\mu}\oplus P_m, \ \ \ 
Z_{\mu_1\mu_2}=Z_{\mu_1\mu_2}\oplus Z^{m_1}_{\mu_1}\oplus Z^{m_1m_2},
\cr
X_{\mu_1\cdots\mu_5}&=&X_{\mu_1\cdots\mu_5}\oplus X^{m_1}_{\mu_1\cdots\mu_4}
\oplus X^{m_1m_2}_{\mu_1\mu_2\mu_3}\oplus X^{m_1m_2m_3}_{\mu_1\mu_2}
\oplus X^{m_1\cdots m_4}_{\mu_1}\oplus X^{m_1\cdots m_5}.
\label{maxsusydec}
\end{eqnarray}
The $N(N-1)$ Lorentz scalar central charges of the 
(maximal) $N$-extended $D<11$ superalgebras originate 
from the Lorentz scalar type terms (under the 
$SO(D-1,1)$ group) on the right hand sides of (\ref{maxsusydec}), 
i.e. $P_m$, $Z^{mm}$ and $X^{m_1\cdots m_5}$.  In this consideration, 
one has to take into account equivalence under the Hodge-duality in 
$D$ dimensions.  $N(N-1)=56$ Lorentz 
scalar central charges of $N=8$ superalgebra in $D=4$ 
originate from the 7 components of $P_m$, ${{7\times 6}\over{2\times 
1}}=21$ terms in $Z^{m_1m_2}$, ${{7\times 6\times 5\times 6\times 5}\over
{5\times 4\times 3\times 2\times 1}}=21$ terms in $X^{m_1\cdots m_5}$ 
and 7 terms in the Hodge-dual of $X^{m_1}_{\mu_1\cdots\mu_4}$.
The similar argument regarding the supertranslational algebras of 
type-IIB and heterotic theories can be made, and details are 
found, for example, in \cite{BAR55,TOW048}.  

We saw that one has to take into account the Lorentz tensor type 
central charges in higher-dimensions to trace the higher-dimensional 
origin of $N(N-1)$ 0-form central charge degrees of freedom in 
$N$-extended supersymmetry in $D<11$.   
This supports the idea that in order for the conjectured string 
dualities (which mix all the electric and magnetic charges associated 
with $N(N-1)$ 0-form central charges in $D<10$ among themselves) 
to be valid, one has to include not only perturbative string states but 
also the non-perturbative branes within 
the full string spectrum.  In lower dimensions, these central 
charges are carried by 0-branes (black holes).  It is a purpose of 
chapter \ref{n4bh} to construct the most general black hole solutions in 
string theories carrying all of 0-form central charges.  In section 
\ref{bprhghemd}, we identify the (intersecting) $p$-branes in 
higher dimensions which reduce to these black holes after 
dimensional reduction. 

\subsubsection{Central Charges and Moduli Fields}

We comment on relation of central charges to 
$U(1)$ charges and moduli fields \cite{ANDdf015,ANDdf105}. 
Except for special cases of $D=4$, $N=1,2$ and $D=5$, $N=2$, 
scalar kinetic terms in supergravity theories 
are described by $\sigma$-model with target space manifold 
given by coset space $G/H$.  Here, a non-compact continuous 
group $G$ is the duality group that acts linearly on field 
strengths $H^{\Lambda}_{\mu_1\cdots\mu_{p+2}}$ and is on-shell 
and/or off-shell symmetry of the action.  The isotropy subgroup 
$H\subset G$ is decomposed into the automorphism group $H_{Aut}$ of 
the superalgebra and the group $H_{matter}$ related to 
the matter multiplets: $H=H_{Aut}\oplus H_{matter}$.  
(Note, the matter multiplets do not exist for $N>4$ in $D=4,5$ and in 
maximally extended supergravities.)   

The properties of supergravity theories are fixed 
by the coset representatives $L$ of $G/H$.  $L$ is a function of the 
coordinates of $G/H$ (i.e. scalars) and is 
decomposed as:
\begin{equation}
L=(L^{\Lambda}_{\Sigma})=(L^{\Lambda}_{AB},L^{\Lambda}_I),\ \ \ \ \ 
L^{-1}=(L_{AB\,\Lambda},L_{I\,\Lambda}),
\label{cosetdecom}
\end{equation}
where $(A,B)$ and $I$ respectively correspond to 
2-fold tensor representation of $H_{Aut}$ and the fundamental 
representation of $H_{matter}$.  Here, $\Lambda$ runs over the 
dimensions of a representation of $G$.  The $(p+2)$-form strength 
$H^{\Lambda}$ kinetic terms are given in terms of the 
following kinetic matrix determined by $L$:
\begin{equation}
{\cal N}_{\Lambda\Sigma}=L_{AB\,\Lambda}L^{AB}_{\Sigma}-
L_{I\,\Lambda}L^I_{\Sigma}.
\label{cskinmetp}
\end{equation}
So, the ``physical'' field strengths of $(p+1)$-form potentials 
in supergravity theories are ``dressed'' with scalars through  
the coset representative and the $(p+2)$-form field strengths 
appear in the supersymmetry transformation laws dressed with the scalars.  
The central charges of extended superalgebra, which is encoded 
in the supergravity transformations rules, are expressed 
in terms of electric $Q_{\Lambda}\equiv\int_{S^{D-p-2}}
{\cal G}_{\Lambda}$ and magnetic $P^{\Lambda}\equiv\int_{S^{p+2}}
{\cal H}^{\Lambda}$ charges of $(p+2)$-form field strengths 
${\cal H}^{\Lambda}=d{\cal A}^{\Lambda}$ (${\cal G}_{\Lambda}=
{{\partial{\cal L}}\over{\partial{\cal H}^{\Lambda}}}$)  
and the asymptotic values of scalars in the form of the coset 
representative, manifesting the geometric structure of moduli space.   

These central charges satisfy the differential equations that follow 
from the Maurer-Cartan equations satisfied by the coset representative.  
One of the consequences of these differential equations is that the 
vanishing of a subset of central charges (resulting from the requirement 
of supersymmetry preserving bosonic background) forces the covariant 
derivative of some other central charges to vanish, i.e. 
``principle of minimal central charge'' \cite{FERk136,FERk54}.   
Furthermore, from defining relations of the kinetic matrix of 
$(p+2)$-form field strengths and the symmetry properties of the 
symplectic section under the group $G$, one obtains the sum rules 
satisfied by the central and matter charges.  

For other cases, in which the scalar manifold cannot be expressed as 
coset space, one can apply the similar analysis as above by using 
techniques of special geometry \cite{STR133,CERdf339,FRE45}.  
For this case, the roles of the coset representative and the Maurer-Cartan 
equations are played respectively by the symplectic sections and the 
Picard-Fuchs equations \cite{CERdfll8}.  

\subsubsection{BPS Supermultiplets}

We discuss massive representations of extended superalgebras 
with non-zero central charges \cite{OLIw78,OSB83,HAAls88,FERsz,STR2}, 
i.e. the BPS states.   
It is convenient to go to the rest frame of states defined 
by $P_{\mu}=(M,0,...,0)$, where $M$ is the rest mass of the state.  
The little group, defined as a set of transformations that leave this 
$P_{\mu}$ invariant, consists of $SO(D-1)$, the automorphism group 
and the supertranslations.  Since central charges $Z^{IJ}$ transform 
as a second rank tensor under the automorphism group, only the subset 
of automorphism group that leaves $Z^{IJ}$ invariant should 
be included in the little group.  
Central charges inactivate some of supercharges, reducing the 
size of supermultiplets.  In the following, we illustrate  
properties of the BPS states for the $D=4$ case with an arbitrary number 
$N$ of supersymmetries.

In the Majorana representation, the central charges $U^{IJ}$ and $V^{IJ}$ 
($I,J=1,...,N$) appear in the $N$-extended superalgebra in $D=4$ in the 
form \cite{FERsz}:
\begin{equation}
\{Q^I_{\alpha},Q^J_{\beta}\}=(\gamma^{\mu}{\cal C})_{\alpha\beta}
P_{\mu}\delta^{IJ}+{\cal C}_{\alpha\beta}U^{IJ} + (\gamma_{5}
{\cal C})_{\alpha\beta}V^{IJ}, 
\label{extsusy}
\end{equation}
where $\alpha,\beta=1,...,4$ are indices of Majorana spinors, 
${\cal C}$ is the charge conjugation matrix ${\cal C}=-{\cal C}^T$, 
and the supercharges $Q^I_{\alpha}$ are in the Majorana representation.  
In the Majorana representation, $U^{IJ}$ and $V^{IJ}$
are hermitian operators and antisymmetric.  

Alternatively, one can express the superalgebra (\ref{extsusy}) 
in the Weyl basis.  In this basis, a 4-component spinor $Q_{\alpha}$ 
($\alpha=1,...,4$) is decomposed into left- and right-handed 2-component 
Weyl spinors: 
\begin{equation}
Q^I_{\alpha}=(Q^I_L)_{\alpha}, \ \ \ \ \ 
(Q^I_R)^{\dot{\alpha}}=\epsilon^{\dot{\alpha}\dot{\beta}}
Q^{*\,I}_{\dot{\beta}}=(i\sigma_2Q^{*\,I}_L)^{\dot{\alpha}}, 
\label{weylspin}
\end{equation}
where $\alpha,\beta=1,2$ and $\dot{\alpha},\dot{\beta}=1,2$ are Weyl spinor 
indices, and $\epsilon^{\alpha\beta}=\epsilon^{\dot{\alpha}\dot{\beta}}=
(i\sigma_2)_{\alpha\beta}=-\epsilon_{\alpha\beta}=-\epsilon_{\dot{\alpha}
\dot{\beta}}$ is the 2-dimensional Levi-Civita symbol.  
Namely, the lower and upper components of a 4-component 
spinor $Q_{\alpha}$ ($\alpha=1,...,4$) are $(Q_L)_{\alpha}$ and 
$\epsilon^{\dot{\alpha}\dot{\beta}}Q^{*}_{L\,\dot{\beta}}$, respectively.  
In this 2-component Weyl basis representation, the anticommutations  
(\ref{extsusy}) become
\begin{eqnarray}
\{Q^I_{\alpha},Q^{*\,J}_{\dot{\beta}}\} &=& 
(\sigma_{\mu})_{\alpha\dot{\beta}}P^{\mu}\delta^{IJ}, 
\cr
\{Q^I_{\alpha},Q^J_{\beta}\} &=& \epsilon_{\alpha\beta}Z^{IJ}, \ \ \ \ \ 
\{Q^{*\,I}_{\dot{\alpha}},Q^{*\,J}_{\dot{\beta}}\} = 
\epsilon_{\dot{\alpha}\dot{\beta}}Z^{IJ}, 
\label{exsuper}
\end{eqnarray}
where $Z^{IJ}\equiv -V^{IJ}+iU^{IJ}$.  

The central charge matrix $Z^{IJ}$ can be brought to the 
block diagonal form by applying the $U(N)$ automorphism group:
\begin{equation}
Z^{IJ} = {\rm diag}(z_1 \epsilon^{ij},z_2 \epsilon^{ij},
\cdots , z_{[{N \over 2}]} \epsilon^{ij})=i\sigma_2\otimes
\hat{Z}_{[{N\over 2}]}, 
\label{block}
\end{equation}
where $z_m$ ($m=1,...,[{N \over 2}]$) are eigenvalues of $Z^{IJ}$ and 
$\hat{Z}_{[{N\over 2}]}\equiv {\rm diag}(z_1,...,z_{[{N\over 2}]})$.  
There are extra 0 entries in the $N$-$th$ row and column in $Z^{IJ}$ 
for an odd $N$.  Further redefining the supercharges and making use of 
the reality condition satisfied by the supercharges, one can simplify  
supersymmetry algebra:
\begin{eqnarray}
\{S^m_{\alpha\,(a)},S^{*\,n}_{\beta\,(b)}\}&=&\delta_{\alpha\beta}\delta^{mn}
\delta_{ab}\,(M-(-)^b z_n), 
\cr
\{S^m_{\alpha\,(a)},S^n_{\beta\,(b)}\}&=&
\{S^{*\,m}_{\alpha\,(a)},S^{*\,n}_{\beta\,(b)}\} =0, 
\label{bogsusy}
\end{eqnarray}
where $\alpha,\beta=1,2$, $a,b=1,2$, and $m,n=1,...,N/2$. 
For odd $N$, there are extra anticommutation relations 
associated with the extra 0 entries in $Z^{IJ}$.  

Since the left-hand-sides of (\ref{bogsusy}) are positive semidefinite 
operators, the rest mass $M$ of the particles in the supermultiplet is 
always greater than or equal to all the eigenvalue of the 
central charge matrix and therefore:
\begin{equation}
M \geq {\rm max}\{|z_m|\}.
\label{bogbound}
\end{equation}
The state that saturates the bound (\ref{bogbound}) is called the BPS state.
The supermultiplet that do not saturate (\ref{bogbound})
is called the long multiplet and is the same as that of extended 
superalgebra without central charges.  The BPS supermultiplet is
called short multiplet, since there are fewer supercharges (or
raising operators) available for building up supermultiplet 
(since the supercharges that annihilate the supersymmetric 
vacuum get projected onto zero norm states).  
The type of supermultiplet that BPS states belong to 
depends on the number of distinct eigenvalues of the central charge 
matrix $Z^{IJ}$.  

In the following, we give examples on all the possible 
BPS multiplets of $N=4$ supersymmetry algebra.
$N=4$ superalgebra has $[{N\over 2}]=2$ eigenvalues $z_1$ and $z_2$. 
There are two types of BPS supermultiplets in the $N=4$ 
superalgebra depending on whether $z_{1,2}$ are the same or different. 

When $z_1 =z_2$, two raising operators $S^1_{\alpha\,(2)}$ and 
$S^2_{\alpha\,(2)}$ in (\ref{bogsusy}) are projected onto the zero-norm 
states; $S^m_{\alpha\,(2)}$ ($m=1,2$) annihilate the 
supersymmetric vacuum state and become zero.  So, $1 \over 2$ 
of supersymmetry is preserved and the remaining raising operators 
$S^1_{\alpha\,(1)}$ and $S^2_{\alpha\,(1)}$ act on the lowest helicity 
state to generate the highest spin 1 multiplet.  

When $z_{1}\neq z_{2}$, say, $z_1>z_2$, the raising operator 
$S^1_{\alpha\,(2)}$ is projected onto the zero-norm states.  Hence, 
$1\over 4$ of supersymmetry is preserved and the remaining 
raising operators  $S^1_{\alpha\,(1)}$, $S^2_{\alpha\,(1)}$ and 
$S^2_{\alpha\,(1)}$ act on the lowest helicity state to generate the 
highest spin $3/2$ multiplet.

\subsection{Positive Energy Theorem and Nester's Formalism}\label{bpsnest}

We discuss the positive mass theorem 
\cite{CHOm,SHOy42,SHOy79,SHOy81,SHOy48,WIT80,NES,GIBhhp,HUL90}
of general relativity. 
The positive mass theorem says that the total energy, i.e. 
the rest mass plus potential energy plus kinetic energy, of the 
gravitating system is always positive 
with a unique zero-energy configuration with appropriate boundary 
conditions at infinity, provided that the matter stress-energy tensor 
$T_{\mu\nu}$ satisfies the dominant energy condition
\begin{equation}
T_{\mu\nu}U^{\mu}V^{\nu}\geq 0,
\label{dominant}
\end{equation}
for any non-space-like vectors $U^{\mu}$ and $V^{\mu}$.  
Here, the unique zero-energy configuration is the Minkowski 
space or anti-de Sitter space, which gravitating configurations approach 
asymptotically at infinity.   The reason why one cannot make 
gravitational energy arbitrarily negative by shrinking the size of the 
gravitating object (as in the case of Newtonian gravity) is that 
when a system collapses beyond certain size, it forms an event horizon, 
which hides the inside region that has singularities and negative 
energy: the system appears to have positive energy to an outside observer.  

In general relativity, there is no intrinsic definition o local energy 
density due to the equivalence principle.   So, one has to define 
energy of a gravitating system as a global quantity defined 
in background (or asymptotic) region \cite{ARNdm122,ABBd}.  
In general, conserved charges of general relativity are associated 
with generators of symmetries of the asymptotic region.  Namely, the 
conserved charge is defined as a surface integral (with the integration 
surface located at infinity) of the time component of the conserved current 
of the symmetry in the asymptotic region.  
Here, the integration surface is taken to be space-like in this section 
and thereby gravitational energy is of the ADM type.  

For a set $\{k^A_{\mu}\}$ of vectors that approach the Killing 
vectors of the asymptotic background, one can define the conserved 
charge
\footnote{At infinity, the current $J^A_{\mu}=\bar{e}\Theta_{\mu\nu}
k^{\nu\,A}$ is conserved, i.e. $\partial^{\mu}J^A_{\mu}
=0$.  Here, $\bar{e}$ is the determinant of the background metric 
and $k^{\nu\,A}$ is now Killing vectors of the background spacetime.} 
$K^A$ through
\footnote{Here, the integration measures for the surface and volume 
integrals are respectively defined as $d\Sigma_{\mu\nu}={1\over 2}
\varepsilon_{\mu\nu\rho\sigma}dx^{\rho}\wedge dx^{\sigma}$ and 
$d\Sigma_{\mu}={1\over 6}\varepsilon_{\mu\nu\rho\sigma}dx^{\nu}\wedge
dx^{\rho}\wedge dx^{\sigma}$.}:
\begin{equation}
K^A=\int_{\Sigma}d^3xe\Theta^{\mu 0}k^A_{\mu}=
\int_{\Sigma}\Theta^{\mu\nu}k^A_{\mu}d\Sigma_{\nu}
={1\over 4}\int_{S=\partial\Sigma}\delta^{\sigma\tau\lambda}_{\mu\nu\rho}
\Gamma^{\nu a}_{\ \ b}k^{A\,\rho}e^{\mu}_ae^b_{\lambda}
d\Sigma_{\sigma\tau},
\label{conscharg}
\end{equation}
where $\Theta_{\mu\nu}$ is the total stress-energy tensor \cite{LANl}
including the pure gravity contributions and $\Gamma^a_{\ b}=
\Gamma^a_{\mu b}dx^{\mu}$ is the connection 1-form for the metric 
$g_{\mu\nu}$.  For a time-like Killing vector $k^A_{\mu}$, $K^A$ is the 
ADM energy relative to the zero-energy background state.  For a generic 
asymptotically flat spacetime, the set of conserved charges $K^A$ 
consists of the ADM 4-momentum $P^{\mu}$ and the angular-momentum tensor 
$J^{\mu\nu}$, which satisfy the Poincar\'e algebra.  
(The ADM mass $M$ is the norm of the ADM 4-momentum: 
$M=\sqrt{-P^{\mu}P_{\mu}}$.)  
For a supersymmetric bosonic background, one can define the (conserved) 
supercharge \cite{DESks16,TEI69} $Q^m$ for the conserved current 
$J^m_{\mu}=\alpha^mR^{\mu}$ ($\partial_{\mu}(e\alpha^mR^{\mu})=0$), 
which is defined in terms of a Killing spinor $\alpha^m$ and a gravitino 
$\psi_{\mu}$ satisfying the Rarita-Schwinger equation:
\begin{eqnarray}
Q^m&=&\int_{\Sigma}d^3xe\alpha^mR^0=\int_{\Sigma}\alpha^m
\varepsilon^{\mu\nu\rho\sigma}\gamma_{\nu}\nabla_{\rho}\psi_{\sigma}
d\Sigma_{\mu}
\cr
&=&{1\over 2}\int_{S=\partial\Sigma}\alpha^m
\varepsilon^{\mu\nu\rho\sigma}\gamma_5\gamma_{\nu}\psi_{\sigma}
d\Sigma_{\mu\rho}
\label{superchks}
\end{eqnarray}
where $R^{\mu}=\varepsilon^{\mu\nu\rho\sigma}\gamma^5\gamma_{\nu}
\nabla_{\rho}\psi_{\sigma}$ and $\sigma^{\mu\nu}\equiv{1\over 4}
[\gamma^{\mu},\gamma^{\nu}]$, etc..
The generators $\tilde{K}_A$ and $\tilde{Q}_m$ of the asymptotic spacetime 
symmetries and the supersymmetry transformation satisfy the following 
algebras:
\begin{equation}
[\tilde{K}_A,\tilde{K}_B]=C^C_{AB}\tilde{K}_C,\ \ \ \ \ 
\{\tilde{Q}_m,\tilde{Q}_n\}=f^M_{mn}\tilde{K}_M,
\label{asygeneralg}
\end{equation}
where $C^C_{AB}$ are the same structure constants that appear in the 
commutator of the Killing vectors $k_A$ and $f^A_{mn}$ are the 
same constants
\footnote{For flat spacetime, one can choose the basis of $\alpha^m$ 
so that $f^A_{mn}=\gamma^A_{mn}$.} 
in the following relation between the Killing vectors 
$k^{\mu}_A$ and the Killing spinors $\alpha_m$:
\begin{equation}
\bar{\alpha}_m\gamma^{\mu}\alpha_n=f^A_{mn}k^{\mu}_A.
\label{kilspvecrel}
\end{equation}
So, these conserved charges satisfy the supersymmetry algebras 
\begin{equation}
[P^{\mu},Q^m]=0,\ \ \ \ 
[J^{\mu\nu},Q^m]=\hbar\sigma^{\mu\nu\,m}_{\ \ \,n}Q^n,\ \ \ \ 
\{Q^m,Q^n\}=\hbar\gamma^{mn}_{\mu}P^{\mu}.
\label{asymalg}
\end{equation}

The third equation in (\ref{asymalg}) leads to the simple proof 
\cite{DESt39} of the positivity of energy in quantum supergravity.  
Since the left hand side of this equation is a positive 
semidefinite operator, one has 
\begin{equation}
\hbar^{-1}\langle s|\{Q^m,Q^{n\dagger}\}|s\rangle =
(\gamma^{\mu}\gamma^0)^{mn}\langle s|P_{\mu}|s\rangle \geq 0, 
\label{simppmth}
\end{equation}
where $|s\rangle$ is a physical state vector.  
For this inequality to be satisfied, the eigenvalues 
$P^0\pm |\vec{P}|$ of $\gamma^{\mu}\gamma^0P_{\mu}$ have to be 
non-negative, leading to proof of the positivity of 
gravitational energy.  

Rigorous proof of the positive energy theorem based on original 
Witten's proof \cite{WIT80} involves the following antisymmetric 
tensor (the Nester's 2-form \cite{NES}):
\begin{equation}
E^{\mu\nu}=\varepsilon^{\mu\nu\rho\sigma}(\bar{\varepsilon}\gamma^5
\gamma_{\rho}\nabla_{\sigma}\varepsilon-\nabla_{\sigma}\bar{\varepsilon}
\gamma^5\gamma_{\rho}\varepsilon),
\label{nesterwit}
\end{equation}
where a Dirac spinor $\varepsilon$ is assumed to approach a constant spinor
\footnote{This is a necessary condition \cite{CHOc,PARt84,REU23} 
for a spinor $\varepsilon$ to satisfy the Witten's condition.}  
$\varepsilon_0$ asymptotically ($\varepsilon\to\varepsilon_0+
{\cal O}(r^{-1})$) and $\nabla_{\mu}$ is the gravitational 
covariant derivative on a spinor.  The ADM 4-momentum $P_{\mu}$ is 
related to the surface integral of $E^{\mu\nu}$ over the 
surface $S$ at the space-like infinity in the following way:
\begin{equation}
P_{\mu}\bar{\varepsilon}_0\gamma^{\mu}\varepsilon_0=
{1\over 2}\int_SE^{\mu\nu}d\Sigma_{\mu\nu}.
\label{fmomnestint}
\end{equation}

Proof of the positive energy theorem involves the surface and the volume 
integrals of the Nester's 2-form, which are related by the Gauss 
divergence theorem: 
$\oint_{S=\partial\Sigma}\,\,{1\over 2}E^{\mu\nu}\,{\rm d}S_{\mu\nu}
=\int_{\Sigma}\,\,\nabla_{\mu}E^{\mu\nu}\,{\rm d}\Sigma_{\nu}$.  
This leads to 
\begin{equation}
P_{\mu}\bar{\varepsilon}_0\gamma^{\mu}\varepsilon_0=
\int_{\Sigma}G_{\mu\nu}\bar{\varepsilon}\gamma^{\mu}\varepsilon 
d\Sigma^{\nu} +\int_{\Sigma}\nabla_{\mu}\bar{\varepsilon}
(\gamma^{\nu}\sigma^{\mu\rho}+\sigma^{\mu\rho}\gamma^{\nu})
\nabla_{\rho}\varepsilon d\Sigma_{\nu},
\label{surfvolnest}
\end{equation}
where $G_{\mu\nu}$ is the Einstein tensor of the metric $g_{\mu\nu}$.  
From the Einstein's equations $G_{\mu\nu}=T_{\mu\nu}$, one sees that 
the first term on the right hand side of (\ref{surfvolnest}) is positive 
if $T_{\mu\nu}$ satisfies the dominant energy condition (\ref{dominant}).  
In the coordinate system in which the $x^0$-direction is normal to 
$\Sigma$, the integrand of the second term on the right hand side of 
(\ref{surfvolnest}) simplifies to
\begin{equation}
2\nabla_i\bar{\varepsilon}\gamma^0\sigma^{ij}\nabla_j\varepsilon=
2|\nabla_j\varepsilon|^2-2\left|\sum^3_{i=1}\gamma^i\nabla_i\varepsilon
\right|^2.
\label{relforwitcon}
\end{equation}
So, if the spinor $\varepsilon$ satisfies the ``Witten condition'' 
\cite{WIT80}: 
\begin{equation}
\sum^3_{i=1}\,\gamma^i\nabla_i\varepsilon=0,
\label{wittcon}
\end{equation}
the second term on the right hand side of (\ref{surfvolnest}) is also 
positive.  Thus, the left hand side of (\ref{surfvolnest}) is always 
non-negative for any $\varepsilon_0$, leading to proof that 
energy $P^0$ of a gravitating system is non-negative.  

The above proof of the positive energy theorem based on the Nester's 
formalism can be readily generalized to gravitating configurations 
in extended supergravities.  In this case, the gravitational 
covariant derivative $\nabla_{\mu}$ on spinors in the 
Nester's 2-form (\ref{nesterwit}) is replaced by the super-covariant 
derivative $\hat{\nabla}_{\mu}$ on spinors.  
So, the Nester's 2-form generalizes to:
\begin{equation}
\hat{E}^{\mu\nu}\equiv 2(\overline{\hat{\nabla}_{\rho}\epsilon}
\Gamma^{\mu\nu\rho}
\epsilon -\bar{\epsilon}\Gamma^{\mu\nu\rho}\hat{\nabla}_{\rho}\epsilon)
=E^{\mu\nu}+H^{\mu\nu}, 
\label{nestextsusy}
\end{equation}
where $E^{\mu\nu}$ is the original Nester's 2-form (\ref{nesterwit}) 
and $H^{\mu\nu}$ denotes the remaining terms, 
which are usually expressed in terms of gauge fields of extended 
supergravities.  Here, the supercovariant derivative on a supersymmetry 
parameter $\varepsilon$ is given by the gravitino supersymmetry 
transformation in bosonic background, i.e. $\delta\psi_{\mu}=\hat{\nabla}_{\mu}
\varepsilon$.  The lower bound for mass is given in terms of central 
charges (coming from the $H^{\mu\nu}$ term in (\ref{nestextsusy})) of the 
extended supergravity and the bound is saturated when the gravitating 
configuration is a bosonic configuration which preserves some of 
supersymmetry.  

In the following, we discuss proof of the positive mass theorem in 
pure $N=2$, $D=4$ supergravity as an example \cite{GIBh82}.   
The $N=2$, $D=4$ supergravity is first obtained in \cite{FERv37} 
by coupling the (2,3/2) gauge action to the (3/2,1) matter multiplet 
by means of the Noether procedure.  The theory unifies electromagnetism 
(spin 1 field) with gravity (spin 2 field).  The theory has 
a manifest invariance under the $O(2)$ symmetry, which rotates 
2 gravitino into each other.  In the bosonic background, the supergravity 
transformation of the gravitino, which define the supercovariant 
derivative $\hat{\nabla}_{\mu}$, is 
\begin{equation}
\delta\psi_{\mu} ={1\over\sqrt{2\pi G_N}}\hat{\nabla}_{\mu}\epsilon
={1\over\sqrt{2\pi G_N}}[\nabla_{\mu}-{\sqrt{G_N}\over 4}F_{\nu\rho}
\gamma^{\nu}\gamma^{\rho}\gamma_{\mu}]\epsilon.
\label{n2susy}
\end{equation}

Substituting the supercovariant derivative $\hat{\nabla}_{\mu}\varepsilon$ 
from (\ref{n2susy}) into the Nester's two-form (\ref{nestextsusy}), 
one has the following expression for $H^{\mu\nu}$:
\begin{equation}
H^{\mu\nu}\equiv 4\bar{\epsilon}(F^{\mu\nu}+
i\gamma_5\,\star F^{\mu\nu})\epsilon.
\label{hn2susy}
\end{equation} 
The integrand of the volume integral takes the form:
\begin{equation}
\nabla_{\nu}\hat{E}^{\mu\nu}=16\pi G_{N}\,T^{{\rm matter}\,\,
\mu}_{\ \ \ \ \ \ \ \ \rho}U^{\rho}
+16\pi \sqrt{G_N}\bar{\epsilon}(J^{\mu}+i\gamma_5
\tilde{J}^{\mu})\epsilon +4\overline{\hat{\nabla}_{\nu}\epsilon}
\Gamma^{\mu\nu\rho}\hat{\nabla}_{\rho}\epsilon, 
\label{volnes}
\end{equation}
where $U^{\mu}\equiv\bar{\epsilon}\gamma^{\mu}\epsilon$ is a 
non-spacelike 4-vector, provided $\epsilon$ is the Killing spinor, and 
the electric and magnetic 4-vector currents $J^{\mu}$ and 
$\tilde{J}^{\mu}$ are defined through the Maxwell's equations and 
the Bianchi identities as $\nabla_{\nu}F^{\mu\nu}=4\pi J^{\mu}$ and
$\nabla_{\nu}\star F^{\mu\nu}=4\pi\tilde{J}^{\mu}$. 

If we assume that the Killing spinor $\epsilon$ approaches 
a constant value $\epsilon_0$ as $r\to\infty$, then the surface and 
the volume integrals are
\begin{eqnarray}
\bar{\epsilon}_0[-P_{\mu}\gamma^{\mu}+\sqrt{G_N}(Q+i\gamma_5 P)]
\epsilon_0
&=&\int_{\Sigma}[T^{{\rm matter}\,\,\mu}_{\ \ \ \ \ \ \ \ \nu}U^{\nu}+ 
{1\over {G_N}}\bar{\epsilon}(J^{\mu}+i\gamma_5\tilde{J}^{\mu})\epsilon]
{\rm d}\Sigma_{\mu}
\cr 
& &+{1\over{4\pi G_N}}\int_{\Sigma}\,
\overline{\hat{\nabla}_{\nu}\epsilon}\Gamma^{\mu\nu\rho}
\hat{\nabla}_{\rho}\epsilon\,{\rm d}\Sigma_{\mu},
\label{n2surfvol}
\end{eqnarray}
where $Q$ and $P$ are the electric and magnetic charges of the gauge 
field $A_{\mu}$.  

The first term on the right hand side of (\ref{n2surfvol}) is always 
non-negative, if $T^{\rm matter}_{\mu\nu}$ satisfies the generalized 
dominant energy condition:  
\begin{equation}
T^{\rm matter}_{\mu\nu}U^{\mu}V^{\nu} \geq 
G^{-1/2}_N[(J_{\mu}V^{\mu})^2 +(\tilde{J}_{\mu}V^{\mu})^2], 
\label{domin}
\end{equation}
for any pairs of non-spacelike four-vectors $U^{\mu}$ and $V^{\mu}$.  
(This says that the local energy density of the matter exceeds the local 
charge density for any observer.)  
The second term on the right hand side of (\ref{n2surfvol}) is non-negative 
provided the following generalized Witten's condition is satisfied:
\begin{equation}
\gamma^a\hat{\nabla}_a\epsilon =0,
\label{genwit}
\end{equation}
where $\hat{\nabla}_a$ is the Israel and Nester's projection of 
$\hat{\nabla}_{\mu}$ on the surface $S=\partial\Sigma$. 
The second term vanishes iff $\hat{\nabla}_a\epsilon =0$ on $S$.  
So, the left hand side of (\ref{n2surfvol}) is always positive 
semidefinite for $\epsilon_0$, provided the generalized Witten's 
condition (\ref{genwit}) and dominant energy condition (\ref{domin}) are 
satisfied.  This is possible iff all the eigenvalues of the 
Hermitian matrix sandwiched between $\epsilon_0$ are non-negative. 
So, we have the following Bogomol'nyi bound for the ADM mass:
\begin{equation}
M\geq G^{-1/2}_N\sqrt{Q^2+P^2}. 
\label{bogon2}
\end{equation}

\section{Duality Symmetries}\label{dual}

Past few years have been an active period for studying string  
dualities \cite{WIT443,WIT121,VAF201,SCH201,POL68,DUF11}.   
Five different string theories  ($E_8\times E_8$ and $SO(32)$ 
heterotic strings, type-IIA and type-IIB strings, and type-I string 
with $SO(32)$ symmetry), which were previously regarded as 
independent perturbative theories, are now understood as 
being related via web of dualities.   

String Dualities are classified into $T$-duality, $S$-duality 
and $U$-duality.   $T$-duality (or target space duality) \cite{GIVpr244} 
is a perturbative duality (i.e. duality that relates theories with the 
same string coupling) that transforms the theory with large [small] 
target space volume to one with small [large] target space volume 
\cite{KIKy149,SAKs75,DINhs322} or connects different points in 
(target-space) moduli space.   Under $T$-duality, type-IIA and type-IIB 
strings \cite{DINhs322,DAIlp}, and $E_8\times E_8$ and $SO(32)$ heterotic 
strings \cite{NAR169,NARsw279,GIN35} are interchanged.  
Another examples of $T$-duality are the $O(10-D,26-D,{\bf Z})$ symmetry 
\cite{GIVrv322} of heterotic string on $T^{10-D}$ and the 
$O(10-D,10-D,{\bf Z})$ symmetry \cite{GIVrv322,SHAw320,GIVmr220,GIVmr238} 
of type-II string on $T^{10-D}$.  
$S$-duality (or strong-weak coupling duality) is a non-perturbative 
duality that transforms string coupling to its inverse (while moduli fields 
remain fixed) and interchanges perturbative string states and 
non-perturbative branes.  Duality that relates Type-I string and $SO(32)$ 
heterotic string \cite{WIT443,POLw460} is an example of $S$-duality.
Another examples are ($i$) the $SL(2,{\bf Z})$ symmetry of type-IIB 
string \cite{SCH226,HOWw238,SCH360,SCH49,SCH367,BERbo53}; 
($ii$) the $D=6$ string-string duality between the heterotic string on 
$T^4$ [on $K3$] and the type-II string on $K3$ [on a Calabi-Yau-threefold] 
\cite{DUFk411,WIT443,VAFw447,SEN450,DUFlm,DUF442,DUFm,FERhsv361,KACv450,BILcdffrsv13,KLElm357,BEHbj467}; 
($iii$) the $SL(2,{\bf Z})$ symmetry of $N=4$ heterotic string in $D=4$ 
\cite{FONil249,SCH125,SCHs312,SCHs411,SEN93,SEN303,SEN038,SEN057,SENint,HARms52}.  
$U$-duality \cite{HULt438,BAR52,BARy53,BAR49}, which is closely related 
to the $D=11$ theory ($M$-theory), is regarded as a consequence of the 
$SL(2,{\bf Z})$ $S$-duality of type-IIB string and $T$-dualities of type-II 
strings on a torus.  
Thus, $U$-duality is a non-perturbative duality of type-II strings 
which necessarily interchanges NS-NS charged state and R-R charged state.  

String dualities require existence of non-perturbative 
states within string spectrum, as well as the  
well-understood perturbative states.  
These non-perturbative states include smooth solitons and new types of 
topological objects called $D$-branes \cite{POL75}.  
Such non-perturbative states are extended objects, which in a low energy 
limit correspond to $p$-branes of the effective field theories.  
So, string theories, which are previously known as theories of 
(perturbative) strings, are no longer theories of strings only, but contain 
objects of higher/lower spatial extends.  These perturbative and 
non-perturbative states are interrelated via string dualities.

One of important discoveries of string dualities is the conjecture that 
there exists more fundamental theory in higher dimensions ($D>10$),  
which reduces to all of 5 perturbative string theories 
in different limits in moduli space when the theory is compactified to 
lower dimensions ($D\leq 10$).   Such fundamental theories include 
$M$-theory \cite{WIT443,SCH367,DUF11,SCH201} in $D=11$, $F$-theory 
in $D=12$ \cite{VAF469}, and $S$-theories in $D=12,13$ 
\cite{BAR54,BAR200,BAR55,BAR061}.  

$M$-theory is defined as an unknown theory in $D=11$  (with 1 time-like 
coordinate) whose low energy effective theory is the $D=11$  
supergravity \cite{CREj78} and which becomes type-IIA theory when 
the extra 1 spatial coordinate is compactified on $S^1$ of radius 
$R$.  Since the radius $R$ of $S^1$ is related to the string coupling 
$\lambda$ of type-IIA theory as $R^3=\lambda^2$, the strong coupling limit 
($\lambda\gg 1$) \cite{WIT443} of type-IIA theory is $M$-theory, 
i.e. the decompactification limit ($R\to\infty$) of $M$-theory on $S^1$.  
Furthermore, the evidence was given in \cite{HORw475} for the conjecture 
that $M$-theory compactified on $S^1/{\bf Z}_2$ is $E_8\times E_8$ 
heterotic string.  We mentioned in the previous paragraph that 
type-IIA and type-IIB strings, and $E_8\times E_8$ and $SO(32)$ 
heterotic strings are related via $T$-duality, and $SO(32)$ heterotic 
string and type-I string are related via $S$-duality.
Thus, all of the 5 different perturbative string theories 
are obtained from $M$-theory by compactifying on $S^1$ or 
$S^1/{\bf Z}_2$, and applying dualities.  

$F$-theory is a conjectured theory in $D=12$ (with 2 time-like coordinates) 
which is proposed in an attempt to find geometric interpretation 
of the $SL(2,{\bf Z})$ $S$-duality of type-IIB theory.  
Namely, the complex scalar formed by the dilaton and the R-R 0-form  
transforms linear-fractionally under the 
$SL(2,{\bf Z})$ transformation, just like the transformation of modulus of 
$T^2$ under the $T$-duality of a string theory compactified 
on $T^2$.  $F$-theory is, therefore, roughly defined as a $D=12$ theory  
which reduces to type-IIB theory upon compactification on 
$T^2$, with the modulus of $T^2$ given by the complex scalar.  
Note, since type-IIB theory on $S^1$ is equivalent to $M$-theory on 
$T^2$, $F$-theory on $T^2\times S^1$ is the same as $M$-theory on $T^2$.  

The essence of string dualities is that strong coupling limit of one 
theory is dual to weak coupling limit of another theory with the strongly 
coupled string states (dual to perturbative string states) identified 
with branes.  So, branes play an important role in understanding 
non-perturbative aspects of and dualities in string theories.   It is 
one of purposes of this review to summarize the recent development in 
solitons and black holes in string theories.   
The purpose of this chapter is to give basic facts on dualities 
in supersymmetric field theories and superstring theories that are 
necessary in understanding the rests of chapters of this review.  
Therefore, readers are referred to other literatures, e.g. 
\cite{WIT443,WIT121,SCH49,SCH367,SCH201,DUF11,DUF203,POL75,POL123,POL123,POL050,POLcj,ASP137,GRE155,TOW121},  
for complete understanding of the subject.  

In the first section, we discuss the symplectic transformations in 
extended supergravities and moduli spaces spanned by scalars 
in the supermultiplets.  
In section \ref{dualn4}, we summarize $T$-duality and $S$-duality of 
heterotic string on tori. In this section, we also discuss solution 
generating transformations that induce electric/magnetic charges of 
$U(1)$ gauge fields of heterotic string on tori when applied to a 
charge neutral solution.  These dualities are basic transformations 
for constructing most general black hole solutions in heterotic string 
on tori.  
In section \ref{dualsix}, we discuss string-string duality between 
heterotic string on $T^4$ and type-II string on $K3$, and 
string-string-string triality among type-IIA, type-IIB and heterotic 
strings in $D=6$.  These dualities transform black hole solutions 
discussed chapter \ref{n4bh} to type-II black holes which carry 
R-R charges \cite{BEH080,BEHd370}, thereby enabling 
interpretation of our black hole solutions in terms of $D$-brane 
picture.  In the final section, we summarize some aspects of $M$-theory 
and $U$-duality.  The generating black hole solutions of heterotic 
string on tori are the generating solutions for type-II string on tori, 
as well.  Namely, when the generating solutions of heterotic strings 
are embedded to type-II theories (note, such generating solutions carry 
only charges of NS-NS sector, which is common to both heterotic and type-II 
theories), subsets of $U$-dualities of type-II theories on tori induce 
the remaining $U(1)$ charges of type-II theories on tori \cite{CVEh}.  

\subsection{Electric-Magnetic Duality}\label{dualem}

The electric-magnetic duality in electromagnetism was conjectured by 
Dirac \cite{DIR} based on the observation that when electric charge and 
current are nonzero the Maxwell's equations lack symmetry under the 
exchange of the electric and magnetic fields, or in other words under 
the Hodge-duality transformation of the electromagnetic field strength. 
Dirac conjectured the existence of magnetic charges \cite{DIR74} to remedy 
the situation.  Magnetic charges are due to the topological defect of 
spacetime and are given by the first Chern class of the $U(1)$ 
principal fiber bundle with the base manifold given by $S^2$ surrounding 
the monopole.  The requirement of the continuity of the transition 
function that patches the 2 covers of the northern and southern 
hemispheres of $S^2$ or the requirement of the unobservability of Dirac 
string singularity restricts magnetic charge $q_m$ to be quantized 
\cite{ZWA1,ZWA2,SCH66,SCH68,WUy14,WUy107,WIT86}
relative to electric charge $q_e$ in the following way through the 
Dirac-Schwinger-Zwanzinger (DSZ) quantization rule:
\begin{equation}
{{q_e q_m}\over {4\pi\hbar}} ={n\over 2} \ \ \ \ (n\in {\bf Z}). 
\label{dirquan}
\end{equation}
Under the duality transformation, electric and magnetic charges 
are interchanged and correspondingly the coupling of the electromagnetic 
interactions is inverted due to the relation (\ref{dirquan}).  
So, the weak [strong] coupling limit of one theory is described 
by the strong [weak] coupling limit of its dual theory.  

The extension of the duality idea to non-Abelian gauge theories  
was made possible by the discovery of the 't Hooft-Polyakov monopole solution 
\cite{HOO79,POL20} in non-Abelian gauge theory.  In the 't Hooft-Polyakov 
monopole configuration, the non-Abelian gauge group is spontaneously broken 
down to the Abelian one at spatial infinity by Higgs fields that transform 
as the adjoint representation of the gauge group.  The magnetic charge 
of the 't Hooft-Polyakov monopole is determined by the second 
homotopy group of $S^2$ formed by the symmetry breaking Higgs vacuum, 
i.e. the winding number around $S^2$ as one wraps around  
$S^2_{\infty}$ surrounding the monopole.  Within this context, Montonen and 
Olive \cite{MONo72} conjectured that the spontaneously broken electric 
non-Abelian gauge group is dual to the spontaneously broken  magnetic
non-Abelian gauge group.  Under this duality, the gauge coupling of one 
theory is inverted in its dual theory, leading to the prediction 
that the strong coupling limit of a gauge theory is the weak 
coupling limit of its dual theory \cite{SEIw426,SEIw431,SEI49}.

Note, the hub of the Montonen-Olive conjecture lies in the existence 
of symmetry breaking Higgs fields which transform in the adjoint 
representation of the non-Abelian gauge group.  
It is a generic feature of extended supersymmetries that 
scalars live in the same supermultiplet as vector fields. 
So, the scalars in vector supermultiplet transform 
as the adjoint representation under the non-Abelian gauge group. 
Furthermore, supersymmetric theories obey the well-known 
non-renormalization theorem (See for example 
\cite{OLIw78,DADhd76,BARc17,MAN213,FLU217,SOHw100,GRIsr159,HOWsw124})  
The extended supersymmetry theories have another nice feature that 
the states preserving some of supersymmetry, i.e. {\it BPS states}, 
are determined entirely by their charges and moduli.   These are 
(degenerate) ground states of the theories (parameterized by moduli).   
Such BPS mass formula is invariant under dualities and 
degeneracy of BPS states remains unchanged under dualities.   
For example, electrically charged BPS states at coupling $g$ have the 
same mass and degeneracy as magnetically charged states BPS states at 
coupling $1/g$.  Furthermore, the supersymmetry algebra prevents the 
number of degeneracy of BPS states from changing as the coupling constant 
is varied.  Thus, it is BPS states that are suitable for testing ideas on 
duality symmetries.

In the following, we discuss generalization of electric-magnetic 
duality of Maxwell's equations to $N$-extended 
supersymmetry theories and study moduli spaces spanned by scalars.   

\subsubsection{Symplectic Transformations in Extended 
Supersymmetries}\label{dualemtran}

In supersymmetry theories, scalars $\phi^I$ are taken as 
coordinates of the target space manifold ${\cal M}_{scalar}$ of  
non-linear $\sigma$-model, which we write in general in the form 
\cite{FRE45}:
\begin{equation}
{\cal L}_{scalar}= g_{IJ}(\phi)g^{\mu\nu}\nabla_{\mu}\phi^I
\nabla_{\nu}\phi^J, 
\label{susysigma}
\end{equation}
where the covariant derivative $\nabla_{\mu}$ on  
$\phi^I$ is with respect to the gauge group $G_{gauge}$ 
that the vector fields ${\cal A}^{\Lambda}_{\mu}$ in the theory 
belong to:
\begin{equation}
\nabla_{\mu}\phi^I=\partial_{\mu}\phi^I+g{\cal A}^{\Lambda}
k^I_{\Lambda}(\phi),
\label{covderiv}
\end{equation}
with Killing vector fields $\vec{\bf k}_{\Lambda}\equiv k^I_{\Lambda}
(\phi){{\partial}\over{\partial\phi^I}}$ satisfying the Lie algebra 
$g_{gauge}$ of $G_{gauge}$:
\begin{equation}
\left[\vec{\bf k}_{\Lambda},\vec{\bf k}_{\Sigma}\right] 
=f^{\Delta}_{\Lambda\Sigma}\vec{\bf k}_{\Delta}.
\label{killalg}
\end{equation}
Note, $g_{gauge}$ is a subalgebra of the isometry 
algebra of ${\cal M}_{scalar}$.

Here, $g_{IJ}$ is the metric of ${\cal M}_{scalar}$.   In other words, 
a scalar is regarded as a map from the spacetime manifold to 
${\cal M}_{scalar}$. It turns out that the types of allowed target space 
manifolds formed by scalars are fixed by the number $m$ of supercharge 
degrees of freedom in $N$-extended superalgebra \cite{SALs}.  
When $m$ exceeds 8, the target space manifold is fixed as a symmetric 
space specified by the number $n$ of vector multiplets.  For example, 
$D=4$, $N=8$ supergravity, for which $m=32$, has the target space 
manifold $E_7/SU(8)$ \cite{CREj80,CREj79}; the $D=4$, $N=4$ theory, 
for which $m=16$, has target space manifold ${{SO(6,n)}\over {SO(6)
\otimes SO(n)}}\otimes {{SU(1,1)}\over {U(1)}}$ 
\cite{ROO,SEN388,SEN303,SCHs411,SCHs312}.  A special case is 
the effective action of the heterotic string on $T^6$, which is 
described by the $N=4$ supergravity coupled to the $N=4$ super-Yang-Mills 
theory with the gauge group $U(1)^{22}$ ($n=22$ case).  
Here, ${{SO(6,12)}\over {SO(6)\otimes SO(12)}}$ describes (classical) 
moduli space of Narain torus \cite{NAR169,NARsw279}, and ${{SU(1,1)}\over 
{U(1)}}$ is parameterized by the dilaton-axion field.  
(In section \ref{dualn4}, we discuss the Sen's approach 
\cite{SENint} of realizing such target space manifolds within the effective 
supergravity through the dimensional reduction of the heterotic string 
effective action.)  
For $m\leq 8$, the target space manifolds are less restrictive.  
For a $D=4$, $N=2$ theory, corresponding to $m=8$, 
the scalar manifold is factorized into a quaternionic one and a  
special K\"ahler manifold \cite{STRw,STR133}, which are respectively spanned 
by the scalars in the hypermultiplets and the vector multiplets.  
For $m=4$ case, e.g. $D=4$, $N=1$ theory, 
the target space manifold is the K\"ahler manifold.   

Within the extended supersymmetry theories described above, one can 
generalize \cite{GAIz193} the electric-magnetic duality transformations,  
which preserve equations of motion for the $U(1)$ field strengths.  
For this purpose, only relevant part of scalars is from vector 
supermultiplets. 
Such generalized electric-magnetic duality transformation is realized 
as follows.
  
We consider the general form of the diffeomorphism of the 
scalar manifold:  
\begin{equation}
t:\,{\cal M}^v_{scalar} \to {\cal M}^v_{scalar}:\,
\phi^I \mapsto t^I(\phi).
\label{diffscal}
\end{equation}
The map that corresponds to the isometry of the scalar manifold, 
i.e. $t^{*} g_{IJ}=g_{IJ}$, becomes the candidate for the symmetry 
of the theory.   

General form of kinetic term for vector fields in vector 
supermultiplets is \cite{FRE45}
\begin{equation}
{\cal L}_{vec}={1\over 2}\gamma_{\Lambda\Sigma}(\phi){\cal F}^{\Lambda}
\wedge\star{\cal F}^{\Sigma} + {1\over 2}\theta_{\Lambda\Sigma}(\phi)
{\cal F}^{\Lambda}\wedge {\cal F}^{\Sigma}, 
\label{veckin}
\end{equation}
where the field strengths ${\cal F}^{\Lambda}_{\mu\nu}$ of  
gauge fields ${\cal A}^{\Lambda}_{\mu}$ are
\begin{equation}
{\cal F}^{\Lambda}_{\mu\nu}\equiv {1\over 2}(\partial_{\mu}
{\cal A}^{\Lambda}_{\nu}-\partial_{\nu}{\cal A}^{\Lambda}_{\mu}
+gf^{\Lambda}_{\Sigma\Delta}A^{\Sigma}_{\mu}A^{\Delta}_{\nu}), 
\label{flstrth}
\end{equation}
and $\star{\cal F}^{\Lambda}_{\mu\nu}\equiv {1\over 2}
\epsilon_{\mu\nu\rho\sigma}{\cal F}^{\Lambda\,\rho\sigma}$ is the 
Hodge-dual of ${\cal F}^{\Lambda}_{\mu\nu}$.  
Here, the $n\times n$ matrix $\gamma_{IJ}(\phi)$ generalizes the
coupling constant of the conventional gauge theory and the antisymmetric 
matrix $\theta_{IJ}(\phi)$ is the generalization of the $\theta$-term 
\cite{WIT86}.   

The transformation properties of the gauge fields and the 
complex symmetric matrix ${\cal N}_{\Lambda\Sigma}(\phi)\equiv 
\theta_{\Lambda\Sigma}(\phi)-i\gamma_{\Lambda\Sigma}(\phi)$  
are determined by the symplectic embedding of the isometry group $G_{iso}$ 
of the scalar manifold ${\cal M}^{v}_{scalar}$ as follows.  

We consider the following homomorphism from the group ${\rm Diff}
({\cal M}_{scalar})$ of diffeomorphisms $t:\,{\cal M}^v_{scalar}
\to {\cal M}^v_{scalar}$ 
to the general linear group $GL(2n,{\bf R})$:
\begin{equation}
\iota_{\delta}:\,{\rm Diff}({\cal M}^v_{scalar})\to 
GL(2n,{\bf R}).
\label{homdg}
\end{equation}
One introduces a $2n\times 1$ matrix $V=(\star{\cal F},
\star{\cal G})^T$, where $\star{\cal G}\equiv -{{\partial{\cal L}}\over
{\partial{\cal F}^T}}$.  Then, the map $\iota_{\delta}$ in (\ref{homdg}) 
is defined by assigning, for each element $\xi$ of ${\rm Diff}
({\cal M}^v_{scalar})$, a $2n\times 2n$ matrix $\iota_{\delta}(\xi)
=\left(\matrix{A_{\xi}& B_{\xi}\cr C_{\xi}& D_{\xi}}\right)$ in 
$GL(2n,{\bf R})$ which transforms $V$ as   
\begin{eqnarray}
\left(\matrix{\star{\cal F} \cr \star{\cal G}}\right) &\mapsto&  
\left(\matrix{\star{\cal F} \cr \star{\cal G}}\right)^{\prime} = 
\left(\matrix{A_t&B_t\cr C_t&D_t}\right) 
\left(\matrix{\star{\cal F} \cr \star{\cal G}}\right), \ \ \ {\rm or}
\cr
\left(\matrix{{\cal F}^+\cr {\cal G}^+}\right) &\mapsto&  
\left(\matrix{{\cal F}^+\cr {\cal G}^+}\right)^{\prime} = 
\left(\matrix{A_t&B_t\cr C_t&D_t}\right)
\left(\matrix{{\cal F}^+\cr {\cal N}{\cal F}^+}\right),
\label{vectran}
\end{eqnarray}
where ${\cal F}^+\equiv {\cal F}-i\star{\cal F}$ and ${\cal G}^+\equiv 
{\cal N}{\cal F}^+$ \cite{DEWv245,DEWv293}.  The transformation law 
(\ref{vectran}) on $V$ is dictated by the requirement that the Bianchi 
identities and the field equations for vector fields remain invariant. 

Under the action of the diffeomorphism $\xi$ on $\phi^I$, 
${\cal N}(\phi)$ also transforms.  If we further require that transformed 
field ${\cal G}^{\prime}$ 
to be defined as $\star{\cal G}^{\prime} =-{{\partial{\cal L}^{\prime}}\over
{\partial{\cal F}^{\prime\,T}}}$, the transformation property of 
${\cal N}_{\Lambda\Sigma}(\phi)$ under the diffeomorphism $t$ on 
$\phi^I$ is fixed as the fractional linear form:
\begin{equation}
{\cal N}(\phi) \mapsto {\cal N}^{\prime}(t(\phi)) = 
[C_t +D_t{\cal N}(\phi)][A_t +B_t{\cal N}(\phi)]^{-1}, 
\label{fracmat}
\end{equation}
with $\iota_{\delta}(\xi)$ now restricted to belong to  
$Sp(2n,{\bf R})\subset GL(2n,{\bf R})$.  

When $B_t\neq 0\neq C_t$, it is a symmetry of equations of motion only.  
When $B_t =0\neq C_t$, the Lagrangian is invariant up to four-divergence.
When $B_t=0=C_t$, the Lagrangian is invariant.  In particular, the 
symplectic transformations (\ref{vectran}) and (\ref{fracmat}) with $B_t\neq 
0$ are non-perturbative, since they necessarily induce magnetic charge from 
purely electric configuration and invert ${\cal N}$, which plays the role 
of the gauge coupling constant.  

When electric/magnetic charges are quantized,  
$Sp(2n,{\bf R})$ gets broken down to $Sp(2n,{\bf Z})$ so that the 
charge lattice spanned by the quantized electric and magnetic charges 
is preserved under the transformation (\ref{vectran}).  This is 
the generalization of the electric-magnetic duality symmetry to the case 
of $D=4$ supersymmetry theory with $n$ vector fields.   

\subsubsection{Symplectic Embedding of Homogeneous Spaces}\label{dualemhom}

When the number of supercharge degrees of freedom exceeds 8, 
e.g. $N\geq 3$ in $D=4$, ${\cal M}_{scalar}$ is a homogeneous space $G/H$ 
with the isometry group $G$.  The supersymmetry restricts the dimension 
of ${\cal M}_{scalar}$ and the number $n$ of vector multiplets to be related 
to each other and, therefore, ${\cal M}_{scalar}$ is determined 
uniquely by $n$.  In the following, we discuss the symplectic embedding 
of the homogeneous space and show how the gauge kinetic matrix 
${\cal N}_{\Lambda\Sigma}$ is determined.

We consider the following embedding of the isometry group $G$ of  
$G/H$ into $Sp(2n,{\bf R})$:
\begin{equation}
\iota_{\delta}:\,G\to Sp(2n,{\bf R}):\, 
L(\phi)\mapsto \iota_{\delta}(L(\phi)).
\label{emhomsym}
\end{equation}
Applying the following isomorphism from the real symplectic group 
$Sp(2n,{\bf R})$ to the complex symplectic group $Usp(n,n)\equiv Sp(2n,C)
\cap U(n,n)$:
\begin{equation}
\mu:\,\left(\matrix{A&B\cr C&D}\right) \mapsto 
\left(\matrix{T&V^*\cr V&T^*}\right), 
\label{isomo}
\end{equation}
where
\begin{equation}
T\equiv {1\over 2}(A-iB)+{1\over 2}(C+iD), \ \ \ 
V\equiv{1\over 2}(A-iB)-{1\over 2}(C+iD),
\label{isodef}
\end{equation}
one can define the complex symplectic matrix ${\cal O}(\phi)$ 
($\in Usp(n,n)$), for each coset representative $L(\phi)$  
of $G/H$, as follows:
\begin{equation}
\mu\cdot\iota_{\delta}:\,G \to Usp(n,n):\, L(\phi)\mapsto {\cal O}(\phi)=
\left(\matrix{U_0(\phi)&U^*_1(\phi) \cr U_1(\phi)&U^*_0(\phi)}\right),  
\label{cxsymp}
\end{equation}
where
\begin{equation}
U_0(\phi)^{\dagger}U_0(\phi)-U_1(\phi)^{\dagger}U_1(\phi)={\bf 1}, 
\ \ 
U_0(\phi)^{\dagger}U_1(\phi)^*-U_1(\phi)^{\dagger}U_0(\phi)^*={\bf 0}.
\label{cpxembd}
\end{equation}

From ${\cal O}(\phi)$ in (\ref{cxsymp}), which is defined  
for each coset representative $L(\phi)$ of $G/H$, 
one can define the following scalar matrix \cite{GAIz193} which has all 
the right properties for the gauge kinetic matrix ${\cal N}_{\Lambda\Sigma}=
\theta_{\Lambda\Sigma}-i\gamma_{\Lambda\Sigma}$:
\begin{equation}
{\cal N}\equiv i[U^{\dagger}_0+U^{\dagger}_1]^{-1}[U^{\dagger}_0-
U^{\dagger}_1],
\label{symgaug}
\end{equation}
namely, ${\cal N}^T={\cal N}$ and ${\cal N}$ transforms fractional 
linearly under $Sp(2n,{\bf R})$.

Specifically, we consider the homogeneous space of the form
\footnote{${\cal ST}[2,n]$ is the only special K\"ahler manifold 
with direct product structure \cite{FERv6} of this form.} 
${\cal ST}[m,n]\equiv {{SU(1,1)}\over{U(1)}}\otimes {{SO(m,n)}\over
{SO(m)\otimes SO(n)}}$, where $m$ is the number of graviphotons and 
$n$ the number of vector multiplets.  Here, ${{SU(1,1)}
\over{U(1)}}$ is parameterized by the axion-dilaton field $S$ and 
${{SO(m,n)}\over{SO(m)\otimes SO(n)}}$ is parameterized by 
a $m\times n$ real matrix $X$.  

In the real basis, the $SO(m,n)$ $T$-duality and $SL(2,{\bf R})$ 
$S$-duality groups of ${\cal ST}[m,n]$ are respectively embedded 
into the symplectic group as:
\begin{eqnarray}
\iota_{\delta}&:&\ L\in SO(m,n) \mapsto \left(\matrix{L&O\cr O&(L^T)^{-1}}
\right)\in Sp(2m+2n,{\bf R})
\cr
\iota_{\delta}&:&\ \left(\matrix{a&b\cr c&d}\right)\in SL(2,{\bf R})\mapsto 
\left(\matrix{a{\bf 1}&b\eta\cr c\eta&d{\bf 1}}\right)\in Sp(2m+2n,{\bf R}),
\label{restdualsym}
\end{eqnarray}
where $\eta$ is an $SO(m,n)$ invariant metric, ${\bf 1}$ is the $(m+n)
\times (m+n)$ identity matrix, and $a,b,c,d\in{\bf R}$ satisfy  
$ad-bc=1$.  
In the complex basis, the embeddings are:
\begin{eqnarray}
\iota_{\delta}&:&\ L\in SO(m,n) \mapsto
\cr
& &\ \ \left(\matrix{{1\over 2}
(L+\eta L\eta)& {1\over 2}(L-\eta L\eta)\cr {1\over 2}(L-\eta L\eta)&
{1\over 2}(L+\eta L\eta)}\right)\in Usp(m+n,m+n),
\cr 
\iota_{\delta}&:&\ \left(\matrix{t&v^*\cr v&t^*}\right)\in SU(1,1)\mapsto 
\cr
& &\ \ \left(\matrix{{\rm Re}\,t\,\,{\bf 1}+i{\rm Im}\,
t\,\,\eta &{\rm Re}\,v\,\,{\bf 1}-i{\rm Im}\,v\,\,\eta \cr 
{\rm Re}\,v\,\,{\bf 1}+i{\rm Im}\,v\,\,\eta &{\rm Re}\,t\,\,
{\bf 1}-i{\rm Im}\,t\,\,\eta}\right)\in Usp(m+n,m+n). 
\label{cxstdualsym}
\end{eqnarray}

The symplectic embeddings (\ref{cxstdualsym}) make it possible 
to express the gauge kinetic matrix ${\cal N}$ in terms of the scalars 
$S$ and $X$, which parameterize ${\cal ST}[m,n]$,  as follows.  
The coset representatives of $SU(1,1)/U(1)$ and $SO(m,n)/[SO(m)
\times SO(n)]$ are respectively 
\begin{eqnarray}
L(S)&\equiv& {1\over{n(S)}}\left(\matrix{{\bf 1}&{{i-S}\over{i+S}}\cr 
{{i+\bar{S}}\over{i-\bar{S}}}&{\bf 1}}\right), 
\cr
L(X)&\equiv&\left(\matrix{({\bf 1}+XX^T)^{1/2}&X\cr X^T&({\bf 1}+X^TX)^{1/2}}
\right), 
\label{stsymp}
\end{eqnarray}
where $n(S)\equiv\sqrt{{4{\rm Im}\,S}\over{1+|S|^2+2{\rm Im}\,S}}$.   
Note, $M\equiv L(X)L^T(X)$ is a symmetric $SO(m,n)$ 
matrix, studied by Sen \cite{SENint}, which will be discussed 
in section \ref{dualn4}.  

Applying the transformations (\ref{cxstdualsym}), one obtains the 
following symplectic embedding of the coset representations of 
${\cal ST}[m,n]$:
\begin{equation}
\iota_{\delta}(L(S))\circ\iota_{\delta}(L(X)) = 
\left(\matrix{U_0(S,X)&U^*_1(S,X)\cr U_1(S,X)&U^*_0(S,X)}\right) 
\in Usp(n+m,n+m), 
\label{homsymp}
\end{equation}
where the explicit expressions for $\iota_{\delta}(L(S))$ and 
$\iota_{\delta}(L(X))$ are obtained by applying the transformations 
(\ref{cxstdualsym}).  
Substituting this expression into the general formula in  
(\ref{symgaug}), one obtains the following gauge kinetic matrix:
\begin{equation}
{\cal N}=i{\rm Im}\,S\eta L(X)L^T(X)\eta +{\rm Re}\,S\eta = 
i{\rm Im}\,S\eta M\eta +{\rm Re}\,S\eta. 
\label{gagkinmat}
\end{equation}
Then, the Lagrangian ${\cal L}_{scalar}+
{\cal L}_{vec}$ (Cf. see (\ref{susysigma}) and (\ref{veckin}))  
takes the following form that corresponds to ${\cal ST}[m,n]$:
\begin{eqnarray}
{\cal L}&=&\sqrt{-g}\left[{\cal R}_g +{1\over{4({\rm Im}\,S)^2}}
\partial_{\mu}S\partial^{\mu}\bar{S}-{1\over 4}{\rm Tr}(\partial_{\mu}
M\partial^{\mu}M)\right. 
\cr
& &\ \ \ \ \ \ \ \left.-{1\over 4}{\rm Im}\,S{\cal F}^I_{\mu\nu}
(\eta L\eta)_{IJ}{\cal F}^{J|\mu\nu} +{1\over{8\sqrt{-g}}}{\rm Re}
S{\cal F}^I_{\mu\nu}\eta_{IJ}{\cal F}^J_{\rho\sigma}
\epsilon^{\mu\nu\rho\sigma}\right]. 
\label{lagstmn}
\end{eqnarray}

The $T$-duality and $S$-duality of the heterotic 
string on $T^6$ \cite{SCH125,SCHs312,MAHs390,SENint}
are special cases of the symplectic transformations 
(\ref{restdualsym}) with $(m,n)=(6,22)$.  In general, under the 
$SO(m,n)$ and $SL(2,{\bf R})$ transformations (\ref{restdualsym}), 
the gauge fields and the gauge kinetic matrix transform, 
respectively, as 
(Cf. (\ref{vectran}) and (\ref{fracmat})):
\begin{eqnarray}
{\cal F}^+&\mapsto&{\cal F}^{+\,\prime}=L{\cal F}^+, \ \ \  
{\cal N}\mapsto {\cal N}^{\prime}=(L^T)^{-1}{\cal N}L^{-1},  
\cr
{\cal F}^+&\mapsto&{\cal F}^{+\,\prime}=
a{\cal F}^+ +b\eta{\cal N}{\cal F}^+, \ \ \ 
{\cal N}\mapsto {\cal N}^{\prime}=(d{\cal N}+c\eta)(b\eta{\cal N}+a)^{-1}. 
\label{sttrans}
\end{eqnarray}
Note, $O(m,n)$ [$SL(2,{\bf R})$] is a perturbative [non-perturbative] 
symmetry, since ${\cal N}$ is not inverted [gets inverted].  
$SL(2,{\bf R})$ is the symmetry of the equations of motion only, 
since electric and magnetic charges get mixed,  and since this 
corresponds to the transformations (\ref{vectran}) and 
(\ref{fracmat}) with $B_t\neq 0\neq C_t$.

\subsubsection{Target Space Manifolds of $N=2$ Theories}\label{dualemman}

Contrary to $D=4$, $N\geq 3$ theories, the scalar 
manifolds of $N=2$ theories are not necessarily expressed as 
homogeneous symmetric coset manifolds
\footnote{But there is a subclass of homogeneous special manifolds, 
which are classified in \cite{CREv2}.  These are ${{SU(1,1)}\over 
{U(1)}}$, ${{SU(1,n_v)}\over{SU(n_v)\times U(1)}}$, ${{SU(1,1)}\over{U(1)}}
\otimes {{SO(2,n_v)}\over{SO(2)\times SO(n_v)}}$, ${{Sp(6,{\bf R})}\over 
{SU(3)\times U(1)}}$, ${{SU(3,3)}\over{SU(3)\times SU(3)}}$, 
${{SO^*(12)}\over{SU(6)\times U(1)}}$, and ${{E_{7(-6)}}\over 
{E_6\times SO(2)}}$ with the corresponding symplectic groups $Sp(2n_v+2)$ 
respectively given by $Sp(4)$, $Sp(2n_v+2)$, $Sp(2n_v+4)$, $Sp(14)$, 
$Sp(20)$, $Sp(32)$, and $Sp(56)$.}.  
Scalar manifold of the $D=4$, $N=2$ theory has the generic form: 
\begin{equation}
{\cal M}_{scalar}={\cal SM}_n \otimes {\cal HM}_m,
\label{n2manif}
\end{equation}
where ${\cal SM}_n$ is a special K\"ahler manifold of the complex 
dimension $n=$ ``the number of the vector supermultiplets'', and 
the manifold ${\cal HM}_m$ spanned by the scalars in the 
hypermultiplets has the dimension $4m$, where $m=$ ``the number of the 
hypermultiplets''.  So, the metric $g_{IJ}(\phi)$ of ${\cal M}_{scalar}$ 
has the form:
\begin{equation}
g_{IJ}(\phi)d\phi^I\otimes d\phi^J = 
g_{ab^*}dz^a\otimes d\bar{z}^{b*} +
h_{uv}dq^u\otimes dq^v. 
\label{n2met}
\end{equation}
Here, $g_{ab^*}$ [$h_{uv}$] is the special K\"ahler metric 
on ${\cal SM}_n$ [the quaternionic metric on ${\cal HM}_n$].  

\paragraph{Special K\"ahler Manifolds}\label{dualemmansp}

$N=2$ super-Yang-Mills theory is described by a chiral superfield 
$\Phi$, which is defined by $\bar{D}^{\dot{\alpha}i}\,\Phi=0$ 
(like chiral superfield in $N=1$ theory),  
with an additional constraint:
\begin{equation}
D^{\alpha}_{(i}D^{\beta}_{j)}\,\Phi\epsilon_{\alpha\beta} = 
\epsilon_{ik}\epsilon_{j\ell}\bar{D}^{\dot{\alpha}(k}\bar{D}^{\ell)
\dot{\beta}}\,\bar{\Phi}\epsilon_{\dot{\alpha}\dot{\beta}}, 
\label{n2chiral}
\end{equation}
where $i=1,2$ labels supercharges of $N=2$ theory, $\alpha,
\dot{\alpha}=1,2$ are indices of chiral spinors, and 
$\bar{D}^{\dot{\alpha}}$ is a covariant chiral superspace derivative.
The component fields of a $N=2$ chiral superfield $\Phi^A$ are a 
scalar $X^A$, spinors $\lambda^{iA}$, $U(1)$ gauge field strength 
$F^A_{\mu\nu}$, and auxiliary scalars $Y^A_{ij}$ satisfying a reality 
constraint $Y_{ij}=\epsilon_{ik}\epsilon_{j\ell}\bar{Y}^{k\ell}$ (due to 
the constraint (\ref{n2chiral})).  The action of $N=2$ chiral superfields 
$\Phi^A$ is determined by an arbitrary holomorphic function $F(\Phi^A)$ of 
$\Phi^A$ as $\int d^4x\int d^4\theta\,F(\Phi^A)\,+\,
c.c.$, and is given, in terms of the component fields, by:
\begin{equation}
{\cal L}^{vec}=g_{A\bar{B}}\partial_{\mu}X^A\partial^{\mu}\bar{X}^B 
+g_{A\bar{B}}\bar{\lambda}^{iA}\gamma^{\mu}\partial_{\mu}
\lambda^{\bar{B}}_i 
+{\rm Im}(F_{AB}{\cal F}^{-A}_{\mu\nu}{\cal F}^{-B\,\mu\nu}) + ..., 
\label{chiract}
\end{equation}
where the dots denote the interaction terms involving fermions, 
$g_{A\bar{B}}=\partial_A\partial_{\bar{B}}\,K$ is a K\"ahler 
metric, and $F_{AB}\equiv\partial_A\partial_B F$.  
Note, this action is a special case of the most general coupling 
of $N=1$ chiral superfields to $N=1$ Abelian vector superfields in which the 
K\"ahler potential $K$ and the holomorphic kinetic term function $F_{AB}$ 
take the following special forms \cite{SIEt,GAT238} 
\begin{equation}
K(X,\bar{X})=i[\bar{F}_A(\bar{X})X^A -F_A(X)
\bar{X}^A]\ \ (F_A\equiv\partial_A F), \ \ 
F_{AB} = \partial_A\partial_B F. 
\label{kahl}
\end{equation}
The K\"ahler manifold with the K\"ahler potential $K(X,\bar{X})$ 
determined by the prepotential $F(X)$ \cite{DEWv245,CREkvdfddg250,DEWlv255}
through (\ref{kahl}) is called the {\it special K\"ahler manifold} 
\cite{DEWlpsv134,DEWv245,STR133,DIXkl329,CANd355,CASdf241,DAUff359}.  

When $N=2$ chiral fields $\Phi^{\Lambda}$ ($\Lambda=0,1,...,n_v$) 
are coupled to the Weyl multiplet (with components given by vierbein, 
2 gravitinos and auxiliary fields) \cite{DEWlpsv134,DEWv245}, the 
invariance under the dilatation requires $F(X)$ to be a homogeneous 
function of degree 2 (so that $F(X)$ has Weyl weight 2) 
\cite{DEWv245,DEWlpsv134}.  Furthermore, the requirement of 
canonical gravitino kinetic term imposes one constraint on the set 
of scalars $X^{\Lambda}$ as 
\begin{equation}
i(\bar{X}^{\Lambda}F_{\Lambda}-\bar{F}_{\Sigma}X^{\Sigma})=1, 
\label{scalconst}
\end{equation}
leading to gauge fixing for dilations and the special K\"ahler manifold 
of the dimension $n_v$.  
Note, the extra chiral superfield $\Phi^0$ is introduced to fix 
the dilatation gauge, to break the $S$-supersymmetry, and to introduce the 
physical $U(1)$ gauge field in the $N=2$ supergravity multiplet 
(the scalar and the spinor components of the superfield $\Phi^0$ 
do not become additional physical particles).
The final form of bosonic action describing $n_v$ numbers of $N=2$ vector 
multiplets coupled to $N=2$ supergravity is
\begin{equation}
e^{-1}{\cal L}=-{1\over 2}{\cal R}+g_{ab^*}\partial_{\mu}z^a
\partial^{\mu}\bar{z}^{b^*} - {\rm Im}\,({\cal N}_{\Lambda\Sigma}
(z,\bar{z}){\cal F}^{+\,\Lambda}_{\mu\nu}
{\cal F}^{+\,\Sigma\,\mu\nu}), 
\label{grachicat}
\end{equation}
where $z^a$ ($a=1,...,n_v$) are the coordinates of a K\"ahler space 
spanned by the scalars $X^{\Lambda}$ ($\Lambda=0,1,...,n_v$) 
which satisfy one constraint (\ref{scalconst}) (therefore, the manifold 
spanned by $X^{\Lambda}$ has $n_v$ complex dimensions).  
A convenient choice for $z^a$ is the inhomogeneous coordinates 
called the {\it special} coordinates: $z^a=X^a(z)/X^0(z)$, $a=1,...,n_v$.  
(Note, $X^a(z)=z^a$ in special coordinates in which ${{\partial (X^a/X^0)}
\over{\partial z^b}}=\delta^a_b$ \cite{CANd355,CASdf241,STR133,CERdfll8}.)   
Here, $K$ and ${\cal N}_{\Lambda\Sigma}$ (Cf. the scalar matrix ${\cal N}$ 
in (\ref{fracmat})) are determined by $F(X)$ to be of the 
forms \cite{DEWv245,CREkvdfddg250,DEWlv255,CECfg4,DAUff359}:
\begin{eqnarray}
e^{-K(z,\bar{z})}&=& i\bar{Z}^{\Lambda}(\bar{z})
F_{\Lambda}(Z(z))-iZ^{\Sigma}(z)\bar{F}_{\Sigma}
(\bar{Z}(\bar{z})), 
\cr
{\cal N}_{\Lambda\Sigma}&=& \bar{F}_{\Lambda\Sigma}+
2i{{{\rm Im}(F_{\Lambda\Delta}){\rm Im}(F_{\Sigma\Pi})X^{\Delta}X^{\Pi}}
\over {{\rm Im}(F_{\Delta\Pi})X^{\Delta}X^{\Pi}}}, 
\label{kahmet}
\end{eqnarray}
where $Z^{\Lambda}(z)=e^{-K/2}X^{\Lambda}$ and 
$\bar{Z}^{\Lambda}(\bar{z})=e^{-K/2}\bar{X}^{\Lambda}$ 
($\Lambda=0,1,...,n_v$) are holomorphic sections of the projective space 
$P{\bf C}^{n+1}$ \cite{CASdf241,CASdf7,DAUff359}, and $F_{\Lambda\Sigma}
\equiv \partial_{\Lambda}F_{\Sigma}(X)$.  
We give some examples of the holomorphic function $F(X)$ 
of $N=2$ theories and the corresponding special K\"ahler manifold 
target spaces:
\begin{eqnarray}
F=iX^0X^1 \ \ \ \ \ \ \  & &{{SU(1,1)}\over {U(1)}}
\cr
F=(X^1)^3/X^0 \ \ \ \ \ \ \ & &{{SU(1,1)}\over {U(1)}}
\cr
F=-4\sqrt{X^0(X^1)^3} \ \ \ \ \ \ \ & &{{SU(1,1)}\over {U(1)}}
\cr
F=iX^{\Lambda}\eta_{\Lambda\Sigma}X^{\Sigma} \ \ \ \ \ \ \ 
& &{{SU(1,n)}\over {SU(n)\otimes U(1)}}
\cr
F={{d_{\Lambda\Delta\Sigma}X^{\Lambda}X^{\Delta}X^{\Sigma}}\over X^0} 
\ \ \ \ \ \ \ & &{\rm Calabi-Yau}. 
\cr
F=-i{X^s\over X^0}\left((X^1)^2-\sum^n_{a=2}(X^a)^2\right)\ \ \ \ \ \ \ 
& & {{SO(2,1)}\over{SO(2)}}\times{{SO(2,n)}\over{SO(2)\times SO(n)}}.
\label{spkahex}
\end{eqnarray}

So far, we defined the {\it special K\"ahler manifold} as the K\"ahler 
manifold with the special form of the K\"ahler metric given by 
(\ref{kahmet}), which depends on the holomorphic prepotential $F$.  
Now, we discuss the symplectic formalism of the 
special K\"ahler manifolds of $N=2$ supergravity coupled to 
$n_v$ vector supermultiplets.   

For the symplectic formalism \cite{CERdf339,DEWv097,CERdf46}
of the special K\"alher manifold ${\cal M}$, one considers the tensor 
bundle of the type ${\cal H}={\cal SV}\otimes {\cal L}$.  Here, 
${\cal SV}\to{\cal M}$ denotes a holomorphic flat vector bundle of 
rank $2n_v+2$ with structural group $Sp(2n_v+2,{\bf R})$
\footnote{The additional two dimensions in ${\cal SV}$ come from 
a vector field in the supergravity multiplet.}, 
and ${\cal L}\to {\cal M}$ denotes the complex line bundle 
whose first Chern class equals the K\"ahler form of the 
$n_v$-dimensional Hodge-K\"ahler manifold ${\cal M}$.

A holomorphic section of the bundle $\cal H$ has the form 
\cite{CASdf241,CANd355,CANdgp258,CANdgp359,DAUff359,CERdfll8}:
\begin{equation}
\Omega =\left(\matrix{X^{\Lambda} \cr F_{\Sigma}}\right)
\ \ \ \ \Lambda,\Sigma=0,1,...,n_v,
\label{holsec}
\end{equation}
which is defined for each coordinate patch 
$U_i\subset {\cal M}$ of the (Hodge-K\"ahler) manifold ${\cal M}$  
and transforms as a vector under the symplectic transformation 
$Sp(2n_v+2,{\bf R})$.  
The Hodge-K\"ahler manifold ${\cal M}$ with a bundle ${\cal H}$ 
described above is called {\it special K\"ahler}, if 
the K\"ahler potential is expressed in terms of the holomorphic section 
$\Omega$ as
\footnote{Alternatively, one can define {\it special K\"ahler} manifold 
by introducing the symplectic section $V$ (\ref{sympsec}) satisfying 
the constraint (\ref{secconst}).  Then, the K\"ahler potential is determined 
in terms of the holomorphic section $\Omega$ as in (\ref{algkahl}) 
through the constraint (\ref{secconst}) with (\ref{sympsec}) substituted.}
\begin{equation}
K=-{\rm log}(i\langle\Omega|\bar{\Omega}\rangle)=
-{\rm log}\left[i\left(\bar{X}^{\Lambda}F_{\Lambda}-
\bar{F}_{\Sigma}X^{\Sigma}\right)\right],
\label{algkahl}
\end{equation}
where $\langle\Omega|\bar{\Omega}\rangle\equiv -\Omega^{\dagger}
\left(\matrix{0&I\cr-I&0}\right)\Omega$ denotes a symplectic inner product.

One further introduces the symplectic section of the bundle ${\cal H}$ 
according to 
\begin{equation}
V=\left(\matrix{L^{\Lambda}\cr M_{\Sigma}}\right)
\equiv e^{K/2}\Omega^T.
\label{sympsec}
\end{equation}  
Then, by definition, $V$ satisfies the 
constraint \cite{DEWv245,CREkvdfddg250,DEWlv255,CECfg4,CASdf241}:
\begin{equation}
1=i\langle V|\bar{V}\rangle = 
i(\bar{L}^{\Lambda}M_{\Lambda}-\bar{M}_{\Sigma}L^{\Sigma}), 
\label{secconst}
\end{equation}
and is covariantly holomorphic:
\begin{equation}
\nabla_{a^*}\,V=(\partial_{a^*}-{1\over 2}\partial_{a^*}K)\,V=0, 
\label{covconst}
\end{equation}
where $\partial_a \equiv \partial/\partial z^a$ and 
$\partial_{a^*} \equiv \partial/\partial \bar{z}^{a^*}$.  

One further introduces the matrix of the following form:
\begin{equation}
U_a =\nabla_a V =(\partial_a +{1\over 2}\partial_a K)V \equiv 
\left(\matrix{f^{\Lambda}_a\cr h_{\Sigma|a}}\right)\ \ \ 
(a=1,...,n_v). 
\label{defhol}
\end{equation}
Then, {\it period matrix} ${\cal N}_{\Lambda\Sigma}$ (which corresponds 
to the gauge kinetic matrix in the $N=2$ theory) is defined via 
the relations
\begin{equation}
\bar{M}_{\Lambda}=\bar{\cal N}_{\Lambda\Sigma}\bar{L}^{\Sigma}, 
\ \ \ \ 
h_{\Lambda|a}=\bar{\cal N}_{\Lambda\Sigma}f^{\Sigma}_a.
\label{peridef}
\end{equation}
Therefore, the period matrix has the following explicit form
\begin{equation}
\bar{\cal N}_{\Lambda\Sigma} =h_{\Lambda|I}\circ (f^{-1})^I_{\Sigma}, 
\label{perihol}
\end{equation}
where the $(n_v+1)\times(n_v+1)$ matrices $f^{\Lambda}_I$ and 
$h_{\Lambda|I}$ in the above are defined as
\begin{equation}
f^{\Lambda}_I=\left(\matrix{f^{\Lambda}_a\cr \bar{L}^{\Lambda}}\right), 
\ \ \ \ \ 
h_{\Lambda|I}=\left(\matrix{h_{\Lambda|a}\cr \bar{M}_{\Lambda}}\right).
\label{dfsymp}
\end{equation}
As a consequence, under the diffeomorphism of the base manifold $\cal M$, 
${\cal N}_{\Lambda\Sigma}$ transforms fractional 
linearly, like the gauge kinetic matrix (Cf. (\ref{fracmat})). 

Note, in the above symplectic formalism of the special K\"ahler 
manifold no reference was made on the prepotential.  In fact, for 
some cases, the existence of the prepotential is not even guaranteed
\footnote{For electrically neutral theories, one can 
always rotate to bases where a prepotential exists \cite{CRArtv073}.} 
\cite{CERdfv444}.  
We now discuss how the concept of the prepotential emerges 
within the framework of symplectic formalism.  

Under the coordinate transformations of ${\cal M}$, $\Omega$ transforms as:
\begin{equation}
\Omega \to \Omega^{\prime}=e^{-f}M\Omega, 
\label{holsectran}
\end{equation}
where the factor $e^{-f}$ corresponds to a $U(1)$ K\"ahler transformation 
(i.e. $K$ transforms as $K\to K+{\rm Re}\,f(z)$) and 
$M\in Sp(2n+2,{\bf R})$, and ${\cal N}_{\Lambda\Sigma}$ 
transforms fractional linearly just like a gauge kinetic matrix 
(Cf. (\ref{fracmat})).
From the transformation law  (\ref{holsectran}) with $M=I$, one can infer 
that $X^{\Lambda}$ can be regarded as homogeneous 
coordinates of a $(n_v+1)$-dimensional projective space at least locally 
\cite{DEWlpsv134,GAT238,DEWv245}, since $X^{\Lambda}$ and 
$e^{-f}X^{\Lambda}$ are identified under the K\"ahler transformations.  
This is possible provided the Jacobian matrix $\partial_a\left({{X^b}
\over{X^0}}\right)$ ($a,b=1,...,n_v$) is invertible \cite{CERdf339}.  
In this case, due to the integrability condition following from 
(\ref{secconst}) and (\ref{covconst}), the lower components $F_{\Sigma}$ 
of $\Omega$ are  expressed as 
\begin{equation}
F_{\Sigma}={\partial\over{\partial X^{\Sigma}}}F,  
\label{prepot}
\end{equation}
in terms of a homogeneous function $F(X)={1\over 2}X^{\Lambda}F_{\Lambda}$ 
of degree 2 in $X^{\Lambda}$.  
Then, one can use $z^a\equiv{{X^a}\over{X^0}}$ ($a=1,...,n_v$) as the 
special coordinates and the holomorphic prepotential is  
${\cal F}(z)\equiv (X^0)^{-2}F(X)$.  In terms of ${\cal F}$ and $z^a$, 
the K\"ahler potential $K$ is expressed as \cite{DEWlpsv134}
\begin{equation}
K(z,\bar{z})=-{\rm log}\,i\left[2({\cal F}-\bar{\cal F})-
(\partial_a{\cal F}+\partial_{a^*}\bar{F})(z^a-
\bar{z}^{a^*})\right]. 
\label{holseckah}
\end{equation}          

\paragraph{Hypergeometry}\label{dualemmanhyp}

$N=2$ hypermultiplet consists of a doublet of 0-form spinors with 
left and right chiralities, and 4 real scalars, which 
can be locally regarded as the 4 components of a quaternion.  
The scalars $q^v$ ($v=1,...,4n_H$) in $n_H$ hypermultiplets 
form a $4n_H$-dimensional real manifold ${\cal HM}$ 
\cite{BAGw222,HITklr108,GAL108,DAUff359,DEWlv255,CASdf7} 	
with a metric
\begin{equation}
ds^2 =h_{uv}(q)dq^u \otimes dq^v.
\label{hypmet}
\end{equation}
This manifold is endowed with 3 complex structures $J^x:\,T({\cal HM}) 
\to T({\cal HM})$ ($x=1,2,3$) satisfying the quaternionic algebra 
$J^xJ^y =-\delta^{xy}{\bf 1}+\epsilon^{xyz}J^z$.  
The metric $h_{uv}(q)$ is hermitian with respect to $J^x$:
\begin{equation}
h(J^x{\bf X},J^x{\bf Y})=h({\bf X},{\bf Y}), \ \ \ 
{\bf X}, {\bf Y} \in T\,{\cal HM}. 
\label{hermqrtmet}
\end{equation}

From $J^x$, one can define triplet of $SU(2)$ Lie-algebra valued 
HyperK\"ahler forms as
\begin{equation}
K^x=K^x_{uv}dq^u\wedge dq^v,
\label{hypkah}
\end{equation}
where $K^x_{uv}=h_{uw}(J^x)^w_v$.  Supersymmetry requires the existence of a 
principal $SU(2)$-bundle ${\cal SU}\to {\cal HM}$ with a connection 
$\omega^x$.  The manifold ${\cal HM}$ is defined by requiring that 
$K^x$ is covariantly closed with respect to the connection $\omega^x$: 
\begin{equation}
\nabla K^x\equiv dK^x +\epsilon^{xyz}
\omega^y\wedge K^z=0.
\label{defhm}
\end{equation}

There are two types of hypergeometry: 
rigid [local] hypergeometry corresponding to global [local] 
$N=2$ supersymmetry is called {\it HyperK\"ahler} [{\it quaternionic}].
The only difference between the two manifolds are the 
structure of the $\cal SU$-bundle.   
A {\it HyperK\"ahler manifold} has the flat $\cal SU$-bundle, 
and a {\it quaternionic manifold} has the curvature of the 
$\cal SU$-bundle proportional to the HyperK\"ahler 2-form.  
Here, the ${\cal SU}$-curvature is defined as
\begin{equation}
\Omega^x \equiv d\omega^x +{1\over 2}\epsilon^{xyz}\omega^y\wedge\omega^z.
\label{sucurv}
\end{equation} 
In the quaternionic case, the curvature is:
\begin{equation}
\Omega^x ={1\over\lambda}K^x, 
\label{hypcurv}
\end{equation}
where $\lambda$ is a real number related to the scale of the quaternionic 
manifold.  In the limit $\lambda\to\infty$, quaternionic manifold becomes  
Hyperk\"ahler manifold \cite{ANDbcdffm032}.

The manifold ${\cal HM}$ has the following holonomy group:
\begin{eqnarray}
{\rm Hol}({\cal HM}) &=& SU(2)\otimes {\cal H}\ \ \ \ \ {\rm for\ 
quaternionic\ manifold},
\cr
{\rm Hol}({\cal HM}) &=& {\bf 1}\otimes {\cal H}\ \ \ \ \ 
{\rm for\ HyperK\ddot{a}hler\ manifold},
\label{holo}
\end{eqnarray}
where ${\cal H}\subset Sp(2n_H,{\bf R})$.  We denote the flat indices that 
run in the fundamental representation of $SU(2)$ [$Sp(2n_H,{\bf R})$] as 
$i,j=1,2$ [$\alpha,\beta=1,...,2n_H$].  Then, the metric  
of the quaternionic manifold is expressed in terms of the vielbein 
1-form ${\cal U}^{i\alpha}={\cal U}^{i\alpha}_u(q)dq^u$ as:
\begin{equation}
h_{uv}={\cal U}^{i\alpha}_u{\cal U}^{j\beta}_v C_{\alpha\beta}\epsilon_{ij}, 
\label{quaviel}
\end{equation}
where $C_{\alpha\beta}=-C_{\beta\alpha}$ [$\epsilon_{ij}=-\epsilon_{ji}$] 
is the flat $Sp(2n_H)$ [$Sp(2)\sim SU(2)$] invariant metric.  
The vielbein ${\cal U}^{i\alpha}$ is covariantly constant with respect to 
the $SU(2)$-connection $\omega^x$ and some $Sp(2n_H,{\bf R})$ Lie 
algebra valued connection $\Delta^{\alpha\beta}=\Delta^{\beta\alpha}$:
\begin{equation}
\nabla {\cal U}^{i\alpha}\equiv d{\cal U}^{i\alpha} +{1\over 2}\omega^x
(\epsilon\sigma_x\epsilon^{-1})^i_j\wedge{\cal U}^{j\alpha} 
+\Delta^{\alpha\beta}\wedge {\cal U}^{i\gamma}C_{\beta\gamma} =0, 
\label{quvielconst}
\end{equation}
where $\sigma^x$ ($x=1,2,3$) are the Pauli spin matrices. 
Also, ${\cal U}^{i\alpha}$ satisfies the reality condition:
\begin{equation}
{\cal U}_{i\alpha}\equiv ({\cal U}^{i\alpha})^* = 
\epsilon_{ij}C_{\alpha\beta}{\cal U}^{j\beta}. 
\label{qureal}
\end{equation}
The curvature 2-form $\Omega^x$ forms the representation of the quaternionic 
algebra: 
\begin{equation}
h^{st}\Omega^x_{us}\Omega^y_{tw}=-\lambda^2\delta^{xy}h_{uw}+
\lambda\epsilon^{xyz}\Omega^z_{uw}, 
\label{curep}
\end{equation}
and can be written in terms of the vielbein ${\cal U}^{i\alpha}$ as
\begin{equation}
\Omega^x =i\lambda C_{\alpha\beta}(\sigma^x\epsilon^{-1})_{ij} 
{\cal U}^{\alpha i}\wedge{\cal U}^{\beta j}. 
\label{curviel}
\end{equation}

\subsection{Target Space and Strong-Weak Coupling Dualities of 
Heterotic String on a Torus}\label{dualn4}

\subsubsection{Effective Field Theory of Heterotic 
String}\label{dualn4d10}

The effective field theory of massless states in heterotic string 
is $D=10$ $N=1$ supergravity coupled to $N=1$ super-Maxwell 
theory \cite{CHA185,BERddv195,CHAm120}.  
The massless bosonic fields of heterotic string   
at a generic point of Narain lattice \cite{NAR169,NARsw279}
are metric $\hat{G}_{MN}$, 2-form field $\hat{B}_{MN}$, 
gauge fields $\hat{A}^I_M$ of $U(1)^{16}$ and  
dilaton field $\Phi$, where $0 \le M,N \le 9$ and $1 \le I \le 16$.  
The field strengths of $\hat{A}^I_M$ and $\hat{B}_{MN}$ are
defined as $\hat{F}^I_{MN} = \partial_M \hat{A}^I_N - \partial_N
\hat{A}^I_M$ and $\hat{H}_{MNP} = \partial_M \hat{B}_{NP} -
{1\over 2}\hat{A}^I_M \hat{F}^I_{NP} + {\rm cyc.\ perms.}$, respectively.
The $D=10$ effective action \cite{CHA185,BERddv195,CHAm120,CALmpf}
of these massless bosonic modes is
\begin{equation}
{\cal L}={1\over{16\pi G_{10}}}\sqrt{-\hat{G}}\,[{\cal R}_{\hat{G}} 
+ \hat{G}^{MN}\partial_M \Phi \partial_N \Phi - {\textstyle {1\over {12}}}
\hat{H}_{MNP}\hat{H}^{MNP}-{\textstyle{1\over 4}}\hat{F}^I_{MN}
\hat{F}^{I\,MN}],
\label{10daction}
\end{equation}
where $\hat G \equiv {\rm det}\,\hat{G}_{MN}$, ${\cal R}_{\hat G}$ is
the Ricci scalar of $\hat{G}_{MN}$, and $G_{10}$ is the 
$D=10$ gravitational constant
\footnote{In this section, we fix the $D=10$ gravitational 
constant to be $G_{10}=8\pi^2$.}.  
We choose the mostly positive signature
convention $(-++\cdots +)$ for the metric $\hat{G}_{MN}$.
For the spacetime vector index convention, the characters ($A,B,...$) 
and ($M,N,...$) denote flat and curved indices, respectively.

The supersymmetry transformations of the fermionic 
fields, i.e. gravitino $\psi_{M}$, dilatino $\lambda$ and 
gaugini $\chi^I$, are
\begin{eqnarray}
\delta\psi_M &=& \nabla_M \varepsilon - {1\over 8}
\hat{H}_{MNP}\hat{\Gamma}^{NP}\varepsilon, 
\nonumber \\
\delta \lambda &=& (\hat{\Gamma}^M \partial_M \hat{\Phi})\varepsilon 
-{1\over 6}\hat{H}_{MNP}\hat{\Gamma}^{MNP}\varepsilon, 
\nonumber \\
\delta \chi^I &=& \hat{F}^I_{MN}\hat{\Gamma}^{MN}\varepsilon ,
\label{hetsusy}
\end{eqnarray}
where $\nabla_M \varepsilon = \partial_M \varepsilon + {1\over 4}
\Omega_{MAB}\hat{\Gamma}^{AB}\varepsilon$ is the gravitational covariant 
derivative on a spinor $\varepsilon$.  Here, $\Omega_{MAB}$ is 
the spin-connection defined in terms of a Zehnbein $\hat{E}^A_{\Lambda}$ 
(defined as $\hat{E}^A_{M}\eta_{AB}\hat{E}^B_{N}=\hat{G}_{MN}$): 
$\Omega_{ABC} \equiv -\tilde{\Omega}_{AB,C} + \tilde{\Omega}_{BC,A} - 
\tilde{\Omega}_{CA,B}$, where $\tilde{\Omega}_{AB,C} \equiv 
\hat{E}^{M}_{[A}\hat{E}^{N}_{B]}\partial_{N}\hat{E}_{M C}$, and curved indices 
are obtained by contracting with Zehnbein.
$\hat{\Gamma}^A$ are gamma matrices of $SO(1,9)$ Clifford algebra 
$\{\hat{\Gamma}^A,\hat{\Gamma}^B\}=2\eta^{AB}$ (those with 
several indices are defined as the antisymmetrized products of 
gamma matrices, e.g. $\hat{\Gamma}^{AB} \equiv 
{1\over 2}(\hat{\Gamma}^A\hat{\Gamma}^B-\hat{\Gamma}^B\hat{\Gamma}^A)$).

\subsubsection{Kaluza-Klein Reduction and Moduli Space}\label{dualn4kk}

The effective field theory of massless bosonic fields in heterotic
string on a Narain torus \cite{NAR169,NARsw279} at a generic point of 
moduli space is obtained by compactifying $D=10$ effective field 
discussed in the previous section on $T^{10-D}$ \cite{MAHs390,SENint}.

Before we discuss the compactification Ansatz, we fix our notation 
for indices.  General indices running over $D=10$ are denoted by 
upper-case letters ($A,B,...; M,N,...$).  The lower-case Greek letters 
($\alpha,...,\beta; \mu,\nu,...$) are for $D<10$ spacetime 
coordinates and the lower-case Latin letters ($a,b,...; m,n,...$) are for 
the internal coordinates.  The flat indices are denoted by the letters 
in the beginning of alphabets ($A,B,...; \alpha,\beta,...; a,b,...$) 
and curved indices are denoted by the letters at the latter parts of 
alphabets ($M,N,...; \mu,\nu,...; m,n,...$).

The compactification \cite{KAL,KLE,CHOf,SCHs82,SCHs153,MAHs390,DUFnp130,DUF046}
on  $T^{10-D}$ is achieved by choosing the following Abelian Kaluza-Klein (KK) 
Ansatz for the $D=10$ metric
\begin{equation}
\hat{G}_{MN}=\left(\matrix{e^{a\varphi}
g_{{\mu}{\nu}}+
G_{{m}{n}}A^{(1)\,m}_{{\mu}}A^{(1)\,n}_{{\nu}} & A^{(1)\,m}_{{\mu}}
G_{{m}{n}}  \cr  A^{(1)\,n}_{{\nu}}G_{{m}{n}} & G_{{m}{n}}}\right),
\label{4dkk}
\end{equation}
where $A^{(1)\,m}_{\mu}$ ($\mu = 0,1,...,D-1$; $m=1,...,10-D$) are 
KK $U(1)$ gauge fields, $\varphi \equiv \hat{\Phi} - {1\over 2}{\rm ln}\,
{\rm det}\, G_{mn}$ is the $D$-dimensional dilaton and $a\equiv 
{2\over{D-2}}$.  Then, the affective action is specified by the 
following massless bosonic fields: the (Einstein-frame) graviton 
$g_{\mu\nu}$, the dilaton $\varphi$, $(36-2D)$ $U(1)$ gauge fields 
${\cal A}^i_{\mu}\equiv (A^{(1)\,m}_{\mu},A^{(2)}_{\mu\,m},
A^{(3)\,I}_{\mu})$ defined as $A^{(2)}_{\mu\,m} \equiv \hat{B}_{\mu m}
+\hat{B}_{mn}A^{(1)\,n}_{\mu}+{1\over 2}\hat{A}^I_mA^{(3)\,I}_{\mu}$ and 
$A^{(3)\,I}_{\mu} \equiv \hat{A}^I_{\mu} - \hat{A}^I_m
A^{(1)\,m}_{\mu}$, the 2-form field $B_{\mu\nu}$ with the 
field strength $H_{\mu\nu\rho}= \partial_{\mu} B_{\nu\rho} - 
{1\over 2} {\cal A}^i_{\mu}L_{ij}{\cal F}^j_{\nu\rho} + {\rm cyc. perms.}$,
and the following symmetric $O(10-D,26-D)$
matrix of scalars (moduli) \cite{MAHs390,SENint}:
\begin{equation}
M=\left ( \matrix{G^{-1} & -G^{-1}C & -G^{-1}a^T \cr
-C^T G^{-1} & G + C^T G^{-1}C +a^T a & C^T G^{-1} 
a^T
+ a^T \cr -aG^{-1} & aG^{-1}C + a & I + aG^{-1}a^T}
\right ),
\label{modulthree}
\end{equation}
where $G \equiv [\hat{G}_{mn}]$, $C \equiv [{1\over 2}
\hat{A}^{(I)}_{{m}}\hat{A}^{(I)}_{n}+\hat{B}_{mn}]$ and
$a \equiv [\hat{A}^I_{{m}}]$ are defined in terms of the
internal parts of $D=10$ fields.  $M$ can be expressed
in terms of the following $O(10-D,26-D)$ matrix as $M = V^T V$
\cite{MAHs390}:
\begin{equation}
V =\left(\matrix{E^{-1} & -E^{-1}C & -E^{-1}a^T \cr
0 & E & 0 \cr 0 & a & I_{16}}\right),
\label{mviel}
\end{equation}
where $E \equiv [e^a_m]$, $C \equiv [{1\over 2}
\hat{A}^I_m \hat{A}^I_n + \hat{B}_{mn}]$ and $a \equiv [\hat{A}^I_m]$.
$V$ plays a role of a Vielbein in the $O(10-D,26-D)$ target space.
Note, $M$ parameterizes the quotient space
$O(10-D,26-D)/[O(10-D)\times O(26-D)]$ with dimensions $26-36D+D^2$.  
The dimensionality precisely matches the number of scalar fields in the 
matrix $M$: $(11-D)(10-D)/2$ scalars $\hat{G}_{mn}$, $(10-D)(9-D)/2$ scalars 
$\hat{B}_{mn}$, and $16\cdot (10-D)$ scalars $\hat{A}^I_m$.  

The resulting theory in $D<10$ corresponds to $26-D$ vector 
multiplets coupled to $D<10$, $N$-extended supergravity.  
The supergravity multiplet consists of graviton $g_{\mu\nu}$, 2-form 
potential $B_{\mu\nu}$, $10-D$ graviphotons $A^{(R)\,a}_{\mu}$ 
($a=1,...,10-D$), dilaton $\varphi$, gravitinos $\psi^{\alpha}_{\mu}$ 
($\alpha=1,...,N$) and dilatinos $\lambda^{\alpha}$.  
The field content in the $26-D$ vector multiplets is 
$26-D$ vector fields $A^{(L)\,I}_{\mu}$ ($I=1,...,26-D$), 
$(10-D)\times(26-D)$ scalars $\phi^{aI}$ parameterizing the coset 
$O(10-D,26-D)/[O(10-D)\times O(26-D)]$ and gauginos $\chi^{\alpha I}$.  
At the string level, the $10-D$ graviphotons originate from the right 
moving sector of the heterotic string and the $26-D$ photons in 
the vector multiplets originate from the left moving sector.  
In terms of the field strengths ${\cal F}^i_{\mu}$ of the 
$U(1)^{36-2D}$ gauge group, the graviphoton field strengths 
$F^{(R)\,a}_{\mu\nu}$ and matter photon field strengths 
$F^{(L)\,I}_{\mu\nu}$ are expressed as
\begin{equation}
F^{(R)}_{\mu\nu}=V_RL{\cal F}_{\mu\nu}, \ \ \ \ \ 
F^{(L)}_{\mu\nu}=V_LL{\cal F}_{\mu\nu},
\label{garvmatphorel}
\end{equation}
where $V=(\matrix{V_R& V_L})^T$ and the $O(10-D,26-D)$ invariant 
metric $L$ is defined in (\ref{4dL}).

Then, the effective $D<10$ action (in the Einstein 
frame) takes the form \cite{MAHs390,SENint}:
\begin{eqnarray}
{\cal L}&=&{1\over{16\pi G_D}}\sqrt{-g}[{\cal R}_g-{1\over (D-2)}g^{\mu\nu}
\partial_{\mu}\varphi\partial_{\nu}\varphi+{1\over 8}g^{\mu\nu}
{\rm Tr}(\partial_{\mu}ML\partial_{\nu}ML)
\cr
& &-{1\over{12}}e^{-2a\varphi}g^{\mu\mu^{\prime}}
g^{\nu\nu^{\prime}}g^{\rho\rho^{\prime}}H_{\mu\nu\rho}
H_{\mu^{\prime}\nu^{\prime}\rho^{\prime}}
-{1\over 4}e^{-a\varphi}g^{\mu\mu^{\prime}}g^{\nu\nu^{\prime}}
{\cal F}^{i}_{\mu\nu}(LML)_{ij}{\cal F}^{j}_{\mu^{\prime}\nu^{\prime}}],
\label{effaction}
\end{eqnarray}
where $g\equiv {\rm det}\,g_{\mu\nu}$, ${\cal R}_g$ is the Ricci
scalar of $g_{\mu\nu}$, and ${\cal F}^i_{\mu\nu} = \partial_{\mu}
{\cal A}^i_{\nu}-\partial_{\nu} {\cal A}^i_{\mu}$ are the
$U(1)^{36-2D}$ gauge field strengths.  Here, the $D<10$  
gravitational constant $G_D$ is defined in terms of the $D=10$  
one $G_{10}$ as $G_{10}=(2\pi\sqrt{\alpha^{\prime}})^{10-D}G_D$
\footnote{We choose $\alpha^{\prime}=1$ in most of cases.}, 
where $\sqrt{\alpha^{\prime}}$ is the radius of internal circles.  
The Einstein-frame metric $g_{\mu\nu}$ is related to the 
string-frame metric $g^{str}_{\mu\nu}$ through Weyl-rescaling  
$g^{str}_{\mu\nu}=e^{\alpha\varphi}g_{\mu\nu}$.  
In terms of the graviphotons $A^{(R)\,a}_{\mu}$ and photons 
$A^{(R)\,I}_{\mu}$ in vector multiplets, the gauge kinetic terms 
take the form: ${\cal F}_{\mu\nu}(LML){\cal F}^{\mu\nu}=
F^{(R)\,T}_{\mu\nu}F^{(R)\,\mu\nu}+F^{(L)\,T}_{\mu\nu}F^{(L)\,\mu\nu}$, 
due to the relation $M=V^TV=V^T_RV_R+V^T_LV_L$.  

In particular for $D=4$, the supersymmetry transformations (\ref{hetsusy}) 
of fermionic fields in the bosonic background take the following 
simplified form in terms of $D=4$ fields \cite{DUFlr}:
\begin{eqnarray}
\delta \psi_\mu &=& [\nabla_\mu-{1\over4}i\gamma^5 {\partial_\mu S_1\over S_2}
-{1\over 8\sqrt{2}}\sqrt{S_2} F^{(R)a}_{\alpha\beta}\gamma^{\alpha\beta}
\gamma_\mu \Gamma^a+{1\over4} Q_\mu^{ab}\Gamma^{ab}]\epsilon,
\nonumber\\
\delta\lambda&=&{1\over4\sqrt{2}}[\gamma^\mu
{\partial_\mu(S_2-i\gamma^5S_1)\over S_2}-{1\over2\sqrt{2}}\sqrt{S_2}
F_{\mu\nu}^{(R)a}\gamma^{\mu\nu}\Gamma^a]\epsilon,
\nonumber\\
\delta\chi&=&{1\over\sqrt{2}}[\gamma^\mu V_L^{\vphantom{T}}
L\partial_\mu V_R^T\cdot\Gamma
-{1\over2\sqrt{2}}\sqrt{S_2}F_{\mu\nu}^{(L)}\gamma^{\mu\nu}]\epsilon,
\label{4dimhetsusytr}
\end{eqnarray}
where $Q_\mu^{ab} = (V_R^{\vphantom{T}}L\partial_\mu V_R^T)^{ab}$ is the
composite $SO(6)$ connection and $S$ is the axion-dilaton field defined 
in section \ref{dualn4ts}.  Here, the $D=10$ gamma matrices 
$\Gamma^A$ ($A=0,1,...,9$) are decomposed into the $D<10$ 
spacetime parts $\gamma^{\mu}$ ($\mu=0,1,...,D-1$) and the 
internal space parts $\Gamma^a$ ($a=1,...,10-D$).

\subsubsection{Duality Symmetries}\label{dualn4ts}

The $D<10$ effective action (\ref{effaction}) is
invariant under the $O(10-D,26-D)$ transformations ($T$-duality)
\cite{MAHs390,SENint}:
\begin{equation}
M \to \Omega M \Omega^T ,\ \  {\cal A}^i_{\mu} \to \Omega_{ij}
{\cal A}^j_{\mu}, \ \  g_{\mu\nu} \to g_{\mu\nu}, \ \ 
\varphi \to \varphi, \ \ B_{\mu\nu} \to B_{\mu\nu},
\label{tdual}
\end{equation}
where $\Omega\in O(10-D,26-D)$, i.e. with the property:
\begin{equation}
\Omega^T L \Omega = L ,\ \ \ L =\left ( \matrix{0 & I_{10-D}& 0\cr
I_{10-D} & 0& 0 \cr 0 & 0 &  I_{26-D}} \right ),
\label{4dL}
\end{equation}
where $I_n$ denotes the $n\times n$ identity matrix.

When electric/magnetic charges are quantized according to the 
Dirac-Schwinger-Zwanzinger-Witten (DSZW) quantization rule 
\cite{DIR74,SCH66,SCH68,ZWA1,ZWA2,WUy14,WUy107,WIT86}, the quantized, 
conserved electric $\vec{\alpha}$ and magnetic $\vec{\beta}$ charge vectors 
live on the even, self-dual, Lorentzian lattice $\Lambda$ 
\cite{KALo48,SEN303}. 
The subset of $O(10-D,26-D,{\bf R})$ symmetry that preserves the lattice 
$\Lambda$ is $O(10-D,26-D,{\bf Z})$, the so-called $T$-duality group of 
heterotic string on a torus.  $T$-duality symmetry is a 
perturbative symmetry, which is proven to be exact to order by order in 
string coupling \cite{GIVrt409}.  Under the $T$-duality, 
the charge lattice vectors transform as 
\begin{equation}
\vec{\alpha} \to L \Omega L \vec{\alpha}, 
\ \ \ \ \ \ 
\vec{\beta} \to L \Omega L \vec{\beta}.
\label{tuallat}
\end{equation} 

In addition, the effective field theory has an on-shell symmetry 
called strong-weak coupling duality ($S$-duality) 
\cite{FONil249,SCH125,SCHs312,SCHs411,SEN93,SEN303,SEN038,SEN057,SENint}.  
The equations of motion for (\ref{effaction}) 
are invariant under the $SO(1,1)$ [$SL(2,{\bf R})$] transformation 
for $5\leq D\leq 10$ [$D=4$].  
Such transformations mix electric and magnetic charges, while transforming 
the dilaton in a nontrivial way.  When the DSZW quantization is taken into 
account, such duality groups break down to integer-valued subgroups 
${\bf Z}_2$ and $SL(2,{\bf Z})$ for $5\leq D\leq 10$ and $D=4$, 
respectively.  These are the conjectured $S$-dualities in heterotic string.  

As an example, we discuss the $D=4$ $SL(2,{\bf R})$ symmetry.
$D=4$ case is special for the following reason.  
Since the 2-form field strengths are self-dual under the 
Hodge-duality, $U(1)$ gauge fields obey electric-magnetic 
duality transformations, which leave the Maxwell's equations and 
Bianchi identities invariant.  
Also, the field strength $H_{\mu\nu\rho}$ of the 2-form potential  
$B_{\mu\nu}$ is Hodge-dualized to a pseudo-scalar $\Psi$ (axion):  
$H^{\mu\nu\rho}=-{{e^{2\varphi}}\over \sqrt{-g}}\varepsilon^{\mu
\nu\rho\sigma}\partial_{\sigma}\Psi$, forming a complex scalar 
$S=\Psi +ie^{-\varphi}$ with dilaton $\varphi$.  
The (Einstein-frame) $D=4$ theory has the on-shell symmetry under 
the $SL(2,{\bf R})$ transformations ($S$-duality) 
\cite{CREsf,CREs74,SENint}:
\begin{eqnarray}
S &\to& { {{aS+b}\over{cS+d}}},\ \ M\to M ,\ \ g_{\mu\nu}\to
g_{\mu\nu},
\cr
{\cal F}^i_{\mu\nu} &\to& (c\Psi + d)
{\cal F}^i_{\mu\nu} + ce^{-2\varphi} (ML)_{ij}
\star{\cal F}^j_{\mu\nu},
\label{sdual}
\end{eqnarray}
where  $\star{\cal F}^{i\,\mu\nu} = {1\over 2\sqrt{-g}}
\varepsilon^{\mu\nu\rho\sigma}{\cal F}^i_{\rho\sigma}$, and
$a,b,c,d \in {\bf R}$ satisfy $ad-bc=1$. 

The instanton effect breaks the $SL(2,{\bf R})$ symmetry
down to $SL(2,{\bf Z})$ \cite{SHAtw,SEN303}.  
The electric and magnetic ``lattice charge vectors''
\cite{SENint} $\vec{\alpha}$ and $\vec{\beta}$ that live on an even, 
self-dual, Lorentzian lattice $\Lambda$ with signature $(6,22)$ are 
given in terms of the physical electric and magnetic charges 
$\vec{Q}$ and $\vec{P}$ (defined as ${\cal F}^i_{tr}\approx{Q^i\over r^2}$ 
and $\star{\cal F}^i_{tr}\approx{P^i\over r^2}$) as 
$\vec{\beta} \equiv L\vec{P}$ and $\vec{\alpha} \equiv
e^{-\phi_{\infty}}M^{-1}_{\infty}\vec{Q}-\Psi_{\infty}
\vec{\beta}$ \cite{SEN303}. Under the $S$-duality, $\vec{\alpha}$ and 
$\vec{\beta}$ transform as \cite{KALo48,SEN303}
\begin{equation}
\left(\matrix{\vec{\alpha} \cr \vec{\beta}}\right) \to 
\left(\matrix{a&-b\cr -c&d}\right) 
\left(\matrix{\vec{\alpha} \cr \vec{\beta}}\right), 
\label{latsdual}
\end{equation}
where $a,b,c,d$ are integers satisfying $ad-bc=1$.

\subsubsection{Solution Generating Symmetries}\label{dualn4sol}

For stationary solutions, which have the Killing time coordinate, 
one can further perform Abelian KK compactification of 
the time coordinate on $S^1$.  The $T$-duality transformation of such 
$(D-1)$-dimensional action can be applied to a known $D$-dimensional 
solution to generate new types of solutions in $D$ dimensions with 
different spacetime structure 
\cite{EHL,HAR9,MAI10,DOBm,LEU,CLE18,CLE118,CHOd,BREmg,SEN271,SEN274,HASs375,SEN69,SEN440,JOHm50,GALk50}.  

The basic idea on solution generating symmetry is as follows. 
If the background configuration is time-independent, then under the 
(time-independent) general coordinate transformations,  
$G_{t\breve{\mu}}$ and $B_{t\breve{\mu}}$ transform as vectors, 
where $\breve{\mu}=1,...,D-1$.  In addition, $B_{t\breve{\mu}}$ transforms  
as a vector under the (time-independent) gauge transformation of  
the 2-form field.   So, one can add 2 new $U(1)$ gauge fields 
associate with $G_{t\breve{\mu}}$ and $B_{t\breve{\mu}}$ to the existing 
$D$-dimensional $36-2D$ $U(1)$ gauge fields, forming a new 
multiplet of vectors $\breve{\cal A}^{\breve{i}}_{\breve{\mu}}$ 
($\breve{i}=1,...,38-2D$) \cite{SEN440}.  
In addition, since $G_{tt}$ and ${\cal A}^i_t$ transform as scalars under 
the transformations mentioned above, the scalar matrix of moduli  
is enlarged to a $(38-2D)\times (38-2D)$ matrix \cite{SEN440}.  

Under the $T$-duality of the $(D-1)$-dimensional effective 
action, the $(t,t)$-component of the $D$-dimensional metric 
$g_{\mu\nu}$ mixes with scalars in the moduli matrix $M$ and the 
$t$-component of the $U(1)^{36-2D}$ gauge fields ${\cal A}^i_{\mu}$, and  
the $(t,\breve{\mu})$-components of $g_{\mu\nu}$ mix with the 
${\cal A}^i_{\mu}$ and the $(t,\breve{\mu})$ components of $B_{\mu\nu}$.  
So, unlike the $D$-dimensional $T$-duality transformation, which leaves the 
$D$-dimensional spacetime intact, the $(D-1)$-dimensional $T$-duality 
transformation can be applied to a known $D$-dimensional solution 
to generate new solutions with different spacetime structure.   
In particular, such transformations can be imposed on charge neutral solutions 
to generate electrically charged (under the $D$-dimensional $U(1)^{36-2D}$ 
gauge group) solutions: $36-2D$ $SO(1,1)$ boosts in the 
$(D-1)$-dimensional $T$-duality group generate electric charges of 
$U(1)^{36-2D}$ gauge field when acted on charge neutral solutions
\footnote{When acted on magnetically charged solutions, such transformations 
induce unphysical Taub-NUT charge \cite{SEN440}.  This is due to the 
singularity of Dirac monopole.}

The $(D-1)$-dimensional effective action is 
\cite{SEN434,SEN440,PEE456}:
\begin{eqnarray}
{\cal L}&=&\sqrt{\breve{g}}e^{-\breve{\varphi}}\left[{\cal R}_{\breve{g}}
+\breve{g}^{\breve{\mu}\breve{\nu}}\partial_{\breve{\mu}}\breve{\varphi}
\partial_{\breve{\nu}}\breve{\varphi}+{1\over 8}
\breve{g}^{\breve{\mu}\breve{\nu}}{\rm Tr}(\partial_{\breve{\mu}}
\breve{M}\breve{L}\partial_{\breve{\nu}}\breve{M}\breve{L})\right.
\cr
& &\ \ \ -\left.
{1\over {12}}\breve{g}^{\breve{\mu}\breve{\mu}^{\prime}}
\breve{g}^{\breve{\nu}\breve{\nu}^{\prime}}
\breve{g}^{\breve{\rho}\breve{\rho}^{\prime}}
\breve{H}_{\breve{\mu}\breve{\nu}\breve{\rho}}
\breve{H}_{\breve{\mu}^{\prime}\breve{\nu}^{\prime}\breve{\rho}^{\prime}} 
-\breve{g}^{\breve{\mu}\breve{\mu}^{\prime}}
\breve{g}^{\breve{\nu}\breve{\nu}^{\prime}}
\breve{\cal F}^{\breve{i}}_{\breve{\mu}\breve{\nu}}
(\breve{L}\breve{M}\breve{L})_{\breve{i}\breve{j}}
\breve{\cal F}^{\breve{j}}_{\breve{\mu}^{\prime}
\breve{\nu}^{\prime}}\right],
\label{genlag}
\end{eqnarray}
where $U(1)$ gauge fields $\breve{\cal A}^{\breve{i}}_{\breve{\mu}}$ 
($\breve{i}=1,...,38-2D$), dilaton $\breve{\varphi}$, 2-form field 
$\breve{B}_{\breve{\mu}\breve{\mu}}$ and the metric $\breve{g}_{\breve{\mu}
\breve{\nu}}$ are defined as
\begin{eqnarray}
\breve{\cal A}^i_{\breve{\mu}}&\equiv&{\cal A}^i_{\breve{\mu}}-
(g_{tt})^{-1}g_{t\breve{\mu}}{\cal A}^i_t ,
\ \ \ \ \ 1\leq i \leq 36-2D,\ 1\leq \breve{\mu}\leq D-1, 
\cr
\breve{\cal A}^{37-2D}_{\breve{\mu}}&\equiv&\textstyle{1\over 2}
(g_{tt})^{-1}g_{t\breve{\mu}},\ \ \
\breve{\cal A}^{38-2D}_{\breve{\mu}} \equiv \textstyle{1\over 2}
B_{t\breve{\mu}}+{\cal A}^i_tL_{ij}\breve{\cal A}^j_{\breve{\mu}}, 
\cr
\breve{\varphi}&\equiv&\varphi-\textstyle{1\over 2}{\rm ln}(-g_{tt}),\ \ \
\breve{g}_{\breve{\mu}\breve{\nu}}=g_{\breve{\mu}\breve{\nu}}-
(g_{tt})^{-1}g_{t\breve{\mu}}g_{t\breve{\nu}}, 
\cr
\breve{B}_{\breve{\mu}\breve{\nu}}&\equiv&B_{\breve{\mu}\breve{\nu}}+
(g_{tt})^{-1}(g_{t\breve{\mu}}{\cal A}^i_{\breve{\nu}}-
g_{t\breve{\nu}}{\cal A}^i_{\breve{\mu}})L_{ij}{\cal A}^j_t 
+\textstyle{1\over 2}(g_{tt})^{-1}(B_{t\breve{\mu}}g_{t\breve{\nu}}-
B_{t\breve{\nu}}g_{t\breve{\mu}}),
\label{lowfield}
\end{eqnarray}
and the symmetric $O(11-D,27-D)$  moduli matrix is given by
\begin{equation}
\breve{M}=\left(\matrix{M+4(g_{tt})^{-1}{\cal A}_t{\cal A}^T_t &
-2(g_{tt}){\cal A}_t & 2ML{\cal A}_t\cr 
& & + 4(g_{tt})^{-1}{\cal A}_t({\cal A}_tL{\cal A}_t)\cr 
-2(g_{tt})^{-1}{\cal A}^T_t & (g_{tt})^{-1} &
-2(g_{tt})^{-1}{\cal A}^T_t L{\cal A}_t \cr 2{\cal A}^T_tLM&
-2(g_{tt})^{-1}{\cal AA}^T_t L{\cal A}_t & g_{tt}+4{\cal A}^T_t LML
{\cal A}_t \cr +4(g_{tt})^{-1}{\cal A}^T_t({\cal A}^T_t L{\cal A}_t)
& &+ 4(g_{tt})^{-1}({\cal A}^T_t L{\cal A}_t)^2}\right).
\label{lowmod}
\end{equation}
Here, $\breve{L}\equiv \left(\matrix{L&0&0\cr0&0&1\cr0&1&0}\right)$ is
an $O(11-D,27-D)$ invariant matrix and ${\cal A}_t\equiv
[{\cal A}^i_t]$.

This action has invariance under the $O(11-D,27-D)$ $T$-duality 
\cite{SEN440,PEE456}:
\begin{equation}
\breve{M}\to \breve{\Omega}\breve{M}\breve{\Omega}^T, \ \
\breve{\cal A}^{\breve{i}}_{\breve{\mu}} \to 
\breve{\Omega}_{\breve{i}\breve{j}}\breve{\cal A}^{\breve{j}}_{\breve{\mu}}, 
\ \ \breve{\varphi}\to \breve{\varphi}, \ \
\breve{g}_{\breve{\mu}\breve{\nu}}\to \breve{g}_{\breve{\mu}\breve{\nu}}, 
\ \ \breve{B}_{\breve{\mu}\breve{\nu}}\to \breve{B}_{\breve{\mu}\breve{\nu}},
\label{lowtran}
\end{equation}
where $\breve{\Omega}\in O(11-D,27-D)$, i.e.
$\breve{\Omega}\breve{L}\breve{\Omega}^T=\breve{L}$.

$D=3$ case is special since a 2-form field strength is dual to a scalar.  
So, the scalar moduli space is enlarged from $O(7,23,{\bf Z})\backslash 
O(7,23)/[O(7)\times O(23)]$ to $O(8,24,{\bf Z})\backslash O(8,24)/[O(8)
\times O(24)]$ \cite{SEN434}. 
The $O(8,24,{\bf Z})$ duality symmetry are generated by 
$D=4$ $SL(2,{\bf Z})$ $S$-duality and $D=3$ $O(7,23,{\bf Z})$ $T$-duality  
\cite{SEN434}, just as $U$-duality in type-II string is generated by 
$S$-duality in $D=10$ and $T$-duality in $D<10$ 
\cite{HULt438}.  The $O(8,24,{\bf Z})$ symmetry transformation puts the 
axion-dilaton field on the same putting as the other moduli fields and, 
therefore, is non-perturbative in nature. 

Since we consider stationary solution, i.e. a solution 
with isometry in the time direction, we compactify the time coordinate 
as well as other internal space coordinates on $T^7$ to obtain 
$D=3$ effective action ((\ref{effaction}) with $D=3$ and now 
$\bar{\mu}=r,\theta,\phi$; $m=t,1,...,6$).  Such action has an off-shell 
symmetry under the $O(7,23)$ transformation (\ref{tdual}).  
The DSZW quantization condition breaks this 
symmetry to integer valued $O(7,23,{\bf Z})$ subset. 

In $D=3$, one can perform the following Hodge-duality 
transformations to trade the $D=3$ $U(1)$ fields 
${\cal A}^{i}_{\bar{\mu}}$ with a set of scalars 
$\psi \equiv [\psi^i]$ \cite{SEN434}:
\begin{equation}
\sqrt{-h}e^{-2\varphi}h^{\bar{\mu}\bar{\mu}^{\prime}}
h^{\bar{\nu}\bar{\nu}^{\prime}}(ML)_{ij}
{\cal F}^j_{\bar{\mu}^{\prime}\bar{\nu}^{\prime}} =
\epsilon^{\bar{\mu}\bar{\nu}\bar{\rho}}\partial_{\bar{\rho}}\psi^{i}, 
\label{sen3ddual}
\end{equation}
where $h_{\bar{\mu}\bar{\nu}}$ is the $D=3$ space metric and 
$i,j=1,...,30$.  Here, $M$ is a symmetric $O(7,23)$ matrix defined 
as in (\ref{modulthree}) but now the time-component is included, and 
$L$ is an $O(7,23)$ invariant metric defined in (\ref{4dL}).    
So, the $D=3$ effective theory is described only  
in terms of graviton and scalars.  The $D=3$ effective 
action has the form \cite{SEN434}:
\begin{equation}
{\cal L} = {\textstyle {1\over 4}}\sqrt{-h}\,[{\cal R}_h +
{\textstyle {1\over 8}}h^{\bar{\mu}\bar{\nu}}{\rm Tr}
(\partial_{\bar{\mu}}{\cal M}{\bf L}
\partial_{\bar{\nu}}{\cal M}{\bf L})],
\label{3daction}
\end{equation}
where $h\equiv {\rm det}\,h_{\bar{\mu}\bar{\nu}}$, ${\cal R}_h$
is the Ricci scalar of $h_{\bar{\mu}\bar{\nu}}$.  ${\cal M}$ 
is a symmetric $O(8,24)$ matrix of $D=3$ scalars defined as 
\cite{SEN434}
\begin{equation}
{\cal M} = \left(\matrix{M-e^{2\varphi}\psi\psi^T &e^{2\varphi}\psi &
ML\psi
\cr & & -{1\over 2}e^{2\varphi}\psi(\psi^TL \psi)\cr 
e^{2\varphi}\psi^T & -e^{2\varphi} & {1\over 2}e^{2\varphi}\psi^T L\psi 
\cr
\psi^T LM &{1\over 2}e^{2\varphi}\psi^T L\psi & -e^{-2\varphi}
+ \psi^T \bar{L}\bar{M}\bar{L}\psi\cr 
-{1\over 2}e^{2\varphi}\psi^T(\psi^TL\psi)& &
-{1\over 4}e^{2\varphi}(\psi^TL\psi)^2}\right).
\label{modultwo}
\end{equation}
The action is manifestly invariant under the $O(8,24)$
transformations \cite{SEN434}:
\begin{equation}
{\cal M} \to {\bf \Omega} {\cal M} {\bf \Omega}^T, \ \ \ \
h_{\bar{\mu}\bar{\nu}} \to h_{\bar{\mu}\bar{\nu}},
\label{o824}
\end{equation}
where ${\bf \Omega} \in O(8,24)$, i.e.
\begin{equation}
{\bf \Omega}{\bf L}{\bf\Omega}^T={\bf L } , \ \  \
 {\bf L} = \left (\matrix{\bar{L} & 0 & 0 \cr 0 & 0 & 1
\cr 0 & 1 & 0}\right )\  .
\label{3dbfo}
\end{equation}
When electric and magnetic charges are quantized according to the 
DSZW quantization condition, the 
$O(8,24)$ is broken down to $O(8,24,{\bf Z})$.  
Since $O(8,24,{\bf Z})$ is generated by the conjectured $S$-duality in 
$D=4$ and $T$-duality in $D=3$ (which is proven to 
hold order by order in string coupling), the establishing $O(8,24,{\bf Z})$ 
invariance of the full string theory is equivalent to proving the 
$S$-duality in $D=4$ \cite{SEN434}.

\subsection{String-String Duality in Six Dimensions}\label{dualsix}

Six dimensions is special in the duality of $(d-1)$-branes \cite{DUFkl}. 
A $(d-1)$-brane in $D$ dimensions is dual to a 
$(\tilde{d}-1)$-brane ($\tilde{d}\equiv D-d-2$) under the Hodge-dual 
transformation of field strengths. 
So, in particular the heterotic string (1-brane) in $D=6$ is dual to 
another string ($\tilde{d}=6-2-2=2$) \cite{DUF442}.  
In fact, it was found out by Duff et. al. \cite{DUFn,TOW283} 
that the type-IIA string compactified on $K3$ surface has the same moduli 
space as that of the heterotic string compactified on $T^4$, i.e. 
$O(4,20,{\bf Z})/O(4,20,{\bf R})\backslash [O(4,{\bf R})\times 
O(20,{\bf R})]$.  Based on these observations, it is conjectured 
\cite{WIT443} that the heterotic string on $T^4$ is dual to 
the Type-IIA string on $K3$ surface, the so-called string-string duality in 
$D=6$ \cite{WIT443,HULt438,DUF442,DUFfk,DUFlm,HARs449}.  

The effective action of heterotic string compactified on 
$T^4$ in the string-frame is \cite{WIT443,SEN450}
\begin{eqnarray}
S&=&{1\over{16\pi G_6}}\int d^6 x\sqrt{-G}e^{-\Phi}[{\cal R}_G G^{\bar{\mu}
\bar{\nu}}\partial_{\bar{\mu}}\Phi\partial_{\bar{\nu}}\Phi 
-{1\over {12}}G^{\bar{\mu}\bar{\mu}^{\prime}}
G^{\bar{\nu}\bar{\nu}^{\prime}}G^{\bar{\rho}\bar{\rho}^{\prime}}
H_{\bar{\mu}\bar{\nu}\bar{\rho}}H_{\bar{\mu}^{\prime}\bar{\nu}^{\prime}
\bar{\rho}^{\prime}}
\cr
& &\ \ \ -G^{\bar{\mu}\bar{\mu}^{\prime}}G^{\bar{\nu}\bar{\nu}^{\prime}}
{\cal F}^i_{\bar{\mu}\bar{\nu}}(LML)_{ij}{\cal F}^j_{\bar{\mu}^{\prime}
\bar{\nu}^{\prime}}+{1\over 8}G^{\bar{\mu}\bar{\nu}}{\rm Tr}
(\partial_{\bar{\mu}}ML\partial_{\bar{\nu}}ML)],
\label{sixhet}
\end{eqnarray}
where $\bar{\mu}=0,...,5$, $i=1,...,24$, $L$ is an 
$O(4,20)$ invariant metric and $M$ is an symmetric $O(4,20)$ matrix, 
i.e. $M^T=M$ and $MLM^T=L$.  
(Definitions of $D=6$ fields in terms of the $D=10$ fields  
are given in section \ref{dualn4kk}.)  
The field strengths of the $U(1)$ gauge fields and the 2-form potential 
are
\begin{equation}
{\cal F}^i_{\bar{\mu}\bar{\nu}}=
\partial_{\bar{\mu}}{\cal A}^i_{\bar{\nu}}-
\partial_{\bar{\nu}}{\cal A}^i_{\bar{\mu}}, \ \ \ 
H_{\bar{\mu}\bar{\nu}\bar{\rho}}=
(\partial_{\bar{\mu}}B_{\bar{\nu}\bar{\rho}}
+2{\cal A}^i_{\bar{\mu}}L_{ij}
{\cal F}^j_{\bar{\nu}\bar{\rho}}) + {\rm cyc.\ perms.}. 
\label{hetfielst}
\end{equation}

The effective action for type-IIA string compactified on 
a $K3$ surface is \cite{WIT443,SEN450}
\begin{eqnarray}
S^{\prime}&=&{1\over{16\pi G_6}}\int d^6 x\left(\sqrt{-G^{\prime}}
\left[e^{-\Phi^{\prime}}\{{\cal R}_{G^{\prime}}+G^{\prime\,\bar{\mu}
\bar{\nu}}\partial_{\bar{\mu}}\Phi^{\prime}
\partial_{\bar{\nu}}\Phi^{\prime}\right.\right.
\cr
& &\ \ \ -{1\over{12}}G^{\prime\,\bar{\mu}\bar{\mu}^{\prime}}
G^{\prime\,\bar{\nu}\bar{\nu}^{\prime}}
G^{\prime\,\bar{\rho}\bar{\rho}^{\prime}}
H^{\prime}_{\bar{\mu}\bar{\nu}\bar{\rho}}
H^{\prime}_{\bar{\mu}^{\prime}\bar{\nu}^{\prime}
\bar{\rho}^{\prime}} 
+{1\over 8}G^{\prime\,\bar{\mu}\bar{\nu}}{\rm Tr}
(\partial_{\bar{\mu}}M^{\prime}L\partial_{\bar{\nu}}M^{\prime}
L)\}
\cr
& &\ \ \ \left.\left.-G^{\prime\,\bar{\mu}
\bar{\mu}^{\prime}}G^{\prime\,\bar{\nu}\bar{\nu}^{\prime}}
{\cal F}^{\prime\,i}_{\bar{\mu}\bar{\nu}}(LM^{\prime}L)_{ij}
{\cal F}^{\prime\,j}_{\bar{\mu}^{\prime}\bar{\nu}^{\prime}}\right]
-{1\over 4}\varepsilon^{\bar{\mu}\bar{\nu}\bar{\rho}
\bar{\sigma}\bar{\tau}\bar{\varepsilon}}
B^{\prime}_{\bar{\mu}\bar{\nu}}
{\cal F}^{\prime\,i}_{\bar{\rho}\bar{\sigma}}
L_{ij}{\cal F}^{\prime\,j}_{\bar{\tau}\bar{\varepsilon}}\right), 
\label{sixtype2}
\end{eqnarray}
where now the corresponding $D=6$ fields in type-IIA 
theory are denoted with primes.  
Note, the field strength of $B^{\prime}_{\bar{\mu}\bar{\nu}}$ is defined 
without Chern-Simmons term involving $U(1)$ gauge fields  
\begin{equation}
H^{\prime}_{\bar{\mu}\bar{\nu}\bar{\rho}}=
\partial_{\bar{\mu}}B^{\prime}_{\bar{\nu}\bar{\rho}}+{\rm cyc.\ perms.}. 
\label{flstrtyp2}
\end{equation}
The scalars and metric are defined similarly as those in 
the effective action (\ref{sixhet}) of heterotic string on $T^4$.  
But since the $K3$ surface does not have a continuous isometry, there are 
no KK $U(1)$ gauge fields, instead there are additional 
$U(1)$ gauge fields arising from the 1-form $A^{(10)}_M$ and the 
3-form $A^{(10)}_{MNP}$ in the R-R sector.  

These two string-frame effective actions are described by the same field 
degrees of freedom and have the same modular space.  So, they can be 
identified as the same action, provided we perform the following conformal 
transformation of the metric and the Hodge-duality transformation of the 
2-form field \cite{WIT443,SEN450}:
\begin{eqnarray}
\Phi^{\prime} = -\Phi, \ \ \ \  
G^{\prime}_{\bar{\mu}\bar{\nu}}&=&
e^{-\Phi}G_{\bar{\mu}\bar{\nu}}, \ \ \ \ 
M^{\prime} = M, \ \ \ \  A^{\prime(a)}_{\bar{\mu}}=
A^{(a)}_{\bar{\mu}}, 
\nonumber \\
\sqrt{-G}e^{-\Phi}H^{\bar{\mu}\bar{\nu}\bar{\rho}}&=&
{1\over 6}\varepsilon^{\bar{\mu}\bar{\nu}\bar{\rho}\bar{\sigma}\bar{\tau}
\bar{\varepsilon}}H^{\prime}_{\bar{\sigma}\bar{\tau}
\bar{\varepsilon}}. 
\label{ststdual}
\end{eqnarray}
Under the string-string duality, the dilaton 
changes its sign, indicating that the string coupling 
$\lambda=e^{-\langle\Phi\rangle}$ of the dual theory is inverse of the 
original theory.  So, a perturbative string state (weak string coupling 
$\lambda \ll 1$) in one theory is mapped to a non-perturbative string 
state (strong string coupling $\lambda \gg 1$) under the string-string 
duality.  For example, perturbative, singular, fundamental string 
in one theory is mapped to non-perturbative soliton string in the other 
theory \cite{SEN450}.

\subsubsection{String-String-String Triality}\label{dualsixtri}

Upon toroidal compactification to $D=4$, 
the $D=6$ string-string duality (\ref{ststdual}) interchanges 
the $D=4$ $S$-duality and the ($T^2$ part of) $T$-duality 
\cite{DUF442}, while the dilaton-axion field and the K\" ahler 
structure of $T^2$ are interchanged.  
So, the axion-dilaton field of the string-string duality 
transformed theory is given by the K\" ahler structure of the original 
theory.  Note, the $T^2$ part of the full $D=4$  
$T$-duality group, i.e. the $O(2,2,{\bf Z})\cong SL(2,{\bf Z})\times 
SL(2,{\bf Z})$ subgroup, 
contains not only the $SL(2,{\bf Z})$ factor parameterized by 
the K\" ahler structure of $T^2$ but also the other $SL(2,{\bf Z})$ 
factor parameterized by the complex structure of $T^2$ 
\cite{DIJvv88,SHAw320,DUFfk}.  
Namely, the effective $D=4$ theory has the 
$SL(2,{\bf Z})\times SL(2,{\bf Z})\times SL(2,{\bf Z})$ symmetry with each 
$SL(2,{\bf Z})$ factor respectively parameterized by the dilaton-axion 
field, the K\" ahler structure and the complex structure.  So, 
on the ground of symmetry argument, one expects another string theory 
whose axion-dilaton field is given  by the complex structure of the original 
theory \cite{DUFlr}.  In fact, mirror symmetry 
\cite{GREp338,HOSkt096,GREp014,ASPgm416,GRE155} exchanges the complex 
structure and the K\" ahler structures of an internal manifold. 
In particular, the mirror symmetry exchanges the type-IIA and type-IIB 
strings, and transforms heterotic string into itself.     
Thus, combining the $D=6$ string-string duality (which 
interchanges the dilaton-axion field and the K\" ahler structure) and 
the Mirror symmetry (which interchanges the K\" ahler structure and the 
complex structure), we establish the ``triality'' \cite{DUFlr} 
among the heterotic string on $K3\times T^2$ and the type-IIA and 
type-IIB strings on the Calabi-Yau-threefold.  

For the purpose of illustrating the triality among these three theories, 
we consider only the $T^2$ part and the NS-NS sector (which is common 
to the three theories) described by the following $D=6$ effective action:
\begin{equation}
{\cal L}={1\over{16\pi G_6}}\sqrt{-G}e^{-\Phi}[{\cal R}_G +
G^{\bar{\mu}\bar{\nu}}\partial_{\bar{\mu}}\Phi
\partial_{\bar{\nu}}
\Phi-{1\over{12}}G^{\bar{\mu}\bar{\mu}^{\prime}}
G^{\bar{\nu}\bar{\nu}^{\prime}}G^{\bar{\rho}\bar{\rho}^{\prime}}
H_{\bar{\mu}\bar{\nu}\bar{\rho}}H_{\bar{\mu}^{\prime}\bar{\nu}^{\prime}
\bar{\rho}^{\prime}}]. 
\label{trunsix}
\end{equation}
All the three theories with such truncation have the effective actions 
of this form. 
We label these three $D=4$ theories as $F_{XYZ}$, where $F=H,A,B$ 
respectively denoting the heterotic theory, the type-IIA theory and 
the type-IIB theory, and the subscripts $X,Y,Z$ respectively are the 
axion-dilaton field, the K\"ahler structure and the complex structure 
of the theory.  
We can take any of these three theories as the starting point, but for 
the purpose of definiteness we start with the heterotic string and impose 
the string-string duality and the Mirror symmetry to obtain all other 5 
theories.  

Compactification on $T^2$ is achieved by the following KK 
Ansatz for the $D=6$ metric:
\begin{equation}
G_{\bar{\mu}\bar{\nu}}=\left(\matrix{e^{\eta}g_{\mu\nu}+
A^m_{\mu}A^n_{\nu}
G_{mn}& A^m_{\mu}G_{mn}\cr A^n_{\nu}G_{mn} &  G_{mn}}\right), 
\label{kk64}
\end{equation}
where $\mu,\nu=0,...,3$ are the $D=4$ spacetime indices 
and $m,n=1,2$ are the internal space indices.
The $D=6$ 2-form field is decomposed as:
\begin{equation}
B_{\bar{\mu}\bar{\nu}}=
\left(\matrix{B_{\mu\nu}+{1\over 2}(A^m_{\mu}B_{m\nu}
-B_{\mu n}A^n_{\nu}) & B_{\mu n}+A^m_{\mu}B_{mn} \cr 
B_{m\nu}+B_{mn}A^n_{\nu} & B_{mn}}\right).
\label{twof64}
\end{equation}
Here, $\eta$, $g_{\mu\nu}$, $A^m_{\mu}$, $B_{\mu\nu}$ and $G_{mn}$ are 
respectively the $D=4$ dilaton (defined below), Einstein-frame 
metric, the KK $U(1)$ gauge field, the 2-form field and the 
internal metric.    

To express the $D=4$ effective action in an $SL(2,{\bf Z})\times 
SL(2,{\bf Z})$ $T$-duality invariant form, we parameterize 
the internal metric and the 2-form field as: 
\begin{equation}
G_{mn}=e^{\rho-\sigma}\left(\matrix{e^{-2\rho}+c^2 &-c\cr -c&1}\right), 
\ \ \ \ \ \ \ \ 
B_{mn}=b\epsilon_{mn}, 
\label{internal}
\end{equation}
and define the $D=4$ dilaton $\eta$ and axion $a$ as: 
\begin{equation}
e^{-\eta}=e^{-\Phi}\sqrt{{\rm det}\,G_{mn}}=e^{-(\Phi+\sigma)},\ \ \ \ 
\epsilon^{\mu\nu\rho\sigma}\partial_{\sigma}a=\sqrt{-g}e^{-\eta}
g^{\mu\sigma}g^{\nu\lambda}g^{\rho\tau}H_{\sigma\lambda\tau}, 
\label{axidil}
\end{equation}
where $H_{\mu\nu\rho}$ is the field strength of $B_{\mu\nu}$.  Then, 
from the above real scalars we define the following complex scalars 
\cite{DIJvv88}
\begin{eqnarray}
S&=&S_1+iS_2\equiv a+ie^{-\eta}, 
\cr
T&=&T_1+iT_2\equiv b+ie^{-\sigma}, 
\cr
U&=&U_1+iU_2\equiv c+ie^{-\rho}, 
\label{cpxscals}
\end{eqnarray}
which (within the framework of the heterotic string) are respectively 
the dilaton-axion field, the K\" ahler structure and the complex structure. 

Then, the final form of the $D=4$ effective action is 
\cite{DUFlr}:
\begin{eqnarray}
{\cal L}&=&{1\over{16\pi G_4}}\sqrt{-g}[{\cal R}_g
-{1\over 4}S_2g^{\mu\mu^{\prime}}g^{\nu\nu^{\prime}}
{\cal F}^T_{\mu\nu}({\cal M}_{T} \otimes {\cal M}_{U})
{\cal F}_{\mu^{\prime}\nu^{\prime}}
\nonumber \\
& &+{1\over 4}g^{\mu\nu}{\rm Tr}(\partial_{\mu}{\cal M}_{T}{\cal L}
\partial_{\nu}{\cal M}_{T}{\cal L})
+{1\over 4}g^{\mu\nu}{\rm Tr}(\partial_{\mu}{\cal M}_{U}{\cal L}
\partial_{\nu}{\cal M}_{U}{\cal L})
\nonumber \\
& &\left.-{1\over{2(S_2)^2}}g^{\mu\nu}\partial_{\mu}S\partial_{\mu}
\bar{S}\right],  
\label{hetslac}
\end{eqnarray}
where ${\cal M}_{T},{\cal M}_{U}\in SL(2,{\bf R})$ are  
defined as 
\begin{equation}
{\cal M}_{T} \equiv {1\over T_2}\left(\matrix{1 & T_1 \cr 
T_1 & |T|^2}\right), \ \ \ \ \ 
{\cal M}_{U} \equiv {1\over U_2}\left(\matrix{1 & U_1 \cr 
U_1 & |U|^2}\right),
\label{tumod}
\end{equation}
and the $U(1)$ gauge fields ${\cal A}^i_{\mu}$ ($i=1,...,4$) are given by 
${\cal A}^1_{\mu} = B_{4\mu}$, ${\cal A}^2_{\mu} = B_{5\mu}$, 
${\cal A}^3_{\mu} = A^{1}_{\mu}$, ${\cal A}^4_{\mu} = A^{2}_{\mu}$. 
Here, ${\cal L}$ is an $SL(2,{\bf Z})$ invariant metric. 
The action is manifestly invariant under the $SL(2,{\bf R})\times 
SL(2,{\bf R})$ $T$-duality:
\begin{equation}
{\cal M}_{T} \to \omega^T_{T}{\cal M}_{T}\omega_{T}, \ \ \ 
{\cal M}_{U} \to \omega^T_{U}{\cal M}_{U}\omega_{U}, \ \ \ 
{\cal F}_{\mu\nu} \to (\omega^{-1}_{T} \otimes \omega^{-1}_{U})
{\cal F}_{\mu\nu}, 
\label{tutran}
\end{equation}
where $\omega_{T,U}\in SL(2,{\bf R})$  
and the rest of the fields are inert.  
In addition, the theory has an on-shell $S$-duality symmetry:
\begin{equation}
S\to {{aS+b}\over{cS+d}}, \ \ \ \ \ 
\left(\matrix{{\cal F}^i_{\mu\nu}\cr \star{\cal F}^i_{\mu\nu}}\right)
\to \omega^{-1}_S \left(\matrix{{\cal F}^i_{\mu\nu}\cr 
\star{\cal F}^i_{\mu\nu}}\right); \ \ \ 
\omega=\left(\matrix{a&b\cr c&d}\right),  
\label{stran}
\end{equation}
where $a,b,c,d\in{\bf Z}$ satisfy $ad-bc=1$, and $\star{F}^i_{\mu\nu}$ 
is the Hodge-dual (defined from the action (\ref{hetslac})) of the field 
strength ${\cal F}^i_{\mu\nu}$.  

We denote the theory described by (\ref{hetslac}) as  
$H_{STU}$, meaning the heterotic theory with the dilaton-axion field, 
the K\" ahler structure and the complex structure given respectively by 
$S,T,U$ defined in (\ref{cpxscals}).  Under the Mirror symmetry, the 
K\" ahler structure and the complex structure are interchanged, and therefore 
we obtain $H_{SUT}$ theory, i.e. the heterotic string with the 
K\" ahler structure and the complex structure now respectively given by 
$U$ and $T$ defined in (\ref{cpxscals}); the effective action is 
(\ref{hetslac}) with $T$ and $U$ fields interchanged.  
We call $H_{STU}$ and $H_{SUT}$ as the $S$-strings, meaning the 
string theories with the dilaton-axion field given by $S$ defined in 
(\ref{cpxscals}).   

Under the $D=6$ string-string duality (\ref{ststdual}), 
the $H_{STU}$ is transformed to $A_{TSU}$. 
Under the Mirror symmetry, $A_{TSU}$ is transformed to 
$B_{TUS}$. 
So, the $A_{TSU}$ and $B_{TUS}$ are the $T$-strings. 

We apply string-string duality to $B_{TUS}$ to obtain $B_{UTS}$.
Under the Mirror symmetry, $B_{UTS}$ is transformed to $A_{UST}$.  
Therefore, we have the $U$-strings given by $B_{UTS}$ and $A_{UST}$. 

We comment on relation of the $D=4$ $S$-duality to the $D=6$ 
string-string duality.  Since effect of the $D=6$ string-string 
duality on the $D=4$ theory is to interchange the complex structure 
and the dilaton-axion field, the $SL(2,{\bf Z})$ subset $T$-duality 
of one theory accounts for the $S$-duality of the string-string duality 
transformed theory \cite{DUFk411,DUF442}. 
Namely, the large-small radius $T$-duality  
($R\to\alpha^{\prime}/R$) of one theory corresponds to the strong-weak 
coupling duality ($g^2/2\pi\to 2\pi/g^2$) of the dual theory.  
In terms of transformation of $U(1)$ gauge fields, one can understand 
this as follows \cite{DUFlr}.  Under the string-string duality, electric 
[magnetic] charges of 2-form $U(1)$ gauge fields of one theory is 
transformed to magnetic [electric] charges of 2-form $U(1)$ gauge 
fields of the dual theory, while those of KK $U(1)$ fields remain 
inert.  Since the $T$-duality interchanges KK $U(1)$ gauge fields and 
 2-form $U(1)$ gauge fields (associated with the same internal 
coordinates), under the combined action of the $D=6$ string-string duality 
and the $D=4$ $T$-duality, electric [magnetic] charges of KK 
$U(1)$ gauge fields and magnetic [electric] charges of 2-form 
$U(1)$ gauge fields are exchanged, which is exactly the $D=4$ $S$-duality.  
Thus, since the $T$-duality is proven to hold order 
by order in string coupling, proof of the conjectured $D=4$ 
$S$-duality amounts to proof of the $D=6$ string-string duality, and 
vice versa.  

Since the string-string duality interchanges the dilaton-axion field with 
the K\"ahler structure, the string coupling $g^2/2\pi$ of one theory 
is transformed to the worldsheet coupling $\alpha^{\prime}/R^2$ of the 
dual theory \cite{DUF442}. So, string quantum corrections controlled 
by the string coupling in one theory correspond to the stringy classical 
corrections ($\alpha^{\prime}$ corrections) controlled by the worldsheet 
coupling in the dual theory;  
the $\alpha^{\prime}$ [quantum] corrections in one theory can 
be understood in terms of the quantum [$\alpha^{\prime}$] corrections of 
the dual theory.

\subsection{$U$-duality and Eleven-Dimensional Supergravity}\label{dualu}

Hull and Townsend \cite{HULt438} conjectured that the type-II superstring 
theories on a torus has full symmetry of low-energy effective field theory, 
which is larger than the direct product of the $D=10$ $SL(2,{\bf Z})$ 
$S$-duality and the $O(10-d,10-d,{\bf Z})$ $T$-duality 
in $D=d<10$ \cite{GIVpr244}.  
For example, the effective action of type-II string on $T^6$, 
which is $D=4$, $N=8$ supergravity \cite{CREj80,CREj79}, has 
an on-shell $E_{7(7)}$ symmetry \cite{CREj79}, which contains 
$SL(2,{\bf R})\times O(6,6)$ as the maximal subgroup.
Hull and Townsend \cite{HULt438} conjectured that the subgroup $E_7({\bf Z})$ 
(broken due to the DSZW charge quantizations) extends to the full string 
theory as a new unified duality group, called $U$-duality.  
$U$-duality unifies the $S$ and $T$ dualities and mixes $\sigma$-model 
and string coupling constants.  

The discrete subgroup $E_7({\bf Z})$ is the intersection 
of the continuous $E_{7(7)}$ group and the discrete symplectic 
$Sp(28,{\bf Z})$ group, which transforms 28 $U(1)$ gauge fields 
of the effective theory linearly:  
$E_7({\bf Z})=E_{7(7)}\cap Sp(28,{\bf Z})$ \cite{HULt438}.  
Under $U$-duality, a set of $28\times 2$ electric and magnetic 
charges transform as a vector, and all the scalars in the theory mix 
among themselves.  Unlike other types of duality, which assigns a 
special role to the dilaton, under $U$-duality the dilaton is on 
the same footing as moduli.  
Thus, unlike $T$-duality which is perturbative in nature, 
$U$-duality, like $S$-duality \cite{SENint}, is non-perturbative in nature, 
so cannot be tested within perturbative spectrum of string theories.
  
We list the conjectured $U$-duality groups in various dimensions 
\cite{HULt438}.  Note, the duality group in higher dimensions is a 
subgroup of lower dimensional duality group, since it survives  
compactification (Cf. The duality group does not act 
on the Einstein-frame metric.).  The $SO(10-d,10-d,{\bf Z})$ $T$-duality 
in $D=d<10$ and the $SL(2,{\bf Z})$ $S$-duality in $D=10$ are unified 
to $U$-duality for $d<8$.  The $U$-duality groups are, in the descending 
order starting from $D=7$: $SO(5,5,{\bf Z})$, 
$SL(5,{\bf Z})$, $E_{6(6)}({\bf Z})$, $E_{7(7)}({\bf Z})$, $E_{8(8)}
({\bf Z})$, $E_{9(9)}({\bf Z})$, $E_{10(10)}({\bf Z})$.  These 
$U$-duality groups in $D=d$ are generated by the $T$-duality in 
$D=d$ and $n$ numbers of $U$-duality groups in $D=d+1$  
\cite{HULt438}, where $n$ is the possible numbers of ways in which 
one can compactify from $D=10$ to $D=d+1$ and then down to $D=d$.  

In comparison, the heterotic string on $T^{10-d}$ with $d>3$ maintains  
the duality group in the form ($T$-duality group) $\times$ ($S$-duality 
group), i.e. $SO(10-d,26-d,{\bf Z})\times {\bf Z}_2$ for $10\geq d\geq 5$ or 
$SO(10-d,26-d,{\bf Z})\times SL(2,{\bf Z})$ for $d=4$.  
For $d=3$ \cite{SEN434} and $d=2$ \cite{SEN447}, the $T$- and $S$-dualities 
are unified to $U$-duality given by $SO(8,24,{\bf Z})$ and $SO(8,24)^{(1)}
({\bf Z})$, respectively.  

As we will see in section \ref{entstr}, BPS electric states are within 
perturbative spectrum of heterotic string \cite{DUFr};   
all the 28 electric charges in the heterotic string on $T^6$ are related 
through the ``perturbative'' $O(6,22,{\bf Z})$ $T$-duality.  
Also, there is non-perturbative spectrum carrying the remaining 28 magnetic 
charges, related to perturbative spectrum via the ${\bf Z}_2$ subset 
of the ``non-perturbative'' $SL(2,{\bf Z})$ $S$-duality 
\cite{SEN038,SCHs312,ORT47}.  These magnetic charges are carried by 
solitons.   
The mass of a state in heterotic string on $T^6$ carrying electric [magnetic] 
charges of the $U(1)^{28}$ gauge group behaves as $\sim 1$ [$\sim 1/g^2_s$] 
in the string frame, as expected for a fundamental string [a soliton].  

For the type-II string, only 12 of the 28 electric charges 
couple to perturbative string states, since the remaining 16 
electric charges are R-R charges, which 
cannot be coupled to perturbative string states.   The ${\bf Z}_2$ subgroup 
of the $SL(2,{\bf Z})$ $S$-duality group relates these perturbative states to 
solitonic states carrying 12 magnetic charges of the same 12 $U(1)$ gauge 
fields.  The remaining 16 electric and 16 magnetic charges of the 
$U(1)^{16}$ gauge group are carried by another type of non-perturbative 
states, whose mass behaves as $\sim 1/g_s$.   Thus, under the 
($T$-duality) $\times$ ($S$-duality) subgroup, 
i.e. $SO(6,6,{\bf Z})\times SL(2,{\bf Z}) \subset E_7({\bf Z})$, the 
fundamental representation $\bf 56$ (representing 28 electric and 28 
magnetic charges of the $D=4$ $U(1)^{28}$ gauge group) is 
decomposed as ${\bf 56}=({\bf 12},{\bf 2}) \times ({\bf 32},{\bf 1})$.  
Here, the first factor $({\bf 12},{\bf 2})$ corresponds to the 12 numbers 
of $SL(2,{\bf Z})$ doublets of perturbative and solitonic states in the 
NS-NS sector and the second factor $({\bf 32},{\bf 1})$ denotes the 
remaining non-perturbative states, which are singlets under the 
$SL(2,{\bf Z})$ group and carries 16 electric and 16 magnetic charges of 
16 $U(1)$ gauge fields in the R-R sector.
 
Since the $O(6,6,{\bf Z})$ $T$-duality group of the type-II string 
on $T^6$ mixes only NS-NS charges among themselves, it is not inconsistent 
that string states carry only NS-NS charges.  However, it is not consistent 
with the conjectured $U$-duality, since $U$-duality puts all the 
$28\times 2$ electric and magnetic charges of $D=4$ $U(1)^{28}$ gauge group 
on the same putting.  In addition, as we saw in the decomposition 
of the $\bf 56$ representation, the $U$-duality requires 
existence of additional $16+16$ electric and magnetic charges in the 
RR-sector that transform irreducibly under the $O(6,6,{\bf Z})$ $T$-duality 
group.  Hence, one leads to the conclusion that all the RR charges, 
which cannot be carried by perturbative string states or solitons, 
should be carried by another type of non-perturbative states.  The 
low-energy or long-distance description of these non-perturbative string 
states is R-R $p$-branes or black holes.  
In the original work by Hull and Townsend \cite{HULt438}, they show that 
all the R-R charged black holes can arise from extreme $p$-branes of 
the $D=10$ effective supergravity via dimensional reduction.  

In \cite{POL75}, Polchinski shows that the states carrying R-R charges 
can be realized within string theories without introducing exotic extended 
objects like $p$-branes.  Such objects are $D$-branes \cite{DAIlp,LEI}, 
boundaries to which the ends of open strings (with Dirichlet boundary 
condition) are attached.  $D$-branes carry one unit of R-R charges.  
$D$-branes are dynamic objects and the open string states describe their 
fluctuations.  In the strong coupling limit, $D$-branes become black holes.  
Such identification made it possible to give precise statistical explanation 
of  black hole entropy.  
In chapter \ref{dbr}, we will summarize aspects of $D$-branes and 
the recent development in $D$-brane calculation of black hole entropy.   

In the following sections, we discuss the $S$-duality of the 
type-IIB string and the $T$-duality of type-II string on a torus.  
The $U$-duality of type-II string on a torus is generated by these 
two dualities.  In particular, starting from generating black holes 
of type-II theories on a torus, one obtains black holes with the 
general charge configuration by applying the $S$-duality 
and the $T$-duality transformations.  For $p$-branes in type-II 
theories, one can generate $p$-branes of the general charge configuration 
by first imposing $SO(1,1)$ boost on a charge neutral solution to 
induce KK electric charge and then applying the $T$- and the $S$-duality 
transformations and/or another $SO(1,1)$ boosts sequentially.  

\subsection{$S$-Duality of Type-IIB String}\label{dualus}

The type-IIB string \cite{GREs109,SCH226} has the $SL(2,{\bf Z})$ 
symmetry \cite{HULt438}. 
The ${\bf Z}_2\subset SL(2,{\bf Z})$ transformation 
exchanges the NS-NS 2-form potential $\hat{B}^{(1)}$ (coupled to a 
perturbative string state) and the R-R 2-form potential $\hat{B}^{(2)}$ 
(coupled to a non-perturbative $D$-brane) while changing the sign of the 
dilaton (or inverting the string coupling). So, the $SL(2,{\bf Z})$ 
symmetry is a strong-weak coupling duality.  

It is well-known that a covariant effective action for type-IIB string 
does not exist, while only the field equations \cite{SCH226,HOWw238} 
exist.   The only problem with construction of the covariant effective 
action is the R-R 4-form potential $\hat{D}$ (with the self-dual 5-form 
field strength $\hat{F}$), whose equation of motion $\hat{F}=\star\hat{F}$ 
cannot be derived from the covariant action.  So, to construct the 
covariant effective action for type-IIB string, one is forced to set 
$\hat{D}$ to zero.   However, it is found in \cite{BERbo53} that one can 
construct the type-IIB effective action which gives rise to the correct 
field equations and compactifies to the correct action for the dimensionally 
reduced type-IIB theories without setting $\hat{D}$ to zero.  
In this approach, one keeps $\hat{F}$ different from zero in the 
effective action but eliminates the self-duality constraint.   
After the field equations are obtained from this effective action, the 
self-duality constraint is imposed in order to get the correct field 
equations for the type-IIB theory.  

In the string-frame, such effective action for type-IIB string has the 
form \cite{BERbo53}:
\begin{eqnarray}
S^{string}_{IIB}&=&{1\over 2}\int d^{10}x\sqrt{-\hat{G}^{str}}
\left[e^{-2\Phi}\left\{-\hat{\cal R}^{str}+4(\partial\Phi)^2-
{3\over 4}(\hat{H}^{(1)})^2\right\}\right.
\cr
& &\ \ \left. -{1\over 2}(\partial\chi)^2-{3\over 4}(\hat{H}^{(2)}-
\chi\hat{H}^{(1)})^2-{5\over 6}\hat{F}^2-{1\over{96\sqrt{-\hat{G}^{str}}}}
\epsilon^{ij}\epsilon\hat{D}\hat{H}^{(i)}\hat{H}^{(j)}\right],
\label{striibact}
\end{eqnarray}
The field strengths of the 2-form potentials $\hat{B}^{(i)}$ 
and the 3-form potential $\hat{D}$, and their gauge transformation rules are
\begin{eqnarray}
\hat{H}^{i}&=&\partial\hat{B}^{(i)}, 
\ \ \ \ \ \ \ \ \ \ \ \ \ \ \ \ \ \ \ \ 
\delta\hat{B}^{(i)}=\partial\hat{\Sigma}^{(i)},
\cr
\hat{F}&=&\partial\hat{D}+{3\over 4}\epsilon^{ij}\hat{B}^{(i)}\hat{B}^{(j)}, 
\ \ \ \ 
\delta\hat{D}=\partial\hat{\Lambda}-{3\over 4}\epsilon^{ij}\partial
\hat{\Sigma}^{(i)}\hat{B}^{(j)},
\label{thfouformtran}
\end{eqnarray}
where $\hat{\Sigma}^{(i)}$ and $\hat{\Lambda}$ are infinitesimal gauge 
transformation parameters.   

The $SL(2,{\bf Z})$ symmetry of the type-IIB theory is manifest 
in the effective action in the Einstein-frame.   
To go to the Einstein-frame, one Weyl-rescales the metric $\hat{G}_{MN}=
e^{-{1\over 2}\Phi}\hat{G}^{str}_{MN}$.  The resulting Einstein-frame 
action has the form:
\begin{eqnarray}
S^{Einstein}_{IIB}&=&{1\over 2}\int d^{10}x\sqrt{-\hat{G}}\left[-\hat{\cal R}
+{1\over 4}{\rm Tr}(\partial_M\hat{\cal M}\partial^M\hat{\cal M}^{-1})
-{3\over 4}\hat{H}^{(i)}\hat{\cal M}_{ij}\hat{H}^{(j)}\right.
\cr
& &\ \ \ \ \ 
\left.-{5\over 6}\hat{F}^2-{1\over{96\sqrt{-\hat{G}}}}\epsilon^{ij}
\epsilon\hat{D}\hat{H}^{(i)}\hat{H}^{(j)}\right],
\label{einiibact}
\end{eqnarray}
where $\hat{\cal M}$ is a $2\times 2$ real matrix formed by the complex 
scalar $\hat{\lambda}=\chi+ie^{-\Phi}$:
\begin{equation}
\hat{\cal M}={1\over{{\cal I}m\hat{\lambda}}}\left(\matrix{|\hat{\lambda|}^2&
-{\cal R}e\,\hat{\lambda} \cr-{\cal R}e\,\hat{\lambda} & 1}\right).
\label{iibslmatscal}
\end{equation}
(\ref{einiibact}) is manifestly invariant under the 
$SL(2,{\bf R})$ transformation \cite{SCH226,HOWw238}:
\begin{equation}
\left(\matrix{\hat{H}^{(1)}\cr\hat{H}^{(2)}}\right) \to 
\omega \left(\matrix{\hat{H}^{(1)}\cr\hat{H}^{(2)}}\right), \ \ \ 
\hat{\cal M} \to (\omega^{-1})^T\hat{\cal M}\omega^{-1}, \ \ \ 
\omega
\in SL(2,{\bf R}).
\label{iibsl2}
\end{equation}
The $SL(2,{\bf R})$ transformation on $\hat{\lambda}$ has the usual 
fractional-linear form $\hat{\lambda}\to{{a\hat{\lambda}+b}\over
{c\hat{\lambda}+d}}$.  When the DSZW type charge quantization is taken 
into account, the $SL(2,{\bf R})$ symmetry breaks down to the 
$SL(2,{\bf Z})$ subset.  

\subsection{$T$-Duality of Toroidally Compactified Strings}\label{dualut}

Closed strings in $D$ dimensions in target space background with 
$d$ commuting isometries have $O(d,d,{\bf Z})$ $T$-duality symmetry
\cite{GIVrv322,SHAw320,GIVmr220,GIVmr238}.  
The $O(d,d,{\bf Z})$ symmetry is a perturbative symmetry proven to hold 
order by order in string coupling.  
For heterotic string, the symmetry is enlarged to $O(d,d+16,{\bf Z})$ due 
to the additional rank 16 background gauge fields.  
Under the ${\bf Z}_2$ subset that inverts the radius of $S^1$  
(i.e. $R_i\to\alpha^{\prime}/R_i$) and interchanges winding and momentum 
modes (i.e. $m_i\leftrightarrow n_i$),  the type-IIA and the type-IIB 
theories are interchanged if odd number of circles are acted on by the 
${\bf Z}_2$ transformations, while  heterotic string transforms to itself.  
The ${\bf Z}_2$ transformation between the type-IIA and the type-IIB 
strings at the effective field theory level is understood as field 
redefinition between type-IIA and type-IIB theories, since the 
compactification of the type-IIA and the type-IIB supergravities leads to 
the same supergravity theory.  

We consider the bosonic string worldsheet $\sigma$-model with only 
NS-NS sector fields (target space metric $G_{\mu\nu}(X)$, 2-form potential  
$B_{\mu\nu}(X)$ and dilaton $\Phi(X)$), which are common to both 
type-II and heterotic strings, turned on.  The action 
with the (curved) $D$-dimensional target space has form
\begin{equation}
S={1\over{2\pi}}\int d^2z\left[\left(G_{\mu\nu}(X)+
B_{\mu\nu}(X)\right)\partial X^{\mu}\partial X^{\nu}-{1\over 4}\phi(X)
{\cal R}^{(2)}\right].
\label{strsigmod}
\end{equation}

Let us assume that (\ref{strsigmod}) is invariant under the 
$d$ commuting, compact Abelian isometries \cite{GIVr380}
$\delta X^{\mu}=\varepsilon{\bf k}^{\mu}_i$ ($i=1,...,d$) along 
the $X^i$-direction, where $[{\bf k}_i,{\bf k}_j]=0$.  
Then, the background fields become independent of $X^i$.  

First, we consider the case where the background fields have only one  
isometry ($d=1$) along, say, the direction $\theta=X^0$.  
The $T$-dual pair $\sigma$-model actions are obtained by gauging the 
Abelian isometry.  Following the procedures discussed in 
\cite{BUS159,BUS194,BUS201,BERek49,BERko51}, one obtains 
the action dual to (\ref{strsigmod}) with the dual background fields 
(with primes) related to the original ones (without primes) as
\footnote{More general transformations with non-zero R-R fields are 
derived in \cite{BERho451}.}  
\begin{eqnarray}
G^{\prime}_{00}&=&{1\over{G_{00}}}, \ \ 
G^{\prime}_{0a}={{B_{0a}}\over{G_{00}}}, \ \  
G^{\prime}_{ab}=G_{ab}-{{G_{a0}G_{0b}+B_{a0}B_{0b}}\over{G_{ab}}}, 
\cr
B^{\prime}_{0a}&=&{{G_{0a}}\over{G_{00}}}, \ \  
B^{\prime}_{ab}=B_{ab}-{{G_{a0}B_{0b}+B_{a0}G_{0b}}\over{G_{00}}}, \ \ 
\Phi^{\prime}=\Phi+\log G_{00},
\label{circtdual}
\end{eqnarray}
where $X^{\mu}=(\theta,X^a)$ ($a=1,...,D-1$).  
This is the curved background generalization of 
the $R\to\alpha^{\prime}/R$ $T$-duality of closed strings 
on $S^1$. 

Alternatively, the dual pair $\sigma$-models are obtained by the method 
of chiral currents \cite{ROCv373}.  
One starts with a $(D+d)$-dimensional $\sigma$-model 
with $d$ Abelian (left-handed) chiral currents $J^i$ and 
(right-handed) anti-chiral currents $\bar{J}^i$ \cite{GIVr380}:  
\begin{eqnarray}
S_{D+d}&=&S_1+S_a+S[X], 
\cr
S_1&=&{1\over{2\pi}}\int d^2z\left[\partial\theta^i_L\bar{\partial}\theta^i_L
+\partial\theta^i_R\bar{\partial}\theta^i_R+2\Sigma_{ij}(X)\partial\theta^i_R
\bar{\partial}\theta^j_L\right.
\cr
& &\ \ \ \ \ \ \ \left.+\Gamma^L_{ai}(X)\partial X^a\bar{\partial}\theta^i_L
+\Gamma^R_{ia}(X)\partial\theta^i_R\bar{\partial}X^a\right],
\cr
S_a&=&{1\over{2\pi}}\int d^2z\left[\partial\theta^i_L\bar{\partial}\theta^i_R
-\partial\theta^i_R\bar{\partial}\theta^i_L\right],
\cr
S[X]&=&{1\over{2\pi}}\int d^2z\left[\Gamma_{ab}(X)\partial X^a\bar{\partial}
X^b-{1\over 4}\Phi(X){\cal R}^{(2)}\right], 
\label{chiralddact}
\end{eqnarray}
where $i,j=1,...,d$ and $a,b=d+1,...,D$.  
Here, chiral and anti-chiral currents, corresponding to the 
$U(1)^d_L\times U(1)^d_R$ affine symmetries $\delta\theta^i_{L,R}
=\alpha^i_{L,R}(z)$, are
\begin{equation}
J^i=\partial\theta^i_L+\Sigma_{ji}\theta^j_R+{1\over 
2}\Gamma^L_{ai}\partial X^a, \ \ 
\bar{J}^i=\bar{\partial}\theta^i_R+\Sigma_{ij}\bar{\theta}^j_L+{1\over 
2}\Gamma^R_{ia}\bar{\partial}X^a.
\label{chircurrlr}
\end{equation}

By gauging either a vector or an axial subgroup of $U(1)^d_L\times 
U(1)^d_R$, one has the following $D$-dimensional dual pair 
$\sigma$-model actions:
\begin{eqnarray}
S^{\pm}_D&=&{1\over{2\pi}}\int d^2z\left[{\cal E}^{\pm}_{\mu\nu}(X^a)
\partial X^{\mu}_{\pm}\bar{\partial}X^{\nu}_{\pm}-
{1\over 4}\phi^{\pm}(X^a){\cal R}^{(2)}\right]
\cr
&=&{1\over{2\pi}}\int 
d^2z\left[E^{\pm}_{ij}(X^a)\partial\theta^i_{\pm}\bar{\partial}\theta^j_{\pm}+
F^{R\,\pm}_{ia}(X^a)\partial\theta^i_{\pm}\bar{\partial}X^a+
F^{L\,\pm}_{ai}(X^a)\partial X^a
\bar{\partial}\theta^i_{\pm}\right.
\cr
& &\ \ \ \ \ \ \ \ \ \left.+F^{\pm}_{ab}(X^a)\partial X^a\bar{\partial}X^b
-{1\over 4}\phi^{\pm}(X^a){\cal R}^{(2)}\right],
\label{finalaxvecac}
\end{eqnarray}
where $(X^{\mu}_{\pm})=(\theta^i_{\pm},X^a)$ with $\mu,\nu=1,...,D$,  
$i=1,...,d$, $a=d+1,...,D$.  Here, $\theta^i_{\pm}\equiv\theta^i_R\pm
\theta^i_L$  and the upper [lower] signs in $\pm$ and 
$\mp$ correspond to the axial [vector] gauged $\sigma$-model.  
The background fields are
\begin{eqnarray}
{\cal E}^{\pm}_{\mu\nu}&=&{\cal G}^{\pm}_{\mu\nu}+{\cal B}^{\pm}_{\mu\nu}=
\left(\matrix{E^{\pm}_{ij}&F^{R\,\pm}_{ib}\cr F^{L\,\pm}_{aj}&
F^{\pm}_{ab}}\right),  \ \ \ \ 
\phi^{\pm}=\Phi+\log\,{\rm det}(I\pm\Sigma), 
\cr
E^{\pm}_{ij}&=&(I\pm\Sigma)_{ik}(I\mp\Sigma)^{-1}_{kj}, \ \ \ \ 
F^{\pm}_{ab}=\Gamma_{ab}\pm{1\over 2}\Gamma^L_{ai}(I\mp\Sigma)^{-1}_{ij}
\Gamma^R_{jb},
\cr
F^{R\,\pm}_{ia}&=&(I\mp\Sigma)^{-1}_{ij}\Gamma^R_{ja},\ \ \ \ \ \ \ \ 
F^{L\,\pm}_{ai}=\pm\Gamma^L_{aj}(I\pm\Sigma)^{-1}_{ji},
\label{vecaxibackfld}
\end{eqnarray}
with ${\cal G}^{\pm}$ [${\cal B}^{\pm}$]  denoting the 
symmetric [the antisymmetric] part of ${\cal E}^{\pm}$.  
The action $S^{\pm}_D$ has an isometry under the translation in the 
$\theta^i_{\mp}$-direction.

The dual pair actions (\ref{finalaxvecac}) are the most 
general $\sigma$-model with $d$ commuting compact Abelian symmetries.  
$S^{+}_D$ and $S^{-}_D$  are related under the combined operations 
of the sign reversal of $\Sigma$ and $\Gamma^L$, and the coordinate 
transformation $\theta^i_L\to-\theta^i_L$ (i.e. $\theta^i_+\leftrightarrow
\theta^i_-$), implying equivalence of two gaugings at least locally.   
To achieve global equivalence of the two gaugings, one has to 
impose the same periodicity conditions on both $\theta^i_{+}$ and 
$\theta^i_{-}$, i.e. $\theta^i_{\pm}\equiv\theta^i_{\pm}+2\pi$.  
Then, one establishes the equivalence  of vector and axial gauged 
$\sigma$-model actions $S^{\pm}_D$ (the so-called ``vector-axial 
duality'' \cite{KIR6}).   

One can relate $S^{\pm}_D$ to the bosonic string $\sigma$-model  
action $S$ in (\ref{strsigmod}) by identifying $X^{\mu}=X^{\mu}_{\pm}$.  
Then, the background fields in $S^{\pm}_D$ are related to those in 
$S$ in the following way:
\begin{eqnarray}
G_{ij}&=&{1\over 2}(E^{\pm}_{ij}+E^{\pm}_{ji}),\ \ \ \ \ \ \ \ 
B_{ij}={1\over 2}(E^{\pm}_{ij}-E^{\pm}_{ji}), 
\cr
G_{ia}&=&{1\over 2}(F^{R\,\pm}_{ia}+F^{L\,\pm}_{ai}),\ \ \ \ \ \ 
B_{ia}={1\over 2}(F^{R\,\pm}_{ia}-F^{L\,\pm}_{ai}), 
\cr
G_{ab}&=&{1\over 2}(F^{\pm}_{ab}+F^{\pm}_{ba}), \ \ \ \ \ \ \ \ 
B_{ab}={1\over 2}(F^{\pm}_{ab}-F^{\pm}_{ba}).
\label{strconfbackrel}
\end{eqnarray}
From this, transformation rule of background 
fields under $T$-duality that relates the dual pair bosonic string 
$\sigma$-model actions (one related to $S^+_D$ and the 
other related to $S^-_D$) follows.  When $S^{\pm}_D$ have isometry along only 
one coordinate direction ($d=1$), one recovers the factorized 
duality transformation (\ref{circtdual}).  

We discuss transformations \cite{GIVmr238,GIVr380}
that relate the different backgrounds (of the same action) describing   
the equivalent conformal field theory.   
$S^{\pm}_D$ have the manifest invariance under the following 
$O(d,d,{\bf Z})$ transformation
\begin{eqnarray}
{\cal E}^{\pm}&\to&{\cal E}^{\pm\,\prime}=(\hat{a}{\cal E}^{\pm}+\hat{b})
(\hat{c}{\cal E}^{\pm}+\hat{d})^{-1}
\cr
& &\ \ \ \ \ \ =\left(\matrix{E^{\pm\,\prime}&
(a-E^{\pm\,\prime}c)F^{R\,\pm}
\cr F^{L\,\pm}(cE^{\pm}+d)^{-1}&F^{\pm}-F^{L\,\pm}(cE^{\pm}+d)^{-1}c
F^{R\,\pm}}\right), 
\cr
\phi^{\pm}&\to&\phi^{\pm\,\prime}=\phi^{\pm}+{1\over 2}\log
\left({{{\rm det}{\cal G}^{\pm}}\over{{\rm det}{\cal G}^{\pm\,\prime}}}
\right),\ \ \ \ \ g=\left(\matrix{a&b\cr c&d}\right)\in O(d,d,{\bf Z}),
\label{oddbacktran}
\end{eqnarray}
where $D\times D$ blocks $\hat{a}$, $\hat{b}$, $\hat{c}$ and $\hat{d}$ 
of the $O(D,D,{\bf Z})$ matrix are
\begin{equation}
\hat{a}=\left(\matrix{a&0\cr 0&I_{D-d}}\right), \ \ 
\hat{b}=\left(\matrix{b&0\cr 0&0}\right), \ \ 
\hat{c}=\left(\matrix{c&0\cr 0&0}\right), \ \ 
\hat{d}=\left(\matrix{d&0\cr 0&I_{D-d}}\right). 
\label{embeddd}
\end{equation}
Here, $I_n$ is the $n\times n$ identity matrix.  
The $d\times d$ matrices $E^{\pm}$ transform fractional linearly under 
$O(d,d,{\bf Z})$:
\begin{equation}
E^{\pm}\to E^{\pm\,\prime}=(aE^{\pm}+b)(cE^{\pm}+d)^{-1}.
\label{fraclinetran}
\end{equation}
The constant ${\cal E}^{\pm}$ case corresponds 
to the transformation in the toroidal background.  

The  $O(d,d,{\bf Z})$ transformation is generated by the following 
transformations.  
\begin{itemize}
\item Integer ``$\Theta$''-parameter shift of $E$, i.e. $E_{ij}\to E_{ij}+
\Theta_{ij}$ ($\Theta_{ij}=-\Theta_{ji}$):
\begin{equation}
\left(\matrix{a&b\cr c&d}\right)=\left(\matrix{I_d&\Theta\cr 0&I_d}
\right)\ \ \ {\rm s.t.}\ \ \ \Theta=-\Theta^T. 
\label{thetatran}
\end{equation}
\item Homogeneous transformations of $E^{\pm}$, $F^{L\,\pm}$, $F^{R\,\pm}$ 
under $GL(d,{\bf Z})$, i.e. $E^{\pm}\to A^TE^{\pm}A$, $F^{L\,\pm}\to 
A^TF^{L\,\pm}$, $F^{R\,\pm}\to F^{R\,\pm}A$ ($A\in GL(2,{\bf Z})$):
\begin{equation}
\left(\matrix{a&b\cr c&d}\right)=\left(\matrix{A^T&0\cr 0&A^{-1}}
\right)\ \ \ {\rm s.t.}\ \ \ A\in GL(2,{\bf Z}). 
\label{homotran}
\end{equation}
\item Factorized dualities $D_i$, corresponding to the inversion of the 
radius $R_i$ of the $i$-th circle, i.e. $R_i\to 1/R_i$:
\begin{equation}
\left(\matrix{a&b\cr c&d}\right)=\left(\matrix{I-e_i&e_i\cr e_i&I-e_i}
\right)\ \ {\rm s.t.}\ \ e_i={\rm diag}(\underbrace{\overbrace{0,...,
0}^{i-1},1,\overbrace{0,...,0}^{d-i-2}}_d).
\label{factdualtrn}
\end{equation}
\end{itemize}

The maximal compact subgroup of $O(d,d,{\bf Z})$ is $O(d,{\bf Z})\times 
O(d,{\bf Z})$ with a group element having the form:
\begin{equation}
\left(\matrix{a&b\cr c&d}\right)={1\over 2}
\left(\matrix{O_L+O_R&O_L-O_R\cr O_L-O_R&O_L+O_R}\right)\ \ \ {\rm s.t.}
\ \ \  O_L,O_R\in O(d,{\bf Z}).
\label{maxsggrpodd}
\end{equation}
The $O(d,{\bf Z})\times O(d,{\bf Z})$ transformation naturally acts on 
the action $S_{D+d}$ as
\begin{eqnarray}
\theta^i_L&\to& O_L\theta^i_L, \ \ 
\theta^i_R\to O_R\theta^i_R; \ \ \ 
\left(\matrix{\theta_-\cr\theta_+}\right)\to 
{1\over 2}\left(\matrix{O_L+O_R&O_L-O_R\cr O_L-O_R&O_L+O_R}\right)
\left(\matrix{\theta_-\cr\theta_+}\right), 
\cr
\Sigma&\to&O_R\Sigma O^T_L, \ \ \ \ \ 
\Gamma^L\to\Gamma^LO^T_L, \ \ \ \ \ 
\Gamma^R\to O_R\Gamma^R, \ \ \ \ \ 
\Gamma\to\Gamma,  
\label{sddmaxsbtran}
\end{eqnarray}
meanwhile on the actions $S^{\pm}_D$ as
\begin{eqnarray}
E^{\pm}&\to&E^{\pm\,\prime}=\left[(O_L+O_R)E^{\pm}+(O_L-O_R)\right]
\left[(O_L-O_R)E^{\pm}+(O_L+O_R)\right]^{-1}, 
\cr
F^{L\pm}&\to& F^{L\pm\,\prime}=2F^{L\pm}
\left[(O_L-O_R)E^{\pm}+(O_L+O_R)\right]^{-1},
\cr
F^{R\pm}&\to&F^{R\pm\,\prime}={1\over 2}\left[(O_L+O_R)-
E^{\pm\,\prime}(O_L-O_R)\right]F^{R\pm}, 
\cr
F^{\pm}&\to& F^{\pm\,\prime}=F^{\pm}-F^{L\pm}\left[(O_L-O_R)E^{\pm}
+(O_L+O_R)\right]^{-1}(O_L-O_R)F^{R\pm}.  
\label{spmdmaxtran}
\end{eqnarray}
   
We now specialize to strings in flat background (i.e. toroidal 
compactification) to understand properties of perturbative string 
spectrum under $T$-duality.  The relevant part of 
the worldsheet action is the toroidal ($T^d$) part:
\begin{equation}
S={1\over{4\pi}}\int^{2\pi}_0d\sigma\int d\tau\left[
\sqrt{g}g^{\alpha\beta}G_{ij}\partial_{\alpha}X^i\partial_{\beta}X^j
+\epsilon^{\alpha\beta}B_{ij}\partial_{\alpha}X^i\partial_{\beta}X^j
-{1\over 2}\sqrt{g}\phi{\cal R}^{(2)}\right],
\label{torowshact}
\end{equation}
where $X^i\sim X^i+2\pi m^i$ and $i,j=1,...,d$.  Here, $m^i$ is a string 
`winding number' along the $X^i$-direction.  $G_{ij}$ and $B_{ij}$ 
can be collected into the ``background matrix'' $E=G+B$.  
The matrix $E$ is a special case of $E^{\pm}$ in (\ref{vecaxibackfld}) 
where $E^{\pm}$ do not depend on $X^a$.  The lattice $\Lambda^d$, which 
defines $T^d={\bf R}^d/(\pi\Lambda^d)$, is spanned by basis vectors 
${\bf e}_i$ satisfying $\sum^d_{a=1}e^a_ie^a_j=2G_{ij}$.   

The mode expansions of $X^i$ and conjugate 
momenta $2\pi P_i=G_{ij}\dot{X}^j+B_{ij}X^{j\,\prime}$ are
\begin{eqnarray}
X^i(\sigma,\tau)&=&x^i+m^i\sigma+\tau G^{ij}(p_j-B_{jk}m^k)
\cr
& &\ \ +{i\over{\sqrt{2}}}\sum_{n\neq 0}{1\over n}[\alpha^i_n(E)
e^{-in(\tau-\sigma)}+\tilde{\alpha}^i_n(E)e^{-in(\tau+\sigma)}],
\cr
2\pi P_i(\sigma,\tau)&=&p_i+{1\over{\sqrt{2}}}\sum_{n\neq 0}
[E^T_{ij}\alpha^j_n(E)e^{-in(\tau-\sigma)}+E_{ij}\tilde{\alpha}^j_n(E)
e^{-in(\tau+\sigma)}],
\label{modeexptor}
\end{eqnarray}
where we made analytic continuation $\tau\to -i\tau$ and the 
momentum zero modes $p_i$ take integer values, i.e. $p_i=n_i\in {\bf Z}$. 
The equal-time canonical commutation relations $[X^i(\sigma,0),
P_j(\sigma^{\prime},0)]=i\delta^i_j\delta(\sigma-
\sigma^{\prime})$ lead to commutation relations 
among the oscillator modes:
\begin{equation}
[x^i,p_j]=i\delta^i_j,\ \ \ 
[\alpha^i_n(E),\alpha^j_m(E)]=[\tilde{\alpha}^i_n(E),
\tilde{\alpha}^j_m(E)]=mG^{ij}\delta_{m+n,0}. 
\label{comrelosciltor}
\end{equation}

The Hamiltonian takes the form:
\begin{eqnarray}
H&=&L_0+\tilde{L}_0={1\over{4\pi}}\int^{2\pi}_0d\sigma(P^2_L+P^2_R)
={1\over 2}Z^TM(E)Z+N+\tilde{N};
\cr
M(E)&=&\left(\matrix{G-BG^{-1}B&BG^{-1}\cr -G^{-1}B&G^{-1}}\right), 
\ \ \ \ 
Z=\left(\matrix{m_a\cr n_a}\right),
\label{hamildtor}
\end{eqnarray}
where $P_{L,R}$ are the left- and the right-moving momenta defined as
\begin{equation}
P_{L\,a}=[2\pi P_i+(G-B)_{ij}X^{j\,\prime}]e^{i\,*}_a, \ \ \ 
P_{R\,a}=[2\pi P_i-(G+B)_{ij}X^{j\,\prime}]e^{i\,*}_a,	
\label{lefrigmovmom}
\end{equation}
and the number operators of the left- and the right-moving modes are
\begin{equation}
N_L=\sum_{n>0}\alpha^i_{-n}(E)G_{ij}\alpha^j_n(E), \ \ \ \ 
N_R=\sum_{n>0}\tilde{\alpha}^i_{-n}(E)G_{ij}\tilde{\alpha}^j_n(E).
\label{lefrignumop}
\end{equation}
Here, ${\bf e}^{i\,*}$ are dual basis vectors, satisfying 
$\sum^d_{a=1}e^a_ie^{j\,*}_a=\delta^j_i$ and $\sum^d_{a=1}e^{i\,*}_a
e^{j\,*}_a={1\over 2}(G^{-1})^{ij}$.  
The left- and the right-moving momenta zero modes 
\begin{equation}
p_R=[n^T+m^T(B-G)]e^*,\ \ \ \  p_L=[n^T+m^T(B+G)]e^*,
\label{zerolefrigmom}
\end{equation}
transforming as a vector under $O(d,d,{\bf R})$, form an even self-dual 
Lorentzian lattice $\Gamma^{(d,d)}$ \cite{NAR169,NARsw279}, i.e. 
$p^2_L-p^2_R=2m^in_i\in 2{\bf Z}$.  

While $\Gamma^{(d,d)}$ is preserved under $O(d,d,{\bf R})$, 
the Hamiltonian zero mode $H_0={1\over 2}(p^2_L+p^2_R)$ 
is invariant only under its maximal compact subgroup 
$O(d,{\bf R})\times O(d,{\bf R})$.  So, the zero-mode spectrum is 
unchanged under the $O(d,{\bf R})\times O(d,{\bf R})$ subgroup, only.  
Note, from (\ref{zerolefrigmom}) one sees that $(p_L,p_R)$ 
and hence $\Gamma^{(d,d)}$ are in one-to-one correspondence 
with a particular background $E=G+B$.  Thus, the moduli space is 
isomorphic to $O(d,d,{\bf R})/[O(d,{\bf R})\times O(d,{\bf R})]$.  

Under the $O(d,d,{\bf Z})$ transformation (\ref{fraclinetran}) 
\cite{KUGz87}, 
\begin{eqnarray}
\ \ \ \ \ \ \ \ \ \ \ & & M(E)\to gM(E)g^T,  
\cr
\alpha_n(E)&\to&(d-cE^T)^{-1}\alpha_n(E^{\prime}), \ \ \ 
\tilde{\alpha}_n(E)\to (d+cE)^{-1}\tilde{\alpha}_n(E^{\prime}).
\label{oddtrancomps}
\end{eqnarray}
So, $N_{L,R}$ are manifestly invariant under $O(d,d,{\bf Z})$; 
the spectrum is $O(d,d,{\bf Z})$ invariant.  
The $O(d,d,{\bf Z})$ transformation is generated 
\cite{GIVrv322,SHAw320,GIVmr220,GIVmr238} by integer $\Theta$-parameter 
shift of $E$, the $GL(2,{\bf Z})$ transformation and the factorized 
duality $D_i$, as discussed above.  
Particularly, under $GL(d,{\bf Z})$, which changes the basis of 
$\Lambda^d$, $E\to AEA^T$ and $(m,n)\to(A^Tm,A^{-1}n)$ 
($A\in GL(d,{\bf Z})$).  
In addition, the spectrum is invariant under the worldsheet parity 
\cite{GIVmr238} $\sigma\to-\sigma$, which acts on $E$ as $B\to -B$.  
The effect of the worldsheet parity on the spectrum is to interchange 
the left-handed and the right-handed modes: 
$p_L\leftrightarrow p_R$ and $\alpha_n\leftrightarrow\tilde{\alpha}_n$.  
The above transformations generate the full spectrum preserving symmetry group 
${\cal G}_d$.  

A particular element $g$ of $O(d,d,{\bf Z})$ with $a=d=0$ and $b=c=I$, 
i.e. $E\to E^{-1}$ \cite{GIVrv322,SHAw320}, corresponds to vector-axial 
duality symmetry (\ref{vecaxibackfld}).  Under this transformation, 
$n\leftrightarrow m$ and the Hamiltonian (\ref{hamildtor}) is 
manifestly invariant.  When $B=0$, the transformation becomes  
$G\to G^{-1}$, i.e. the volume inversion of $T^d$.  
A significance of $E\to E^{-1}$ is that the gauge symmetry is enhanced 
to the affine algebra $SU(2)^d_L\times SU(2)^d_R$ at a single fixed point 
$G=I$ and $B=0$.  The gauge symmetry is maximally enhanced \cite{GIVrv322} at 
fixed points under $E\to E^{-1}$ modulo $SL(d,{\bf Z})$ and $\Theta({\bf Z})$ 
transformations, i.e. $E$ such that $E^{-1}=M^T(E+\Theta)M$ ($M\in 
SL(d,{\bf Z})$).  
Hence, an enhanced symmetry point corresponds to an orbifold singularity 
point \cite{DIXhvw261,DIXhvw274} in the moduli space under some non-trivial 
$O(d,d,{\bf Z})$ transformation.  At the fixed point, $E$ takes the 
following form in terms of the Cartan matrix $C_{ij}$ of the rank $d$, 
semi-simple, simply-laced symmetry group \cite{ELIgrs283}:
\begin{equation}
E_{ij}=C_{ij}\ \ (i>j),\ \ \ \ E_{ii}={1\over 2}C_{ii},\ \ \ \ 
E_{ij}=0\ \ (i<j). 
\label{maxgausymfx}
\end{equation}
Non-maximally enhanced symmetry points correspond to fixed points under 
factorized dualities $D_i$ instead of the full inversion $E\to E^{-1}$. 
A simplest but non-trivial example is the $d=2$ case (i.e. 
compactification on $T^2$), which we discuss in section \ref{n4bh4dsphsing}.  
At the fixed point $E=I$ under $E\to E^{-1}$, the gauge symmetry is enhanced 
to $\left(SU(2)\times SU(2)\right)_L\times\left(SU(2)\times SU(2)\right)_R$.  
The gauge symmetry is maximally enhanced to $SU(3)_L\times SU(3)_R$ at the 
point $E=\left(\matrix{1&1\cr 0&1}\right)$.  These fixed points correspond 
to orbifold singularities \cite{SHAw320} in the fundamental domain of the 
moduli space (parameterized by two complex coordinates of the moduli space
\footnote{Note, $O(2,2,{\bf R})\cong SL(2,{\bf R})\times 
SL(2,{\bf R})$.  Therefore, $E$, which parameterizes 
the moduli space $O(2,2,{\bf R})/[O(2,{\bf R})\times O(2,{\bf 
R})]$, is reparameterized by the complex coordinates $\rho$ and 
$\tau$, each parameterizing $SL(2,{\bf R})/U(1)$.} 
$SL(2,{\bf R})/U(1)\times SL(2,{\bf R})/U(1)$).

\subsection{$M$-Theory}\label{dualm}

In this section, we discuss some aspects of $M$-theory.  We illustrate how 
different superstring theories emerge from different moduli space of 
compactified $M$-theory and discuss the $M$-theory origin of string dualities.
  
In this picture, each of 5 different string theories  
represents a perturbative expansion about different points in moduli space 
of the compactified $M$-theory.  Namely, 5 perturbative string theories 
and uncompactified $M$-theory are located at different subsets of moduli 
space, and it is dualities that map one subset of moduli space to 
another, thereby making transition between different theories.  
In the following, we illustrate this idea by showing how different 
theories are achieved by taking different limits of parameters of 
moduli space and how dualities are realized as transformations 
of parameters of moduli space.  

First, we discuss the connection between the type-II theories and M-theory.  
Type-IIA theory is obtained from $M$-theory by compactifying the 
extra 1 spatial coordinate on $S^1$ of radius $R_{11}$ 
\cite{HUQn,CAMw243,DUFhis,TOW350,WIT443}.    
The type-IIA and the type-IIB theories are related via $T$-duality 
\cite{DINhs322,DAIlp}.  
Namely, the type-IIA theory on $S^1$ of radius 
$R_A$ is perturbatively equivalent (under $T$-duality) 
to the type-IIB theory on $S^1$ of radius 
$R_B=1/R_A$.  So, one can think of the $S^1$-compactified 
type-IIB theory as $T^2$ compactified $M$-theory
\footnote{Note, due to the no-go theorem for KK compactification 
of the $D=11$ supergravity \cite{WIT85}, it might be impossible 
to obtain a chiral theory like type-IIB supergravity through  
dimensional reduction.  
This no-go theorem can be circumvented to obtain the (chiral) type-IIB 
theory by compactifying on orbifolds (rather than manifolds) 
\cite{HORw475,DASm465,WIT463,SEN11}.  In the case of compactification 
of $M$-theory on $T^2$ (which is relevant to our discussion), when  
the size of $T^2$ goes to zero at the fixed shape, 
one obtains ``chiral'' type-IIB theory, due to additional massive 
`wrapping' modes (of membrane) which become massless 
\cite{BERho451,ASP46,SCH367,SCH077}.}.  

To understand the connection between type-II theories and $M$-theory, 
one has to compactify $M$-theory on $T^2=S^1\times S^1$ 
(with the radii of each circle given by $R_{11}$ and $R_{10}$ from the 
$D=11$ point of view), and compactify the type-IIA and the 
type-IIB theories on circles of radii $R_A$ and $R_B$, respectively.  
Here, the radius $R_A$ [$R_B$] is measured with $D=10$ string 
frame metric of the type-IIA [the type-IIB] theory.  

We first relate parameters of the type-II theories (i.e. 
the radii $R_{A,B}$ and the string couplings
\footnote{The string coupling $g_s$ is defined as $g_s=
e^{\langle\phi\rangle}$.} 
$g^{(A,B)}_s$ in the type-IIA/B 
theories) to parameters $R_{11}$ and $R_{10}$ of $T^2$ moduli space 
before we understand the various limits in the moduli space.  
$R_{11}$ is related to $g^{(A)}_s$ as $R_{11}=(g^{(A)}_s)^{2\over 3}$.
As for the second circle of $T^2=S^1\times S^1$, which is also the 
circle upon which the type-IIA theory is compactified, the radius is 
measured differently depending on the dimensionality of spacetime.  
Note, we denoted the radius measured in $D=11$ [in $D=10$ 
by the type-IIA string-frame metric] as $R_{10}$ [$R_A$].
Namely, since the string-frame metric $g^{IIA}_{\mu\nu}$ ($\mu,\nu=
0,1,...,9$) of the type-IIA theory is related to the $D=11$ metric 
$G^{(11)}_{MN}$ ($M,N=0,1,...,10$) as $G^{(11)}_{\mu\nu}\sim 
e^{-{2\over 3}\phi_A}g^{IIA}_{\mu\nu}$, where $\phi_A$ is the dilaton 
of the type-IIA theory, one sees that $R_{10}=R_A/(g^{(A)}_s)^{1\over 3}$.
Furthermore, one can relate the string coupling $g^{(B)}_s$ of the type-IIB 
theory to $R_{11}$ and $R_{10}$ as follows. 
Under the $T$-duality between the type-IIA and the type-IIB theories, the 
string couplings are related as $g^{(B)}_s=g^{(A)}_s/R_A$.   
By using other relations among parameters, one finds that $g^{(B)}_s=
R_{11}/R_{10}$.   

We discuss the various limits in the $M$-theory moduli space of $T^2$ 
in terms of $R_{11}$ and $R_{10}$ \cite{ASP46}: 
$M$-theory and the type-IIA,B theories are located at 
various limiting points in the $T^2$-moduli space.  
First, $M$-theory is located at $(R_{10},R_{11})=(\infty,\infty)$ (i.e. 
the decompactification limit), which is also the strong coupling limit 
($g^{(A)}_s=(R_{11})^{3\over 2}\to\infty$) of the type-IIA theory 
\cite{WIT443,TOW350}. 
The (uncompactified) type-IIA theory, defined as $R_A\to\infty$ and finite 
string coupling $g^{(A)}_s$,  is located at $(R_{10},R_{11})=
(\infty,finite)$, i.e. $M$-theory on $S^1$ of radius $R_{11}<\infty$.  
The (uncompactified) type-IIB theory can be defined as the limit 
$R_B\to\infty$, $R_{11}\to 0$ and finite string coupling $g^{(B)}_s$.  
In this limit, $R_{10}=R_A/(g^{(A)})^{1/3}=R_A/(R_Ag^{(B)}_s)^{1/3}=
R^{2/3}_A(g^{(B)}_s)^{-1/3}=R^{-2/3}_B(g^{(B)}_s)^{-1/3}\to 0$.  
So, in terms of parameters of $T^2$, the (uncompactified) type-IIB 
theory corresponds to the limit in which $(R_{10},R_{11})
=(0,0)$ while keeping the ratio $g^{(B)}_s=R_{11}/R_{10}$ finite.  
The value of $g^{(B)}_s$ depends on how the limit $(R_{10},R_{11})\to(0,0)$ 
is taken and, therefore, $(R_{10},R_{11})=(0,0)$ is not really a point 
in the moduli space.  

Note, when 2 circles in $T^2=S^1\times S^1$ are exchanged 
(i.e. $R_{11}\leftrightarrow R_{10}$), $g^{(B)}_s=R_{11}/R_{10}$ 
is inverted: $g^{(B)}_s\to 1/g^{(B)}_s$.  Such interchange of 
2 circles is a subset of more general $SL(2,{\bf Z})$ 
reparameterization of $T^2$, which acts on the complex 
modulus $\tau$ of $T^2$ fractional linearly.  Thus, the reparameterization 
symmetry of $T^2$, upon which $M$-theory is compactified, manifests 
in the type-IIB theory as the $SL(2,{\bf Z})$ $S$-duality \cite{HULt438}, 
which acts on the complex scalar $\rho=\chi+ie^{-\phi}$ (formed by 0-form 
$\chi$ and the dilaton $\phi_B$) fractional linearly.   

Next, we discuss string theories with $N=1$ supersymmetry, 
i.e. the $E_8\times E_8$ and $SO(32)$ heterotic strings and type-I string.  
To understand the connection between $M$-theory and these 
$N=1$ string theories, one has to consider the moduli space of $M$-theory 
compactified on $S^1/{\bf Z}_2\times S^1$, i.e. a cylinder of length $L$ 
and radius $R$.  

We first comment on the relation of type-I string theory to $M$-theory.  
One can think of the type-I theory as an `orientifold' of the type-IIB 
theory, namely a theory of unoriented closed string (gauged under the 
worldsheet parity transformation $\Omega$ \cite{HOR327,HOR231,PRAs216}) 
and open string with $SO(32)$ Chan-Paton factor \cite{GREs149}.  
To see the direct relation to $M$-theory, it is convenient to first 
compactify one spatial coordinate, which we call $Y$, 
of the type-I theory on $S^1$ and then $T$-dualize along the 
$S^1$-direction, inverting the radius of $S^1$.  We call such 
theory as the type-I$^{\prime}$ theory \cite{DAIlp,POLw460}.  
Since the dual coordinate $\tilde{Y}$ is pseudo-scalar under the worldsheet 
parity transformation (i.e. $\Omega[\tilde{Y}](\tau,\sigma)=-\tilde{Y}
(\tau,-\sigma)$), $S^1$ is mapped under this $T$-duality to the orbifold 
$S^1/{\bf Z}_2$ with fixed points at $\tilde{Y}=0,\pi$.  So, the 
type-I$^{\prime}$ theory is effectively described by the type-IIA theory 
on $S^1/{\bf Z}_2$; closed strings wrapped around 
$S^1/{\bf Z}_2$ look like open string stretched between two 
8-plane boundaries located at fixed points of $S^1/{\bf Z}_2$.  
(These (parallel) boundaries corresponds to $D\,8$-branes.)  

Second, the $E_8\times E_8$ heterotic string theory is obtained by 
compactifying $M$-theory on the orbifold $S^1/{\bf Z}_2$ 
\cite{HORw460,HORw475}.  Namely, $M$-theory on $S^1/{\bf Z}_2$ of 
length $L$ gives rise to spacetime with two $D=10$ faces (the so-called 
``end-of-the-world 9-branes'') which are separated by a distance $L$.  
Each of the two faces carries an $E_8$ gauge field of the $E_8\times E_8$ 
heterotic string \cite{HORw460,HORw475}.  
In this picture, a fundamental string of the $E_8\times E_8$ theory is 
interpreted as a cylindrical $M\,2$-brane attached between the two faces.  
(So, the intersection of the cylindrical $M\,2$-brane with the faces is a 
circle.)  The string coupling is $g^{E_8}_s=L^{3/2}$ and, therefore, 
in the strong coupling limit ($g^{E_8}_s\gg 1$) the two faces move apart far 
away from each other, revealing the extra 11-th space dimension 
\cite{TOW350,WIT443,HORw460,HORw475}.  
When the separation is very small ($L\approx 0$), the cylindrical 
$M\,2$-brane is well approximated by a closed string in $D=10$.  
(This is the weak coupling limit $g^{E_8}_s=L^{3/2}\approx 0$ of the 
$E_8\times E_8$ theory.)  Finally, the $SO(32)$ heterotic theory 
is related to the $E_8\times E_8$ heterotic theory via $T$-duality 
\cite{NAR169,NARsw279,GIN35}, and to the type-I theory via $S$-duality 
\cite{WIT443,DAB357,HUL357,POLw460}.  A corollary of the these dualities 
is the duality between $M$-theory on a cylinder $S^1/{\bf Z}_2\times S^1$ 
and the $SO(32)$ theories on $S^1$ \cite{SCH077}.  

Now, we discuss the various limits in the moduli space of $M$-theory on 
$S^1/{\bf Z}_2\times S^1$ in terms of parameters $L$ and $R$ 
\cite{HORw460,HORw475}.  An obvious limit in the moduli space is the 
small $L$ and $R\to\infty$ limit, which is the uncompactified 
$E_8\times E_8$ heterotic string.  
Here, $L$ is the size of $S^1/{\bf Z}_2$ upon which $M$-theory is 
compactified to lead to the $E_8\times E_8$ heterotic theory.  
The string coupling of the $E_8\times E_8$ heterotic 
string is $g^{E_8}_s=L^{3\over 2}$.  The second obvious limit is $R\to 0$.  
In this limit, the $M\,2$-brane wrapped around the cylinder looks like 
open string stretched between the interval $S^1/{\bf Z}_2$ of the length $L$, 
i.e. the (uncompactified) type-I$^{\prime}$ open string with $SO(16)$ 
Chan-Paton factor attached at each end located at the 9-plane boundary.  

In general, a point in the moduli space with small $L$ and finite $R$ 
corresponds to the $E_8\times E_8$ heterotic string on $S^1$ of radius 
$R$.  A non-vanishing `Wilson line' around $S^1$ breaks $E_8\times E_8$ 
down to subgroups \cite{WIT258}, e.g. $U(1)^{16}$ 
or $SO(16)\times SO(16)$ depending on the choice of the Wilson line.  
In particular, the $E_8\times E_8$ heterotic string on $S^1$ with 
gauge group $SO(16)\times SO(16)$ is obtained from the 
$SO(32)$ heterotic string on $S^1$ with gauge group 
$SO(16)\times SO(16)$ by inverting the radius of $S^1$.  
As $R$ is decreased to a small value, one can switch to 
the $SO(32)$ heterotic string on $S^1$ of inverse radius $1/R$ 
by using the $T$-duality between the $E_8\times E_8$ and $SO(32)$ 
heterotic strings.  For this case, the string coupling of the 
$SO(32)$ heterotic string is $g^{het}_s=L/R$, which 
is small as long as $R\gg L$.  As the radius $R$ approaches smaller 
value so that $R$ becomes much smaller than $L$, one can switch to 
the type-I theory by applying the $S$-duality between the $SO(32)$ 
heterotic string and the type-I string.  The string coupling of the 
type-I theory is then given by $g^I_s=1/g^{het}_s=R/L$.  Note, under 
the ${\bf Z}_2$ transformation that exchanges two moduli $R$ and $L$ 
of the cylinder, the string couplings of the type-I and the $SO(32)$ 
heterotic theories are inverted, thereby manifesting as the non-perturbative 
type-I/heterotic duality.  

In the limit $(R,L)\to (0,0)$ with fixed small $g^I_s=R/L$, one has the 
uncompactified type-I theory.  This is understood as follows.  
We saw that the type-I$^{\prime}$ theory, which is obtained from 
the type-I theory on $S^1$ by inverting the radius, 
is the $R\to 0$ limit of $M$-theory on the cylinder.  
If we further let the length $L$ of the cylinder approach zero, 
then in the type-I side the radius of $S^1$, upon which the 
type-I theory is compactified, approach infinity (i.e. the 
decompactification limit of the type-I theory).  

So far, we discussed connections among $M$-theory and string theories 
with either $N=1$ or $N=2$ supersymmetry.  
One can further relate $N=1$ and $N=2$ theories.  
For this purpose, one breaks $1/2$ of supersymmetry in  
$N=2$ theories by compactifying on a manifold with non-trivial holonomy.  
An obvious example is the $D=6$ string-string duality between 
type-IIA theory on $K3$ and heterotic string on $T^4$ 
\cite{HULt451,WIT443}.  Both of the $D=6$ theories have (non-chiral) 
$N=2$ supersymmetry.  A corollary of this duality is that $M$-theory on 
$K3$ is equivalent to heterotic string on $T^3$ \cite{WIT443}, since 
type-IIA theory is $M$-theory on $S^1$.  The fundamental string in 
heterotic string on $T^3$ is nothing but $M\,5$-brane wrapped 
around a 4-cycle of $K3$ surface \cite{DUF442,HARs449,TOW354}.  
Furthermore, $K3$-compactified $M\,2$-brane through direct dimensional 
reduction is solitonic 5-brane in the heterotic theory wrapped around 
a 3-cycle of $T^3$.  Thus, it leads to the conjecture that the strong 
coupling limit of heterotic string on $T^3$ is $K3$-compactified 
supermembrane in $D=11$.  

We point out that $M$-theory and string theories that are connected 
within the moduli space (of either $T^2$ for the $N=2$ string theories 
or $S^1\times S^1/{\bf Z}_2$ for the $N=1$ string theories) are on an 
equal putting if one includes ``non-perturbative'' branes, as well as 
perturbative string states, within the spectra of the 5 superstring 
theories.  Namely, a brane that appears in one theory is 
necessarily related to branes of the other theories through the 
dimensional reduction and/or dualities \cite{SCH367,SCH077}.   
In particular, all the branes in the 5 string theories should have 
interpretations in terms of $M$-branes through dimensional reductions 
and string dualities.  It turns out that $p$-branes obtained 
in this way have the right property as $p$-branes of string theories 
\cite{TOW350} (e.g. the tension $T$ of branes in string-frame depend on 
the string coupling $g_s$ as $\sim 1$, $\sim 1/g_s$ and $\sim 1/g^2_s$ 
for a fundamental string, $D\,p$-branes and solitonic 5-brane, 
respectively) and the worldvolume actions (derived from those of 
$M$-theory \cite{BERst189,BERst185,BERdps224,BERdo386}) have the right 
forms \cite{TOW373,BERt173,DUFl390,SCH467}.  In the following, we discuss 
the $M$-theory origin of branes in string theories.   

First, we discuss branes in the type-IIA theory.  
Since the type-IIA theory is $M$-theory compactified on $S^1$, all 
of $p$-branes in type-IIA theory (i.e. a fundamental string and a 
solitonic 5-brane in the NS-NS sector, and $D\,p$-branes with $p=0,2,4,6,8$ 
in the R-R sector) should be obtained in this way.  

In $D=11$, there are $M\,2$-brane \cite{BERst189,DUFs253} and 
$M\,5$-brane \cite{GUE276} which are elementary and solitonic branes 
carrying electric and magnetic charges of the 3-form potential, 
respectively. Starting from $M\,2$-brane \cite{BERst189,BERst185}, 
one obtains either fundamental string \cite{DUFhis,DUFs253,DUFgt332}  
in the NS-NS sector or the $D\,2$-brane in the R-R sector, through 
double or direct dimensional reduction.  
The fundamental string and $D\,4$-brane obtained, respectively, from 
the $M\,2$- and $M\,5$-branes via double dimensional reduction have 
the right dependence of the tensions on $g_s$, i.e. $\sim 1$ and $\sim 
1/g_s$, respectively.

Next, a $D\,0$-brane can be thought of as the KK 
momentum mode of the $D=11$ theory on $S^1$ \cite{TOW350}.  
This state with the momentum number $n$ along the $S^1$-direction has mass 
(measured in the string-frame) given by $M={n\over {g_s}}$, which is the 
right dependence of the mass on the string coupling for BPS states 
carrying R-R charges \cite{WIT443}.  
The integer $n$ is the electric charge of the KK $U(1)$ gauge field 
associated with the $S^1$-direction, indicating that such KK state is 
electrically charged under the 1-form potential in the R-R sector.  
The $n=1$ KK state is interpreted as a single $D\,0$-brane and the $n>1$ 
case corresponds to the (marginal) bound state of $n$ $D\,0$-branes.   
The $D\,6$-brane is regarded as the KK monopole \cite{TOW350}, which is 
magnetically charged under the 1-form potential in the R-R sector.  
For the $D\,8$-brane, currently there is no interpretation in terms of 
the $D=11$ theory available yet.  

Second, we discuss branes in type-IIB theory.  In general,  
branes in type-IIB theory can be obtained from those in type-IIA 
theory by applying the $T$-duality between type-IIA and type-IIB 
theories \cite{BAC374,BERd380,GREht382,DEA388,BERho451}.  
For example, starting from $D\,2$-brane of type-IIA theory, one 
obtains $D\,1$-brane of type-IIB theory by 
$T$-dualizing along one of coordinates with the Neumann boundary 
condition (i.e. along a longitudinal direction of the $D\,2$-brane).   
The fundamental string and the solitonic 5-brane in the NS-NS sector 
are obtained from the $M\,2$- and the $M\,5$-branes by dimensional 
reduction similarly as in the type-IIA case.  

On the other hand, one can directly relate branes in the compactified 
type-IIB theory to those in compactified $M$-theory by applying 
equivalence between type-IIB theory on $S^1$ and  $M$-theory on $T^2$.  
In this relation, one identifies complex modulus $\tau$ of $T^2$ with 
the complex scalar $\rho$ of (uncompactified) type-IIB theory, 
i.e. $\tau=\rho$ \cite{SCH360,ASP46}.  
(This is motivated by the observation that the non-perturbative 
$SL(2,{\bf Z})$ symmetry of type-IIB theory is interpreted as the 
$T^2$ moduli group of $M$-theory on $T^2$.)  

First, we discuss 1-branes in (uncompactified) type-IIB theory.  
These carry (integer valued) electric charges
\footnote{These electric charges $(q_1,q_2)$ transform linearly under 
$SL(2,{\bf Z})$, while the complex scalar $\rho$ 
transforms fractional linearly, i.e. $\rho\to{{a\rho+b}\over{c\rho+d}}$ 
($ad-bc=1$).} 
$q_1$ and $q_2$ of 2-form potentials in the NS-NS and the R-R sectors 
\cite{SCH360}, respectively, and are bound states of $q_1$ fundamental 
strings and $q_2$ $D$-strings.  This bound state is absolutely stable 
against decay into individual strings iff the integers $q_1$ and $q_2$ 
are relatively prime \cite{WIT460}, due to the ``tension gap'' and 
charge conservation.  
In the string-frame with the vacuum expectation value 
$\langle\rho\rangle=i(g^{(B)}_s)^{-1}$, the tension of the $(q_1,q_2)$ 
string \cite{SCH360} is $T_{(q_1,q_2)}=\sqrt{q^2_1+(g^{(B)}_s)^{-2}
q^2_2}T^{(B)}_1$, where $T^{(B)}_1$ is the tension of the fundamental 
string.  This tension formula has the right limiting behavior:  
$T_{(1,0)}\sim 1$ for a fundamental string and $T_{(0,1)}\sim 
(g^{(B)}_s)^{-1}$ for a $D$-string \cite{POL75}.  

When these 1-branes are wrapped around $S^1$ of radius $R_B$, 
one has 0-branes in $D=9$ with the momentum mode $m$ and the 
winding mode $n$ around $S^1$.  This 0-brane of the $D=9$ type-IIB theory 
is identified with the $M\,2$-brane wrapped around $T^2$.  Namely, the 
momentum mode $m$ [winding mode $n$] of the type-IIB string is interpreted 
in the $M\,2$-brane language as the wrapping [the KK modes] of the 
$M\,2$-brane on $T^2$.  Through these identifications, one has 
relations between the tension of the fundamental string of 
(uncompactified) type-IIB theory and the tension $T^{(M)}_2$ of  
$M\,2$-brane: $(T^{(B)}_1L^2_B)^{-1}={1\over{(2\pi)^2}}T^{(M)}_2
A^{3/2}_M$, which is consistent with string dualities.  
Here, $L_B=2\pi R_B$ is the circumference of $S^1$, upon which  
type-IIB theory is compactified, and $A_M$ is the area of $T^2$ measured 
in the $D=11$ metric.

Direct dimensional reduction of 1-branes in type-IIB theory gives 
rise to strings with charges $(q_1,q_2)$ and the tension $T_{(q_1,q_2)}$ 
in $D=9$.  This type-IIB string in $D=9$ is identified with $M\,2$-brane 
wrapped around a $(q_1,q_2)$ homology cycle of $T^2$ with the minimal 
length $L_{(q_1,q_2)}=2\pi R_{11}|q_1-q_2\tau|$.  
Such string of the compactified $M$-theory has the tension (measured by 
the $D=11$ metric) $T^{(11)}_{(q_1,q_2)}=L_{(q_1,a_2)}T^{(M)}_2$, 
which is consistent with relation between $T^{(B)}_1$ and 
$T^{(M)}_2$ in the previous paragraph.  

Second, we discuss the $D=9$ $p$-branes related to  
$D\,3$-brane \cite{DUFl273} of (uncompactified) type-IIB theory.  
By wrapping $D\,3$-brane around $S^1$, one obtains  
2-brane with tension $L_BT^{(B)}_3$ in $D=9$.  Here, 
$T^{(B)}_3$ is the tension of $D\,3$-brane in $D=10$.  
This 2-brane of type-IIB theory is identified with 2-brane 
in the $T^2$-compactified $M$-theory obtained by direct dimensional 
reduction of $M\,2$-brane.  Such identification of the two 2-branes 
of type-IIB and $M$-theory leads to relation between the tensions 
$T^{(B)}_1$ and $T^{(B)}_3$ of fundamental string and $D\,3$-brane:  
$T^{(B)}_3={1\over{2\pi}}(T^{(B)}_1)^2$.  When $D\,3$-brane of 
type-IIB theory is compactified via direct dimensional reduction on 
$S^1$, one has 3-brane in $D=9$.  This 3-brane is identified with 
$M\,5$-brane wrapped around $T^2$.  Such an identification leads to the 
correct DSZ quantization relation $T^{(M)}_5={1\over{2\pi}}(T^{(M)}_2)^2$ 
on tensions of $M\,2$- and $M\,5$-branes.  

Third, 5-branes in type-IIB theory carry magnetic charges 
$(p_1,p_2)$ of 2-form potentials in the R-R and the NS-NS sectors, with 
the tension $T^{(B)}_{5(p_1,p_2)}$ given similarly as that of 
1-branes.  4-branes with the tension $L_BT^{(B)}_{5(p_1,p_2)}$ are obtained 
by wrapping this 5-branes around $S^1$.  The corresponding 4-branes in 
the $M$-theory side is $M\,5$-brane wrapped around a $(p_1,p_2)$ cycle 
of $T^2$.  This identification leads to the correct expression for 
the tension of the type-IIB 5-brane (in the string-frame) given by 
$T^{(B)}_{5(p_1,p_2)}={{(g^{(B)}_s)^{-1}}\over{(2\pi)^2}}|p_1-p_2\langle\rho
\rangle|(T^{(B)}_1)^3$.  In the limit where the 5-brane carries 
only either the R-R or the NS-NS magnetic charge, the tension behaves as 
$T^{(B)}_{5(1,0)}\sim (g^{(B)}_s)^{-1}$ and $T^{(B)}_{5(0,1)}\sim 
(g^{(B)}_s)^{-2}$, as expected for solitons and $D$-branes.  

5-branes in $D=9$ have different interpretations.  
First, a singlet 5-brane of the type-IIB theory on $S^1$ is 
magnetically charged under the KK $U(1)$ gauge field associated with the 
$S^1$-direction.  Second, the $SL(2,{\bf Z})$ family of 5-branes of the 
type-IIB theory on $S^1$ is charged under the doublet of 2-form $U(1)$ 
gauge fields.  The corresponding singlet 5-brane in the $T^2$-compactified 
$M$-theory is magnetically charged under the 3-form $U(1)$ gauge field.  
The $SL(2,{\bf Z})$ multiplet of 5-branes that are matched with those of 
the type-IIB theory on $S^1$ is magnetically charged under the doublet of 
KK $U(1)$ fields of $M$-theory on $T^2$.  

As for the $D=9$ branes associated with $D\,7$-brane 
(magnetically charged under the 0-form potential), 
the $M$-theory interpretation is not well-understood 
yet.  In \cite{SCH367}, it is argued that $p$-branes with $p=7,8,9$ in 
$M$-theory that would give rise to 7-brane in $D=9$ do not 
exist, and 6-brane in $D=9$ cannot be obtained from 
$D\,7$-brane of type-IIB theory by the periodic 
array along the compact direction and also is not consistent with 
the $D=9$ type-IIB theory.  

Finally, we comment on branes in the $SO(32)$ theories, i.e. 
the type-I and the $SO(32)$ heterotic strings.  These 2 theories 
are related by $S$-duality, which inverts the string couplings 
(i.e. $g^{het}_s=1/g^I_s$) and exchanges the 2-form potentials 
of the 2 theories (the 2-form potential is in the NS-NS sector [the R-R 
sector] for the $SO(32)$ heterotic theory [the type-I theory]).  
The electric [magnetic] charge of the 2-form potential is carried by 1-branes 
[5-branes].  When this 1-brane [5-brane] is compactified on $S^1$, 
one has either 0-brane or 1-brane [4-brane or 5-brane] in $D=9$  
depending on whether or not these branes are wrapped around $S^1$.  
The $M$-theory origin of these $D=9$ branes is understood from 
the observation that $M$-theory on $S^1/{\bf Z}_2\times S^1$ 
with length $L$ and radius $R$ is related to $SO(32)$ theory on $S^1$.  
Note, while $M\,2$-brane can wrap on $S^1/{\bf Z}_2$, $M\,5$-brane can 
wrap around $S^1$, only.  So, $M\,5$-brane compactified on 
$S^1/{\bf Z}_2\times S^1$ give rise to either 4-brane or 5-brane 
in $D=9$, depending on whether $M\,5$-brane is wrapped around 
$S^1$ or not.  These branes are identified with those of the $SO(32)$ 
theory.  Similarly, 0-brane and 1-brane of the $SO(32)$ theories on 
$S^1$ are identifies with the $M\,2$-brane which is first wrapped around 
$S^1/{\bf Z}_2$ and then either wrapped around $S^1$ or not.  Note, 
under the exchange of parameters  $L$ and $R$ of $S^1/{\bf Z}_2\times S^1$, 
the $SO(32)$ heterotic theory and the type-I theory is exchanged, while 
the string couplings are inverted (i.e. $g^{het}_s=1/g^I_s=L/R$).  
Thus, the roles of $R$ and $L$ are interchanged when one identifies 
$p$-branes of the $S$-dual theory on $S^1$ with those of the 
$M$-theory on $S^1/{\bf Z}_2\times S^1$.  These identifications yield the 
tensions for 1-brane and 5-brane of the $SO(32)$ theories with consistent 
limiting behavior, i.e. $T^{het}_5\sim (g^{het}_s)^{-2}$ and 
$T^{het}_1\sim 1$ for the heterotic theory, and $T^I_5\sim (g^I_s)^{-1}$ 
and $T^I_1\sim (g^I_s)^{-1}$ for the type-I theory.  

\section{Black Holes in Heterotic String on Tori}\label{n4bh}
\subsection{Solution Generating Procedure}\label{n4bhgen}

The primary goal of this chapter is to generate the general 
black hole solutions in the effective theories of the heterotic string 
on tori by applying the solution generating transformations described 
in section \ref{dualn4}.  

In principle, $D<10$ black hole solutions which have 
the most general electric/magnetic charge configurations and 
are compatible with the conjectured no-hair theorem 
\cite{ISR164,ISR8,CAR26,HAW71,HAW25} can be constructed by imposing 
the $SO(1,1)$ boosts on charge neutral solutions, i.e. 
Schwarzschield or Kerr solution.  Here, the $SO(1,1)$ boosts 
that generate electric charges of $U^{36-2D}(1)$ gauge group in 
$D<10$ are contained in the $O(11-D,27-D)$ symmetry group  
(\ref{lowtran}) of the $(D-1)$-dimensional Lagrangian.  Meanwhile, 
magnetic charges of the $U^{36-2D}(1)$ gauge group and the $D=5$  
NS-NS 2-form potential are generated by the $SO(1,1)$ 
boosts in the $O(8,24)$ symmetry group (\ref{o824}) of the $D=3$ action.  

However, it is not necessary to generate all the electric/magnetic charges by 
applying the $SO(1,1)$ boosts, as we explain in the following.  
The $D$-dimensional dualities, which leave the (Einstein-frame) 
metric intact, can be used to remove some of charge degrees of 
freedom (associated with the $U^{36-2D}(1)$ gauge group in $D<10$) of 
black holes.  The general black hole solution with all the 
redundant charge degrees of freedom removed by the $D$-dimensional dualities 
is called the ``generating solution'', since the most general solution in 
the class is obtained by applying the $D$-dimensional dualities. 
Thus, one only needs to generate electric [and magnetic] charges of the 
generating solutions by applying the $SO(1,1)$ boosts in $O(11-D,27-D)$ 
[$O(8,24)$] duality group on the Schwarzschield or the Kerr solution 
\cite{SEN440}.  The generating solution is equivalent to the solution 
with the most general charge configuration due to the conjectured 
string dualities.  
This is a reminiscence of automorphism transformations of 
$N$-extended superalgebra discussed in section \ref{bpssusy}, which brings 
the algebra in a simple form in which only $[N/2]$ eigenvalues of central 
charge matrix appear in the algebra rather than whole $N(N-1)$ central 
charges.  The charge assignments for the generating solution for each 
dimensions are: 
\begin{itemize}
\item dyonic black holes in $D=4$ \cite{CVEy127}: 
5 charge degrees of freedom associated with gauge fields in the 
$T^2$ part.
\item black holes in $D=5$ \cite{CVEy476}: 
a magnetic charge of the NS-NS 3-form field strength (or an electric 
charge of its Hodge-dual), and 2 electric charges of KK 
and 2-form $U(1)$ gauge fields associated with the same internal 
coordinate. 
\item  black holes in $D\geq 6$ \cite{CVEy477}:
2 electric charges of KK and 2-form $U(1)$ gauge 
fields associated with the same internal coordinate. 
\end{itemize}

For the purpose of constructing solutions, it is convenient to choose  
scalar asymptotic values in the ``canonical forms'' \cite{SEN440}:  
$M_{\infty}=I_{10-D,26-D}$ and $\varphi_{\infty}=0$ [and $\Psi_{\infty}=0$ 
for the $D=4$ case].  This is not an arbitrary choice since 
one can bring arbitrary scalar asymptotic values to the 
canonical forms by applying the following $O(10-D,26-D,{\bf R})$ 
transformation: 
\begin{equation}
M_{\infty}\to \hat{M}_{\infty}=\Omega M_{\infty}\Omega^T =I_{10-D,26-D}, 
\ \ \ \Omega\in O(10-D,26-D,{\bf R}), 
\label{canmodaym}
\end{equation}
and the $S$-duality, for example for the $D=4$ 
case, given by the $SL(2,{\bf R})$ transformation
\begin{equation}
S_{\infty}\to \hat{S}_{\infty}=(aS_{\infty}+b)/d =i, \ \ \ ad=1.
\label{cancxasym}
\end{equation}
The $D=3$ $O(8,24,{\bf R})$ transformation that brings an asymptotic 
value of modulus ${\cal M}$ (\ref{modultwo}) to the form 
${\cal M}_{\infty}=I_{4,28}$ is equivalent \cite{SEN434} to the $D=4$
$SL(2,{\bf R})$ transformation plus the $D=3$ $O(7,23,{\bf R})$ 
transformation.  
Furthermore, one brings asymptotic values of $U(1)$ gauge 
fields ${\cal A}^i_{\mu}$ to zero by applying global $U(1)$ gauge 
transformations.  

Then, the subset of $O(11-D,27-D)$ [$O(8,24)$] that preserves the 
canonical asymptotic value $M_{\infty}=I_{10-D,26-D}$ [${\cal M}_{\infty}
=I_{4,28}$] is $SO(26-D,1)\times SO(10-D,1)$ [$SO(22,2)\times SO(6,2)$] 
\cite{SEN440}.  There are $36-2D$ [$2\times 28$] $SO(1,1)$ boosts in  
$SO(26-D,1)\times SO(10-D,1)$ [$SO(22,2)\times SO(6,2)$].  
When applied to a charge neutral solution, these boosts in 
$SO(26-D,1)\times SO(10-D,1)$ [$SO(22,2)\times SO(6,2)$] induce electric 
charges of the $U(1)^{36-2D}$ gauge group in $D<10$  
[electric and magnetic charges of the $U(1)^{28}$ gauge group in $D=4$]
\footnote{While the $SO(1,1)$ boosts in $SO(22,1)\times SO(6,1) \subset 
SO(22,2)\times SO(6,2)$  induce electric charges of the $D=4$ 
$U(1)^{28}$ gauge group, the remaining $SO(1,1)$ boosts in $SO(22,2)\times 
SO(6,2)-SO(26-D,1)\times SO(10-D,1)$ induce magnetic charges of the 
$U(1)^{28}$ gauge group.}.  

The starting point of constructing the generating solution is 
the $D$-dimensional Kerr solution, parameterized by 
the ADM mass and $[{{D-1}\over 2}]$ angular momenta.  
The solution in the ``Boyer-Lindquist'' coordinate has the  
form \cite{MYEp172}: 
\begin{eqnarray}
ds^2&=&-{{(\Delta-2N)}\over \Delta}dt^2+ {\Delta \over
{\prod^{[{{D-1}\over 2}]}_{i=1}(r^2+l^2_i)-2N}}dr^2
\cr
& &+(r^2+l^2_1\cos^2\theta+K_1\sin^2\theta)d\theta^2 
\cr
& &+(r^2+l^2_{i+1}\cos^2\psi_i+K_{i+1}\sin^2\psi_i)
\cos^2\theta\cos^2\psi_1\cdots\cos^2\psi_{i-1}d\psi^2_i 
\cr
& &-2(l^2_{i+1}-K_{i+1})\cos\theta\sin\theta\cos^2\psi_1\cdots
\cos^2\psi_{i-1}\cos\psi_i\sin\psi_i d\theta d\psi_i 
\cr
& &-2\sum_{i<j}(l^2_j-K_j)\cos^2\theta\cos^2\psi_1\cdots
\cos^2\psi_{i-1}
\cr
& &\ \ \ \ \times\cos\psi_i\sin\psi_i\cdots\cos^2\psi_{j-1}
\cos\psi_j\sin\psi_jd\psi_i d\psi_j \cr
& &+{\mu^2_i \over \Delta}[(r^2+l^2_i)\Delta+2l^2_i\mu^2_iN ]
d\phi^2_i - {{4l_i\mu^2_iN}\over \Delta} dtd\phi_i 
\cr
& &+\sum_{i<j}{{4l_i l_j\mu^2_i\mu^2_jN}\over \Delta}d\phi_i d\phi_j, 
\label{kerr}
\end{eqnarray}
where for 
\begin{itemize}
\item  Even dimensions:
\begin{eqnarray}
\Delta &\equiv& \alpha^2\prod^{{D-2}\over 2}_{i=1}(r^2+l^2_i)
+r^2\sum^{{D-2}\over 2}_{i=1}\mu^2_i(r^2+l^2_1)\cdots(r^2+l^2_{i-1})
\cr
& &\ \ \ \ \times(r^2+l^2_{i+1})\cdots(r^2+l^2_{{D-2}\over 2}), 
\cr
K_i&\equiv&l^2_{i+1}\sin^2\psi_i+\cdots+l^2_{{D-2}\over 2}
\cos^2\psi_i\cdots\cos^2\psi_{{D-6}\over 2}\sin^2\psi_{{D-4}\over 2},
\cr
N&=&mr ,
\label{edef1}
\end{eqnarray}
and
\begin{eqnarray}
\mu_1 &\equiv&\sin\theta,\ \ \ \  \mu_2\equiv\cos\theta\sin\psi_1,
\ \ \ \ \cdots ,
\cr
\mu_{{D-2}\over 2}&\equiv&\cos\theta\cos\psi_1\cdots
\cos\psi_{{D-6}\over 2}\sin\psi_{{D-4}\over 2},
\cr
\alpha&\equiv&\cos\theta\cos\psi_1\cdots\cos\psi_{{D-4}\over 2}, 
\label{edef2}
\end{eqnarray}
\item Odd dimensions:
\begin{eqnarray}
\Delta &\equiv& r^2\sum^{{D-1}\over 2}_{i=1}\mu^2_i(r^2+l^2_1)\cdots
(r^2+l^2_{i-1})(r^2+l^2_{i+1})\cdots (r^2+l^2_{{D-1}\over 2}),
\cr
K_i&\equiv&l^2_{i+1}\sin^2\psi_i+\cdots+l^2_{{D-3}\over 2}
\cos^2\psi_i\cdots\cos^2\psi_{{D-7}\over 2}\sin^2\psi_{{D-5}\over 2}
\cr
& &+l^2_{{D-1}\over 2}\cos^2\psi_i\cdots\cos^2\psi_{{D-5}\over 2}, 
\cr
N&=&mr^2,
\label{odef1}
\end{eqnarray}
and
\begin{eqnarray}
\mu_1&\equiv&\sin\theta,\ \ \ \  \mu_2\equiv\cos\theta\sin\psi_1,
\ \ \ \  \cdots , 
\cr
\mu_{{D-3}\over 2}&\equiv&\cos\theta\cos\psi_1\cdots
\cos\psi_{{D-7}\over 2}\sin\psi_{{D-5}\over 2},\cr
\mu_{{D-1}\over 2}&\equiv&\cos\theta\cos\psi_1\cdots
\cos\psi_{{D-5}\over 2}.
\label{odef2}
\end{eqnarray}
\end{itemize}
Here, the repeated indices are summed over: $i,j$ in
$\psi$  [$\phi$]  run from 1 to $[{{D-4}\over 2}]$    
[from 1 to $[{{D-1}\over 2}]$].  
The ADM mass and the angular momenta $J_i$ are
\begin{equation}
M={{(D-2)\Omega_{D-2}}\over {8\pi G_D}}m,  \ \ \ \ \ \ 
J_i={\Omega_{D-2}\over {4\pi G_D}}ml_i = {2\over{D-2}}Ml_i, 
\label{kerrpar}
\end{equation}
where $G_D$ is the $D$-dimensional Newton's constant and 
$\Omega_{D-2}={{2\pi^{(D-1)/2}}\over {\Gamma({{D-1}\over 2})}}$ is the 
area of $S^{D-2}$.   

When compactified to $D-1$ dimensions [3 dimensions (for the 4-dimensional 
Kerr solution)], the transformation that generates inequivalent solutions 
from the Kerr solution is $(SO(26-D,1)\times SO(10-D,1))/
(SO(26-D)\times SO(10-D))$ [$(SO(22,2)\times SO(6,22))/(SO(22)\times 
SO(6)\times SO(2))$], which has $(9-D)+(25-D)$ [$2\times 28 +1$] 
parameters; these parameters are interpreted as $(9-D)+(25-D)$ electric 
charge degrees of freedom [$2\times 28$ electric and magnetic charge 
degrees of freedom plus unphysical Taub-NUT charge] introduced to the 
Kerr solution \cite{SEN440}.  For the $D=5$ case, an 
additional charge associated with the NS-NS 2-form field is generated 
by an $SO(1,1)$ boost in $O(8,24)$ \cite{CVEy476}. 
 
After the generating solutions are constructed from the $SO(1,1)$  
boosts, the remaining charge degrees of freedom are induced 
(without changing the Einstein-frame spacetime) from subsets 
of $D$-dimensional (continuous) duality transformations that generate 
new charge configurations from the generating ones while keeping the 
canonical scalar asymptotic values intact.  This is 
$[SO(10-D)\times SO(26-D)]/[SO(9-D) \times SO(25-D)]$, which 
introduces $(9-D)+(25-D)$ new charge degrees of freedom, and, for the 
$D=4$ case, $SO(1)\subset SL(2,{\bf R})$, which introduces 
one more charge degree of freedom.  Subsequently, to obtain 
solutions with arbitrary scalar asymptotic values, one has to undo 
the transformations (\ref{canmodaym}) and (\ref{cancxasym}).  In the 
following sections, we discuss the generating solutions in each dimensions.  
Note, due to the conjectured string-string duality between heterotic 
string on $T^4$ and type-II string on $K3$, these also correspond to the 
generating solutions of general black holes in type-II strings on 
$K3\times T^n$ for $n=6-D=0,1,2$.  For this case, some of charges of the 
generating solutions can be dualized to R-R charges, rendering 
interpretation in terms of $D$-branes.
It turns out that by applying $U$-dualities of type-II string on 
tori to such generating solutions, one can generate the general class 
of solutions of the effective type-II string on tori as well \cite{CVEh} 
(see section \ref{bprhghclas}) for discussions).

\subsection{Static, Spherically Symmetric Solutions in 
Four Dimensions}\label{n4bh4dsph}

\subsubsection{Supersymmetric Solutions}\label{n4bh4dsphsusy}

In this section, we derive a general BPS  
spherically symmetric solution with a diagonal moduli \cite{CVEy672}.  
Such a solution, after subsets of $O(6,22)$ and 
$SL(2,{\bf R})$ transformations are applied, satisfies one $U(1)$ 
charge constraint, missing one parameter to describe the most general 
BPS solutions. 
The solution generalizes the previously known black hole solutions in 
heterotic string on tori as special cases, and are shown to be exact 
to all orders in expansions of $\alpha^{\prime}$ \cite{CVEt366}.  
At particular points in moduli space, such a solution becomes massless, 
enhancing not only gauge symmetry but also supersymmetry \cite{CVEy674}.  

\paragraph{Generating Solutions}\label{n4bh4dsphsusysol}

A general BPS non-rotating black hole solution with a diagonal 
moduli matrix is obtained by solving the Killing spinor equations.  
With spherically symmetric Ans\" atze for fields and a diagonal form of 
moduli $M$, the Killing spinor equations $\delta\psi_{M}=0$, 
$\delta\lambda=0$ and $\delta\chi^I=0$ (Cf. (\ref{hetsusy})) are  
satisfied by restricted charge configurations (see \cite{CVEy672} 
for details on allowed charge configurations), which we choose without 
loss of generality to be $P^{(1)}_1, P^{(2)}_1, Q^{(1)}_2, Q^{(2)}_2$.  
The explicit BPS non-rotating solution with such a charge configuration has 
the form \cite{CVEy672}:
\begin{eqnarray}
\lambda &=& r^2/[(r-\eta_P P^{(1)}_{1})(r-\eta_P P^{(2)}_{1})
(r- \eta_Q Q^{(1)}_{2})(r-\eta_Q Q^{(2)}_{2})]^{1\over 2},
\nonumber\\
R &=& [(r-\eta_P P^{(1)}_{1})(r - \eta_P P^{(2)}_{1})
(r - \eta_Q Q^{(1)}_{2})(r-\eta_Q Q^{(2)}_{2})]^{1\over 2},
\nonumber\\
e^{\varphi}&=&\left [{(r-\eta_P P^{(1)}_{1})
(r- \eta_P P^{(2)}_{1})} \over {(r- \eta_Q Q^{(1)}_{2})
(r- \eta_Q Q^{(2)}_{2})}\right]^{1\over 2},
\nonumber\\
g_{11}&=&\left ({{r- \eta_P P^{(2)}_{1}} \over {r-\eta_P P^{(1)}_{1}}}
\right ), \
g_{22}=\left ({{r- \eta_Q Q^{(1)}_{2}} \over {r- \eta_Q Q^{(2)}_{2}}}
\right ),\
g_{mm}=1\ \ \ (m \neq 1,2),
\label{gensol}
\end{eqnarray}
where $\lambda$ and $R$ are components of the metric $g_{\mu\nu}dx^{\mu}
dx^{\nu}=-\lambda dt^2+\lambda^{-1}dr^2+R(d\theta^2+{\rm sin}^2\theta 
d\phi^2)$, $\eta_{P,Q}=\pm 1$ and the radial coordinate is chosen so that 
the horizon is at $r=0$.  

The requirement that the ADM mass saturates 
the Bogomol'nyi bound restricts choice of $\eta_{P,Q}$
such that $\eta_P {\rm sign}(P^{(1)}_1+P^{(2)}_1)=-1$ and $\eta_Q
{\rm sign}(Q^{(1)}_2+Q^{(2)}_2)=-1$, thus yielding non-negative 
ADM mass of the form
\begin{equation}
M_{BPS} = |P^{(1)}_{1}+ P^{(2)}_{1}|+|Q^{(1)}_{2}+ Q^{(2)}_{2}|.
\label{bpsmass}
\end{equation}
Note, the relative signs for the pairs $(Q^{(1)}_2,Q^{(2)}_2)$ and 
$(P^{(1)}_1,P^{(2)}_1)$ are not restricted but in this 
section we consider the case where both of pairs have 
the same relative signs so that the solution (\ref{gensol}) has 
regular horizon.  

The Killing spinor $\varepsilon$ of the above BPS solution satisfies the 
following constraints:
\begin{itemize}
\item 
$P^{(1)}_1 \ne 0$ and/or $P^{(2)}_1 \ne 0$: 
$\hat{\Gamma}^5\hat{\Gamma}^{a=1}\varepsilon = i\eta_P\varepsilon$ 
\item
$Q^{(1)}_2 \ne 0$ and/or $Q^{(2)}_2 \ne 0$: 
$\hat{\Gamma}^0\hat{\Gamma}^{a=2}\varepsilon = \eta_Q\varepsilon$.
\end{itemize}
From these constraints, one sees that purely electric (or magnetic) 
solutions preserve $1/2$, while dyonic solutions preserve 
$1/4$ of $N=4$ supersymmetry.  The former and the latter 
configurations fall into vector- and hyper-supermultiplets, respectively.  

Since the Killing spinor equations are invariant under the $O(6,22)$ and 
$SL(2,{\bf R})$ transformations, one can generate new BPS solutions 
by applying the $O(6,22)$ and $SL(2,{\bf R})$ transformations to a 
known BPS solution.   The $[SO(6)/SO(4)]\times [SO(22)/SO(20)]$ 
transformation with ${{6\cdot 5 - 4\cdot 3}\over 2}+{{22\cdot 21 
-20\cdot 19} \over 2}=50$ parameters to the solution (\ref{gensol}) 
leads to a general solution with zero axion and $4+50=54=56-2$ charges.  
28 electric ${\vec Q}$ and 28 magnetic ${\vec P}$ charges of such a 
solution satisfy the two constraints
\begin{equation}
{\vec P}^T{\cal M}_{\pm}{\vec Q}=0\ \ \ \
({\cal M}_{\pm} \equiv (LML)_{\infty}\pm L).
\label{gencon}
\end{equation}
The subsequent $SO(2) \subset SL(2,{\bf R})$ transformation introduces one
more parameter (along with a non-trivial axion), which replaces 
the two constraints (\ref{gencon}) with the following one $SL(2,{\bf R})$ 
and $O(6,22)$ invariant constraint on charges:
\begin{equation}
{\vec P}^T{\cal M}_{-}{\vec Q}\,[{\vec Q}^T{\cal M}_{+}{\vec Q} -
{\vec P}^T{\cal M}_{+}{\vec P}] - (+ \leftrightarrow -) = 0.
\label{fgencon}
\end{equation}
Thus, general solution in this class has $4+50+1=55=2\cdot 28-1$ charge 
degrees of freedom.  

By applying the $O(6,22)$ and $SL(2,{\bf R})$ transformations to 
(\ref{bpsmass}), one obtains the following ADM mass for general 
solutions preserving $1/4$ of supersymmetry:
\begin{eqnarray}
M^2_{BPS}&=&e^{-\varphi_\infty} \left\{{\vec P}^T {\cal M}_+ {\vec P} +
{\vec Q}^T {\cal M}_+ {\vec Q}\right.
\cr
& &\left.\ \ + 2\left [({\vec P}^T{\cal M}_+{\vec P})
({\vec Q}^T{\cal M}_+{\vec Q})-({\vec P}^T{\cal M}_+
{\vec Q})^2\right]^{1\over 2}\right\}. 
\label{Bogmass}
\end{eqnarray}
This agrees with the expression (\ref{massbound}) obtained \cite{DUFlr} 
by the Nester's procedure.
When magnetic $\vec{P}$ and electric $\vec{Q}$ charges
are parallel in the $SO(6,22)$ sense, i.e. 
$\vec{P}^T{\cal M}_{+}\vec{Q}=0$, (\ref{Bogmass}) becomes 
the ADM mass of configurations preserving $1/2$ of $N=4$ 
supersymmetry 
\cite{HARl,BANddf212,KHU259,KHU294,KHU387,SENint,SEN10,GIBp84,PEE456}: 
\begin{equation}
M^2_{BPS}=e^{-\phi_{\infty}}(\vec{P}^T{\cal M}_{+}\vec{P} + 
\vec{Q}^T{\cal M}_{+}\vec{Q}),
\label{admsen}
\end{equation}
whose corresponding generating solution is purely electric or magnetic  
subset of (\ref{gensol}).  

\paragraph{General Supersymmetric Solution with Five Charges}

The BPS solution (\ref{gensol}) has a charge 
configuration satisfying one constraint (\ref{fgencon}) when 
further acted on by the $[SO(6)/SO(4)]\times[SO(22)/SO(20)]$ and $SO(2)$ 
transformations.  So, to construct the generating solution for the 
most general BPS solution, which conforms to the conjectured classical 
``no-hair'' theorem, one has to introduce one more charge degree of 
freedom into (\ref{gensol}). 

Such a generating solution was constructed in \cite{CVEt53} 
by using the chiral null model approach, and has the following 
charge configuration
\begin{eqnarray}
(Q^{({1})}_1, P^{(1)}_1)&=&(q,P_1), \ \  \ \
(Q^{({1})}_2, P^{(1)}_2)=(Q_1,0),
\cr
(Q^{({2})}_1, P^{(2)}_1)&=&(-q,P_2), \ \  \ \
(Q^{({2})}_2, P^{(2)}_2)=(Q_2,0) .
\label{chnullchr}
\end{eqnarray}
Explicitly, the solution has the form:
\begin{eqnarray}
\lambda&=&{{r^2}\over{\bigg[(r+Q_1)(r+Q_2)(r+P_1)(r+P_2)
-q^2 [r + {1\over 2}(P_1 +P_2)]^2\bigg]^{1\over 2}}},
\cr
e^{2\varphi}&=&{{(r+P_1)(r+P_2)}\over{\bigg[(r+Q_1)(r+Q_2)(r+P_1)(r+P_2)
-q^2 [r + {1\over 2}(P_1 +P_2)]^2\bigg]^{1\over 2}}},
\cr
\Psi&=&{q (P_2 - P_1)\over{2(r+P_1)(r + P_2)}},
\cr
G_{11}&=&{{r+P_2}\over{r+P_1}}, \ 
G_{22}={{r+Q_1}\over{r+Q_2}}, \ 
G_{12}=-B_{12}=
{{q [r+{1\over 2} (P_1+P_2)]}\over{(r+Q_2)(r+P_1)}}.
\label{chnulsol}
\end{eqnarray}
For this solution to have a regular horizon, the charges have to satisfy 
the constraints
\begin{eqnarray}
P_1&>&0, \ \ \ P_2>0, \ \ \ Q_1>0,\ \ \ Q_2>0, 
\cr
Q_1Q_2-q^2&>&0,\ \ \ \  \ \ (Q_1Q_2-q^2)P_1P_2- 
{1\over 4} q^2(P_1-P_2)^2>0.
\label{chnullchcons}
\end{eqnarray}
The ADM mass has the same form as that of the 4-parameter 
solution (\ref{gensol}):
\begin{equation}
M_{ADM} = Q_1 +Q_2 +P_1 +P_2,
\label{chnullbhmass}
\end{equation}
independent of the additional parameter $q$.
Meanwhile, the horizon area, i.e. 
${\bf A}\equiv 4\pi(\lambda^{-1}r^{2})_{r=0}$, is 
modified in the following way due to $q$:
\begin{equation}
{\bf A}= 4\pi \bigg[(Q_1Q_2-q^2)P_1P_2-
{1\over 4}q^2(P_1-P_2)^2\bigg]^{1\over 2}.  
\label{chnullbhmes}
\end{equation}

The following ADM mass and horizon area of BPS non-rotating 
black hole with general charge configuration are obtained by 
applying the $[SO(6)/SO(4)]\times [SO(22)/SO(20)]$ and $SO(2)\subset 
SL(2,{\bf R})$ transformations to (\ref{chnullbhmass}) and 
(\ref{chnullbhmes}):  
\begin{eqnarray}
M_{ADM}^2&=&{1\over 2}{\vec{\tilde\alpha}^T}\mu_+{\vec{\tilde\alpha}}
+{1\over 2}e^{-4\varphi_{\infty}}{\vec \beta}^T 
\mu_+{\vec \beta}
\cr
& &\ \ \ +e^{-2\varphi_{\infty}} 
\left[({\vec \beta}^T\mu_+{\vec \beta})({\vec \alpha}^T\mu_+
{\vec \alpha})-({\vec \beta}^T\mu_+{\vec \alpha})^2\right]^{1\over 2},
\cr
{\bf A}&=&\pi e^{2\varphi_{\infty}}\bigg[({\vec \beta}^T 
L{\vec \beta})({\vec {\alpha}}^TL{\vec {\alpha}})-({\vec \beta}^TL
{\vec{\alpha}})^2\bigg]^{1\over 2}, 
\label{chnullmassarea}
\end{eqnarray}
where the charge lattice vectors $\vec{\alpha}$ and $\vec{\beta}$ live 
on the even self-dual Lorentzian lattice $\Lambda_{6,22}$ with signature 
$(6,22)$, ${\vec{\tilde \alpha}}\equiv{\vec \alpha}+\Psi_{\infty}
{\vec\beta}$ and $\mu_{\pm}\equiv M_{\infty}\pm L$. 
Here, $\vec{\alpha}$ and $\vec{\beta}$ are related to the physical $U(1)$ 
charges $\vec Q$ and $\vec P$ as:
\begin{equation}
\sqrt{2}Q_i=e^{2\varphi_{\infty}}
M_{ij\,\infty}(\alpha_j+\Psi_{\infty}\beta_j),
\ \ \ \ \ \ \ 
\sqrt{2}P_i=L_{ij}{\beta}_j,  
\label{chnulllatchrel}
\end{equation}
where we assumed that $\alpha'=2,\ G_N={1 \over 8}\alpha^{\prime} 
e^{2\varphi_{\infty}}={1\over 4}e^{2\varphi_{\infty}}$.  

The ADM mass and horizon area in (\ref{chnullmassarea}) 
can be put in the $SL(2,{\bf Z})$ $S$-duality, as well as the $O(6,22)$ 
$T$-duality, invariant forms by expressing them with new charge 
lattice vector ${\vec v}^T=({\vec v}^{1\,T},{\vec v}^{2\, T})\equiv 
({\vec \alpha}^T,{\vec \beta}^T)$ and by introducing the following 
$SL(2,{\bf R})$ invariant matrices:
\begin{equation}
{\cal M}={\rm e}^{2\varphi}\left(\matrix{1&\Psi\cr
\Psi& \Psi^2+{\rm e}^{-4\varphi}} \right ) \ \ ,\ \
{\cal L}=\left(\matrix{0&1\cr -1&0}\right).
\label{mlsl}
\end{equation}
The final forms are 
\begin{eqnarray}
M_{ADM}^2&=&(8G_N)^{-1}\left({\cal M}_{\infty\,ab}({\vec v}^{a\,T}\mu_+ 
{\vec v}^b)+\left[2{\cal L}_{ac}{\cal L}_{bd}({\vec v}^{a\,T}\mu_+
{\vec v}^b)({\vec v}^{c\, T}\mu_+{\vec v}^d)\right]^{1\over 2}\right),
\cr
{\bf S}&=&{{\bf A}\over{4G_N}}=\pi\bigg[{1\over 2}{\cal L}_{ac}
{\cal L}_{bd}({\vec v}^{a\,T}L{\vec v}^{b})
({\vec v}^{c\,T}L{\vec v}^{d})\bigg]^{1\over 2}.
\label{chnullbhgenentp}
\end{eqnarray}
These are manifestly $SL(2,{\bf R})$ invariant, since ${\cal M}$  
and ${\vec v}$ transform under $SL(2,{\bf R})$ as \cite{SENint}
\begin{equation}
{\cal M }\to \omega {\cal M }\omega^T ,\ \ \ 
{\vec v}\to {\cal L}\omega{\cal L}^T{\vec v},\ \ \ \ 
\omega \in SL(2,{\bf R}).
\label{slmattrans}
\end{equation}
An important observation is that while for fixed values of 
$\vec\alpha$ and $\vec \beta$, mass changes under the variation of 
moduli and string coupling, {\it entropy remains the same 
as one moves in the moduli and coupling space} \cite{FERks52,FERk136,FERk54}. 
The fact that entropy is independent of coupling constants and 
moduli is consistent with the expectation that 
degeneracy of BPS states is a topological quantity which is 
independent of vacuum scalar expectation values and the fact that 
entropy measures {\it the number of generate microscopic states}, 
which should be independent of continuous parameters. 

\paragraph{Bogomol'nyi Bound}\label{n4bh4dsphsusybog}

We derive the Bogomol'nyi bound on the ADM mass
of asymptotically flat configurations within the effective theory of 
heterotic string on $T^6$ \cite{DUFlr}.  For this 
purpose, we introduce the Nester-like 2-form \cite{NES}:
\begin{equation}
\hat{E}_{\mu\nu} \equiv \bar{\varepsilon}\gamma^{\mu\nu\rho}
\delta_{\varepsilon}\tilde{\psi}_{\rho}, 
\label{nester}
\end{equation}
where $\delta_{\varepsilon}\tilde{\psi}_{\mu}$ is the supersymmetry 
transformation of physical gravitino in $D=4$.
Given supersymmetry transformations (\ref{hetsusy}) of fermionic fields 
expressed in terms of $D=4$ fields \cite{CVEy672}, 
the Nester's 2-form reduces to the form:
\begin{equation}
\hat{E}^{\mu\nu}= \bar{\varepsilon}^{\mu\nu\rho}\delta_{\rho}
\varepsilon + {1\over{2\sqrt{2}}}e^{-\varphi/2}\bar{\varepsilon}
(V_R L({\cal F}-i\gamma^5 \tilde{\cal F})^{\mu\nu})^a\Gamma^a\varepsilon 
+ \cdots ,
\label{nest}
\end{equation}
where $V$ is a vielbein defined in (\ref{mviel}) and $L$ is an 
invariant metric of $O(6,22)$ given in (\ref{4dL}).

Derivation of the Bogomol'nyi bound consists of evaluating the surface 
integral of the Nester's 2-form (\ref{nest}), which is related through 
the Stokes theorem to the volume integral of its covariant derivative. 
The surface integral yields
\begin{equation}
{1\over {4\pi G_4}}\int_{\partial\Sigma} dS_{\mu\nu}e\hat{E}^{\mu\nu}= 
\bar{\varepsilon}_{\infty}\left[ P^{ADM}_{\mu}\gamma^{\mu}
+{1\over{2\sqrt{2}G_4}}e^{-\varphi_{\infty}/2}\{V_{R\,\infty} L(\vec{Q}-
i\gamma^5\vec{P})\}^a\Gamma^a\right]\varepsilon_{\infty}, 
\label{surfnest}
\end{equation}
where $P^{ADM}_{\mu}$ is the ADM 4-momentum \cite{ARNdm122} of the 
configuration, and $\vec{Q}$ and $\vec{P}$ are physical electric and 
magnetic charges of $U(1)^{28}$ gauge group. 

The integrand of the volume integral is a positive semidefinite 
operator, provided spinors $\varepsilon$ satisfy the (modified) 
Witten's condition \cite{WIT80} $n\cdot\hat{\nabla}\varepsilon=0$ 
($n$ is the 4-vector normal to a space-like hypersurface $\Sigma$).  
Thus, the bilinear form on the right hand side of (\ref{surfnest}) 
is positive semidefinite, which requires that the ADM mass $M$ has to be 
greater than or equal to the largest of the following  
2 eigenvalues of central charge matrix:
\begin{equation}
|Z_{1,2}|^2 = {1\over{(4G_4)^2}}e^{-\varphi_{\infty}}
\left[\vec{Q}^2_R + \vec{P}^2_R \pm 2\{\vec{Q}^2_R \vec{P}^2_R - 
(\vec{Q}^T_R\vec{P}_R)^2\}^{1\over 2}\right],
\label{eigenvals}
\end{equation}
where $\vec{Q}_R \equiv \sqrt{2}(V_{R\,\infty} L\vec{Q})$ and similarly for 
$\vec{P}_R$.  This yields the Bogomol'nyi bound: 
\begin{equation}
M^2_{ADM} \geq {1\over{(4G_4)^2}}e^{-\varphi_{\infty}}
\left[\vec{Q}^2_R + \vec{P}^2_R + 2\{\vec{Q}^2_R \vec{P}^2_R - 
(\vec{Q}^T_R\vec{P}_R)^2\}^{1\over 2}\right]. 
\label{massbound}
\end{equation}
This bound is saturated iff supersymmetric variations of fermionic fields 
are zero, i.e. BPS configurations.  The Bogomol'nyi bound 
(\ref{massbound}) can be expressed explicitly in terms of electric 
$\vec{Q}$ and magnetic $\vec{P}$ charges, and asymptotic values 
$M_{\infty}$ and $\varphi_{\infty}$ of scalars, by using the 
identity: $LV^T_R V_RL = {1\over 2}[L(M+L)L]$. 
For example, $\vec{Q}^2_R = \vec{Q}^T L(M_{\infty}+L)L\vec{Q}$.

\subsubsection{Singular Black Holes and Enhancement of 
Symmetry}\label{n4bh4dsphsing}

In perturbative heterotic string theories, gauge symmetry is enhanced to 
non-Abelian ones through the Halpern-Frenkel-Ka\v c (HFK) mechanism 
\cite{FRE151,GROhmr54,GROhmr256,GROhmr267,ENGn163,CASent162,NAR169}.  
The HFK mechanism is due to extra spin one string states which are 
normally massive at generic points in moduli space but become massless 
at particular points. These points are the fixed 
points under discrete subgroups of $T$-duality of the worldsheet theory. 

As shown in \cite{HULt451}, BPS states in $N=4$ theories become 
massless at particular points in the moduli space.  
Since BPS multiplets in $N=4$ theories generically contain massive 
spin one states, gauge symmetry is enhanced to non-Abelian ones 
when the BPS states become massless.  Note, the BPS states carry magnetic, 
as well as electric, charges and, therefore, are non-perturbative in 
character.  When BPS multiplets with highest spin $3/2$ state become 
massless, supersymmetry as well as gauge symmetry is enhanced, a 
phenomenon that is never observed within perturbative string theories.  
In this section, we illustrate the enhancement of symmetries 
in the BPS states of $N=4$ theories by studying massless black holes 
in heterotic string on $T^6$ \cite{CVEy674,CARclmr464,CHAc375}.  

\paragraph{Massless Black Holes and Symmetry Enhancement}
\label{n4bh4dsphsingsym}

First, we consider the subset of BPS states with diagonal $M_{\infty}$ 
and purely imaginary $S_{\infty}$ \cite{CVEy674}.  We 
rewrite the corresponding generating solution (\ref{gensol}) 
with explicit dependence on scalar asymptotic values 
\begin{eqnarray}
\lambda &=& r^2/[(r-\eta_P{\bf P}^{(1)}_{1\,\infty})
(r-\eta_P {\bf P}^{(2)}_{1\,\infty})
(r- \eta_Q {\bf Q}^{(1)}_{2\,\infty})
(r-\eta_Q {\bf Q}^{(2)}_{2\,\infty})]^{1\over 2},
\cr
R &=& [(r-\eta_P {\bf P}^{(1)}_{1\,\infty})
(r - \eta_P {\bf P}^{(2)}_{1\,\infty})
(r - \eta_Q {\bf Q}^{(1)}_{2\,\infty})
(r-\eta_Q {\bf Q}^{(2)}_{2\,\infty})]^{1\over 2},
\label{gensolasym}
\end{eqnarray}
where the solution now depends on the following ``screened'' charges:  
\begin{eqnarray}
& &({\bf  P}^{(1)}_{1\,\infty}, {\bf P}^{(2)}_{1\,\infty},
{\bf Q}^{(1)}_{2\,\infty}, {\bf Q}^{(2)}_{2\,\infty}) 
\cr
& &\ \ \ \equiv 
e^{-{\varphi_{\infty}\over 2}}(g^{1\over 2}_{11\,\infty}P_1^{(1)},
g^{-{1\over 2}}_{11\,\infty}P_1^{(2)}, g^{-{1\over 2}}_{22\,\infty}
Q_2^{(1)}, g^{1\over 2}_{22\,\infty}Q_2^{(2)}) 
\cr
& &\ \ \ =(e^{-{\varphi_{\infty}
\over 2}}g^{1\over 2}_{11\,\infty}\beta_1^{(2)}, e^{-{\varphi_{\infty}
\over 2}} g^{-{1\over 2}}_{11\,\infty}\beta_1^{(1)}, e^{{\varphi_{\infty}
\over 2}} g^{{1\over 2}}_{22\,\infty}\alpha_2^{(1)}, e^{{\varphi_{\infty}
\over 2}}g^{-{1\over 2}}_{22\,\infty}\alpha_2^{(2)}). 
\label{screen}
\end{eqnarray}
Here, the quantized charge lattice vectors $\vec\alpha$ and $\vec\beta$ 
live on the even, self-dual, Lorentzian lattice $\Lambda$ \cite{SEN303}.  

When $\eta_{P,Q}$ are chosen to satisfy $\eta_P\,{\rm sign}(P_1^{(1)}
+P_1^{(2)})= -1$ and $\eta_Q\, {\rm sign}(Q_2^{(1)}+Q_2^{(2)})=-1$, 
the ADM mass takes the following form that saturates the BPS bound:
\begin{equation}
M_{\rm BPS} = e^{-{\varphi_{\infty} \over 2}}|g^{1\over 2}_{11\,\infty}
\beta^{({2})}_1 + g^{-{1\over 2}}_{11\,\infty}\beta^{(1)}_1| +
e^{{\varphi_\infty}\over 2}|g^{{1\over 2}}_{22\,\infty}\alpha^{(1)}_2+
g^{-{1\over 2}}_{22\,\infty}\alpha^{(2)}_2|.
\label{ADMmass}
\end{equation}

Only electric states ($\vec{\beta}=0$) with $\vec{\alpha}^T
L\alpha \geq 2$ are matched onto perturbative string states \cite{DUFr}.  
As for dyonic states that break $1/4$ of supersymmetry 
(i.e. those with non-parallel electric and magnetic charge vectors, i.e.  
$\vec{Q}^T{\cal M}_{+}\vec{P}\neq 0$, and therefore cannot be related 
to electric solution via the $SL(2,{\bf Z})$ transformations), 
the consistency with the $SL(2,{\bf Z})$ symmetry and 
consistent electric limit require that the electric 
and the magnetic lattice vectors separately satisfy 
the constraints $\vec{\alpha}^T L\vec{\alpha} \geq -2$ and 
$\vec{\beta}^T L\vec{\beta} \geq -2$.  These subsets of BPS states 
(\ref{gensolasym}) become massless \cite{BEH455,KALl52,CVEy674} at the 
fixed points under $T$-duality $R \to 1/R$ (i.e. at the $T^2$ 
self-dual point $g_{22\,\infty}=1$ [$g_{11\,\infty}=1=
g_{22\,\infty}$]), when $\alpha^{(1)}_2 =-\alpha^{(2)}_2=\pm 1$ 
[$\alpha^{(1)}_2 =-\alpha^{(2)}_2=\pm 1$ and $\beta^{(1)}_1 =
-\beta^{(2)}_1=\pm 1$] for the case $\vec{\beta}=0$ [$\vec{\alpha}\neq 0 
\neq\vec{\beta}$].  There are also additional infinite number of 
$SL(2,{\bf Z})$ related massless BPS states.  

The extra massless spin 1 states associated with $\alpha^{(1)}_2 =
-\alpha^{(2)}_2=\pm 1$ [$\alpha^{(1)}_2 =-\alpha^{(2)}_2=\pm 1$ and 
$\beta^{(1)}_1 =-\beta^{(2)}_1=\pm 1$] at the self-dual points of 
$T^2$ contribute to enhancement of Abelian gauge symmetry to 
$SU(2)$ [$SU(2)\times SU(2)$] non-Abelian symmetry.  The extra massless 
spin 1 states together with generic massless $U(1)$ gauge fields form 
the adjoint representations of the enhanced non-Abelian gauge groups.  

The BPS multiplet which preserves $1/4$ of $N=4$ supersymmetry 
contains spin $3/2$ state.  Thus, additional 4 massless gravitino 
associated with $\alpha^{(1)}_2=-\alpha^{(2)}_2=\pm 1$ 
and $\beta^{(1)}_1 =-\beta^{(2)}_1=\pm 1$ contribute to enhancement 
of supersymmetry from $N=4$ to $N=8$.  

Note, the infinite number of $SL(2,{\bf Z})$ related massless states and 
enhancement of supersymmetry are not realized within perturbative string 
theories.  These are new non-perturbative phase of string theories that 
are required by non-perturbative string dualities.  

\paragraph{Maximal Gauge Symmetry Enhancement in the Moduli Space 
of Two-Torus}\label{n4bh4dsphsingmax}

In this subsection, we study maximal symmetry enhancement in full 
moduli space of $T^2$ parameterized by arbitrary scalar asymptotic 
values \cite{CARclmr464,CHAc375}.   We consider the general BPS 
mass formula (\ref{Bogmass}).  

The moduli space of $T^2$ is parameterized by the following real matrix:
\begin{equation}
M=\left(\matrix{G^{-1} & -G^{-1}B \cr -B^TG^{-1} & 
G+B^TG^{-1}B}\right), 
\label{twomod}
\end{equation}
where $G\equiv [G_{mn}]$ and $B\equiv [B_{mn}]$ with $(m,n)=1,2$ 
being indices associated with $T^2$. Thus,  
the effective theory has the $O(2,2,{\bf R})$ $T$-duality symmetry. 
Since $O(2,2,{\bf R})\cong SL(2,{\bf R})\times SL(2,{\bf R})$ 
\cite{DIJvv88,SHAw320}, $M$ is reparameterized \cite{DIJvv88} by the 
following 2 complex scalars, which separately parameterize each 
$SL(2,{\bf R})$ moduli space: 
\begin{equation}
\tau=\tau_1+i\tau_2\equiv{\sqrt{G}\over G_{11}}-i{G_{12}\over G_{11}},\ \ \ 
\rho=\rho_1+i\rho_2\equiv\sqrt{G}+iB_{12},
\label{slmodul}
\end{equation}
where $G \equiv G_{11}G_{22}-G^2_{12}$.  Here, $\tau$ [$\rho$] is the  
complex [K\" ahler] structure of $T^2$.
Each $SL(2,{\bf Z})$ factor \cite{DIJvv88,SHAw320,GIVmr238} is generated 
by 2 transformations, which are given, for the $SL(2,{\bf Z})$ factor 
associated with $\tau$, by
\begin{equation}
S_{\tau}:\ \ \tau \to {1\over \tau}, \ \ \ \ \
T_{\tau}:\ \ \tau \to \tau + i,
\label{tdualsltran}
\end{equation}
where $\rho$ remains intact, and similarly for
the $SL(2,{\bf Z})$ factor associated with $\rho$.
In addition, the $\sigma$-model corresponding to $T^2$ is symmetric 
under the ``mirror symmetry'' ($S_{mirr}:\,\tau\leftrightarrow\rho$), 
the worldsheet parity symmetry $(\tau,\rho)\to (\tau,-\bar{\rho})$ 
corresponding to $\sigma\to -\sigma$, and the symmetry $(\tau,\rho)\to 
(-\bar{\tau},-\bar{\rho})$ associated with the reflection $X^1\to -X^1$. 

In accordance with the above reparameterization of moduli space, 
one can express the central charges (\ref{eigenvals}) of the $N=4$ 
theory in the following $SL(2,{\bf R})_{\tau}\times 
SL(2,{\bf R})_{\rho}\times {\bf Z}^{\tau\leftrightarrow\rho}_2$ 
invariant form \cite{CARclmr464} 
(the subscript $\infty$ is omitted in ADM mass and central charges)
\begin{eqnarray}
|Z_{1,2}|^2 &=& {1\over {4(S+\bar{S})(\tau +\bar{\tau})(\rho +\bar{\rho})}}
|{\cal M}_{1,2}|^2,
\nonumber \\
{\cal M}_{1} &\equiv& (\hat{\alpha}_j +iS\hat{\beta}_j)\hat{P}^j, \ \ \
{\cal M}_{2} \equiv (\hat{\alpha}_j -i\bar{S}\hat{\beta}_j)\hat{P}^j,
\label{slcentral}
\end{eqnarray}
where $\vec{\hat{\alpha}}\equiv (\alpha^{(1)}_2,\alpha^{(2)}_2,
\alpha^{(2)}_1,-\alpha^{(1)}_1)^T$, $\vec{\hat{\beta}}\equiv
(\beta^{(1)}_2,\beta^{(2)}_2,\beta^{(2)}_1,-\beta^{(1)}_1)^T$, and
$\hat{P} \equiv (1,-\tau\rho,i\tau,i\rho)^T$.
Here, the charge lattice vectors $\vec{\hat{\alpha}}$ and $\vec{\hat{\beta}}$ 
transform under $SL(2,{\bf R})_{\tau}\times SL(2,{\bf R})_{\rho}\times 
{\bf Z}^{\tau\leftrightarrow\rho}_2$ as
\begin{eqnarray}
SL(2,{\bf R})_{\tau}&:&\ \ 
\left(\matrix{\hat{\alpha}_3 \cr \hat{\alpha}_1}\right) \to 
\omega_{\tau}\left(\matrix{\hat{\alpha}_3 \cr \hat{\alpha}_1}\right), \ \ 
{\rm similary\ for}\ 
\left(\matrix{\hat{\alpha}_2 \cr \hat{\alpha}_4}\right), 
\left(\matrix{\hat{\beta}_3 \cr \hat{\beta}_1}\right), 
\left(\matrix{\hat{\beta}_2 \cr \hat{\beta}_4}\right), 
\nonumber \\
SL(2,{\bf R})_{\rho}&:&\ \ 
\left(\matrix{\hat{\alpha}_3 \cr \hat{\alpha}_0}\right) \to 
\omega_{\rho}\left(\matrix{\hat{\alpha}_3 \cr \hat{\alpha}_0}\right), \ \ 
{\rm similary\ for}\ 
\left(\matrix{\hat{\alpha}_1 \cr \hat{\alpha}_2}\right), 
\left(\matrix{\hat{\beta}_3 \cr \hat{\beta}_0}\right), 
\left(\matrix{\hat{\beta}_1 \cr \hat{\beta}_2}\right), 
\nonumber \\
{\bf Z}^{\tau\leftrightarrow\rho}_2&:&\ \ 
\hat{\alpha}_2 \leftrightarrow \hat{\alpha}_3, \ \ \ 
\hat{\beta}_2 \leftrightarrow \hat{\beta}_3.
\label{slchtran}
\end{eqnarray}
In addition, the central charges (\ref{slcentral}) are invariant under 
the $SL(2,{\bf R})_S$ $S$-duality:
\begin{equation}
SL(2,{\bf R})_S :\ \ \left(\matrix{\vec{\hat{\alpha}}\cr 
\vec{\hat{\beta}}}\right) 
\to \left(\matrix{a&c\cr b&d}\right)\left(\matrix{\vec{\hat{\alpha}}\cr 
\vec{\hat{\beta}}}\right). 
\label{slsch}
\end{equation}
Note, the ADM mass of BPS states is given by the largest of $|Z_1|$ 
and $|Z_2|$.   

First, we consider the short multiplet, i.e. the BPS multiplet 
with 2 central charges $Z_{1,2}$ equal in magnitude.  It has the 
highest spin 1 state and preserves $1/2$ of the $N=4$ supersymmetry.  
The charge lattice vectors $\vec{\alpha}$ and $\vec{\beta}$ 
of the short multiplet live on the  $S$-orbit satisfying $p\hat{\alpha}_i 
= s\hat{\beta}_i$ ($s,p \in {\bf Z}$).  Explicitly, we 
write $\vec{\hat{\alpha}}$ and $\vec{\hat{\beta}}$ in the $S$-orbit as
\begin{eqnarray}
\hat{\alpha}_1 &=& sm_2, \ \ \  \hat{\alpha}_2 = sn_2, \ \ \ 
\hat{\alpha}_3 = sn_1, \ \ \  \hat{\alpha}_4 = -sm_1,
\nonumber \\
\hat{\beta}_1 &=& pm_2, \ \ \  \hat{\beta}_2 = pn_2, \ \ \ 
\hat{\beta}_3 = pn_1, \ \ \  \hat{\beta}_4 = -pm_1,
\label{shortchvec}
\end{eqnarray}
where $m_i$ [$n_i$] corresponds to momentum [winding] number of 
perturbative string states when $p=0$ and $s$ [$p$] denotes electric 
[magnetic] quantum number associated with the $S$-modulus.  
Then, the central charge (\ref{slcentral}) becomes
\begin{equation}
{\cal M}=(s+ipS)(m_2-im_1\tau +in_1\rho -n_2\tau\rho). 
\label{shortbps}
\end{equation}

In the complex moduli space parameterized by $(\tau,\rho)$, 
$M_{BPS}$ can vanish at fixed lines \cite{DINhs322,IBAllt352,CARlm455} 
under the Weyl reflections $w_1\equiv S_{mirr}$, $w_2\equiv S_{\rho}
S_{mirr}S_{\rho}$, $w_3\equiv T_{\rho}S_{mirr}T^{-1}_{\rho}$ and 
$w_4\equiv w^{-1}_3W_2w_3$ in the $T$-duality group.  Along these lines, 
the Abelian gauge group $U(1)_a\times U(1)_b\times U(1)_c\times 
U(1)_d\equiv U(1)^{(1)}_1 \times U(1)^{(1)}_2 \times U(1)^{(2)}_1 
\times U(1)^{(2)}_2$ is enhanced to the non-Abelian $U(1)^3\times SU(2)$ 
group.   The BPS states (labeled by $\vec{\alpha}$) which become massless 
along the fixed lines ${\cal L}_i$ under $w_i$ are as follows 
\cite{CARlm455,CARlm45}:
\begin{itemize}
\item ${\cal L}_1=\{\tau =\rho$\}: $\vec{\alpha}=\vec{\lambda}_{1\pm}=
\pm(1,0,-1,0)$ contributing to $U(1)_b \times U(1)_d \times U(1)_{a+c} 
\times SU(2)_{a-c}$
\item ${\cal L}_2=\{\tau =\rho^{-1}\}$: $\vec{\alpha}=\vec{\lambda}_{2\pm}=
\pm(0,1,0,-1)$ contributing to $U(1)_a \times U(1)_c \times U(1)_{b+d} 
\times SU(2)_{b-d}$
\item ${\cal L}_3=\{\tau =\rho -i\}$: $\vec{\alpha}=\vec{\lambda}_{3\pm}=
\pm(1,1,-1,0)$ contributing to $U(1)_d \times U(1)_{a+c} \times 
U(1)_{a-2b-c} \times 
SU(2)_{a+b-c}$
\item ${\cal L}_4=\{\tau = {\rho\over{i\rho +1}}\}$: $\vec{\alpha}=
\vec{\lambda}_{4\pm}=\pm(1,0,-1,1)$ contributing to $U(1)_b\times 
U(1)_{a+c} \times U(1)_{a-c-2d}\times SU(2)_{a-c+d}$.  
\end{itemize}
At points where the lines ${\cal L}_i$ intersect 
\cite{CARlm450,CARlm45}, there are additional massless states, resulting 
in the maximal enhancements of gauge symmetries:
\begin{itemize}
\item ${\cal L}_1 \cap {\cal L}_2=\{\tau =\rho =1\}$: 
$\vec{\alpha}=\vec{\lambda}_{1\pm}$ or $\vec{\lambda}_{2\pm}$ contributing 
to $U(1)_{a+c}\times U(1)_{b+d}\times SU(2)_{a-c}\times SU(2)_{b-d}$ 
\item ${\cal L}_2 \cap {\cal L}_3\cap {\cal L}_4=\{\bar{\tau}=\rho=
e^{i{\pi\over 6}}\}$: $\vec{\alpha}=\vec{\lambda}_{2\pm}$ or 
$\vec{\lambda}_{3\pm}$ or $\vec{\lambda}_{4\pm}$ contributing to 
$U(1)_{a+c}\times U(1)_{a-2b-c-2d}\times SU(3)_{b-d,2a+b-2c+d}$. 
\end{itemize}
Along with the above perturbative massless states, there are accompanying 
infinite massless dyonic states, so-called $S$-orbit 
$p\hat{\alpha}_i=s\hat{\beta}_i$, related via $SL(2,{\bf Z})_S$ $S$ duality.  

Second, we consider the intermediate multiplets, i.e. the BPS 
multiplets with $|Z_1|\neq |Z_2|$.  They have the highest spin $3/2$ 
states and preserve $1/4$ of the $N=4$ supersymmetry.  In this case, 
$\vec{\hat{\alpha}}$ and $\vec{\hat{\beta}}$ are 
not proportional, i.e. $\hat{\alpha}_i\hat{\beta}_j - \hat{\alpha}_j
\hat{\beta}_i\neq 0$.   The requirement that the ADM mass is 
zero, i.e. $|Z_1|=0$ and $\Delta Z^2=|Z_1|-|Z_2|=0$, leads to the relations 
\cite{CARclmr464}:
\begin{equation}
\alpha^{(1)}_2-\alpha^{(2)}_2\tau\rho +i\alpha^{(2)}_1\rho 
-i\alpha^{(1)}_1=0,\ \ \  
\beta^{(1)}_2-\beta^{(2)}_2\tau\rho +i\beta^{(2)}_1\rho
-i\beta^{(1)}_1=0.
\label{interm}
\end{equation}
These relations are satisfied by the following fixed points 
\cite{CARclmr464,CHAc375}:
\begin{itemize}
\item $\tau=\rho=i$: $\ \ (\vec{\alpha},\vec{\beta})=(\vec{\lambda}_{1\pm}, 
\vec{\lambda}_{2\pm})$
\item $\bar{\tau}=\rho=e^{i{\pi\over 6}}$: $\ \ (\vec{\alpha},\vec{\beta})=
(\vec{\lambda}_{i\pm},\vec{\lambda}_{j\pm})$, $2\leq i<j\leq 4$.
\end{itemize}
In addition to the above massless dyonic states, there are infinite number 
of $SL(2,{\bf Z})_S$ related dyonic states.  Since these additional 
massless states belong to the highest spin $3\over 2$ supermultiplet, 
supersymmetry as well as gauge symmetry are enhanced \cite{CVEy674}.

\paragraph{Properties of Massless Black Holes}\label{n4bh4dsphsingprop}

When both of the pairs $(Q^{(1)}_2,Q^{(2)}_2)$ and $(P^{(1)}_1,P^{(2)}_1)$ 
have the same relative signs \cite{CVEy672}, the singularity of the solution 
(\ref{gensolasym}) is always behind or located at the event horizon 
at $r=0$, corresponding to the Reissner-Nordstr\" om-type horizon or 
null singularity, respectively.
However, the moment at least one of the pairs has the opposite relative 
signs \cite{BEH455,KALl52,CVEy674}, there is a singularity outside of 
the event horizon, i.e. naked singularity $r_{\rm sing} > 0$.  
Explicitly, the curvature singularity is at 
$r=r_{\rm sing}\equiv{\rm max}\{{\rm min}
[|{\bf P}^{(1)}_{1\,\infty}|,|{\bf P}^{(2)}_{1\,\infty}|],{\rm min}
[|{\bf Q}^{(1)}_{2\,\infty}|,|{\bf Q}^{(2)}_{2\,\infty}|]\} >0$.

These singular solutions have an unusual property of repelling massive 
test particles \cite{KALl52}.  (Note, the BPS black holes 
need not be massless to be able to repel massive test particles 
\cite{CVEy674}.)  
There is a stable gravitational equilibrium point for a test particle at 
$r=r_c$ where the gravitational force is attractive for $r>r_c$ and 
repulsive for $r<r_c$ \cite{KALl52}.  
This can also be seen by calculating the traversal time of the geodesic 
motion for a test particle with energy $E$, mass $m$ and zero angular 
momentum along the radial coordinate $r$, as measured by an asymptotic 
observer \cite{CVEy674}:
\begin{equation}
t(r)=\int^r_{r_{\infty}} {{E\,dr}\over{\lambda(r)\sqrt{E^2-m^2\lambda(r)}}}.
\label{time}
\end{equation}
The minimum radius that can be reached by a test particle corresponds to
$r_{\rm min}>r_{\rm sing}$ for which $\lambda(r=r_{\rm min})=E^2/m^2$, since 
it takes infinite amount of time to go beyond $r=r_{min}$.  
Here, $r=r_{sing}$ is the singularity. 
Massive test particles cannot reach the singularity of singular black 
holes in finite time and are reflected back.   
On the other hand, classical massless particles with zero angular
momentum do not feel the repulsive gravitational potential due to
increasing $\lambda(r)$, and they reach the singularity
in a finite time.

Note, for regular solutions, studied in section 
\ref{n4bh4dsphsusy}, $\lambda\le 1$, while for singular solutions 
studied in this section, $\lambda \ge 1$ for $r$ small enough.   
Thus, for regular solutions, particles are always attracted 
toward the singularity.  When only one charge is non-zero, 
the regular solution has a naked singularity at $r=0$;  
$t(r=r_{sing}=0)$ is finite.

\subsubsection{Non-Extreme Solutions}\label{n4bh4dsphnon}

The following non-extreme generalization \cite{CVEy675} of the BPS 
solution (\ref{gensol}) is obtained by solving 
the Einstein and Euler-Lagrange equations:
\begin{eqnarray}
\lambda &=& r(r+2\beta)/[(r+P^{(1)\,\prime}_1)
(r+P^{(2)\,\prime}_1)(r+Q^{(1)\,\prime}_2)
(r+Q^{(2)\,\prime}_2)]^{1\over 2},
\nonumber\\
R(r) &=& [(r+P^{(1)\,\prime}_1)(r+P^{(2)\,\prime}_1)
(r+Q^{(1)\,\prime}_2)(r+Q^{(2)\,\prime}_2)]^{1\over 2},
\nonumber\\
e^{\varphi} &=& \left [ {{(r+P^{(1)\,\prime}_1)
(r+P^{(2)\,\prime}_1)} \over {(r+Q^{(1)\,\prime}_2)
(r+Q^{(2)\,\prime}_2)}}\right ]^{1\over 2},\ \Psi=0,
\nonumber\\
g_{11} &=& {{r+P^{(2)\,\prime}_1}\over
{r+P^{(1)\,\prime}_1}}, \ \
g_{22} = {{r+Q^{(1)\,\prime}_2}\over
{r+Q^{(2)\,\prime}_2}},\ \
g_{mm} = 1\ \ (m \neq 1,2),\nonumber\\
 g_{mn}&=&B_{mn}=0\ \  (m\ne n),\ \ \ a^I_m=0,
\label{hetnonex}
\end{eqnarray}
where $\beta>0$ measures deviation from the corresponding BPS  
solution and $P^{(1)\,\prime}_1 \equiv \beta \pm \sqrt{(P^{(1)}_1)^2
+ \beta^2}$, etc..  The ADM mass is
\begin{equation}
M= P^{(1)\,\prime}_1 + P^{(2)\,\prime}_1 +
Q^{(1)\,\prime}_2 + Q^{(2)\,\prime}_2-4\beta .
\label{nonexmass}
\end{equation}

The signs $\pm$ in the expressions for $P^{(1)\,\prime}_1$,
etc. should be chosen so that $M\to M_{BPS}$ as $\beta\to 0$.  
To have a regular horizon, one has to choose the relative signs of both 
pairs $(Q^{(1)}_2,Q^{(2)}_2)$ and $(P^{(1)}_1,P^{(2)}_1)$ 
to be the same \cite{CVEy672}.  
In this case, the non-extreme solution is (\ref{hetnonex}) 
with {\it positive } signs in the expressions 
for $P^{(1)\,\prime}_1$, etc. and have the ADM mass 
\begin{equation}
M=\sqrt{(P^{(1)}_1)^2+\beta^2}+
\sqrt{(P^{(2)}_1)^2 +\beta^2}+
\sqrt{(Q^{(1)}_2)^2+\beta^2}+
\sqrt{(Q^{(2)}_2)^2+\beta^2},
\label{regnonmass}
\end{equation}
which is always compatible with the Bogomol'nyi bound:
\begin{equation}
M_{\rm BPS}=|P^{(1)}_1|+|P^{(2)}_1|+|Q^{(1)}_2|
+|Q^{(2)}_2|.
\label{regexmass}
\end{equation}
Such solutions always have nonzero mass.  

\paragraph{Space-Time Structure and Thermal Properties}
\label{n4bh4dsphnongl} 

We now study spacetime properties 
\cite{CVEy672,CVEy675} of the regular $D=4$ 4-charged black hole 
discussed in the previous subsections.  
 
There is a spacetime singularity, i.e. the Ricci scalar 
$\cal R$ blows up, at the point $r=r_{sing}$ where $R=0$. 
The event horizon, defined as a location where the $r={\rm constant}$ 
surface is null, is at $r=r_{H}$ where $g^{rr}=\lambda=0$.  
The horizon(s) forms, provided $r_H > r_{sing}$ (time-like singularity).  
In some cases, singularity and the event horizon coincide: $r_{sing}=r_H$.  
In this case, the singularity is ($i$) naked (space-like singularity) 
when the singularity is reachable by an outside observer (at $r=r_0>r_H$) 
in a finite affine time $\tau = \int^{r_0}_{r_{sing}}dr \sqrt{g_{rr}\over 
g_{tt}} = \int^{r_0}_{r_H}dr \lambda^{-1}(r)$ and ($ii$) 
also an event horizon (null singularity) when $\tau = \infty$. 

Thermal properties of the solution (\ref{gensol}) are specified 
by spacetime at the event horizon.  
The Hawking temperature \cite{HAW248,HAW75} is defined by the surface 
gravity $\kappa$ at the event horizon: 
\begin{equation}
T_H = \kappa/(2\pi) = |\partial_r\lambda(r=r_H)|/(4\pi). 
\label{tempform}
\end{equation}
Entropy $S$ is given by the Bekenstein's formula 
\cite{BEK73,BEK74,HAW71,HAW13,BARch31}:
\begin{equation}
S = {1\over 4}\times({\rm the\ surface\ area\ of\ the\ event\ horizon}) 
= \pi R(r=r_H).
\label{entform}
\end{equation}

We classify thermal and spacetime properties according to 
the number of non-zero charges: 
\begin{itemize}
\item
All the 4 charges non-zero:
There are 2 horizons at $r=0,-2\beta$ and a 
time-like singularity is hidden behind the inner horizon, i.e. the
global spacetime is that of the non-extreme Reissner-Nordstr\" om black
hole.  The Hawking temperature is $T_H=\beta/(\pi
\sqrt{P^{(1)\,\prime}_1P^{(2)\,\prime}_1
Q^{(1)\,\prime}_2 Q^{(2)\,\prime}_2})$ and the entropy
is $S=\pi\sqrt{P^{(1)\,\prime}_1 P^{(2)\,\prime}_1
Q^{(1)\,\prime}_2 Q^{(2)\,\prime}_2}$.  When $\beta \to 0$, 
spacetime is that of extreme Reissner-Nordstr\" om black holes.
\item
3 nonzero charges: 
A space-like singularity is located at the inner
horizon ($r=-2\beta$).  For example when $P^{(1)}_1=0$, $T_H=
\beta^{1\over 2}/(\pi\sqrt{2P^{(2)\,\prime}_1 Q^{(1)\,\prime}_2
Q^{(2)\,\prime}_2})$ and $S=\pi\sqrt{2\beta P^{(2)\,\prime}_1
Q^{(1)\,\prime}_2 Q^{(2)\,\prime}_2}$.  When $\beta \to 0$, 
the singularity coincides with the horizon at $r=0$.
\item
2 nonzero charges: 
A space-like singularity is at $r=-2\beta$.  For example when 
$P^{(1)}_1\ne 0 \ne P^{(2)}_1$, $T_H=1/(2\pi\sqrt{P^{(1)\,
\prime}_1 P^{(2)\,\prime}_1})$ and $S=\pi\sqrt{4\beta^2 P^{(1)\,\prime}_1 
P^{(2)\,\prime}_1}$.
As $\beta \to 0$, the singularity coincides with the horizon at $r=0$.
\item
1 nonzero charge: 
A space-like singularity is at $r=-2\beta$.  
For example when $P^{(1)}_1 \neq 0$, $T_H=1/(2\pi\sqrt{2\beta 
P^{(1)\,\prime}_1})$ and $S=\pi\sqrt{8\beta^3 P^{(1)\,\prime}_1}$.  
As $\beta \to 0$, the singularity becomes naked.
\end{itemize}

\subsubsection{General Static Spherically Symmetric Black Holes in Heterotic 
String on a Six-Torus}\label{n4bh4dsphgen}

In section \ref{n4bh4dsphsusy}, we saw that a general solution obtained 
by applying subsets of $O(6,22)$ and $SL(2,{\bf R})$ transformations 
on the 4-charged solution has 1 charge degree of freedom missing for  
describing non-rotating black holes with the most general charge 
configuration.   It is a purpose of this section to introduce such 1 
missing charge degree of freedom by applying 2 $SO(1,1)$ boosts 
(in the $D=3$ $O(8,24)$ duality group) along a $T^2$ direction 
(associated with 4 non-zero charges) with a zero-Taub-NUT constraint
\footnote{An additional $SO(1,1)$ boost along a $T^2$ direction 
on the 4-charged black hole solution necessarily induces Taub-NUT term, 
since the metric components $g_{\mu\nu}$ get mixed with the 
$\phi$-component of the $U(1)$ gauge potential, which is singular 
\cite{SEN440}.} 
to construct the ``generating solution'' for non-extreme, non-rotating 
black holes in heterotic string on $T^6$ with 
the most general charge configuration of $U(1)^{28}$ gauge group 
\cite{CVEy127}.  (See \cite{JATmp484} for an another attempt.)  
So, the generating solution is parameterized by the non-extremality 
parameter (or the ADM mass) and 6 $U(1)$ charges with one zero-Taub-NUT 
constraint.  

\paragraph{Explicit Form of the Generating Solution}\label{n4bh4dsphgensol}

For the purpose of constructing the generating solution in a simplest 
possible form, it is convenient to first generate the 4-charged 
non-extreme solution (\ref{hetnonex}) with the following non-zero charges 
by applying 4 $SO(1,1)$ boosts to the Schwarzschield solution:
\begin{eqnarray}
P^{(1)}_1&=&2m{\rm sinh}\delta_{p1}{\rm cosh}\delta_{p1}\equiv P_1,
\ \ \ \ \, 
P^{(2)}_1=2m{\rm sinh}\delta_{p2}{\rm cosh}\delta_{p2}\equiv P_2,
\cr
Q^{(1)}_2&=&2m{\rm cosh}\delta_{q1}{\rm sinh}\delta_{q1} \equiv
Q_1, \ \ \ \ \, 
Q^{(2)}_2 = 2m{\rm cosh}\delta_{q2}{\rm sinh}\delta_{q2} \equiv Q_2. 
\label{charge}
\end{eqnarray}
Only non-extreme solutions compatible with the Bogomol'nyi bound and, 
therefore, within spectrum of states, are those with the same 
relative signs for both pairs $(Q_1,Q_2)$ and $(P_1,P_2)$.  For this case,  
$\hat{P}_1\equiv 2m\cosh^2\delta_{p1}-m=\pm\sqrt{(P_1)^2+m^2}$, etc. 
are given with plus signs.

As the next step, one introduces one missing charge degree of freedom 
by applying 2 $SO(1,1)\subset O(8,24)$ boosts with parameters $\delta_{1,2}$
\footnote{One can induce any 2 of the remaining charges in the 
$U(1)^{(1)}_1\times U(1)^{(1)}_2\times U(1)^{(2)}_1\times U(1)^{(2)}_{2}$ 
gauge group.  But we here choose to induce $P^{(2)}_2$ and $Q^{(1)}_1$.} 
satisfying the zero Taub-NUT condition:
\begin{equation}
P_1{\rm tanh}\delta_1 -  Q_2{\rm tan}\delta_2 = 0.
\label{notaubnut}
\end{equation}
Assuming, without loss of generality, that $Q_2 \geq P_1$, 
one has, from (\ref{notaubnut}), $\delta_2$ in terms of the other parameters 
\footnote{For the case $Q_2 \leq P_1$, the role of $\delta_1 $ and 
$\delta_2$ are interchanged.}: 
\begin{equation}
{\rm cosh}\,\delta_2 = Q_2 {\rm cosh}\,\delta_1 /\Delta, \ \ \ \ \
{\rm sinh}\,\delta_2 = P_1 {\rm sinh}\,\delta_1 /\Delta,
\label{boostrel}
\end{equation}
where $\Delta \equiv {\rm sign}(Q_2) \sqrt{(Q_2)^2 {\rm cosh}^2
\delta_1 - (P_1)^2 {\rm sinh}^2 \delta_1}$.

The final form of the generating solution \cite{CVEy127} 
(with zero Taub-NUT charge) is
\footnote{The BPS limit ($m=0$ and $\delta_1\to\infty$) of this solution 
is related to the solution (\ref{chnulsol}) 
via subsets of $SO(2)\times SO(2)\subset O(2,2)$ ($T^2$ $T$-duality) 
and $SO(2)\subset SL(2,{\bf R})$ transformations.} 
\begin{eqnarray}
\lambda &=& { {{(r+m)(r-m)}\over
{(XY-Z^2)^{1\over 2}}}},\ \ \
R=(XY-Z^2)^{1\over 2},\ \ \ \ \ \ \
e^{2\varphi}= { {W^2 \over {XY-Z^2}}},
\cr
\partial_r \Psi &=& { {1\over {\Delta^3\,W}}}\left[
\Delta^2 (P_1 Q_1 + P_2 Q_2)+ P_1 Q_2 [(P_1)^2(r+\hat{Q_2})-
(Q_2)^2(r+\hat{P}_1)] \right.\cr
&\times&\left.{{{P_1 P_2 (r-\hat{Q}_1){\rm sinh}^2 \delta_1
+ Q_1 Q_2 (r+\hat{P}_2){\rm cosh}^2 \delta_1 } \over {XY-Z^2}}}
\right]{{\rm sinh}\delta_1
{\rm cosh}\delta_1},
\cr
G_{11}&=&{{X\over {(r+\hat{P}_1)(r+\hat{Q}_2)}}}, \ \ \
G_{22}={ {Y\over {(r+\hat{P}_1)(r+\hat{Q}_2)}}}, 
\cr
G_{12}&=&-{ {Z\over {(r+\hat{P}_1)(r+\hat{Q}_2)}}},
\cr
B_{12}&=&-{ {{[(Q_2)^2(r+\hat{P}_1)-(P_1)^2(r+
\hat{Q}_2)]{\rm cosh}\delta_1 {\rm sinh}\delta_1} \over
{\Delta (r+\hat{P}_1)(r+\hat{Q}_2)}}},\cr
G_{ij}&=&\delta_{ij}, \  \ \  B_{ij}=0, \ \ \ (i,j \neq 1, 2),
\ \ \ a^I_m=0\
\label{gentsol}
\end{eqnarray}
with
\begin{eqnarray}
X&=&r^2+[(\hat{P}_2 +\hat{Q}_2){\rm cosh}^2 \delta_1
+(\hat{Q}_1 - \hat{P}_1){\rm sinh}^2 \delta_1]r 
\cr
& &+(\hat{P}_1\hat{Q}_1{\rm sinh}^2 \delta_1
+ \hat{Q}_2 \hat{P}_2{\rm cosh}^2 \delta_1),
\cr
Y&=&r^2+{\textstyle {1\over \Delta^2}}[(P_1)^2 (\hat{P}_2 -\hat{Q}_2)
{\rm sinh}^2 \delta_1 + (Q_2)^2 (\hat{P}_1 +\hat{Q}_1)
{\rm cosh}^2 \delta_1]r 
\cr
& &+{\textstyle{1\over \Delta^2}}[(P_1)^2 \hat{Q}_2
\hat{P}_2\sinh^2\delta_1 +(Q_2)^2\hat{Q}_1\hat{P}_1
{\rm cosh}^2 \delta_1],
\cr
Z&=&{\textstyle {1\over \Delta}}[(P_1 P_2 +
Q_1 Q_2)r +(\hat{P}_1 Q_1 Q_2 +
\hat{Q}_2 P_1 P_2)]{{\rm cosh}\delta_1 {\rm sinh}\delta_1},
\cr
W&=& r^2 +{\textstyle {1\over \Delta^2}}[(Q_2)^2 (\hat{P}_1 +
\hat{P}_2){\rm cosh}^2 \delta_1 + (P_1)^2 (\hat{Q}_1 - \hat{Q}_2)
{\rm sinh}^2 \delta_1]r\cr
& &+ {\textstyle {1\over \Delta^2}}[(Q_2)^2\hat{P}_1
\hat{P}_2 {\rm cosh}^2 \delta_1 + (P_1)^2 \hat{Q}_1
\hat{Q}_2 {\rm sinh}^2 \delta_1]. 
\label{def}
\end{eqnarray}
For the sake of simplification, the coordinate is chosen 
so that the outer horizon is at $r=m$.

This solution has the following non-zero charges:
\begin{eqnarray}
P^{(1)}_1 &=& P_1 Q_2/\Delta, \ \ \ \ \ \ \ \ \ \ \ \
\ \ \ \,
Q^{(1)}_1 = (\hat{P}_1 - \hat{P}_2 - \hat{Q}_1 - \hat{Q}_2)
{\rm cosh}\delta_1 {\rm sinh}\delta_1,\cr
P^{(1)}_2 &=& 0, \ \  \ \ \ \ \ \ \ \ \ \ \ \ \ \
\ \ \ \ \ \ \ \ \
Q^{(1)}_2 = (Q_1 Q_2 {\rm cosh}^2 \delta_1 + P_1 P_2 {\rm sinh}^2
\delta_1)/\Delta, \cr
P^{(2)}_1 &=& (Q_2 P_2 {\rm cosh}^2 \delta_1 +
 Q_1 P_1 {\rm sinh}^2 \delta_1)/\Delta,
\ \ \ \ \ \ \ \ \ \ \ \ \ \ \ \ \ \ Q^{(2)}_1 = 0, \cr
P^{(2)}_2&=& P_1 Q_2
(Q_2-Q_1-P_1-P_2){\rm sinh}\delta_1{\rm cosh}\delta_1 /
\Delta^2, \ \ \ \ \
Q^{(2)}_2 = \Delta,
\label{ch}
\end{eqnarray}
and the ADM mass, compatible with the  BPS bound
\cite{CVEy672,DUFlr}, is
\begin{eqnarray}
M_{ADM}&=&{\textstyle {1\over \Delta^2}}
[(P_1)^2(\hat{P}_2-\hat{Q}_2){\rm sinh}^2 \delta_1
+ (Q_2)^2(\hat{P}_1 + \hat{Q}_1){\rm cosh}^2 \delta_1]
\cr
& &+(\hat{P}_2+\hat{Q}_2){\rm cosh}^2 \delta_1 +
(\hat{Q}_1-\hat{P}_1){\rm sinh}^2 \delta_1.
\label{adm}
\end{eqnarray}

\paragraph{$S$- and $T$-Duality Transformations}\label{n4bh4dsphgentr}

The additional $51$ charge degrees of freedom needed for parameterizing 
the most general $U(1)^{28}$ charge configuration are introduced by 
$[O(6)\times O(22)]/[O(4)\times O(20)]$ and $SO(2)$ transformations.  
The resulting general solution has the charge configuration
\begin{equation}
\vec{\cal Q}^{\prime}=\textstyle{1\over \sqrt{2}} {\cal U}^T \left 
( \matrix{U_6({\bf e}_u-{\bf e}_d)\cr U_{22}\left(\matrix{{\bf e}_u
+{\bf e}_d \cr 0_{16}}\right)}\right),\ 
\vec{\cal P}^{\prime}=\textstyle{1\over \sqrt{2}} {\cal U}^T \left(
\matrix{U_6({\bf m}_u-{\bf m}_d)\cr U_{22}\left(\matrix{{\bf m}_u
+{\bf m}_d \cr 0_{16}}\right)}\right),
\label{totch}
\end{equation}
where
\begin{eqnarray}
{\bf e}^T_u &\equiv& (Q^{(1)}_1 {\rm cos}\gamma +
P^{(2)}_1{\rm sin}\gamma , Q^{(1)}_2 {\rm cos}\gamma +
P^{(2)}_2{\rm sin}\gamma ,\overbrace{0,...,0}^{4}),
\nonumber \\
{\bf e}^T_d &\equiv& (P^{(1)}_1{\rm sin}\gamma , Q^{(2)}_2
{\rm cos}\gamma , \overbrace{0,...,0}^{4}),
\nonumber \\
{\bf m}^T_u &\equiv& (P^{(1)}_1{\rm cos}\gamma , Q^{(2)}_2
{\rm sin}\gamma , \overbrace{0,...,0}^{4}),
\nonumber \\
{\bf m}^T_d &\equiv& (P^{(2)}_1 {\rm cos}\gamma +
Q^{(1)}_1{\rm sin}\gamma , P^{(2)}_2 {\rm cos}\gamma +
Q^{(1)}_2{\rm sin}\gamma ,\overbrace{0,...,0}^{4}),
\label{fourdsubmat}
\end{eqnarray}
$\gamma$ is the $SO(2) \subset SL(2,{\bf R})$ rotational angle, 
$U_6 \in SO(6)$, $U_{22} \in SO(22)$, $0_{16}$ is a
$(16\times 1)$-matrix with zero entries and ${\cal U}\in O(6,22,{\bf R})$ 
brings to the basis where the $O(6,22)$ invariant metric (\ref{4dL}) 
is diagonal.  And the complex scalar $S$ and the moduli $M$ transform to
\begin{eqnarray}
S^{\prime}&=&{{(\Psi{\rm cos}\gamma - {\rm sin}\gamma)+ie^{-\varphi}
{\rm cos}\gamma}\over {(\Psi{\rm sin}\gamma + {\rm cos}\gamma)+ie^{-\varphi}
{\rm sin}\gamma}}, 
\cr
M^{\prime}&=&{\cal U}^T \left(\matrix{U_6 &0 \cr 0 & U_{22}}\right){\cal U}
M{\cal U}^T \left (\matrix{U^T_6 & 0 \cr 0 & U^T_{22}}\right){\cal U},
\label{4dgenscalar}
\end{eqnarray}
where $\Psi$, $e^{-\varphi}$ and $M$ are the axion, the dilaton and
the moduli of the generating solution (\ref{gentsol}).
The ``Einstein-frame'' metric $g_{\mu\nu}$ in (\ref{gentsol}) remains 
unchanged, but the ``string-frame'' metric $g^{string}_{\mu\nu}$ is 
transformed to the most general form $g^{string}_{\mu\nu}= 
g_{\mu\nu}/{\rm Im}\,(S)$.

\paragraph{Special Cases of the General Solution}\label{n4bh4dsphgeneg}

The generating solution (\ref{gentsol}), when supplemented by  
appropriate subsets of $S$- and $T$-dualities, reproduces all the 
previously known spherically symmetric black holes in heterotic 
string on $T^6$.  Here, we give some examples. 
\begin{itemize}
\item Non-rotating black holes in Einstein-Maxwell-dilaton 
system with the gauge kinetic term ${1\over 4}e^{-a\varphi}F_{\mu\nu}
F^{\mu\nu}$ \cite{GIBm,GARhs43,HOLw380,DUFmr418} :\\
(1) $P_1=P_2=Q_1=Q_2\neq 0$ case:  the Reissner-Nordstr\"om black hole, 
i.e. $a=0$ \\
(2) any of 3 charges non-zero and equal:  the $a=1/\sqrt{3}$ case 
\cite{DUFlr} \\ 
(3) only 2 magnetic (or electric) charges non-zero and equal: 
the $a=1$ case \cite{KALlopp46} \\
(4) only 1 charge non-zero:  the $a=\sqrt{3}$ case, which contains 
in the extreme limit the followings $i$) $P_1 \neq 0$ case: 
KK monopole \cite{GROp226,GIBp84,SOR51}, and $ii$) 
$P_2 \neq 0$ case: H-monopole  
\cite{KHU259,KHU294,KHU387,GAUhl,BEHk468}.
\item $P_1=P_2$ and $Q_1=Q_2$ solution with subsets of $S$- and $T$-dualities 
applied becomes general axion-dilaton black holes found in 
\cite{KALo48,ORT47,KALkot50,BERko478}.  
\item The solution with $Q_1\neq 0\neq Q_2$, when supplemented
by $S$- and $T$-dualities, is the general electric black hole 
in heterotic string \cite{SEN440,SEN10}.  The $S$-dual counterpart is 
the general magnetic solution  \cite{BEHk53}.
\item The {\it non-BPS extreme solution} (i.e.
$m\to 0$, $P_2=Q_1=0$, $|Q_2|-|P_1|\to 0$ and $\delta_1 \to\infty$, 
while keeping $me^{2|\delta_1|}$ and $(|Q_2|-|P_1|)e^{2|\delta_1|}$ as 
finite constants) is related by $S$- and $T$-dualities to 
the non-BPS extreme KK black hole studied in \cite{GIBw}. 
\end{itemize}

\paragraph{Global Space-Time Structure and Thermal Properties}
\label{n4bh4dsphgenpro}

We classify all the possible spacetime and thermal
properties of non-rotating black holes in heterotic 
string on $T^6$.  These properties are determined by the 6 
parameters $P_{1,2}$, $Q_{1,2}$, $\delta_1$ and $m$ of the generating 
solution (\ref{gentsol}), since the $D=4$ $T$- and $S$-dualities, 
which introduce the remaining charge degrees of freedom, do not affect 
the ``Einstein-frame'' spacetime.  We separate the solutions into 
non-extreme ($m>0$) and extreme ($m=0$) ones.  Within each class, 
we analyze their properties according to the range of the other 5 
parameters
\footnote{In the following analysis, it is understood that
$Q_2 \neq 0$ when $\delta_1 \neq 0$.  When $Q_2=0$, $P_1=0$ 
due to initial assumption $|Q_2| \geq |P_1|$.  
Then, $\delta_{1,2}$ are not constrained by (\ref{notaubnut}).  
In this case, we have a non-extreme 4-charged solution with  
charges $P^{(2)}_1, P^{(2)}_2, Q^{(1)}_1$ and $Q^{(1)}_2$.  
Such a solution is related to (\ref{gentsol}) 
through subsets of $SO(2)\times SO(2) \subset O(2,2)
\subset O(6,22)$ and $SO(2) \subset SL(2,{\bf R})$ transformations.}
$P_{1,2}$, $Q_{1,2}$ and $\delta_1$.
\\
\\
\noindent{\it Global Space-Time Structure} 
\\
There is a spacetime singularity at $r=r_{sing}$ where 
$R(r)=0$.  
The event horizon(s) is located at $r=r_{hor\, \pm}$ 
where $\lambda=0$, provided $r_{hor\, \pm} \geq r_{sing}$.  

($A$) {\it Non-extreme solutions} ($m> 0$)

By analyzing roots of $XY-Z^2$, one sees that a singularity is 
always at $r_{sing} \leq -m$.  Thus, global spacetime is either 
that of non-extreme Reissner-Nordstr\"om black hole when $r_{sing}<-m$ 
(case with 2 horizons at $r=\pm m$) or that of Schwarzschield black 
hole when $r_{sing}=-m$.

$XY-Z^2$ has a single root at $r_{sing}=-m$, in which case a 
singularity and the inner horizon coincide at $r=-m$, when
($a$) $\delta_1 \neq 0$ and  $P_1=0$, or 
($b$) $\delta_1 =0$ and at least one of $P_{1,2}$ and $Q_{1,2}$ is zero.
$XY-Z^2$ has a double root at $r_{sing}=-m$, in 
which case the inner horizon disappears and a singularity forms at 
$r=-m$, when ($a$) $\delta_1 \neq 0$ and only $Q_2$ is non-zero, or 
($b$) $\delta_1 = 0$ and at least 2 of $P_{1,2},Q_{1,2}$ are zero.

($B$) {\it Extreme solutions} ($m \to 0$) 

When $\delta_1$ is finite, the ADM mass of the 
generating solution always saturates the Bogomol'nyi bound 
as $m\to 0$, i.e. becomes {\it BPS extreme solution}. 

When both pairs $(P_1,P_2)$ and $(Q_1,Q_2)$ have the same relative signs, 
the singularity is always at $r_{sing}\leq 0$.   Global spacetime 
is, therefore, that of the extreme Reissner-Nordstr\"om black hole when 
$r_{sing}<0$ (time-like singularity), or the singularity and the horizon 
coincide (null singularity) when $r_{sing}=r_{hor\,\pm}=0$.  The latter 
case happens when at least one out of $P_{1,2}, Q_{1}$ (and $Q_2$) is 
zero with $\delta_1 \neq 0$ (with $\delta_1 =0$).  The horizon at 
$r_{hor}=0$ disappears (naked-singularity) when ($i$) only $Q_2$ is non-zero 
with $\delta_1 \neq 0$, or ($ii$) only one out of $P_{1,2},Q_{1,2}$ is 
non-zero with $\delta_1 =0$.  

When at least one of the pairs $(P_1,P_2)$ and $(Q_1,Q_2)$ has the 
opposite relative sign, the singularity is outside of the horizon, 
i.e. $r_{sing}>0$ (singularity is naked) \cite{BEH455,KAL13,KALl52,CVEy674}.

In the case of {\it non-BPS extreme solutions} 
\cite{CVEy649,CVEy127}, 
the singularity is always behind the event horizon 
($r_{sing}<r_{hor}=0$), i.e. the global spacetime of the extreme 
Reissner-Nordstr\" om black hole (time-like singularity).
\\ 
\\
\noindent{\it Thermal Properties} 
\\
Thermal properties are specified by spacetime at the (outer) 
horizon.  So, we consider only regular solutions, which 
include non-extreme solutions compatible with the Bogomol'nyi bound 
and extreme solutions with the same relative sign 
for both pairs ($P_1,P_2$) and ($Q_1,Q_2$).   

The entropy $S$ is of the form
\begin{eqnarray}
S&=&{\textstyle {\pi\over { |\Delta|}}}\left | \left
[(\hat Q_2+m)(\hat P_2+m){\rm cosh}^2 \delta_1 +
(\hat P_1+m)(\hat Q_1-m){\rm sinh}^2 \delta_1\right ]
\right.
\cr
& &\times\left[(Q_2)^2(\hat Q_1+m)(\hat P_1+m)
{\rm cosh}^2 \delta_1  + (P_1)^2 (\hat Q_2+m)(\hat P_2-m)
{\rm sinh}^2 \delta_1\right ]
\cr
& &\left.-\left[P_1P_2 (\hat Q_2+m)+ Q_1Q_2 (\hat P_1+
m)\right]^2 {\rm cosh}^2 \delta_1 {\rm sinh}^2 \delta_1
\right |^{{1\over 2}},
\label{4dstgensolent}
\end{eqnarray}
where $\hat{P_1}=+\sqrt{P_1^2+m^2}$, etc.  Entropy increases with
$\delta_1$, approaching infinity [finite value] as $\delta_1 \to
\infty$ [non-BPS extreme limit is reached].
For BPS extreme solutions, entropy is
\noindent{($a$) non-zero and finite, approaching infinity as
$\delta_1 \to \infty$, when $P_{1,2}$ and $Q_{1,2}$ are non-zero, and}
\noindent{($b$) always zero when at least one of $P_{1,2}$,
$Q_{1}$ (and $Q_2$) is zero with $\delta_1 \neq 0$ (with $\delta_1 =0$).}

The Hawking temperature $T_H = |\partial_r \lambda (r=m)|/4\pi$ is 
\begin{equation}
T_H={\textstyle {m\over \sqrt 2}} S^{-1}.
\label{temp}
\end{equation}
As $\delta_1$ increases, $T_H$ decreases, approaching zero.   
In the BPS extreme limit with at least 3 of 
$P_{1,2},Q_{1,2}$ non-zero, $T_H$ is always zero.  
With 2 of them non-zero, $T_H$ is non-zero and finite, 
approaching zero as $\delta_1 \to \infty$.  When only one of them 
(only $Q_2$) is non-zero (for the case $\delta_1 \neq 0$), 
$T_H$ becomes infinite.  In the non-BPS extreme limit, $T_H$ is zero.

\paragraph{Duality Invariant Entropy}

We discuss the duality invariant form of entropy of near-extreme, 
non-rotating black hole in heterotic string on $T^6$ 
\cite{CVEg134}.  

The $(t,t)$-component of metric for general $N=4$ spherically 
symmetric solutions has the form $g_{tt}=\pi (r+m)(r-m)S^{-1}(r)$ with 
$S(r)$ given by
\begin{equation}
S(r)=\pi \ \prod_{i=1}^{4} \sqrt{(r +\lambda_{i})}. 
\end{equation}
Entropy $S$ of non-extreme solutions is given by 
$S(r)$ at the outer horizon, i.e. $S\equiv S(m)$.  

Generally, $\lambda_i$ are functions of $28+28$ electric and magnetic 
charges (\ref{totch}) and $m$ (through $m^2$), and 
their duality invariant forms are hard to obtain.  However, for the 
near-extreme case, in which $\lambda_i$ are expressed to leading order 
in $m^2$ around their BPS values $\lambda^{(0)}_i$, one can obtain the 
$T$- and $S$-duality invariant entropy expression, which reads
\begin{eqnarray}
\label{proposal}
S &=& \pi \ \prod_{i=1}^{4} \sqrt{\lambda_{i}^{[0]}} \ + \ 
\frac{\pi}{2} \ m\ \prod_{i=1}^{4} \sqrt{1 / \lambda_{i}^{[0]}} \ 
\sum_{i<j<k} 
\lambda_{i}^{[0]} \lambda_{j}^{[0]} \lambda_{k}^{[0]}
\ + \ {\cal O}(m^{2}).  
\end{eqnarray}
Here, the $T$- and $S$-duality invariants are 
\begin{eqnarray}
\prod_{i=1}^{4}\lambda_{i}^{[0]}&\equiv&S^{2}_{BPS}/\pi
\ = \  \frac{1}{4} \ F(L,\Gamma)F(L,-\Gamma),
\nonumber\\
\sum_{i=1}^{4} \lambda_{i}^{[0]}&\equiv&M_{BPS}
\ = \  \sqrt{F({\cal M}_{+},\Gamma)}+\sqrt{F({\cal M}_{+},-\Gamma)},
\cr
\sum_{i<j}\lambda_{i}^{[0]}\lambda_{j}^{[0]}&=&
\frac{1}{2g^2_s} 
\left(\vec{\cal Q}^{T}L\vec{\cal Q}+\vec{\cal P}^{T}L\vec{\cal P}\right)
+\sqrt{ F({\cal M}_{+},\Gamma) F({\cal M}_{+},-\Gamma)},
\cr
\sum_{i<j<k}\lambda_{i}^{[0]}\lambda_{j}^{[0]}\lambda_{k}^{[0]}&=& 
\frac{1}{4g^2_sM_{BPS}}\left\{M_{BPS}^{2}\left( \vec{\cal Q}^{T}L
\vec{\cal Q}+\vec{\cal P}^{T}L\vec{\cal P}\right)\right.
\cr
& &\left.-(\vec{\cal Q}^{T}{\cal M}_{+}\vec{\cal Q}-\vec{\cal P}^{T} 
{\cal M}_{+}\vec{\cal P})(\vec{\cal Q}^{T}L\vec{\cal Q}-\vec{\cal P}^{T}
L\vec{\cal P})\right.
\cr
& &\left.-4(\vec{\cal Q}^{T} LM_{\infty}L\vec{\cal P})(\vec{\cal Q}^{T}
{\cal M}_{+}\vec{\cal P})\right\}, 
\end{eqnarray}
where
\begin{eqnarray}
F({\cal M}_{+},\pm\Gamma)&=&\frac{1}{2g^2_s}\ 
\left(\vec{\cal Q}^{T} {\cal M}_{+} \vec{\cal Q}\ +\ 
\vec{\cal P}^{T}{\cal M}_{+}\vec{\cal P}\ \pm \ \Gamma({\cal M}_{+})\right)
\cr
\Gamma({\cal M}_{+})&=&\sqrt{4\ (\vec{\cal P}^{T}{\cal M}_{+} 
\vec{\cal Q})^{2}\ +\ (\vec{\cal Q}^{T} {\cal M}_{+}\vec{\cal Q}
-\ \vec{\cal P}^{T}{\cal M}_{+}\vec{\cal P})^{2}}.
\end{eqnarray}

\subsection{Rotating Black Holes in Four Dimensions}\label{n4bh4drot}

We generalize the 4-charged non-extreme  
solution (\ref{gensol}) to include an angular momentum 
\cite{CVEy54}.  (For an another attempt, see \cite{JATmp384}.  
But this solution has only 3 charge degrees of freedom and is 
a special case of a general solution to be discussed in this section.)

\subsubsection{Explicit Solution}\label{n4bh4drotsol}

By applying the solution generating technique discussed in the beginning 
this chapter, one obtains the following $D=4$, 
non-extreme, rotating black hole solution \cite{CVEy54}:  
\begin{eqnarray}
g_{11}&=&{{(r+2m{\rm sinh}^2 \delta_{p2})(r+2m{\rm sinh}^2 \delta_{e2})
+l^2{\rm cos}^2 \theta}\over {(r+2m{\rm sinh}^2 \delta_{p1})(r+2m
{\rm sinh}^2 \delta_{e2})+l^2{\rm cos}^2\theta}},  
\cr
g_{12}&=&{2ml{\rm cos}\theta({\rm sinh}\delta_{p1}{\rm cosh}\delta_{p2}
{\rm sinh}\delta_{e1}{\rm cosh}\delta_{e2}-{\rm cosh}\delta_{p1}
{\rm sinh}\delta_{p2}{\rm cosh}\delta_{e1}{\rm sinh}\delta_{e2})\over 
{(r+2m{\rm sinh}^2 \delta_{p1})(r+2m{\rm sinh}^2 \delta_{e2})+
l^2{\rm cos}^2\theta}},  
\cr
g_{22}&=&{{(r+2m{\rm sinh}^2 \delta_{p1})(r+2m{\rm sinh}^2 \delta_{e1})
+l^2{\rm cos}^2 \theta}\over {(r+2m{\rm sinh}^2 \delta_{p1})(r+2m
{\rm sinh}^2 \delta_{e2})+l^2{\rm cos}^2\theta}},  
\cr
B_{12}&=&-{{2ml{\rm cos}\theta({\rm sinh}\delta_{p1}{\rm cosh}
\delta_{p2}{\rm cosh}\delta_{e1}{\rm sinh}\delta_{e2}-{\rm cosh}
\delta_{p1}{\rm sinh}\delta_{p2}{\rm sinh}\delta_{e1}{\rm cosh}
\delta_{e2})}\over{(r+2m{\rm sinh}^2 \delta_{p1})(r+2m{\rm sinh}^2 
\delta_{e2})+l^2{\rm cos}^2\theta}},  
\cr
e^{\varphi}&=&{{(r+2m{\rm sinh}^2 \delta_{p1})(r+2m{\rm sinh}^2 
\delta_{p2})+l^2{\rm cos}^2 \theta}\over \Delta^{1\over 2}},\cr
ds^2_{E}&=&\Delta^{1\over 2}[-{{r^2-2mr+l^2{\rm cos}^2\theta}\over 
\Delta}dt^2+{{dr^2}\over{r^2-2mr+l^2}} + d\theta^2 
\cr
& &+{{{\rm sin}^2\theta}\over \Delta}\{(r+2m{\rm sinh}^2 
\delta_{p1})(r+2m{\rm sinh}^2 \delta_{p2})(r+2m{\rm sinh}^2 \delta_{e1})
\cr
& &\ \ \times(r+2m{\rm sinh}^2 \delta_{e2})+l^2(1+{\rm cos}^2\theta)r^2+W 
+2ml^2r{\rm sin}^2\theta\}d\phi^2
\cr
& &-{{4ml}\over \Delta}\{({\rm cosh} \delta_{p1}{\rm cosh}
\delta_{p2}{\rm cosh} \delta_{e1}{\rm cosh} \delta_{e2}
-{\rm sinh} \delta_{p1}{\rm sinh} \delta_{p2}
{\rm sinh} \delta_{e1}{\rm sinh} \delta_{e2})r
\cr
& &\ \ +2m{\rm sinh}\delta_{p1}{\rm sinh}\delta_{p2}{\rm sinh}\delta_{e1}
{\rm sinh}\delta_{e2}\}{\rm sin}^2 \theta dtd\phi],
\label{4dsol}
\end{eqnarray}
where
\begin{eqnarray}
\Delta &\equiv& (r+2m{\rm sinh}^2 \delta_{p1})
(r+2m{\rm sinh}^2 \delta_{p2})(r+2m{\rm sinh}^2 \delta_{e1})
(r+2m{\rm sinh}^2 \delta_{e2})
\cr
& &+(2l^2r^2+W){\rm cos}^2\theta, 
\cr
W &\equiv& 2ml^2({\rm sinh}^2\delta_{p1}+{\rm sinh}^2\delta_{p2}+
{\rm sinh}^2\delta_{e1}+{\rm sinh}^2\delta_{e2})r 
\cr
& &+4m^2l^2(2{\rm cosh}\delta_{p1}{\rm cosh}\delta_{p2}{\rm cosh}
\delta_{e1}{\rm cosh}\delta_{e2}{\rm sinh}\delta_{p1}{\rm sinh}
\delta_{p2}{\rm sinh}\delta_{e1}{\rm sinh}\delta_{e2}
\cr 
& &-2{\rm sinh}^2 \delta_{p1}{\rm sinh}^2 \delta_{p2}{\rm sinh}^2 
\delta_{e1}{\rm sinh}^2 \delta_{e2}-{\rm sinh}^2 \delta_{p2}
{\rm sinh}^2 \delta_{e1}{\rm sinh}^2 \delta_{e2} 
\cr
& &-{\rm sinh}^2 \delta_{p1}{\rm sinh}^2 \delta_{e1}
{\rm sinh}^2 \delta_{e2}-{\rm sinh}^2 \delta_{p1}{\rm sinh}^2 
\delta_{p2}{\rm sinh}^2 \delta_{e2}
\cr
& &-{\rm sinh}^2 \delta_{p1}{\rm sinh}^2 \delta_{p2}{\rm sinh}^2 
\delta_{e1})+l^4{\rm cos}^2 \theta.
\label{4ddef}
\end{eqnarray}
The axion $\Psi$ also varies with spatial coordinates, but 
since its expression turns out to be cumbersome, we shall not 
write here explicitly. 

The ADM mass, $U(1)$ charges, and angular momentum are
\begin{eqnarray}
M&=&2m({\rm cosh}2\delta_{e1}+{\rm cosh}2\delta_{e2}+
{\rm cosh}2\delta_{p1}+{\rm cosh}2\delta_{p2}), \cr
Q^{(1)}_2 &=&2m{\rm sinh}2\delta_{e1},\ \ \ \ \ \ \ 
Q^{(2)}_2=2m{\rm sinh}2\delta_{e2}, 
\cr 
P^{(1)}_1&=&2m{\rm sinh}2\delta_{p1},\ \ \ \ \ \ \ 
P^{(2)}_1=2m{\rm sinh}2\delta_{p2}, 
\cr
J&=&8lm({\rm cosh}\delta_{e1}{\rm cosh}\delta_{e2}{\rm cosh}\delta_{p1}
{\rm cosh}\delta_{p2}-{\rm sinh}\delta_{e1}{\rm sinh}\delta_{e2}
{\rm sinh}\delta_{p1}{\rm sinh}\delta_{p2}),  
\label{4dphys}
\end{eqnarray}
where $G_N^{D=4}={1\over 8}$ and the convention of  
\cite{MYEp172} is followed.  

When $Q_2^{(1)}=Q_2^{(2)}=P_1^{(1)}=P_1^{(2)}$, all the scalars are constant, 
and thus the solution becomes the Kerr-Newman solution.  
The $\delta_{p1}=\delta_{p2}=0$ case is the generating solution 
of a general electric rotating solution \cite{SEN440}.  
The case with $Q_2^{(1)}=P_1^{(1)}$ and $Q_2^{(2)}=P_1^{(2)}$ is 
constructed in \cite{JATmp384}.

The solution (\ref{4dsol}) has the inner $r_-$ and the outer $r_+$ 
horizons at
\begin{equation}
r_{\pm}=m\pm \sqrt{m^2-l^2},
\label{4dhorizon}
\end{equation}
provided $m\ge |l|$.  In this case, the solution has the 
global spacetime of the Kerr-Newman black hole with the 
ring singularity at $r={\rm min}\{Q^{(1)}_2,Q^{(2)}_2,P^{(1)}_1,
P^{(2)}_1\}$ and $\theta={\pi\over 2}$.  

The extreme solution ($r_+=r_-$) is obtained by taking the limit 
$m\to |l|^+$.  In this case, the global spacetime is that of the extreme 
Kerr-Newman solution. 

The BPS limit is reached by taking $m\to 0$ and 
$\delta_{e1,e2,p1,p2}\to\infty$ while keeping $me^{2\delta_{e1,e2,p1,p2}}$ 
as finite constants so that the charges remain non-zero.  When $J$ is 
non-zero, i.e. $l\neq 0$, the singularity is naked since the condition 
$m\geq |l|$ for existence of regular horizon (\ref{4dhorizon}) is not 
satisfied
\footnote{See \cite{TSE381} for the same result from the conformal 
$\sigma$-model perspective.}.   
To have a BPS solution with regular horizon, one has to take $l\to 0$, 
leading to a solution with $J=0$.  
Thus, the only regular BPS solution in $D=4$ is the non-rotating solution, 
with global spacetime of the extreme Reissner-Nordstr\" om black hole.
This is in contrast with the $D=5$ 3-charged solution 
\cite{CVEy476,BREmpv}, where one can take $l_{1,2}$ to zero (so that 
the BPS solution has regular horizon) but the angular momenta $J_{1,2}$ 
can be non-zero.  For $D>5$, the regular BPS limit with non-zero angular 
momentum is achieved without taking $l_i$ to zero if only {\it one} 
angular momentum is non-zero \cite{HORs53}.

\subsubsection{Entropy of General Solution}
\label{n4bh4drotent}

The thermal entropy of the solution (\ref{4dsol}) is \cite{CVEy54}
\begin{eqnarray}
S&=&\textstyle{1\over 4G_N} A=16\pi [m^2(\prod^4_{i=1}\cosh\delta_i+
\prod^4_{i=1}\sinh\delta_i)
\cr
& &\ \ \ \ \ +m\sqrt{m^2-l^2}(\prod^4_{i=1}
\cosh\delta_i-\prod^4_{i=1}\sinh\delta_i)]\cr
&=& 16\pi\left[m^2(\prod^4_{i=1}\cosh\delta_i+\prod^4_{i=1}
\sinh\delta_i)\right.
\cr
& &\ \ \ \ \ +\left.\sqrt{m^4(\prod^4_{i=1}\cosh\delta_i
-\prod^4_{i=1}\sinh\delta_i)^2-J^2}\right], 
\label{4dent}
\end{eqnarray}
where $\delta_{1,2,3,4}\equiv \delta_{e1,e2,p1,p2}$ and 
$A=\int d\theta d\phi \left.\sqrt{g_{\theta\theta}g_{\phi\phi}}
\right|_{r=r_+}$ is the outer-horizon area.  

Note, the thermal entropy has the form which 
is {\it sum} of `left-moving' and `right-moving' contributions.  
Each term is {\it symmetric} in $\delta_i$, i.e. in the 4 
charges, manifesting $U$-duality symmetry \cite{HULt438}.  On the other 
hand, (\ref{4dent}) is asymmetric in $J$: only the 
{\it right-moving} term has $J$, which reduces the right-moving 
contribution to the entropy.  This reflects right-moving worldsheet 
supersymmetry of the corresponding $\sigma$-model.  

When $J=0$, the entropy becomes \cite{CVEy675}:
\begin{equation}
S=32\pi m^2\prod^4_{i=1}\cosh\delta_i, 
\end{equation}
which again has $U$-duality symmetry under the exchange of 4 charges.  

In the regular BPS limit as well as the extreme limit, the `right-moving' 
term in (\ref{4dent}) becomes zero, however entropy has different form 
in each case.  In the regular BPS limit ($J=0$) \cite{CVEy672}: 
\begin{equation}
S=32\pi m^2\prod^4_{i=1}\cosh\delta_i=
2\pi\sqrt{P^{(1)}_1P^{(2)}_1Q^{(1)}_2Q^{(2)}_2}, 
\label{4dbpsent}
\end{equation}
while in the extreme limit:
\begin{equation}
S=16\pi m^2(\prod^4_{i=1}\cosh\delta_i+\prod^4_{i=1}
\sinh\delta_i)=2\pi\sqrt{J^2+P^{(1)}_1P^{(2)}_1Q^{(1)}_2
Q^{(2)}_2}.
\label{4dexarea}
\end{equation}

Entropy of a black hole with general charge configuration 
in the class and with arbitrary scalar asymptotic values  
is independent of scalar asymptotic values when 
expressed in terms of the charge lattice vectors $\vec{\alpha}$ and 
$\vec{\beta}$, and has the $S$- and $T$-duality invariant form 
\cite{CVEy54}: 
\begin{equation}
S=2\pi\sqrt{J^2+\{(\vec{\cal \alpha}^TL\vec{\cal \alpha})
(\vec{\cal \beta}^TL\vec{\cal \beta})-(\vec{\cal \alpha}^TL
\vec{\cal \beta})^2\}}.   
\label{4dvecexarea}
\end{equation}

\subsection{General Rotating Five-Dimensional Solution}
\label{n4bh5d}

We construct the most general rotating black 
hole in heterotic string on $T^5$ \cite{CVEy476}.  In $D=5$, 
black holes carry only electric charges of $U(1)$ gauge fields.  
Since the NS-NS 3-form field strength $H_{\mu\nu\rho}$ is Hodge-dual to 
a 2-form field strength in $D=5$ in the following way
\begin{equation}
H^{\mu\nu\rho}=-{e^{4\varphi/3}\over{2!\sqrt{-g}}}
\varepsilon^{\mu\nu\rho\lambda\sigma}F_{\lambda\sigma},
\end{equation}
where $F_{\mu\nu}$ is the field strength of a new $U(1)$ gauge field $A_\mu$, 
black holes in $D=5$ carry an additional charge associated with 
the NS-NS 2-form field $B_{\mu\nu}$ as well as 26 electric charges of 
the $U(1)^{26}$ gauge group.  Thus, the most general 
black hole in heterotic string on $T^5$, compatible with 
the conjectured ``no-hair theorem'' \cite{ISR164,ISR8,CAR26,HAW71,HAW25}, 
is parameterized by 27 electric charges, 2 angular momenta and the 
non-extremality parameter.

\subsubsection{Generating Solution}\label{n4bh5dsol}

We choose to parameterize the ``generating solution'' in terms of  
electric charges $Q$, $Q^{(1)}_1$ and $Q^{(2)}_1$ associated with 
$H_{\mu\nu\rho}$, $A^{(1)\,1}_{\mu}$ and $A^{(2)}_{\mu\,1}$, respectively.  
These charges are induced through solution generating 
procedure described in section \ref{n4bhgen}.  

The final form of the generating solution is \cite{CVEy476}  
\begin{eqnarray}
g_{11}&=&{{r^2+2m{\rm sinh}^2 \delta_{e1}+l^2_1{\rm cos}^2\theta+
l^2_2{\rm sin}^2\theta}\over{r^2+2m{\rm sinh}^2 \delta_{e2}+l^2_1
{\rm cos}^2\theta+l^2_2{\rm sin}^2\theta}}, \cr
e^{2\varphi}&=&{{(r^2+2m{\rm sinh}^2 \delta_{e}+l^2_1{\rm cos}^2
\theta+l^2_2{\rm sin}^2\theta)^2}\over{\prod^2_{i=1}
(r^2+2m{\rm sinh}^2\delta_{ei}+l^2_1{\rm cos}^2\theta
+l^2_2{\rm sin}^2\theta)}},
\cr
A^{(1)}_{t\,1}&=&{{m{\rm cosh}\delta_{e1}{\rm sinh}\delta_{e1}}
\over{r^2+2m{\rm sinh}^2\delta_{e1}+l^2_1{\rm cos}^2\theta +
l^2_2{\rm sin}^2 \theta}}, \cr
A^{(1)}_{\phi_1\,1}&=&m{\rm sin}^2\theta
{{l_1{\rm sinh}\delta_{e1}{\rm sinh}\delta_{e2}{\rm cosh}\delta_{e}
-l_2{\rm cosh}\delta_{e1}{\rm cosh}\delta_{e2}{\rm sinh}\delta_{e}}
\over{r^2+2m{\rm sinh}^2\delta_{e1}+l^2_1{\rm cos}^2\theta +
l^2_2{\rm sin}^2 \theta}}, \cr
A^{(1)}_{\phi_2\,1}&=&m{\rm cos}^2\theta
{{l_1{\rm cosh}\delta_{e1}{\rm sinh}\delta_{e2}{\rm sinh}\delta_{e}
-l_2{\rm sinh}\delta_{e1}{\rm cosh}\delta_{e2}{\rm cosh}\delta_{e}}
\over{r^2+2m{\rm sinh}^2\delta_{e1}+l^2_1{\rm cos}^2\theta +
l^2_2{\rm sin}^2 \theta}}, \cr
A^{(2)}_{t\,1}&=&{{m{\rm cosh}\delta_{e2}{\rm sinh}\delta_{e2}}
\over{r^2+2m{\rm sinh}^2\delta_{e2}+l^2_1{\rm cos}^2\theta +
l^2_2{\rm sin}^2 \theta}}, \cr
A^{(2)}_{\phi_1\,1}&=&m{\rm sin}^2\theta
{{l_1{\rm cosh}\delta_{e1}{\rm sinh}\delta_{e2}{\rm cosh}\delta_{e}
-l_2{\rm sinh}\delta_{e1}{\rm cosh}\delta_{e2}{\rm sinh}\delta_{e}}
\over{r^2+2m{\rm sinh}^2\delta_{e2}+l^2_1{\rm cos}^2\theta +
l^2_2{\rm sin}^2 \theta}}, \cr
A^{(2)}_{\phi_2\,1}&=&m{\rm cos}^2\theta
{{l_1{\rm sinh}\delta_{e1}{\rm cosh}\delta_{e2}{\rm sinh}\delta_{e}
-l_2{\rm cosh}\delta_{e1}{\rm sinh}\delta_{e2}{\rm cosh}\delta_{e}}
\over{r^2+2m{\rm sinh}^2\delta_{e2}+l^2_1{\rm cos}^2\theta +
l^2_2{\rm sin}^2 \theta}}, \cr
\hat{B}^{(10)}_{t\phi_1}&=&-2m\sin^2\theta{{l_1\sinh\delta_{e1}
\sinh\delta_{e2}\cosh\delta_e-l_2\cosh\delta_{e1}\cosh\delta_{e2}
\sinh\delta_e}\over
{r^2+l^2_1\cos^2\theta+l^2_2\sin^2\theta+2m\sinh^2\delta_{e2}}},
\cr
\hat{B}^{(10)}_{t\phi_2}&=&-2m\cos^2\theta{{l_2\sinh\delta_{e1}\sinh
\delta_{e2}\cosh\delta_e-l_1\cosh\delta_{e1}\cosh\delta_{e2}\sinh\delta_e}
\over{r^2+l^2_1\cos^2\theta+l^2_2\sin^2\theta+2m\sinh^2\delta_{e2}}},
\cr
\hat{B}^{(10)}_{\phi_1\phi_2}&=&-{{2m\cosh\delta_e\sinh\delta_e\cos^2\theta
(r^2+l^2_1+2m\cosh^2\delta_{e2})} \over 
{r^2+l^2_1\cos^2\theta+l^2_2\sin^2\theta+
2m\sinh^2\delta_{e2}}},
\cr
ds^2_E&=& \bar{\Delta}^{1\over 3} \left [-{{(r^2+l^2_1{\rm cos}^2
\theta +l^2_2{\rm sin}^2\theta)(r^2+l^2_1{\rm cos}^2\theta 
+l^2_2{\rm sin}^2\theta-2m)}\over \bar{\Delta}} dt^2 \right .
\cr
& &+{r^2 \over {(r^2+l^2_1)(r^2+l^2_2)-2mr^2}} dr^2 +d\theta^2
\cr
& &+{{4m{\rm cos}^2\theta {\rm sin}^2\theta}\over \bar{\Delta}}
[l_1 l_2\{(r^2+l^2_1{\rm cos}^2\theta +l^2_2{\rm sin}^2\theta) 
\cr
& &\ \ -2m({\rm sinh}^2\delta_{e1}{\rm sinh}^2\delta_{e2}+
{\rm sinh}^2\delta_{e}{\rm sinh}^2\delta_{e1}
+{\rm sinh}^2\delta_{e}{\rm sinh}^2\delta_{e2})\}
\cr
& &\ \ +2m\{(l^2_1+l^2_2){\rm cosh}\delta_{e1}{\rm cosh}
\delta_{e2}{\rm cosh}\delta_{e}
{\rm sinh}\delta_{e1}{\rm sinh}\delta_{e2}{\rm sinh}\delta_{e}
\cr
& &\ \ -2l_1l_2{\rm sinh}^2 \delta_{e1}{\rm sinh}^2 \delta_{e2}
{\rm sinh}^2 \delta_{e}\}] d\phi_1 d\phi_2 
\cr
& &-{{4m{\rm sin}^2\theta}\over \bar{\Delta}}[(r^2+l^2_1
{\rm cos}^2\theta+l^2_2{\rm sin}^2\theta)(l_1{\rm cosh}
\delta_{e1}{\rm cosh}\delta_{e2}{\rm cosh}\delta_{e}
\cr
& &\ \ -l_2{\rm sinh}\delta_{e1}{\rm sinh}\delta_{e2}{\rm sinh}\delta_{e})
+2ml_2{\rm sinh}\delta_{e1}{\rm sinh}\delta_{e2}{\rm sinh}
\delta_{e}]d\phi_1 dt 
\cr
& &-{{4m{\rm cos}^2\theta}\over\bar{\Delta}}
[(r^2+l^2_1{\rm cos}^2\theta+l^2_2{\rm sin}^2\theta)
(l_2{\rm cosh}\delta_{e1}{\rm cosh}\delta_{e2}
{\rm cosh}\delta_{e}
\cr
& &\ \ \ -l_1{\rm sinh}\delta_{e1}{\rm sinh}
\delta_{e2}{\rm sinh}\delta_{e})+2ml_1{\rm sinh}\delta_{e1}
{\rm sinh}\delta_{e2}{\rm sinh}\delta_{e}] d\phi_2 dt 
\cr
& &+{{{\rm sin}^2\theta}\over \bar{\Delta}} 
[(r^2+2m{\rm sinh}^2\delta_e+l^2_1)
\cr
& &\ \ \ \times\prod^2_{i=1}(r^2+2m{\rm sinh}^2\delta_{ei}
+l^2_1{\rm cos}^2\theta+l^2_2{\rm sin}^2\theta)
\cr
& &\ \ \ +2m{\rm sin}^2 \theta\{(l^2_1{\rm cosh}^2\delta_e
-l^2_2{\rm sinh}^2\delta_e)(r^2+l^2_1{\rm cos}^2\theta +l^2_2{\rm sin}^2
\theta)
\cr
& &\ \ \ +4ml_1 l_2{\rm cosh}\delta_{e1}
{\rm cosh}\delta_{e2}{\rm cosh}\delta_{e}{\rm sinh}\delta_{e1}
{\rm sinh}\delta_{e2}{\rm sinh}\delta_{e}
\cr
& &\ \ \ -2m{\rm sinh}^2\delta_{e1}
{\rm sinh}^2\delta_{e2}(l^2_1{\rm cosh}^2\delta_e+l^2_2{\rm sinh}^2\delta_e)
\cr
& &\ \ \ -2ml^2_2{\rm sinh}^2\delta_e({\rm sinh}^2\delta_{e1}+
{\rm sinh}^2\delta_{e2})\}]d\phi^2_1 
\cr
& &+{{{\rm cos}^2\theta}\over \bar{\Delta}}
[(r^2+2m{\rm sinh}^2\delta_e+l^2_2)
\cr
& &\ \ \ \times\prod^2_{i=1}(r^2+2m{\rm sinh}^2\delta_{ei}+
l^2_1{\rm cos}^2\theta+l^2_2{\rm sin}^2\theta)
\cr
& &\ \ \ +2m{\rm cos}^2 \theta\{(l^2_2{\rm cosh}^2\delta_e 
-l^2_1{\rm sinh}^2\delta_e)(r^2+l^2_1{\rm cos}^2\theta 
+l^2_2{\rm sin}^2\theta)
\cr
& &\ \ \ +4ml_1 l_2 {\rm cosh}\delta_{e1}
{\rm cosh}\delta_{e2}{\rm cosh}\delta_{e}{\rm sinh}\delta_{e1}
{\rm sinh}\delta_{e2}{\rm sinh}\delta_{e}
\cr
& &\ \ \ -2m{\rm sinh}^2\delta_{e1}{\rm sinh}^2\delta_{e2}
(l^2_1{\rm sinh}^2\delta_e+l^2_2{\rm cosh}^2\delta_e)
\cr
& &\ \ \ \left.-2ml^2_1{\rm sinh}^2\delta_e({\rm sinh}^2\delta_{e1}
+{\rm sinh}^2\delta_{e2})\}]d\phi^2_2 \right], 
\label{5daxisol}
\end{eqnarray}
where
\begin{eqnarray}
\bar{\Delta} &\equiv& (r^2+2m{\rm sinh}^2\delta_{e1}+
l^2_1{\rm cos}^2\theta+l^2_2{\rm sin}^2\theta)(r^2+2m
{\rm sinh}^2\delta_{e2}+l^2_1{\rm cos}^2\theta+l^2_2
{\rm sin}^2\theta)\cr
& &\times(r^2+2m{\rm sinh}^2\delta_{e}+l^2_1{\rm cos}^2\theta
+l^2_2{\rm sin}^2\theta), 
\label{5def}
\end{eqnarray}
and the subscript $E$ in the line element denotes the Einstein-frame.  
The $U(1)$ charges, the ADM mass and the angular momenta  
of the generating solution (\ref{5daxisol}) (with 
$G_N^{D=5}={\pi\over 4}$) are
\begin{eqnarray}
Q^{(1)}_1&=&m{\rm sinh}2\delta_{e1},\ \ \ 
Q^{(2)}_1=m{\rm sinh}2\delta_{e2},\ \ \ 
Q=m{\rm sinh}2\delta_{e}, 
\cr
M&=&m({\rm cosh}2\delta_{e1}+{\rm cosh}2\delta_{e2}+
{\rm cosh}2\delta_{e})
\cr
&=&\sqrt{m^2+(Q^{(1)}_1)^2}+\sqrt{m^2+(Q^{(2)}_1)^2}+
\sqrt{m^2+Q^2}, 
\cr
J_1&=&4m(l_1{\rm cosh}\delta_{e1}{\rm cosh}\delta_{e2}
{\rm cosh}\delta_{e}-l_2{\rm sinh}\delta_{e1}{\rm sinh}\delta_{e2}
{\rm sinh}\delta_{e}), 
\cr
J_2&=&4m(l_2{\rm cosh}\delta_{e1}{\rm cosh}\delta_{e2}
{\rm cosh}\delta_{e}-l_1{\rm sinh}\delta_{e1}{\rm sinh}\delta_{e2}
{\rm sinh}\delta_{e}).
\label{5ddef}
\end{eqnarray}
The solution has the outer and inner horizons at:
\begin{equation}
r_{\pm}^2=m-{1\over 2}l_1^2-{1\over 2}l_2^2 \pm
{1\over 2}\sqrt{(l^2_1-l^2_2)^2+4m(m-l^2_1-l^2_2)},
\label{oihori}
\end{equation}
provided $m\ge (|l_1|+|l_2|)^2$.

When $Q=Q_{1}^{(1)}=Q_1^{(2)}$, the generating solution becomes the 
$D=5$ Kerr-Newman solution, since $g_{11}$ and $\varphi$ become constant.  
The generating solution with $Q=Q_1^{(2)}$ corresponds to the case where 
the $D=6$ dilaton $\varphi_6=\varphi+{\textstyle{1\over 2}}\log
{\rm det} g_{11}$ is constant.  In this case, with a subsequent rescaling 
of scalar asymptotic values one obtains the static 
solution of \cite{HORs77} and rotating solution of \cite{BRElmpsv}. 

The BPS limit with $J_{1,2}\neq 0$ and regular event horizon 
is defined as the limit in which $m\to 0$, $l_{1,2}\to 0$ and 
$\delta_{e1,e2,e}\to\infty$ while keeping ${1\over 2}me^{2\delta_{e1}}=
Q^{(1)}_1,\,{1\over 2}me^{2\delta_{e2}}=Q^{(2)}_1,\,{1\over 2}m
e^{2\delta_{e}}=Q$,\,$l_1/m^{1/2}=L_1$ and $l_2/m^{1/2}=L_2$ constant.  
In this limit, the generating solution becomes  
\begin{eqnarray}
A^{(1)}_{t\,1}&=&{{{1\over 2}Q^{(1)}_1} \over {r^2+Q^{(1)}_1}}, 
\ \ \ \ 
A^{(1)}_{\phi_1\,1}={{{1\over 4}J\sin^2\theta} \over 
{r^2+Q^{(1)}_1}}, \ \ \ \ 
A^{(1)}_{\phi_2\,1}={{{1\over 4}J\cos^2\theta} \over 
{r^2+Q^{(1)}_1}}, \cr
A^{(2)}_{t\,1}&=&{{{1\over 2}Q^{(2)}_1} \over {r^2+Q^{(2)}_1}}, 
\ \ \ \ 
A^{(2)}_{\phi_1\,1}={{{1\over 4}J\sin^2\theta} \over 
{r^2+Q^{(2)}_1}}, \ \ \ \ 
A^{(2)}_{\phi_2\,1}={{{1\over 4}J\cos^2\theta} \over 
{r^2+Q^{(2)}_1}}, 
\cr
\hat{B}^{(10)}_{t\phi_1}&=&
-{{{1\over 2}J\sin^2\theta}\over {r^2+Q^{(2)}_1}}, \ \ \ 
\hat{B}^{(10)}_{t\phi_2}={{{1\over 2}J\cos^2\theta}\over 
{r^2+Q^{(2)}_1}}, \ \ \ 
\hat{B}^{(10)}_{\phi_1\phi_2}=-Q\cos^2\theta, 
\cr
g_{11}&=&{{r^2+Q^{(1)}_1}\over{r^2+Q^{(2)}_1}}, 
\ \  e^{\varphi}={{(r^2+Q)}\over{[(r^2+Q^{(1)}_1)
(r^2+Q^{(2)}_1)]^{1\over 2}}}, 
\cr
ds^2_E&=&\bar{\Delta}^{1\over 3}\left[-{r^4\over \bar{\Delta}}dt^2 
+{{dr^2} \over r^2} +d\theta^2+{{J^2{\rm cos}^2\theta
{\rm sin}^2\theta}\over {2\bar{\Delta}}}d\phi_1 d\phi_2\right.
\cr
& &-{{2Jr^2{\rm sin}^2\theta}\over 
\bar{\Delta}}dtd\phi_1+{{2Jr^2{\rm cos}^2\theta}\over\bar{\Delta}}
dtd\phi_2 
\cr
& &+{{{\rm sin}^2\theta} \over \bar{\Delta}}\{(r^2+Q^{(1)}_1)
(r^2+Q^{(2)}_1)(r^2+Q)-{1\over 4}J^2{\rm sin}^2\theta\}
d\phi^2_1 \cr
& &+\left.{{{\rm cos}^2\theta} \over \bar{\Delta}}\{(r^2+Q^{(1)}_1)
(r^2+Q^{(2)}_1)(r^2+Q)-{1\over 4}J^2{\rm cos}^2
\theta\}d\phi^2_2 \right], 
\label{5dhetBPS}
\end{eqnarray}
where
\begin{equation}
\bar{\Delta}\equiv (r^2+Q^{(1)}_1)(r^2+Q^{(2)}_1)(r^2+Q).
\label{5dBPSdef}
\end{equation}
The solution is specified by 3 charges and {\it only 1} angular 
momentum $J$
\footnote{When 1 or 3 boost parameters are negative, one has the 
BPS limit with $J_1=J_2$.}:
\begin{equation}
J_1=-J_2\equiv J=(2Q^{(1)}_1Q^{(2)}_1Q)^{1\over 2}
(L_1-L_2),
\end{equation}
while its ADM mass saturates the Bogomol'nyi bound:
\begin{equation}
M_{BPS}=Q^{(1)}_1+Q^{(2)}_2+Q.
\end{equation}

\subsubsection{$T$-Duality Transformation}\label{n4bh5dtr}

The remaining 27 electric charges (needed for parameterizing the most 
general charge configuration) are introduced by the $[SO(5)\times 
SO(21)]/[SO(4)\times SO(20)]$ transformation on the generating solution 
(\ref{5daxisol}). The final expression for electric charges is
\begin{equation}
\vec{\cal Q} = {1\over \sqrt{2}} {\cal U}^T \left ( 
\matrix{U_5 ({\bf e}_u-{\bf e}_d)\cr U_{21}\left(\matrix{{\bf e}_u
+{\bf e}_d \cr 0_{16}}\right)}\right ), 
\label{newch}
\end{equation}
where 
\begin{equation}
{\bf e}^T_u \equiv (Q^{(1)}_1 ,\overbrace{0,...,0}^{4}),\ \ \ 
{\bf e}^T_d \equiv (Q^{(2)}_1, \overbrace{0,...,0}^{4}),
\label{fivdsubmat}
\end{equation}
$U_5\in SO(5)$, $U_{21}\in SO(21)$, $0_{16}$ is a $(16\times 1)$-matrix 
with zero entries and ${\cal U}\in O(5,21,{\bf R})$ 
brings to the basis where the $O(5,21)$ invariant metric $L$ 
(\ref{4dL}) is diagonal.  And the charge $Q$ associated with $B_{\mu\nu}$ 
remains unchanged.  The moduli $M$ is transformed to
\begin{equation}
M^{\prime}={\cal U}^T \left(\matrix{U_5 &0 \cr 0 & U_{21}}\right)
{\cal U}M{\cal U}^T \left (\matrix{U^T_5 & 0 \cr 0 & U^T_{21}}
\right){\cal U},
\label{5dgenscalar}
\end{equation}
where $M$ is the moduli of the generating solution (\ref{5daxisol}).  
The subsequent $O(5,21)\times SO(1,1)$ transformation 
leads to the solution with arbitrary asymptotic values $M_\infty$ 
and $\varphi_\infty$.

\subsubsection{Entropy of General Solution}\label{n4bh5dent}

The thermal entropy of the generating solution (\ref{5daxisol}) 
is \cite{CVEy54}
\begin{eqnarray}
S&=&{1\over {4G_N}}A=4\pi \left[m\{2m- (l_1-l_2)^2\}^{1/2}
(\prod^3_{i=1}\cosh\delta_i +\prod^3_{i=1}\sinh\delta_i)\right.\cr
& &\ \ \ \ +\left. m\{2m- (l_1+l_2)^2\}^{1/2}(\prod^3_{i=1}\cosh
\delta_i-\prod^3_{i=1}\sinh\delta_i)\right]
\cr
&=&4\pi\left[\sqrt{2m^3(\prod^3_{i=1}\cosh\delta_i+\prod^3_{i=1}
\sinh\delta_i)^2-\textstyle{1\over 16}(J_1-J_2)^2}
\right.\cr
& &\ \ \ \ +\left.\sqrt{2m^3(\prod^3_{i=1}\cosh\delta_i-\prod^3_{i=1}
\sinh\delta_i)^2-\textstyle{1\over 16}(J_1+J_2)^2}
\right],
\label{5dent}
\end{eqnarray}
where $\delta_{1,2,3}\equiv \delta_{e1,e2,e}$, $G_N={\pi\over 4}$ and 
the outer horizon area is defined as 
$A=\int d\theta d\phi_1 d\phi_2 \left.\sqrt{g_{\theta\theta}(g_{\phi_1\phi_1}
g_{\phi_2\phi_2}-g_{\phi_1\phi_2}^2)}\right |_{r=r_+}$. 

Note, each term is {\it symmetric} under the permutation of $\delta_i$ 
(i.e. 3 charges), manifesting the conjectured $U$-duality symmetry 
\cite{HULt438}.  Again, as in the $D=4$ case, the entropy (\ref{5dent}) 
is cast in the form as {\it sum} of `left-moving' and 
`right-moving' contributions, hinting at the possibility of statistical 
interpretation of each term  
as left- and right-moving ($D$-brane worldvolume) 
contributions to microscopic degrees of freedom.  Each term now carries 
left- or right-moving angular momentum that could be interpreted 
as left- or right-moving $U(1)$ charge \cite{BANdfm299,BANd307} 
of the $N=4$ superconformal field theory when the generating solution 
(\ref{5daxisol}) is transformed to a solution of type-IIA string on 
$K3\times S^1$ through the conjectured string-string duality in 
$D=6$ \cite{WIT443}.  When $J_i=0$, the entropy rearranges itself as 
a single term \cite{HORms383,CVEy54}:
\begin{equation}
S=8\sqrt{2}\pi m^{3/2}\prod^3_{i=1}\cosh\delta_i, 
\end{equation}
which again has manifest symmetry under permutation of charges.  

We discuss duality invariant forms of the entropy and the ADM mass of 
non-extreme, rotating black hole with general charge configuration 
(\ref{newch}).  The entropy and the ADM mass are expressed in terms of 
the following $T$-duality invariants (obtained by applying  
$T$-duality to charges of the generating solution):
\begin{eqnarray}
\label{coordinates_inv_5}
Q_1^{(1)} &\to&X={1\over 2}\sqrt{\vec{\cal Q}^{T}{\cal M}_{+}\vec{\cal Q}}
+{1\over2}\sqrt{\vec{\cal Q}^{T}{\cal M}_{-}\vec{\cal Q}}
\cr
Q_1^{(2)}&\to&Y={1\over 2}\sqrt{\vec{\cal Q}^{T}{\cal M}_{+}\vec{\cal Q}} 
-{1\over 2}\sqrt{\vec{\cal Q}^{T}{\cal M}_{-}\vec{\cal Q}}, 
\end{eqnarray}
while $Q$ remains intact under $T$-duality.  
From these 3 $T$-duality invariant ``coordinates''  $X,Y,Q$, 
one defines the following duality invariant ``non-extreme hatted'' 
quantities $\hat{X}_i$: 
\begin{equation}
\hat X_{i} \equiv\sqrt{ X_{i}^{2}+m^{2}}, \ \ \   X_{i}=(X,Y,Q).
\end{equation} 

Duality invariant forms of the entropy and the ADM mass are
\begin{eqnarray}
S&=&2\pi\left[\sqrt{\prod_i\hat{X}_i+m^2\sum_i\hat{X}_i
+\sqrt{\prod_i(\hat{X}^{2}_i-m^2)}-(J_1-J_2)^{2}}\right.
\cr
& &+\left.\sqrt{\prod_i\hat{X}_i+m^2\sum_i\hat{X}_i
-\sqrt{\prod_i(\hat{X}^{2}_i-m^2)}-(J_1+J_2)^{2}}\right], 
\cr
M&=&\hat{X}+\hat{Y}+\hat{Q}.
\end{eqnarray}

When $J_i=0$, the duality invariant expression for entropy 
is
\begin{equation}
S=4\pi\sqrt{(\hat{X}+m^2)(\hat{Y}+m^2)(\hat{Q}+m^{2})}.
\end{equation}

\paragraph{BPS limit}\label{n4bh5dentbps}

In the regular BPS limit, the event horizon area (\ref{5dent}) becomes 
\cite{CVEy476}
\begin{equation}
A_{BPS}=4\pi^2[(Q^{(1)}_1Q^{(2)}_1Q)(1-\textstyle{1\over 2}(L_1-
L_2)^2)]^{1\over 2}
=4\pi^2[Q^{(1)}_1Q^{(2)}_1Q-\textstyle{1\over 4}J^2]^{1\over 2}.
\label{5dbpsarea}
\end{equation}
Entropy of BPS black hole with general charge configuration 
(\ref{newch}) and with arbitrary scalar asymptotic values depends 
only on (quantized) charge lattice vectors $\vec{\alpha}$ and $\beta$ 
\cite{CVEy476}, being a statistical quantity 
\cite{FERks52,STR383,FERk136,FERk54,CVEy672,LARw375,CVEt53}:
\begin{equation}
{\bf S}_{BPS}=4\pi\sqrt{\beta({\vec\alpha}^T{ L}{\vec \alpha})-
{1\over 4}J^2}.  
\label{genarea}
\end{equation}
Here, $\vec{\alpha}$ and $\beta$ are related to 
the physical charges $\vec{\cal Q}$ and $Q$ as
\begin{equation}
\vec{\cal Q}=e^{2\varphi_{\infty}/3}M_{\infty}\vec{\alpha},\ \ \ \ 
Q=e^{-4\varphi_{\infty}/3}\beta.
\label{5dlatvecchrel}
\end{equation}

In the regular BPS limit, the ADM mass (\ref{5ddef}) becomes 
\begin{equation}
M_{ADM}=Q^{(1)}_1+Q^{(2)}_1+Q.
\label{5dbhbpsmass}
\end{equation}
A subset of the $SO(5,21)$ $T$-duality transformation on  
(\ref{5dbhbpsmass}) leads to the following 
$T$-duality invariant expression for ADM mass of the general 
$D=5$ black hole:
\begin{equation}
M_{ADM}=e^{2\varphi_{\infty}/3}[\vec{\alpha}^T({\cal M}_{\infty}+L)
\vec{\alpha}]^{1/2}+e^{-4\varphi_{\infty}/3}\beta,
\label{gen5dadm}
\end{equation}
which has dependence on scalar asymptotic values as well as 
charge lattice vectors.  

\paragraph{Near-Extreme Limit}\label{n4bh5dentnear}

Infinitesimal deviation from the BPS limit is achieved by taking 
the limit in which $m$ and $l_{1,2}$ are very close 
to zero, and $\delta$'s are very large such that charges and 
$l_{1,2}/m^{1/2}=L_{1,2}$ remain as finite, non-zero 
constants, and then keeping only the leading order terms in $m$ 
\cite{CVEy476}.    

To the leading order in $m$, the inner and the outer horizons 
are located at
\begin{equation}
r^2_{\pm}\approx m\left(1-{1\over 2}(L^2_1+L^2_2)\pm{1\over 2}
\sqrt{[2-(L_1+L_2)^2]
[2-(L_1-L_2)^2]}\right).
\label{devhorizon}
\end{equation}
The outer horizon area to the leading order in $m$ is 
\cite{CVEy476,BRElmpsv}
\begin{eqnarray}
A&\approx&4\pi^2\left [(Q^{(1)}_1Q^{(2)}_1Q)\sqrt{1-{1\over 2}(L_1-L_2)^2}
+m(Q^{(1)}_1Q^{(2)}_1+Q^{(1)}_1Q+Q^{(2)}_1Q)\right.\cr
& &\ \ \ \times\left.\sqrt{[1-{1\over 2}(L_1+L_2)^2]}\right]^{1\over 2}. 
\label{devarea}
\end{eqnarray}
$J_1$ and $J_2$ are no longer equal in magnitude and opposite in sign 
anymore:
\begin{eqnarray}
J\equiv\textstyle{1\over 2}(J_1-J_2)&=& 
(2Q^{(1)}_1Q^{(2)}_1Q)^{1\over 2}(L_1-L_2)+{\cal O}(m^2),
\cr
\Delta J\equiv{\textstyle{1\over 2}}(J_2+J_1)&=& 
m(2Q^{(1)}_1Q^{(2)}_1Q)^{1\over 2}\left(
{1\over{Q^{(1)}_1}}+{1\over{Q^{(2)}_1}}+{1\over{Q}}\right)(L_1+L_2)
+{\cal O}(m^2), 
\end{eqnarray}
while the ADM mass still has the form
\begin{equation}
M=\sqrt{m^2+(Q^{(1)}_1)^2}+\sqrt{m^2+(Q^{(2)}_1)^2}+
\sqrt{m^2+Q^2}.
\end{equation}
Note, when one of the charges is taken small, 
e.g. $Q^{(1)}_1\to 0$, as in study of the 
microscopic entropy near the BPS limit 
\cite{HORs77,BRElmpsv}, the ADM mass is $M=M_{BPS} +{\cal O}(m)$, 
while the area is $A=A_{BPS}+{\cal O}(m^{1/2})$.  However, 
when all the charges are non-zero, the deviation from 
the BPS limit is of the forms $M=M_{BPS} +{\cal O}(m^2)$ 
and $A=A_{BPS}+ {\cal O}(m)$.  

\subsection{Rotating Black Holes in Higher Dimensions}
\label{n4bhhigh}

We discuss rotating black holes in  
heterotic string on $T^{10-D}$ ($4\leq D\leq 9$) 
with general $U(1)^{36-2D}$ electric charge configurations   
\cite{CVEy477,LLA397}.   The generating solution is 
parameterized by the ADM mass $M_{BH}$ (or alternatively the 
non-extremality parameter $m$), $[{{D-1}\over 2}]$ angular momenta 
$J_i$ ($i=1,...,\left[{{D-1}\over 2}\right]$), and 2 electric charges 
of the KK and the 2-form $U(1)$ gauge fields associated with the 
same compactified direction, which we choose without loss of generality to 
be $Q^{(1)}_1$ and $Q^{(2)}_1$, i.e. those associated with the 
first compactified direction, as well as asymptotic values of a  
toroidal modulus $G_{11\,\infty}$ and the dilaton $\varphi_\infty$.  

The non-trivial fields of the generating solution are \cite{CVEy477}
\begin{eqnarray}
A^{(1)\,1}_t&=&{{N\sinh\delta_1\cosh\delta_1}\over
{2N\sinh^2\delta_1 + \Delta}}, \ \ \ \ \ \ \ \ \ \
A^{(2)}_{t\,1}={{N\sinh\delta_2\cosh\delta_2}\over
{2N\sinh^2\delta_2 + \Delta}},\cr
A^{(1)\,1}_{\phi_i}&=&{{Nl_i\mu^2_i\sinh\delta_1\cosh\delta_2}\over
{2N\sinh^2\delta_1 + \Delta}}, \ \ \ \ \
A^{(2)}_{\phi_i\,1}={{Nl_i\mu^2_i\sinh\delta_2\cosh\delta_1}\over
{2N\sinh^2\delta_2 + \Delta}},\cr
e^{2\varphi}&=&{\Delta^2 \over W},
\ \ \ \ \ \ \ \ \ \ \ \ \ \ \ \ \ \ \ \ \ \ \ \ \ \ \ \
G_{11}={{2N\sinh^2\delta_1 + \Delta}\over
{2N\sinh^2\delta_2 + \Delta}}, \cr
B_{t\phi_i}&=&-{{2Nl_i\mu^2_i\sinh\delta_1\sinh\delta_2
[m(\sinh^2\delta_1+\sinh^2\delta_2)r+\Delta]} \over W}, \cr
B_{\phi_i\phi_j}&=&-4N^2l_il_j\mu^2_i\mu^2_j\sinh\delta_1
\sinh\delta_2\cosh\delta_1\cosh\delta_2
\cr
& &\times[N(\sinh^2\delta_1+\sinh^2\delta_2) +\Delta]
[2N^2\sinh^2\delta_1\sinh^2\delta_2
\cr
& &+N\Delta(\sinh^2\delta_1+\sinh^2\delta_2-1)+\Delta^2]/
[(\Delta-2N)W^2],
\cr
ds^2&=&\Delta^{{D-4}\over{D-2}} W^{1\over{D-2}}
\left[-{{\Delta-2N}\over W}dt^2+
{dr^2 \over {\prod^{[{{D-1}\over 2}]}_{i=1}(r^2+l^2_i)-2N}}\right.
\cr
& &+{{r^2+l^2_1\cos^2\theta+K_1\sin^2\theta}\over \Delta}
d\theta^2 
\cr
& &+{{\cos^2\theta\cos^2\psi_1\cdots\cos^2\psi_{i-1}}
\over \Delta}(r^2+l^2_{i+1}\cos^2\psi_i+K_{i+1}\sin^2\psi_i)
d\psi^2_i 
\cr
& &-2\sum_{i<j}{{l^2_j-K_j}\over \Delta}
\cos^2\theta\cos^2\psi_1\cdots\cos^2\psi_{i-1}
\cr
& &\ \ \ \ \times\cos\psi_i\sin\psi_i
\cdots\cos^2\psi_{j-1}\cos\psi_j\sin\psi_j d\psi_i d\psi_j 
\cr
& &-2{{l^2_{i+1}-K_{i+1}}\over \Delta}
\cos\theta\sin\theta\cos^2\psi_1\cdots\cos^2\psi_{i-1}
\cos\psi_i\sin\psi_i d\theta d\psi_i 
\cr
& &+{\mu^2_i \over {\Delta W}}[(r^2+l^2_i)(2N\sinh^2\delta_1+\Delta)
(2N\sinh^2\delta_2+\Delta)
\cr
& &\ \ \ \ +2l^2_iN(\Delta-2N\sinh^2\delta_1\sinh^2\delta_2)]
d\phi^2_i
\cr
& &-{{2Nl_i\mu^2_i\cosh\delta_1\cosh\delta_2}\over W}dtd\phi_i
\cr
& &\left.+\sum_{i<j}{{4Nl_il_j\mu^2_i\mu^2_j(\Delta-2N\sinh^2\delta_1
\sinh^2\delta_2)}\over {\Delta W}}d\phi_i d\phi_j\right],
\label{ebh}
\end{eqnarray}
where
\begin{equation}
W \equiv (2N\sinh^2\delta_1 +\Delta)(2N\sinh^2\delta_2 +\Delta)
\label{ebhdef}
\end{equation}
and $\Delta,K_i,N,\mu_i,\alpha$ are defined separately for even and odd 
$D$ in (\ref{edef1})$-$(\ref{odef2}).  

The ADM mass, angular momenta and electric charges 
of the generating solution are
\footnote{We use the convention of \cite{MYEp172}, 
keeping in mind that matter Lagrangian in 
(\ref{effaction}) has $1/(16\pi G_D)$ prefactor.} 
\begin{eqnarray}
M_{BH}&=&{{\Omega_{D-2}m}\over {16\pi G_D}}
[(D-3)(\cosh 2\delta_1+\cosh 2\delta_2)+2],\cr
J_i&=&{\Omega_{D-2} \over {4\pi G_D}}ml_i\cosh\delta_1
\cosh\delta_2, 
\cr
Q^{(1)}_1&=&{{\Omega_{D-2}\over {16\pi G_D}}} (D-3)m\sinh 2\delta_1, 
\cr
Q^{(2)}_1&=&{{\Omega_{D-2}\over {16\pi G_D}}}(D-3)m\sinh 2\delta_2.
\label{para}
\end{eqnarray}
For the canonical choice of asymptotic values 
$G_{ij}=\delta_{ij}$, i.e. compactification on $(10-D)$ self-dual circles 
with radius $R=\sqrt{\alpha^{\prime}}$, the $D$-dimensional gravitational 
constant is $G_{D}=G_{10}/(2\pi\sqrt{\alpha^{\prime}})^{10-D}$.  
Also, the KK and the 2-form field $U(1)$ charges $Q^{(1)}_1$ and 
$Q^{(2)}_1$ are quantized as $p/\sqrt{\alpha^{\prime}}$ and 
$q/\sqrt{\alpha^{\prime}}$, respectively, where $p,q\in {\bf Z}$.

The outer horizon area of the generating solution is \cite{HORs53,CVEy477} 
\begin{equation}
A_D=2mr_+\Omega_{D-2}\cosh\delta_1\cosh\delta_2,
\label{area}
\end{equation}
where the outer horizon $r_+$ is determined by 
\begin{equation}
[\prod^{[{{D-1}\over 2}]}_{i=1}(r^2+l^2_i)-2N]_{r=r_+}=0.
\label{highhorizon}
\end{equation}

The surface gravity $\kappa$ at the (outer) event horizon is defined as 
$\kappa^2 = {\rm lim}_{r\to r_+}\nabla_{\mu}\lambda\nabla^{\mu}\lambda$, 
where $\xi^{\mu}\xi_{\mu} \equiv -\lambda^2$ and $\xi \equiv \partial/
\partial t+\Omega_i\partial/\partial\phi_i$. Here, $\Omega_i$ is the 
angular velocity at the (outer) horizon and is defined by the condition 
that $\xi$ is null on the (outer) horizon.  The surface gravity 
and angular velocity at the outer-horizon of the generating solution are
\begin{equation}
\kappa={1\over {{\rm cosh}\delta_1 {\rm cosh}\delta_2}} 
{{\partial_r (\Pi -2N)}\over {4N}}|_{r=r_+},\ \ \ 
\Omega_i={1\over{{\rm cosh}\delta_1 {\rm cosh}\delta_2}}
{l_i\over {r^2_+ + l^2_i}}, 
\label{axitemp}
\end{equation}
where $\Pi\equiv \prod^{[{{D-1}\over 2}]}_{i=1}(r^2+l^2_i)$.  

The generating solution has a ring-like singularity 
at $(r,\theta)=(0,{\pi\over 2})$ and thus spacetime 
is that of the Kerr solution. 

The BPS limit of (\ref{ebh}), where the ADM mass 
$M_{BH}$ saturates the Bogomol'nyi bound 
\begin{equation}
M_{BH}\geq |Q^{(1)}_1 + Q^{(2)}_1|,
\label{ddbogbound}
\end{equation} 
is defined as the limits $m\to 0$ and $\delta_{1,2}\to 
\infty$ such that $Q^{(1),(2)}_1$  remain as finite constants.  
For $D\geq 6$ with only one of $l_i$ non-zero, 
the BPS limit is also the extreme limit \cite{HORs53}, 
i.e. all the horizons collapse to $r=0$ as $m\to 0$.  
However, with more than one $l_i$ non-zero, the singularity at $r=0$ 
becomes naked, i.e. horizons disappear. 

\section{Black Holes in $N=2$ Supergravity Theories}\label{n2bh}
\subsection{$N=2$ Supergravity Theory}\label{n2bhsg}
\subsubsection{General Matter Coupled $N=2$ Supergravity}\label{n2bhsglag}

We consider the general $N=2$ supergravity 
\cite{DEWlv255,CASdf241,DAUff359,ANDbcdff476,ANDbcdffm032,FRE182}
coupled to $n_v$ vector multiplets and $n_H$  hypermultiplets. 
The field contents are as follows.  
The $N=2$ supergravity multiplet contains the graviton, the $SU(2)$ 
doublet of gravitinos $\psi^i_{\mu}$ (the $SU(2)$ index $i=1,2$ labels two 
supercharges of $N=2$ supergravity and $\mu=0,1,2,3$ is a spacetime 
vector index), and the graviphoton.  
The $N=2$ vector multiplets contain $U(1)$ gauge fields, doublets of 
gauginos $\lambda^a_i$ and scalars $z^a$ ($a=1,...,n_v$), which span 
the $n_v$-dimensional special K\"ahler manifold.  
The hypermultiplets consist of hyperinos $\zeta_{\alpha}$, 
$\zeta^{\alpha}$ ($\alpha=1,...,2n_H$) with left and right chiralities 
and real scalars $q^u$ ($u=1,...,4n_H$), which span the $4n_H$-dimensional 
quaternionic manifold.  
The general form of the bosonic action is \cite{ANDbcdff476}
\begin{eqnarray}
{\cal L}^{N=2}&=&\sqrt{-g}\left[-\textstyle{1\over 2}{\cal R}+g_{ab^*}
(z,\bar{z})\nabla^{\mu}z^a\nabla_{\mu}\bar{z}^{b^*} + 
h_{uv}(q)\nabla^{\mu}q^u\nabla_{\mu}q^v \right.
\cr
& &\ \ \ \ \ \ +\left.i(\bar{\cal N}_{\Lambda\Sigma}
{\cal F}^{-\,\Lambda}_{\mu\nu}
{\cal F}^{-\,\Sigma\,\mu\nu}-{\cal N}_{\Lambda\Sigma}
{\cal F}^{+\,\Lambda}_{\mu\nu}{\cal F}^{+\,\Sigma\,\mu\nu})\right], 
\label{n2act}
\end{eqnarray}
where $g_{ab^*}=\partial_a\partial_{b^*}K(z,\bar{z})$ is the K\"ahler 
metric
\footnote{The K\"ahler potential $K(z,\bar{z})$ and the period matrix 
${\cal N}_{\Lambda\Sigma}$ are defined in terms of the holomorphic 
prepotential $F(X)$ and the scalar fields $X^{\Lambda}$ as in 
(\ref{kahmet}).}, 
$h_{uv}(q)$ is the quaternionic metric, 
${\cal F}^{\pm\,\Lambda}_{\mu\mu}\equiv {1\over 2}({\cal F}^{\Lambda}_{\mu\nu}
\pm{i\over 2}\epsilon^{\mu\nu\rho\sigma}{\cal F}^{\Lambda}_{\rho\sigma})$ 
are the  (anti-)self-dual parts of the field strengths 
${\cal F}^{\Lambda}_{\mu\nu}=\partial_{\mu}{\cal A}^{\Lambda}_{\mu}
-\partial_{\nu}{\cal A}^{\Lambda}_{\mu}+gf^{\Lambda}_{\Sigma\Delta}
{\cal A}^{\Sigma}_{\mu}{\cal A}^{\Delta}_{\nu}$ 
of the $U(1)$ gauge fields ${\cal A}^{\Lambda}_{\mu}$ 
($\Lambda=0,1,...,n_v$) in the $N=2$ 
supergravity and $N=2$ vector multiplets, and $g$ is the gauge coupling.  
Here, the gauge covariant differentials on the scalars are defined 
as:
\begin{eqnarray}
\nabla_{\mu}\,z^a&\equiv&\partial_{\mu}z^a +gA^{\Lambda}_{\mu}k^a_{\Lambda}(z)
\cr
\nabla_{\mu}\,\bar{z}^{a*}&\equiv&\partial_{\mu}\bar{z}^{a*}
+gA^{\Lambda}_{\mu}k^{a*}_{\Lambda}(\bar{z})
\cr
\nabla_{\mu}q^u&\equiv&\partial_{\mu}q^u +gA^{\Lambda}_{\mu}k^u_{\Lambda}(q),
\label{gaugcov}
\end{eqnarray}  
where $k^a_{\Lambda}(z)$ [$k^u_{\Lambda}(q)$] are the holomorphic 
[triholomorphic] Killing vectors of the K\"ahler [quaternionic] manifold 
(Cf. see (\ref{killalg})).  

We introduce a symplectic vector of the anti-self-dual field strengths:
\begin{equation}
{\cal Z}^{-}\equiv \left(\matrix{{\cal F}^{-\,\Lambda}\cr
{\cal G}^{-}_{\Sigma}}\right), 
\label{sympvec}
\end{equation}
where ${\cal G}^{-}_{\Lambda}\equiv \bar{\cal N}_{\Lambda\Sigma}
{\cal F}^{-\,\Sigma}$ \cite{DEWv245}.  The symplectic vector ${\cal Z}^+$ 
of self-dual field strengths is the complex conjugation of 
(\ref{sympvec}).  It is convenient to redefine field strengths 
${\cal F}^{\Lambda}$ as \cite{CERdf339,BILcdffrsv13}:
\begin{eqnarray}
T^{-}&\equiv& \langle V|{\cal Z}^{-}\rangle = 
(M_{\Lambda}{\cal F}^{-\,\Lambda}-L^{\Sigma}{\cal G}^{-}_{\Sigma}),
\cr
F^{-\,a}&\equiv&g^{ab^*}\langle \bar{U}_{b^*}|{\cal Z}^{-}\rangle 
=g^{ab^*}(\nabla_{b^*}\bar{M}_{\Lambda}{\cal F}^{-\,\Lambda}-
\nabla_{b^*}\bar{L}^{\Lambda}{\cal G}^{-}_{\Lambda}). 
\label{photons}
\end{eqnarray}
Then, $T^-$ and $F^{-\,a}$ ($a=1,...,n_v$), respectively, correspond to 
the field strengths of the gravi-photon of the supergravity multiplet 
and the gauge fields of $n_v$ super-Yang-Mills multiplets.  

The supersymmetry transformation laws for the gravitinos, the gauginos 
and hyperinos in the bosonic field background are 
\begin{eqnarray}
\delta\psi_{i\,\mu}&=&{\cal D}_{\mu}\varepsilon_i+
[igS_{ij}\eta_{\mu\nu}+\epsilon_{ij}T^{-}_{\mu\nu}]
\gamma^{\nu}\varepsilon^{j},
\cr
\delta\lambda^{a\,i}&=&i\gamma^{\mu}\nabla_{\mu}z^a\varepsilon^i 
+\epsilon^{ij}(F^{-\,a}_{\mu\nu}\gamma^{\mu\nu}+k^a_{\Lambda}
\bar{L}^{\Lambda})\varepsilon_j,
\cr
\delta\zeta_{\alpha}&=&i{\cal U}^{j\beta}_{u}\nabla_{\mu}q^u\gamma^{\mu}
\epsilon_{ij}C_{\alpha\beta}\varepsilon^i+gN^i_{\alpha}\varepsilon_i,
\label{n2susytr}
\end{eqnarray}
where $\epsilon_{ij}$ [$C_{\alpha\beta}$] is the flat $Sp(2)$ [$Sp(2n_H)$] 
invariant matrix and ${\cal U}^{j\beta}_u$ is the quaternionic vielbein 
\cite{BAGw222}. 
Here, $S_{ij}$ and $N^i_{\alpha}$ are mass-matrices given by
\begin{equation}
S_{ij}={i\over 2}(\sigma_x)^{\,\,k}_i\epsilon_{jk}
{\cal P}^x_{\Lambda}L^{\Lambda}, \ \ \ \ \ 
N^i_{\alpha}=2{\cal U}^i_{\alpha u}k^u_{\Lambda}\bar{L}^{\Lambda}, 
\label{massmat}
\end{equation}
where ${\cal P}^x_{\Lambda}$ is a triplet of real 0-form prepotentials 
on the quaternionic manifold.

\subsubsection{BPS States}\label{n2bhsgbps}

The BPS states of the $N=2$ theory have mass equal to the central charge, 
which is just the graviphoton charge given by:
\begin{equation}
Z\equiv -{1\over 2}\oint_{S^2}\,T^{-}.
\label{central}
\end{equation}
Thus, central charge $Z$ is characterized by the vacuum expectation 
of the moduli in the symplectic vector $V$ and the symplectic charge vector 
given by
\begin{equation}
{\cal Q}=\left(\matrix{P^{\Lambda}\cr Q_{\Sigma}}\right); \ \ \ \ 
P^{\Lambda}\equiv \oint_{S^2}\,{\rm Re}\,{\cal F}^{-\,\Lambda}, \ \ 
Q_{\Sigma}\equiv \oint_{S^2}\,{\rm Re}\,{\cal G}^{-}_{\Sigma},
\label{chvec}
\end{equation}
in the following way \cite{SEIw426,SEIw431,DEWv245,CERdfv444,CERdf46}
\begin{equation}
Z=\langle V|{\cal Q}\rangle =(L^\Lambda Q_{\Lambda}-M_{\Sigma}P^{\Sigma})
=e^{{K(z,\bar{z})}\over 2}(X^{\Lambda}(z)Q_{\Lambda}-F_{\Sigma}(z)P^{\Sigma}).
\label{invcentr}
\end{equation}
Note, since two vectors $V$ and $\cal Q$ transform covariantly under 
the symplectic transformation $Sp(2n_v+2)$, the central charge and, therefore, 
the ADM mass $M=|Z|$ have manifest symplectic covariance.

\subsection{Supersymmetric Attractor and Black Hole Entropy}\label{n2bhent}

Since entropy is a statistical quantity defined as the degeneracy of 
microscopic states, the horizon area, which defines the 
thermal entropy, should be independent of 
continuous quantities like scalar asymptotic values.  
This is an another illustration of no-hair theorem where properties 
of black holes are independent of scalar hairs; all the information of 
scalar asymptotic values get lost at the event horizon.  
It was discovered in \cite{FERks52} within $N=2$ theories that this is 
a generic property of BPS solutions in supersymmetric theory and can 
be derived from supersymmetry alone 
\cite{FERk136,FERk54,ANDdf105}. 

To illustrate this idea, we consider general magnetic, 
spherically symmetric solution in $N=2$ theory coupled to vector 
superfields \cite{FERks52}.  Spherically symmetric Ans\"atze are 
\cite{TOD121}: 
\begin{equation}
ds^2=g_{\mu\nu}dx^{\mu}dx^{\nu}=-e^{2U}dt^2 +e^{-2U}d\vec{x}^2, \ \ \ \ \ 
\hat{\cal F}^{\Lambda}_r = {{q^{\Lambda}_{(m)}}\over {r^2}}e^{U(r)}, 
\label{n2sph}
\end{equation}
and the scalars (moduli) $z^{\Lambda}\equiv X^{\Lambda}/X^0$ 
($\Lambda=0,1,...,n_v$) depend on $r$, only.  

With these Ans\"atze, the Killing spinor equations 
$\delta\,\psi_{i\,\mu}=0$ and $\delta\,\lambda^{a\,i}=0$ yield the 
following coupled first order differential equations \cite{FERks52}:
\begin{eqnarray}
4U^{\prime}&=&-\sqrt{{(\bar{z}{\cal N}q_{(m)})(z{\cal N}q_{(m)})
(\bar{z}{\cal N}z)}\over{(z{\cal N}z)(\bar{z}{\cal N}\bar{z})}} e^U, 
\cr
(z^{\Lambda})^{\prime}&=&-{{e^U}\over 4}\sqrt{{(z{\cal N}z)(\bar{z}{\cal N}
q_{(m)})(\bar{z}{\cal N}z)}\over{(\bar{z}{\cal N}\bar{z})(z{\cal N}q_{(m)})}}
(z^{\Lambda} q^0_{(m)}-q^{\Lambda}_{(m)}), 
\label{n2first}
\end{eqnarray}
where the prime denotes the differentiation with respect to $\rho\equiv 
1/r$, and $(z{\cal N}q_{(m)})\equiv z^{\Lambda}{\cal N}_{\Lambda\Sigma}
q^{\Sigma}_{(m)}$, etc.. 
From (\ref{n2first}), one obtains the following second order ordinary 
differential equations for the moduli fields $z^{\Lambda}$:
\begin{eqnarray}
(z^{\Lambda})^{\prime\prime}&-&\left({{(z{\cal N}q_{(m)})}\over{(z{\cal N}z)}}
+q^0_{(m)}\right)
{{((z^{\Lambda})^{\prime})^2}\over{z^{\Lambda}q^0_{(m)}-q^{\Lambda}_{(m)}}}
\cr 
& &\ \ \ \ +{1\over 2}\left({\rm ln}{{(z{\cal N}z)(\bar{z}{\cal N}q_{(m)})
(\bar{z}{\cal N}z)}\over{(\bar{z}{\cal N}\bar{z})(z{\cal N}q_{(m)})}}
\right)^{\prime}(z^{\Lambda})^{\prime}=0. 
\label{geomod}
\end{eqnarray}

(\ref{geomod}) can be viewed as a geodesic equation for moduli 
fields $z^{\Lambda}$ that determines how $z^{\Lambda}$ evolves as $\rho$ 
varies from 0 to $\infty$.  The initial conditions $z^{\Lambda}|_{\rho=0}$ 
and $(z^{\Lambda})^{\prime}|_{\rho=0}$ for the geodesic motion 
in the phase space (with coordinates $z^{\Lambda}$ and 
$(z^{\Lambda})^{\prime}$) are the asymptotic values 
($r=1/\rho\to\infty$) of $z^{\Lambda}$ and their derivatives 
(determined by  $z^{\Lambda}|_{r\to\infty}$ and $q^{\Lambda}_{(m)}$ 
through the second equation in (\ref{n2first})).  Given initial conditions 
$z^{\Lambda}|_{\rho=0}$ and $(z^{\Lambda})^{\prime}|_{\rho=0}$, 
$z^{\Lambda}$ evolve with $\rho$, following damped geodesic motion in the 
phase space until they run into an attractive fixed 
point, i.e. a point where the velocities ${{dz^{\Lambda}}\over{d\rho}}$ 
of $z^{\Lambda}$ vanish.  
For the special example under consideration, as we see (\ref{n2first}), 
the fixed point is located at 
\begin{equation}
z^{\Lambda}_{fixed}=q^{\Lambda}_{(m)}/q^0_{(m)}.
\label{fixedch}
\end{equation} 
At the fixed point, moduli depend on $U(1)$ charges only, loosing 
all their information on the initial conditions at infinity.   
From this observation, one arrives at stronger version of {\it no-hair 
theorem} for black holes in supersymmetry theories: black holes lose 
all their scalar hairs near the horizon and are characterized by discrete 
$U(1)$ charges (and angular momenta), only.  

Nearby the horizon, the black hole approximates to the 
Bertotti-Robinson geometry \cite{LEV26,BER116,ROB7}
with the topology  $AdS_2\times S^{D-2}$.  This geometry 
is conformaly flat and the graviphoton field strength is 
covariantly constant.  Thus, in this region the ADM mass 
reduces to the Bertotti-Robinson mass and the supersymmetry is 
completely restored 
\cite{GIB207,KAL282,KALp46,LOWs73,GIBt71,CHAfg55}.   In the 
asymptotic region ($r\to\infty$), the spacetime is flat and, therefore,  
supersymmetry is unbroken.  In between these regions, solutions break 
fraction of supersymmetries, indicating the BPS nature.

Note, the supersymmetric configuration under consideration  
is a bosonic configuration, i.e. a solution to  
supergravity theory with all the fermionic fields set equal to zero.  
However, the supersymmetry parameters associated with the unbroken 
supersymmetries, called ``anti-Killing spinors'', generate a whole 
supermultiplets of solutions, i.e. the superpartners to black hole 
solution.  To generate such solutions, one start with a bosonic 
configuration and applies supersymmetry transformations iteratively 
with the supersymmetry parameters given by the anti-Killing spinors.  
Such a procedure induces fermionic fields, as well as corrections to 
bosonic fields.  It was shown in \cite{KALl161} that when this 
procedure is performed on double-extreme black solutions, i.e.  
extreme solutions with constant scalars, in the $N=2$ supergravity 
coupled to vector and hyper multiplets, there are no corrections to 
the fields in the vector and hyper multiplets.  This implies that 
although the metric, graviphoton and gravitino receive corrections, 
the moduli at the fixed attractor point as functions of $U(1)$ charges, 
only, remain intact under the supersymmetry transformations which 
generate the fermionic partners of the supersymmetric black holes.

\subsection{Explicit Solutions}\label{n2bhentsol}

\subsubsection{General Magnetically Charged Solutions}\label{n2bhentsolmag}

We discuss the general spherically symmetric, magnetic 
($q^{(e)}_{\Sigma}=0$) solutions in $N=2$ supergravity coupled to 
$n_v$ vector multiplets with scalar fields varying with the 
radial coordinate $r$ \cite{FERks52}.  The Ans\"atze for the fields are 
given in (\ref{n2sph}) with the scalar fields depending on 
the radial coordinate $r$, only.  The scalar fields and the metric 
components satisfy the differential equations  (\ref{n2first}) 
$-$ (\ref{geomod}).  

For the purpose of solving the differential equations (\ref{n2first}) 
$-$ (\ref{geomod}), we consider the simple case with $q^0_{(m)}=0$. 
In this case, the solutions are given by 
\begin{eqnarray}
e^{2U(\rho)}&=&e^{K(z,\bar{z})-K_{\infty}}
\cr 
z^a&=&\left\{\matrix{z^a_{\infty}+{{q^a_{(m)}}\over 4}\rho e^{-K_{\infty}/2}, 
\ \ \ \  {\rm for}\ z^a_{\infty}=\bar{z}^a_{\infty} \cr 
z^a_{\infty}+i{{q^a_{(m)}}\over 4}\rho e^{-K_{\infty}/2}, 
\ \ \ \  {\rm for}\ z^a_{\infty}=-\bar{z}^a_{\infty}}\right. 
\cr
ds^2&=&-e^{K(z^a,\bar{z}^a=z^a)-K_{\infty}}dt^2+ 
e^{-K(z^a,\bar{z}^a=z^a)+K_{\infty}}d\vec{x}^2. 
\label{n2sol}
\end{eqnarray}
The explicit solutions for $N=2$ theories with specific prepotentials $F$ 
are obtained by substituting the corresponding K\"ahler potential $K$ 
into the general solution (\ref{n2sol}) \cite{FERks52}.  

\subsubsection{Dyonic Solutions}\label{n2bhentsoldy}

We generalize the magnetic black holes in section 
\ref{n2bhentsolmag} to include electric charges as well 
\cite{STR383}.  Since it would be hard to solve the resulting 
differential equations with non-zero electric and magnetic charges, 
we take all the moduli fields $z^I$ to be constant.  In fact, as we 
discuss in the subsequent sections, such class of solutions corresponds 
to the minimum energy configurations among extreme solutions and, 
therefore, is physically interesting. 

Assuming that the moduli $z^I=X^I/X^0$ are constant, 
from $\delta\,\lambda^{a\,i}=0$ 
one obtains the following electric and magnetic charges 
of dyonic solutions:
\begin{equation}
(q^I_{(m)},q^{(e)}_I)= {\rm Re}(CX^I_f,-C{i\over 2}F_I(X_f)), 
\label{dyfxdch}
\end{equation}
where $C$ is a constant and the subscript $f$ denotes the fixed point.  
With a suitable choice of K\"ahler gauge, which eliminates the redundant 
degrees of freedom in $X^I$ by half, one can solve the $2n+2$ equations 
in (\ref{dyfxdch}) to find the expressions for $z^I=X^I/X^0$ in terms of 
quantized charges $q^I_{(m)}$ and $q^{(e)}_I$.  

From $\delta\,\psi_{i\,\mu}=0$ with $U(1)$ field strengths
\footnote{These are obtained by solving $\delta\,\lambda^{a\,i}=0$.} 
$\hat{\cal F}^I=CX^I\epsilon^+$ and ${\cal G}^+_I ={C\over 2}F_I\epsilon^+$ 
($\epsilon^+_{\mu\nu}$ obeys $*\epsilon^+_{\mu\nu}=i\epsilon^+_{\mu\nu}$ and 
is normalized to give $2\pi$ after being integrated over $S^2$) substituted, 
one obtains the following solution for the metric: 
\begin{equation} 
e^{-U}=1+{\sqrt{C\bar{C}}\over {4r}}. 
\label{fxddysol}
\end{equation}
This solution has the surface area given by
\begin{equation}
A={\pi\over 4}C\bar{C}.
\label{dyarea}
\end{equation}

\subsection{Principle of a Minimal Central Charge}\label{n2bhentcen}

At the fixed attractor point in phase space, the central charge 
eigenvalue is extremized with respect to moduli fields, 
so-called ``principle of a minimal central charge'' 
\cite{FERk136,FERk54,KALsw,KALrw,FERgk103}.   For $N>2$ theories, the 
largest eigenvalue is extremized and the smaller central charges 
become zero \cite{FERk54} at the fixed point
\footnote{Thus, for $N\geq 4$ theories, one can 
determine moduli fields (at the fixed point) in terms of 
$U(1)$ charges, by minimizing the largest eigenvalue and  
setting the remaining eigenvalues equal to zero.}.  
Since scalar asymptotic values are expressed only in terms of 
$U(1)$ charges at the fixed point, the extremized (largest) central 
charge depends only on $U(1)$ charges, thereby becoming a 
candidate for describing black hole entropy. 
It turns out that entropy of extreme black holes for each 
dimension has the following universal dependence on the extremal value 
$Z_{fix}$ of the (largest) central charge eigenvalue regardless of the number 
$N$ of supersymmetries \cite{FERk136,FERk54}:
\begin{equation}
S={A\over 4}=\pi|Z_{fix}|^{\alpha}, 
\label{cenent}
\end{equation}
where $\alpha=2\,[3/2]$ for $D=4$ [$D=5$].  

As an example, we consider the BPS dilatonic dyon 
\cite{GIB207,KALlopp46} in $D=4$ with the mass:
\begin{equation}
M_{BPS}=|Z|={1\over 2}(e^{-\varphi_{\infty}}|p|+e^{\varphi_{\infty}}|q|). 
\label{dilbh}
\end{equation}
The minimum of the central charge $|Z|$ is located at $g^2_{fix}=
e^{2\varphi_0}=\left|{p\over q}\right|$, which leads 
to the following correct expression for the entropy which is 
{\it independent of dilaton asymptotic value}: 
\begin{equation}
S={A\over 4}=\pi|Z_{fix}|^2=\pi|pq|. 
\label{dilent}
\end{equation}

One can prove the principle of a minimal central charge as follows.  
We consider the ungauged $N=2$ supergravity coupled to Abelian vector 
multiplets and hypermultiplets, defined by the Lagrangian (\ref{n2act}) 
and the supersymmetry transformations (\ref{n2susytr}) with $g=0$.  
Since we are interested in configurations at the fixed point, 
the derivatives of scalars are zero, i.e. $\partial_{\mu}z^a=0$ and 
$\partial_{\mu}q^u=0$, at the horizon.  So, from $\delta\lambda^{a\,i}=0$, 
one has ${\cal F}^{-\,a}_{\mu\nu}=0$. 
Note, the covariant derivative of the central charge defined in 
(\ref{invcentr}) is
\begin{equation}
Z_a\equiv \nabla_a\,Z=-{1\over 2}\oint_{S^2}g_{ab^*}{\cal F}^{+\,b^*}
=(Q_{\Lambda}\nabla_aL^{\Lambda}-P^{\Sigma}\nabla_aM_{\Sigma})
=\langle U_a|{\cal Q}\rangle.
\label{covcentch}
\end{equation}
Since ${\cal F}^{-\,a}=0$ is equivalent to ${\cal F}^{+\,a}=0$, 
the central charge $Z$ is covariantly constant at the fixed point 
of the moduli space:
\begin{equation}
Z_a=\nabla_a Z=\langle U_a|{\cal Q}\rangle=0.
\label{cencvr}
\end{equation}
It can be shown \cite{FERk136} that the condition (\ref{cencvr}) is 
equivalent to the statement that the central charge takes extremum 
value at the fixed point:
\begin{equation}
\partial_a\,|Z|=0.
\label{cenext}
\end{equation}
Thus, within ungauged general Abelian $N=2$ supergravity we 
establish that {\it central charge is minimized at the fixed point of 
geodesic motion of moduli evolving with $\rho=1/r$.}  

At the fixed point in the moduli space,  
the central charge is expressed in terms of the 
symplectic $U(1)$ charge vector $\cal Q$ and the moduli as \cite{FERk136}:
\begin{equation}
|Z|^2=-{1\over 2}{\cal Q}^T\cdot {\bf M}({\cal N})\cdot {\cal Q},
\label{ctchfx}
\end{equation}
\begin{equation}
{\bf M}({\cal N})\equiv\left(
\matrix{{\rm Im}\,{\cal N}+({\rm Re}\,{\cal N})({\rm Im}\,{\cal N})^{-1}
({\rm Re}\,{\cal N}) & -({\rm Re}\,{\cal N})({\rm Im}\,{\cal N})^{-1} \cr
-({\rm Im}\,{\cal N})^{-1}({\rm Re}\,{\cal N})& ({\rm Im}\,{\cal N})^{-1}}
\right),
\label{matdef}
\end{equation}
with the moduli in the matrix ${\bf M}({\cal N})$ taking values at the 
fixed attractor point.  

The central charge minimization condition (\ref{cencvr}) fixes the 
asymptotic values of the moduli in terms of $\cal Q$. 
By using the relations $\langle U_a|V\rangle =0=\langle U_a|\bar{V}
\rangle$ satisfied by the general symplectic section $V$, one can solve 
(\ref{cencvr}) to express $\cal Q$ in terms of the moduli as \cite{FERk136}
\begin{equation}
{\cal Q}=i(\bar{Z}V-Z\bar{V}),
\label{chsecrel}
\end{equation}
or in component form:
\begin{equation}
P^{\Lambda}=2{\rm Im}(\bar{Z}L^{\Lambda}), \ \ \ \ \ 
Q_{\Sigma}=2{\rm Im}(\bar{Z}M_{\Sigma}).
\label{chsecfx}
\end{equation}
(\ref{chsecfx}) can be solved to express the moduli (at the fixed point) 
in terms of $U(1)$ charges:
\begin{equation}
V=-{1\over{2\bar{\bf Z}}}\left[
\left(\matrix{{\bf 0}& {\bf I}\cr -{\bf I}& {\bf 0}}\right)\cdot 
{\bf M}({\cal F}) + 
i\left(\matrix{{\bf I}& {\bf 0}\cr {\bf 0}& {\bf I}}\right)
\right]\cdot {\cal Q},
\label{modchrel}
\end{equation}
or in terms of components:
\begin{eqnarray}
-2\bar{\bf Z}L^{\Lambda}&=&
\left[iP-({\rm Im}\,{\cal N})^{-1}({\rm Re}\,{\cal N})P
+({\rm Im}\,{\cal N})^{-1}Q\right]^{\Lambda}, 
\cr
-2\bar{\bf Z}M_{\Sigma}&=&
\left[iQ-\left(({\rm Im}\,{\cal N})+({\rm Re}\,{\cal N})({\rm Im}\,
{\cal N})^{-1}({\rm Re}\,{\cal N})\right)P\right.
\cr
& &\left.\ \ \ \ +({\rm Re}\,{\cal N})
({\rm Im}\,{\cal N})^{-1}Q\right]_{\Sigma}.
\label{mdchrelcomp}
\end{eqnarray}
Here, ${\bf M}({\cal F})$ is defined as in (\ref{matdef}) with 
${\cal N}_{\Lambda\Sigma}$ replaced by ${\cal F}_{\Lambda\Sigma}\equiv
\partial_{\Lambda}F_{\Sigma}(X)$.

Alternatively, one can rederive the relations (\ref{chsecrel}) obeyed by 
the moduli and $U(1)$ charges by a variational principle \cite{BEHcdk488} 
associated with a potential
\begin{equation}
{\cal V}_{\cal Q}(Y,\bar{Y})\equiv 
-i\langle \bar{\Pi}|\Pi\rangle -\langle \bar{\Pi}+\Pi|{\cal Q}
\rangle,
\label{varpot}
\end{equation}
where 
\begin{equation}
\Pi\equiv\left(\matrix{Y^{\Lambda}\cr F_{\Sigma}(Y)}\right); \ \ \ \ 
Y^{\Lambda}\equiv \bar{Z}X^{\Lambda}.
\label{defpot}
\end{equation}
Namely, at the minimum of ${\cal V}_{\cal Q}(Y,\bar{Y})$, the relation 
(\ref{chsecrel}) for the minimal central charge, which can be expressed 
in terms of $\Pi$ as $\Pi-\bar{\Pi}=i{\cal Q}$, is satisfied.  
In particular, the entropy for $D=4$ black holes is rewritten as 
\begin{equation}
{S\over{\pi}}=|Z_{fix}|^2=i\langle\bar{\Pi}|\Pi\rangle =
|Y^0|^2\exp[-K(z,\bar{z})]|_{fix}.
\label{newent}
\end{equation}

Many black holes are uplifted to intersecting $p$-branes.  
In this case, energy of black holes is sum of energies 
of the constituent $p$-branes.  The minimal energy of 
$p$-branes corresponds to the ADM mass of 
the corresponding double-extreme black holes in lower dimensions.  
In taking variation of moduli to find the minimum energy 
configuration, one has to keep the gravitational constant of 
lower dimensions
\footnote{Note, lower-dimensional gravitational constant is expressed 
in terms of the $D=10$ gravitational constant and the volume of the 
internal space, i.e. a modulus.} 
as constant.  The minimum energy of $p$-brane is achieved when 
energy contributions from each constituent $p$-brane are equal 
\cite{KAL55}. 

\subsubsection{Generalization to Rotating Black Holes}\label{n2bhentrot}

Generally, rotating black holes have naked singularity 
in the BPS limit.   $D=5$ rotating black holes with 3 charges 
has regular BPS limit (thereby the horizon area can be defined), 
if 2 angular momenta have the same absolute values
\footnote{It is argued in \cite{GAUt55} that singular $D=4$ heterotic BPS 
rotating black holes can be described by regular $D=5$ BPS 
rotating black holes which are compactified through generalized dimensional 
reduction including massive Kaluza-Klein modes.} 
\cite{CVEy476,BREmpv}.  We discuss generalization of the principle of 
minimal central charge to the rotating black hole case \cite{KALrw}.
 
We consider the following truncated theory of 11-dimensional supergravity 
compactified on a Calabi-Yau three-fold 
\cite{GUNst84,GUNst85,CADcdf357,PAPt357}:
\begin{eqnarray}
{\cal L}&=&\sqrt{-g}\left[-{1\over 2}{\cal R}-{1\over 4}e^{{2\over 3}\varphi}
F_{\mu\nu}F^{\mu\nu}-{1\over 4}e^{-{4\over 3}\varphi}G_{\mu\nu}G^{\mu\nu}
+{1\over 6}(\partial_{\mu}\varphi)^2\right]
\cr
& &\ \ \ \ \ -{1\over{4\sqrt{2}}}\epsilon^{\mu\nu\rho\sigma\lambda}
F_{\mu\nu}F_{\rho\sigma}B_{\lambda}.
\label{11dcy}
\end{eqnarray}
This corresponds to the $N=2$ theory with $F={1\over 6}
C_{\Lambda\Sigma\Delta}X^{\Lambda}X^{\Sigma}X^{\Delta}$. 
The supersymmetry transformations of the gravitino $\psi_{\mu}$ and 
the gaugino $\chi$ in the bosonic background are
\begin{eqnarray}
\delta\psi_{\mu}&=&\nabla_{\mu}\,\varepsilon+{1\over{12}}
(\Lambda^{\ \rho\sigma}_{\mu}-4\delta^{\ \rho}_{\mu}\Gamma^{\sigma})
(e^{{\varphi}\over 3}F_{\rho\sigma}-{1\over{\sqrt{2}}}e^{-{2\over 3}\varphi}
G_{\rho\sigma})\varepsilon, 
\cr
\delta\chi&=&-{1\over{2\sqrt{3}}}\Gamma^{\mu}\partial_{\mu}\varphi\,
\varepsilon+{1\over{4\sqrt{3}}}\Gamma^{\rho\sigma}(e^{{\varphi}\over 3}
F_{\rho\sigma}+\sqrt{2}e^{-{2\over 3}\varphi}G_{\rho\sigma})\varepsilon.
\label{11dcysusy}
\end{eqnarray}
The model corresponds to the $N=2$ supergravity with the graviphoton 
$(e^{{\varphi}\over 3}F_{\rho\sigma}-{1\over{\sqrt{2}}}e^{-{2\over 3}\varphi}
G_{\rho\sigma})$ coupled to one vector multiplet with the vector field 
component $(e^{{\varphi}\over 3}F_{\rho\sigma}+\sqrt{2}e^{-{2\over 3}\varphi}
G_{\rho\sigma})$.  

We consider the following $D=5$ BPS rotating black hole solution  
\cite{BREmpv,TSE11,CVEy476} (Cf. (\ref{5dhetBPS}))  
to the above theory:
\begin{eqnarray}
ds^2&=&\left(1-{{r^2_0}\over{r^2}}\right)^2\left[dt-
{{4J\sin^2\theta}\over{\pi(r^2-r^2_0)}}d\phi+{{4J\cos^2\theta}\over
{\pi(r^2-r^2_0)}}d\psi\right]^2
\cr
& &-\left(1-{{r^2_0}\over{r^2}}\right)^{-2}dr^2
-r^2(d\theta^2+\sin^2\theta d\phi^2+\cos^2\theta d\psi^2).
\label{5drotreg}
\end{eqnarray}

For this solution, the scalar $\varphi$ is constant everywhere
(double-extreme): $e^{2\varphi}=\lambda^6$.  
So, from $\delta\chi=0$, one sees that the vector field in the vector 
multiplet vanishes, i.e. $B=-{{\lambda^3}\over{\sqrt{2}}}A$, from which 
one can express $\varphi$ in terms of $U(1)$ charges as:
\begin{equation}
e^{2\varphi}={{8Q^2_F}\over{\pi^2Q^2_H}}\equiv \lambda^6; \ \ 
Q_H\equiv{{\sqrt{2}}\over{4\pi^2}}\int_{S^3}e^{-4\varphi/3}\star G, \  
Q_F\equiv{{\sqrt{2}}\over{16\pi}}\int_{S^3}e^{2\varphi/3}\star F.
\label{fxsc5d}
\end{equation}
Furthermore, the entropy is still expressed in terms of the central 
charge $Z_{fix}$ at the fixed point, but modified by the non-zero 
angular momentum $J$:
\begin{equation}
S=\pi\sqrt{Z^3_{fix}-J^2}.
\label{5drotfxent}
\end{equation}

The argument can be extended to more general rotating solutions 
in the $N=2$ supergravity coupled to $n_v$ vector multiplets with 
the gaugino supersymmetry transformations
\begin{equation}
\delta\lambda_a=-{i\over 2}g_{ab}(\varphi)\Gamma^{\mu}\partial_{\mu}
\varphi^b\,\varepsilon 
+{1\over 4}\left({3\over 4}\right)^{2/3}t_{\Lambda,a}\Gamma^{\mu\nu}
F^{\Lambda}_{\mu\nu}\,\varepsilon.
\label{n2sgngau}
\end{equation}
From $\delta\lambda_a=0$, one sees that at the fixed point 
($\partial_{\mu}\varphi^a=0$) $F^{\Lambda}_{\mu\nu}=0$.  
So, the central charge is extremized at the fixed point: 
$\partial_a Z=0$.  

We discuss the enhancement of supersymmetry near the horizon 
\cite{KALp46,CHAfg55}.  
Since the vector field in the vector multiplet is zero for 
(\ref{5drotreg}), the solution is effectively described by the 
{\it pure} $N=2$ supergravity \cite{FERv37} with the graviphoton 
$\tilde{F}\equiv {{\sqrt{3}}\over 2}\lambda F$.  
The supersymmetry transformation for the gravitino is
\begin{equation}
\delta\psi_{\mu}=\hat{\nabla}_{\mu}\varepsilon=
\nabla\varepsilon+{1\over{4\sqrt{3}}}(\Gamma^{\rho\sigma}\Gamma_{\mu}
+2\Gamma^{\rho}\delta^{\ \sigma}_{\mu})\tilde{F}_{\rho\sigma}\,\varepsilon.
\label{n2sgr}
\end{equation}
The integrability condition for the Killing spinor equation $\delta
\psi_{\mu}=0$ is:
\begin{equation}
[\hat{\nabla}_a,\hat{\nabla}_b]\,\varepsilon=
\hat{\cal R}_{ab}\,\varepsilon =0, 
\label{intn2bh}
\end{equation}
where the super-curvature $\hat{\cal R}_{ab}$ for the solution 
(\ref{5drotreg}) is defined as
\begin{equation}
\hat{\cal R}_{ab}={{(r^2-r^2_0)}\over{r^6}}X_{ab}(1+\Gamma^0).
\label{spcurv}
\end{equation}
Here, the explicit forms of matrices $X_{ab}$, which 
can be found in \cite{KALrw}, are unimportant for our purpose.
At the horizon ($r=r_0$) and at infinity ($r\to\infty$), 
$\hat{\cal R}_{ab}=0$, thereby (\ref{intn2bh}) does not constraint 
the spinor $\varepsilon$, i.e.  supersymmetry is not broken.   
However, for finite values of $r$ outside of the event horizon, 
for which $\hat{\cal R}_{ab}\neq 0$, $\varepsilon$ is constrained by 
the relation:
\begin{equation}
(1+\Gamma^0)\,\varepsilon=0, 
\label{n2spcnst}
\end{equation}
indicating that $1/2$ of supersymmetry is broken.

\subsubsection{Generalization to $N>2$ Case}\label{n2bhentng2}

The principle of minimal central charge is generalized to the $N=4,8$ 
cases by reducing $N=4,8$ theories to $N=2$ theories, and then 
by applying the formalism of $N=2$ theories \cite{FERk136}.  
For $N\geq 4$ theories, there are more than 1 central charge eigenvalues 
$Z_i$ ($i=1,...,[{N\over 2}]$).  The ADM mass of the BPS 
configuration is given by the max$\{|Z_i|\}$.  
When the principle of minimal central charge is applied to this 
eigenvalue, the smaller eigenvalues vanish and all the 
scalar asymptotic values are expressed in terms of $U(1)$ charges, 
only \cite{FERk54}.  So, the extremum of the largest central charge 
continues to depend on integer-valued $U(1)$ charges, only.  
The entropy of extreme black holes in each $D$ has the universal 
dependence on the extreme value of the largest central charge eigenvalue: 
$S={A\over 4}=\pi|Z_{fix}|^{\alpha}$, where $\alpha=2$ $[3/2]$ for $D=4$ 
[$D=5$], regardless of the number $N$ of supersymmetry.

\paragraph{Pure $N=4$ Supergravity}\label{n2bhentng24}

Pure $N=4$ theory can be regarded as $N=2$ supergravity coupled to
one $N=2$ vector multiplet.  This can also be regarded as either 
$SU(2)\times SO(4)$ or $SU(2)\times SU(4)$ invariant truncation of 
$N=8$ theory.  The former corresponds to the $N=2$ theory with 
$F(X)=-iX^0X^1$ and the latter has no prepotential. 
These two theories are related by the symplectic transformation 
\cite{CERdfv444}:
\begin{equation}
\hat{X}^0=X^0,\ \ \ \  \hat{F}_0=F_0,\ \ \ \ 
\hat{X}^1=-F_1,\ \ \ \ \hat{F}_1=X^1,
\label{n4simpl}
\end{equation}
where the hat denotes the $SU(4)$ model \cite{CREsf}.   
The $SO(4)$ version \cite{DAS15,CREsf68,CREs127} 
of the $D=4$, $N=4$ supergravity action without axion is
\begin{equation}
I={1\over{16\pi}}\int d^4x\sqrt{-g}\left[
-{\cal R}+2\partial_{\mu}\varphi\partial^{\mu}\varphi
-{1\over 2}(e^{-2\varphi}F^{\mu\nu}F_{\mu\nu}
+e^{2\varphi}\tilde{G}^{\mu\nu}\tilde{G}_{\mu\nu})
\right],
\label{so4sg}
\end{equation}
where the field strength $\tilde{G}_{\mu\nu}$ of the $SO(4)$ theory 
is related to that $G_{\mu\nu}$ of the $SU(4)$ theory as 
$\tilde{G}^{\mu\nu}={i\over 2}{1\over{\sqrt{-g}}}e^{-2\varphi}
\epsilon^{\mu\nu\rho\lambda}G_{\rho\lambda}$.
 
The dilatino supersymmetry transformation is
\begin{equation}
{1\over 2}\delta\Lambda_I = -\gamma^{\mu}\partial_{\mu}\phi\,\varepsilon_I 
+{1\over{\sqrt{2}}}\sigma^{\mu\nu}(e^{-\phi}F_{\mu\nu}\alpha_{IJ}-
e^{\phi}\tilde{G}_{\mu\nu}\beta_{IJ})^{-}\,\varepsilon^J.  
\label{n4dil}
\end{equation}
At the fixed point ($\partial_{\mu}\phi=0$), the Killing spinor 
equations $\delta\Lambda_I=0$ fix $\phi$ in terms of electric and 
magnetic charges:  
$e^{-2\phi_{fix}}={{|q|}\over{|p|}}$.
Then, writing $\delta\Lambda_I =0$ at the fixed point in 
the form $(Z_{IJ})_{fix}\,\varepsilon^J=0$, one learns \cite{KALlopp46} 
that
\begin{itemize}
\item $pq>0$ case:  $\varepsilon^{3,4}$ non-vanishing, $Z_{34}=0$ 
and $M_{ADM}=|Z_{12}|$. 
\item $pq<0$ case:  $\varepsilon^{1,2}$ non-vanishing, $Z_{12}=0$ 
and $M_{ADM}=|Z_{34}|$.
\end{itemize}
So, smaller eigenvalues, which correspond to broken 
supersymmetries, vanish and entropy is given by the largest  
eigenvalue at the fixed point. 

\paragraph{$N=4$ Supergravity Coupled to $n_v$ Vector 
Multiplets}\label{n2bhentng2vec}

The target space manifold of $N=4$ supergravity coupled to $n_v$ vector 
multiplets is ${{SU(1,1)}\over {U(1)}}\times {{O(6,n_v)}\over 
{O(6)\times O(n_v)}}$ with the first factor parameterized by the 
axion-dilaton field $S$ and the second factor by the coset 
representatives $L^A_{\Lambda}=(L^{ij}_{\Lambda},L^a_{\Lambda})$ 
($i,j=1,2,3,4$, $\Lambda=1,...,6+n_v$ and $a=1,...,n_v$) \cite{BERks155}.  
The central charge is 
\begin{equation}
Z_{ij}=e^{K/2}[L^{\Lambda}_{ij}q_{\Lambda}-SL_{ij\,\Lambda}p^{\Lambda}], 
\label{n4cent}
\end{equation}
where $K=-{\rm ln}\,i(S-\bar{S})$ is the K\"ahler potential for $S$. 

At the fixed point, the gaugino Killing spinor equations 
$\delta\lambda^a_i=0$ require that
\begin{equation}
SL^a_{\Lambda}p^{\Lambda} - L^{a\Lambda}q_{\Lambda}=0. 
\label{n4req}
\end{equation}
The dilatino Killing spinor equation $\delta \chi^i=0$ requires the 
following smaller of the central charge eigenvalues to vanish:
\begin{equation}
|Z_2|^2 ={1\over 4}\left(Z_{ij}\bar{Z}^{ij}-\sqrt{(Z_{ij}\bar{Z}^{ij})^2 
-{1\over 4}|\epsilon^{ijkl}Z_{ij}Z_{kl}|^2}\right), 
\label{n2fixs}
\end{equation}
which fixes $S$ at the fixed point.  
At the fixed point, the difference between two eigenvalues
\begin{equation}
|Z_1|^2-|Z_2|^2={1\over 2}\sqrt{(Z_{ij}\bar{Z}^{ij})^2-{1\over 4}
|\epsilon^{ijkl}Z_{ij}Z_{kl}|^2} 
\label{diffeigs}
\end{equation}
becomes independent of scalars and gives rise to the horizon area 
\cite{CVEy672,DUFlr}
\begin{equation}
A=4\pi(M_{ADM})_{fix}=4\pi|Z_{1\,fix}|^2
=2\pi\sqrt{q^2 p^2-(q\cdot p)^2}.
\label{n4vec}
\end{equation}

\paragraph{$N=8$ Supergravity}\label{n2bhentng28}

The consistent truncation of $N=8$ down to $N=2$ is achieved by choosing 
$H\subset SU(8)$ such that 2 residual supersymmetries are 
$H$-singlet.  Such theory corresponds to $N=2$ supergravity couple to 15  
vector multiplets ($n_v=15$) and 10 hypermultiplets ($n_H=10$).  
(This is the upper limit on the number of matter multiplets 
that can coupled to $N=2$ supergravity.)  
Under $N=2$ reduction of $N=8$, $SU(8)$ group breaks down to 
$SU(2)\times SU(6)$, leading to the decomposition of 26 central charges 
$Z_{AB}$ of $N=8$ into $(1,1)+(2,6)+(1,15)$ under $SU(2)\times SU(6)$.  
The $SU(2)$ invariant part $(1,1)+(1,15)$ is $(Z,D_iZ)$, where $Z$ is the 
$N=2$ central charge.  So, the horizon area is again $A\sim |Z|^2$.  
For example, for type-IIA theory on $T_6/{\bf Z}_3$ 
truncated so that only 2 electric and 2 magnetic 
charges are nonzero, the central charge at the fixed point 
is product of $U(1)$ charges, which is black hole horizon area.  

We consider the following truncation of $N=8$ supergravity:
\begin{eqnarray}
S&=&{1\over{16\pi G}}\int d^4 x\sqrt{-g}\left({\cal R}-
{1\over 2}[(\partial\eta)^2+(\partial\sigma)^2+(\partial\rho)^2]\right.
\cr
& &-\left.{{e^{\eta}}\over 4}[e^{\sigma+\rho}(F_1)^2+e^{\sigma-\rho}(F_2)^2
+e^{-\sigma-\rho}(F_3)^2+e^{-\sigma+\rho}(F_4)^2]\right). 
\label{stulag}
\end{eqnarray}
This is a special case of $STU$ model
\footnote{This model also corresponds to $T^2$ part of type-IIA theory 
on $K_3\times T^2$ or heterotic theory on $T^4\times T^2$.  
See section \ref{dualn4kk} for the explicit action.} 
\cite{DUFlr} with the real parts of complex scalars zero
\begin{equation}
e^{-\eta}={\rm Im}\,S \equiv s, \ \ \ \ \ 
e^{-\sigma}={\rm Im}\,T \equiv t, \ \ \ \ \ 
e^{-\rho}={\rm Im}\,U \equiv u. 
\label{imstu}
\end{equation}
The following black hole solution to this model is reparameterization 
\cite{RAH372} of the general solution obtained in \cite{CVEy672}:
\begin{eqnarray}
d^2s&=&-e^{2U}dt^2 +e^{-2U}d\vec{x}^2, 
\ \ \ \ \ \ \ \ 
e^{4U}=\psi_1\psi_3\chi_2\chi_4, 
\cr
e^{-2\eta}&=& {{\psi_1\psi_3}\over{\chi_2\chi_4}}, \ \ \ \ \ \ \ \ \ 
e^{-2\sigma}={{\psi_1\chi_4}\over{\chi_2\psi_3}}, \ \ \ \ \ \ \ \ \ 
e^{-2\rho}={{\psi_1\chi_2}\over{\psi_3\chi_4}}, 
\cr
F_1&=&\pm d\psi_1\wedge dt, \ \ \ \ \ 
\tilde{F}_2=\pm d\chi_1\wedge dt, 
\cr
F_3&=&\pm d\psi_3\wedge dt, \ \ \ \ \  
\tilde{F}_4=\pm d\chi_4\wedge dt, 
\label{stusol}
\end{eqnarray}
where
\begin{eqnarray}
\psi_1 &=&\left(e^{{\eta_{\infty}+\sigma_{\infty}+\rho_{\infty}}\over 2}
+{{|q_1|}\over r}\right)^{-1}, \ \ \ \ \ 
\chi_2=\left(e^{{-\eta_{\infty}-\sigma_{\infty}+\rho_{\infty}}\over 2}
+{{|p_2|}\over r}\right)^{-1},
\cr
\psi_3 &=&\left(e^{{\eta_{\infty}-\sigma_{\infty}-\rho_{\infty}}\over 2}
+{{|q_3|}\over r}\right)^{-1}, \ \ \ \ \ 
\chi_4=\left(e^{{-\eta_{\infty}+\sigma_{\infty}-\rho_{\infty}}\over 2}
+{{|p_4|}\over r}\right)^{-1},
\label{stuharm}
\end{eqnarray}
and $\chi_{2,4}$ are magnetic potentials related to 
$\tilde{F}_{2,4}=e^{\eta\pm(\sigma-\rho)}\star F_{2,4}$.  
The ADM mass of (\ref{stusol}) is 
\begin{equation}
M_{ADM}={1\over 4}\left(stu|q_1|+{s\over{tu}}|q_3|+{u\over{st}}|P_2|
+{t\over{su}}|p_4|\right). 
\label{stuadm}
\end{equation}
By minimizing (\ref{stuadm}) with respect to $s,t,u$, one 
obtains the ADM mass at the fixed point 
\cite{CVEy672}:
\begin{equation}
(M_{ADM})_{fix}=|q_1p_2q_3p_4|^{1/4}, 
\label{stufxm}
\end{equation}
and finds that the smaller central charges are zero at the fixed point. 

This result is proven in general setting as follows. 
We consider the $N=8$ supersymmetry transformations \cite{CREj79}
of gravitinos $\Psi_{\mu\,A}$ and fermions $\chi_{ABC}$ at the fixed point:
\begin{equation}
\delta \Psi_{\mu\,A}=D_{\mu}\varepsilon_A + Z_{AB\,\mu\nu}
\gamma^{\nu}\varepsilon^B, \ \ \ \ \ 
\delta\chi_{ABC}= Z_{[AB\,\mu\nu}\sigma^{\mu\nu}\varepsilon_{C]}, 
\label{n8susy}
\end{equation}

where $A=1,...,8$ labels supercharges of $N=8$ theory.  
We truncate the Killing spinors as
\begin{equation}
\varepsilon_a =0, \ \ \ \ 
\varepsilon_i = \{\varepsilon_1,\varepsilon_2\neq 0, 
\varepsilon_3=\varepsilon_4=0\}, 
\label{killtrunc}
\end{equation}
where 6 supersymmetries are projected onto null states. 
Here, we splitted the index $A$ as $A=(i,a)$ in accordance with the 
breaking of $SU(8)$ to $SU(4)\times SU(4)$.   By bringing $Z_{AB}$ to a 
block diagonal form \cite{FERsz} through $SU(8)$ transformation 
(See section \ref{bpssusy}, for details.), one finds that the 
supersymmetry variations of $\psi_{\mu\,a}$, $\chi_{abc}$ and $\chi_{aij}$ 
vanish due to (\ref{killtrunc}) and the block diagonal choice of $Z_{AB}$.  
From $\delta\chi_{iab}=0$, one finds that $Z_{ab}=0$ (i.e. $Z_3=Z_4=0$) 
and from $\delta\chi_{ijk}=0$, one finds that $Z_2=0$.  
So, we proved within the class of configurations characterized by 
truncation (\ref{killtrunc}) that the condition for unbroken 
supersymmetry requires the smaller central charges to vanish. 
And the largest central charge at the fixed point gives the ADM mass 
and the horizon area.  

\paragraph{Five-Dimensional Theories}\label{n2bhentng2d5}

$N=1$, $N=2$ and $N=4$ theories in $D=5$  
\cite{GUNst84,GUNst85,AWAt255,DEWv293,CADcdf357,PAPt357,ANTft460}
have 1, 2 and 3 central charges, respectively.  At the fixed 
point, the largest central charge is minimized and the smaller 
central charges vanish.  The horizon area is given in terms of the 
central charge at the fixed point by $A=4\pi Z^{3/2}_{fix}$. 
The general expressions for the (largest) central charge at the fixed 
point for each case are as follows:
\begin{itemize}
\item $N=1$ theory: 
$Z_{fix} =\sqrt{d^{AB}(q)^{-1}q_A q_B}$, 
where $d^{AB}(q)^{-1}$ is the inverse of $d_{AB}=d_{ABC}t^C(z)$ evaluated 
at the fixed point \cite{FERk136}.  
\item $N=2$ theory: 
$Z_{fix} =\left(Q_HQ^2_F\right)^{1/3}$, 
where $Q_H$ is a charge of the 2-form potential 
and $Q^2_F$ is the Lorentzian $(5,n_v)$ norm of other $5+n_v$ 
charges \cite{STRv}.
\item $N=4$ theory: 
$Z_{fix}=\left(q_{ij}\Omega^{jl}q_{lm}\Omega^{mn}q_{np}\Omega^{pi}
\right)^{1/3}$, where $q_{ij}$ is 27 quantized charges transforming 
under $E_6({\bf Z})$ and $\Omega^{ij}\in Sp(8) $ is traceless.  
\end{itemize}

\subsection{Double Extreme Black Holes}\label{n2bhdb}

We discuss the most general extreme spherically symmetric 
black holes in $N=2$ theories in which all the scalars are {\it frozen} 
to be constant all the way from the horizon ($r=0$) to infinity 
($r\to\infty$) \cite{KALsw}, called {\it double-extreme} black holes
\footnote{After the first draft of this chapter is finished, more general 
class of $N=2$ supergravity black hole solutions 
\cite{SAB101,SAB147,BEHls169,BEHls065,SAB103}, which include general 
rotating black holes and Eguchi-Hanson instantons, are constructed.  
These solutions are entirely determined in terms of the Kahler
potential and the Kahler connection of the underlying 
special geometry, where also the holomorphic sections 
are expressed in terms of harmonic functions.  Such general class of 
solutions turns out to be very important in addressing questions related 
to the conifold transitions in type II superstrings on Calabi-Yau spaces, 
when they become massless \cite{BEHls065}.}.  
For this case, the ADM mass (or the largest central charge) takes the 
minimum value (related to the horizon area) and, therefore, is equal to 
the Bertotti-Robinson mass.  Whereas all the scalars are restricted to 
take values determined by $U(1)$ charges, all the $U(1)$ charges can take 
on arbitrary values.  Double-extreme black holes are also of interests 
since they are the minimum-energy extreme configurations in a moduli 
space for given charges.   

The general double-extreme solution is obtained by starting 
from the spherically symmetric Ansatz for metric
\begin{equation}
ds^2=e^{2U}dt^2 - e^{-2U}d\vec{x}^2, \ \ \ \ \ \  
U(r)\to 0, \ \ {\rm as}\ \ r\to\infty, 
\label{doubmet}
\end{equation}
and assuming that all the scalars are constant 
everywhere ($\partial_{\mu}z^i=0$ and $\partial_{\mu}q^u=0$) and that  
consistency condition ${\cal F}^{-\,i}=0$ for unbroken 
supersymmetry is satisfied. 
Since all the scalars are constant, the spacetime is 
that of extreme Reissner-Nordstrom solution: 
\begin{equation}
e^{-U}= 1+{M\over r}. 
\label{doubmettt}
\end{equation}
By solving the equations of motion following from (\ref{n2act}), 
one obtains
\begin{equation}
{\cal F}^{\Lambda} = e^{2U}{{2Q^{\Lambda}}\over {r^2}}dt\wedge dr - 
{{2P^{\Lambda}}\over{r^2}}
rd\theta\wedge r{\rm sin}\theta d\phi.   
\label{doubgauge}
\end{equation}

From equations of motion along with (\ref{doubmet})$-$(\ref{doubgauge}), 
one obtains the ADM mass $M$ in terms of the electric $Q^{\Lambda}$ and 
magnetic $P^{\Sigma}$ charges of ${\cal F}^{\Lambda}$: 
\begin{equation}
M^2 =-2{\rm Im}\,{\cal N}_{\Lambda\Sigma}(Q^{\Lambda}Q^{\Sigma}+
P^{\Lambda}P^{\Sigma}).
\label{doubadm}
\end{equation}
The $U(1)$ charges $(P^{\Sigma},Q_{\Lambda})$ are related to the 
symplectic charges ${\cal Q}=(q^{\Lambda}_{(m)},q^{(e)}_{\Sigma})=
(\int{\cal F}^{\Lambda},\int{\cal G}_{\Sigma})$ as:
\begin{equation}
\left(\matrix{q^{\Lambda}_{(m)}\cr q^{(e)}_{\Sigma}}\right)=
\left(\matrix{2P^{\Lambda}\cr 
2({\rm Re}\,{\cal N}_{\Lambda\Sigma})P_{\Lambda}-
2({\rm Im}\,{\cal N}_{\Lambda\Sigma})Q^{\Lambda}}\right).
\label{symchtran}
\end{equation}
In terms of $(q^{\Lambda}_{(m)},q^{(e)}_{\Sigma})$, $M$ is expressed as
\footnote{The other sum rule for $Z$ and $Z_a$ is 
$|Z|^2-|Z_a|^2=-{1\over 2}{\cal Q}^T{\cal M}({\cal F}){\cal Q}$.} 
\begin{eqnarray}
M^2&=&-{1\over 2}\left(q^{\Lambda}_{(m)},q^{(e)}_{\Lambda}\right)
\left(\matrix{({\rm Im}{\cal N}+{\rm Re}{\cal N}{\rm Im}{\cal N}^{-1}{\rm Re}
{\cal N})_{\Lambda\Sigma} & (-{\rm Re}{\cal N}{\rm Im}
{\cal N}^{-1})^{\ \Sigma}_{\Lambda} 
\cr
(-{\rm Im}{\cal N}^{-1}{\rm Re}{\cal N})^{\ \Lambda}_{\Sigma} & 
({\rm Im}{\cal N}^{-1})^{\Lambda\Sigma}}\right) 
\cr
& &\ \ \times\left(\matrix{q^{\Sigma}_{(m)}\cr q^{(e)}_{\Sigma}}\right) 
=|Z|^2 + |\nabla_a Z|^2. 
\label{doubsymadm}
\end{eqnarray}  
From the consistency condition ${\cal F}^{-\,a}=0$ for unbroken 
supersymmetry, one has $\nabla_a Z=0$, which is equivalent to 
$\partial_a Z=0$.  So, the ADM mass of double extreme black holes is 
$M=|Z|_{\nabla_aZ=0}$ with scalars constrained to 
take values defined by $\nabla_aZ=0$.  
By solving $\nabla_a Z=0$, one obtains the following 
relation between $(q^{(e)},q_{(m)})$ and the 
holomorphic section $(L^{\Lambda}, M_{\Sigma})$:
\begin{equation}
\left(\matrix{q^{\Lambda}_{(m)}\cr q^{(e)}_{\Sigma}}\right)=
{\rm Re}\left(\matrix{2i\bar{Z}L^{\Lambda}\cr 2i\bar{Z}M_{\Sigma}}\right),
\label{cenqpcon}
\end{equation}
which can be solved to express $(L,M)$ in terms of $(q^{(e)},q_{(m)})$.  
Since the ADM mass $M$ for double-extreme solutions obeys the 
stabilization equations (\ref{cenqpcon}), the entropy is related to the 
ADM mass as:
\begin{equation}
S=\pi M^{\alpha},
\label{entdoub}
\end{equation}
where $\alpha=2\,[3/2]$ for the $D=4$ [$D=5$] solutions.  

\subsubsection{Moduli Space and Critical Points}

We have seen that the BPS condition requires scalars at the event 
horizon take their fixed point values expressed in terms of quantized 
electric/magnetic charges and, thereby, the (largest) central charge 
at the event horizon is related to the black hole entropy.  In this 
subsection, we point out that such property of extreme black holes 
at the fixed point can be derived from bosonic field equations 
and regularity requirement of configurations near 
the event horizon without using supersymmetry \cite{FERgk103}.  
For non-extreme configurations, the horizon area has non-trivial 
dependence on (continuous) scalar asymptotic values.  

We consider the following general form of Bosonic Lagrangian 
\begin{eqnarray}
{\cal L}&=&\sqrt{-g}\left[-{1\over 2}{\cal R}
+{1\over 2}G_{ab}\partial_{\mu}\phi^a\partial_{\nu}
\phi^bg^{\mu\nu}-{1\over 4}\mu_{\Lambda\Sigma}
{\cal F}^{\Lambda}_{\mu\nu}{\cal F}^{\Sigma}_{\lambda\rho}
g^{\mu\lambda}g^{\nu\rho}\right.
\cr
& &\ \ \ \ \ \ \ \,\left. -{1\over 4}\nu_{\Lambda\Sigma}
{\cal F}^{\Lambda}_{\mu\nu}\star{\cal F}^{\Sigma}_{\lambda\rho}
g^{\mu\lambda}g^{\nu\rho}\right],
\label{critbosact}
\end{eqnarray}
where ${\cal F}^{\Lambda}_{\mu\nu}\equiv\partial_{\mu}{\cal A}^{\Lambda}_{\nu}
-\partial_{\nu}{\cal A}^{\Lambda}_{\mu}$ are Abelian field strengths 
with charges $(p^{\Lambda},q_{\Lambda})=
({1\over{4\pi}}\int{\cal F}^{\Lambda},{1\over{4\pi}}\int[\mu_{\Lambda
\Sigma}\star{\cal F}^{\Sigma}+\nu_{\Lambda\Sigma}{\cal F}^{\Sigma}])$
and $\mu_{\Lambda\Sigma}$, $\nu_{\Lambda\Sigma}$ are moduli dependent 
matrices.  We restrict our attention to static Ansatz for the metric
\begin{equation}
ds^2=e^{2U}dt^2-e^{-2U}\gamma_{mn}dx^mdx^n,
\label{statansatz}
\end{equation}
where for spherically symmetric configurations
\begin{equation}
\gamma_{mn}dx^mdx^n={{c^4d\tau^2}\over{\sinh^4c\tau}}+
{{c^2}\over{\sinh^2c\tau}}(d\theta^2+\sin^2\theta d\varphi), 
\label{sphsymspacemet}
\end{equation}
where $\tau$ runs from $-\infty$ (horizon) to 0 (spatial infinity).  
The function $U$ satisfies the boundary conditions $U\to c\tau$ as $\tau
\to -\infty$ and $U(0)=1$.  

The equations of motion for $U(\tau)$ and $\phi^a(\tau)$ can be derived 
from the following 1-dimensional action 
\begin{equation}
{\cal L}_{geod}=\left({{dU}\over{d\tau}}\right)^2+G_{ab}{{d\phi^a}\over
{d\tau}}{{d\phi^b}\over{d\tau}}+e^{2U}V(\phi,(p,q)),
\label{1dgeodeffact}
\end{equation}
describing geodesic motion in a potential $V=(p\,\,q)\left(\matrix{
\mu+\nu\mu^{-1}\nu&\nu\mu^{-1}\cr \mu^{-1}\nu&\mu^{-1}}\right)
\left(\matrix{p\cr q}\right)$ and with the constraint
\begin{equation}
\left({{dU}\over{d\tau}}\right)^2+G_{ab}{{d\phi^a}\over
{d\tau}}{{d\phi^b}\over{d\tau}}-e^{2U}V(\phi,(p,q))=c^2. 
\label{geomtnconstraint}
\end{equation}
where a constant $c$ is related to the entropy $S$ and temperature $T$ 
as $c^2=2ST$.  

For non-extreme configurations, where scalars $\phi^a$ 
have non-zero scalar charges $\Sigma^a$ ($\phi^a\sim\phi^a_{\infty}+
{{\Sigma^a}\over r}$), the first law of thermodynamics is 
modified \cite{GIBkk77} to 
\begin{equation} 
dM={{\kappa dA}\over{8\pi}}+\Omega dJ+\psi^{\Lambda}dq_{\Lambda}
+\chi_{\Lambda}dp^{\Lambda}-G_{ab}(\phi_{\infty})\Sigma^bd\phi^a,
\label{modfstthermlaw}
\end{equation}
whereas the Smarr formula remains in a standard form.  $\Sigma^a$ 
vanish iff $\phi^a$ take the values which extremize $M$, i.e. 
double extreme solutions (i.e. $\phi^a(\tau)=\phi^a_{\infty}$), provided 
$V_{ab}=\nabla_a\nabla_b V$ is non-negative (convexity condition).  
Here, $\nabla_a$ is the Levi-Civita covariant derivative with respect 
to the scalar manifold metric $G_{ab}$.  (This can also be directly seen 
from $\left({{\partial M}\over{\partial\phi^a}}\right)_{A,J,p,q}=
-G_{ab}(\phi_{\infty})\Sigma^b$.)  
For this case, $\phi^a_{\infty}$ have to extremize $V$,  
i.e. $\left({{\partial V}\over{\partial\phi^a}}\right)_{\phi^a_{\infty}}=0$, 
so that $\phi^a$ are regular near the horizon and have fixed values in 
terms of conserved charges $(p^{\Lambda},q_{\Lambda})$.  
The bound $A\leq 4\pi V(p,q,\phi_{horizon})$, which is derived from the 
requirement of finite event horizon area $A$ together with the constraint 
(\ref{geomtnconstraint}), is saturated for the (double) extreme case.  

Since $U\to M\tau$ as $\tau\to 0$, one obtains the following relation 
from (\ref{geomtnconstraint}):
\begin{equation}
M^2+G_{ab}\Sigma^a\Sigma^b-V(\phi^a_{\infty})=4S^2T^2,
\label{genrelfornonex}
\end{equation}
which states that the total self-force on black holes due to the 
attractive forces of gravity and scalars is not exceeded by 
the repulsive self-force due to vectors. 
The net force on black holes vanishes only in the extreme case 
($c^2=0$).  In the double-extreme case, the ADM mass is given by 
$V$ at the fixed point, i.e. $M^2=V(p,q,\phi^a_{fix})$ with the 
fixed values $\phi^a_{fix}$ of scalars determined by 
$\left({{\partial V(\phi,p,q)}\over{\partial\phi^a}}
\right)_{fix}=0$, since $c=0=\Sigma^a$.  From this, one obtains the 
bound on the ADM mass $M(S,\phi_{\infty},(p,q))\geq M(S,\phi_{fix},(p,q))$. 
Note, these results are derived only from the requirement of 
regularity of configurations near the event horizon 
without using supersymmetry.  

We specialize to the case where the target space manifold 
is a K\"ahler manifold spanned by complex scalars $z^i$ with 
K\"ahler potential $K$: $G_{ab}d\phi^ad\phi^b={{\partial^2K}\over
{\partial z^i\partial\bar{z}^j}}dz^id\bar{z}^j$.  We consider the 
bosonic action of $N=2$ supergravity coupled to vector multiplets
\begin{eqnarray}
{\cal L}_{N=2}&=&-{1\over 2}{\cal R}+G_{i\bar{j}}\partial_{\mu}z^i
\partial_{\nu}\bar{z}^{\bar{j}}g^{\mu\nu}+{\rm Im}\,
{\cal N}_{\Lambda\Sigma}{\cal F}^{\Lambda}_{\mu\nu}
{\cal F}^{\Sigma}_{\lambda\rho}g^{\mu\lambda}g^{\nu\rho}
\cr
& &\ \ \ \ \ \ +{\rm Re}\,{\cal N}_{\Lambda\Sigma}
{\cal F}^{\Lambda}_{\mu\nu}\star{\cal F}^{\Sigma}_{\lambda\rho}
g^{\mu\lambda}g^{\nu\rho}.
\label{n2sigmodel}
\end{eqnarray}
The moduli dependent matrices $\mu_{\Lambda\Sigma}$ and 
$\nu_{\Lambda\Sigma}$ in (\ref{critbosact}) are given  by $\nu+i\mu=
-{\cal N}$.  So, $V$ in (\ref{1dgeodeffact}) has the form 
$V(p,q,\phi^a)=|Z(z,p,q)|^2+|D_iZ(z,p,q)|^2$, where $Z$ is the central 
charge and $D_iZ$ is its K\"ahler covariant derivative.  By applying 
properties of special geometries, one obtains the following ADM mass and 
scalar charges
\begin{equation}
M=|Z|(z_0,p,q),\ \ \ \ \ \Sigma^i=G^{i\bar{j}}\bar{D}_{\bar{j}}\bar{Z}. 
\label{n2masscalchrg}
\end{equation}
By applying the general results in the previous paragraph, one can 
see that at the critical points of $V$ ($\partial_iV=0$), 
where $\Sigma^i=0$, $Z$ is extremized: $D_iZ=0=\bar{D}_{\bar{k}}\bar{Z}$.  
Since the second covariant derivative of $|Z|$ at the critical point 
coincides with the partial (non-covariant) second derivative, one has 
$(\bar{\partial}_{\bar{i}}\partial_j|Z|)_{\rm cr}={1\over 2}G_{\bar{i}j}
|Z|_{\rm cr}$.  So, when $G_{\bar{i}j}$ is positive [negative] at the 
critical point, $M$ at the critical point reaches its minimum [maximum].  
Generally, when $G_{\bar{i}j}$ changes its sign and becomes negative, 
some sort of a phase transition occurs and the effective Lagrangian 
breaks down unless new massless states appear.

\subsubsection{Examples}\label{n2bhdbeg}

In the following, we discuss the explicit expression for 
$M$ in the metric component (\ref{doubmettt}) with specific 
prepotentials.

\paragraph{Axion Dilaton Black Holes}\label{n2bhdbegaxd}

The axion-dilaton black holes in the $SO(4)$ [$SU(4)$] formulation 
of pure $N=4$ supergravity can be regarded as black holes  
in $N=2$ supergravity coupled to one vector multiplet with the prepotential 
$F=-iX^0X^1$ [without prepotential]. 
The holomorphic sections and $U(1)$ charges of $SU(4)$ theory 
\cite{CREsf} (with hats) are related to those of $SO(4)$ theory 
\cite{DAS15,CREsf68,CREs127} (without hats) as 
\cite{DEWv245,CERdfv444,CERdf46}:
\begin{eqnarray}
\hat{X}^0&=&X^0, \ \ \ \ \hat{F}_0=F_0, \ \ \ \ 
\hat{X}^1=-F_1, \ \ \ \ \hat{F}^1=X^1, 
\cr
\hat{q}^0_{(m)}&=&q^0_{(m)}, \ \ \ \ \hat{q}^{(e)}_0=q^{(e)}_0, \ \ \ \ 
\hat{q}^1_{(m)}=-q^{(e)}_1, \ \ \ \ \hat{q}^{(e)}_1=q^1_{(m)}. 
\label{susotran}
\end{eqnarray}

First, we discuss the $SO(4)$ case. 
We choose the gauge $X^0=1$.  Then, the prepotential $F=-iX^0X^1$ yields 
the K\"ahler potential $e^K={1\over {2(z+\bar{z})}}$ ($z\equiv {{X^1}
\over{X^0}}$)
\footnote{From this expression for $K$, one sees that the real part of 
the moduli $z$ has to be positive, leading to the constraint 
${\rm Re}\,z=|q^{(e)}_0q^{(e)}_1+q^0_{(m)}q^1_{(m)}|$.} 
and (\ref{cenqpcon}) can be solved to fix the moduli $z$ in terms 
charges:
\begin{equation}
z={{q^{(e)}_0-iq^1_{(m)}}\over {q^{(e)}_1-iq^0_{(m)}}}. 
\label{modch}
\end{equation}
So, one has central charge in terms of  $U(1)$ charges:
\begin{equation}
Z=L^Iq^{(e)}_I-M_Iq^I_{(m)}
=\left({{q^{(e)}_0q^{(e)}_1+q^0_{(m)}q^1_{(m)}}\over {(q^{(e)}_1)^2+
(q^0_{(m)})^2}}\right)^{1/2}(q^{(e)}_1+iq^0_{(m)}), 
\label{cench}
\end{equation}
by solving (\ref{cenqpcon}).  This leads to the ADM of the 
double-extreme black hole:
\begin{equation}
M^2 =|Z|^2 = |q^{(e)}_0q^{(e)}_1 + q^0_{(m)}q^1_{(m)}|. 
\label{admso4}
\end{equation}

The corresponding expressions in the $SU(4)$ theory are obtained by 
applying the transformations (\ref{susotran}).  The moduli field 
and the ADM mass are \cite{KALo48,KAL54}
\begin{equation}
z={{\hat{q}^{(e)}_1+i\hat{q}^{(e)}_0}\over
{\hat{q}^0_{(m)}-i\hat{q}^1_{(m)}}}, \ \ \ \ \ \ 
M^2=|Z|^2=|\hat{q}^0_{(m)}\hat{q}^{(e)}_1-\hat{q}^{(e)}_0\hat{q}^1_{(m)}|.
\label{quansu4}
\end{equation}

\paragraph{$N=2$ Heterotic Vacua}\label{n2bhdbeghet}

The effective field theory of $N=2$ heterotic string is described by 
the $N=2$ theory with a prepotential 
\cite{ANDbcdff476,ANDbcdffm032}
\begin{equation}
F=-{{X^1}\over{X^0}}\left[X^2X^3-\sum^{n+1}_{i=4}(X^i)^2\right].
\label{prehet}
\end{equation}
This prepotential corresponds to the manifold 
${{SU(1,1)}\over{U(1)}}\times {{SO(2,n)}\over{SO(2)\times SO(n)}}$ 
\cite{FERv6,DEWkll451} with the first factor parameterized by the 
axion-dilaton field $S=-iX^1/X^0=-iz^1$ and the second factor being  
the special K\"ahler manifold parameterized by $n$ complex moduli 
$z^i=X^i/X^0$ ($i=2,...,n+1$).  
$S$ belongs to a vector multiplet and the remaining 
vector multiplets with the scalar components $z^i$ are associated 
with the $U(1)$ gauge fields in the left moving sector of heterotic string.  
In particular, the $n=2$ case is the $STU$-model 
\cite{BEHkrsw54,CARclmr464} with the complex scalars $S$, $T$ and $U$ 
parameterizing each $SL(2,R)$ factor of the moduli group.  
$S$, $T$ and $U$ are related to $z^i$ as
\begin{equation}
z^1=iS, \ \ \ \ \ z^2=iT, \ \ \ \ \  z^3=iU,
\label{modstu}
\end{equation}
and, therefore, the prepotential takes the form:
\begin{equation}
F=-STU.
\label{prepstu}
\end{equation}

It is convenient to apply a singular symplectic transformation 
\cite{CERdfv444} (defined as $X^1\to -F_1$ and $F_1\to -X^1$) on 
$(X^{\Lambda},F_{\Sigma})$ to bring it to the form \cite{CERdfv200}: 
\begin{equation}
\left(\matrix{X^{\Lambda}\cr F_{\Sigma}}\right) = 
\left(\matrix{X^{\Lambda}\cr S\eta_{\Sigma\Lambda}X^{\Lambda}}\right). 
\label{althet}
\end{equation}
In this basis, theory has a uniform weak coupling behavior as ${\rm Im}\,
S\to\infty$ and the holomorphic section satisfies the constraints
\begin{equation} 
\langle X|X\rangle=\langle F|F\rangle=X\cdot F=0.
\label{hetconst}
\end{equation}   
Here, $\langle A|B\rangle \equiv A^{\Lambda}\eta_{\Lambda\Sigma}B^{\Sigma}
=A_{\Lambda}\eta^{\Lambda\Sigma}B_{\Sigma}$ and $A\cdot B\equiv 
A^{\Lambda}B_{\Lambda}$ with 
\begin{equation}
\eta_{\Lambda\Sigma}=\left(\matrix{{\cal L}&0&0\cr 0&{\cal L}&0\cr 0&0&
-{\bf I}}\right);\ \ \ \ \ 
{\cal L}\equiv\left(\matrix{0&1\cr 1&0}\right).
\label{etamat}
\end{equation}

By solving (\ref{cenqpcon}), along with (\ref{althet}), one  
fixes $S$ in terms of $U(1)$ symplectic charges and 
obtains the ADM mass of double extreme solution  
\cite{CVEy672,DUFlr}:
\begin{eqnarray}
S&=&{{q_{(m)}\cdot q^{(e)}}\over {\langle q_{(m)}|
q_{(m)}\rangle}}+i{\sqrt{\langle q_{(m)}|q_{(m)}\rangle
\langle q^{(e)}|q^{(e)}\rangle -
(q_{(m)}\cdot q^{(e)})^2} \over {\langle q_{(m)}|
q_{(m)}\rangle}}, 
\cr
M^2&=&|Z|^2 =\sqrt{\langle q_{(m)}|q_{(m)}\rangle
\langle q^{(e)}|q^{(e)}\rangle 
-(q_{(m)}\cdot q^{(e)})^2}
=({\rm Im}\,S)\langle q_{(m)}|q_{(m)}\rangle.
\label{cenhet}
\end{eqnarray}

\paragraph{Cubic Prepotential}\label{n2bhdbegcub}

We consider the following general form of cubic prepotential 
\cite{CREkvdfddg250}:
\begin{equation}
F=d_{abc}{{X^aX^bX^c}\over{X^0}},\ \ \ a,b,c=1,...,n_v. 
\label{cyprep}
\end{equation}
The $n_v=3$ case is the $STU$ model \cite{DEWclmr481,BEHkrsw54}.

By solving (\ref{cenqpcon}), along with (\ref{cyprep}), one obtains 
the ADM mass at the fixed point in the moduli space \cite{SHM076}:
\begin{equation}
M^2={1\over{3q^0_{(m)}}}\sqrt{{4\over 3}(\Delta_a\tilde{x}^a)^2-
9[q^0_{(m)}(\vec{q}_{(m)}\cdot \vec{q}^{(e)})-2D]^2},
\label{admcy}
\end{equation}
where $D\equiv d_{abc}q^a_{(m)}q^b_{(m)}q^c_{(m)}$, 
$D_a\equiv d_{abc}q^b_{(m)}q^c_{(m)}$ and 
$\Delta_a\equiv 3D_a-q^0_{(m)}q^{(e)}_a$.  
Here, $\tilde{x}^a$ in (\ref{admcy}) are real solutions 
to the system of equations:
\begin{equation}
d_{abi}\tilde{x}^a\tilde{x}^b=\Delta_i.
\label{algsys}
\end{equation}
The moduli are fixed in terms of the symplectic $U(1)$ charges as
\begin{equation}
z^a={3\over 2}{{\tilde{x}^a}\over{q^0_{(m)}(\Delta_c\tilde{x}^c)}}
[q^0_{(m)}(\vec{q}_{(m)}\cdot\vec{q}^{(e)})-2D]+{{q^a_{(m)}}\over
{q^0_{(m)}}}-i{3\over 2}{{\tilde{x}^a}\over{(\Delta_c\tilde{x}^c)}}
M^2.
\label{modchcy}
\end{equation}

When $q^0_{(m)}=0$, (\ref{cenqpcon}) can be solved 
explicitly to yield the ADM mass:
\begin{equation}
M^2=\sqrt{{D\over 3}(q^{(e)}_aD^a+12q^{(e)}_0)},
\label{sympcyadm}
\end{equation}
where $D_{ab}\equiv d_{abc}q^c_{(m)}$, $D^{ab}\equiv[D^{-1}]_{ab}$ and 
$D^a\equiv D^{ab}q^{(e)}_b$.  
And the moduli $z^a$ take the following fixed point value in 
terms of the symplectic charges:
\begin{equation}
z^a={{D^a}\over 6}-i{{q^a_{(m)}}\over 2}DM^2.
\label{symchcy2}
\end{equation}

When the prepotential (\ref{cyprep}) 
has an extra topological term \cite{HOSly182,LOUst480,BEHcdk488}
$\sum^{n_v}_{a=1}{{c_2\cdot J_a}\over{24}}z^a$, one only needs to apply 
the symplectic transformation \cite{CERdfv444,BEHcdk488} with 
the matrix $\left(\matrix{{\bf 1}&0\cr W&{\bf 1}}\right)\in Sp(2n_v+2)$, 
where the non-zero components of $W_{\Lambda\Sigma}$ are 
$W_{0a}={{c_2\cdot J_a}\over{24}}$.  Then, the ADM mass is given by 
(\ref{admcy}) or (\ref{sympcyadm}) with the symplectic charges 
$(\vec{q}_{(m)},\vec{q}^{(e)})$ replaced by
\begin{equation}
\tilde{q}^{(e)}_{\Lambda}=q^{(e)}_{\Lambda}-W_{\Sigma\Lambda}
q^{\Sigma}_{(m)};\ \ \ \ 
\left\{\matrix{\tilde{q}^{(e)}_0=q^{(e)}_0-{{c_2J_a}\over{24}}q^a_{(m)}\cr
\tilde{q}^{(e)}_a=q^{(e)}_a-{{c_2J_a}\over{24}}q^0_{(m)}}\right..
\label{sympcy}
\end{equation}  

\paragraph{$CP(n-1,1)$ Model}\label{n2bhdbegcp}

The $CP(n-1,1)\equiv{{SU(1,n)}\over{SU(n)\times U(1)}}$ model has the 
isometry group $SU(1,n)$.  The $n=1$ case is the axion-dilaton black 
holes \cite{KALlopp46,KALkot50,BERko478,KALsw}.  
The form of prepotential depends on the way in which $SU(1,n,{\bf Z})$ 
is embedded into $Sp(2n+2,{\bf Z})$ \cite{SAB486}.  

For the $Sp(2n+2,{\bf Z})$ embedding $\Omega=\left(\matrix{A&B\cr C&D}
\right)$ of $M=U+iV\in SU(1,n)$ given by:
\begin{equation}
A=U, \ \ \  C=\eta V,\ \ \ 
B=V\eta,\ \ \  D=\eta U\eta ,
\label{embed1}
\end{equation}
where $\eta$ is an $SU(1,n)$ invariant metric, 
the holomorphic prepotential is 
\begin{equation}
F={i\over 2}X^t\eta X.
\label{prep1}
\end{equation}
For this case, $X$ transforms as a vector under $SU(1,n)$, i.e. 
$X\to MX$, and the moduli space is parameterized by 
$\phi=(\phi^0,...,\phi^{n+1})^T$ as $\phi^0={1\over{\sqrt{Y}}}$ 
and $\phi^a={{z^a}\over{\sqrt{Y}}}$ with $Y=1-\sum_az^a\bar{z}^a$. 
In general, the ADM mass of the extreme solution has the form 
\cite{SAB210}:
\begin{equation}
M^2_{BPS}={{|m_c-n_ct-Q_{ic}{\cal A}^i|^2}\over 
{2(1-t\bar{t}-\sum_i\,{\cal A}^i\bar{\cal A}^i)}}, 
\label{cpadm}
\end{equation}
where 
\begin{eqnarray}
m_c&\equiv&q^{(e)}_0+iq^0_{(m)}, \ \ 
n_c\equiv iq^1_{(m)}-q^{(e)}_1, \ \ 
Q_{ic}\equiv iq^i_{(m)}-q^{(e)}_i, 
\cr
t&\equiv&{{X^1}\over {X^0}}, \ \ \ \ \ \ 
{\cal A}^i\equiv {{X^i}\over {X^0}}.
\label{defcomp}
\end{eqnarray}
For the following fixed-point values of the moduli fields, which satisfy 
(\ref{cenqpcon}), 
\begin{equation}
\bar{t}={{n_c}\over{m_c}}, \ \ \ \ \ \ 
\bar{\cal A}^i={{Q_{ic}}\over {m_c}},
\label{cpmodch}
\end{equation}
the ADM mass $M$ takes the minimum value \cite{SAB210}
\begin{equation}
M^2={1\over 2}\left(|m_c|^2-|n_c|^2-|Q_{ic}|^2\right) 
=\pi\left(\matrix{q^{(e)}& q_{(m)}}\right) 
\left(\matrix{\eta&0\cr 0&\eta}\right)
\left(\matrix{q^{(e)}\cr q_{(m)}}\right).
\label{cpadm1}
\end{equation}

For other embedding $\Omega^{\prime}=S\Omega S^{-1}$ of  
$SU(1,n)$ into $Sp(2n+2)$ related via the symplectic transformation 
$S\in Sp(2n+2)$, the ADM mass is given in terms of new 
symplectic $U(1)$ charges $(\vec{q}^{(e)\,\,\prime}\,  
\vec{q}^{\,\prime}_{(m)})^T=(\vec{q}^{(e)}\,\,\vec{q}_{(m)})S^{-1}$ 
by \cite{SAB210}
\begin{equation}
M^2=\pi\left(\matrix{\vec{q}^{(e)\,\prime}& \vec{q}^{\,\prime}_{(m)}}\right)S
\left(\matrix{\eta&0\cr 0&\eta}\right)S^T
\left(\matrix{\vec{q}^{(e)\,\prime}\cr \vec{q}^{\,\prime}_{(m)}}\right).
\label{cpadm2}
\end{equation}

\paragraph{General Quadratic Prepotential}\label{n2bhdbegqd}

We discuss the case where the lower-component of $V$ is proportional 
to the upper component \cite{BEHs010}:
\begin{equation}
\left(\matrix{L^I\cr M_J}\right)=
\left(\matrix{L^I\cr \Sigma_{JK}L^K}\right),
\label{proprep}
\end{equation}
where $\Sigma_{IJ}=\alpha_{IJ}-i\beta_{IJ}$ with real matrices 
$\alpha_{IJ}$ and $\beta_{IJ}$.  Note, it is sufficient to consider the 
case where $\Sigma_{IJ}=-i\beta_{IJ}$, since the most general case with 
$\alpha_{IJ}\neq 0$ is achieved by applying the symplectic transformation
\footnote{The ADM mass and entropy transform under this symplectic 
transformation similarly as (\ref{cpadm2}).}  
$\left(\matrix{1&0\cr \alpha_{IJ}&1}\right)\in Sp(2n+2)$ 
on the configuration with $\alpha_{IJ}=0$.  

By solving (\ref{cenqpcon}) with (\ref{proprep}), one obtains the 
ADM mass
\begin{eqnarray}
M^2&=&{i\over 2}\left(\matrix{\vec{q}^{(e)}&\vec{q}_{(m)}}\right)
\left(\matrix{(\bar{\Sigma}-\Sigma)^{-1}& (\Sigma-\bar{\Sigma})^{-1}\Sigma\cr 
\bar{\Sigma}(\Sigma-\bar{\Sigma})^{-1}& \bar{\Sigma}(\bar{\Sigma}-\Sigma)^{-1}
\Sigma}\right)
\left(\matrix{\vec{q}^{(e)}\cr \vec{q}_{(m)}}\right)
\cr
&=&{1\over 2}\left(\matrix{\vec{q}^{(e)}&\vec{q}_{(m)}}\right)
\left(\matrix{\beta^{-1}& -\beta^{-1}\alpha \cr 
-\alpha\beta^{-1}& \beta+\alpha\beta^{-1}\alpha}\right)
\left(\matrix{\vec{q}^{(e)}\cr \vec{q}_{(m)}}\right).
\label{quadmass}
\end{eqnarray}
The case $\Sigma=-i\eta$, where $\eta$ is an $SU(1,n)$ invariant 
metric, is the $CP(n-1,1)$ model, i.e. (\ref{quadmass}) reduces to 
(\ref{cpadm1}).

The most general extreme solution to this model has the form \cite{TOD121}:
\begin{eqnarray}
ds^2&=&-e^{-2U}dt^2+e^{2U}d\vec{x}\cdot d\vec{x},
\cr
F^I_{\mu\nu}&=&\epsilon_{\mu\nu\rho}\partial_{\rho}\,\tilde{H}^I, \ \ \ \ \ 
G_{J\,\mu\nu}=\epsilon_{\mu\nu\rho}\partial_{\rho}\,H_J, 
\cr
Y&\equiv&\bar{Z}L=i(\Sigma-\bar{\Sigma})^{-1}(\bar{\Sigma}\tilde{H}-H),
\label{genn2bh}
\end{eqnarray}
where 
\begin{eqnarray}
e^{2U}&=&i\left(\matrix{H^T&\tilde{H}^T}\right)
\left(\matrix{(\bar{\Sigma}-\Sigma)^{-1}& (\Sigma-\bar{\Sigma})^{-1}\Sigma\cr 
\bar{\Sigma}(\Sigma-\bar{\Sigma})^{-1}& \bar{\Sigma}(\bar{\Sigma}-\Sigma)^{-1}
\Sigma}\right)
\left(\matrix{H\cr \tilde{H}}\right), 
\cr
\tilde{H}^I&=&(\tilde{h}^I+{{q^I_{(m)}}\over r}), \ \ \ \ \ 
H_J=(h_J+{{q^{(e)}_J}\over r}). 
\label{defgenn2bh}
\end{eqnarray}
The asymptotically flatness condition leads to 
the following constraint on $\tilde{h}^I$ and $h_J$ in (\ref{defgenn2bh}):
\begin{equation}
\left(\matrix{h^T&\tilde{h}^T}\right)
\left(\matrix{(\bar{\Sigma}-\Sigma)^{-1}& (\Sigma-\bar{\Sigma})^{-1}\Sigma\cr 
\bar{\Sigma}(\Sigma-\bar{\Sigma})^{-1}& \bar{\Sigma}(\bar{\Sigma}-\Sigma)^{-1}
\Sigma}\right)
\left(\matrix{h\cr\tilde{h}}\right)=-i.
\label{cnstgenn2}
\end{equation}

\subsection{Quantum Aspects of $N=2$ Black Holes}\label{n2bhq}

Supersymmetric field theories respect remarkable 
perturbative non-renormalization theorems.  In $N=1$ theories,  
superpotentials are not renormalized in perturbation theory 
\cite{GRIsr159,HOWsw124}.  
$N\geq 4$ theories are finite to all orders in perturbative 
quantum corrections \cite{SOHw100,MAN213}.  So, the classical BPS 
solutions in $N\geq 4$ theories are exact to all orders 
in perturbative corrections.  (Cf. Some classical solutions of $N=4$ 
theories are also exact solutions 
\cite{TSE4,TSE12,HORt50,HORt73,HORt51,RUSt449,BEH348,BEH455,CVEt366,CVEt53} 
of conformal $\sigma$-model of string theory and, therefore, exact to all 
orders in $\alpha^{\prime}$-corrections.)   
For $N=2$ theories, prepotentials, which determine the Lagrangians, 
receive perturbative quantum corrections up to one-loop level 
\cite{SEI206,GRIsr159,DEWkll451,ANTfgnt447}.  
Hence, one has to study quantum effect on prepotentials for 
complete understanding of solutions in $N=2$ theories.  

In the following, we study the quantum aspects of black holes 
in the effective $N=2$ theories of compactified superstring theories.   
It is conjectured that the $E_8\times E_8$ heterotic string 
on $K3\times T^2$ and the type-II string on a Calabi-Yau manifold 
are a $N=2$ string dual pair 
\cite{KACv450,FERhsv361,KAPlt357,ANTgnt455,ASPl369,KACklmv459,ANTp460,CARclmr464}.  
Since the dilaton-axion field $S$ belongs to a vector multiplet 
[hyper multiplet] of the heterotic [type-II] theory, moduli space 
of hyper multiplet [prepotential for a vector multiplet] is exact at 
the tree level, due to the absence of neutral perturbative couplings 
between vector multiplets and hypermultiplets 
\cite{KLElm357,KAPlt357,ANTgnt455,CUR366,CUR368,CARclm382}.  
Thus, applying the duality between heterotic 
and type-II strings, one can compute the exact prepotential 
for vector multiplets [hyper multiplet superpotential] of the 
heterotic [type-II] theory.  

The prepotentials of the $N=2$ effective field theories of 
these string theories contain the cubic terms:
\begin{equation}
F(X)=d_{abc}{{X^{a}X^{b}X^{c}}\over{X^0}}, 
\label{cubpreppo}
\end{equation}
plus  correction terms that include part of quantum corrections, 
instanton and topological terms which cannot be included in 
(\ref{cubpreppo}).  
For the type-IIA string on a Calabi-Yau three-fold, real coefficients 
$d_{abc}$ are the topological intersection numbers  
$d_{abc}\equiv\int\,J_a\wedge J_b\wedge J_c$, where 
$J_a\in H^{1,1}(Y,{\bf Z})$ are the K\"ahler cone generators.  
For the heterotic string on $K3\times T^2$, 
$d_{abc}$ describe the classical parts and the non-exponential 
parts of perturbative corrections to the prepotential.  
The K\"ahler potential associated with (\ref{cubpreppo}) is
\begin{equation}
K(z,\bar{z})=-\log\left(-id_{abc}(z-\bar{z})^a(z-\bar{z})^b(z-\bar{z})^c
\right).
\label{cubkahl}
\end{equation}
General double-extreme black holes and a special class of extreme 
black holes with non-constant scalars of the $N=2$ theory with 
(\ref{cubpreppo}) are discussed in \cite{BEH232}.  
In the following, we study the effect of the quantum correction 
terms of the prepotential on the classical solutions 
\cite{CARlm388,BEHcdk488,REY157,BEH232,BEH053}. 

\subsubsection{$E_8\times E_8$ Heterotic String on $K3\times 
T^2$}\label{n2bhqhet}

At generic points in moduli space, the $E_8\times E_8$ heterotic 
string on $K3\times T^2$ is characterized by 65 gauge-neural hypermultiplets
\footnote{The scalar components of the remaining hypermultiplets in 
$\bf 56$ of $E_7$ are not spectrum-preserving moduli, since their non-zero 
vacuum expectation values induce mass for some of the non-Abelian 
fields, resulting in change in the spectrum of light particles 
in the effective theories.} 
(20 from the $K3$ surface and 45 from the gauge bundle) and 
19 vectors (18 from vector multiplets and 1 from the gravity 
multiplet).   So, the moduli $z^a$ ($a=1,...,n_v$) in the 
vector multiplets consist of the axion-dilaton $S$, the $T^2$ moduli $T$ and 
$U$, and Wilson lines $V^i$ ($i=1,...,n_v-3$):
\begin{equation}
z^1=iS, \ \ \ z^2=iT, \ \ \ z^3=iU, \ \ \ 
z^{i+3}=iV^i.
\label{hetmodul}
\end{equation}
We denote the moduli other than $S$ as $T^m$ ($m=2,...,n_v$).  
The number of Wilson lines $V^i$ (or the generic unbroken 
Abelian gauge group $U(1)^{n_v+1}$ with one of $U(1)$ factors 
coming from the gravity multiplet) depends on the choice of 
$SU(2)$ bundles with instanton numbers $(d_1,d_2)=(12-n,12+n)$ 
\cite{KACv450,CANf170,ALDfiq461}. 
For example, the $STU$ model (i.e. the complete Higgsing of the 
$D=6$ gauge group $E_7\times E_7$) is possible for $n=0,1,2$.  

Since prepotentials of $N=2$ theories are not renormalized beyond 
one-loop perturbative levels, the prepotentials are written in the 
form \cite{CERdfv444,DEWkll451,ANTfgnt,ANTfgnt447}:
\begin{equation}
F=F^{(0)}+F^{(1)}+F^{(NP)}, 
\label{n2prepot}
\end{equation}
where $F^{(0)}$ is the tree-level prepotential and $F^{(1)}$  
[$F^{(NP)}$] is the one-loop [non-perturbative] correction.  
The classical moduli space of the $E_8\times E_8$ heterotic 
string on $K3\times T^2$ is 
${{SU(1,1)}\over{U(1)}}\otimes {{SO(2,n_v-1)}\over{SO(2)\times SO(n_v-1)}}$ 
\cite{FERv6,FERgkp192,FERp206,FREs371,CERdf339,CERdfv200,CERdfv444} 
with $S$ residing in the separate moduli space ${{SU(1,1)}\over{U(1)}}$.  
The tree-level prepotential is 
\begin{equation}
F^{(0)}=-{{X^1}\over{X^0}}\left[X^2X^3-\sum^{n_v}_{i=4}(X^i)^4\right]
=-S\left[TU-\sum_i(V^i)^2\right],
\label{treeprep}
\end{equation}
with the corresponding tree-level K\"ahler potential:
\begin{equation}
K^{(0)}=-\log(S+\bar{S})-\log\left[(T+\bar{T})(U+\bar{U})
-\sum_i(V^i+\bar{V}^i)^2\right].
\label{treekahl}
\end{equation}
Since the dilaton is the loop-counting parameter 
of the heterotic string, $F^{(1)}$ is independent of $S$: $F^{(1)}=h(T^m)$  
with the infinite series terms exponentially suppressed in the 
decompactification limit of large moduli.  Non-perturbative 
corrections arise from the spacetime gauge or gravitational instantons 
\cite{SEIw426,SEIw431,KLElty344,ARGf74}, and give rise to holomorphic 
but exponentially suppressed corrections.  
So, the heterotic prepotential has the form:
\begin{equation}
F=-S\left[TU-\sum_i(V^i)^2\right]+h(T^m)+f^{NP}(e^{-2\pi S},T^m).
\label{hetprep}
\end{equation}

Since the $T$-duality is an exact symmetry of the heterotic string 
to all orders in perturbation, the one-loop correction $h(T^m)$ 
\cite{DEWkll451,ANTfgnt447,HARm463,HENm145,KIRk442,KIRk456,KIRkpr091,KIRkpr483,CARclm382,CARcl154} 
has to have an well-defined $T$-duality transformation properties 
\cite{DEWkll451,ANTfgnt447}. 
For models with one Wilson line
\footnote{This is possible for the instanton embedding with $n=0,1,2$.} 
($STUV$-model) \cite{CARcl154}, 
\begin{equation}
h(T,U,V)=p_n(T,U,V)-c-{1\over{4\pi^3}}\sum_{k,l,b\in {\bf Z} \atop 
(k,l,b)>0}\,c_n(4kl-b^2)Li_3({\bf e}[ikT+ilU+ibV]), 
\label{oneloop}
\end{equation}
where $c={{c_n(0)\zeta(3)}\over{8\pi^3}}$ and ${\bf e}[x]=
\exp 2\pi ix$.  Here, $c_n(4kl-b^2)$ are the expansion coefficients 
of particular Jacobi modular form \cite{CARcl154} and $p_n(T,U,V)$ 
is the one-loop cubic polynomial, which depends on the particular 
instanton embedding $n$, given by 
\cite{BERkk456,LOUst480,CARcl154}
\begin{equation}
p_n(T,U,V)=-{1\over 3}U^3-({4\over 3}+n)V^3+(1+{1\over 2}n)UV^2
+{1\over 2}nTV^2.
\label{polynom}
\end{equation}

\paragraph{Perturbative Corrections}\label{n2bhqhetper}

In section \ref{n2bhdbeg}, we obtained the general expression 
for entropy (or the ADM mass of the double extreme black hole) 
in the {\it tree level} effective theory of the heterotic 
string on $K3\times T^2$.  (See (\ref{cenhet}) for the tree level 
expression.)  The entropy depends on the symplectic charge vector 
${\cal Q}=(q_{(m)}\,\,q^{(e)})$ through the full triality 
invariant form $D=\langle q_{(m)}|q_{(m)}\rangle
\langle q^{(e)}|q^{(e)}\rangle -(q_{(m)}\cdot q^{(e)})^2$ 
\cite{CARlm388} and is invariant under $T$-duality since the 
dilaton ${\rm Im}\,S$ and the combination $\langle q_{(m)}|q_{(m)}\rangle$ 
remain intact under $T$-duality \cite{CERdfv444,DEWkll451,CARlm388}.  

Once perturbative quantum corrections are considered, 
$T$-duality transformations get modified due to the 
one-loop corrections to the prepotential:
\begin{equation}
F=F^{(0)}+F^{(1)}=-S\left[TU-\sum_i(V^i)^2\right]+h(T^m),
\label{hetpert}
\end{equation}
where the one-loop correction term $h(T^m)$ is independent of 
the axion-dilaton field $S=-iX^1/X^0$.  For the purpose of discussing 
duality transformations, it is convenient to go to the symplectic 
basis, applying the transformations 
\cite{FERgkp192,FREs371,CERdf339,CERdfv444,DEWkll451} 
\begin{equation}
X^1\to \hat{X}^1=F_1,\ \ \ \ 
F_1\to \hat{F}_1=-X^1,
\label{singsimp}
\end{equation}
with the corresponding tree-level holomorphic section taking the 
form (\ref{althet}) in the new basis.  (In this basis, the theory 
has a uniform weak coupling behavior as Im$S\to\infty$.)
Since one-loop corrections are independent of $X^1$, the relation  
$\hat{X}^1=F_1$ is not modified at one loop level and, as a result,  
$\hat{X}^{\Lambda}$ still satisfies the constraint $\langle\hat{X}|\hat{X}
\rangle=0$ at one-loop level.  However, $\hat{F}_{\Lambda}$ gets 
modified due to the perturbative correction $F^{(1)}$ in the following 
way \cite{ANTgnt413} (Cf. See (\ref{althet}) for the tree-level expression):
\begin{equation}
\hat{F}_{\Lambda}=S\eta_{\Lambda\Sigma}\hat{X}^{\Sigma}+F^{(1)}_{\Lambda}, 
\label{pertfi}
\end{equation}
where $F^{(1)}_{\Lambda}\equiv \partial F^{(1)}/\partial X^{\Lambda}$.  
Note, $F^{(1)}_1$ keeps the classical value.  

Whereas $\hat{X}^{\Lambda}$ transforms exactly the same way as in  
classical theory, the $T$-duality transformation rules of 
$\hat{F}_{\Lambda}$ get modified at one-loop level due to the 
modification of prepotential \cite{DEWkll451,HARm463,ANTfgnt447,CARlm455}:
\begin{equation}
\hat{X}^{\Lambda}\to \hat{U}^{\Lambda}_{\Sigma}\hat{X}^{\Sigma}, 
\ \ \ \ \ \ 
\hat{F}_{\Lambda}\to [(\hat{U}^T)^{-1}]^{\ \Sigma}_{\Lambda}\hat{F}_{\Sigma}
+[(\hat{U}^T)^{-1}{\cal C}]_{\Lambda\Sigma}\hat{X}^{\Sigma}, 
\label{pertdual}
\end{equation}
where $U\in SO(2,2+n,{\bf Z})$ and the symmetric integral matrix 
$\cal C$ encodes the quantum corrections.  
From the relation $X^1=-\hat{F}_1=-iS\hat{X}^0$, one sees that 
$S$ is no longer invariant under the perturbative $T$-duality 
transformation (\ref{pertdual}), but transforms as \cite{DEWkll451}
\begin{equation}
S\to S+{{i[(\hat{U}^T)^{-1}]^{\ \Sigma}_1
(H^{(1)}_{\Sigma}+{\cal C}_{\Sigma\Delta}\hat{X}^{\Delta})}\over
{\hat{U}^0_{\Lambda}\hat{X}^{\Lambda}}}.
\label{pertdil}
\end{equation}

Note, $\langle q_{(m)}|q_{(m)}\rangle$ in the classical expression 
(\ref{cenhet}) is still invariant under the perturbative $T$-duality, but 
the dilaton ${\rm Im}\,S$ transforms  under the perturbative $T$-duality
(\ref{pertdil}).  Since superstring theories are exact under $T$-duality 
order by order in perturbative corrections, one expects entropy to be 
invariant under $T$-duality.  One way of making entropy to be manifestly 
invariant under $T$-duality is to introduce the invariant 
dilaton-axion field $S_{pert}$ \cite{DEWkll451,HARm463} which do not 
transform under $T$-duality.  This motivates the 
following conjectured expression for entropy at one-loop level 
\cite{CARlm388}:
\begin{equation}
S_{pert}=\pi|Z_{pert}|^2=\pi({\rm Im}\,S_{pert})
\langle p_{(m)}|p_{(m)}\rangle.  
\label{pertent}
\end{equation}
The perturbative dilaton ${\rm Im}\,S_{pert}$ is understood as follows.  
The perturbative prepotential (\ref{hetpert}) leads to the  
following perturbative K\"ahler potential \cite{DEWkll451}
\begin{equation}
K=-\log\left[(S+\bar{S})+V_{GS}(T^m,\bar{T}^m)\right]
-\log\left[(T^m+\bar{T}^m)\eta_{mn}(T^n+\bar{T}^n)\right], 
\label{pertkahl}
\end{equation}
where 
\begin{equation}
V_{GS}(T^m,\bar{T}^m)={{2(h+\bar{h})-(T^m+\bar{T}^m)
(\partial_{T^m}h+\partial_{\bar{T}^m}\bar{h})}\over 
{(T^m+\bar{T}^m)\eta_{mn}(T^n+\bar{T}^n)}}
\label{grsch}
\end{equation}
is the Green-Schwarz term \cite{LOU91,DERfkg372,CARo392} describing 
the mixing of the dilaton with the other moduli $T^m$.  
From this expression for $K$, one infers that the true string perturbative  
coupling constant $g_{pert}$ is of the modified form:
\begin{equation}
{{4\pi}\over{g^2_{pert}}}=
{1\over 2}\left(S+\bar{S}+V_{GS}(T^m,\bar{T}^m)\right).  
\label{pertcoup}
\end{equation}
One can prove this conjectured form (\ref{pertent}) of the 
perturbative entropy by solving (\ref{cenqpcon}) with (\ref{hetpert}) 
substituted.  

In the following, as examples of quantum corrections to $N=2$ black holes, 
we find explicit expressions of  
entropy for the axion-free (${\rm Re}\,z^a=0$) solution with  
special forms of the perturbative prepotential.  
We assume that $2Y^0-ip^0=Y^0+\bar{Y}^0\equiv\lambda\neq 0$.  
Note, the symplectic magnetic charge in the perturbative 
basis defined by(\ref{singsimp}) is expressed in terms 
of the charges $p^{\lambda}$ and $q_{\Sigma}$ in the original basis as 
$\hat{q}_{(m)}=(p^0,q_1,p^2,...)$.  
By solving the stability equation $\Pi-\bar{\Pi}=i{\cal Q}$
with the following perturbative prepotential 
\begin{equation}
F(Y)={{d_{abc}Y^aY^bY^c}\over{Y^0}}+ic(Y^0)^2,\ \ \ \ 
Y^{\Lambda}\equiv\bar{Z}X^{\Lambda},
\label{specprep}
\end{equation}
one obtains the following expression for the entropy 
(Cf. See (\ref{newent}).):
\begin{equation}
{S\over{\pi}}=-2(q_0-2c\lambda)\left[\lambda+
{{(p^0)^2}\over{\lambda}}\right].
\label{spechetent}
\end{equation}

Here, $\lambda$ in (\ref{spechetent}) is 
obtained by solving the following equation derived 
from (\ref{cenqpcon}):
\begin{equation}
q_0={{d_{abc}p^ap^bp^c}\over{\lambda^2}}+2c\lambda, \ \ \ \ \ 
q_a=-{{3p^0}\over{\lambda^2}}d_{abc}p^bp^c.
\label{chlamrel}
\end{equation}

For the case $cp^0\neq 0$, one can express $\lambda$ in terms of 
charges as
\begin{equation}
\lambda={{3p^0q_0+p^aq_a}\over{6cp^0}},
\label{cpneq0}
\end{equation}
from which one sees that charges satisfy the following 
constraint when $cp^0=0$:
\begin{equation}
3p^0q_0+p^aq_a=0.
\label{cpeq0}
\end{equation}

For the case $c=0$ and $q_0\neq 0$, $\lambda$ is
\begin{equation}
\lambda=\pm\sqrt{{d_{abc}p^ap^bp^c}\over{q_0}}, 
\label{c0q0neq0}
\end{equation}
with the sign $\pm$ determined by the condition $q_0\lambda<0$ 
that the entropy should be positive.  
As a special case, when $c=p^0=0$ the entropy is
\begin{equation}
{S\over{\pi}}=2\sqrt{q_0d_{abc}p^ap^bp^c},
\label{cp0eq0}
\end{equation}
with the solution having only $n_v+1$ independent charges, i.e. 
$q_a=0$.  In particular, with the cubic prepotential 
$F(Y)=-b{{Y^1Y^2Y^3}\over{Y^0}}+a{{(Y^3)^3}\over{Y^0}}$, the 
entropy (\ref{cp0eq0}) becomes
\begin{equation}
{S\over{\pi}}=2\sqrt{-q_0\left(bp^1p^2p^3-a(p^3)^3\right)}.
\label{cubcp0}
\end{equation}

The explicit solution with non-constant scalar fields is obtained 
by just substituting the symplectic charges by the associated  
harmonic functions in the stabilization equations.  For the case 
where the prepotential is the general cubic prepotential, i.e. 
(\ref{specprep}) with $c=0$, the solution has the form \cite{BEH232}:
\begin{eqnarray}
ds^2&=&-e^{-2U}dt^2+e^{2U}d\vec{x}\cdot d\vec{x}, \ \ \ \ \ 
e^{2U}=\sqrt{H_0d_{abc}H^aH^bH^c}, 
\cr
F^a_{mn}&=&\epsilon_{mnp}\partial_pH^a,\ \ \ 
F_{0\ 0m}=\partial_m\,(H_0)^{-1},\ \ \ 
z^a=iH_0H^ae^{-2U},
\label{gencubsol}
\end{eqnarray}
where the harmonic functions in the solution are
\begin{equation}
H^a=\sqrt{2}\left(h^a+{{p^a}\over r}\right), \ \ \ \ 
H_0=\sqrt{2}\left(h_0+{{q_0}\over r}\right).
\label{cubharm}
\end{equation}
Here, the constants $h$'s are constrained to satisfy the 
asymptotically flat spacetime condition:
\begin{equation}
4h_0{1\over 6}C_{abc}h^ah^bh^c=1,
\label{flatcon}
\end{equation}
and, therefore, the ADM mass of the solution is 
\begin{equation}
M={{q_0}\over{4h_0}}+{1\over 2}p^ah_0C_{abc}h^bh^c.
\label{cubadm}
\end{equation}
When $h$'s take the values at the horizon expressed 
in terms of charges as $h_0=q_0/c$ and $h^a=p^a/c$, the ADM mass 
(\ref{cubadm}) reduces to the entropy ${\cal S}=\pi M^2|_{min}$ 
in (\ref{cp0eq0}).  

As a special case, we consider the following prepotential corresponding 
to the $STU$ model with the a cubic quantum correction:
\begin{equation}
F={{X^1X^2X^3}\over{X^0}}+a{{(X^3)^3}\over{X^0}}.
\label{stuqc}
\end{equation}
For this case, the metric component has the form
\begin{equation}
e^{4U}=4(h_0+{{q_0}\over r})\left((h^1-{{p^1}\over r})
(h^2-{{p^2}\over r})(h^3+{{p^3}\over r})
+a(h^3+{{p^3}\over r})^3\right).
\label{spcubsol}
\end{equation}
Note, the quantum correction term acts as a regulator, smoothing 
out singularity, for example those of massless black holes 
\cite{STR451,GREms451,BEH455,BEH080,CVEy674,KALl52,KALl53,ORT76}, 
of classical solutions.

\subsubsection{Type-II String on Calabi-Yau Manifolds}\label{n2bhqii}

Type-IIA string on a Calabi-Yau three-fold \cite{CANdgp359,HOSkty167}
gives rise to $N=2$ theory with $h_{1,1}+1$ vector fields and 
$h_{1,2}+1$ hypermultiplets ($h_{1,1}$ and $h_{1,2}$ being the Hodge 
numbers of the three-fold), 
with the additional hyper multiplet and vector field coming 
from those associated with the dilaton and the gravity multiplet, 
respectively.  The moduli in the vector multiplets consist of 
the K\"ahler-class moduli $t^a$ ($a=1,...,n_v=h_{1,1}$), where 
$h_{1,1}={\rm dim}\left(H^{1,1}(M,{\bf Z})\right)$. 
The general form of type-IIA prepotential is 
\cite{CANdgp359,HOSkty167}:
\begin{equation}
F^{II}=-{1\over 6}C_{abc}t^at^bt^c-{{\chi\zeta(3)}\over{2(2\pi)^3}}
+{1\over{(2\pi)^3}}\sum_{d_1,...,d_h}n^r_{d_1,...,d_h}
Li_3({\bf e}[i\sum_ad_at^a]),
\label{typeIIprep}
\end{equation}
where $n^r_{d_1,...,d_h}$ are the rational instanton numbers of genus 0 
and $\chi$ is the Euler number.  Here, the prepotential is defined inside 
of the K\"ahler cone $\sigma(K)=\{\sum_a\,t^aJ_a|t^a>0\}$, where 
$J_a$ are the $(1,1)$-forms of the Calabi-Yau 3-fold $M$.   
In the large K\"ahler class moduli limit ($t^a\to\infty$), 
only the classical part in the prepotential, which is related to the 
intersection numbers, survives. 
To the general form (\ref{typeIIprep}) of the prepotential, one can add 
an extra topological term which is determined by the second Chern class 
$c_2$ of the Calabi-Yau 3-fold:
\begin{equation}
F^{II}_{top}=\sum^h_a{{c_2\cdot J_a}\over{24}}t^a; \ \ \ \ 
c_2\cdot J_a \equiv \int_M\,c_2\wedge J_a.
\label{preptop}
\end{equation}
The effect of adding such term to the prepotential is the 
symplectic transformation corresponding to
constant shift of the $\theta$-angle \cite{WIT86} (Cf. See the 
paragraph below (\ref{symchcy2})).  

In particular, the type-IIA model dual to the Heterotic string 
on $K3\times T^2$ with $n_v=4$ and $n=2$ (an $STUV$ model) \cite{CARcl154}
corresponds to the compactification on the Calabi-Yau three-fold 
$P_{1,1,2,6,10}(20)$ \cite{KACv450} with $h_{1,1}=4$ and Euler number 
$\chi=-372$ \cite{BERkkm483}.  The transformations between the 
moduli in the pair of these type-IIA and  heterotic theories are
\begin{equation}
t^1=U-2V,\ \ \ t^2=S-T,\ \ \ 
t^3=T-U,\ \ \ t^4=V, 
\label{tranhetII}
\end{equation}
and some of the instanton numbers \cite{BERkkm483,CARcl154} and the 
Euler number of the three-fold are given in terms of the quantities 
of the heterotic string by:
\begin{equation}
n^r_{l+k,0,k,2l+2k+b}=-2c_2(4kl-b^2),\ \ \ \ \ \ 
\chi=2c_2(0).
\label{numhetII}
\end{equation}
The cubic intersection-number part of the prepotential is
\begin{eqnarray}
-F^{II}_{cubic}&=&t^2\left((t^1)^2+t^1t^3+4t^1t^4+2t^3t^4+3(t^4)^2\right)
\cr
& &+{4\over 3}(t^1)^3+8(t^1)^2t^4+t^1(t^3)^2+2(t^1)^2t^3+8t^1t^3t^4
\cr
& &+2(t^3)^2t^4+12t^1(t^4)^2
+6t^3(t^4)^2+6(t^4)^3.
\label{cubprep}
\end{eqnarray}
In the limit $t^4=V=0$ (i.e. the $STU$ model of the heterotic string 
with $n=2$), the model reduces to the type-IIA string compactified on 
$P_{1,1,2,8,12}(24)$ with $h_{1,1}=3$ and $\chi=-480$.  
The linear topological term, for this case, takes the form:
\begin{equation}
\sum^3_A{{c_2\cdot J_A}\over{24}}t^A=
{{23}\over{6}}t^1+t^2+2t^3=S+T+{{11}\over 6}U.
\label{topstu}
\end{equation}

\paragraph{Entropy Formula}\label{n2bhqiient}

For black holes in type-IIA string on a Calabi-Yau 
3-fold, entropy depends not only on $U(1)$ charges, but also on the 
topological quantities of the 3-fold.  

In the following, we consider the double-extreme black holes 
in the type-IIA superstring on a Calabi-Yau 3-fold with 
the following special form of prepotential:
\begin{equation}
F^{II}(Y)=-{{C_{abc}Y^aY^bY^c}\over{6Y^0}}+{{c_2\cdot J_a}\over{24}}
Y^0Y^a,
\label{speiiaprep}
\end{equation}
where $Y^a\equiv \bar{Z}X^a$.  Note, the prepotential is determined 
by the classical intersection numbers $C_{abc}=-6d_{abc}$ and the 
expansion coefficients $c_2\cdot J_a=24W_{0a}$ of the 2nd Chern 
class $c_2$ of the 3-fold.  

Note, the above form of prepotential can be obtained by imposing 
the symplectic transformation of the following form on the prepotential 
without 2nd Chern class terms:
\begin{equation}
\left(\matrix{\tilde{p}^{\Lambda}\cr \tilde{q}_{\Sigma}}\right)=
\left(\matrix{{\bf 1}&0\cr W&{\bf 1}}\right)
\left(\matrix{p^{\Lambda}\cr q_{\Sigma}}\right).
\label{typiisymp}
\end{equation}
The general expression for the entropy with $p^0=q_a=0$ and $W_{0a}=0$ 
is obtained in (\ref{cp0eq0}).  By imposing the symplectic transformation 
(\ref{typiisymp}), one obtains the following entropy for the type-IIA 
theory with the prepotential (\ref{speiiaprep}) and 
the charge configuration $p^0=q_a=0$:
\begin{equation}
{S\over{\pi}}=2\sqrt{(\tilde{q}_0-W_{0a}\tilde{p}^a)d_{bcd}
\tilde{p}^b\tilde{p}^c\tilde{p}^d}.
\label{typiient}
\end{equation}

\subsubsection{Higher-Dimensional Embedding}\label{n2bhqemb}

The above $D=4$ black holes in string theories with the cubic 
prepotential arise from the compactification of the following 
intersecting $M$-brane solution
\begin{equation}
ds^2_{11}={1\over{({1\over 6}C_{abc}H^aH^bH^c)^{1\over 3}}}\left[
dudv+H_0du^2+{1\over 6}C_{abc}H^aH^bH^cd\vec{x}^2+H^a\omega_a
\right], 
\label{cymbran}
\end{equation}
due to the duality \cite{WIT443} among the heterotic string on 
$K3\times T^2$, the type-II string on a Calabi-Yau 3-fold ($CY$) and 
$M$-theory on $CY\times S^1$ \cite{CADcdf357,FERkm375,FERms474}.  
This solution corresponds to 3 $M\,5$-branes (with the corresponding 
harmonic functions $H^a$, $a=1,2,3$) intersecting over a 3-brane 
(with the spatial coordinates $\vec{x}$), and the momentum (parameterized 
by the harmonic function $H_0$) flowing along the common string.  
Here, the intersection of 4-cycles that each $M\,5$-brane wraps around 
is determined by the parameters $C_{abc}$ and each pair of such 4-cycles 
intersect over the 2-dimensional line element $\omega_a$.  

Compactifying the internal coordinates and the common string direction, 
one obtains the following $D=4$ solution with spacetime 
of the extreme Reissner-Nordstrom black hole:
\begin{equation}
ds^2_4=-{1\over{\sqrt{-{1\over 6}H^0C_{abc}H^aH^bH^c}}}dt^2+
\sqrt{-{1\over 6}H^0C_{abc}H^aH^bH^c}d\vec{x}^2.
\label{cyfdsol}
\end{equation} 

\section{$p$-Branes}\label{pbr}

The purpose of this chapter is to review the recent development in 
$p$-branes and other higher-dimensional configurations in string 
theories.  (See also \cite{GAU011} for another review on this subject.) 
Study of $p$-branes plays an important part in 
understanding non-perturbative aspects of string theories in string 
dualities.  The recently conjectured string dualities require the 
existence of $p$-branes within string spectrum along with well-understood 
perturbative string states.  
Furthermore, microscopic interpretation of entropy, absorption and radiation 
rates of black holes within string theories involves embedding of black 
holes in higher dimensions as intersecting $p$-brane.  

This chapter is organized as follows.  In the first section, we 
summarize properties of single-charged $p$-branes.  In the second 
section, we systematically study multi-charged $p$-branes, which include 
dyonic $p$-branes and intersecting $p$-branes.  In the final section, we 
review the lower-dimensional $p$-branes and their classification.  
We also discuss various $p$-brane embeddings of black holes. 

\subsection{Single-Charged $p$-Branes}\label{pbrsing}

In this section, we discuss $p$-branes which carry one 
type of charge.   Such single-charged $p$-branes are  
basic constituents from which ``bound state'' multi-charged $p$-branes, 
such as dyonic $p$-branes and intersecting $p$-branes, are constructed.    
$p$-branes in $D$ dimensions are defined as $p$-dimensional 
objects which are localized in $D-1-p$ spatial coordinates and 
independent of the other $p$ spatial coordinates, thereby having 
$p$ translational spacelike isometries.   
Note, the allowed values of $(D,p)$ for which supersymmetric 
$p$-branes exist are limited and can be determined by the bose-fermi 
matching condition \cite{ACHetw198}.  (The details are discussed in 
section \ref{bpssusykp}.) 

The effective action for a single-charged $p$-brane has the form:
\begin{equation}
I_D(p)={1\over{2\kappa^2_D}}\int d^Dx\sqrt{-g}
[{\cal R}-{1\over 2}(\partial\phi)^2-{1\over{2(p+2)!}}
e^{-a(p)\phi}F^2_{p+2}],
\label{pbranaction}
\end{equation}
where $\kappa_D$ is the $D$-dimensional gravitational constant, 
$\phi$ is the $D$-dimensional dilaton and 
$F_{p+2}\equiv dA_{p+1}$ is the field strength of $(p+1)$-form 
potential $A_{p+1}$.  Here, the parameter $a(p)$ given below is 
determined by the requirement that the effective action (\ref{pbranaction}) 
and the $\sigma$-model action (\ref{pbransigmact})  
scale in the same way \cite{DUFl416} under the rescaling of fields
\begin{equation}
a^2(p)=4-{{2(p+1)(\tilde{p}+1)}\over{p+\tilde{p}+2}}
=4-{{2(p+1)(\tilde{p}+1)}\over{D-2}}, 
\label{apdef}
\end{equation}
where $\tilde{p}\equiv D-p-4$ corresponds to spatial dimensions of 
the dual brane. 

\subsubsection{Elementary $p$-Branes}\label{bprsingelm}

Electric charge of $(p+1)$-form potential $A_{p+1}$ 
in $I_D(p)$ is carried by the ``elementary'' $p$-brane.  
The ``elementary'' $p$-brane has a $\delta$-function singularity at 
the core, requiring existence of singular electric charge source 
for its support so that equations of motion are satisfied everywhere.   
Namely, the electric charge carried by the ``elementary'' $p$-brane 
is a Noether charge with the Noether current associated with the $p$-brane 
worldvolume $\sigma$-model action:
\begin{eqnarray}
S_p&=&T_p\int d^{p+1}\xi[-{1\over 2}\sqrt{-\gamma}\gamma^{ij}
\partial_iX^M\partial_jX^Ng_{MN}e^{a(p)\phi/(p+1)}
+{{(p-1)}\over 2}\sqrt{-\gamma}
\cr
& &\ \ \ \ -{1\over{(p+1)!}}\varepsilon^{i_1\cdots i_{p+1}}
\partial_{i_1}X^{M_1}\cdots\partial_{i_{p+1}}X^{M_{p+1}}
A_{M_1\cdots M_{p+1}}],
\label{pbransigmact}
\end{eqnarray}
where $X^M(\xi^i)$ ($M=0,1,...,D-1; i=0,1,...,p$) is the spacetime 
trajectory of $p$-branes, $T_p$ is the elementary $p$-brane tension, 
$\gamma_{ij}(\xi)$ [$g_{MN}(X)$] is the worldvolume [spacetime] metric 
and $A_{p+1}=A_{M_1\cdots M_{p+1}}dX^1\wedge\cdots dX^{p+1}$.  
The metric $g_{MN}$ in (\ref{pbransigmact}) is related to the 
canonical metric $g^{can}_{MN}$ in the Einstein-frame effective action 
(\ref{pbranaction}) through the Weyl-rescaling $g_{MN}=e^{a(p)\phi/(p+1)}
g^{can}_{MN}$.  
The ``elementary'' $p$-brane is a solution to the equations of motion 
of the combined action $I_D(p)+S_p$.  In particular, field equation 
and Bianchi identity of $A_{p+1}$ are
\begin{equation}
d\star(e^{-a(p)\phi}F_{p+2})=2\kappa^2_D(-1)^{(p+1)^2}\,\star J_{p+1}, 
\ \ \ \ dF_{p+2}=0,
\label{pformeqns}
\end{equation}
where the electric charge source current $J_{p+1}=J^{M_1\cdots M_{p+1}}
dX_{M_1}\wedge\cdots\wedge dX_{M_{p+1}}$ is
\begin{equation}
J^{M_1\cdots M_{p+1}}(x)=T_p\int d^{p+1}\xi\,\varepsilon^{i_1\cdots i_{p+1}}
\partial_{i_1}X^{M_1}\cdots\partial_{i_{p+1}}X^{M_{p+1}}{{\delta^D(x-X)}
\over\sqrt{-g}}.
\label{pformeleccurrent}
\end{equation}
Here, $\star$ denotes the Hodge-dual operator in $D$ dimensions, 
i.e. $(\star V)^{M_1\cdots M_{D-d}}\equiv{1\over{d!}}
\varepsilon^{M_1\cdots M_D}V_{M_{D-d+1}\cdots M_D}$ with the alternating 
symbol $\varepsilon^{M_1\cdots M_D}$ defined as 
$\varepsilon^{01\cdots D-1}=1$. The Noether electric charge is
\begin{equation}
Q_p=\int_{M^{D-p-1}}\,(\star J)_{D-p-1}=
{1\over{\sqrt{2}\kappa_D}}\int_{S^{\tilde{p}+2}}
e^{-a(p)\phi}\,\star F_{p+2},
\label{pbreleccharg}
\end{equation}
where $S^{\tilde{p}+2}$ surrounds the elementary $p$-brane.  

In solving the Euler-Lagrange equations of the combined action $I_D(p)+S_p$ 
to obtain the elementary $p$-brane solution, one assumes 
the $P_{p+1}\times SO(D-p-1)$ symmetry for the configuration.  Here, 
$P_{p+1}$ is the $(p+1)$-dimensional Poincare group of the $p$-brane 
worldvolume and $SO(D-p-1)$ is the orthogonal group of the transverse space. 
Accordingly, the spacetime coordinates are splitted into 
$x^M=(x^{\mu},y^{m})$, where $\mu=0,...,p$ and $m=p+1,...,D-1$.  
Due to the $P_{p+1}$ invariance, fields are independent of $x^{\mu}$, and  
$SO(D-p-1)$ invariance further requires that this dependence is only 
through $y=\sqrt{\delta_{mn}y^my^n}$.   
In solving the equations, it is convenient to make the following static 
gauge choice for the spacetime bosonic coordinates $X^M$ of the $p$-brane:
\begin{equation}
X^{\mu}=\xi^{\mu}, \ \ \ \ Y^m={\rm constant}, 
\label{statgaugeboscoord}
\end{equation}
where $\mu=0,1,...,p$ [$m=p+1,...,D-1$] corresponds to directions  
internal [transverse] to the $p$-brane.  The general Ansatz for 
metric with $P_{p+1}\times SO(D-p-1)$ symmetry is
\begin{equation}
ds^2=e^{2A(y)}\eta_{\mu\nu}dx^{\mu}dx^{\nu}+e^{2B(y)}\delta_{mn}dy^mdy^n.
\label{pbransptimemet}
\end{equation}
By solving the Euler-Lagrange equations with these Ans\"atze, one 
obtains the following solution for the elementary $p$-brane:
\begin{eqnarray}
ds^2&=&f(y)^{-{{\tilde{p}+1}\over{p+\tilde{p}+2}}}\eta_{\mu\nu}dx^{\mu}
dx^{\nu}+f(y)^{{p+1}\over{p+\tilde{p}+2}}\delta_{mn}dy^mdy^n, 
\cr
e^{\phi}&=&e^{\phi_{0}}f(y)^{-{{a(p)}\over 2}},\ \ \ \ 
A_{\mu_1\cdots\mu_{p+1}}=-{{e^{{{a(p)}\over 2}\phi_0}}\over{^pg}}
\varepsilon_{\mu_1\cdots\mu_{p+1}}f(y)^{-1}, 
\label{pbransolinddim}
\end{eqnarray}
where $^pg$ is the determinant of the metric $g_{\mu\nu}$ and 
$f(y)$ is given by
\begin{equation}
f(y)=\left\{\matrix{1+{{2e^{{{a(p)}\over 2}\phi_0}\kappa^2_DT_p}\over
{(\tilde{p}+1)\Omega_{\tilde{p}+2}}}{1\over{y^{\tilde{p}+1}}}, \ \ \ \ 
\tilde{p}>-1 \cr
1-{{e^{{{a(p)}\over 2}\phi_0}\kappa^2_DT_p}\over \pi}\ln y, \ \ \ \ 
\tilde{p}=-1}\right..
\label{fyforms}
\end{equation}

By solving the Killing spinor equations with the 
$P_{p+1}\times SO(D-p-1)$ symmetric field Ans\"atze, one 
sees that the extreme ``elementary'' $p$-brane  
preserves $1/2$ of supersymmetry with   
the Killing spinors satisfying the constraint: 
\begin{equation}
(1-\bar{\Gamma})\varepsilon=0,
\label{elempbranspinconst}
\end{equation}
where 
\begin{equation}
\bar{\Gamma}\equiv{1\over{(p+1)!\sqrt{-\gamma}}}
\varepsilon^{i_0i_1\cdots i_p}\partial_{i_0}X^{M_0}
\partial_{i_1}X^{M_1}\cdots\partial_{i_p}X^{M_p}
\Gamma_{M_0M_1\cdots M_p},
\label{defbargamma}
\end{equation}
which has properties $\bar{\Gamma}^2=1$ and ${\rm Tr}\,\bar{\Gamma}=0$ that
make ${1\over 2}(1\pm\bar{\Gamma})$ a projection operator.
This originates from the fermionic $\kappa$-symmetry of the super-$p$-brane 
action. The extreme ``elementary'' $p$-brane (\ref{pbransolinddim}) 
saturates the following Bogomol'nyi bound for the mass per unit area 
${\cal M}_p=\int d^{D-p-1}y\,\theta_{00}$:
\begin{equation}
\kappa_D{\cal M}_p\geq {1\over\sqrt{2}}|Q_p|e^{a(p)\phi_0/2},
\label{elepbranbogbd}
\end{equation} 
where $\theta_{MN}$ is the total energy-momentum pseudo-tensor of the 
gravity-matter system. 

\subsubsection{Solitonic $\tilde{p}$-Branes}\label{bprsingsol}

The magnetic charge of $A_{p+1}$ is carried by a solitonic 
$\tilde{p}$-brane, which is topological in nature and free of spacetime 
singularity.  Since magnetic charges can be supported without source at 
the core, solitonic $\tilde{p}$-branes are solutions to 
the Euler-Lagrange equations of the effective action $I_D(p)$ alone.      
The ``topological'' magnetic charge $P_{\tilde{p}}$ of $A_{p+1}$ 
is defined as
\begin{equation}
P_{\tilde{p}}={1\over{\sqrt{2}\kappa_D}}\int_{S^{p+2}}
F_{p+2},
\label{pformmagcharge}
\end{equation}
where $S^{p+2}$ surrounds the solitonic $\tilde{p}$-brane.   
The magnetic charge $P_{\tilde{p}}$ is quantized 
relative to the electric charge $Q_p$ via a Dirac quantization 
condition:
\begin{equation}
{{Q_pP_{\tilde{p}}}\over{4\pi}}={n\over 2}, \ \ \  n\in{\bf Z}.
\label{pbrandiracquant}
\end{equation}

Solitonic $\tilde{p}$-brane solution has the form:
\begin{eqnarray}
ds^2&=&g(y)^{-{{p+1}\over{p+\tilde{p}+2}}}\eta_{\mu\nu}dx^{\mu}dx^{\nu}
+g(y)^{{\tilde{p}+1}\over{p+\tilde{p}+2}}\delta_{mn}dy^mdy^n,
\cr  
e^{\phi}&=&e^{\phi_0}g(y)^{{a(p)}\over 2},\ \ \ \ 
F_{p+2}=\sqrt{2}\kappa_DP_{\tilde{p}}\varepsilon_{p+2}/\Omega_{p+2},
\label{solpbransol}
\end{eqnarray}
where  $\varepsilon_{p+2}$ is the volume form on $S^{p+2}$ and 
$g(y)$ is given by
\begin{equation}
g(y)=1+{{\sqrt{2}e^{-{{2(p+1)}\over{a(p)(p+\tilde{p}+2)}}\phi_0}\kappa_D
P_{\tilde{p}}}\over{(p+1)\Omega_{p+2}}}{1\over{y^{p+1}}}.
\label{gydefintion}
\end{equation}

The ``solitonic'' $\tilde{p}$-brane (\ref{solpbransol}) preserves $1/2$ 
of supersymmetry and saturates the following Bogomol'nyi bound 
for the ADM mass per unit $\tilde{p}$-volume:
\begin{equation}
{\cal M}_{\tilde{p}}\geq {1\over\sqrt{2}}|P_{\tilde{p}}|e^{-a(p)\phi_0/2}.
\label{solpbrnbogbnd}
\end{equation}

Note, the sign difference in dependence of the mass densities 
(\ref{elepbranbogbd}) and (\ref{solpbrnbogbnd}) on  
the dilaton asymptotic value $\phi_0$.  
So, in the limit of large $\phi_0$, the mass density ${\cal M}_p$ 
[${\cal M}_{\tilde{p}}$] is large [small], and vice versa.

\subsubsection{Dual Theory}\label{bprsingdual}

We consider the theory whose actions $\tilde{I}_D(\tilde{p})$  and 
$\tilde{S}_{\tilde{p}}$ are given by (\ref{pbranaction}) and 
(\ref{pbransigmact}) with $p$ replaced by $\tilde{p}=D-p-4$.  
So, in this new theory, $\tilde{p}$-branes carry electric 
charge $\tilde{Q}_{\tilde{p}}$ and $p$-branes carry magnetic charge 
$\tilde{P}_p$ of $(\tilde{p}+1)$-form potential $\tilde{A}_{\tilde{p}+1}$.

Since a $p$-brane from one theory and a $\tilde{p}$-brane from 
the other theory are both ``elementary'' (or ``solitonic''), 
it is natural to assume that these branes are dual pair describing 
the same physics.  
One assumes that the graviton and the dilaton in 
the pair actions are the same, but the field strengths 
$F_{p+2}$ and $\tilde{F}_{\tilde{p}+2}$ are related by 
$\tilde{F}_{\tilde{p}+2}=e^{-a(p)\phi}\,\star F_{p+2}$.  (Thereby, 
the role of field equations and Bianchi identities are interchanged.)   
Then, it follows that since $Q_p=\tilde{P}_p$ and $P_{\tilde{p}}=
\tilde{Q}_{\tilde{p}}$ the Dirac quantization conditions for 
electric/magnetic charge pairs ($Q_p$,$P_{\tilde{p}}$) and 
($\tilde{Q}_{\tilde{p}}$,$\tilde{P}_p$) lead to the following 
quantization condition for the tensions of the dual pair ``elementary'' 
$p$-brane and $\tilde{p}$-brane:
\begin{equation}
\kappa^2_D T_pT_{\tilde{p}}=|n|\pi.
\label{tensionquant}
\end{equation}

By performing the Weyl-rescaling of metrics to the string-frame, 
one sees that the $p$-brane [$\tilde{p}$-brane] loop counting 
parameter $g_p$ [$g_{\tilde{p}}$] is
\begin{equation}
g_p=e^{{{(D-2)a(p)}\over{4(p+1)}}\phi_0}, \ \ \ \ \ \ \ 
g_{\tilde{p}}=e^{{{(D-2)a(\tilde{p})}\over{4(\tilde{p}+1)}}\phi_0}, 
\label{pbranloopcount}
\end{equation}
with $a(\tilde{p})=-a(p)$.  It follows that the brane 
loop counting parameters of the dual pair are related by 
\begin{equation}
(g_p)^{p+1}=1/(g_{\tilde{p}})^{\tilde{p}+1}.
\label{looppararel}
\end{equation}
Thus, the strongly [weakly] coupled $p$-branes are  
the weakly [strongly] coupled $\tilde{p}$-branes.  
In particular, a string ($p=1$) in $D=6$ are dual to another string 
($\tilde{p}=6-1-4=1$), thereby strongly [weakly] coupled string theory 
being equivalent to weakly [strongly] coupled dual string theory.  
Other interesting examples are membrane/fivebrane 
dual pair in $D=11$, string/fivebrane dual pair in $D=10$,  
self-dual 3-branes in $D=10$ and self-dual 0-branes in $D=4$.  
Note, in $D=11$ supergravity strong and weak coupling 
limits do not have meaning due to absence of the dilaton.

\subsubsection{Blackbranes}\label{bprsingblk}

We discuss non-extreme generalization of BPS ``solitonic'' 
$\tilde{p}$-brane in section \ref{bprsingsol}.  Such solution is obtained 
\cite{LU313} by solving the Euler-Lagrange equations following from 
$I_D(p)$ with ${\bf R}\times SO(p+3)\times E(\tilde{p})$ symmetric 
field Ans\"atze.  The non-extreme $\tilde{p}$-brane solution is
\begin{eqnarray}
ds^2&=&-\Delta_+\Delta^{-(\tilde{p}+1)/(p+\tilde{p}+2)}_-dt^2+
\Delta^{-1}_+\Delta^{{{a(p)^2}\over{2(p+1)}}-1}_-dr^2
\cr
& &+r^2\Delta^{{a(p)^2}\over{2(p+1)}}_-d\Omega^2_{p+2}+
\Delta^{{p+1}\over{p+\tilde{p}+2}}_-dx^idx_i,
\cr
e^{-2\phi}&=&\Delta^{a(p)}_-,\ \ \ \ \ 
F_{p+2}=(p+1)(r_+r_-)^{{p+1}\over 2}\varepsilon_{p+2},
\label{backpbransol}
\end{eqnarray}
where $\Delta_{\pm}\equiv 1-({r_{\pm}\over r})^{p+1}$ and $i=1,...,\tilde{p}$. 
The magnetic charge $P_{\tilde{p}}$  and the ADM mass per unit 
$\tilde{p}$-volume ${\cal M}_{\tilde{p}}$ of (\ref{backpbransol}) are
\begin{equation}
P_{\tilde{p}}={\Omega_{p+2}\over{\sqrt{2}\kappa_D}}(p+1)
(r_+r_-)^{{p+1}\over 2},\ \ \ \ 
{\cal M}_{\tilde{p}}={{\Omega_{p+2}}\over{2\kappa^2_D}}
[(p+2)r^{p+1}_+-r^{p+1}_-].
\label{magchmassblackp}
\end{equation}
The solution (\ref{backpbransol}) has the event horizon at $r=r_+$ and 
the inner horizon at $r=r_-$, and therefore is alternatively called 
``black'' $\tilde{p}$-brane.  The requirement of the regular event horizon, 
i.e. $r_+\geq r_-$, leads to the Bogomol'nyi bound $\sqrt{2}
{\cal M}_{\tilde{p}}\geq |P_{\tilde{p}}|e^{-a(p)\phi_0/2}$.  

In the limit $r_-=0$, $\phi$ and $A_{p+1}$ become trivial, i.e. 
$P_{\tilde{p}}=0$, and spacetime reduces to the product of 
$(D-\tilde{p})$-dimensional Schwarzschield spacetime and flat 
${\bf R}^{\tilde{p}}$.  In the extreme limit ($r_+=r_-$), the 
symmetry is enhanced to that of the BPS $\tilde{p}$-brane, i.e. 
$P_{\tilde{p}+1}\times SO(p+3)$, since $g_{tt}=g_{x_ix_i}$.  
Such extreme solution is related to the BPS solution (\ref{solpbransol}) 
through the change of variable $y^{p+1}=r^{p+1}-r^{p+1}_-$.  
In the extreme limit, the event-horizon and the singularity completely 
disappear, i.e. becomes soliton with geodesically complete 
spacetime in the region $r>r_+=r_-$.

\subsection{Multi-Charged $p$-Branes}\label{bprmlt}

In this section, we discuss $p$-branes carrying more than one 
types of charges.  These $p$-branes are ``bound states'' 
of single-charged $p$-branes.  Multi-charged $p$-branes are classified 
into two categories, namely ``marginal'' and ``non-marginal'' configurations.  
The ``marginal'' (BPS) bound states have zero binding energy and, therefore, 
the mass density $M$ is sum of the charge densities $Q_i$ of 
the constituent $p$-branes, i.e. $M=\sum_iQ_i$.   Such bound states 
with $n$ constituent $p$-branes preserve at least $({1\over 2})^n$ of 
supersymmetry.  The marginal bound states include intersecting and 
overlapping $p$-branes.  
The ``non-marginal'' (BPS) bound states have non-zero binding energy 
and the mass density of the form $M^2=\sum_iQ^2_i$.  The quantized charges 
$Q_i$ of ``non-marginal'' bound states take relatively prime integer values.  
In general, non-marginal $p$-brane bound states are obtained from 
single-charged $p$-branes or marginal $p$-brane bound states by applying 
the $SL(2,{\bf Z})$ electric/magnetic duality transformations and, 
therefore, preserve the same amount of supersymmetries as the initial 
$p$-brane configurations (before the $SL(2,{\bf Z})$ 
transformations).  In particular, intersecting $p$-branes are further 
categorized into orthogonally intersecting $p$-branes and $p$-branes 
intersecting at angles.  

\subsubsection{Dyonic $p$-Branes}\label{bprmltdyn}

In $D=2p+4$, i.e. dimensions for which $p=\tilde{p}$, $p$-branes can 
carry both electric and magnetic charges of $A_{p+1}$.  
Examples are dyonic black holes ($p=0$) in $D=4$; 
dyonic strings ($p=1$) in $D=6$; dyonic membranes ($p=2$) in $D=8$.   
Such dyonic $p$-branes can be constructed by applying the $D=2(p+2)$ 
$SL(2,{\bf Z})_{EM}$ electric/magnetic duality transformations on 
single-charged $p$-branes.  
These dyonic $p$-branes have $P_{p+1}\times SO(D-p-1)$ symmetry and 
are characterized by {\it one} harmonic function, just like single-charged 
$p$-branes, since the $SL(2,{\bf Z})$ transformations leave the 
Einstein-frame metric intact.  
In particular, in $D=2$ mod 4, the $(p+2)$-form field strengths satisfy 
a real self-duality condition $F_{p+2}=-\star F_{p+2}$ and, thereby, 
electric and magnetic charges are identified, i.e. $Q_p=-P_p$.  

The $SL(2,{\bf Z})_{EM}$ electric/magnetic duality transformations  
of $2k$-form field strength $F_{2k}$ in $D=4k$ can generally be 
understood as the $T^2$ moduli transformations 
of $D=(4k+2)$ theory compactified on $T^2$ \cite{GRElpt384}.  
We consider the following $D=(4k+2)$ action
\begin{equation}
I_{(4k+2)}=\int d^{4k+2}x\sqrt{-g}{\cal R}+
\alpha\int[dC\wedge H+{1\over 2}H\wedge\,\star H],
\label{4kpl2act}
\end{equation}
where $C$ is a $(4k+2)$-form potential with the field strength 
$H=dC$.   We compactify the action (\ref{4kpl2act}) on $T^2$ 
with the following Ans\"atze for the fields
\begin{eqnarray}
ds^2(M_{4k+2})&=&ds^2(M_{4k})+{1\over{{\rm Im}\tau}}
(|\tau|^2dy^2+2{\rm Re}\tau\,dxdy+dx^2),
\cr
C&=&Bdy+Adx,\ \ \ \ \ \ H=Gy+Fdx,
\label{t2compansats}
\end{eqnarray}
where $\tau$ is the moduli parameter of $T^2$, 
$(x,y)$ are coordinates of $T^2$ (i.e. $x\sim x+1$ and $y\sim 
y+1$), and $A$, $B$ [$F$, $G$] are $(2k-1)$-forms [$2k$-forms] 
in $D=4k$.  By applying the self-duality condition 
\cite{VER455} of $H$, one finds that $2k$-form $G$ is an auxiliary 
field, which can be eliminated by its field equation as 
$G={\rm Im}\tau\,\star F-{\rm Re}\tau\,F$, and one finds that $F=dA$.  

The final expression for the $D=4k$ action is
\begin{equation}
I_{4k}=\int d^{4k}x\sqrt{-g}\left[{\rm R}-{1\over 2}
{{d\tau d\bar{\tau}}\over{({\rm Im}\tau)^2}}\right]
+\alpha\int F\wedge G,
\label{4kdimsl2act}
\end{equation}
where a real constant $\alpha$ is, in the convention of 
\cite{IZQlpt}, given by $\alpha=2[(2k)!]^{-1}$.  
The complex scalar $\tau$, which is expressed in terms of 
real scalars $\rho$ and $\sigma$ as $\tau=2\rho+ie^{-2\sigma}$, 
parameterizes the target space manifold ${\cal M}=
SL(2,{\bf Z})\backslash SL(2,{\bf R})/U(1)$, 
which is the fundamental domain of $SL(2,{\bf Z})$ in the upper 
half $\tau$ complex plane.  Here, $U(1)$ is a subgroup of 
$SL(2,{\rm R})$ which preserves the vacuum expectation value 
$\langle\tau\rangle$.  The field equations are invariant under 
the following $SL(2,{\bf R})_{EM}$ electric/magnetic duality 
transformation of $F$:
\begin{equation}
(F,G)\to (F,G)A^{-1},\ \ \ 
\tau\to{{a\tau+b}\over{c\tau+d}}; \ \ \ \ 
A=\left(\matrix{a&b\cr c&d}\right)\in SL(2,{\bf R}).  
\label{slemdualtran}
\end{equation}
Note, this $SL(2,{\bf Z})_{EM}$ transformation is not 
$S$-duality of string theories, since $\sigma$ 
is not the $D=4k$ dilaton.  (In (\ref{4kpl2act}), the dilaton 
is set to zero.)  The following electric $Q$ and magnetic $P$ charge 
densities form an $SL(2,{\bf R})_{EM}$ doublet:
\begin{equation}
Q={1\over{\Omega_{2k}}}\oint G,\ \ \ \ \ \ 
P={1\over{\Omega_{2k}}}\oint F.
\label{2kformemcharge}
\end{equation}

This $SL(2,{\bf R})_{EM}$ transformation on single-charged 
$(2k-2)$-branes yields $(2k-2)$-branes which 
carry both electric and magnetic charges of $A_{2k-1}$.  
Such dyonic $p$-branes preserve $1/2$ of supersymmetry.  
Charges $(Q,P)$ and $(Q^{\prime},P^{\prime})$ of two dyonic $(2k-2)$-branes 
satisfy the generalized Nepomechie-Teitelboim quantization condition 
\cite{NEP31,TEI167}:
\begin{equation}
QP^{\prime}-Q^{\prime}P\in{\bf Z}.
\label{dypbrnquant}
\end{equation}

When such dyonic $(2k-2)$-branes are uplifted to $D=(4k+2)$  
through (\ref{t2compansats}), the solutions become self-dual 
$(2k-1)$-branes \cite{DUFl273,DUFfkl356}.  The electric and 
magnetic charges of the dyonic $(2k-2)$-branes are interpreted 
as winding numbers of the self-dual $(2k-1)$-branes around the $x$ and 
$y$ directions of $T^2$.   The dyonic $(2k-2)$-branes ($k=1,2$) uplifted 
to $D=11$ describe $p$-brane which interpolates between the 
$M\,2$-brane and the $M\,5$-brane, i.e. a membrane within a 
5-brane $(2|2_M,5_M)$.  

In the following, we specifically discuss the $k=2$ case 
\cite{IZQlpt,GRElpt384}.  The associated action ((\ref{4kdimsl2act}) 
with $k=2$) is type-IIB effective action (\ref{striibact}) 
consistently truncated and compactified to $D=8$.  
The $N=2$, $D=8$ supergravity has an $SL(3,{\bf R})\times SL(2,{\bf R})$ 
on-shell symmetry, whose $SL(3,{\bf Z})\times SL(2,{\bf Z})$ subset 
is the conjectured $U$-duality symmetry of $D=8$ type-II string.  
The R-R 4-form field strength and its dual field strength 
transform as $({\bf 1},{\bf 2})$ under $SL(3,{\bf R})\times SL(2,{\bf R})$.  
This $U$-duality group contains as a subset the $SO(2,2,{\bf Z})\equiv
[SL(2,{\bf Z})\times SL(2,{\bf Z})]/{\bf Z}_2$ $T$-duality group.  
The $SL(2,{\bf Z})$ factor (in $SL(3,{\bf Z})\times 
SL(2,{\bf Z})$) is the electric/magnetic duality 
(\ref{slemdualtran}), which is a subset of ``perturbative'' 
$T$-duality group.  All the ``non-perturbative'' transformations are 
contained in the $SL(3,{\bf Z})$ factor.  

The following dyonic membrane solution with $\langle\tau\rangle=i$ is 
obtained by applying the $U(1)\subset SL(2,{\bf R})_{EM}$ transformation 
with a parameter $\xi$ to purely magnetic membrane:
\begin{eqnarray}
ds^2_{(8)}&=&H^{-{1\over 2}}ds^2({\bf M}^3)+H^{1\over 2}ds^2({\bf E}^5), 
\cr
F_4&=&{1\over 2}\cos\xi\,(\,\star dH)+{1\over 2}\sin\xi\,dH^{-1}\wedge
\epsilon({\bf M}^3),
\cr
\tau&=&{{\sin(2\xi)(1-H)+2iH^{1\over 2}}\over{2(\sin^2\xi+H\cos^2\xi)}},
\label{8ddyonmebran} 
\end{eqnarray}
where $ds^2({\bf M}^3)$ [$ds^2({\bf E}^5)$] is the metric of 
$D=3$ Minkowski space ${\bf M}^3$ [the 5-dimensional Euclidean 
space ${\bf E}^5$], $\epsilon({\bf M}^3)$ is the volume form of 
${\bf M}^3$, $\star$ is the Hodge-dual operator in ${\bf E}^5$ and 
$H$ is a harmonic function given by (\ref{gydefintion}) with $p=2$.  
Here, $U(1)$ is the subgroup of $SL(2,{\bf R})_{EM}$ that preserves 
$\langle\tau\rangle=i$.  In the quantum theory, the $U(1)$ group  
breaks down to ${\bf Z}_2$ due to Dirac quantization condition, resulting 
in either electric or magnetic solutions when applied 
to purely magnetic solution.  (But the solution in 
(\ref{8ddyonmebran}) with an arbitrary $\xi$ satisfies the 
Euler-Lagrange equations following from (\ref{4kdimsl2act}) and, 
therefore, can be taken as an initial solution to which the 
``integer-valued'' duality transformations are applied.)  

To generate dyonic solutions with an arbitrary 
$\langle\tau\rangle$ and are relevant to the quantum theory, 
one has to apply the full $SL(2,{\bf R})_{EM}$ transformation to 
(\ref{8ddyonmebran}).  
The pair of electric and magnetic charges of such dyonic solutions 
take co-prime integer values.  The ADM mass density of this extreme 
dyonic membrane saturates the Bogomol'nyi bound:
\begin{equation}
M^2\geq {1\over 4}\left[e^{2\langle\sigma\rangle}
(Q+2\langle\rho\rangle P)^2+e^{-2\langle\sigma\rangle}P^2
\right],
\label{8dimdymembog}
\end{equation}
and therefore $1/2$ of supersymmetry is preserved.  

When uplifted to $D=10$, this dyonic membrane becomes 
the following self-dual 3-brane of type-IIB theory \cite{DUFl273}:
\begin{eqnarray}
ds^2_{(10)}&=&H^{-{1\over 2}}[ds^2({\bf M}^3)+dv^2]+H^{1\over 2}
[ds^2({\bf E}^5)+du^2],
\cr
F_5&=&{1\over 2}(\,\star dH)\wedge du+{1\over 2}dH^{-1}\wedge
\epsilon({\bf M}^3)\wedge dv,
\label{slfdl3brn}
\end{eqnarray}
where the coordinates $(u,v)$ are related to the coordinates $(x,y)$ 
in (\ref{t2compansats}) through $(y,x)=(v,u)A^{-1}$ ($A\in 
SL(2,{\bf Z})$).  The electric and magnetic charges of the above 
$D=8$ dyonic membrane are respectively interpreted as the 
winding numbers of this $D=10$ self-dual 3-brane around 
$x$ and $y$ directions of $T^2$.  

The $D=8$ dyonic membrane (\ref{8ddyonmebran}) uplifted to 
$D=11$ is a special case of orthogonally intersecting $M$-brane 
interpreted as a membrane within a 5-brane $(2|2_M,5_M)$:
\begin{eqnarray}
ds^2_{(11)}&=&H^{1\over 3}(\sin^2\xi+H\cos^2\xi)^{1\over 3}
\left[H^{-1}ds^2({\bf M}^3)\right.
\cr
& &+\left.(\sin^2\xi+H\cos^2\xi)^{-1}
ds({\bf E}^3)+ds^2({\bf E}^5)\right], 
\cr
F^{(11)}_4&=&{1\over 2}\cos\xi\,\star dH+{1\over 2}\sin\xi\,dH^{-1}\wedge
\epsilon({\bf M}^3)
\cr
& &+{{3\sin 2\xi}\over{2[\sin^2\xi+H\cos^2\xi]^2}}dH\wedge 
\epsilon({\bf E}^3).
\label{11dsolof8dmem}
\end{eqnarray}
This $M$-brane bound state interpolates
\footnote{The above procedure can be applied to intersecting $p$-branes 
to generate $p$-brane bound states which interpolate between 
different intersecting $p$-branes, e.g. the interpolation between the 
intersecting two $(p+2)$-branes and the intersecting two $p$-branes; the 
interpolation between the intersecting $p$-brane and $(p+2)$-brane and the 
intersecting $(p+2)$-brane and $p$-brane \cite{COSc204}.} 
between the $M\,5$-brane and the $M\,2$-brane as $\xi$ is varied from 
$0$ to $\pi\over 2$.  As long as magnetic charge is non-zero 
($\xi\neq 0$), (\ref{11dsolof8dmem}) is non-singular, 
thereby singularity of the $D=8$ dyonic membrane  
(\ref{8ddyonmebran}) is resolved by its interpretation in $D=11$.  
By compactifying an extra spatial isometry direction of (\ref{11dsolof8dmem})
on $S^1$, one obtains 3 different types of dyonic $p$-branes in type-IIA 
theory: ($i$) a membrane within a 4-brane $(2|2,4)$, ($ii$) 
a membrane within a 5-brane $(2|2,5)$ and ($iii$) a string within a 
4-brane $(1|1,4)$.  

One can construct dyonic $p$-branes in $D\neq 2(p+2)$ by 
compactifying purely electric or purely magnetic $p$-branes  
down to $D=2(p+2)$ (or $D=2(\tilde{p}+2)$), 
applying electric/magnetic duality transformations of $p$-branes 
(or $\tilde{p}$-branes) in $D=2(p+2)$ (or $D=2(\tilde{p}+2)$), 
and then uplifting the dyonic solution to the original dimensions.   
In particular, the ${\bf Z}_2$ subset of the $D=2(p+2)$ 
(or $D=2(\tilde{p}+2)$) electric/magnetic duality 
transformations relates electric $p$-brane 
and magnetic $p$-brane.  
Thus, the elementary $p$-brane in $D\neq 2(p+2)$ is interpreted
as the electrically charged partner of magnetic $p$-brane,  
establishing electric/magnetic duality between electric 
``elementary'' $p$-brane and magnetic ``solitonic'' $p$-brane 
in $D\neq 2(p+2)$.  
To enlarge this ${\bf Z}_2$ electric/magnetic duality 
symmetry in $D=2(p+2)$ (or $D=2(\tilde{p}+2)$) to the 
$SL(2,{\bf Z})$ symmetry so that one can generate dyonic $p$-branes 
from single-charged $p$-branes, one has to turn on a pseudo-scalar field.  
For example, the dyonic 5-brane in $D=10$ type-IIB theory is constructed 
in \cite{BERbo53} by applying the $SL(2,{\bf Z})_{IIB}\times 
SL(2,{\bf Z})_{EM}$ transformation of the truncated type-IIB theory in 
$D=6$.  (This $SO(2,2)\equiv SL(2,{\bf Z})_{IIB}\times SL(2,{\bf Z})_{EM}$ 
group is a subgroup of $SO(5,5)$ $U$-duality group of type-II string 
on $T^4$.)  There, it is found out that non-zero R-R fields (which are 
related to the pseudo-scalar field) are needed for the solution 
to have both electric and magnetic charges.  

The type-IIB $SL(2,{\bf Z})_{IIB}$ $S$-duality transformation 
leads to dyonic $p$-brane whose electric and magnetic charges 
coming from different sectors (NS-NS/R-R) of string theory.  
General dyonic solutions where form fields carry both electric and 
magnetic charges are generated by additionally applying the 
$SL(2,{\bf Z})_{EM}$ electric/magnetic transformation in $D=6$ 
\cite{BERbo53}.  In particular, the electric/magnetic duality 
transformation that relates the solitonic 5-brane and elementary 
5-brane is the product of $({\bf Z}_2)_{IIB}$ and $({\bf Z}_2)_{EM}$ 
transformations.  Such dyonic 5-brane solutions preserve $1/2$ of 
supersymmetry.  

We comment on generalization of dyonic 
$p$-branes discussed in this section.  In \cite{DUFfkl356}, general 
dyonic $p$-branes within consistently truncated heterotic 
string on $T^6$, where truncated moduli fields are parameterized by 3 
complex modulus parameters $T^{(i)}$ ($i=1,2,3$) of 3 $T^2$ in $T^6$, is 
constructed.  Thereby, the $O(6,22,{\bf Z})$ $T$-duality symmetry of 
heterotic string on $T^6$ is broken down to $SL(2,{\bf Z})^3$.  The general 
class of multi-charged $p$-brane solution is then characterized by 
harmonic functions each associated with $T^{(i)}$ and the dilaton-axion 
scalar $S$.  Such solutions break more than $1/2$ of supersymmetry.  
With trivial $S$, $({1\over 2})^n$ of supersymmetry is preserved for 
$n$ non-trivial $T^{(i)}$.  With non-trivial $S$, additional $1/2$ of 
supersymmetries is broken unless all of $T^{(i)}$ are non-trivial.  
With all $T^{(i)}$ non-trivial, $({1\over 2})^3$ or none of supersymmetry is 
preserved, depending on the chirality choices.  In particular, a special 
case of general class of solution with $S$ and only one of $T^{(i)}$  
non-trivial corresponds to generalization of $D=6$ self-dual 
dyonic string, when such solution is uplifted to $D=6$.  
This dyonic solution is parameterized by 2 harmonic functions, which 
are respectively associated with electric and magnetic charges of 
2-form potential.  In the self-dual limit, i.e. when the electric 
and magnetic charges are equal, the solution becomes the $D=6$  
self-dual dyonic string.  Within the context of non self-dual theory, 
the solution preserve only $1/4$ of supersymmetry, whereas as a solution 
of self-dual theory it preserve $1/2$ of supersymmetry.  

\subsubsection{Intersecting $p$-Branes}\label{bprmltint}

\paragraph{Constituent $p$-Branes}\label{bprmltintcon} 

Before we discuss intersecting $p$-branes, we summarize various 
single-charged $p$-branes in $D=10,11$, which 
are constituents of intersecting and overlapping $p$-branes.  
These $p$-branes are special cases of ``elementary'' $p$-branes and 
``solitonic'' $\tilde{p}$-branes discussed in section \ref{pbrsing}.
They are characterized by a harmonic function $H(y)$ in the 
transverse space (with coordinates $y_{p+1},...,y_{D-1}$) 
and break $1/2$ of supersymmetry.

$D=11$ supergravity has 3-form potential.  So, 
the basic $p$-branes (called $M\,p$-branes) are an electric 
``elementary'' membrane and a magnetic ``solitonic'' fivebrane:
\begin{eqnarray}
ds^2&=&H^{(p+1)/9}_p[H^{-1}_p(-dt^2+dx^2_1+\cdots dx^2_p)+
(dy^2_{p+1}+\cdots+dy^2_{10})],
\cr
{\cal F}_{tx_1x_2y_{\alpha}}&=&-{c\over 2}\partial_{y_{\alpha}}H^{-1}_p, 
\ \ \ \ \ \ \ \alpha=3,...,10\ \ \ \ \ ({\rm for}\ p=2),
\cr
{\cal F}_{y_{\alpha_1}\cdots y_{\alpha_4}}&=&{c\over 2}
\epsilon_{\alpha_1\cdots\alpha_5}\partial_{y_{\alpha_5}}H_p,\ \ \ 
\alpha_i=6,...,10\ \ \ \ \ ({\rm for}\ p=5),
\label{singlembrane}
\end{eqnarray}
where $c=1 (-1)$ for (anti-) branes and harmonic function $H_p=
1+{{c_pQ_p}\over{|\vec{y}-\vec{y}_0|^{8-p}}}$ is for $M\,p$-brane 
located at the $\{0,1,...,p\}$ hyperplane at $y^i=y^i_0$.    
The Killing spinor $\epsilon$ of these $M\,p$-branes satisfies the 
following constraint:
\begin{equation}
\Gamma_{01\cdots p}\epsilon=c\epsilon, 
\label{spinconstsingmbr}
\end{equation}
where $\Gamma_{01\cdots p}\equiv\Gamma_0\Gamma_1\cdots\Gamma_p$ is 
the product of flat spacetime gamma matrices associated with 
the worldvolume directions.

In $D=10$, there are 3 types of $p$-branes, depending 
on types of charges that $p$-branes carry.  
The electric charge of NS-NS 2-form potential is carried by  
NS-NS strings (or fundamental strings):
\begin{equation}
ds^2=H^{-1}(-dt^2+dx^2)+dy^2_2+\cdots+dy^2_9,\ \ \ \ 
e^{2\phi}=H^{-1}.
\label{fundstringsol}
\end{equation}
The magnetic charge of NS-NS 2-form potential is carried by  
NS-NS 5-branes (or solitons):
\begin{equation}
ds^2=-dt^2+dx^2_1+\cdots+dx^2_5+H(dy^2_6+\cdots+dy^2_9),\ \ \ 
e^{2\phi}=H.
\label{solifivbrsol}
\end{equation}
The charges of R-R $(p+1)$-form potentials are carried by R-R 
$p$-branes:
\begin{eqnarray}
ds^2&=&H^{-1/2}(-dt^2+dx^2_1+\cdots+dx^2_p)+H^{1/2}(dy^2_{p+1}+\cdots+
dy^2_9),
\cr 
e^{2\phi}&=&H^{-{{p-3}\over 2}},\ \ \ \ \ \ \ \ 
{\cal F}_{tx_1\cdots x_py_i}=\partial_{y_i}H^{-1},
\label{pbransol}
\end{eqnarray}
where $p=0,2,4,6$ [$p=1,3,5,7$] for R-R $p$-branes in type-IIA [type-IIB] 
theory.  These R-R $p$-branes of the effective field theory are long 
distance limit of $D\,p$-branes in type-II superstring 
theories \cite{DAIlp}: the transverse [longitudinal] directions of 
R-R $p$-branes correspond to coordinates with Dirichlet [Neumann] boundary 
conditions.  

The Killing spinors of $D=10$ $p$-branes satisfy one 
constraint.  The left-moving and the right-moving Majorana-Weyl spinors 
$\epsilon_L$ and $\epsilon_R$ have the same [opposite] chirality for 
the type-IIB [type-IIA] theory, i.e. $\Gamma_{10}\epsilon_{L,R}=
\epsilon_{L,R}$ [$\Gamma_{10}\epsilon_L=\epsilon_L$ and 
$\Gamma_{10}\epsilon_R=-\epsilon_R$].  So, spinor constraints are 
different for type-IIA/B theories:
\begin{itemize}
\item NS-NS strings:  
$\epsilon_L=\Gamma_{01}\epsilon_L$,  $\epsilon_R=-\Gamma_{01}\epsilon_R$
\item IIA NS-NS fivebranes:  
$\epsilon_L=\Gamma_{01\cdots 5}\epsilon_L$, 
$\epsilon_R=\Gamma_{01\cdots 5}\epsilon_R$
\item IIB NS-NS fivebranes:  
$\epsilon_L=\Gamma_{01\cdots 5}\epsilon_L$, 
$\epsilon_R=-\Gamma_{01\cdots 5}\epsilon_R$
\item R-R $p$-branes:  
$\epsilon_L=\Gamma_{01\cdots p}\epsilon_R$.
\end{itemize}

Whereas in type-II superstring theories there are $D\,p$-branes with 
$p=-1,0,...,9$, R-R $p$-branes in {\it massless} effective field 
theories cover only range $p\leq 6$.  So, there is no place in 
the {\it massless} type-II supergravities for R-R 7- and 8-branes
\footnote{R-R 9-brane of type-IIB theory is interpreted as $D=10$ 
spacetime.}.  
Although the R-R 7-brane in type-IIB theory can be related to 
8-form dual of pseudo-scalar of type-IIB theory \cite{GIBgp370}, 
it cannot be $T$-dualized to R-R $p$-branes in type-IIA theory 
since it is specific to the {\it uncompactified} type IIB theory, only.  
In \cite{BERdgpt470}, it is proposed that $D\,p$-branes of type-II 
superstring theories with $p>6$ can be realized as R-R $p$-branes 
of {\it massive} type-II supergravity theories.  In the following, 
we discuss R-R 7- and 8-branes in some detail, since their properties 
and $T$-duality transformation rules are different from other R-R 
$p$-branes.  

The R-R 8-brane is coupled to 9-form potential.  The introduction of 
10-form field strength into the theory does not increase the bosonic 
degrees of freedom (therefore, is not ruled out by supersymmetry 
consideration), but leads to non-zero cosmological constant.  In fact, 
the {\it massive} type-IIA supergravity constructed in \cite{ROM169} 
contains such cosmological constant term.  It is argued 
\cite{POL75} that the existence of the massive type-IIA supergravity with 
the cosmological constant is related to the existence of the 9-form potential 
of type-IIA theory.  In \cite{BERdgpt470}, new {\it massive} type-IIA 
supergravity is formulated by introducing 9-form potential $A_9$ whose 
10-form field strength $F_{10}=10dA_9$ is interpreted as the cosmological 
constant once Hodge-dualized.  (Note, the cosmological constant $m$ in 
the massive type-IIA supergravity is independent of the dilaton in the 
string-frame, which is typical for terms in R-R sector.)  
In this new formulation, the cosmological constant $m$ is promoted to 
a field $M(x)$ by introducing $A_9$ as a Lagrange multiplier for the 
constraint $dM=0$ that $M(x)=m$ is a constant: the field equation for 
$M(x)$ simply determines the new field strength $F_{10}$, while the 
field equation for $A_9$ implies that $M(x)=m$.  It is conjectured in 
\cite{COWlpst} that the massive type-IIA supergravity is related to 
hypothetical $H$-theory \cite{BAR55} in $D=13$ through the Scherk-Schwarz 
type dimensional reduction (see below for the detailed discussion), 
rather than to $M$-theory.  
The conjectured type-IIB $Sl(2,{\bf Z})$ duality requires the pseudo-scalar 
$\chi$ to be periodically identified ($\chi\sim\chi+1$), which 
together with $T$-duality between massive type-IIA and type-IIB 
supergravities implies that the cosmological constant $m$ is quantized 
in unit of the radius of type-IIB compactification circle, i.e. 
$m={n\over{R_B}}$ ($n\in{\bf Z}$).  

The massive type-IIA supergravity admits the following 8-brane 
(or domain wall) as a natural ground state solution:
\begin{equation}
ds^2=H^{-{1\over 2}}\eta_{\mu\nu}dx^{\mu}dx^{\nu}+H^{1\over 2}dy^2,\ \ \ \ \ 
e^{-4\phi}=H^5,
\label{8branesoliia}
\end{equation}
and the Killing spinor $\epsilon$ satisfies one constraint 
$\bar{\Gamma}_y\epsilon=\pm\epsilon$.  The form of harmonic function $H(y)$, 
which is linear in $y$, depends on $M(x)$.  When $M(x)=m$ {\it everywhere}, 
$H=m|y-y_0|$, with a kink singularity at $y=y_0$.  When $M(x)$ is 
{\it locally} constant, the corresponding solution is interpreted as 
a domain wall separating regions with different values of $M$ (or 
cosmological constant).  An example is $H=-Q_{-}y+b$ [$H=Q_{+}y+b$] for 
$y<0$ [$y>0$], where  8-brane charges $Q_{\pm}$ are defined as the values 
of $M$ as $y\to\pm\infty$ and $b$ is related to string coupling constant 
$e^{\phi}$ at the 8-brane core.  
This 8-brane solution is asymptotically left-flat [right-flat] 
when $Q_{-}=0$ [$Q_{+}=0$].  The multi 8-brane generalization can be 
accomplished by allowing kink singularities of $H$ at ordered points 
$y=y_0<y_1<\cdots <y_n$.   

The R-R 8-brane of {\it massive} type-IIA supergravity can be interpreted 
as the KK 6-brane of $D=11$ supergravity \cite{BERdgpt470}.  
Namely, after the R-R 8-brane is compactified to 6-brane in $D=8$, the 
6-brane can be lifted as the R-R 6-brane of {\it massless} type-IIA 
supergravity, which is interpreted as the KK monopole of 
$M$-theory on $S^1$ \cite{TOW350}.  

Under the $T$-duality, the R-R 8-brane is expected to transform 
to R-R 7-brane or 9-brane in type-IIB theory.  First, $T$-duality 
transformation of massless type-II supergravity along a transverse 
direction of the R-R 8-brane leads to the product of $S^1$ and $D=9$ 
Minkowski spacetime, which is 9-brane.  (Note, the direct dimensional 
reduction requires $H$ to be constant.)  Second, $T$-duality 
transformation leading to type-IIB 7-brane is much involved and we 
discuss in detail in the following.  

$T$-duality transformation involving massive type-IIA supergravity 
requires construction of $D=9$ massive type-IIB supergravity.  
$D=9$ massive type-IIB supergravity is obtained from massless type-IIB 
supergravity through the Scherk-Schwarz type dimensional 
reduction procedure \cite{SCHs82}, i.e. fields are allowed to depend 
on internal coordinates.   This is motivated by the observation that 
the `St\"uckelberg' type symmetry, which fixes the $m$-dependence of 
field strengths in type-IIA supergravity, is realized within type-IIB 
supergravity as a general coordinate transformation in the internal 
direction, which requires some of R-R fields to depend on the internal 
coordinates.  Namely, an axionic field $\chi(x,z)$ (i.e. R-R 0-form 
field) is allowed to have an additional linear dependence on the internal 
coordinate $z$, i.e. $\chi(x,z)\to mz+\chi(x)$, where $x$ is the 
lower-dimensional coordinates.  Since $\chi$ appears always through 
$d\chi$ in the action, the compactified action has no dependence 
on the internal coordinate $z$.  The result is the  
massive supergravity with cosmological term.   

The {\it massive} type-IIA supergravity compactified on $S^1$  
through the standard KK procedure is related via $T$-duality 
to the {\it massless} type-IIB supergravity compactified on $S^1$  
through the Scherk-Schwarz procedure.   This {\it massive} $T$-duality 
transformation generalizes those of {\it massless} type-II supergravity 
in \cite{BERho451}.  The explicit expression for $m$-dependent 
correction to the $T$-duality transformation can be found in 
\cite{BERdgpt470}.  Under the {\it massive} $T$-duality, the type-IIA 
8-brane transforms to the type-IIB 7-brane, which is the field 
theory realization of $D\,7$-brane of type-IIB superstring theory.  
By applying the $T$-duality of {\it massless} type-II supergravities 
to this 7-brane, one obtains a 6-brane of type-IIA theory, 
whereas transformation to 8-brane of type-IIA theory requires 
the application of {\it massive} $T$-duality.  

This generalized compactification Ansatz for $\chi$ is a special case of 
the generalized compactification on a $d$-dimensional manifold $M_d$ where 
an $(n-1)$-form potential $A_{n-1}$ ($n\leq d$) for which $n$-th cohomology 
class $H^n(M_d,{\bf R})$ of $M_d$ is non-trivial is allowed 
to have an additional linear dependence on the $(n-1)$-form 
$\omega_{n-1}(z)$, i.e. $A_{n-1}(x,z)=m\omega_{n-1}(z)+{\rm standard\ terms}$ 
\cite{COWlpst,LAVp492}.  (All the other fields are reduced by the standard 
KK procedure.)  Here, $d\omega_{n-1}$ represents non-trivial $n$-th 
cohomology of $M_d$.  (In the case of compactification of type-IIB theory 
on $S^1$, the cohomology $dz$ is the volume form on $S^1$.)  Since 
$A_{n-1}$ appears in the action always through $dA_{n-1}$, the 
lower-dimensional action depends on $A_{n-1}$ only through its zero mode 
harmonics on $M_d$, only.  The constant $m$ manifests in lower dimensions 
as a cosmological constant and, thereby, the compactified 
$D$-dimensional action admits domain wall, i.e. $(D-2)$-brane, solutions.  
The general pattern for mass generation is as follows.  
First, the KK vector potentials always become massive.  
Second, a field that appears in a bilinear term in the Chern-Simons 
modification of a higher rank field strength acquires mass if it 
is multiplied by  $A_{n-1}$ (with general dimensional reduction Ansatz).  
Third, when axionic field $A^{(ijk)}_0$ associated with $D=11$ 
3-form potential $A_{MNP}$ is used for the Scherk-Schwarz reduction, 
the lower-dimensional theory contains a topological mass term.  
In these mechanisms, the fields associated with the St\"uckelburg 
symmetry (under which the eaten fields undergo pure non-derivative 
shift symmetries) get absorbed by other fields to become mass terms 
for the potentials that absorb them.  The consistency of the theory 
requires that the fields that are eaten should not appear in the Lagrangian.  
Note, whereas the original Scherk-Schwarz mechanism \cite{SCHs82} is 
designed to give a mass to the gravitino, thus breaking supersymmetry, 
and do not generate scalar potentials, in our case a cosmological constant 
is generated and the full supersymmetry is preserved
\footnote{But the Scherk-Schwarz reduced theories have smaller symmetry than 
those reduced via the standard KK procedure \cite{COWlpst}.}.   

In addition to single-charged $p$-branes, there are other supersymmetric 
configurations which are basic building blocks of $p$-brane bound states. 
These are the gravitational plane fronted wave (denoted $0_w$), 
called `pp-wave', and the KK monopole (denoted $0_m$).  
Their existence within $p$-brane bound states are required by 
duality symmetries.  

First, the KK monopole in $D=11$ is introduced \cite{TOW350} in an attempt 
to give $D=11$ interpretation of type-IIA $D\,6$-brane 
\cite{HORs360}.  The KK monopole is the magnetic dual of the KK modes of 
$D=11$ theory on $S^1$, which is identified as R-R 0-brane 
(electrically charged under the KK $U(1)$ gauge field) \cite{WIT443}.  
The $D=11$ KK monopole, which preserves $1/2$ of supersymmetry, has the form: 
\cite{TOW350}:
\begin{equation}
ds^2_{11}=-dt^2+d{\bf y}\cdot d{\bf y}+Vd{\bf x}\cdot d{\bf x}
+V^{-1}(dx^{11}-{\bf A}\cdot d{\bf x})^2,
\label{elvenkkmonop}
\end{equation}
where $\bf y$ [$\bf x$] is the coordinates of the Euclidean space 
${\bf R}^6$ [${\bf R}^3$], $V=1+{\mu\over\rho}$ ($\rho\equiv\sqrt{{\bf x}
\cdot{\bf x}}$) and a 1-form potential ${\bf A}$, satisfying $\nabla\times
{\bf A}=\nabla V$, has the field strength ${\bf F}=\mu\varepsilon_2$, 
where $\varepsilon_2$ is the volume form on $S^2$.  
The singularity of (\ref{elvenkkmonop}) at $\rho=0$ is a coordinate 
singularity, which is removed once the coordinate $x^{11}$ is periodically 
identified with the period $4\pi\mu$.  By compactifying the solution 
(\ref{elvenkkmonop}) along $x^{11}$ on $S^1$, one obtains R-R 6-brane 
in type-IIA theory (carrying magnetic charge $\mu$), which is completely 
non-singular.  When (\ref{elvenkkmonop}) is reduced along a direction in 
${\bf R}^6$, one has the KK monopole in $D=10$.  

Second, the pp-wave, which preserves $1/2$ of supersymmetry, plays as 
important role as $p$-branes.  First of all, string dualities require 
the existence of pp-wave within the string spectrum.  
Namely, $T$-duality on type-IIA [type-IIB] fundamental 
string along the longitudinal direction yields a pp-wave in type-IIB 
[type-IIA] theory \cite{HORhs68,HORt73}.   Furthermore, the type-IIA 
$D\,0$-brane has pp-wave as its image in $D=11$, with the momentum of 
pp-wave identified as the $D\,0$-brane charge.  When 
pp-wave is compactified along a transverse direction, one has 
pp-wave in $D=10$.  The pp-wave propagating in $y$-direction 
has the form:
\begin{equation}
ds^2=dudv+W(u,{\bf x})du^2+d{\bf x}\cdot d{\bf x},\ \ \ 
\partial^2_xW=0,\ \ \ u,v\equiv y\pm t.
\label{ppwavesol}
\end{equation}
The pp-wave is constructed in the following way.  One imposes 
a Lorentz boost $(t,y)\to(t\cosh\beta+y\sinh\beta,
x\cosh\beta+t\sinh\beta)$ on the Schwarzschield solution and 
takes the extreme limit, i.e. infinite boost (boost with the speed of 
light) and zero Schwarzschield mass limit \cite{GIB207}. Then, one obtains 
pp-wave solution with $W={{\cal Q} \over {|{\bf x}|^n}}$, where 
$n\in{\bf Z}^+$ depends on $D$ and the numbers of isometry directions 
of the configuration.  More general solution is obtained by replacing 
$W={{\cal Q}\over {|{\bf x}|^n}}$ by $W(u,{\bf x})$.  
In particular, the choice $W=f_i(u)x^i$ corresponds to wave with the 
profile $f_i$ propagating along $y$; an asymptotic observer, however, 
observes pp-wave with $W={{\cal Q}\over {|{\bf x}|^n}}$ and ${\cal Q}
\sim\langle[\int duf_i(u)]^2\rangle$.  When the Lorentz boost 
is imposed on a black $p$-brane and extreme limit is taken, 
one has a bound state of a $p$-brane and pp-wave.  

\paragraph{Orthogonally Intersecting $p$-Branes}\label{bprmltintort}

From single-charged $p$-branes, one can construct 
``marginal'' bound states of $p$-branes by applying general intersection 
rules.  The ``marginal'' bound states are called orthogonally ``overlapping'' 
[``intersecting''] $p$-branes if the constituent $p$-branes are separated 
[located at the same point] in a direction transverse to all of the 
$p$-branes.  We denote the configuration where a $(p+r)$-brane intersects 
with a $(p+s)$-brane over a $p$-brane as $(p|p+r,p+s)$ \cite{PAPt393}.  
We add subscripts ($NS$, $R$, $M$) in this notation to specify 
types of charges that the constituent $p$-branes carry, e.g. $2_M$, 
$1_{NS}$ and $3_R$ respectively denote $M\,2$-brane, NS-NS string 
(or fundamental string) and R-R $3$-brane.   

The intersection rules are first studied in \cite{PAPt380} 
in an attempt to interpret G\"uven solutions \cite{GUE276}, 
and general harmonic superposition rules (which prescribe how 
products of powers of the harmonic functions of intersecting 
$p$-branes occur) of intersecting $p$-branes are 
formulated in \cite{TSE475}.  (Such harmonic superposition rules of 
intersecting $M$-branes are already manifest in general 
overlapping $M$-branes constructed in \cite{GAUkt478}.)  
Intersection rules can be independently 
derived from the `no force' condition on a $p$-brane probe moving in another 
$p$-brane background \cite{TSE487}.  
Alternatively, one can derive  general intersection rules for ``marginal'' 
bound state $p$-branes in diverse dimensions from equations of motion, only 
\cite{ARGeh398}.  Namely, from the equations of motion following from general 
$D$-dimensional action with one dilaton $\phi$ and arbitrary numbers of 
$n_A$-form field strengths $F_{n_A}$ ($A=1,...,N$) with kinetic terms 
$\sum_A{1\over{2n_A!}}e^{a_A\phi}F^2_{n_A}$, one obtains the 
following intersection rule prescribing $(\bar{p}|p_A,p_B)$:
\begin{equation}
\bar{p}+1={{(p_A+1)(p_B+1)}\over{D-2}}-{1\over 2}\epsilon_Aa_A
\epsilon_Ba_B,
\label{ddidmprrintrule}
\end{equation}
where $\epsilon_A=+(-)$ when the brane $A$ is electric (magnetic).  
It is interesting that the relation (\ref{ddidmprrintrule}), 
derived from the equations of motion of the effective action 
alone, predicts $D$-branes $(0|1_{NS},p_{R})$ and $D$-branes for higher 
branes.  In the following, we discuss intersection rules for a pair of 
branes.  Intersecting configurations with more than two constituents are 
constructed by applying the intersection rules to all the possible pairs 
of branes.  

There are 3 types of intersecting $p$-branes: 
self-intersections, branes ending on branes and branes within branes.  
First, $p$-branes of the same type intersect only over 
$(p-2)$-brane (so-called $(p-2)$ self-intersection rule), denoted as 
$(p-2|p,p)$.  
This can be understood \cite{PAPt380} from the fact that 
$p$-brane worldvolume theory contains a scalar (interpreted as 
a Goldstone mode of spontaneously broken translational invariance by the 
$p$-brane), which is Hodge-dualized to a worldvolume $(p-1)$-form 
potential that the $(p-2)$-brane couples to.
The second type, denoted as $(p-1|p,q)$ with $q>p$, is interpreted as 
a $q$-brane ending on a $p$-brane with $(p-1)$-branes being the ends 
of $q$-branes on the $p$-branes.  This type of intersecting $p$-branes 
can be constructed by applying $S$- and $T$-duality transformations on 
$M\,2$-brane ending on $M\,5$-brane $(1|2_M,5_M)$ or fundamental string 
ending on R-R $p$-brane $(0|1_{NS},p_R)$.  
The third type, denoted as $(p|p,q)$ with $p<q$, is interpreted as 
$p$-branes inside of the worldvolume of $q$-brane \cite{DOU077}.  
This type of intersecting $p$-branes preserves fraction of  
supersymmetry when the projection operators $P_p$ and $P_q$ (defining spinor 
constraints associated with the constituent $D\,p$-branes) either commutes 
or anticommutes \cite{PAPt393}.   If $P_p$ and $P_q$ commute, then 
$(p|p,q)$ preserves $1/4$ of supersymmetry.  This happens iff $p=q$ mode 4.  
When $P_p$ and $P_q$ anticommute, one has configurations 
that preserve $1/2$ of supersymmetry.  An example is 0-brane within membrane 
$(0|0_R,2_R)$, which can be obtained from $(2|2_M,5_M)$ (\ref{11dsolof8dmem}) 
through dimensional reduction on $S^1$ and $T$-duality transformations.

The spacetime coordinates of intersecting branes are divided into 3  
parts: $(i)$ the overall worldvolume coordinates $\xi^{\mu}$  
($\mu=0,1,...,d-1$), which are common to all the constituent 
branes; $(ii)$ the relative transverse coordinates $x^a$ ($a=1,...,n$), 
which are transverse to part of the constituent branes; 
$(iii)$ the overall transverse coordinates $y^i$ ($i=1,...,\ell$), which are 
transverse to all of the constituent branes.   Since a transverse 
[longitudinal] coordinate of R-R $p$-branes corresponds to a coordinate 
with Dirichlet [Neumann] boundary condition, these 3 types of 
coordinates respectively correspond to coordinates of open strings of 
the NN-, ND- (or DN-) and DD-types.  

Intersecting $p$-branes are divided into 3 types, according 
to the dependence of harmonic functions on these coordinates.  
The first type, for which the general intersection rules are formulated 
in \cite{PAPt380,TSE475}, has all the harmonic 
functions depending on the overall transverse coordinates, only.  
For the second type, one harmonic function depends on the overall 
transverse coordinates and the other on the relative transverse coordinates.  
The third type has both harmonic functions depending on the relative 
transverse coordinates.  For the first [third] type, constituent $p$-branes 
are, therefore, localized in the overall [relative] transverse 
directions but delocalized in the relative [overall] transverse directions.  
We will be mainly concerned with intersection rules for the first 
type and later we comment on the rest of types.  

One can construct supersymmetric (BPS) intersecting 
$p$-branes when spinor constraints associated with constituent 
$p$-branes (given in the previous paragraphs) are compatible with 
one another with non-zero Killing spinors.  
When none of spinor constraint is expressed as a combination of 
other spinor constrains, intersecting $N$ number of $p$-branes preserve 
$({1\over 2})^N$ of supersymmetry.  (Note, from now on $N$ stands 
for the number of constituent $p$-branes, not the number of extended 
supersymmetries.)  One can introduce an additional $p$-brane without 
breaking any more supersymmetry, if some combination of spinor 
constraints of existing constituent $p$-branes gives rise to spinor 
constraint of the added $p$-brane.  

First, we discuss intersecting $M\,p$-branes.  Intersecting rules, 
studies in \cite{PAPt380,TSE475}, are:
\begin{itemize}
\item 
Two $M\,2$-branes [$M\,5$-branes] intersect over a 0-brane 
[a 3-brane], i.e. $(0|2_M,2_M)$ [$(3|5_M,5_M)$].  
\item 
$M\,2$-brane and $M\,5$-brane intersect over a string, i.e.  
$(1|2_M,5_M)$, interpreted as $M\,2$-brane ending on 
$M\,5$-brane over a string. The worldvolume theory is described by 
$D=6$ $(0,2)$ supermultiplet with bosonic fields given by 5 scalars and 
a 2-form that has self-dual field strength, and $M\,2$-brane charge 
is carried by a self-dual string inside the worldvolume theory.  
\item
One can add momentum along an isometry direction by applying 
an $SO(1,1)$ boost transformation.  
\end{itemize}
All the possible intersecting $M$-branes are determined by 
these intersection rules.  One can intersect up to 8 $M\,p$-branes 
by applying intersection rules to each pair of constituent 
$M\,p$-branes: the complete classification up to $T$-duality 
transformations is given in \cite{BERdejv095}.

Intersecting $M$-brane solutions can be constructed from the 
harmonic superposition rules.  These are first formulated in 
\cite{TSE475} for BPS $M$-branes and are generalized to the 
non-extreme case in \cite{DUFlp382,CVEt478} and 
to the rotating case in \cite{CVEy229}.  
First, the harmonic superposition rules for the BPS case are 
as follows:
\begin{itemize}
\item 
The overall conformal factor of the metric is the product of 
the appropriate powers of the harmonic functions associated with 
constituent $M\,p$-branes:  
\begin{equation}
ds^2=\prod_iH^{(p_i+1)/9}_{p_i}(y)[\cdots],
\label{bpsconfintmbran}
\end{equation}
where $p_i=2,5$, and $H_{p_i}(y)$ are harmonic functions in overall 
transverse space (with dimension $\ell$), namely of $(i)$ the form 
$H_{p_i}(y)=1+{{c_{p_i}}\over{|y-y_0|^{\ell-1}}}$ for $\ell>2$,  $(ii)$ 
logarithmic form for $\ell=2$, and $(iii)$ linear form for $\ell=1$.  Here, 
$y=\sqrt{y^2_1+\cdots y^2_{\ell}}$ is the radial coordinate of the overall 
transverse space.  So, the spacetime is asymptotically flat when $\ell>2$.  
\item
The metric is diagonal (unless there is a momentum along an 
isometry direction) and each component inside of $[\cdots]$ in 
(\ref{bpsconfintmbran}) is the product of the inverse of harmonic 
functions associated with the constituent $p_i$-branes whose worldvolume 
coordinates include the corresponding coordinate. 
\item 
When momentum is added to an isometry direction, say $\xi$-direction, 
the above rules are modified by the harmonic function $K(y)$ associated 
with the momentum as follows:
\begin{equation}
-dt^2+d\xi^2\to -K^{-1}(y)dt^2+K(y)\widehat{d\xi}^2,\ \ 
\widehat{d\xi}\equiv d\xi+[K^{-1}(y)-1]dt.
\label{bstbpsmbrn}
\end{equation}
\item
When $\ell>3$, one can add a KK monopole in a 3-dimensional subspace 
of the overall transverse space.  All the harmonic functions depend only 
on these 3 transverse coordinates and the metric is modified 
in the overall transverse components as follows
\begin{eqnarray}
dy^idy^i&\to&dy^2_1+\cdots +dy^2_{\ell-4}
\cr
& &+H^{-1}(dy_{\ell-3}\pm\alpha_{KK}
\cos\theta d\phi)^2+H(dr^2+r^2d\Omega^2_2), 
\label{addkkmontombr}
\end{eqnarray}
where the harmonic function $H=1+{{\alpha_{KK}}\over r}$ is associated 
with the KK monopole charge $P_{KK}=\pm\alpha_{KK}\Omega_2$.  The 
2-form field strength associated with the KK monopole has the form 
${\cal F}_2=P_{KK}\epsilon_2$.  
\item
Non-zero components of 4-form field strength ${\cal F}$ are given 
by (\ref{singlembrane}) for each constituent $M\,p$-branes.  
\end{itemize}

Now, we discuss generalization of harmonic superposition rules 
to the non-extreme, rotating case.  The solutions 
get modified by the harmonic functions $g_i(y)=1+{{l^2_i}\over{y^2}}$ 
associated with angular momentum parameters $l_i$ and the following 
combinations of $g_i$:
\begin{equation}
{\cal G}_{\ell}\equiv\left\{\matrix{\alpha^2+\sum^{{\ell-2}\over 2}_{i=1}
\mu^2_ig^{-1}_i& {\rm for\ even\ } \ell \cr 
\sum^{{\ell-1}\over 2}_{i=1}\mu^2_ig^{-1}_i& {\rm for\ odd\ }\ell}\right., 
\ \ \ \ 
f^{-1}_{\ell}\equiv{\cal G}_{\ell}\prod^{[{{\ell-1}\over 2}]}_{i=1}g_i. 
\label{angparharmfun}
\end{equation}
Here, $\alpha$ and $\mu_i$ are defined in (\ref{edef2}) and 
(\ref{odef2}).   The harmonic superposition rules are modified in the 
following way:
\begin{itemize}
\item
Harmonic functions are modified as
\begin{equation}
H=1+{{2m\sinh\delta\cosh\delta}\over{y^{\ell-2}}}\ \ \to \ \ 
H=1+f_{\ell}{{2m\sinh^2\delta}\over{y^{\ell-2}}},
\label{modfharmang}
\end{equation}
where the boost parameter $\delta$ is associated with electric/magnetic 
charge or momentum $2m\sinh\delta\cosh\delta$.
\item
The metric components get further modified as
\begin{equation}
dt^2\to fdt^2,\ \ \ \ \ \ \ 
\delta_{ij}dy^idy^j\to f^{\prime\,-1}dy^2+y^2\widehat{d\Omega}^2_{\ell-1},
\label{angmodfofmetrc}
\end{equation}
where $f\equiv 1-f_{\ell}{{2m}\over{y^{\ell-2}}}$ and $f^{\prime}\equiv 
{\cal G}^{-1}_{\ell}-f_{\ell}{{2m}\over{y^{\ell-2}}}$ are 
non-extremality functions.  
Here, $\widehat{d\Omega}^2_{\ell-1}$ denotes the angular parts of the metric.  
In the limit $l_i=0$, $\widehat{d\Omega}^2_{\ell-1}$ becomes flat 
metric $d\Omega^2_{\ell-1}$ of $S^{\ell-1}$.  
Generally, $\widehat{d\Omega}^2_{\ell-1}$ is given by the angular 
metric components of rotating black hole in (compactified) $D=\ell+1$ 
and the general rules for constructing these components 
are unknown yet.  (The explicit forms of $\widehat{d\Omega}^2_{\ell-1}$ up to 
3 $M\,p$-brane intersections with momentum along an isometry 
direction are given in \cite{CVEy229}.)    
\item
The harmonic superposition rule (\ref{bstbpsmbrn}) with a momentum along 
an isometry direction $\xi$ is modified as
\begin{equation}
-dt^2+d\xi^2\to K^{-1}fdt^2+K\widehat{d\xi}^2,\ \ \ \ 
\widehat{d\xi}\equiv d\xi+[K^{\prime\,-1}-1]dt,
\label{momharmmbrang}
\end{equation}
where $\delta$ is a boost parameter associated with 
momentum $2m\sinh\delta\cosh\delta$ and $K^{\prime\,-1}\equiv 
1-f_{\ell}{{2m\sinh^2\delta}\over{r^{\ell-2}}}K^{-1}$.  
\item
The non-zero components of the 4-form field strength are 
the same as the BPS case, when $l_i=0$.  For $l_i\neq 0$, there are 
additional non-zero components associated with the induced electric/magnetic 
fields due to rotations.  The general construction rules are unknown yet; 
the explicit expressions up to 3 intersections with momentum along an 
isometry direction are given in \cite{CVEy229}.
\end{itemize}

Next, we discuss intersecting $p$-branes in $D=10$.  
The intersection rules are as follows:
\begin{itemize}
\item Two fundamental strings can only be parallely oriented, 
i.e. $1_{NS}\parallel 1_{NS}$.
\item Two solitonic 5-branes orthogonally intersect over 3-spaces, i.e. 
$(3|5_{NS},5_{NS})$.
\item A fundamental string and a solitonic 5-brane can only be parallely 
oriented, i.e. $1_{NS}\parallel 5_{NS}$ or $(1|1_{NS},5_{NS})$.  
\item An R-R $p$-brane and an R-R $q$-brane intersect over $n$-spaces 
$(n|p_{RR},q_{RR})$ such that $p+q-2n=4$.  
\item A fundamental string orthogonally intersects an R-R $p$-brane over 
a point, i.e. $(0|1_{NS},p_R)$.
\item A solitonic 5-brane intersects an R-R $p$-brane over $n$-spaces 
$(n|5_{NS},p_R)$ such that $p-n=1$.
\end{itemize}
These intersecting $p$-branes preserve $1/4$ of supersymmetry except 
the multi-centered configuration $1_{NS}\parallel 1_{NS}$.  
These intersection rules are derived in \cite{TSE487} by applying 
`no force' condition and in \cite{ARGeh398} from the equations of motion.
Alternatively, one can derive these rules from the intersection rules 
of $M$-branes by applying KK procedure and duality 
transformations, which we discuss in the following in details.  

Intersecting $p$-branes in $D=10$ can be obtained 
from intersecting $M$-branes through compactification on $S^1$ and 
duality transformations.   
We have the following $p$-branes in type-IIA theory from the 
compactifications $R_{\parallel}$ and $R_{\perp}$ of $M$-branes 
along a longitudinal and a transverse directions, i.e. the double 
and direct dimensional reductions, respectively:
\begin{equation}
2_M {\buildrel R_{\parallel}\over\longrightarrow}1_{NS},\ \ \ \ \ 
2_M {\buildrel R_{\perp}\over\longrightarrow}2_{R},\ \ \ \ \ 
5_M {\buildrel R_{\parallel}\over\longrightarrow}4_{R},\ \ \ \ \ 
5_M {\buildrel R_{\perp}\over\longrightarrow}5_{NS},
\label{mbrtoiiapbrkk}
\end{equation}
which can be understood from the relations of type-IIA form fields 
to the $D=11$ 3-form field under the standard KK procedure.  
Dimensional reduction of $0_w$ and $0_m$ in $D=11$ yields the following 
type-IIA configurations:
\begin{equation}
0_w {\buildrel R_{\parallel}\over\longrightarrow} 0_R,\ \ \ \ 
0_w {\buildrel R_{\perp}\over\longrightarrow} 0_w,\ \ \ \ 
0_m {\buildrel R_{\parallel}\over\longrightarrow} 6_R,\ \ \ \ 
0_m {\buildrel R_{\perp}\over\longrightarrow} 0_m,
\label{othereldimreduc}
\end{equation}
where $R_{\parallel}$ [$R_{\perp}$] on $0_m$ denotes the reduction 
in the direction associated with Taub-NUT term [the other directions] 
of the metric.  

Duality transformations further relate different types of $p$-branes.  

First, we consider $T$-duality between type-IIA and type-IIB theories 
on $S^1$.  
$T$-duality (\ref{circtdual}) on type-IIA/B R-R $p$-branes 
along  a tangent [transverse] direction leads to type-IIB/A R-R 
$(p-1)$-branes [$(p+1)$-branes].  
Note, $T$-duality on a longitudinal direction 
introduces an additional overall transverse coordinate that 
harmonic functions have to depend on, which is not always guaranteed.  
So, the $T$-duality on a longitudinal direction is called 
``dangerous'', whereas the $T$-duality on a transverse direction 
is ``safe'' since the resulting configurations are guaranteed to be 
solutions to the equations of motion.  

$T$-duality transformation rules of NS-NS $p$-branes can be inferred 
from $T$-duality transformation of fields \cite{BERho451} 
(see also (\ref{circtdual})) as follows.  Among other things,  
$T$-duality interchanges off-diagonal metric components 
$g_{\mu\alpha}$ and the same components $B_{\mu\alpha}$ 
of the NS-NS 2-form potential, where $\alpha$ is the $T$-duality 
transformation direction.  
So, when $p$-brane has non-trivial $(\mu=t,\alpha)$-component of the   
metric or NS-NS 2-form potential, $T$-duality transformation is 
reminiscent of interchange of momentum mode (electric charge of KK $U(1)$  
field $g_{\mu\alpha}$) and winding mode (electric charge of NS-NS 2-form 
$U(1)$ field $B_{\mu\alpha}$) under $T$-duality.  So, pp-wave 
(whose linear momentum is identified with KK electric charge) and 
NS-NS string (carrying electric charge of the NS-NS 2-form field, 
identified as string winding mode) are interchanged when $T$-duality 
is performed along the longitudinal direction of string or the 
direction of pp-wave propagation.  $T$-duality along the other directions  
yields the same type of solutions.  
Second, the $(\mu=\phi_i,\alpha)$-component of metric [NS-NS 2-form 
potential], where $\phi_i$ are angular coordinates associated with 
rotational symmetry, corresponds to the Taub-NUT term [is associated with 
magnetic charge of a solitonic 5-brane $5_{NS}$].  
So, magnetic monopole (or Taub-NUT solution) and NS-NS 5-brane 
are interchanged when $T$-duality transformation is applied along 
the direction transverse to NS-NS 5-brane or along the coordinate 
associated with Taub-NUT term.  $T$-duality along the other directions  
yields the same type of solutions.  

Second, the type-IIB $SL(2,{\bf Z})$ $S$-duality transformation 
can be used to relate NS-NS charged solutions and 
R-R charged solutions which are coupled to 2-form potentials.  
Under the ${\bf Z}_2$ subset transformation, NS-NS string and NS-NS 5-brane 
transform to R-R $1$-brane and R-R $5$-brane, respectively.  
Full $SL(2,{\bf Z})$ transformations on NS-NS string [NS-NS 5-brane] 
yields ``non-marginal'' BPS bound states of $p$ NS-NS strings and 
$q$ R-R strings [$p$ NS-NS 5-brane and $q$ R-R $5$-brane] with the pair 
of integers $(p,q)$ relatively prime.  

The duality transformation rules are, therefore, summarized as follows:
\begin{eqnarray}
p_R&{\buildrel T_{\parallel}\over\longrightarrow}&(p-1)_R, \ \ \ 
p_R{\buildrel T_{\perp}\over\longrightarrow}(p+1)_R, \ \ \ 
1_{NS}{\buildrel T_{\parallel}\over\longrightarrow}0_w,\ \ \ \ 
1_{NS}{\buildrel T_{\perp}\over\longrightarrow}1_{NS}, 
\cr
0_w&{\buildrel T_{\parallel}\over\longrightarrow}&1_{NS},\ \ \ \ \ \ \ \ \ 
0_w{\buildrel T_{\perp}\over\longrightarrow}0_w,\ \ \ \ \ \ \ \ \ \ \,
5_{NS}{\buildrel T_{\parallel}\over\longrightarrow}5_{NS},\ \ \   
5_{NS}{\buildrel T_{\perp}\over\longrightarrow}0_m,
\cr
0_m&{\buildrel T_{\parallel}\over\longrightarrow}&5_{NS},\ \ \ \ \ \ \ \ \,
0_m{\buildrel T_{\perp}\over\longrightarrow}0_m,\ \ \ \ \ \ \ \ \ \ 
1_{NS}{\buildrel S\over\longleftrightarrow}1_{R},\ \ \ \  
5_{NS}{\buildrel S\over\longleftrightarrow}5_{R}.
\label{dualtypeiirule}
\end{eqnarray}

We now discuss various intersecting $p$-branes in $D=10$.  
  
First, we consider intersecting R-R $p$-branes in type-II theories.  
$D$-brane configurations are supersymmetric if the number $\nu$ of 
coordinates of DN or ND type is the multiple of 4 \cite{POL050}.  
(See section \ref{dbrentint} for details on this point.)  
At the level of low-energy intersecting R-R $p$-branes of the 
effective field theories, this means that solutions are supersymmetric 
when the number of relative transverse coordinates is the multiple of 4, 
i.e. $n=4,8$.   This can also be derived from the condition that 
the Killing spinor constraints $\epsilon_L=\Gamma_{01\cdots p}\epsilon_R$ 
of the constituent R-R $p$-branes are compatible with one another.  
When both of the harmonic functions depend only on the relative 
transverse coordinates, BPS configurations are possible for 
$n=8$ case only, and otherwise only $n=4$ configurations are 
BPS.   Since our main concern is the intersecting R-R 
$p$-branes with all the harmonic functions depending only on the overall 
transverse coordinates, we concentrate on the $n=4$ case.   
Since $T$-duality preserves the 
total number $n$ of the relative transverse coordinates, one can obtain 
all the intersecting 2 R-R $p$-branes with $n=4$ by applying ``safe'' 
$T$-duality transformations to intersecting R-R 0-brane and R-R 4-brane, 
i.e. $(0|0_R,4_R)$.   One can further add R-R $p$-branes in such a way that 
$n=4$ for each pair of constituent R-R $p$-branes.   It is shown  
\cite{BERdejv095} that one can intersect up to 8 R-R $p$-branes which 
satisfy the $n=4$ rule for each pair:  the complete classification up 
to $T$-dualities is given in \cite{BERdejv095}.  The explicit solutions 
for BPS intersecting R-R $p$-branes can be constructed by the following 
harmonic superposition rules:
\begin{itemize}
\item
The metric is diagonal, with each component having 
the multiplicative contribution of $H^{-1/2}_p$ [$H^{1/2}_p$] from each 
constituent R-R $p$-brane whose worldvolume [transverse] coordinates 
include the associated coordinate.  
\item 
Dilaton is given by the product of harmonic functions associated 
with the constituent R-R $p$-branes: $e^{-2\phi}=\prod_{k}
H^{{p_k-3}\over 2}_{p_k}$.
\item 
Non-zero components of $(p+2)$-form field strengths are given by 
(\ref{pbransol}) for each constituent R-R $p$-branes.
\end{itemize}
As an example, solution for intersecting R-R $(p+r)$-brane and R-R 
$(p+s)$-brane over a $p$-brane, i.e. $(p|p+r,p+s)$, is of the form:
\begin{eqnarray}
ds^2_{10}&=&(H_{p+r}H_{p+s})^{-1/2}\eta_{\mu\nu}d\xi^{\mu}d\xi^{\nu}
+\left({{H_{p+r}}\over{H_{p+s}}}\right)^{1/2}\sum^{s}_{a=1}(dx^a)^2
\cr
& &+\left({{H_{p+s}}\over{H_{p+r}}}\right)^{1/2}\sum^{s+r}_{a=s+1}(dx^a)^2
+(H_{p+r}H_{p+s})^{1/2}\delta_{ij}dy^idy^j,
\cr
e^{-2\phi}&=&H^{{p+r-3}\over 2}_{p+r}H^{{p+s-3}\over 2}_{p+s},
\cr
{\cal F}_{tx^1\cdots x^sy^i}&=&\partial_{y^i}H^{-1}_{p+s}, \ \ \ \ \ \ \ 
{\cal F}_{tx^{s+1}\cdots x^{s+r}y^i}=\partial_{y^i}H^{-1}_{p+r}.
\label{intertwodbrnsol}
\end{eqnarray}

We discuss intersecting $p$-branes which contain NS-NS $p$-branes.  
First, $(0|1_{NS},p_R)$ with $0\leq p\leq 8$, is nothing but open strings 
that end on $D$-brane.  This type of configurations can be obtained by 
first compactifying $(0|2_M,2_M)$ on $S^1$ along a 
longitudinal direction of one of $M\,2$-brane (resulting in $(0|1_{NS},
2_R)$), and then by sequentially applying $T$-duality transformations 
along the directions transverse to the NS-NS string.   
Second, $(p-1|5_{NS},p_R)$ with $1\leq p\leq 6$, are interpreted as 
$D\,p$-brane ending on NS-NS 5-brane.  Namely, NS-NS 5-branes act as a 
$D$-brane for $D$-branes.  This interpretation is consistent with the 
observation \cite{BECb472} that $M\,5$-branes are boundaries of 
$M\,2$-branes.  This type of intersecting branes can be 
constructed by first compactifying $(1|2_M,5_M)$ on $S^1$ along an 
overall transverse direction (resulting in $(1|5_{NS},2_R)$)), and then by 
sequentially applying $T$-duality transformations along the longitudinal 
directions of the NS-NS 5-brane.   Third, intersecting NS-NS $p$-branes 
can be obtained in the following ways: $(i)$ compactification of 
$(3|5_M,5_M)$ along an overall transverse direction leads to 
$(2|5_{NS},5_{NS})$, $(ii)$ compactification of $(1|2_M,5_M)$ along a 
relative transverse direction which is longitudinal to $M\,2$-brane 
leads to $(1|2_{NS},5_{NS})$, $(iii)$ the type-IIB $S$-duality on 
$(-1|1_R,1_R)$ yields $(-1|1_{NS},1_{NS})$.  

We comment on the case some or all of harmonic 
functions depend on the relative transverse coordinates \cite{BERdejv095}.  
These types of intersecting $p$-branes can be constructed by applying 
the general harmonic superposition rules, taking into account of dependence 
of harmonic functions on the relative transverse coordinates.  
In particular, the metric components associated with 
the relative coordinates (that harmonic functions depend on) have to 
be the same so that the equations of motion are satisfied. 
First, the second type of intersecting $p$-branes, i.e. one harmonic 
function depends on the relative transverse coordinates, can be 
constructed from the first type of intersecting $p$-branes, i.e. all the 
harmonic functions depend on the overall transverse coordinates, by letting 
one of harmonic functions depend on the relative transverse coordinates.  
Thus, the classification of the second type is the same as that of the 
first type.  The third type of $p$-branes, i.e. all the harmonic functions 
depend on the relative transverse coordinates, have 8 relative 
transverse coordinates ($n=8$) for a pair of $p$-branes.  It is impossible 
to have more than two $p$-branes with each pair having $n=8$.  In $D=11$,  
the only configuration of the third type is $(1|5_M,5_M)$
\footnote{$M\,2$-brane with its longitudinal coordinates given by the 
overall longitudinal and overall transverse coordinates of $(1|5_M,5_M)$ 
can be further added without breaking any more supersymmetry.  The added 
$M\,2$-brane intersects the $M\,5$-branes over strings and is interpreted 
as an $M\,2$-brane stretched between two $M\,5$-branes.}  
\cite{GAUkt478}.   This $M$-brane preserves $({1\over 2})^2$ 
of supersymmetry, since the Killing spinor satisfies two constraints of 
the form (\ref{spinconstsingmbr}), each corresponding to a constituent 
$M\,5$-brane.  By compactifying an overall transverse direction of 
$(1|5_M,5_M)$ on $S^1$, one obtains $(1|5_{NS},5_{NS})$
\footnote{One can further add a fundamental string along the string 
intersection without breaking any more supersymmetry.  $T$-duality along 
the fundamental string direction leads to type-IIB 
$(1|5_{NS},5_{NS})$ with pp-wave propagating along the string 
intersection.  Note, the former configuration preserves only $1/8$ of 
supersymmetry, rather than $1/4$, if regarded as a solution of type-IIB 
theory.}, 
which was first constructed in \cite{KHU143}.  
Further application of the type-IIB $SL(2,{\bf Z})$ transformation 
leads to $(1|5_R,5_R)$.  The series of application of 
$T$-duality transformations, then,  yield a set of overlapping 2  
R-R $p$-branes with $n=8$.  (Complete list can be found in 
\cite{GAUkt478}.)  These overlapping $p$-branes correspond, at string 
theory level, to $D$-brane bound states with 8 ND or DN directions, 
and therefore should be supersymmetric.  
When 2 R-R $p$-branes intersect in a point, one can add a 
fundamental string without breaking any more supersymmetry.  This type of 
configurations is interpreted as a fundamental string stretching between 
two $D$-branes.  For the case where 2 R-R $p$-branes intersect in a 
string, one can add pp-wave along the string intersection without 
breaking anymore supersymmetry.  
Another third type of intersecting $p$-branes in $D=10$ can 
be constructed by compactifying $(1|5_M,5_M)$ along 
a relative transverse direction, resulting in $(1|4_R,5_{NS})$, 
followed by series of $T$-duality transformations along the 
longitudinal directions of the NS-NS 5-brane, resulting in $(p-3|p_R,5_{NS})$ 
with $3\leq p\leq 8$.  One can further add R-R $(p-2)$-branes to these 
configurations; these configurations are interpreted as 
a $D\,(p-2)$-brane stretching between $D\,p$-brane and NS-NS 5-brane.  

\paragraph{Other Variations of Intersecting $p$-Branes}\label{bprmltintoth}

So far, we discussed intersecting $p$-branes with $p-p^{\prime}=0$ mod 4.  
Existence of such classical intersecting 
$p$-branes that preserve fraction of supersymmetry is expected 
from the perturbative $D$-brane argument \cite{POLcj,POL050}.  
One can construct such solutions by applying the harmonic superposition 
rules discussed in the above.  

In this subsection, we discuss another type of $p$-brane bound states 
which do not follow the intersection rules discussed in the previous 
subsection.  Such $p$-brane bound states contain pair of constituent 
$p$-branes with $p-p^{\prime}=2$ and still preserve fraction of 
supersymmetry
\footnote{It is argued \cite{POLcj,POL050} that $D$-brane bound 
state with $p-p^{\prime}=6$ is not supersymmetric and is unstable 
due to repulsive force.  Although a solution that may be interpreted 
as $(0|0_{RR},6_{RR})$ is constructed in \cite{COSp04}, its 
interpretation is ambiguous due to abnormal singularity structure of 
harmonic functions, and it cannot be derived from $(2|2_M,5_M)$ though 
a chain of $T$-duality transformations.}.  
These $p$-brane configurations can be generated by applying dimensional 
reduction or $T$-duality along a direction {\it at angle} with a 
transverse and a longitudinal directions of the constituent $p$-branes 
\cite{RUSt047,BREmm174,COSp04}.    
(Hereafter, we call them as `tilted' reduction and `tilted' 
$T$-duality.)  
These $p$-brane configurations can also be 
constructed by applying `ordinary' dimensional reduction and sequence of 
`ordinary' duality transformations on $(2|2_M,5_M)$ 
(\ref{11dsolof8dmem}), which preserves $1/2$ of supersymmetry. 

In fact, it conflicts with diffeomorphic invariance of the underlying theory 
that one has to choose specific directions (which are either transverse or 
longitudinal to the constituent $p$-branes) for dimensional reduction or 
$T$-duality transformations \cite{COSp04}.  So, the existence of 
such new $p$-brane configurations is required by ($M$-theory/IIA string 
and IIA/IIB string) duality symmetries and diffeomorphism invariance of 
the underlying theories.  

We now discuss the basic rules of `tilted' $T$-duality and 
`tilted' dimensional reduction on constituent $p$-branes.  
Before one applies `tilted' $T$-duality and dimensional reduction to 
$p$-branes, one rotates a pair of a longitudinal $x$ and a transverse 
$y$ coordinates of a (constituent) $p$-brane by an angle $\alpha$ 
(a diffeomorphism that mixes $x$ and $y$):
\begin{equation}
\left(\matrix{x^{\prime}\cr y^{\prime}}\right)=
\left(\matrix{\cos\alpha& -\sin\alpha\cr \sin\alpha& \cos\alpha}
\right)\left(\matrix{x\cr y}\right).
\label{rotlotrcoord}
\end{equation}
(Note, $x$ and $y$ have to be diffeomorphic directions so that 
one can compactify these directions on $S^1$.)  
Then, one compactifies or applies $T$-duality along 
the $x^{\prime}$-direction.  These procedures preserve supersymmetry.  
When $\alpha=0$ [$\alpha=\pi/2$], such compactification or $T$-duality 
transformation is the compactification or $T$-duality transformation 
along a longitudinal direction $x$ [a transverse direction $y$].   
Thus, as the angle $\alpha$ is varied from $0$ to $\pi/2$, the 
resulting bound state interpolates between the corresponding two limiting 
configurations.  

First, we discuss the `tilted' reduction $R_{\alpha}$ 
of solutions in $D=11$.  
The `tilted' reduction on $M$-branes leads to type-IIA $p$-brane bound 
states, interpreted as `brane within brane':
\begin{equation}
2_M {\buildrel R_{\alpha}\over\longrightarrow} (1|1_{NS},2_R)_A,
\ \ \ \ \ \ \ \ 
5_M {\buildrel R_{\alpha}\over\longrightarrow} (4|4_R,5_{NS})_A, 
\label{tilkkmpbran}
\end{equation}
where the subscript $A$ means type-IIA configuration. 
From $0_m$ and $0_w$ in $D=11$, one obtains the following type-IIA 
bound states:
\begin{equation}
0_m {\buildrel R_{\alpha}\over\longrightarrow} (0_w|6_R)_A, 
\ \ \ \ \ \ \ \ \ 
0_w {\buildrel R_{\alpha}\over\longrightarrow} (0_m|0_R)_A,
\label{tillkkmonwav}
\end{equation}
where $(0_w|0_R)_A$ is interpreted as a `boosted' $D\,0$-brane.  

Second, we discuss `tilted' $T$-duality transformations $T_{\alpha}$.   
The off-diagonal metric component $g_{x^{\prime}y^{\prime}}$ 
induced by the coordinate rotation (\ref{rotlotrcoord}) 
is transformed to the same component $B_{x^{\prime}y^{\prime}}$ of 
2-form potential under $T$-duality.  So, the $T$-transformed solution has 
diagonal metric and non-zero ${\cal F}_{\mu\nu}=
B_{\mu\nu}+2\pi\alpha^{\prime}F_{\mu\nu}$, where $F_{\mu\nu}$ is the 
worldvolume gauge field strength \cite{LEI}.  For R-R $p$-brane bound 
states, the corresponding perturbative $D$-brane now, 
therefore, satisfies the modified boundary condition 
$\partial_nX^{\mu}-i{\cal F}^{\mu}_{\ \nu}\partial_lX^{\nu}=0$.  
The induced flux ${\cal F}_{x^{\prime}y^{\prime}}$ 
is related to $\alpha$ as ${\cal F}_{x^{\prime}
y^{\prime}}=-\tan\alpha$.  The ADM mass of the transformed 
configuration is of the form $M^2\sim Q^2_1+Q^2_2$, a characteristic of 
non-threshold bound states.  First, the `tilted' $T$-duality on type-IIA/B
$p$-branes leads to the following type-IIB/A bound states:
\begin{equation}
1_{NS}{\buildrel T_{\alpha}\over\longrightarrow}(0_w|1_{NS}),\ \ \,  
5_{NS}{\buildrel T_{\alpha}\over\longrightarrow}(0_m|5_{NS}),\ \ \,   
(p+1)_R{\buildrel T_{\alpha}\over\longrightarrow}(p|p_R,(p+2)_R),
\label{tendpbrtiltdual}
\end{equation}
where $(0_w|1_{NS})$ is simply a boosted fundamental string.  
The existence of the $D$-brane bound states $(p|p_R,(p+2)_R)$ preserving 
$1/2$ of supersymmetry is also expected from the 
perturbative $D$-brane considerations.  
One can apply `tilted' $T$-duality transformation more than once to 
obtain new $p$-brane bound state configurations.  For example, by applying 
`tilted' $T$-duality transformations to $D\,2$-brane in two different 
directions, one obtains $D\,(4,2,2,0)$-brane bound state \cite{BREmm174}.   
Second, `tilted' $T$-duality on $0_m$ and $0_w$ 
in type-IIA/B theory yields the following type-IIB/A bound states:
\begin{equation}
0_m {\buildrel T_{\alpha}\over\longrightarrow}(0_m|5_{NS}),\ \ \ \ \ 
0_w {\buildrel T_{\alpha}\over\longrightarrow}(0_w|1_{NS}).  
\label{tendmompptiltdual}
\end{equation}

Next, we discuss $p$-brane bound states obtained by 
first imposing a Lorentz boost along a transverse direction and 
then applying $T$-duality transformation or reduction along 
the direction of the boost: {\it $T$-duality along a 
boost} and {\it reduction along a boost}, respectively denoted as 
$T_v$ and $R_v$.  The Lorentz boost yields  
non-threshold bound state of a $p$-brane and pp-wave. 
This bound state interpolates between extreme $p$-brane and plane wave, 
as the boost angle $\alpha$ (see below for its definition) varies from zero 
(no boost) to $\pi/2$ (infinite boost).  The ADM energy $E$ of such 
non-threshold bound state of extreme $p$-brane (with mass $M$) and  
pp-wave (with momentum $p$) has the form $E^2=M^2+p^2$, a reminiscent of 
relativistic kinematic relation of a particle of the rest mass $M$ with 
the linear momentum $p$, rendering the interpretation of such bound state 
as `boosted' $p$-brane.  The Lorentz boost with velocity 
$v/c=\sin\alpha$ along a transverse direction $y$ 
has the form \cite{RUSt047}:
\begin{equation}
t\to t^{\prime}=(\cos\alpha)^{-1}(t+y\sin\alpha),\ \ \ \ \ 
y\to y^{\prime}=(\cos\alpha)^{-1}(y+t\sin\alpha).
\label{lorenbstcoordtrn}
\end{equation}
In the above, the angle
\footnote{This boost angle $\alpha$ can be identified with the angle 
$\alpha$ of coordinate rotation in (\ref{rotlotrcoord}).  Namely, 
the non-threshold type-IIA $(q_1,q_2)$ string bound state obtained from 
$D=11$ pp-wave through tilted dimensional reduction at an 
angle $\alpha$, followed by $T$-duality, can also be obtained by 
reduction along a boost of $M\,2$-brane with the same boost angle 
$\alpha$, followed by $T$-duality transformation.}
$\alpha$ is related to the boost parameter $\beta$ as 
$\cosh\beta=1/\cos\alpha$.  
After the Lorentz boost (\ref{lorenbstcoordtrn}), one 
compactifies or applies $T$-duality transformation along 
$y^{\prime}$.  

First, we discuss the reduction $R_v$ along a boost.  Since the momentum 
of pp-wave manifests as the KK electric charge after reduction 
along the direction of momentum flow, the resulting bound state always 
involves R-R 0-brane.  
The following type-IIA bound states are obtained from the 
compactification with a boost of configurations in $D=11$
\footnote{One can straightforwardly apply this procedure to intersecting 
$M$-branes, followed by sequence of $T$-duality transformations, to 
construct $p$-brane bound states that interpolate between those that 
preserve $1/4$ of supersymmetry and those that preserve $1/2$ of 
supersymmetry as $\alpha$ is varied from 0 to $\pi/2$ 
\cite{COSc204}.}:
\begin{eqnarray}
2_M&{\buildrel R_v\over\longrightarrow}&(0|0_R,2_R)_A,
\ \ \ \ \ \ \ \ \ \ 
5_M{\buildrel R_v\over\longrightarrow}(0|0_R,5_{NS})_A,
\cr 
0_w&{\buildrel R_v\over\longrightarrow}&(0_R|0_w)_A,
\ \ \ \ \ \ \ \ \ \ \ \ \,\,
0_m{\buildrel R_v\over\longrightarrow}(0|0_R,6_R)_A.
\label{kkwithboost}
\end{eqnarray}
Second, we discuss the $T$-duality $T_v$ along a boost.  
Under the $T$-duality, the linear momentum of a pp-wave is 
transformed to the electric charge of the NS-NS 2-form potential 
\cite{HORhs68}.  
So, the $T$-dualized configurations always involve a 
fundamental string with non-zero winding mode.  
We have the following type-IIB/A bound states from type-IIA/B configurations:
\begin{eqnarray}
1_{NS}&{\buildrel T_v\over\longrightarrow}&1_{NS},\ \ \ \ 
5_{NS}{\buildrel T_v\over\longrightarrow}(0_m|1_{NS}),\ \ \ \ 
p_R{\buildrel T_v\over\longrightarrow}(1|1_{NS},(p+1)_R),
\cr
& &0_w{\buildrel T_v\over\longrightarrow}(0_w|1_{NS}),
\ \ \ \ \ \ \ \ \ \ \ \ 
0_m{\buildrel T_v\over\longrightarrow}(0_m|1_F,5_{NS}).
\label{tdualwithboost}
\end{eqnarray}

The diffeomorphic invariance and duality symmetries require new type of 
bound states in $M$-theory that preserve $1/2$ of supersymmetry.  
These new $M$-theory configurations can be constructed by uplifting new 
type-IIA configuration discussed in this subsection.  For example, 
by uplifting $(0_m|5_{NS})_A$ or $(4|4_R,6_R)_A$, one obtains the 
$M\,5$-brane and the KK monopole bound state in $D=11$.  Another example 
is the $M\,2$-brane and the KK monopole bound state uplifted from 
$(1|1_{NS},6_R)_A$ or $(0_m|1_{NS})_A$.  

Non-threshold type-IIB $(q_1,q_2)$ string, obtained from 
R-R or NS-NS string through $SL(2,{\bf Z})$ duality, 
can be related to $D=11$ pp-wave compactified at angle
\cite{RUSt047}.  This is understood as follows.  
When the compactification torus $T^2$ (parameterized by isometric 
coordinates $(x,y)$) is rectangular, the angle $\alpha$ of coordinate 
rotation defines the direction of $(q_1,q_2)$ cycle in $T^2$, around 
which the $D=11$ pp-wave is wrapped, as $\cos\alpha=q_1/\sqrt{q^2_1+q^2_2}$.  
(So, choice of different angle $\alpha$ corresponds to different choice of 
a cycle in $T^2$.)  Starting from pp-wave propagating along $x$, one 
performs coordinate rotation (\ref{rotlotrcoord}) in the plane $(x,y)$, 
where $y$ is an isometric transverse direction of the pp-wave, and then 
compactifies along $y^{\prime}$-direction, which is the 
direction of $(q_1,q_2)$-cycle of $T^2$ with coordinates $(x,y)$.   
The resulting type-IIA configuration is a non-threshold bound state of  
pp-wave and $D\,0$-brane.  Subsequent $T$-duality transformation along 
$y_1$ leads to type-IIB $(q_1,q_2)$ string solution.  
This is related to the fact that the type-IIB $SL(2,{\bf Z})$ symmetry 
is the modular symmetry of the $D=11$ 
supergravity on $T^2$ \cite{BERho451,SCH360,SCH49} (see section 
\ref{dualm} for detailed discussion).

Had we started from the bound state of $M\,2$-brane and pp-wave 
along a longitudinal direction of the $M\,2$-brane, we would end up 
with boosted type-IIB $(q_1,q_2)$ string 
that preserves $1/4$ of supersymmetry.  (The $M\,2$-brane charge is, 
therefore, interpreted as a momentum of the type-IIB $(q_1,q_2)$ string.)    

On the other hand, the non-threshold type-IIB $(q_1,q_2)$ 5-brane 
can be related to $M\,5$-brane.  Namely, one first compactifies 
$M\,5$-brane at an angle to obtain $(4|4_R,5_{NS})_A$ 
and then applies $T$-duality transformation along the relative transverse 
direction to obtain type-IIB $(q_1,q_2)$ 5-brane bound state.  
Following the similar procedures, one can obtain the non-threshold 
type-IIB $(q_1,q_2)$ string from $M\,2$-brane.  

\paragraph{Branes Intersecting at Angles}\label{bprmltintang}

In \cite{BERdl480}, it is shown that one can construct BPS $D$-brane 
bound states where the constituent $D$-branes intersect at angles 
other than the right angle.  We first summarize formalism of 
\cite{BERdl480}.  Then, we discuss the corresponding classical solutions 
\cite{BREmm041,GAUgpt202,BEHc56,COSc204,BALll143} in the effective 
field theory.  

In the presence of a $D\,p$-brane, two spinors $\varepsilon$ 
and $\tilde{\varepsilon}$ (corresponding to $N=2$ spacetime supersymmetry) 
of type-II string theory satisfy the constraint:
\begin{equation}
\tilde{\varepsilon}=\Gamma_{(p)}\varepsilon:=
\prod^p_{i=0}e^{\mu}_i\Gamma_{\mu}\varepsilon,
\label{dpbrnspnrcnst}
\end{equation}
where $e_i$ is an orthonormal frame spanning $D\,p$-brane worldvolume.  
For $N$ numbers of constituent $D$-branes, it is convenient to define 
the following raising and lowering operators from gamma matrices:  
\begin{equation}
a^{\dagger}_k={1\over 2}(\Gamma_{2k-1}-i\Gamma_{2k}),\ \ \ 
a^k={1\over 2}(\Gamma_{2k-1}+i\Gamma_{2k}), \ \ \ k=1,...,N, 
\label{railowopgamm}
\end{equation}
which satisfy the anticommutation relations:
\begin{equation}
\{a^j,a^{\dagger}_k\}=\delta^j_k,\ \ \ \ \ 
\{a^j,a^k\}=0=\{a^{\dagger}_j,a^{\dagger}_k\}.
\label{anticmreldpbr}
\end{equation} 
The lowering operators $a^k$ define the `vacuum' $|0\rangle$, satisfying
$a^k|0\rangle=0$.
Under an $SU(N)$ rotation $z^i\to R^i_jz^j$ of the complex coordinates 
$z^k\equiv x^{a_k}+ix^{b_k}$ spanned by $D\,p$-branes, the raising and 
the lowering operators transform as
\begin{equation}
a^k\to R^k_{\ j}a^j,\ \ \ \ \ 
a^{\dagger}_k\to R^{\dagger\ j}_{\ k}a^{\dagger}_j.
\label{suntranrailow}
\end{equation}

One can construct intersecting $D\,p$-branes at angles in the 
following way.  One starts from two $D\,p$-branes oriented, say, 
along the directions ${\rm Re}\,z^i$ and applies $SU(p)$ rotation 
$z^i\to R^i_{\ j}z^j$ to one of $D\,p$-branes.   For this type of 
intersecting $D\,p$-branes at angles, the spinor $\varepsilon$ satisfies 
the constraint:
\begin{equation}
\prod^p_{k=1}(a^{\dagger}_k+a^k)\varepsilon=
\prod^p_{k=1}(R^{\dagger\ j}_{\ k}a^{\dagger}_j+R^k_{\ j}a^j)\varepsilon.
\label{spncnstang}
\end{equation}
So, the resulting configuration has unbroken supersymmetries 
$|0\rangle$ and $\prod^p_{k=1}a^{\dagger}_k|0\rangle$.  These two spinors 
have the same [opposite] chirality for $p$ even [odd].   
One can further compactify this intersecting $D\,p$-branes at angles on 
a torus and then apply $T$-duality transformations to obtain other 
types of intersecting $D\,p$-branes at angles.  
Alternatively, one can start from $(q|(p+q)_R,(p+q)_R)$ and rotate one 
$D\,(p+q)$-brane relative to the other by applying the $SO(2p)$ 
transformation.  The resulting configuration is supersymmetric when the 
$SU(p)\subset SO(2p)$ transformation is applied.  When these intersecting 
$D$-branes are further compactified on tori, the consistency 
of toroidal compactification imposes the quantization condition for the 
intersecting angles $\theta$'s in relation to the moduli of tori 
\cite{BERdl480,BALl55}.  

When intersecting $D$-branes at angles are compactified on a manifold 
${\cal M}$, the unbroken supersymmetry should commute with the 
`generalized' holonomy group (defined by a modified connection 
$-$ with torsion $-$ due to non-zero antisymmetric tensor backgrounds) 
of ${\cal M}$.  Here, the spinor constraint 
$\eta=\prod_ie^{\mu}_i\Gamma_{\mu}\eta$ defines an action of discrete 
generalized holonomy.  
Generally, starting from an intersecting $D\,p$-brane at angles with 
$\hat{F}=F+B=0$, where $F$ [$B$] is the worldvolume 2-form field strength 
[the NS-NS 2-form], one obtains a configuration with $\hat{F}\neq 0$ when 
$T$-duality is applied.  
For intersecting $D\,2$-branes at angles, the necessary and sufficient 
condition for preserving supersymmetry 
is that $\hat{F}$ is anti-self-dual \cite{BERdl480}.  

Classical solution realization of intersecting $D$-branes at angles 
is first constructed in \cite{BREmm041}.  
Starting from $n$ parallel $D\,2$-branes (with each constituent 
$D\,2$-brane located at $\vec{x}=\vec{x}_a$, $a=1,...,n$, and its 
charge related to $\ell_a>0$) lying in the $(y^2,y^4)$ plane, one 
rotates each constituent $D\,2$-brane by an $SU(2)$ angle $\alpha_a$ 
in the $(y^1,y^2)$ and $(y^3,y^4)$ planes, i.e. $(z^1,z^2)\to 
(e^{i\alpha_a}z^1,e^{-i\alpha_a}z^2)$ where $z^1=y^1+iy^2$ and 
$z^2=y^3+iy^4$.  The solution in the string frame has the form:
\begin{eqnarray}
ds^2&=&\sqrt{1+X}\left[{1\over{1+X}}\left(-dt^2+\sum^4_{j=1}(dy^j)^2
\right.\right.
\cr
& &\ \ \ \ \ \ \left.\left.+\sum^n_{a=1}X_a\left\{[(R_a)^1_{\ i}
dy^i]^2+[(R_a)^3_{\ j}dy^j]^2\right\}\right)+\sum^9_{i=5}(dx^i)^2\right],
\cr
A^{(3)}&=&{{dt}\over{1+X}}\wedge\left\{\sum^n_{a=1}X_a\,(R_a)^2_{\ i}dy^i
\wedge (R_a)^4_{\ j}dy^j\right.
\cr
& &\ \ \ \ \ \ \left.-\sum^n_{a<b}X_aX_b\sin^2(\alpha_a-\alpha_b)
(dy^1\wedge dy^3-dy^2\wedge dy^4)\right\},
\cr
e^{2\phi_A}&=&\sqrt{1+X};\ \ \ \ \ 
X\equiv\sum^n_{a=1}X_a+\sum^n_{a<b}X_aX_b\sin^2(\alpha_a-\alpha_b),
\label{bmmintdbrang}
\end{eqnarray}
where $R^a$ ($a=1,..,n$) are (block-diagonal) $SO(4)$ matrices that 
correspond to the above mentioned rotation of constituent $D\,2$-branes 
and harmonic functions $X_{a}(\vec{x})={1\over 3}\left({{\ell_a}\over
{|\vec{x}-\vec{x}_a|}}\right)^3$ are associated with constituent 
$D\,2$-branes located at $\vec{x}=\vec{x}_a$.  This configuration preserves 
$1/4$ of supersymmetry and interpolates between previously known 
configurations: $(i)$ $\alpha_a=0,\pi/2$ case is orthogonally 
oriented $D\,2$-branes, $(ii)$ $\alpha_a=\alpha_0$, $\forall a$, case is 
parallel $n$ $D\,2$-branes oriented in different direction through 
$SU(2)$ rotations, etc..   

The ADM mass density of (\ref{bmmintdbrang}) is the sum of those 
of constituent $D\,2$-branes, i.e. $m={{{\cal A}_4}
\over{2\kappa^2}}\sum^n_{a=1}\ell^3_a$, and is independent of the $SU(2)$ 
rotation angles $\alpha_a$.  The physical charge density is also simply 
the sum of those of constituent $D\,2$-branes, although the charge 
densities in different planes $(y^i,y^j)$ of the intersecting 
$D\,2$-branes depend on $\alpha_a$.  

$T$-duality on (\ref{bmmintdbrang}) yields other types of 
$D$-brane bound states.  The $T$-duality along transverse directions 
leads to angled $D\,p$-branes with $p>2$.  The $T$-duality along 
worldvolume directions leads to more exotic bound states 
of $D$-branes.  Namely, since the constituent $D\,2$-branes intersect 
with one another at angles, the worldvolume direction that one chooses for 
$T$-duality transformation is necessarily at angle with some of 
constituent $D\,2$-branes.  Consequently, the resulting configuration is 
exotic bound state of $D\,p$-brane ($p\neq 2$) and bound states of the 
type studied in \cite{BREmm174} (e.g. bound state of $D\,(p+1)$-brane and 
$D\,(p-1)$-brane, and $D\,(4,2,2,0)$-brane bound state)  
obtained by applying the `tilted' $T$-duality transformation(s) 
on a $D\,p$-brane.  

The above intersecting $D$-branes at angles and related configurations 
are alternatively derived by ($i$) applying the `tilted' boost transformation 
on the orthogonally intersecting two $D$-branes, followed by the sequence 
of $T$-$S$-$T$ transformations of type-II strings \cite{BEHc56}, or ($ii$) 
applying `reduction along a boost' followed by $T$-duality transformations 
\cite{COSc204}.  For the former case \cite{COSc204}, the resulting 
configurations are mixed bound states of R-R branes that necessarily 
involves fundamental string.   
It is essential that one has to turn on both $D$-brane charges 
of original orthogonally intersecting $D$-branes and apply the $S$-duality 
between two $T$-duality transformations to have configurations 
where the $D$-branes intersect at angles.  

Intersecting $p$-branes at angles in more general setting, starting from 
$M$-branes with the flat Euclidean transverse space ${\bf E}^{4n}$ 
replaced by the toric hyper-K\"ahler manifold ${\cal M}_n$, are 
studied in \cite{GAUgpt202}.  
We first discuss the general formalism and then specialize to the 
case of intersecting $p$-branes in $D=10,11$. 

$4n$-dimensional toric hyper-K\"ahler manifold ${\cal M}_n$ with a 
tri-holomorphic
\footnote{The manifold ${\cal M}_n$ is tri-holomorphic iff the triplet 
K\"ahler 2-forms ${\bf \Omega}=(d\varphi_i+A_i)d{\bf x}^i-{1\over 2}
U_{ij}d{\bf x}^i\times d{\bf x}^j$ are independent of  
$\varphi_i$, i.e. ${\cal L}_{{\partial}\over{\partial\varphi_i}}
{\bf\Omega}=0$.} 
$T^n$ isometry has the following general form of metric:
\begin{equation}
ds^2_{HK}=U_{ij}d{\bf x}^i\cdot d{\bf x}^j+U^{ij}(d\varphi_i+A_i)
(d\varphi_j+A_j)\ \ \  (i,j=1,...,n),
\label{hyperkaltrihol}
\end{equation}
where $\varphi_i$ (which are periodically identified $-$ 
to remove a coordinate singularity $-$ $\varphi_i\sim\varphi_i+2\pi$, 
thereby parameterizing $T^n$) 
correspond to the $U(1)$ isometry directions of ${\cal M}_n$ and 
${\bf x}^i=\{x^i_r|r=1,2,3\}$ parameterize $n$ copies of  
Euclidean spaces ${\bf E}^3$.   The $n=1$ case is Taub-NUT space.  
The hyper-K\"ahler condition relates $n$ 1-forms $A_i=d{\bf x}^j\cdot 
{\bf \omega}_{ij}$ with field strengths $F_i=dA_i$ to $U_{ij}$ 
through $\varepsilon^{rsl}\partial^l_jU_{ki}=F^{rs}_{jki}=
\partial^r_j\omega^s_{ki}-\partial^s_k\omega^r_{ji}$.  
(So, a toric hyper-K\"ahler metric is specified by $U_{ij}$ alone.)  
This implies that $U_{ij}$ are harmonic functions on ${\cal M}_n$, 
i.e. $U^{ij}\partial_i\cdot\partial_jU=0$.  Generally, a positive definite 
symmetric $n\times n$ matrix $U({\bf x}^i)$ is linear combination 
\begin{equation}
U_{ij}=U^{(\infty)}_{ij}+\sum_{\{p\}}\sum^{N(\{p\})}_{m=1}
U_{ij}[\{p\},{\bf a}_m(\{p\})]
\label{genuijform}
\end{equation}
of the following harmonic functions specified by a set of $n$ real numbers 
$\{p_i|i=1,...,n\}$, called a `$p$-vector', and an arbitrary 3-vector 
${\bf a}$:
\begin{equation}
U_{ij}[\{p\},{\bf a}]={{p_ip_j}\over{2|\sum_kp_k{\bf x}^k-{\bf a}|}}.
\label{harmfuncuij}
\end{equation}

The vacuum hyper-K\"ahler manifold ${\bf E}^{3n}\times T^n$ with  
moduli space $Sl(n,{\bf Z})\backslash Gl(n,{\bf R})/SO(n)$ 
has the metric (\ref{hyperkaltrihol}) with $U_{ij}=
U^{(\infty)}_{ij}$ (so, $A_i=0$).  
Regular non-vacuum hyper-K\"ahler manifold is represented by 
harmonic functions $U_{ij}[\{p\},{\bf a}]$  
associated with a $3(n-1)$-plane in ${\bf E}^{3n}$ defined by 
3-vector equations $\sum^n_{k=1}p_k{\bf x}^k={\bf a}$.   
The hyper-K\"ahler metric (\ref{hyperkaltrihol}) is non-singular,  
provided $\{p\}$ are coprime integers.  
The $Sl(n,{\bf Z})$ transformation $U\to S^TUS$ ($S\in Sl(n,{\bf Z})$) 
on the hyper-K\"ahler metric (\ref{hyperkaltrihol}) leads to another 
hyper-K\"ahler metric with the $Sl(n,{\bf Z})$ transformed 
$p$-vector $S\{p\}$.  The angle $\theta$ between two $3(n-1)$-planes  
defined by two $p$-vectors $\{p\}$ and $\{p^{\prime}\}$ is given by  
$\cos\theta=p\cdot p^{\prime}/\sqrt{p^2p^{\prime\,2}}$.  Here, the 
inner product is defined as $p\cdot q=(U^{(\infty)})^{ij}p_iq_j$, 
which is invariant under $Sl(n,{\bf Z})$.  
The solution (\ref{hyperkaltrihol}) is, therefore, 
specified by angles and distances between mutually intersecting 
$3(n-1)$-planes associated with harmonic functions 
$U_{ij}[\{p\},{\bf a}_m(\{p\})]$.  

The special case where $\Delta U\equiv U-U^{(\infty)}$ is diagonal 
(i.e. the $p$-vectors have the form $(0,...,1,...,0)$ and  
$\Delta U_{ij}=\delta_{ij}{1\over{2|{\bf x}^i|}}$:   
there are only $n$ intersecting $3(n-1)$-planes) describes 
$n$ fundamental BPS monopoles in maximally-broken rank $(n+1)$ gauge 
theories found in \cite{LEEwy54}, thereby called LWY metric.  
When additionally $U^{(\infty)}$ is diagonal (so that $U$ is diagonal), 
$n$ $3(n-1)$-planes intersect orthogonally ($\cos\theta=0$) and ${\cal M}_n
={\cal M}_1\times\cdots\times {\cal M}_1$.  In this case, one can 
always choose ${\bf\omega}_{ij}$ 
such that $F_i=dA_i$ and $U_{ii}$ are related as $F_i=\star dU_{ii}$, 
where $\star$ is the Hodge-dual on ${\bf E}^3$.  

The hyper-K\"ahler manifold ${\cal M}_n$ preserves fraction of 
supersymmetry.  It admits $(n+1)$ covariantly constant $SO(4n)$ spinors 
(in the decomposition of $D$-dimensional Lorentz spinor representation 
under the subgroup $Sl(n,{\bf R})\times SO(4n)$) if the holonomy of 
${\cal M}_n$ is strictly $Sp(n)$, which corresponds to the 
case where $3(n-1)$-planes intersect non-orthogonally, i.e. 
$U^{(\infty)}$ is non-diagonal.  These covariantly constant $SO(4n)$ 
spinors arise as singlets in the decomposition of the spinor representation 
of $SO(4n)$ into representations of holonomy group of ${\cal M}_n$, i.e. 
$Sp(n)$ for this case.  The only toric hyper-K\"ahler manifolds 
whose holonomy is a proper subgroup of $Sp(n)$ are those corresponding to 
the `orthogonally' intersecting or `parallel' $3(n-1)$-planes.  
For this case, ${\cal M}_n={\cal M}_1\times\cdots {\cal M}_1$ 
(i.e. product of $n$ Taub-NUT space) with $Sp(1)^n$ holonomy and 
diagonal $U_{ij}$ (thereby, $3(n-1)$-planes intersecting orthogonally), 
and more supersymmetry is preserved since the non-singlet spinor 
representations of $Sp(n)$ are further decomposed under the proper 
subgroup $Sp(1)^n$.   
A trivial case corresponds to the case $U_{ij}=U^{(\infty)}_{ij}$, 
i.e. the vacuum hyper-K\"ahler manifold: since the holonomy group is 
trivial, all the supersymmetries are preserved.  

The starting point of general class of intersecting $p$-branes is 
the following $D=11$ solution, 
which is the product of the $D=3$ Minkowski space and ${\cal M}_2$:
\begin{equation}
ds^2_{11}=ds^2({\bf E}^{2,1})+U_{ij}d{\bf x}^i\cdot d{\bf x}^j+
U^{ij}(d\varphi_i+A_i)(d\varphi_j+A_j), 
\label{elvnhypkalsol}
\end{equation} 
where $i=1,2$.  
For a general solution with non-diagonal $U_{ij}$, thereby 
with the $Sp(2)$ holonomy for ${\cal M}_2$, (\ref{elvnhypkalsol}) admits 
$(n+1)=3$ covariantly constant spinors.  Namely, 32-component real 
spinor in $D=11$ is decomposed under $Sl(2,{\bf R})\times SO(8)$ as   
${\bf 32}\to ({\bf 2},{\bf 8}_s)\oplus({\bf 2},{\bf 8}_c)$.  Two 
$SO(8)$ spinor representations
\footnote{The subscripts $s$ and $c$ denote two possible $SO(8)$ 
chiralities.} 
${\bf 8}_s$ and ${\bf 8}_c$ are further, respectively, decomposed under 
$Sp(2)\subset SO(8)$ as ${\bf 8}_s\to {\bf 5}\oplus {\bf 1}
\oplus {\bf 1}\oplus {\bf 1}$ and ${\bf 8}_c\to {\bf 4}\oplus {\bf 4}$.   
So, $3/16$ of supersymmetry is preserved.  
When $U_{ij}$ is diagonal (so, ${\cal M}_2={\cal M}_1\times {\cal M}_1$ 
and 3-planes orthogonally intersect), the holonomy group is 
$Sp(1)\times Sp(1)$.  Under the $Sp(1)\times Sp(1)$ subgroup, non-singlet 
$Sp(2)$ spinor representations ${\bf 5}$ and ${\bf 4}$ are, respectively, 
decomposed as ${\bf 5}\to ({\bf 2},{\bf 2})\oplus ({\bf 1},{\bf 1})$ and 
${\bf 4}\to ({\bf 2},{\bf 1})\oplus ({\bf 1},{\bf 2})$.  So,  
$8/32=1/4$ of supersymmetry is preserved.  

One can generalize the solution (\ref{elvnhypkalsol}) to include 
$M$-branes without breaking any more supersymmetry, resulting in 
`generalized $M$-branes', where the transverse  Euclidean space 
is replaced by ${\cal M}_n$.  
The harmonic functions (associated with $M$-branes) are independent 
of the $U(1)$ isometry coordinates $\varphi_i$, thereby $M\,p$-branes 
are delocalized in the $\varphi_i$-directions.  

First, one can naturally include an $M\,2$-brane to the solution 
(\ref{elvnhypkalsol}), since the transverse space of $M\,2$-brane 
has dimensions 8:
\begin{eqnarray}
ds^2_{11}&=&H^{-{2\over 3}}ds^2({\bf E}^{2,1})+H^{1\over 3}
[U_{ij}d{\bf x}^i\cdot d{\bf x}^j+U^{ij}(d\varphi_i+A_i)
(d\varphi_j+A_j)],
\cr
F&=&\pm\omega({\bf E}^{2,1})\wedge dH^{-1},
\label{geneldimhypkal}
\end{eqnarray}
where $\omega({\bf E}^{2,1})$ is the volume form on ${\bf E}^{2,1}$, 
the signs $\pm$ are those of $M\,2$-brane charge and 
$H=H({\bf x}^i)$ is a harmonic function (associated with $M\,2$-brane) 
on ${\cal M}_2$, i.e. $U^{ij}\partial_i\cdot\partial_jH=0$.   
The $SO(1,10)$ Killing spinor of this solution is decomposed into 
the $SO(8)$ spinors of definite chiralities  
${\bf 8}_c$ and ${\bf 8}_s$, which are related to the signs $\pm$.  
So, depending on the sign of $M\,2$-brane charge, either all 
supersymmetries are broken or $3/16$ of supersymmetry 
is preserved.  

Second, one can add $M\,5$-branes to the solution (\ref{elvnhypkalsol}) 
if ${\cal M}_2={\cal M}_1\times {\cal M}_1$, i.e. $U={\rm diag}
(U_1({\bf x}^1),U_2({\bf x}^2))$.  
For this purpose, it is convenient to introduce 2 1-form 
potentials $\tilde{A}_i$ ($i=1,2$) with field strengths $\tilde{F}_i$ 
which can be related to the harmonic functions $H_1({\bf x}^1)$ and 
$H_2({\bf x}^2)$ (associated with 2 $M\,5$-branes) as 
$dH_i=\star\tilde{F}_i$.  (This is analogous to the relations 
$dU_i=\star F_i$ satisfied by the diagonal components of $U$ and 
the field strengths $F_i=dA_i$ of the solution (\ref{elvnhypkalsol}) 
when both $\Delta U$ and $U^{(\infty)}$ are diagonal.)  Here, 
$\star$ is the Hodge-dual on ${\bf E}^3$.  The explicit solution 
has the form:
\begin{eqnarray}
ds^2_{11}&=&(H_1H_2)^{2\over 3}\left[(H_1H_2)^{-1}ds^2({\bf E}^{1,1})
+H^{-1}_1[U_2d{\bf x}^2\cdot d{\bf x}^2+U^{-1}_2(d\varphi_2+A_2)^2]
\right.
\cr
& &\ \left.+H^{-1}_2[U_1d{\bf x}^1\cdot d{\bf x}^1+U^{-1}_1
(d\varphi_1+A_1)^2]+dz^2\right],
\cr
F&=&[\tilde{F}_1\wedge(d\varphi_1+A_1)+
\tilde{F}_2\wedge(d\varphi_2+A_2)]\wedge dz.
\label{mfbrhypkalel}
\end{eqnarray}
Generally with non-constant $U_i$ and $H_i$, (\ref{mfbrhypkalel}) 
preserves $1/8$ of supersymmetry, provided the proper relative 
sign of $M\,5$-brane charges is chosen.  

In the following, we discuss the intersecting (overlapping) $p$-brane 
interpretation of solutions obtained via dimensional reduction and 
duality transformations of the $D=11$ solutions 
(\ref{elvnhypkalsol})--(\ref{mfbrhypkalel}).  
Due to the triholomorphicity of the Killing vector fields $\partial/
\partial\varphi_i$, the Killing spinors survive in these procedures.  
As for the $p$-branes associated 
with the harmonic functions $U_{ij}$, there is a one-to-one correspondence 
between $3(n-1)$-planes and $p$-branes, and the intersection angle of 
$p$-branes is given by the angle between the corresponding $p$-vectors, 
which define $3(n-1)$-planes.  

First, we discuss intersecting (overlapping) $p$-branes related to 
(\ref{elvnhypkalsol}) and (\ref{geneldimhypkal}).  
Since (\ref{elvnhypkalsol}) is a special case of (\ref{geneldimhypkal}) 
with $H=1$, we consider $p$-branes related to 
(\ref{geneldimhypkal}), and then comment on the $H=1$ case.  
First, one compactifies one of the $U(1)$ isometry directions of 
${\cal M}_2$, say $\varphi_2$ without lose of generality, on $S^1$, 
resulting in a type-IIA solution, and then applies the $T$-duality 
transformation along the other $U(1)$ isometry direction, i.e. the 
$\varphi_1$-direction, to obtain the following type-IIB solution:
\begin{eqnarray}
ds^2_E&=&({\rm det}\,U)^{3\over 4}H^{1\over 2}[H^{-1}({\rm det}\,U)^{-1}
ds^2({\bf E}^{2,1})+({\rm det}\,U)^{-1}U_{ij}d{\bf x}^i\cdot d{\bf x}^j
+H^{-1}dz^2],
\cr
B_{(i)}&=&A_i\wedge dz,\ \ \ 
\tau=-{{U_{12}}\over{U_{11}}}+i{\sqrt{{\rm det}\,U}\over{U_{11}}}, \ \ \ 
i_kD=\omega({\bf E}^{2,1})\wedge dH^{-1},
\label{iibhypkalsol}
\end{eqnarray}
where $\phi_B$ is the dilaton, $\tau\equiv\ell+ie^{-\phi_B}$ 
($\ell$ = R-R 0-form field), $B_{(i)}$ ($i=1,2$) are 2-form potentials 
in the NS-NS and R-R sectors, $D$ is the 4-form potential, and 
$z\equiv\varphi_1$.  
This solution is interpreted as $D\,3$-brane (with harmonic function 
$H$) stretching between 5-branes along the $z$-direction. 
The 5-branes in this type-IIB configuration are specified by a set 
of intersecting 3-planes $\sum^n_{k=1}p_k{\bf x}^k={\bf a}$ in ${\bf E}^6$.  
From the expression for $\tau$ in (\ref{iibhypkalsol}), 
one sees that the $Sl(2,{\bf R})$ transformation $U\to (S^{-1})^TUS^{-1}$ 
($S\in Sl(2,{\bf R})$) in ${\cal M}_2$ is realized in this type-IIB 
configuration as the type-IIB $Sl(2,{\bf R})$ symmetry $\tau\to{{a\tau+b}
\over{c\tau+d}}$ of equations of motion, where $S=\left(\matrix{a&b\cr c&d}
\right)$.  
The condition that $Sl(2,{\bf R})$ is broken down to $Sl(2,{\bf Z})$ 
so that ${\cal M}_2$ with the coprime integers $\{p_1,p_2\}$ 
remains non-singular after the transformation is translated into the 
type-IIB language that the $Sl(2,{\bf R})$ symmetry of the equations of 
motion is broken down to the $Sl(2,{\bf Z})$ $S$-duality symmetry of 
type-IIB string theory.  
In the following we discuss particular cases of (\ref{iibhypkalsol}).  

We first consider the solution (\ref{iibhypkalsol}) with 
$U={\rm diag}(H_1({\bf x}^1),H_2({\bf x}^2))$ and 
$H=1$.  In this case, ${\cal M}_2={\cal M}_1\times {\cal M}_1$ 
with holonomy $Sp(1)\times Sp(1)$, thereby preserving 
$1/4$ of supersymmetry.  The corresponding solution is 
`orthogonally' intersecting $(2|5_{NS},5_R)$:
\begin{equation}
ds^2_E=(H_1H_2)^{3\over 4}[(H_1H_2)^{-1}ds^2({\bf E}^{2,1})
+H^{-1}_2d{\bf x}^1\cdot d{\bf x}^1
+H^{-1}_1d{\bf x}^2\cdot d{\bf x}^2+dz^2],
\label{diagiibhypkal}
\end{equation}
where harmonic functions $H_i=1+(2|{\bf x}^i|)^{-1}$ ($i=1,2$) are 
respectively associated with NS-NS 5-brane and R-R 5-brane 
\cite{GAUgpt202}.   

A particular case of (\ref{iibhypkalsol}) with $U^{(\infty)}
=1=H$ and a single 3-plane in ${\bf E}^6$ (defined by $\{p_1,p_2\}$) 
is the bound state of NS-NS 5-brane and R-R 5-brane with charge vector 
$(p_1,p_2)$.  So, the restriction that ${\cal M}_2$ is non-singular, 
i.e. $\{p_1,p_2\}$ are coprime integer, manifests in the type-IIB theory 
that the corresponding $p$-brane configuration is a non-marginal bound 
state of NS-NS 5-brane and R-R 5-brane. 
There is a correlation between the $D=11$ $Sl(2,{\bf Z})$ transformation, 
which rotates a 3-plane in ${\bf E}^6$, and the type-IIB $Sl(2,{\bf Z})$ 
transformation, which rotates the charge vectors of the 2-form field 
doublet $B_{(i)}$.  

The general type-IIB solution (\ref{iibhypkalsol}) with non-diagonal 
$U^{(\infty)}$ is interpreted as an arbitrary number of 5-branes 
intersecting or overlapping at angles.  $p$-vectors and 
$U^{(\infty)}$ specify orientations and charges of 5-branes, and 
${\bf a}$ determines distance of 5-branes from the 
origin.  Since the corresponding ${\cal M}_2$ has the $Sp(2)$ holonomy, 
$3/16$ of supersymmetry is preserved.  

Next, we discuss the intersecting $p$-branes in type-IIA string and 
$M$-theory related to (\ref{geneldimhypkal}).  The intersecting $p$-branes 
in type-IIA theory are constructed in the following way.  
First, one $T$-dualizes the type-IIB solution (\ref{iibhypkalsol}) along a 
direction in ${\bf E}^{2,1}$ to obtain the following type-IIA `generalized 
fundamental string' solution, which can also be obtained from 
(\ref{elvnhypkalsol}) by compactifying on a spatial direction in 
${\bf E}^{2,1}$:  
\begin{eqnarray}
ds^2&=&H^{-1}ds^2({\bf E}^{1,1})+U_{ij}d{\bf x}^i\cdot d{\bf x}^j+
U^{ij}(d\varphi_i+A_i)(d\varphi_j+A_j),
\cr
B&=&\omega({\bf E}^{1,1})H^{-1},\ \ \ \ \ \ 
\phi_A=-{1\over 2}\ln\,H.
\label{genstriiahypkal}
\end{eqnarray}
Subsequent $T$-dualities along $\varphi_1$ and $\varphi_2$ lead to the 
type-IIA solution:
\begin{eqnarray}
ds^2&=&H^{-1}ds^2({\bf E}^{1,1})+U_{ij}(d{\bf x}^i\cdot d{\bf x}^j+
d\varphi^id\varphi^j), 
\cr
B&=&A_i\wedge d\varphi^i+\omega({\bf E}^{1,1})H^{-1},\ \ \ \ \ \ \ \ 
\phi_A={1\over 2}\ln\,{\rm det}\,U-{1\over 2}\ln\,H,
\label{iiahypkahlsol}
\end{eqnarray}
where $B$ is the NS-NS 2-form potential and $\phi_A$ is the dilaton. 
This solution is interpreted as an arbitrary number of NS-NS 5-branes 
intersecting on a fundamental string (with a harmonic function $H$), 
generalizing the solutions in \cite{KHU143}.  The case 
with diagonal $U$ represents orthogonally intersecting NS-NS 
5-branes.  More general case with non-diagonal $U$ represents 
intersecting NS-NS 5-branes at angles and preserves $3/16$ of 
supersymmetry.  The following solution in $D=11$ is  
obtained by uplifting the type-IIA solution (\ref{iiahypkahlsol}):
\begin{eqnarray}
ds^2_{11}&=&H^{1\over 3}({\rm det}\,U)^{2\over 3}[H^{-1}({\rm det}\,U)^{-1}
ds^2({\bf E}^{1,1})
\cr
& &\ \ +({\rm det}\,U)^{-1}U_{ij}(d{\bf x}^i\cdot d{\bf x}^j
+d\varphi^id\varphi^j)+H^{-1}dy^2],
\cr
F&=&[F_i\wedge d\varphi^i+\omega({\bf E}^{1,1})\wedge dH^{-1}]\wedge dy.
\label{mbrhypkalhsol}
\end{eqnarray}
When $U$ is of LWY type, this solution represents parallel $M\,2$-branes 
which intersect intersecting 2 $M\,5$-branes (orthogonally when 
$U^{(\infty)}$ is diagonal as well) over a string.   
For more general form of $U$, the solution represents the $M\,2$-branes 
intersecting arbitrary numbers of $M\,5$-branes at angles 
and preserves $3/16$ of supersymmetry.  

$T$-duality on the type-IIA solution (\ref{iiahypkahlsol}) along 
the fundamental string direction leads to intersecting type-IIB NS-NS 
5-branes with a pp-wave along the common intersection direction.  
Further application of ${\bf Z}_2\subset SL(2,{\bf Z})$ $S$-duality 
transformation leads to the following solution involving R-R 5-branes, 
which preserves $3/16$ of supersymmetry when 5-branes intersect at angles: 
\begin{eqnarray}
ds^2_E&=&({\rm det}\,U)^{1\over 4}[dtd\sigma+Hd\sigma^2+
U_{ij}(d{\bf x}^i\cdot d{\bf x}^j+d\varphi^id\varphi^j)],
\cr
B^{\prime}&=&A_i\wedge d\varphi^i,\ \ \ \ \ \ \ \ 
\tau=i\sqrt{{\rm det}\,U},
\label{iibdbrhypkal}
\end{eqnarray} 
where $B^{\prime}$ is the R-R 2-form potential.  
The $H=1$ case is classical solution realization of intersecting $D$-branes 
at angles of \cite{BERdl480}.  In \cite{BERdl480}, the condition 
for unbroken supersymmetry is given by the holonomy condition 
arising in the KK compactifications.  This corresponds 
to the holonomy condition on the hyper-K\"ahler manifold of 
\cite{GAUgpt202}.  Namely, intersecting R-R $p$-branes preserve 
fraction of supersymmetry if orientations of the constituent 
R-R $p$-branes are related by rotations in the $Sp(2)$ subgroup 
of $SO(8)$.  This is seen by considering spinor constraints of 
intersecting two $D\,5$-branes, where one $D\,5$-brane is oriented in the 
$(12345)$ 5-plane and the other $D\,5$-brane rotated into the $(16289)$ 
5-plane by an angle $\theta$.   
For this configuration, type-IIB chiral spinors $\epsilon^A$ ($A=1,2$) 
satisfy the constraints $\Gamma_{012345}\epsilon^1=\epsilon^2$ and 
$R^{-1}(\theta)\Gamma_{016289}R(\theta)\epsilon^1=\epsilon^2$, where 
$R(\theta)={\rm exp}\{-{1\over 2}\theta(\Gamma_{26}+\Gamma_{37}+\Gamma_{48}
+\Gamma_{59})\}$ is the $SO(1,9)$ spinor representation of the above 
mentioned $SO(8)$ rotation.  (The rotational matrix $R(\theta)$ is 
associated with an element of $U(2,{\bf H})\cong Sp(2)$ that commutes 
with quaternionic conjugation, which is the rotation mentioned above.)  
It can be shown \cite{GAUgpt202} that $(i)$ for $\theta=0,\pi$, $1/2$ of 
supersymmetry is preserved since the second spinor constraint 
is trivially satisfied, $(ii)$ for $\theta=\pm\pi/2$, $1/4$ 
is preserved, and $(iii)$ for all other values of $\theta$, 
$3/16$ is preserved. 

Finally, we discuss intersecting (overlapping) $p$-branes related 
to (\ref{mfbrhypkalel}).  The compactification on one of the $U(1)$ 
isometry directions followed by the $T$-duality along the other 
$U(1)$ isometry direction yields the following type-IIB solution:
\begin{eqnarray}
ds^2_E&=&(H_1H_2U_1U_2)^{3\over 4}\left[(U_1U_2H_1H_2)^{-1}ds^2
({\bf E}^{1,1})+(U_2H_2)^{-1}d{\bf x}^1\cdot d{\bf x}^1\right.
\cr
& &\ \left.+(U_1H_1)^{-1}d{\bf x}^2\cdot d{\bf x}^2+(H_1H_2)^{-1}dz^2
+(U_1U_2)^{-1}dy^2\right], 
\cr
B&=&A_1\wedge dz+\tilde{A}_2\wedge dy, \ \ 
B^{\prime}=A_2\wedge dz+\tilde{A}_1\wedge dy,\ \ 
\tau=i\sqrt{{H_1U_2}\over{H_2U_1}}.
\label{iibfromgenintmf}
\end{eqnarray}
This solution represents 2 NS-NS 5-branes  
in the planes $(1,2,3,4,5)$ and $(1,6,7,8,9)$ and 2 R-R 5-branes 
in the planes $(1,5,6,7,8)$ and $(1,2,3,4,9)$ intersecting orthogonally.  
Since the spinor constraint associated with one of the constituent 
$p$-branes is expressed as a combination of the rest three 
independent spinor constraints, the solution preserves $({1\over 2})^3$ 
of supersymmetry.   Related intersecting $p$-brane is constructed by 
applying $T$-duality, oxidation and dimensional reduction.  One can 
further include additional $p$-branes without 
breaking any more supersymmetry, provided the spinor constraints of 
the added $p$-branes can be expressed as a combination of spinor 
constraints of the existing $p$-branes.

\subsection{Dimensional Reduction and Higher Dimensional Embeddings}
\label{bprhgh}

The lower-dimensional ($D<10$) $p$-branes can be obtained 
from those in $D=10,11$ through dimensional reduction.   
Reversely, most of lower-dimensional $p$-branes are related 
to $D=10,11$ $p$-branes via dimensional reduction and dualities.  
In particular, many black holes in $D<10$ originate from 
$D=10,11$ $p$-brane bound states, which makes it 
possible to find microscopic origin of black hole entropy.  
It is purpose of this section to discuss various $p$-branes 
in $D<10$.  We also discuss various $p$-brane embeddings of black holes.  

There are two ways of compactifying $p$-branes to lower dimensions.  
First, one can compactify along a longitudinal direction.   
It is called the `double dimensional reduction' (since both worldvolume 
and spacetime dimensions are reduced, bringing a $p$-brane 
in $D$ dimensions to $(p-1)$-brane in $D-1$ dimensions diagonally 
in the $D$ versus $p$ brane-scan) or `wrapping' of 
branes (around cycles of compactification manifold).  
Since target space fields are independent of longitudinal coordinates, 
one only needs to require periodicity of fields in the compactification 
directions.   
Second, one can compactify a transverse direction of a $p$-brane.   
It is called the `direct dimensional reduction' 
(since this takes us vertically on the bran-scan, taking a $p$-brane 
in $D$ dimensions to a $p$-brane in $D-1$ dimensions) or 
`constructing periodic arrays' of $p$-branes (along the compactified 
direction).   
Since fields depend on transverse coordinates, direct dimensional reduction 
is more involved \cite{KHU387,GAUhl,LUps481}.  
For this purpose, one takes periodic array of parallel $p$-branes 
(with the period of the size of compact manifold) along the transverse 
direction
\footnote{Such staking-up procedure breaks down for 
$(D-3)$-branes in $D$ dimensions, due to conical asymptotic 
spacetime \cite{LUps481}.  This is also related to the fact that 
$(D-2)$-branes reside in massive supergravity, rather than ordinary 
massless type-II theory.}.   
Then, one takes average over the transverse coordinate, 
integrating over continuum of charges distributed over the  
transverse direction.  The resulting configuration 
is independent of the transverse coordinate,  
making it possible to apply standard Kaluza-Klein dimensional 
reduction
\footnote{Or one can promote such isometry directions as the spatial 
worldvolume of a $p$-brane, leading to an intersecting $p$-brane solution 
\cite{GAUkt478,TSE475,LUps481,KHVklp388}}.  
In the double [direct] dimensional reduction, the values of 
$\tilde{p}$ [$p$] and $\Delta$ are preserved; in the direct dimensional 
reduction, the asymptotic behavior of the fields 
(which goes as $\sim 1/|\vec{y}|^{\tilde{p}}$) changes. 

Conventionally, the direct dimensional reduction uses 
the zero-force property of BPS $p$-branes, which allows 
the construction of multicentered $p$-branes.  
Note, however that it is also possible to apply the vertical dimensional 
reduction even for non-BPS extreme $p$-branes \cite{LUps481} and 
non-extreme $p$-branes \cite{LUpx489}, contrary to the 
conventional lore.   Namely, since the equations of motion 
(of a non-extreme, axially symmetric black $(D-4)$-brane in $D$ dimensions, 
for the non-extreme case) can be reduced to Laplace equations in the 
transverse space with suitable choice of field Ans\"atze, one can still 
construct multi-center $p$-branes for non-BPS and non-extreme 
cases as well.  
For the non-extreme case, when an {\it infinite} number of non-extreme 
$p$-branes are periodically arrayed 
along a line, the net force on each $p$-brane is zero and 
the conical singularities along the axis of periodic array act 
like ``struts'' that hold the constituents in place.  Furthermore, 
since the direction of periodic array is compactified on $S^1$ with 
each $p$-brane precisely separated by the circumference of $S^1$, 
the instability problem of such a configuration can be overcome.  

One can also lift $p$-branes as another $p$-branes in 
higher-dimensions, so-called `dimensional oxidation'.  
First, the oxidation of a $p$-brane in $D$ dimensions to a $p$-brane 
in $D+1$ dimensions 
(i.e. the reverse of the direct dimensional reduction) is never possible in 
the standard KK dimensional reduction, since the oxidized $p$-brane 
in $D+1$ dimensions has to depend on the extra transverse 
coordinate introduced by oxidation
\footnote{Such a $p$-brane in $D$ dimensions should rather be viewed as a 
$p$-brane in $D+1$ dimensions whose charge is uniformly distributed along 
the extra coordinate.  This is interpreted as the limit where one of 
charges of intersecting two $p$-branes in $D+1$ dimensions is zero 
\cite{LUps481}.}.  
(This is analogous to the `dangerous' $T$-duality transformation.)  
Second, the oxidation of a $p$-brane in $D$ dimensions to a $(p+1)$-brane 
in $D+1$ dimensions (i.e. the reverse of the double dimensional reduction)
is classified into two groups.  A $p$-brane in $D$ dimensions is called 
`rusty' if it can be oxidized to a $(p+1)$-brane in $D+1$ 
dimensions.  Otherwise, it is called `stainless'.  
Thus, $p$-branes in bran scan are KK descendants of stainless 
$p$-branes in some higher dimensions
\footnote{Contrary to the conventional lore that all 
the $p$-branes in $D<10$ are obtained from those in $D=10,11$ through 
dimensional reductions, there are stainless $p$-branes in $D<10$ which 
cannot be viewed as dimensional reductions of $p$-branes in 
$D=10,11$. So, the conventional brane scan is modified \cite{LUpss456}.}.   
A $p$-brane in $D$ dimensions is `stainless' when 
($i$) there is not the antisymmetric tensor in $D+1$ 
dimensions that the corresponding $(p+1)$-brane couples to, or 
($ii$) 
the exponential prefactors $a_{D+1}$ and $a_D$ for $(p+2)$-form 
field strength kinetic terms in $D+1$ and $D$ dimensions do not satisfy 
the relation $a^2_{D+1}=a^2_D-{{2(\tilde{p}+1)^2}\over{(D-2)(D-1)}}$.  
(This relation is satisfied by the expression for $a$ in 
(\ref{expprefactor}), provided $\Delta$ remains 
unchanged in the dimensional reduction procedure.)  

In this section, we focus on $p$-branes in $D<10$ with only field strengths 
of the same rank turned on, comprehensively studied in 
\cite{LUpss456,LUps481,LUp465,LUp12,LUpx489,DUFlp382,LUmp224,LUpx11,LUp177,LUps476}.  
A special case is black holes, which are 0-brane bound states.  
We also discuss their supersymmetry properties and interpretations as bound 
states of higher-dimensional $p$-branes.  

\subsubsection{General Solutions}\label{bprhghsol}  

We concentrate on $p$-branes in $D=11$ supergravity on tori.  
Bosonic Lagrangian of $D=11$ supergravity is 
\begin{equation}
{\cal L}_{11}=\sqrt{-\hat{G}}\left[{\cal R}_{\hat{G}}-
{1\over{48}}\hat{F}^2_4\right]+{1\over 6}\hat{F}_4\wedge\hat{F}_4\wedge
\hat{A}_3,
\label{elevendimsglag}
\end{equation}
where 
$\hat{F}_4=d\hat{A}_3$ is the field strength of the 3-form potential 
$\hat{A}_3$.  So, such $p$-branes have interpretation in terms of 
$M$-theory or type-II string theory configurations.  
Although one can directly reduce the $D=11$ action 
down to $D<11$ by compactifying on $T^{11-D}$ in one step, 
it turns out to be more convenient to reduce the action 
(\ref{elevendimsglag}) one dimension at a time iteratively until one reaches 
$D$ dimensions.  Namely, one compactifies $11-D$ times on $S^1$, making use 
of the following KK Ansatz:
\begin{eqnarray}
ds^2_{D+1}&=&e^{2\alpha\varphi}ds^2_D+e^{-2(D-2)\alpha\varphi}
(dz+{\cal A}_1)^2,
\cr
A_n(x,z)&=&A_n(x)+A_{n-1}(x)\wedge dz,
\label{circlekkcompac}
\end{eqnarray}
where $\varphi$ is a dilatonic scalar, ${\cal A}_1={\cal A}_{\mu}
dx^{\mu}$ is a KK 1-form field, $A_n$ is an $n$-form field arising 
from $\hat{A}_3$ and $\alpha\equiv 1/\sqrt{2(D-1)(D-2)}$.  
The explicit form of resulting $D<11$ action can be found elsewhere 
\cite{LUp465}.  The advantage of such compactification procedure is 
that spin-0 fields are manifestly divided into two classes in the 
Lagrangian.  
Namely, only dilatonic scalars $\vec{\phi}=(\phi_1,...,\phi_{11-D})$ 
appear in the exponential prefactors of the $n$-form field kinetic terms.  
The dilatonic scalars originate from the diagonal components 
of the internal metric and are true scalars.  The couplings of 
$\vec{\phi}$ to $n$-form potentials $A^{\alpha}_n$ are characterized by 
the ``dilaton vectors'' $\vec{a}_{\alpha}$ in the following way:
\begin{equation}
e^{-1}{\cal L}_{n-{\rm form}}=-{1\over{2n!}}\sum_{\alpha}
e^{\vec{a}_{\alpha}\cdot\vec{\phi}}(F^{\alpha}_{n+1})^2.
\label{nformkinwithdil}
\end{equation}
The complete expressions for ``dilaton vectors'', which are expressed as 
linear combinations of basic constant vectors, are found in \cite{LUp465}. 
On the other hand, the remaining spin-0 fields coming from the off-diagonal 
components of $\hat{G}_{MN}$ and the internal components of $\hat{A}_3$ are 
axionic, being associated with constant shift symmetries, and should rather 
be called 0-form potentials, which couple to solitonic $(D-3)$-branes. 
(Note, there are no elementary $p$-branes for 1-form field strengths.)  

Up to present time, study of $p$-branes within the above 
described theory has been mostly concentrated on the case where  
only $n$-form potentials of the same rank are turned on, with the 
restrictions that terms related to the last term in (\ref{elevendimsglag}) 
(denoted as ${\cal L}_{FFA}$ from now on) and the ``Chern-Simons'' 
terms in $(n+1)$-form field strengths are zero.  These restrictions 
place constraints on possible charge configurations for $p$-branes.  
These constraints become non-trivial when a $p$-brane  
involves both undualized and dualized field strengths, i.e. when 
the $p$-brane has both electric and magnetic charges coming from 
different field strengths. 
The former
\footnote{In particular, to set the 0-forms $A^{ijk}_0$ 
to zero consistently with their equations of motion, the bilinear 
products of field strengths that occur multiplied by $A^{ijk}_0$ 
in ${\cal L}_{FFA}$ should vanish.} 
[later] type of constraint is satisfied as long as the dualized 
and undualized field strengths have [do not have] common internal 
indices $i,j,k$.  

\paragraph{Supersymmetry Properties}\label{bprhghsolsusy}

The supersymmetry preserved by $p$-branes is determined from  
the Bogomol'nyi matrix ${\cal M}$, which is defined by the commutator of 
supercharges $Q_{\epsilon}=
\int_{\partial\Sigma}\bar{\epsilon}\Gamma^{ABC}\psi_Cd\Sigma_{AB}$ 
per unit $p$-volume:
\begin{equation}
[Q_{\epsilon_1},Q_{\epsilon_2}]=\int_{\partial\Sigma}N^{AB}d\Sigma_{AB}
\sim\epsilon^{\dagger}_1{\cal M}\epsilon_2,
\label{bogmatpbran}
\end{equation}
where $N^{AB}=\bar{\epsilon}_1\Gamma^{ABC}\delta_{\epsilon_2}\psi_C$ 
is the Nester's form defined from the supersymmetry transformation 
rule of $D=11$ gravitino $\psi_A$:
\begin{equation}
N^{AB}=\bar{\epsilon}_1\Gamma^{ABC}D_C\epsilon_2+
\textstyle{1\over 8}\bar{\epsilon_1}\Gamma^{C_1C_2}\epsilon_2
F^{AB}_{\ \ C_1C_2}+\textstyle{1\over{96}}\bar{\epsilon}_1
\Gamma^{ABC_1\cdots C_4}\epsilon_2F_{C_1\cdots C_4}.
\label{11dsgnester}
\end{equation}
The first term in $N^{AB}$ gives rise to the ADM mass density 
and the last two terms respectively contribute 
to the electric and magnetic charge density terms in ${\cal M}$.  
The Bogomol'nyi matrix for the 11-dimensional supergravity on 
$(S^1)^{11-D}$ is in the form of the ADM mass density $m$ term plus 
the electric and magnetic Page charge density \cite{PAG28} (defined 
respectively as ${1\over{4\omega_{D-n}}}\int_{S^{D-n}}\star F_n$ and 
${1\over{4\omega_n}}\int_{S^n}F_n$) terms.   

Since the Bogomol'nyi matrix is obtained from the {\it Hermitian} 
supercharges, its eigenvalues are non-negative.  
The matrix ${\cal M}$ has zero eigenvalues for each component of 
unbroken supersymmetry associated with the Killing spinor 
$\epsilon$ satisfying $\delta_{\epsilon}\psi_A=0$.  Since 
$D=11$ spinor has 32 components, the fraction of preserved supersymmetry 
is ${k\over{32}}$, where $k$ is the number of 0 eigenvalues of 
${\cal M}$ (i.e. the nullity of the matrix ${\cal M}$) or equivalently 
the number of linearly independent Killing spinors.  
The amount of preserved supersymmetry 
is determined as follows.  First, one calculates the ADM mass density 
$m$ from the $p$-brane solutions.  Then, one plugs $m$, together with 
the Page electric and magnetic charge densities of the $p$-branes, 
into the Bogomol'nyi matrix.  
The multiplicity $k$ of 0 eigenvalues of the resulting matrix 
${\cal M}$ determines the fraction of supersymmetry preserved by 
the corresponding $p$-branes.  

\paragraph{Multi-Scalar $p$-Branes}\label{bprhghsolmult}

General $p$-branes with more than one non-zero $p$-brane charges
are called ``multi-scalar $p$-branes'', since such $p$-branes have
more than one non-trivial dilatonic scalars.
The Lagrangian density has the following truncated form:
\begin{equation}
{\cal L}=\sqrt{-g}\left[{\cal R}-{1\over 2}(\partial\vec{\phi})^2
-{1\over{2(p+2)!}}\sum_{\alpha}e^{\vec{a}_{\alpha}\cdot\vec{\phi}}
(F^{\alpha}_{p+2})^2\right],
\label{multscallagdens}
\end{equation}
where field strengths $F^{\alpha}_{p+2}=dA^{\alpha}_{p+1}$ are defined
without ``Chern-Simons'' modifications.
In this action, the rank $p+2$ of field strengths is assumed to not
exceed $D/2$, namely those with $p+2>D/2$ are Hodge-dualized.
This is justified by the fact that the dual of
field strength of an elementary [solitonic] $p$-brane
is identical to the field strength of solitonic [elementary]
$(D-p-2)$-brane, with the corresponding dilaton vector differing only by
sign.

We consider extreme $p$-branes with $N$ non-zero $(p+2)$-form field 
strengths, each of which is either elementary or solitonic,  
but not both.  
For the simplicity of calculations, the $SO(1,p-2)\times SO(D-p+1)$ 
symmetric metric Ansatz (\ref{pbransptimemet}) is assumed to 
satisfy $(p+1)A+(\tilde{p}+1)B=0$ \cite{LUp12}, so that the  
field equations are linear.    The $p$-brane solutions are then 
determined completely by the dot products $M_{\alpha\beta}\equiv
\vec{a}_{\alpha}\cdot\vec{a}_{\beta}$ of the dilaton vectors 
$\vec{a}_{\alpha}$ associated 
with non-zero $(p+1)$-form field strengths $F^{\alpha}_{p+2}$ 
($\alpha=1,...,N$).  In solving the equations, it is assumed that 
$M_{\alpha\beta}$ is invertible
\footnote{When $M_{\alpha\beta}$ is singular, 
analysis depends on the number of rescaling parameters \cite{LUp465}.  
The only new solution is the case $\sum_{\alpha}\vec{a}_{\alpha}=0$, 
which yields solutions with $a=0$ and $(F^{\alpha})^2=F^2/N$, 
$\forall\alpha$.},  
which requires the number $N$ of non-trivial $F^{\alpha}_{p+2}$ 
to be not greater than the number of the components in $\vec{\phi}$, 
i.e. $N\leq 11-D$.  
For such $p$-branes, only $N$ components 
$\varphi_{\alpha}\equiv\vec{a}_{\alpha}\cdot\vec{\phi}$ of $\vec{\phi}$ 
are non-trivial.  
If one further takes the Ansatz $-\epsilon\varphi_{\alpha}+2(p+1)A
\propto\sum_{\beta}(M^{-1})_{\alpha\beta}\varphi_{\beta}$, then   
$M_{\alpha\beta}$ takes the form:
\begin{equation}
M_{\alpha\beta}=4\delta_{\alpha\beta}-{{2(p+1)(\tilde{p}+1)}\over
{D-2}}.
\label{malphabeta}
\end{equation}
The conditions on fields that linearize the field equations 
and lead to $M_{\alpha\beta}$ of the form 
(\ref{malphabeta}) are also dictated by supersymmetry transformation 
rules for the BPS configurations.  
Thus, a necessary condition for ``multi-scalar $p$-branes'' to be  
BPS is for the dilaton vectors $a_{\alpha}$ associated with 
participating field strengths to satisfy (\ref{malphabeta}).  
The following extreme multi-scalar $p$-brane solution is obtained by 
taking further simplifying Ansatz discussed in \cite{LUp12}:
\begin{equation}
e^{{1\over 2}\epsilon\varphi_{\alpha}-(p+1)A}=H_{\alpha},\ \ 
ds^2=\prod^N_{\alpha=1}H^{-{{\tilde{p}+1}\over{D-2}}}_{\alpha}dx^{\mu}dx^{\nu}
\eta_{\mu\nu}+\prod^N_{\alpha=1}H^{{p+1}\over{D-2}}_{\alpha}dy^mdy^m,
\label{mutichargepbran}
\end{equation}
where harmonic functions $H_{\alpha}=1+{{\lambda_{\alpha}}\over
{\tilde{p}+1}}{1\over{y^{\tilde{p}+1}}}$ ($y\equiv\sqrt{y^my^m}$) are 
associated with $p$-branes carrying the Page charges 
$P_{\alpha}=\lambda_{\alpha}/4$, and the field strengths 
are given, respectively, for the electric and magnetic cases by
\begin{equation}
F^{\alpha}_{p+2}=dH^{-1}_{\alpha}\wedge d^{p+1}x,\ \ \ \ \ \ 
F^{\alpha}_{p+2}=\star(dH^{-1}_{\alpha}\wedge d^{p+1}x).
\label{mutscalpbrfidstr}
\end{equation}
The elementary and solitonic $p$-branes are related by 
$\vec{\phi}\to-\vec{\phi}$.  
The ADM mass density is the sum of the mass 
densities of the constituent $p$-branes, i.e. $m=\sum^N_{\alpha=1}
P_{\alpha}$.   Multi-center generalization is 
achieved by replacing harmonic functions by 
$H_{\alpha}=1+\sum_i{{\lambda_{\alpha,\,i}}\over{|\vec{y}-
\vec{y}_{\alpha,\,i}|^{\tilde{p}+1}}}$ \cite{KHVklp388}.  

\paragraph{Single-Scalar $p$-Branes}\label{bprhghsolsing}

We discuss the case where the bosonic Lagrangian 
for 11-dimensional supergravity on $(S^1)^{11-D}$ is consistently 
truncated to the following form with one dilatonic scalar and one  
$(p+2)$-form field strength \cite{LUpss456,LUp465}:
\begin{equation}
{\cal L}=\sqrt{-g}\left[{\cal R}-{1\over 2}(\partial\phi)^2
-{1\over{(p+2)!}}e^{a\phi}(F_{p+2})^2\right],
\label{onescalformlag}
\end{equation}  
where we parameterize the exponential prefactor $a$ in the form:
\begin{equation}
a^2=\Delta-{{2(p+1)(\tilde{p}+1)}\over{D-2}}.
\label{expprefactor}
\end{equation}
This expression for $a$ is motivated from (\ref{apdef}), now with an 
arbitrary parameter $\Delta$ replacing 4.  
By consistently truncating (\ref{multscallagdens}), one has 
(\ref{onescalformlag}) with $(F_{p+2})^2\equiv\sum_{\alpha}
(F^{\alpha}_{p+2})^2$ and $a,\phi$ given by (for the case 
$M_{\alpha\beta}$ is invertible) 
\begin{equation}
a^2=\left(\sum_{\alpha,\beta}(M^{-1})_{\alpha\beta}\right)^{-1},
\ \ \ \ \ 
\phi=a\sum_{\alpha,\beta}(M^{-1})_{\alpha\beta}\vec{a}_{\alpha}
\cdot\vec{\phi}.
\label{aphisingmult}
\end{equation}

By taking Ans\"atze which reduce equations of motion for 
(\ref{onescalformlag}) to the first order, one obtains \cite{LUpss456} 
the ``single-scalar $p$-brane'' solution with the Page charge 
density $P=\lambda/4$: 
\begin{equation}
e^{\phi}=H^{{2a}\over{\epsilon\Delta}},\ \ \ 
ds^2=H^{-{{4(\tilde{p}+1)}\over{\Delta(D-2)}}}dx^{\mu}dx^{\nu}\eta_{\mu\nu}
+H^{{4(p+1)}\over{\Delta(D-2)}}dy^mdy^m,
\label{singscalpbrn}
\end{equation}
where $H=1-{{\sqrt{\Delta}\lambda}\over{2(p+1)}}{1\over{r^{\tilde{p}+1}}}$.   
The mass density is $m={{\lambda}\over{2\sqrt{\Delta}}}$.  Although 
inequivalent charge configurations give rise 
to the same $\Delta$, i.e. the same solution, supersymmetry 
properties depend on charge configurations. 

Note, although one can obtain $p$-brane solutions (\ref{singscalpbrn}) 
for any values of $p$ and $a$, hence for any values of $\Delta$, 
only specific values of $p$ and $a$ can occur in supergravity theories.  
The value of $\Delta$ is preserved 
in the compactification process, provided no fields are truncated.  

For $p$-branes with 1 constituent, $\Delta$ is always 4, as can be 
seen from the form of $a$ in (\ref{apdef}), determined by 
the requirement of scaling symmetry
\footnote{For $p$-branes with $\Delta\neq 4$, the scaling symmetry 
of combined worldvolume and effective supergravity actions is broken.},  
and always $1/2$ of supersymmetry is preserved.    
The value $\Delta=4$ can also be understood from the facts that 
$\Delta=4$ in $D=11$, since there is no dilaton in $D=11$, and 
the value of $\Delta$ is preserved in dimensional reduction  
not involving field truncations. 

When the field strength 
is a linear combination of more than one original field strengths, 
$\Delta<4$.  With all the Page charges $P_{\alpha}=\lambda_{\alpha}/4$ 
of ``multi-scalar $p$-brane'' (\ref{mutichargepbran}) equal, 
one has ``single-scalar $p$-brane'' with the Page charge 
$P=\lambda/4$ ($\lambda=\sqrt{N}\lambda_{\alpha}$).  
By substituting $M_{\alpha\beta}$ in (\ref{malphabeta}) into 
(\ref{aphisingmult}), one has $\Delta=4/N$. 
So, ``single-scalar $p$-branes'' with $\Delta=4/N$ ($N\geq 2$) 
are bound states of $N$ single-charged $p$-branes (with $\Delta=4$) 
with zero binding energy, and preserve the same fraction 
of supersymmetry as their multi-scalar generalizations.  Only ``single-scalar 
$p$-branes'' with $\Delta=4/N$ ($N\in{\bf Z}^+$) and ``multi-scalar 
$p$-branes'' can be supersymmetric. (Non-supersymmetric 
$p$-branes in this class is related to supersymmetric ones by 
reversing the signs of certain charges.)  
And only single-scalar $p$-branes with $\Delta=4/N$ ($N\geq 2$) have 
multi-scalar generalizations. 

\paragraph{Dyonic $p$-Branes}\label{bprhghsoldyo}

In $D=2(p+2)$, $p$-branes can carry both electric and magnetic charges 
of $(p+2)$-form field strengths.  There are two types of dyonic $p$-branes 
\cite{LUp465}: 
$(i)$ the first type has electric and magnetic charges 
coming from different field strengths, $(ii)$ the second type has 
dyonic field strengths.   
As in the multi-scalar $p$-brane case, the requirements that 
${\cal L}_{FFA}=0$ and the Chern-Simons terms are zero place constraints 
on possible dyonic solutions in $D=2(p+2)=4,6,8$.  
Such restrictions rule out dyonic $p$-branes of the first type in 
$D=6,8$.  For the second type, dyonic $p$-brane in $D=8$ is special since 
it has non-zero 0-form potential $A^{(123)}_0$ \cite{IZQlpt}, thereby 
requiring non-zero source term $\hat{F}_{MNPQ}\hat{F}_{RSTU}
\epsilon^{MNPQRSTU}$, and can be obtained from purely electric/magnetic 
membrane by duality rotation, unlike dyonic $D=6$ string and $D=4$ 
0-brane of the second type.  
Dyonic $p$-branes of the second type include self-dual 
3-branes in $D=10$ \cite{HORs360,DUFl273}, self-dual string \cite{DUFl416} 
and dyonic string \cite{DUFfkl356} in $D=6$, and dyonic black hole 
in $D=4$ \cite{LUp465}.

There are two possible dyonic $p$-branes (associate with 
(\ref{onescalformlag})) of the second type
\footnote{The solutions for dyonic $p$-branes of the first type 
have the form (\ref{mutichargepbran}) with Lagrangian 
(\ref{multscallagdens}) containing both Hodge-dualized and undualized 
field strengths.} 
with the Page charge densities $\lambda_i/4$ \cite{LUp465}: 
$(1)$ $a^2=p+1$ case (i.e. $\Delta=2p+2$) with 
the solution
\begin{equation}
e^{-{1\over 2}a\phi-(p+1)A}=1+{{\lambda_1}\over{a\sqrt{2}}}{1\over{r^{p+1}}}
,\ \ \ \ 
e^{+{1\over 2}a\phi-(p+1)A}=1+{{\lambda_2}\over{a\sqrt{2}}}{1\over{r^{p+1}}},
\label{dyonpbr2nd1}
\end{equation}
and $(2)$ $a=0$ case (i.e. $\Delta=p+1$) with the solution
\begin{equation}
\phi=0,\ \ \ \ e^{-(p+1)A}=1+{1\over 2}\sqrt{{\lambda^2_1+\lambda^2_2}\over
{p+1}}{1\over{r^{p+1}}}.
\label{dyonpbr2nd2}
\end{equation}
The ADM mass density for (\ref{dyonpbr2nd1}) is 
$m={1\over{2\sqrt{\Delta}}}(\lambda_1+\lambda_2)$, 
whereas for (\ref{dyonpbr2nd2}) 
$m={1\over{2\sqrt{\Delta}}}\sqrt{\lambda^2_1+\lambda^2_2}$.  
The solution (\ref{dyonpbr2nd2}) is invariant under 
electric/magnetic duality and, therefore, is equivalent to 
the purely elementary ($\lambda_2=0$) or solitonic ($\lambda_1=0$) case. 
For (\ref{dyonpbr2nd1}) with $\lambda_1=\lambda_2$, the field 
strength is self-dual and $\phi=0$, thereby (\ref{dyonpbr2nd1}) and 
(\ref{dyonpbr2nd2}) are equivalent, but for (\ref{dyonpbr2nd2}) 
$\lambda_1$ and $\lambda_2$ are independent.  When $\lambda_1=-\lambda_2$, 
(\ref{dyonpbr2nd1}) corresponds to anti-self-dual massless string with 
enhanced supersymmetry.  

Note, the solutions (\ref{dyonpbr2nd1}) and (\ref{dyonpbr2nd2}) 
are not restricted to those obtained from the $D=11$ supergravity 
on tori.  For $D=8,6$ and 4, which are relevant for the $D=11$ 
supergravity on tori, $\Delta$'s for (\ref{dyonpbr2nd1}) and 
(\ref{dyonpbr2nd2}) are respectively $\{6,3\}$, $\{4,2\}$ 
and $\{2,1\}$.  So, (\ref{dyonpbr2nd1}) and (\ref{dyonpbr2nd1}) with 
$p=2$ (i.e. $D=8$) are excluded. 

\paragraph{Black $p$-Branes}\label{bprhghsolblck}

We discuss non-extreme $p$-branes.  Non-extreme $p$-branes are 
additionally parameterized by the non-extremality parameter $k>0$.  
There are two ways of constructing non-extreme $p$-branes.  

The first method involves a universal prescription for ``blackening'' 
extreme $p$-branes, which deforms extreme solutions with $e^{2f}=1-
{k\over{r^{\tilde{p}+1}}}$ ($r\equiv |\vec{y}|$) \cite{DUFlp382}:
\begin{equation}
dt^2\to e^{2f}dt^2,\ \ \ \ \ \ \ 
dr^2\to e^{-2f}dr^2,  
\label{blackening}
\end{equation}
while modifying harmonic functions associated with $p$-branes as 
$H=1+k\sinh 2\delta_{\alpha}/r^{\tilde{p}+1}\to 1+k\sinh^2
\delta_{\alpha}/r^{\tilde{p}+1}$.  
The resulting non-extreme $p$-branes, called ``type-2 non-extreme 
$p$-branes'', 
have an event horizon at $r=r_+=k^{1/(\tilde{p}+1)}$, which 
covers the singularity at the core $r=0$.  
The ADM mass density has the generic form 
$m\sim\sum_{\alpha}\sqrt{(Q_{\alpha})^2+k^2}$, which is always 
larger than the extreme counterpart,  
and all the supersymmetry is broken since the Bogomol'nyi bound is not 
saturated.  For type-2 non-extreme $p$-branes (with $p\geq 1$), 
the Poincar\'e invariance is broken down to ${\bf R}\times E^p$  
because of the extra factor $e^{2f}$ in the $(t,t)$-component of the metric.  
For 0-branes, the metric remains isotropic but the 
quantity $(p+1)A+(\tilde{p}+1)B$ no longer vanishes.  

In the second method, the metric Ansatz (\ref{pbransptimemet}) remains intact 
but instead general solution to the field equations  
is obtained \cite{LUpx11,LUmp224} without simplifying  
Ans\"atze, e.g. $(p+1)A+(\tilde{p}+1)B=0$, that linearize field 
equations.  (In solving the field equations without 
simplifying Ans\"atze, one encounters an additional integration constant 
interpreted as non-extremality parameter.) 
So, the resulting non-extreme $p$-branes, called ``type-1 non-extreme 
$p$-branes'', preserve the full Poincar\'e invariance (in the worldvolume) 
of extreme $p$-branes.  So, type-1 non-extreme $p>0$ solutions do not 
overlap with the type-2 non-extreme counterparts.  But type-1 non-extreme 
0-branes contain type-2 non-extreme 0-branes as a subset.  

The equations of motion for single-scalar $p$-branes and 
dyonic $p$-branes of the second type are, respectively, casted into 
the forms of the Liouville equation and the Toda-like equations for two 
variables, which are subject to the first integral constraint. 
The equations of motion for dyonic $p$-branes are solvable 
when $a^2=p+1$ (i.e. $\Delta=2(p+1)$)  or $a^2=3(p+1)$ (i.e. $\Delta=
4(p+1)$).  When $a^2=p+1$, the equations of motion are 
reduced to two independent Liouville equations.  Since $\Delta\leq 4$ 
in supergravity theories, only dyonic strings in $D=6$ and 
dyonic black holes in $D=4$ are relevant, with only dyonic strings having 
BPS limit.  When $a^2=3(p+1)$, the equations 
of motion are reduced to $SU(3)$ Toda equations.  Only dyonic black holes 
are possible in supergravity theory for this case.  In the extreme 
limit, such black holes preserve supersymmetry when either electric or 
magnetic charge is zero.

For multi-scalar $p$-branes with $N$ field strengths, the equations of 
motion are Toda-like in general, but when the extreme limit is 
BPS (i.e. dilaton vectors satisfy (\ref{malphabeta})) the 
equations of motion become $N$ independent Liouville equations.  
The requirements that non-extreme $p$-branes are asymptotically 
Minkowskian and dilatons are finite at the event horizon (thereby 
the event horizon is regular) place restrictions on parameters of the 
solutions.  

\paragraph{Massless $p$-Branes}\label{bprhghsolmssles}

For multi-scalar $p$-branes and a dyonic $p$-brane (\ref{dyonpbr2nd2}) 
of the second type, the ADM mass density has the form $m\sim
\sum_{\alpha}\lambda_{\alpha}$.  So, they can be massless when some of 
the Page charges are negative.  In this case, there are additional 0 
eigenvalues of the Bogomol'nyi matrix, enhancing supersymmetry.  
Generally massless $p$-branes are ruled out if one requires 
the Bogomol'nyi matrix to have only non-negative eigenvalues
\footnote{In the exceptional case of 4-scalar solution with 
2-form field strengths in $D=4$, it is possible to have massless 
$p$-branes where the Bogomol'nyi matrices have no negative 
eigenvalues \cite{LUp12}.}, 
since the Bogomol'nyi matrix is obtained from the commutator 
of the {\it Hermitian} supercharges.  Since some of the Page 
charges are negative, the massless $p$-branes have naked singularity.  
On the other hand, if one allows negative eigenvalues, one can have 
$p$-branes preserving more than $1/2$ of supersymmetry and some of 
non-BPS multi-scalar $p$-branes can become supersymmetric 
due to the appearance of 0 eigenvalues with suitable sign choice of 
Page charges (but their single-scalar counterparts are 
non-BPS, since Page charges have to be equal in the single-scalar 
limit) \cite{LUp12}.  

\subsubsection{Classification of Solutions}\label{bprhghclas}

In this subsection, we classify $p$-branes discussed in 
the previous subsection according to their supersymmetry properties.  
Since single-scalar $p$-branes and their multi-scalar generalizations 
preserve the same amount of supersymmetry (except for the special case 
of massless $p$-branes),  
the classification of multi-scalar $p$-branes is along the same line as 
that of single-scalar $p$-branes.  Single-scalar $p$-branes are 
supersymmetric only when $\Delta=4/N$ ($N\in{\bf Z}^+$) and the 
dilaton vectors $\vec{a}_{\alpha}$ (associated with the participating 
field strengths) of their multi-scalar $p$-brane counterparts satisfy 
the relations (\ref{malphabeta}).  

Spin-0 fields, i.e. dilatonic scalars and 0-form fields, form target 
space manifold of $\sigma$-model.  The target space manifold 
has a coset structure $G/H$, where $G\simeq E_{n(+n)}({\bf R})$ ($n=11-D$) 
is the (real-valued) $U$-duality group
\footnote{For $D\leq 6$, this is the case only when all the field 
strengths are Hodge-dualized to those with rank $\leq D/2$ \cite{LUp177}.}  
and $H$ is a linearly-realized maximal subgroup of $G$.  
Under the $G$-transformations, the equations of motion are invariant.  
When the DSZ quantization is taken into account, $G$ and $H$ break down 
to integer-valued subgroups.  The subgroup $G({\bf Z})$ is the 
conjectured $U$-duality group
\footnote{In the following, we also call the real-valued group 
$G$ as the $U$-duality group, but the distinction between $G({\bf R})$ 
and $G({\bf Z})$ will be clear from the context.} 
of type-II string on tori.  

The asymptotic values of spin-0 fields, called ``moduli'', 
define the ``scalar vacuum''.  The asymptotic values of dilatonic 
scalars and 0-form fields are, respectively, interpreted as the ``coupling 
constants'' and ``$\theta$-angles'' of the theory.  
One can parameterize spin-0 fields by a $G$-valued matrix 
$V(x)$ which transforms under rigid $G$-transformation by 
right multiplication and under local $H$-transformation by 
left transformation:  $V(x)\to h(x)V(x)\Lambda^{-1}$, 
$h(x)\in H$ and $\Lambda\in G$.  It is convenient to define 
a new scalar matrix $M\equiv V^TV$, which is 
inert under $H$ but transforms under $G$ as $M\to \Lambda M 
\Lambda^T$.   So, the $U$-duality group $G$ generally changes the 
``vacuum'' of the theory.  By applying a $G$-transformation, one can 
bring the asymptotic value of $M$ to the canonical form 
$M_{\infty}={\bf 1}$.  The subgroup $H$ leaves $M_{\infty}={\bf 1}$ 
intact (i.e. $H$ is the $U$-duality little group of the scalar vacuum), 
thereby acting as solution classifying isotropy group (of the vacuum) 
that organizes the distinct solutions of the theory 
into families of the same energy.   The integer-valued subgroup 
$G({\bf Z})\cap H$ is identified with the Weyl group of $G$ that 
transforms the set of dilaton vectors $\vec{a}_{\alpha}$ associated 
with field strengths of the same rank as weight vectors of the 
irreducible representations of $G({\bf Z})$.  

The $U$ Weyl group in $D\leq 9$ contains a subgroup $S_{11-D}$ 
consisting of the permutations of the internal coordinates 
($i\leftrightarrow j$), corresponding to the permutations of 
field strengths
\footnote{This permutation is a discrete subset of 
$G({\bf R})$ which acts on a field strength multiplet linearly.}, 
and (for $D\leq 8$) the additional discrete symmetries that 
interchange field strengths and the Hodge dualized field strengths 
(namely, the interchange of field strength equations of motion 
and Bianchi identity).  At the same time, the associated dilaton 
vectors transform in such a way that theory is invariant, respectively, 
by permutation or change of signs, forming an irreducible multiplet under 
the Weyl group.    

In particular, $M_{\alpha\beta}$ are invariant under the $U$ Weyl 
transformations and therefore $\Delta$ is also preserved.  Furthermore, 
since the Bogomol'nyi matrix ${\cal M}$ is invariant under the $U$ Weyl 
group, $p$-branes with the same eigenvalues of ${\cal M}$ and, therefore, 
the same supersymmetry property are related by the $U$ Weyl group. 
Starting with a $p$-brane with a set of $\vec{a}_{\alpha}$, one generates 
a $U$ Weyl group multiplet of $p$-branes with the same $M_{\alpha\beta}$ 
(or same $\Delta$) and the same eigenvalues of ${\cal M}$.  
In the case of multi-scalar or dyonic $p$-branes, where 
the $N$ Page charges are independent parameters, the size of 
the $U$ Weyl multiplet is larger than that of single-scalar 
$p$-brane counterparts, since the participating field strengths 
are now distinguishable.  

We classify $p$-branes according to the rank of field strengths that 
$p$-branes couple to \cite{LUp465,LUp12}.  
Particularly, BPS $p$-branes are possible with $N=1$ 
4-form/3-form field strength, $N\leq 4$ 2-form field strengths 
and $N\leq 7$ 1-form field strengths.  BPS $p$-branes with 
$N$ participating field strengths appear in lower dimensions once they 
occur in some higher dimensions; the $p$-branes in those maximal 
dimensions are ``stainless super $p$-branes''.   Generally, 
BPS $p$-branes with $\Delta=4,2,{4\over 3}$ respectively 
preserve ${1\over 2}$, ${1\over 4}$, ${1\over 8}$ of  
supersymmetry, and 0-branes with $\Delta={4\over 5},{2\over 3},{4\over 
7}$, which occur only in $D=4$, all preserve $1\over{16}$.  As for the 
super $p$-branes with 4 field strengths, there are two inequivalent 
solutions: $(i)$ those that preserve $1\over 8$ of supersymmetry 
(denoted $\Delta=1^{\prime}$) and are coupled to 2-form and 1-form field 
strengths $(ii)$ those that preserve $1\over {16}$ (denoted $\Delta=1$) 
and are coupled to 1-form field strengths, only.  
Lastly, we show that $H$-transformations on black holes discussed 
in chapter \ref{n4bh} generate the most general black holes in 
$D=11$ supergravity on tori \cite{CVEh}. 

\paragraph{4-Form Field Strength}\label{bprhghclas4f} 

There is only one 4-form field strength in each dimension, but 
within the supergravity models under consideration in this 
section, the 4-form field strength exists only in $D\geq 8$, since 
those in $D<8$ are Hodge-dualized to lower ranks.  
So, no multi-scalar generalization is possible.  
There is a unique single-scalar $p$-brane, which is either elementary 
membrane or solitonic $(D-6)$-brane.  
In $D=8$, one can construct dyonic membrane of the second type 
(\ref{dyonpbr2nd1}), but it is ruled out by the constraint 
${\cal L}_{FFA}=0$.  

\paragraph{3-Form Field Strengths}\label{bprhghclas3f}

There are $11-D$ 3-forms in $D\geq 6$, except in $D=7$ where 
there is an extra 3-form coming from the Hodge-dualization of 
the 4-form.   The associated dilaton vectors 
satisfy $M_{\alpha\beta}=2\delta_{\alpha\beta}-{{2(D-6)}\over{D-2}}$, 
which are not of the form (\ref{malphabeta}), and, therefore, the 
multi-scalar generalization is not possible.   In fact, this  
expression for $M_{\alpha\beta}$ yields $\Delta=2+{2\over N}$ in the 
limit $F^2_{\alpha}=F^2/N,\forall\alpha$: supersymmetry is completely 
broken unless $N=1$ (i.e. $\Delta=4$), in which case $1/2$ of 
supersymmetry is preserved.  
In $D=6$, one can construct dyonic strings.  
Due to the constraint ${\cal L}_{FFA}=0$, 
only dyonic strings of the second type, which are (\ref{dyonpbr2nd1}) and 
(\ref{dyonpbr2nd2}) with $\Delta=4$ and 2, respectively, are possible.  

\paragraph{2-Form Field Strengths}\label{bprhghclas2f}

Two-form field strengths couple to elementary 0-branes and 
solitonic $(D-4)$-branes.  Analysis of 2-form field strengths and 
1-form field strengths is complicated due to their proliferation in 
lower dimensions.  We therefore discuss only supersymmetric cases; 
complete classification of $p$-branes including non-supersymmetric 
ones can be found in \cite{LUp465,LUp12}.  
The dilaton vectors $\vec{a}_{\alpha}$ associated with $N$ participating 
2-form field strengths satisfy (\ref{malphabeta}) only for $N\leq 4$.  
The dimensions $D$ in which these BPS $p$-branes with 
$N=1,2,3,4$ 2-form field strengths appear are, respectively, 
$D\leq 10,9,5,4$.  
For $N=1,2,3$, the $p$-branes preserve $2^{-N}$ of supersymmetry, and 
for the $N=4$ case, the solutions preserve $1/8$.  Whereas $p$-branes 
with $N\leq 3$ can be either purely electric/magnetic or dyonic 
(of the first type), $p$-branes with $N=4$ are intrinsically dyonic 
(of the first type).  

In $D=4$, there are 4 inequivalent BPS black holes with 
$\Delta={4\over N}$ ($N=1,2,3,4$), corresponding to dilaton couplings 
$a=\sqrt{3},1,{1\over\sqrt{3}},0$, respectively
\footnote{Note, only for these values of $a$, the 0-branes  
have a regular event horizon.}.  
These black holes are interpreted as bound states of $N$ $D=5$ KK 
black holes with $a=\sqrt{3}$ \cite{RAH372,DUFr481}.  

In $D=4$, dyonic 0-branes of both first and second types satisfy the 
constraint ${\cal L}_{FFA}=0$.  Discussion on the first type is along 
the same line as the multi-scalar 0-branes.  As for the second type, 
we have solutions (\ref{dyonpbr2nd1}) and 
(\ref{dyonpbr2nd2}) with $a=1/\sqrt{3}$ (i.e. $N=2$) and $a=0$ 
(i.e. $N=4$), respectively.  First, $a=0$ case is 
intrinsically dyonic of the first type even when $\lambda_1=0$ or 
$\lambda_2=0$.  Although the explicit forms of solutions are 
insensitive to signs of Page charges, their supersymmetry 
properties depend on their relative signs.  
Second, the supersymmetry property of $a=1/\sqrt{3}$ case is insensitive 
to the signs of Page charges.  Supersymmetry is preserved when $(i)$ 
$\lambda_1=0$ or $\lambda_2=0$, corresponding to purely 
solitonic or elementary solution preserving $1/4$ of 
supersymmetry, or $(ii)$ $\lambda_1=-\lambda_2$, corresponding to 
a massless black hole preserving $1/2$.  

\paragraph{1-Form Field Strengths}\label{bprhghclas1f}

1-form field strengths couple to solitonic $(D-3)$-branes, only.  
The dilaton vectors satisfy (\ref{malphabeta}) for $N\leq 7$ 
participating field strengths.   
$N=1,2,3$ cases occur, respectively, in $D\leq 9,8,6$, whereas 
$N=5,6,7$ cases occur in $D=4$, only.  As for the $N=4$ case, 
there are 2 inequivalent BPS solutions: ($i$) those occurring 
in $D\leq 6$, denoted $N=4^{\prime}$ or $\Delta=1^{\prime}$ and 
($ii$) those occurring only in $D=4$, denoted $N=4$ or $\Delta=1$.  
For generic values of Page charges, $p$-branes preserve $2^{-N}$ 
[${1\over{16}}$] of supersymmetry for $N=1,2,3,4$ [$N=5,6,7$].  
The $N=4^{\prime}$ case preserves $1\over 8$.  
In the case $\Delta=1^{\prime},{4\over 5},{2\over 3},{4\over 7}$, 
$p$-branes are BPS or non-BPS, depending on the  
signs of the Page charges.  However, for $\Delta=4,2,{4\over 3},1$,  
their supersymmetry properties are independent of the Page charge signs. 

\paragraph{Black Holes in $4\leq D\leq 9$}\label{bprhghclasbck}

The 0-branes in $D=11$ supergravity on tori 
with the most general charge configurations can be obtained by 
applying subsets of $U$-duality transformations on 
the generating solutions.  
As in the case of black holes in heterotic string theory on tori, 
the set of transformations that generate the general 
black holes with the canonical asymptotic value of scalar matrix 
$M_{\infty}={\bf 1}$ from the generating solutions is of the form $H/H_0$, 
where $H_0$ is the largest subgroup of $H$ that leaves the generating 
solutions intact.  The $H/H_0$ transformation introduces 
${\rm dim}(H)-{\rm dim}(H_0)$ parameters, which together with the 
parameters of the generating solutions form the complete parameters of the 
most general solution.  The number of $U(1)$ charges of the generating 
solutions are 5, 3, 2 for $D=4,5,\geq 6$, respectively.  The charge 
configurations for these generating solutions are the same as the 
heterotic case in chapter \ref{n4bh}, with all the charges coming from 
the NS-NS sector.  To generate solutions with an arbitrary 
asymptotic value of the scalar matrix $M$, one additionally imposes a 
general (real-valued) $U$-duality transformation.   

The ``dressed'' 0-brane charge $\bar{\cal Z}=V_{\infty}{\cal Z}$ 
can be rearranged in an $N\times N$ anti-symmetric complex matrix
\footnote{For 0-branes in $D=4$, the matrix $Z_{4\,AB}$ 
is defined as follows \cite{CREj80,HULt438}.   The electric $q_I$ and 
magnetic $p_I$ charges of the $U(1)^{28}$ gauge group of the $N=8$, $D=4$ 
supergravity are combined into a 56-vector ${\cal Z}^T=(p^I,q_I)$, 
which transforms under $G$ as ${\cal Z}\to \Lambda{\cal Z}$. 
The dressed 0-brane charge $\bar{\cal Z}=V_{\infty}{\cal Z}=
(\bar{p}^I,\bar{q}_I)^T$ is invariant under $G$ but transforms under 
local $SU(8)$.  The central charge matrix $Z_{4\,AB}$, which is the 
complex antisymmetric representation of $SU(8)$, that appears in the $N=8$, 
$D=4$ supersymmetry algebra is related to the ``dressed'' charges 
$\bar{q}_I$ and $\bar{p}^I$ as $Z_{4\,AB}=(\bar{q}_I+i\bar{p}^I)t^I_{AB}$, 
where $t^I_{AB}=-t^I_{BA}$ are the generators of $SO(8)$.}  
$Z_{D\,AB}$ ($A,B=1,...,N$), where $N$ is the number of 
the maximal supersymmetry in $D$ dimensions.  
$\bar{\cal Z}$ and $Z_{D\,AB}$ are invariant under 
the global $G$-transformation but transform under the local 
$H$-transformation.
$Z_{D\,AB}$ appears in the supersymmetry algebra in the form 
$[Q_{A\alpha},Q_{B\beta}]=C_{\alpha\beta}Z_{D\,AB}$.  
In general, $Z_{D\,AB}$ is splitted into blocks of 
${N\over 2}\times {N\over 2}$ submatrices.  Two diagonal blocks 
$Z_{R,L}$ correspond to NS-NS charges and two off-diagonal blocks 
represent R-R charges.  By applying the $H$-transformation 
$Z_D\to Z^0_D=hZ_Dh^T$ ($h\in H$), one can bring the matrix $Z_D$ into 
the skew-diagonal form with complex skew eigenvalues $\lambda_i$ 
($i=1,...,N/2$).  These eigenvalues $\lambda_i$ are related to charges 
of the generating solutions in a simple way, which we show in the 
following.  

We now discuss the subsets of $H$ that generate 0-branes with the most 
general charge configurations from the generating solutions.
\begin{itemize}
\item $D=4$: 
The most general 0-brane carries $28+28$ electric and magnetic charges 
of the $U(1)^{28}$ gauge group.  The $U$-duality group is $G=E_7$ with 
the maximal compact subgroup $H=SU(8)$.  The skew eigenvalues $\lambda_i$ 
($i=1,...,4$) are related to the charges $Q_{1\,R,L}\equiv Q_1\pm Q_2$, 
$P_{2\,R,L}\equiv P_1\pm P_2$ and $q$ of the generating solution as
\begin{eqnarray}
\lambda_1&=&Q_{1\,R}+P_{2\,R},\ \ \ \ \ \ \ \ \ \ 
\lambda_2=Q_{1\,R}-P_{2\,R},
\cr
\lambda_3&=&Q_{1\,L}+P_{2\,L}+2iq,\ \ \ 
\lambda_4=Q_{1\,L}-P_{2\,L}-2iq.
\label{4dgenchtoskew}
\end{eqnarray}
The subset of $H=SU(8)$ that leaves the generating solution unchanged 
is $SO(4)_L\times SO(4)_R$.  The $63-(6+6)=51$ parameters of 
$H/H_0=SU(8)/[SO(4)_L\times SO(4)_R]$ are introduced to 
the generating solution.  
\item $D=5$:  
The ``dressed'' 27 electric charges of the most general 0-brane transform 
as a $\bf 27$ of the $USp(8)$ maximal compact subgroup of $U$-duality 
group $E_6$.  The skew eigenvalues $\lambda_i$ ($i=1,...,4$) with 
a constraint $\sum^4_{i=1}\lambda_i=0$ are related to the charges 
$Q_{1\,R,L}\equiv Q_1\pm Q_2$ and $Q$ of the generating solution as
\begin{equation}
\lambda_1=Q+Q_{1\,R},\  
\lambda_2=Q-Q_{1\,R},\  
\lambda_3=-Q+Q_{1\,L},\ 
\lambda_4=-Q-Q_{1\,L}.
\label{5dgenchtoskew}
\end{equation}
The subset $SO(4)_L\times SO(4)_R\subset USp(8)$ leaves this charge 
configuration intact.  The $USp(8)/[SO(4)_L\times SO(4)_R]$ transformation 
introduces remaining $36-12=24$ charge degrees of freedom into the 
generating solution. 
\item $D=6$:  
The most general 0-brane carries 16 electric charges, which transform 
as a $\bf 16$ (spinor) of the $SO(5,5)$ $U$-duality group, whereas 
the ``dressed'' charges transform as $({\bf 4},{\bf 4})$ under 
the maximal compact subgroup $SO(5)\times SO(5)$.  
The skew eigenvalues $\lambda_i$ ($i=1,2$) are related to the charges 
$Q_{1\,R,L}\equiv Q_1\pm Q_2$ as
\begin{equation}
\lambda_1=Q_{1\,R},\ \ \ \ \ \lambda_2=Q_{1\,L}.
\label{6dgenchtoskew}
\end{equation}
The subgroup $SO(3)_L\times SO(3)_R$ of the maximal compact subgroup 
$SO(5)\times SO(5)$ leaves the generating solution intact.  
The transformation $[SO(5)\times SO(5)]/[SO(3)_L\times SO(3)_R]$ 
introduces remaining $2(10-3)=14$ charge degrees of 
freedom. 
\item $D=7$:  
The most general 0-brane carries 10 electric charges, which transform  
as a $\bf 10$ under the $SL(5,{\bf R})$ $U$-duality group, whereas 
the ``dressed'' charges also transform as ${\bf 10}$ under $SO(5)$.  
The skew eigenvalues $\lambda_i$ ($i=1,2$) are related to the charges 
$Q_{1\,R,L}\equiv Q_1\pm Q_2$ in the same way as the $D=6$ case.  
The subgroup $SO(2)_L\times SO(2)_R$ of the maximal compact subgroup 
$SO(5)$ preserves the generating solution.  $10-2=8$ parameters 
of $SO(5)/[SO(2)_L\times SO(2)_R]$ are introduced into the generating 
solution. 
\item $D=8$:  
6 electric charges of the general 0-brane transform as $({\bf 3},
{\bf 2})$ under the $U$-duality group $SL(3,{\bf R})\times SL(2,{\bf R})$. 
There is no subgroup of the maximal compact subgroup $SO(3)\times U(1)$ 
that leaves the generating solution intact.  The $SO(3)\times U(1)$ 
transformation induces $3+1=4$ remaining charge degrees of freedom 
into the generating solution. 
\item $D=9$:  
4 electric charges of the general 0-brane transform as $({\bf 3},
{\bf 1})$ under the $U$-duality group $SL(2,{\bf R})\times {\bf R}^+$.  
The maximal compact subgroup $U(1)$ introduces an additional electric 
charge into the generating solution. 
\end{itemize}

Since the equations of motion and, especially, the Einstein frame metric  
are invariant under the $U$-duality, it is natural to expect that 
quantities derived from the metric, e.g. the ADM mass and the 
Bekenstein-Hawking entropy, are $U$-duality invariant.  In fact, one 
can express the Bekenstein-Hawking entropy in a manifestly $U$-duality 
invariant form in terms of unique $G$ invariants of $D=11$ 
supergravity on tori.  Such manifestly $U$-duality 
invariant entropy depends only on ``integer-valued'' quantized {\it bare} 
charges \cite{FERks52}.  

In the following, we give the manifestly $U$-duality invariant form 
for the Bekenstein-Hawking entropy 
of black holes with general charge configuration.  
\begin{itemize}
\item $D=4$:  
The quartic $E_{7(7)}$ invariant is given in terms of $Z_{4\,AB}$ as 
\cite{CREj79}
\begin{eqnarray}
J_4&=&Z_{4\,AB}\bar{Z}^{BC}_4Z_{4\,CD}\bar{Z}^{DA}_4-\textstyle{1\over 4}
(Z_{4\,AB}\bar{Z}^{AB}_4)^2
\cr
&+&\textstyle{1\over{96}}(\epsilon_{ABCDEFGH}\bar{Z}^{AB}_4\bar{Z}^{CD}_4
\bar{Z}^{EF}_4\bar{Z}^{GH}_4+\epsilon^{ABCDEFGH}Z_{4\,AB}Z_{4\,CD}
Z_{4\,EF}Z_{4\,GH}).  
\label{4dudualinvar}
\end{eqnarray}
In terms of the skew-eigenvalues $\lambda_i$, $J_4$ takes the form
\begin{equation}
J_4=\sum^4_{i=1}|\lambda_i|^4-2\sum^4_{j>i}|\lambda_i|^2|\lambda_j|^2
+4(\bar{\lambda}_1\bar{\lambda}_2\bar{\lambda}_3\bar{\lambda}_4+
\lambda_1\lambda_2\lambda_3\lambda_4).
\label{4dskiewuinv}
\end{equation}
By substituting  $\lambda_i$ in (\ref{4dgenchtoskew}) into the 
following $E_7$ invariant entropy \cite{KALk53,KALr}, one reproduces the 
Bekenstein-Hawking entropy (\ref{chnullbhmes}) of the generating 
solution:
\begin{equation}
S_{BH}={\pi\over 8}\sqrt{J_4}.
\label{4dudualinvent}
\end{equation}
\item $D=5$:
The cubic $E_{6(6)}$ invariant has the form \cite{CRE80,CRE}:
\begin{equation}
J_3=-\sum^8_{A,...,F=1}\Omega^{AB}Z_{5\,BC}\Omega^{CD}Z_{5\,DE}
\Omega^{EF}Z_{5\,FA},
\label{5dudualinvar}
\end{equation}
which is expressed in terms of the real skew eigenvalues $\lambda_i$ as
\begin{equation}
J_3=2\sum^4_{i=1}\lambda^3_i.
\label{5dskiewuinv}
\end{equation}
Here, $\Omega$ is the $USp(8)$ symplectic invariant.  
The manifestly $E_{6(6)}$ invariant expression for the entropy of 
general solution is of the form \cite{CVEh}:
\begin{equation}
S_{BH}=\pi\sqrt{\textstyle{1\over{12}}J_3},
\label{5dudualinvent}
\end{equation}
which reproduces the entropy (\ref{5dbpsarea}) of the generating 
solution if the expressions for $\lambda_i$ in (\ref{5dgenchtoskew}) 
are substituted. 
\item $6\leq D\leq 9$:
There is no non-trivial $U$-duality invariant 
in $D\geq 6$.  This is consistent with the fact that 
the Bekenstein-Hawking entropy of the general BPS 
black holes in $D\geq 6$ is zero, which is the only $U$-duality 
invariant in $D\geq 6$.  
For near-extreme black holes, which has non-zero 
Bekenstein-Hawking entropy, the entropy can be expressed in a 
duality invariant form in terms of ``dressed'' electric charges 
and, therefore, has dependence on scalar asymptotic values 
\cite{CVEh}.  
\end{itemize}

The ADM mass $M$ of the BPS solution is given by the largest eigenvalue 
${\rm max}\{|\lambda_i|\}$ of $Z_{D\,AB}$.   The $U$-duality invariant 
form of the ADM mass can be expressed in terms 
of the $U$-duality invariant quantities ${\rm Tr}(Y^m_D)$ ($m=1,...,
[N/2]-p+1$; $Y_D\equiv Z^T_DZ_D$) and corresponds to the largest root 
of a polynomial of degree $[N/2]-p+1$ in $M$ with coefficients 
involving ${\rm Tr}(Y^m_D)$.  
The BPS solution preserves $p/N$ of supersymmetry if 
$p$ of the central charge eigenvalues have the same magnitude, i.e. 
$|\lambda_1|=\cdots=|\lambda_p|$.  This depends on charge configurations 
of black holes.  As for the generating solutions, the number of 
identical eigenvalues $|\lambda_i|$ can be determined from  
(\ref{4dgenchtoskew}), (\ref{5dgenchtoskew}) and (\ref{6dgenchtoskew}).  
In the following, we discuss $D=4$ black holes as an example
\footnote{The fraction of supersymmetry preserved by $N=8$, $D=4$ BPS 
black holes can also be determined from the Killing spinor equations 
\cite{CHA717}.} 
\cite{CVEh}. 
\begin{itemize}
\item $p=4$ case:  
The generating solution preserves $1/2$ of supersymmetry 
when only one charge is non-zero.  The $U$-duality invariant ADM mass 
is $M=-{1\over 8}{\rm Tr}(Y_4)$.  
\item $p=3$ case:  
An example is the case where $(Q_1,Q_2,P_1=P_2)\neq 0$ with 
$q={1\over 2}\sqrt{(Q_{1\,R}+P_{2\,R})^2-Q^2_{1\,L}}$.  The $U$-duality 
invariant mass has the form $M^2=-{1\over 8}{\rm Tr}(Y_4)+\sqrt{{1\over
{24}}{\rm Tr}(Y^2_4)-{1\over{192}}({\rm Tr}\,Y_4)^2}$.  
\item $p=2$ case:
An example is the case where only $Q_1$ and $Q_2$ are non-zero.  
The $U$-duality invariant mass $M$ is the largest root of a cubic equation in 
$M$ with coefficients involving $U$-duality invariants ${\rm Tr}(Y^m_4)$ 
($m=1,2,3$).  
\item $p=1$ case:  
Examples are the case where only $Q_1$, $Q_2$ and $P_1$ are non-zero, 
or the case where all the five charges are non-zero and independent.  
The largest root of a quartic equation involving invariants 
${\rm Tr}(Y^m_4)$ ($m=1,...,4$) corresponds to the ADM mass of the 
BPS solution.  
\end{itemize}

\subsubsection{$p$-Brane Embedding of Black Holes}\label{bprhghemd}

We discuss the $D=10,11$ $p$-brane embeddings of black 
holes in $D<10$.  Starting from $p$-branes in $D=10,11$, one obtains  
0-branes in $D<10$ by wrapping all the constituent 
$p$-branes around the cycles of the internal manifold. 
The resulting black hole solution has the following generic form:
\begin{equation}
ds^2_D=h^{1\over{D-2}}(r)[-h^{-1}(r)f(r)dt^2+f^{-1}(r)dr^2
+r^2d\Omega^2_{D-2}],
\label{dbhfrompbrn}
\end{equation}
where $f=1-{{2m}\over{r^{D-3}}}$ [$H_{\alpha}=1+{{2m\sinh^2
\delta_{\alpha}}\over{r^{D-3}}}$] is harmonic function associated 
with non-extremality parameter $m$ [charge $Q_{\alpha}=(D-3)m\sinh 
2\delta_{\alpha}$] and $h(r)=\prod^N_{\alpha=1}H_{\alpha}(r)$. 
The ADM mass and the Bekenstein-Hawking entropy are
\footnote{Generally, for a multi-charged black $p$-brane with 
the Page charges $\lambda_{\alpha}=(\tilde{p}+1)m\sinh 2\delta_{\alpha}$ 
($\alpha=1,...,N$), $M_{ADM}=2m\left[(\tilde{p}+1)
\sum^N_{\alpha=1}\sinh^2\delta_{\alpha}+\tilde{p}+2\right]=
\sum^N_{\alpha=1}\sqrt{\lambda^2_{\alpha}+\mu^2}+2\left({{\tilde{p}+2}\over
{\tilde{p}+1}}-{N\over 2}\right)\mu$ and $S_{BH}={1\over 4}
(2m)^{{\tilde{p}+2}\over{\tilde{p}+1}}\omega_{\tilde{p}+2}
\prod^N_{\alpha=1}\cosh\delta_{\alpha}\sim \mu^{{{\tilde{p}+2}\over
{\tilde{p}+1}}-{N\over 2}}\omega_{\tilde{p}+2}\prod^N_{\alpha=1}
\left(\sqrt{\lambda^2_{\alpha}+\mu^2}+\mu\right)^{1/2}$, 
where $\mu\equiv(\tilde{p}+1)m$ is the rescaled non-extremality parameter.}
\begin{eqnarray}
M_{ADM}&=&2m[(D-3)\sum^N_{\alpha=1}\sinh^2\delta_{\alpha}+D-2]
\cr
&=&\sum^N_{\alpha=1}\sqrt{Q^2_{\alpha}+\mu^2}+2\left({{D-2}\over
{D-3}}-{N\over 2}\right)\mu,
\cr
S_{BH}&=&{1\over 4}(2m)^{{D-2}\over{D-3}}\omega_{D-2}\prod^N_{\alpha=1}
\cosh\delta_{\alpha}
\cr
&\sim&\mu^{{{D-2}\over{D-3}}-{N\over 2}}
\prod^N_{i=1}\left(\sqrt{Q^2_{\alpha}+\mu^2}+\mu\right)^{1/2},
\label{genadmbhqntt}
\end{eqnarray}
where $\mu\equiv (D-3)m$ is the rescaled non-extremality parameter and 
we neglected overall factor related to the gravitational constant, since 
we are interested only in the dependence on $m$, $\delta_{\alpha}$ and 
$Q_{\alpha}$. 

As can be seen from (\ref{dbhfrompbrn}), dimensional reduction of 
single-charged $p$-branes leads to black holes with singular 
horizon and zero horizon area in the BPS limit.   
To construct black holes with regular event horizon and 
non-zero horizon area in the BPS limit, 
one has to start from multi-charged $p$-branes in higher dimensions.  
This is achieved in $D=4,5$ with $N=4,3$, respectively, which can 
be seen from the BPS limit of entropy in (\ref{genadmbhqntt}).  
In fact, it is shown in \cite{BERdejv095} that the number $N$ 
of distinct BPS black hole solutions to the equations 
of motion for the action (\ref{onescalformlag}) that  
have intersecting $p$-brane origins in $D=11,10$ is 
$N=4,3$ and 2 for $D=4,5$ and $D\geq 6$, respectively.   

In the following, we discuss intersecting $p$-branes 
which give rise to regular BPS black holes in $D=4,5$, as well 
as black holes with singular BPS limit.  We concentrate on 
intersecting $M$-branes; intersecting $p$-branes in $D=10$ 
are related to intersecting $M$-branes through dimensional 
reduction and dualities.  All the possible $D=10,11$ BPS, intersecting 
$p$-branes that satisfy intersection rules are classified in 
\cite{BERdejv095}.  
Also, there is a $M$-brane configuration (\ref{11dsolof8dmem}) 
interpreted as a $M\,2$-brane within a $M\,5$-brane ($2\subset 5$), 
which preserves $1/2$ of supersymmetry \cite{IZQlpt}.  
By including $2\subset 5$ configurations, one constructs new type of 
black holes \cite{COS181,COS138} which are mixture of marginal and 
non-marginal bound states.  Namely, the ADM mass and 
horizon area of $p$-branes that contain $2\subset 5$ have the forms  
$M\sim\sum_i\sqrt{\alpha^2_i+\mu^2}+c\mu$ and $A_H\sim \mu^{c^{\prime}}
\prod_i(\mu+\sqrt{\alpha^2_i+\mu^2})^{1/2}$ with $\alpha_i=
\sqrt{Q^2_i+P^2_i}$ for each $2\subset 5$ constituent, where $Q_i$ 
[$P_i$] is charge of $M\,2$-brane [$M\,5$-brane] in the 
$2\subset 5$ constituent and $c,c^{\prime}$ are appropriate constants.  
One can also add KK monopole to intersecting $M$-branes with overall 
transverse space dimensions higher than 3.  
We will not show the explicit intersecting $p$-brane solutions, 
since one can straightforwardly construct them applying harmonic 
superposition rules discussed in the previous section; 
explicit solutions can be found, for example, in 
\cite{KLEt,GAUkt478,TSE475,CVEt478,COS181,COS138}.  

\paragraph{Four-Dimensional Black Holes}\label{bprhghemd4d} 

Intersecting $M$-branes which reduce to $D=4$ black holes with 4 
charges, i.e. (\ref{dbhfrompbrn}) with $N=4$, should have 
4 or 3 (with momentum along a isometry direction) 
$M\,p$-brane constituents and at least 3 overall transverse 
directions.  Such configurations are $(i)$ $2\perp 2\perp 5\perp 5$ 
for $N=4$, and $(ii)$ $5\perp 5\perp 5$, $2\perp 5\perp 5$ and 
$2\perp 2\perp 5$ for $N=3$.  
Additionally, one has the following $D=11$ configurations that 
reduce to $D=4$ black holes preserving $1/8$ of supersymmetry: 
($i$) $2\perp 2\perp 2\perp+KK\ monopole$, ($ii$) $2\perp 5+boost+
KK\ monople$, ($iii$) $(2\subset 5)\perp(2\subset 5)\perp(2\subset 5)$, 
($iv$) $(2\subset 5)\perp 5+boost$, ($v$) $(2\subset 5)\perp 2+KK\ monopole$, 
($vi$) $(2\subset 5)+boost+KK\ monopole$.  Also, the $boost+KK\ monoplole$ 
configuration reduces to $D=4$ black hole that carries KK electric 
and magnetic charges and preserves $1/4$ of supersymmetry.  

\paragraph{Five-Dimensional Black Holes}\label{bprhghemd5d}

Intersecting $M$-branes with 3 or 2 (with a boost along a isometry 
direction) $M\,p$-brane constituents and at least 4 overall transverse 
directions can be reduced to $D=5$ black holes with 3 charges.  
These are $2\perp 2\perp 2$ and $2\perp 5$ with a momentum along a 
isometry direction.   An additional $M$-brane configuration 
that reduces to $D=5$ black hole with 3 charges is 
$(2\subset 5)\perp 2$, which preserves $1/4$ of supersymmetry.  

\paragraph{Black Holes in $D\geq 6$}\label{bprhghemdhigh}

0-branes in $D\geq 6$ can be supersymmetric with up to 2 constituent 
$p$-branes.  Supersymmetric 2 intersecting $M$-branes are $5\perp 5$, 
$2\perp 5$ and $2\perp 2$, which are compactified to 2-charged black holes 
in $D=4,5$ and 7, respectively, after wrapping each constituent 
$M\,p$-brane around cycles of a compact manifold.  One can compactify 
overall transverse directions of these $D=5,7$ 
black holes to obtain black holes with 2 charges in $D=4$ and 
$D\leq 6$, respectively.  One can also construct black holes in 
$D\leq 9$ and $D\leq 5$ by compactifying $M\,2$-brane and $M\,5$-brane
with momentum along a longitudinal direction, respectively.  
Additionally, the $M$-brane configuration $(2\perp 5)+boost$ 
reduces to $D=6$ black hole that preserves $1/4$ of supersymmetry.  

\section{Entropy of Black Holes and Perturbative String States}\label{ent}

One of challenging problems in quantum gravity for past decades 
is the issues related to black hole thermodynamics.  
It was early 1970's \cite{BEK72,BEK73,BEK74,CAR238,BARch31} when 
it was first noticed that the event horizon area $A$ behaves much 
like entropy $S$ of classical thermodynamics.  Namely, it is observed 
\cite{PENf299,HAW71} that the event horizon area tends to grow 
($\delta A \geq 0$), resembling the second law of thermodynamics 
($\delta S \geq 0$).  Furthermore, Bardeen, Carter and Hawking 
\cite{BARch31} proved that the surface gravity $\kappa$ of a stationary 
black hole is constant over the event horizon, resembling the 
zeroth law of thermodynamics, which states that the temperature is 
uniform over a body in thermal equilibrium.  
They also realized the following relation
\footnote{For the Kerr-Newmann black hole, this relation 
is generalized to $dM = {1\over{8\pi G_N}}\kappa dA+\Phi dQ+\Omega dJ$, where 
$\Phi$ [$\Omega$] is the potential [angular velocity] at the event horizon 
and $Q$ [$J$] is a $U(1)$ charge [angular momentum].} 
between the ADM mass $M$ of black holes and the event horizon area $A$:
\begin{equation}
dM = {1\over{8\pi G_N}}\kappa dA,
\label{bhther}
\end{equation}
which resembles the thermodynamic relation between energy $E$ and entropy 
$S$ (the first law of thermodynamics):
\begin{equation}
dE = TdS,  
\label{thermo}
\end{equation}
if one identifies the energy $E$ with the ADM mass $M$ and 
the entropy $S$ with the event horizon area $A$ with some unknown 
constant of proportionally.   Such analogy between horizon area  
and entropy met initially with skepticism, until Hawking  discovered 
\cite{HAW248,HAW75} that black hole is indeed thermal system which radiates 
(quasi-Planckian black body) thermal spectrum with temperature $T_{H}=\hbar
\kappa/2\pi$, due to quantum effect.  Since then, it is widely accepted 
that a black hole, as a thermal system, is endowed with ``thermodynamic'' 
entropy given by a quarter of the event horizon area in Planck 
units, the so-called {\it Bekenstein-Hawking entropy} 
\cite{BEK72,BEK73,BEK74,HAW71,HAW13,BARch31,MOS69,VIS48,KALop}:
\begin{equation}
S_{BH}={A\over{4\hbar G_N}}.
\label{bhtherent}
\end{equation}

Puzzles on black hole entropy stem from the fact that 
entropy is a thermodynamic quantity, which arises from 
the fundamental microscopic dynamics of a large complicated 
system as a universal macroscopic quantity which does not 
depend on the details of the underlying microscopic dynamics.
So, if the correspondence between laws of black hole 
mechanics and thermodynamic laws is to be valid, 
the thermodynamic black hole entropy (\ref{bhtherent}) 
should have a statistical interpretation in terms of the degeneracy of the 
corresponding microscopic degrees of freedom.  Based upon our knowledge
of statistical mechanics, one could guess several possible interpretations
of the statistical origin of black hole entropy: ($i$)
internal black hole states associated with a single black hole exterior
\cite{BEK73,BEK75,BEK76,HAW13}, ($ii$) the number of different ways
the black holes can be formed \cite{BEK73,HAW13}, ($iii$) the number
of horizon quantum states \cite{HOO90,SUStu48},
($iv$) missing information during the black hole evolution
\cite{HAW14,GID49}.
Another difficulty comes from the fact that contrary to 
ordinary thermodynamics, understanding of black hole thermodynamics 
requires the treatment of quantum effects, as we noted the crucial role 
that the Hawking effect plays in establishing black hole thermodynamics.  
Thus, the statistical interpretation of black hole entropy should 
necessarily entail quantum theory of gravity, 
of which we have only rudimentary understanding.  

The early attempts during the 1970's and 1980's were not successful in 
the sense that the most of approaches either ($i$) did not touch upon the 
statistical meaning of entropy, since the calculations were mostly based 
on thermodynamic relations (e.g. calculating entropy using Clausius's 
rule $S=\int dM/T$ given that the black hole temperature $T$ is determined 
by the surface gravity method \cite{BARch31,VIS46}), or ($ii$) is purely 
classical (e.g. in Gibbons-Hawking Euclidean (on-shell functional integral) 
method \cite{GIBh77} involving grand partition function, the black hole 
entropy $A/(4\hbar G_N)$ is reproduced at tree-level of 
quantum gravity calculation).   Another major was the ultraviolet 
quadratic divergence
\footnote{This is closely related to the fact that quantum gravity in 
point-like particle field theory is non-renormalizable, as we will 
discuss below.} 
(related to the divergence of the number of energy levels a particle can 
occupy in the vicinity of black hole horizon)    
in black hole entropy when the Euclidean functional formulation of the 
partition function is evaluated for quantum fields in the black hole 
background.  To avoid such divergences, 't Hooft 
\cite{HOO256} introduced ``cutoff'' at a small distance $\varepsilon$  
just above the event horizon in the path integral of real free scalar 
field
\footnote{'t Hooft proposed that the entropy of black hole is nothing 
but the entropy of particles which are in thermal 
equilibrium with black hole background \cite{HOO79,HOO256,HOO36}.} 
$\phi$, assuming that there are no states at the interval between 
the event horizon and the cutoff (the so-called {\it brick-wall method}).  
The black hole entropy in field theory based on brick-wall 
method in general depends on the cut-off distance $\varepsilon$ in the form 
$S_{\phi}\sim {A\over{4\varepsilon^2}}$, reflecting the quadratic 
divergence.  It was conjectured \cite{HOO90} by 't Hooft that such 
ultraviolet divergence of the statistical entropy might be related to 
Hawking's information paradox \cite{HAW14}, i.e. a black hole is an 
infinite sink of information.  

In \cite{SUSu50}, it is shown that the divergence associated with the 
Euclidean functional formulation of the partition function for canonical 
quantum gravity (of point-like particles) is related to the renormalization 
of the gravitational coupling $G_N$.  When the contributions to entropy 
(obtained from the partition function) 
from the pure gravity and matter fields are added, the entropy takes a 
suggestive form $S={A\over 4}({1\over{G_N}}+{c\over{\varepsilon^2}})$ 
similar to (\ref{bhtherent}) but the bare gravitational constant $G_N$ 
is renormalized.  The explicit calculation of quantum corrections of 
quantum gravity shows \cite{DEMlm52,KAB453,KABss182} that 
the renormalized gravitational constant $G_N$ takes the same form.   
Since superstring theory is  
a promising candidate for a finite theory of quantum gravity, the 
contradiction encountered in the point-like particle field theory has to be 
resolved \cite{SUS145,SUS116,SUSu50,SUSu45}.  
It is indeed shown in \cite{SUSu50,DAB347} 
that theory of superstring propagating in a black hole background 
gives rise to a {\it finite} expression for black hole entropy in 
the calculation of partition function through Euclidean path integral, 
with the {\it finite} renormalized gravitational coupling $G_N$: 
the genus zero contribution gives rise to the classical result 
(\ref{bhtherent}) with a bare Newton's constant $G_N$ and the higher genus 
terms contribute to finite corrections to $G_N$.    

This can be seen intuitively by considering microscopic states near the 
event horizon \cite{SUS49}.  For point-like particles, due to the 
arbitrarily small longitudinal Lorentz contraction near the event horizon, an 
arbitrarily large number of particles can be 
packed close to the event horizon, giving rise to a divergent entropy.  
However, the Lorentz contraction of strings along the longitudinal direction 
is eventually halted to a finite extent and, therefore, only finite number of 
strings can be packed near the horizon, leading to finite entropy.   

Just as only contribution to the first-quantized path integral of the 
point-like particle field theory is from the set of paths that encircle or 
touch the black hole event horizon,  only string graphs which contribute 
to the entropy through the partition functions are those that are somehow 
entangled with the event horizon.  From the point of 
view of an external observer, this kind of closed strings, which are 
partially hidden behind the event horizon, look like open strings frozen 
to the horizon.  
Thus, the black hole entropy can somehow be interpreted as being 
associated with oscillation degrees of freedom of fluctuating open strings 
whose ends are attached to the black hole horizon \cite{SUSu50}.  

This chapter is organized as follows.  
In section \ref{entstr}, we discuss connection between black holes 
and perturbative string states.  Identification of black holes with 
string states makes it possible to do explicit calculations of 
statistical entropy of black holes, based on the conjecture that 
microscopic origin of entropy is from degenerate string 
states with mass given by the corresponding ADM mass of the black hole.  
In section \ref{entsen}, we discuss Sen's original calculation of 
statistical entropy of the BPS static black hole, which was compared to 
Bekenstein-Hawking entropy evaluated at the stretched horizon.  
In section \ref{entnear}, we generalize Sen's result to 
near-extreme rotating black holes.  In section \ref{entfun}, 
we discuss the level matching of black holes to macroscopic 
string states at the core.  This justifies our working hypothesis 
that black holes are perturbative string states.  
In section \ref{entch}, we summarize Tseytlin's method of chiral null 
model for calculating statistical entropy of the BPS black holes that 
carries magnetic charges as well as perturbative NS-NS electric charges.

\subsection{Black Holes as String States}\label{entstr}

It is not a new idea that elementary particles might behave like black 
holes \cite{HAW152,SAL,HOO90}.  A particle whose mass exceeds 
the Plank mass and therefore whose wavelength is less than its 
Schwarzschield radius exhibits an event horizon, a characteristic 
property of black holes. 
Since typical massive excitations of strings have mass 
of the order of the Planck mass, one would expect that massive string 
states become black holes when gravitational coupling 
(or string coupling) 
is large enough.  It has been shown that black holes with 
given charges and angular momenta behave like string states with 
the corresponding quantum numbers.

Before we discuss identification of black holes with string states, 
we summarize \cite{GIVrv322} some aspects of perturbative spectrum, 
moduli space and $T$-duality of heterotic string on a torus.  
Heterotic string \cite{GROhmr54} is a theory of closed string whose 
left- and right-moving modes are respectively described by bosonic and 
super string theories.  So, the critical dimensions, in which 
the conformal anomali is absent, for each mode are different:  
$D=26$ for the left-movers and $D=10$ for the right-movers.  In 
compactifying the extra 16 coordinates of the left-movers on 
$T^{16}$, one obtains a rank 16 non-Abelian gauge group 
which is associated with the even-self-dual lattice of the type 
$E_8 \times E_8$ or $Spin(32)/{\bf Z}_2$.  Thus, the massless bosonic 
modes of heterotic string in $D=10$ at a generic point of moduli space are  
$U(1)^{16}$ gauge fields, as well as graviton, 2-form field and dilaton 
in the NS-NS sector ground state. 

We compactify the extra $d$ spatial coordinates $X^{\mu}$ 
($\mu=1,...,d$) on $T^d$ to obtain a theory in 
$D=10-d$.  The toroidal compactification is defined by the 
periodic identification of each internal coordinate, i.e. 
$X^{\mu} \sim X^{\mu} + 2\pi m^{\mu}$, where $m^{\mu}$ is 
the integer-valued string ``winding mode''.  
Since the holonomy of a torus is trivial, 
all of supersymmetry is preserved in compactification, i.e. 
$N=4$ for the compactification on $T^6$.  
The $T^d$ part of the heterotic string worldsheet action in flat 
background, including the coupling to gauge fields 
$A^{\alpha}_{\mu}$ and a 2-form potential $B_{\mu\nu}$, is
\begin{eqnarray}
S &=& {1\over{2\pi}}\int d^2 z[(G_{\mu\nu}+B_{\mu\nu})\partial X^{\mu}
\bar{\partial}X^{\nu} + A_{\mu\alpha}\partial X^{\mu}\bar{\partial}
X^{\alpha} 
\nonumber \\
& &\ \ + (G_{\alpha\beta}+B_{\alpha\beta})\partial X^{\alpha} 
\bar{\partial}X^{\beta}] + ({\rm ferminionic\ terms}), 
\label{sigma}
\end{eqnarray}
where $\mu , \nu = 1,...,d$ [$\alpha =1,...,16$] correspond to the 
coordinates of $T^d$ [$T^{16}$ of left-movers], and 
the complex worldsheet coordinate and derivative are defined as
\begin{equation}
z = {1\over\sqrt{2}}(\tau + i\sigma), \ \ \ \ \ 
\partial = {1\over\sqrt{2}}(\partial_{\tau}-i\partial_{\sigma}).
\label{cxcoord}
\end{equation}
The internal coordinates $X^{\alpha}$ live on the weight lattice of 
$E_8 \times E_8$ or $Spin(32)/{\bf Z}_2$.  
The background fields $G_{\mu\nu}$, $B_{\mu\nu}$ and $A_{\mu\alpha}$, 
which parameterize the moduli space $O(d+16,d,{\bf Z})/[O(d+16,{\bf Z})
\times O(d,{\bf Z})]$ of $T^d\times T^{16}$, can be organized into 
the ``background matrix'' of the form:
\begin{equation}
E \equiv B+G = \left (\matrix{(G+B+{1\over 4}A^K A_K)_{ij} & A_{iJ} \cr 
0 & (G+B)_{IJ}} \right), 
\label{defmat}
\end{equation}
where $B$ [$G$] is the antisymmetric [symmetric] part of $E$.  
(For the relations between the background fields in (\ref{sigma}) and 
$E$, see the next footnote.) 
Here, the indices $i,j$ [$I,J$], associated with $T^d$ [$T^{16}$],  
run from 1 to $d$ [from 1 to 16].  In particular, the components 
$E_{IJ}=(B+G)_{IJ}$ of $E$ are related to the Cartan matrix $C_{IJ}$ of 
$E_8\times E_8$ or $Spin(32)/{\bf Z}_2$ as
\begin{equation}
E_{IJ}=C_{IJ}\ \ (I>J),\ \ \ 
E_{II}={1\over 2}C_{II},\ \ \ 
E_{IJ}=0\ \ (I<J).  
\label{backmatcart}
\end{equation}

The Narain lattice \cite{NAR169,NARsw279} $\Lambda^{(d+16)}$, 
which defines $T^d\times T^{16}$ 
is spanned by basis vectors $\alpha\equiv(\alpha_i,\alpha_I)$ of the form:
\begin{equation}
\alpha_i = (e_i, {1\over 2}A^K_i E_K), \ \ \ \ \ 
\alpha_I =(0,E_I), 
\label{latbasis}
\end{equation}
where the vectors $\{E^{\alpha}_I|I=1,...,16\}$ and $\{e^{\mu}_i|i=1,...,d\}$ 
are defined through
\begin{equation}
E_I \cdot E_J = 2G_{IJ}, \ \ \ \  
e_i \cdot e_j = 2G_{ij}, \ \ \ \ 
e_i \cdot E_I = 0.  
\label{basis}
\end{equation}

The zero modes $(p_R,p_L)$ of the right- and left-moving momenta 
form an even self-dual lattice  
$\Gamma^{(d,d+16)} = \Gamma^{(d,d)} \oplus \Gamma^{(0,16)}$. 
Quantized 
momentum zero modes $(p_R,p_L)$ are embedded into  
$\Gamma^{(d+16,d+16)}$ as
\begin{equation}
\left(\matrix{p_R \cr p_L} \right ) = 
\left(\matrix{p^i_R & 0 \cr p^j_L & p^J_L}\right ),  
\label{momentum}
\end{equation}
where
\begin{equation}
p_R = [n^t + m^t(B-G)]{\alpha}^* , \ \ \  
p_L = [n^t + m^t(B+G)]{\alpha}^*,\ \ \ m,n\in{\bf R}^{d+16}. 
\label{rl}
\end{equation}
Here, ${\alpha}^*$ is the basis vector of the lattice dual to 
$\Lambda^{(d+16)}$:
\begin{equation}
\alpha^{i\,*}=(e^{i\,*},0), \ \ \ \ 
\alpha^{I\,*}=(-{1\over 2}A^I_ie^{i\,*},E^{I\,*}),
\label{dualvechet}
\end{equation}
where $E^{I\,*}$ [$e^{i\,*}$] are dual 
\footnote{By contracting with $e^{i\,*}_{\mu}$ and 
$E^{I\,*}_{\alpha}$, one obtains the background fields in (\ref{sigma}) 
from $E$.  For example, $G^{\alpha\beta}=2G_{IJ}(E^{I\,*})^{\alpha}
(E^{J\,*})^{\beta}$ and $B^{\mu\nu}=2B_{ij}(e^{i\,*})^{\mu}(e^{j\,*})^{\nu}$, 
etc.}
to $E_I$ [$e_i$], i.e. $\sum^d_{\mu=1}e^{\mu}_ie^{j\,*}_{\mu}=\delta^j_i$ 
[$\sum^{16}_{\alpha=1}E^{\alpha}_IE^{J\,*}_{\alpha}=\delta^J_I$].  

The heterotic string with the action (\ref{sigma}) has an  
$O(d+16,d,{\bf Z})$ $T$-duality symmetry. 
This group is a subgroup of the following $O(d+16,d+16,{\bf Z})$ 
transformation that preserves the triangular form of $E$ in (\ref{defmat}): 
\begin{equation}
E \rightarrow E^{\prime} = (aE+b)(cE+d)^{-1},\ \ \ 
\left(\matrix{a & b \cr c & d}\right)\in O(d+16,d+16,{\bf Z}), 
\label{frac}
\end{equation}
and $(p_R,p_L)$ in $\Gamma^{(d,d+16)}$ transforms as 
a vector.  $T$-duality is proven  
\cite{GIVrt409} to be exact to all orders in string coupling.  

The mass of perturbative states for heterotic string on a torus  
is \cite{GROhmr54} 
\begin{equation}
M^2 = 
{1 \over {8\lambda^{(0)}_2}}\{(p_R)^2 + 2N_R - 1\} =
{1 \over {8\lambda^{(0)}_2}}\{(p_L)^2 + 2N_L - 2\}, 
\label{stmass}
\end{equation}
where $N_{L,R}$ are left- and right-moving oscillator numbers, 
$\lambda^{(0)}_2$ is the vacuum expectation value of the dilaton (or 
string coupling).  

We now identify string states with black holes. 
The mass of the BPS purely electric black holes in heterotic 
string on $T^6$, which preserves $1\over 2$ of the $N=4$ supersymmetry, is 
\cite{HARl,SEN038,SCHs312,SENint,SEN10}:
\begin{equation}
m^2 = {1 \over {16\lambda^{(0)}_2}}\alpha^a(M^{(0)}+L)_{ab}\alpha^b,
\label{bpmass}
\end{equation}
where $\vec{\alpha}$ is the charge lattice vector on an even, 
self-dual, Lorentzian lattice $\Lambda$ with the $O(6,22)$ metric $L$  
and the subscript $(0)$ denotes asymptotic values. 
Here, $M$ is the moduli matrix of $T^6$ defined in (\ref{modulthree}).  
Under the $T$-duality, the moduli matrix and the charge lattice transform 
as \cite{SENint} 
\begin{equation}
M^{(0)} \to \Omega M^{(0)} \Omega^T, \ \ \ 
\Lambda \to L\Omega L \Lambda , \ \ \ \ \Omega\in O(6,22)
\label{tdualtran}
\end{equation}
and the BPS mass (\ref{bpmass}) is invariant.  
With a choice of the asymptotic values $\lambda^{(0)}_2 = 1$ and 
$M^{(0)}=I_{6,22}$, the mass takes a simple form:
\begin{equation}
m^2 = {1\over {16}}\alpha^a(I_{6,22}+L)_{ab}\alpha^b =
{1\over 8}(\vec{\alpha}_R)^2, \ \ \ 
\alpha^a_{R,L} \equiv {1\over 2}(I_{6,22} \pm L)_{ab}\alpha^b.
\label{bhsimp}
\end{equation}  

The string momentum [winding] zero modes are identified 
with the quantized electric charges of KK [2-form] $U(1)$ 
gauge fields, i.e. $\vec{\alpha}_{R,L}= \vec{p}_{R,L}$.   
Then, $m=M$, provided $N_R = {1\over 2}$ \cite{DUFr}:   
the BPS black holes are identified with the ground states of the right 
movers.  
With a further inspection of (\ref{stmass}), one finds $N_L$ in terms of 
$\vec{\alpha}$ \cite{DUFr}:
\begin{equation}
N_L-1 = {1\over 2}\left( (\vec{\alpha}_R)^2 - (\vec{\alpha}_L)^2\right) 
= {1\over 2} \vec{\alpha}^T L \vec{\alpha}, 
\label{leftoscil}
\end{equation}
leading to $\vec{\alpha}^T L \vec{\alpha} \geq -2$.  
So, the various BPS black holes in the heterotic 
string on a torus are identified \cite{DUFr} as string states with the 
corresponding value of $N_L$ [or $\vec{\alpha}^T L \vec{\alpha}$]. 

Non-extreme black holes 
are identified with string states with the right movers excited 
as well.   Identification of black holes in other dimensions and in 
type-II theories with string states is proceeded similarly as above.  
For type-II string theories, the mass of perturbative string state is 
\begin{equation}
M^2 = {1 \over {8\lambda^{(0)}_2}}\{(p_R)^2 + 2N_R - 1\} =
 {1 \over {8\lambda^{(0)}_2}}\{(p_L)^2 + 2N_L - 1\}.
\label{typeiibperpmass}
\end{equation}
Since the type-II strings have supersymmetry in both the right- and 
left-moving sectors, perturbative string states can ($i$) preserve 
supersymmetry in both sectors ($N_R=N_L=1/2$), leading to short 
supermultiplet; ($ii$) preserve supersymmetry in one sector, only 
($N_{R,L}>N_{L,R}=1/2$), leading to intermediate supermultiplet; 
($iii$) break supersymmetry in both sectors ($N_R,N_L>1/2$), leading 
to long supermultiplet.  Further study of equivalence of 
string states and extreme black holes, including spins of string states 
and dipole moments of rotating black holes, is carried out in 
\cite{DUFr481,DUFlr015}.

\subsection{BPS, Purely Electric Black Holes and Perturbative 
String States}\label{entsen}

In the previous section, we showed that BPS electric black holes in the 
string low energy effective actions are identified with perturbative 
string states.  
Thus, it is natural to infer that the microscopic degeneracy of  
black holes originates from the degenerate string states in 
a corresponding level.  

In general, non-extreme black holes are also identified with 
perturbative string states.  
However, non-extreme solutions are plagued with (unknown gravitational) 
quantum corrections and, therefore, the ADM mass cannot be trusted.  
In fact, the number of states in non-extreme black hole grows with the 
ADM mass $M_{bh}$ like $\sim e^{M^2_{bh}}$ \cite{HOO90}, whereas the 
string state level density grows with the string state mass $M_{string}$ 
as $\sim e^{M_{string}}$.  
Thus, if one is to identify string states with 
black hole states, one is force to identify $M^2_{bh}$ with $M_{string}$ 
\cite{SUS145,RUSs437}.  In \cite{SUS145}, Susskind attributes the 
discrepancy to the mass renormalization due to unknown quantum 
corrections.   (See also section \ref{dbrbh}, where it is discussed that the 
Bekenstein-Hawking entropy of non-extreme black holes has to be evaluated 
at the specific string coupling at the black hole and microscopic 
configuration ($D$-branes and fundamental string) transition point.)

As first pointed out by Vafa \cite{SUS145}, 
the BPS solutions do not receive quantum corrections \cite{OLIw78} due 
to renormalization theorem of supersymmetry. 
Such class of solutions
\footnote{It is argued in \cite{HAWhr51} that the Bekenstein-Hawking 
formula for entropy, i.e. (entropy) $\propto$ (horizon area), 
ceases to hold for extreme case and entropy of an extreme black 
hole is always zero, displaying discontinuity in going from non-extreme  
to extreme case.  This is attributed \cite{HAWhr51} as being due to 
difference in Euclidean topologies for the two cases.  The cure for this 
discontinuity is proposed in \cite{GHOm78}, where it is suggested that 
one has to extremize after quantization, rather than quantizing after 
extremization.} 
are, therefore, suitable for testing the hypothesis that the statistical 
origin of black hole entropy is from the degenerate string states with 
mass given by the ADM mass of black hole. 

So, one can calculate the ``statistical'' entropy  
by taking logarithm of the string level density. 
This yields the finite non-zero entropy $\sim\sqrt{N_L}$.  However, 
the ``thermal'' entropy of the BPS purely electric black holes 
in heterotic string is zero. 
In \cite{SEN10}, Sen circumvented with problem by postulating that the 
``thermal'' entropy of the BPS black hole is not the event horizon area, 
but the area of a surface close to the event horizon, a so-called 
``stretched'' horizon \cite{THOpm,SUStu48,SUSt49}.  
Although the BPS electric black hole solutions are 
free of quantum corrections, they receive (classical) 
stringy $\alpha^{\prime}$ corrections due to 
the singularity at the event horizon.  This leads to the shift of the 
event horizon by the amount of an order of $\alpha^{\prime}$.  

Originally, the stretched horizon is defined \cite{SUStu48} as the 
surface where the local Unruh temperature for an observer, who is 
stationary in the Schwarzschield coordinate, is of the order of the 
Hagedorn temperature \cite{HAG}.  
Namely, it is a surface where the string interactions become significant.  
In \cite{KARks3,SUS71,MEZpt50}, it is observed that the 
transverse size of strings diverges logarithmically and 
fill up a region at the stretched horizon, melting to form a single string.  
Thereby, information in the string states is stored and 
thermalized with black hole environment in the region near the stretched 
horizon \cite{SUS71,SUS49,LOWsu327,LOWpstu,MEZpt50}, 
and black hole states are in one-to-one correspondence with single string 
states.  
So, the statistical entropy is due to degenerate 
strings states in equilibrium with the black hole background at the 
stretched horizon \cite{SUSu50}.  

In this section, we summarize \cite{SEN10,PEE456} to illustrate this idea.  
The electric black hole considered in \cite{SEN10} is a special case of 
the general solution \cite{CVEy672} discussed in section 
\ref{n4bh4dsphsusy}.  But for the purpose of illustrating the idea of 
perturbative string state and black hole correspondence, we follow Sen's 
parameterization of solution in terms of left-moving and right-moving 
electric charges, rather than in terms of KK and 2-form electric 
charges.  

\subsubsection{Black Hole Solution}\label{entsensol}

In Sen's notation, the most general non-rotating, electric black 
hole solution in the heterotic string on $T^6$, in the Einstein-frame, 
is \cite{SEN10}
\begin{equation}
g_{\mu\nu}dx^{\mu}dx^{\nu} 
=-{{r(r-2m)}\over \Delta^{1\over 2}}dt^2 + {\Delta^{1\over 2}\over
{r(r-2m)}}dr^2 +\Delta^{1\over 2}(d\theta^2 + {\rm sin}^2\theta d\varphi^2), 
\label{senbh}
\end{equation}
where $\Delta\equiv r^2[r^2+2mr({\rm cosh}\alpha{\rm cosh}\beta -1)+
m^2({\rm cosh}\alpha-{\rm cosh}\beta)^2]$.  

The ADM mass and the electric charges are 
\begin{eqnarray}
M_{BH}&=&{1\over {2G_N}}m(1+{\rm cosh}\alpha{\rm cosh}\beta), 
\nonumber \\
Q^i&=&\left\{\matrix{{m\over\sqrt{2}}g_sn^i{\rm sinh}\alpha{\rm cosh}
\beta \ \ \ \ {\rm for}\ 1\leq i\leq 22 \cr
{m\over\sqrt{2}}g_sp^{i-22}{\rm sinh}\beta{\rm cosh}\alpha \ \ \ \ 
{\rm for}\ 23\leq i\leq 28}\right. , 
\label{senphys}
\end{eqnarray}
where $\vec{n}$ [$\vec{p}$] is an arbitrary 22 [6] component unit vector.  
The left and right handed charges are defined as
\begin{eqnarray}
Q^i_{L}&\equiv&{1\over 2}(I_{6,22}- L)_{ij}Q^j =
{{n^i_L}\over \sqrt{2}}g_sm{\rm sinh}\alpha {\rm cosh}\beta ,  
\nonumber \\
Q^i_{R}&\equiv&{1\over 2}(I_{6,22}+ L)_{ij}Q^j =
{{n^i_R}\over \sqrt{2}}g_sm{\rm sinh}\beta {\rm cosh}\alpha ,  
\label{lrcharge}
\end{eqnarray}
and the 28-component left and right handed unit vectors $\vec{n}_L$ 
and $\vec{n}_R$ are similarly defined.   The solution 
(\ref{senbh}) is in the frame where the $O(6,22)$ invariant metric 
$L$ (\ref{4dL}) is diagonal.  This parameterization of  
black hole solution has a convenient form in which only left [right] 
handed charges are non-zero when $\beta=0$ [$\alpha=0$] with all 
the parameters finite.  

The solution has 2 horizons
at $r=r_{+,-}=2m,0$.  The event horizon area is
\begin{equation}
A=\int d\theta d\varphi \sqrt{g_{\theta\theta}g_{\varphi\varphi}}|_{r=r_+} = 
8\pi m({\rm cosh}\alpha +{\rm cosh}\beta).
\label{senarea}
\end{equation}
The surface gravity at the event horizon is 
\begin{equation}
\kappa = {\rm lim}_{r\to r_+} \sqrt{g^{rr}}\partial_r\sqrt{-g_{tt}}|_{\theta
=0} = {1\over{2m({\rm cosh}\alpha +{\rm cosh}\beta)}}.
\label{sentemp}
\end{equation}

\subsubsection{Extreme Limit and String States}\label{entsenstat}

The extreme limit is defined as a limit where the inner and outer 
horizons coincide, i.e. $m\to 0$.  In taking the ``non-extremality 
parameter'' $m$ to zero, one has to let one (or both) of the boost parameters 
$\alpha$ and $\beta$ go to infinity so that the electric charges 
(\ref{lrcharge}) do not vanish.  Since we are interested in the BPS 
solutions, we let the ADM mass depend only on the right-handed electric 
charge.  
This is achieved by taking the limit $\beta\to\infty$ and $m\to 0$ such that 
$\hat{m}= {1\over 2}me^{\beta}$ remains as a finite non-zero constant, 
while $\alpha$ remaining finite.  In this limit, 
the ADM mass and the electric charges are
\begin{eqnarray}
M_{BPS}&=&{1\over {2G_N}}\hat{m}{\rm cosh}\alpha, 
\cr
Q^i_L&=&{1\over\sqrt{2}}g_sn^i_L\hat{m}{\rm sinh}\alpha, \ \ \ \ 
Q^i_R={1\over\sqrt{2}}g_sn^i_R\hat{m}{\rm cosh}\alpha, 
\label{senbpspar}
\end{eqnarray}
thereby the ADM mass depends on the right-handed electric charge, 
only:
\begin{equation}
M^2_{BPS}={1\over{8g^2_s}}\vec{Q}^2_R, 
\label{senmassch}
\end{equation}
where $G_N=2$.  
In this limit, the solution has the form (\ref{senbh}) with $m=0$ and 
\begin{equation}
\Delta = r^2(r^2+2\hat{m}r{\rm cosh}\alpha +\hat{m}^2).
\label{delbpssen}
\end{equation}

The event horizon area (\ref{senarea}) is zero in the BPS limit.  
However, the string states are degenerate.  One can circumvent such 
problem by calculating entropy at the ``stretched horizon'' right above 
the event horizon.  To find a location of the stretched horizon, one 
considers a region close to the event horizon in the ``string frame'' metric:
\begin{eqnarray}
dS^2 &\equiv& g^{string}_{\mu\nu}dx^{\mu}dx^{\nu} \simeq 
-{{r^2}\over {\hat{m}^2}}g^2_s dt^2 +g^2_sdr^2 +g^2_sr^2(d\theta^2 + 
{\rm sin}^2\theta d\varphi^2) 
\nonumber \\
&=&-\bar{r}^2d\bar{t}^2 +d\bar{r}^2 +\bar{r}^2(d\theta^2 +{\rm sin}^2\theta 
d\varphi^2), 
\label{senhor}
\end{eqnarray}
where $\bar{r}\equiv g_sr$ and $\bar{t}\equiv t/\hat{m}$.  Note, in 
the frame $(\bar{t},\bar{r},\theta,\phi)$, all the 
dependence on the other parameters has disappeared.  One can show that 
the other background fields also become independent of the parameters 
near the event horizon, if one performs a suitable $O(6,22)$ transformation. 
Thus, the location of stretched horizon, i.e. the location where higher 
order stringy corrections become important, is unambiguously estimated to 
be located at $\bar{r}=C$, a distance of order 1 (in unit of string scale) 
from the event horizon.  In terms of the original coordinate, the stretched 
horizon is located at $r=C/g_s \equiv \eta$.  

The stretched horizon area, calculated from (\ref{senbh}) with $m=0$ 
and (\ref{delbpssen}), is 
\begin{equation}
A\simeq 4\pi\eta\hat{m} = 4\pi\hat{m}C/g_s, 
\label{stretch1}
\end{equation}
where only the term leading order in $\eta$ is kept, and therefore 
the thermal entropy is $S_{BH}\equiv{A\over{4G_N}}= 
{\pi\over 2}{{\hat{m}C}\over {g_s}}$.  To compare this expression with the 
statistical entropy, 
one expresses $S_{BH}$ in terms of electric charges by using the 
relation $\hat{m}=4\sqrt{M^2_{BH}-{{\vec{Q}^2_L}\over {8g^2_s}}}$ derived 
from (\ref{senbpspar}):
\begin{equation}
S_{BH}={{2\pi C}\over g}\sqrt{M^2_{BH}-{{\vec{Q}^2_L}\over {8g^2_s}}}. 
\label{stretch2}
\end{equation}

Now we compare the thermal entropy (\ref{stretch2}) 
with the degeneracy of string states.  Since string states 
identified with BPS black holes have the right movers in 
ground state ($N_R={1\over 2}$), the string state degeneracy is 
from the left movers with $N_L$ given in terms of electric charges 
as (details are along the same line as in section \ref{entstr}):
\begin{equation}
N_L \simeq {4\over g^2_s}(M^2_{BPS}-{{\vec{Q}^2_L}\over {8g^2_s}}),  
\label{leftoscilsen}
\end{equation}
for large $\vec{Q}^2_L$.  
So, the statistical entropy associated with the degeneracy of string 
states is
\begin{equation}
S_{STAT} \equiv {\rm ln}\,d(N_L)\simeq 4\pi\sqrt{N_L} 
\simeq {{8\pi}\over {g_s}}\sqrt{M^2_{BH}-{{\vec{Q}^2_L}\over {8g^2_s}}}. 
\label{statent}
\end{equation}
This entropy expression has the same dependence on $M_{BPS}$ and 
$Q_L$ as the thermal entropy (\ref{stretch2}) calculated at the 
stretched horizon, and the two expressions agree if one 
chooses $C=4$ in (\ref{stretch2}).  Note, it is crucial that the constant 
$C$ does not depend on parameters of black holes; 
otherwise, the dependence of $S_{STAT}$ (\ref{statent}) on $\vec{Q}^2_L$ 
and $M^2_{BH}$ cannot trusted because of the unknown dependence of $C$ on 
these parameters.

\subsection{Near-Extreme Black Holes as String States}\label{entnear}

In the previous section, we saw that thermal entropy of the BPS, 
non-rotating, electric black holes agrees (up to numerical factor 
of order one) with statistical entropy associated with the degeneracy of 
string states, if it is evaluated at the ``stretched 
horizon''.  However, the rotating black hole case is problematic for 
the following reasons.  Since the electric, rotating black hole 
(\ref{ebh}) in the BPS limit with all the angular momenta 
non-zero has naked singularity, thermal quantities cannot be defined.  
The BPS limit with a horizon is possible in $D\geq 6$ with at most 
1 non-zero angular momentum \cite{HORs53}.  Even for this case, not 
only the event horizon is singular (i.e. the event horizon and the 
singularity coincide) and has zero surface area, but also the area of 
the stretched horizon (which is assumed to be independent 
of parameters of the black hole) is independent of angular momenta.
We surmise that this is due to the unknown dependence of the location of 
the stretched horizon on physical parameters, unlike the non-rotating 
black hole case.  
The determination of the stretched horizon location may require  
understanding of $\alpha^{\prime}$ corrections with rotating black hole 
as the target space configuration, which is difficult to estimate 
at this point.  

We propose \cite{CVEy477} an alternative way to circumvent 
the problems of the BPS electric black holes.  
Instead of defining the thermal entropy of the BPS black holes {\it at the 
stretched horizon}, we propose to calculate the thermal entropy of 
{\it near extreme black holes} at the event horizon.  
Then, the thermal entropy of near-extreme black holes takes 
suggestive form which can be interpreted in terms of string state 
degeneracy.  We attempt statistical interpretation of 
such thermal entropy expression by using the conformal field theory of 
$\sigma$-model with the near-extreme solution as a target space 
configuration and with angular momenta identified 
with $[{{D-1}\over 2}]$ $U(1)$ left-moving world-sheet currents.

\subsubsection{Thermal Entropy of Near-BPS Black Holes}\label{entnearther}

The proper way of taking the near-BPS limit of rotating black holes 
is to take the limit in such a way that the angular momentum 
contribution to the thermal entropy is not negligible compared with the 
contribution of the other terms, while ensuring the regular horizon so 
that thermal quantities can be evaluated at the {\it event horizon}.   
This is achieved as follows.  First of all, 
the near-BPS limit is defined as the limit in which the non-extremal 
parameter $m>0$ is very small and the boost parameters  
$\delta_i$ are very large such that the combinations 
$me^{2\delta_i}$ ($i=1,2$) remain as finite, non-zero constants. 
Then, as long as $l_i$ are non-zero, 
$J_i$ (\ref{para}) do not vanish.  Second, 
the requirement of the regular event horizon restricts the range of the 
parameters of the solution \cite{MYEp172}, e.g. $m\geq |l_1|$ for $D=4$ and 
$m\geq (|l_1|+|l_2|)^2$ for $D=5$.  
For an arbitrary $D$, we write such a constraint generally as $Q_1^{(1)}
Q_1^{(2)}\gg J^2_{1,\cdots,[{{D-1}\over 2}]}$.  Third, for the 
thermal entropy to be macroscopically non-negligible, the electric charges 
have to be very large, i.e. $Q^{(1),(2)}_1 \gg m={\cal O}(1)$.  This is 
required also 
for the statistical entropy, since the system has to be very large 
so that statistical quantities approach the exact values.  
Fourth, while keeping the angular momenta small so that the regular horizon 
is always ensured as $m\approx 0$, one has to make sure that angular momenta 
contribution to entropy is not macroscopically negligible compared with the 
other terms.  This is achieved by taking the limit $J^2_{1,\cdots,[{{D-1}
\over 2}]}\gg {\sqrt{ Q_1^{(1)}Q_1^{(2)}}}$.  

In such a near-BPS limit, the thermal entropy (\ref{area}) takes 
the form \cite{CVEy477}:
\begin{equation}
S_{thermo}=2\pi \left[{4\over {(D-3)^2}}Q^{(1)}_1
Q^{(2)}_1(2m)^{2\over{D-3}}-{{2}\over{(D-3)}}
\sum_{i=1}^{[{{D-1}\over 2}]} J^2_i\right]^{1\over 2}.
\label{nexarea}
\end{equation}

\subsubsection{Microscopic Interpretation}\label{entnearstat}

In this section, we calculate the {\it statistical entropy} of 
near-extreme, rotating black holes by counting the degenerate  
string states with the specific angular momenta.  

In principle, to calculate the statistical entropy 
of rotating black holes, one has to extract the degenerate 
string states (in a given level) with the specific values of angular momenta.  
This was first attempted in \cite{RUSs437}.  
Note, string states in a given level consist of states with different angular 
momenta (with the maximum angular momentum determined by the level).   
  
Alternatively, one can use the level density formula that has
contribution from all the possible angular momenta in a given level 
and employ the technique of conformal field theory to extract 
the specific contribution of states with given angular momenta.  
The main point is that the $U(1)$ charges of the left-moving worldsheet 
currents are interpreted as target space spins of string states.   
States with non-zero spins are obtained by applying 
the affine $U(1)$ current operators to spin zero states.  
The procedures described in the following paragraphs are also 
applicable to $D$-brane interpretation of entropy of rotating black holes 
\cite{BREmpv,BRElmpsv,MALs77}.  

As in the BPS case, one identifies the KK electric 
charges $Q^{(1)}_i$ [the 2-form electric charges $Q^{(2)}_i$] with the 
(internal) momentum zero modes [string winding modes].  Then, mass of the 
perturbative string states takes the form:
\begin{equation}
M_{string}^2=(Q_1^{(1)}+Q_1^{(2)})^2+{4\over {\alpha^{\prime}}}
(N_R-{1\over 2})=(Q_1^{(1)}-Q_1^{(2)})^2+{4\over
{\alpha^{\prime}}}(N_L-1), 
\label{nearstrmss}
\end{equation}
where each circle in the torus has self-dual radius 
$R=\sqrt{\alpha^{\prime}}$.  
From the second equality in (\ref{nearstrmss}), i.e the Virasoro constraint, 
one has the following relation between $N_L$ and $N_R$:
\begin{equation}
N_L= \alpha^{\prime}Q_1Q_2+N_R+{1\over 2}.
\label{lmodes}
\end{equation}
Note, for the statistical interpretation to be valid, the 
electric charge (quantized in unit $1/\sqrt{\alpha^{\prime}}$) 
has to be very large, i.e. $Q_1^{(1),(2)} \gg {1 \over
{\sqrt{\alpha^{\prime}}}}$. 

In the near extreme limit ($m\approx 0$), the BPS mass (\ref{para}) takes 
the form 
\begin{equation}
M^2_{BH} \approx (Q^{(1)}_1 +Q^{(2)}_2)^2 + {\cal O}(m). 
\label{nearexmass}
\end{equation}
By identifying the ADM mass (\ref{nearexmass}) 
with the mass of string state (\ref{nearstrmss}), one finds that the right 
movers are barely excited: $N_R\approx {1\over 2}+{\cal O}(m)$.  So, 
$N_R$ is negligible compared to $N_L$ and $N_L\approx\alpha^{\prime}
Q^{(1)}_1Q^{(2)}_2$ to a good approximation: $N_L \approx \alpha^{\prime}
Q^{(1)}_1Q^{(2)}_2 + 1+{\cal O}(m) \approx \alpha^{\prime}Q^{(1)}_1
Q^{(2)}_2 \gg N_R$.  Thus, to leading order, the 
logarithm of degeneracy $d(N_L,N_R)$ of the near-BPS states at the 
level $(N_L,N_R)$ takes the form
\begin{equation}
{\rm ln}\, d(N_L,N_R) \approx 2\pi\sqrt{{1\over 6}c_{eff}N_L} = 
4\pi\sqrt{N_L}, 
\label{aplevel}
\end{equation}
with the right-mover contributions neglected, just like  
BPS black holes.  Here, $c_{eff}=26-2=24$ since we are considering left 
moving bosonic string modes, only.  Note, this level density contains 
contribution from all the spin states in the level $(N_L,N_R)$.  

To extract the contribution by the states with particular spins, 
we employ conformal field theory technique.  Recall that the 
$\sigma$-model with the target space configuration 
\cite{CALhs359,CALhs367} given by a rotating black hole is 
described by the WZNW model \cite{NOV37,WIT92,KNIz247,GEPw278} 
with the $U(1)^{[{{D-1}\over 2}]}$ affine Lie algebra (i.e. Cartan 
sub-algebra of the $O(D-1)$ rotational group), or the conformal field 
theory with the $U(1)^{[{{D-1}\over 2}]}$ group manifold.  

The eigenvalues of the left-moving $U(1)$ worldsheet currents 
$j_i=i\partial_{\bar z}H^i$ ($i=1,\cdots,[{{D-1}\over 2}]$) are 
interpreted as the $[{{D-1}\over 2}]$ spins of string states.  
A general state in this WZNW model is labeled by charges of the 
affine Lie algebra as well as by the oscillator numbers.  
The conformal field 
${\bf \Phi}_{J_{1,\cdots,[{{D-1}\over 2}]}}$ with $U(1)$ charges 
$J_{1,\cdots,[{{D-1}\over 2}]}$ can be expressed as
\begin{equation}
{\bf \Phi}_{J_{1,\cdots,[{{D-1}\over 2}]}} =
\prod_{i=1}^{[{{D-1}\over 2}]}e^{{iJ_iH^i}}{\bf \Phi}_0,
\label{spinfields}
\end{equation}
where ${\bf \Phi}_0$ is a conformal field 
without $U(1)$ charges, or without target space spins.  
Thus, the left-moving conformal dimensions $\bar{h}$'s, 
i.e. the eigenvalues of the left-moving Virasoro generator $L_0$, 
of ${\bf \Phi}_{J_{1,\cdots,[{{D-1}\over 
2}]}}$ and ${\bf \Phi}_0$ are related as:
\begin{equation}
{\bar h}_{{\bf \Phi}_{J_{1,\cdots,[{{D-1}\over 2}]}}} =
{1\over {2}} \sum_{i=1}^{[{{D-1}\over 2}]}
J_i^2 + {\bar h}_{{\bf \Phi}_0}.
\label{confdim}
\end{equation}
This implies that the total number $N_{L\,0}$ of left moving 
oscillations of spinless states is reduced by the amount ${1\over {2}} 
\sum_{i=1}^{[{{D-1}\over 2}]}J_i^2$ relative to the total number 
$N_L$ of left moving oscillations of states with the specific spins 
$J_{1,\cdots,[{{D-1}\over 2}]}$:
\begin{equation}
N_L \to N_{L\,0}= N_L - {1\over {2}}
\sum_{i=1}^{[{{D-1}\over 2}]}J_i^2. 
\label{modNL}
\end{equation}

Note, the level density for spinless states in a given level 
$(N_L,N_R)$ differs from the level density $d(N_L,N_R)$ of all the 
states in the level $(N_L,N_R)$ by a numerical factor, which 
can be neglected in the large $(N_L,N_R)$ limit if one takes 
logarithm of the level densities.  So, one can use the formula 
(\ref{aplevel}) as the logarithm of the level density of spinless 
states to a good approximation. 
Then, the statistical entropy associated with degenerate string 
states with particular spins $J_{1,\cdots,[{{D-1}\over 2}]}$ at the  
level $(N_L,N_R)$ is 
\begin{equation}
S_{stat} \equiv \log d(N_{L\,0},N_{R\,0})\approx
4\pi\sqrt{N_{L\,0}} = 4\pi\left(N_L - {1\over {2}}
\sum_{i=1}^{[{{D-1}\over 2}]}J_i^2\right)^{1\over 2},
\label{statentnear}
\end{equation}
in the limit $N_L\gg {1\over {2}}\sum_{i=1}^{[{{D-1}\over 2}]}J_i^2$. 
For the angular momenta contribution to be statistically 
non-negligible, $\sqrt{N_L}\sum_{i=1}^{[{{D-1}\over 2}]}J_i^2\gg 1$ 
has to be satisfied.

In the near-extreme limit, $N_{L}\approx\alpha^{\prime}Q^{(1)}_1Q^{(2)}_1$.  
So, in terms of electric charges and angular momenta,  
the statistical entropy  takes the form:
\begin{equation}
S_{stat}=2\pi\left(4\alpha^{\prime}Q_1Q_2-2
\sum_{i=1}^{[{{D-1}\over 2}]}J_i^2\right)^{1\over 2}.
\label{statentf}
\end{equation}
This qualitatively agrees with the thermal entropy (\ref{nexarea}).  

\subsection{Black Holes and Fundamental Strings}\label{entfun}

In the previous sections, we calculated statistical entropy of 
black holes by assuming that perturbative string states are 
black holes.  Based on this assumption, we equated the mass of 
string states with the ADM mass of black holes, and identified the 
left and right moving momentum zero modes of the string states with 
the left and right handed electric charges of black holes.  This 
fixes $N_{L,R}$ (which determines the microscopic degeneracy of states) 
in terms of the macroscopic parameters 
of black holes, making it possible to calculate the {\it statistical 
entropy} of black holes.

In this section, we justify \cite{DABghw,CALmp475} such identification 
of perturbative string states with microscopic black hole states.  
The starting point is the fundamental string in $5\leq D\leq 10$. 
Here, the fundamental string is defined as a 1-brane solution of 
the combined action $S+S_{\sigma}$ with 
macroscopic string (described by $S_{\sigma}$) as its electric charge source.  
When the fundamental string is compactified on $S^1$ along its longitudinal 
direction, it {\it asymptotically approaches black hole}, as $r\to\infty$, 
with its core having milder singularity (than black hole in $(D-1)$ 
dimensions) of a $D$-dimensional string source.  
With this identification,  microscopic degrees of freedom of the 
{\it asymptotic} black hole in $(D-1)$ dimensions is interpreted as being 
due to oscillating macroscopic string at its core.  
And electric charges and angular momenta of the black hole are 
determined by momentum and winding modes of the core string along 
its longitudinal direction and the frequency of string 
oscillation in the rotational planes.  

\subsubsection{Fundamental String in $D$ Dimensions}\label{entfunsol}

We consider the following worldsheet action of macroscopic string 
moving in a background of the string massless modes: 
\begin{equation}
S_{\sigma}={1\over {4\pi\alpha^{\prime}}}\int d^2\sigma 
[\sqrt{\gamma}\gamma^{\alpha\beta}\partial_{\alpha}X^M \partial_{\beta}
X^N\hat{G}_{MN}+{\epsilon}^{\alpha\beta}\partial_{\alpha}X^M \partial_{\beta}
X^N\hat{B}_{MN}-{1\over 2}\sqrt{\gamma}\hat{\Phi}{\cal R}^{(2)}].
\label{sigmafun}
\end{equation}
The conformal invariance of  $S_{\sigma}$ leads to the 
equations of motion for the massless background fields \cite{TSE3,TSE4}.   
These equations of motion can be reproduced \cite{CALmpf} by the 
Euler-Lagrange equations of the effective action:
\begin{equation}
S={1\over{2\kappa^2_D}}\int d^D x\sqrt{-\hat{G}}e^{-2\hat{\Phi}}
[{\cal R}_{\hat{G}} +4\partial_M\hat{\Phi}\partial^M\hat{\Phi} 
-{1\over{12}}\hat{H}_{MNP}\hat{H}^{MNP}],
\label{effecact}
\end{equation}
where the $D$-dimensional gravitational constant $\kappa_D$ is 
related to the Newton's constant $G_D$
\footnote{$G_D$ is related to the quantities in (\ref{sigmafun}) as 
$G_D={1\over 8}\alpha^{\prime}e^{2\hat{\Phi}_{\infty}}$.} 
as $G_D={{\kappa^2_D}\over{8\pi}}$. 

When the target space is flat, i.e. $\hat{G}_{MN}=\eta_{MN}$ and 
$\hat{B}_{MN}=0$, one can exactly solve the $\sigma$-model (\ref{sigmafun}) 
to construct perturbative string states.  We concentrate on 
compactification of the $\sigma$-model on $S^1$ of radius $R$ 
in flat background.  The string states in this model are characterized 
by string winding number $n$ and quantized momentum zero mode $m/R$ along 
the $S^1$-direction.  The right- and left-moving momenta along the 
$S^1$-direction are
\begin{equation}
p_{R}={m\over{2R}}-{{nR}\over{2\alpha^{\prime}}}, \ \ \ \ 
p_{L}={m\over{2R}}+{{nR}\over{2\alpha^{\prime}}}.
\label{lrmom}
\end{equation}
For BPS states in heterotic string, whose supersymmetry is generated by 
right-moving worldsheet current, all the right movers are in the ground state 
($N_R={1\over 2}$) and the total number of left-moving oscillations 
is determined by the Virasoro constraint to be:
\begin{equation}
N_L=1+\alpha^{\prime}(p^2_R-p^2_L)=1-mn. 
\label{loscill}
\end{equation}
The mass of BPS states depend $p_R$, only:
\begin{equation}
M^2_{string}=4p^2_R.
\label{stringmass}
\end{equation}

From now on, we concentrate on BPS fundamental string solution
\cite{DABh,DABgh,DABghw,DAB357,HUL357} to this theory. 
The background fields of such ``straight'' fundamental string solution 
have the form \cite{DABgh}:
\begin{eqnarray}
\hat{G}_{MN}dx^M dx^N&=&-e^{2\hat{\Phi}}dudv + d\vec{x}\cdot d\vec{x}, 
\ \ \ 
\hat{B}_{uv}={1\over 2}(e^{2\hat{\Phi}}-1), 
\cr 
e^{-2\hat{\Phi}}&=&1+{Q\over r^{D-4}}, \ \ \ \ 
(Q={{\kappa^2_D}\over{\pi\alpha^{\prime}(D-4)\Omega_{D-3}}}), 
\label{straight}
\end{eqnarray}
where $u\equiv x^0-x^{D-1}$ and $v=x^0+x^{D-1}$ are the lightcone
coordinates along the string worldsheet, and $x^m$ ($m=1,...,(D-2)$)
are the transverse coordinates.  
In deriving this solution, we chose the static gauge for $X^M$:
\begin{equation}
X^{\mu}=\xi^{\mu}, \ \ \ X^m={\rm constant}, 
\label{static}
\end{equation}
where $\xi^{\mu}=(\tau,\sigma)$ is the worldsheet coordinate.
This fundamental string solution preserves $1/2$ of the spacetime 
supersymmetry \cite{DABgh}.  When the fundamental string 
is compactified along its longitudinal direction on $S^1$ of radius 
$R$, i.e. $x^{D-1}=x^{D-1}+2\pi R$, one obtains point-like solution in 
$D-1$ dimensions with its charge proportional to the winding number 
$n$ along the $S^1$-direction.  

We now obtain solution that 
also has an arbitrary left-moving oscillation,  
which is a source for microscopic degeneracy.  The 
zero-modes of the left-moving oscillation induce momentum 
$m/R$ in the $S^1$-direction.  Applying the general prescription of 
solution generating transformation discussed in 
\cite{VAC277,VAC238,GAR41,GARv42,GAR46} to the straight fundamental 
string solution (\ref{straight}), one obtains the following 
left-moving oscillating fundamental string solution 
\begin{eqnarray}
\hat{G}_{MN}dx^M dx^N &=& -e^{2\hat{\Phi}}(dudv-T(v,\vec{x})dv^2)+
d\vec{x}\cdot d\vec{x}, 
\nonumber \\
\hat{B}_{uv}&=&{1\over 2}(e^{2\hat{\Phi}}-1), \ \ \ \ \ 
e^{-2\hat{\Phi}}=1+{Q\over r^{D-4}},  
\label{oscil}
\end{eqnarray}
where $T(v,\vec{x})$ is a solution to $\partial^2_{\vec{x}}T(v,\vec{x})=0$.  
The general form of $T(v,\vec{x})$ that can 
be matched onto the string source at the core is
\begin{equation}
T(v,\vec{x}) = \vec{f}(v)\cdot\vec{x} + p(v)r^{-D+4}, 
\label{rellap}
\end{equation}
where the first term corresponds to oscillating string source and the 
second term corresponds to a momentum without oscillations.  

One can bring (\ref{oscil}) to a manifestly asymptotically flat form 
by applying the coordinate transformations 
\begin{equation}
v=v^{\prime}, \ \ \ 
u=u^{\prime}-2\dot{\vec{F}}\cdot\vec{x}^{\prime}+2\dot{\vec{F}}\cdot
\vec{F}-\int^{v^{\prime}} \dot{F}^2 dv, \ \ \ 
\vec{x}=\vec{x}^{\prime} -\vec{F}, 
\label{tranflat}
\end{equation}
where $\dot{}\equiv\partial/\partial v$, 
$\vec{f}(v)=-2\ddot{\vec{F}}$ and $\dot{F}^2 = \dot{\vec{F}}\cdot
\dot{\vec{F}}$.  
In this new coordinates, (\ref{oscil}) takes the form (with primes 
suppressed)
\begin{eqnarray}
\hat{G}_{MN}dx^M dx^N&=&-e^{2\hat{\Phi}}dudv +[e^{2\hat{\Phi}}p(v)r^{-D+4}
-(e^{2\hat{\Phi}}-1)\dot{F}^2]dv^2 
\nonumber \\
& &+2(e^{2\hat{\Phi}}-1)\dot{\vec{F}}\cdot d\vec{x}dv +d\vec{x}
\cdot d\vec{x}, 
\nonumber \\
\hat{B}_{uv}&=&{1\over 2}(e^{2\hat{\Phi}}-1), \ \ \ 
\hat{B}_{vi}=\dot{F}_i(e^{2\hat{\Phi}}-1), 
\nonumber \\
e^{-2\hat{\Phi}}&=&1+{Q\over{|\vec{x}-\vec{F}|^{D-4}}}.  
\label{oscilflat}
\end{eqnarray}
This solution has the ADM mass $\pi^{0,0}_{ADM}$  
and the ADM momentum per unit length $\pi^{0,i}_{ADM}$, 
and the $D$-momentum flow along the string $\pi^{D-1,M}_{ADM}$ 
given by
\begin{equation}
\left(\matrix{\pi^{0,M}_{ADM}\cr \pi^{D-1,M}_{ADM}}\right) = 
{{(D-4)\Omega_{D-3}}\over{2\kappa^2_D}}\left(
\matrix{Q+Q\dot{F}^2+p & Q\dot{F}^i & -Q\dot{F}^2-p \cr
-Q\dot{F}^2-p & -Q\dot{F}^i & -Q+Q\dot{F}^2+p}\right). 
\label{admfmom}
\end{equation}

\subsubsection{Level Matching Condition}\label{entfunmat}

In section \ref{entfunsol}, we constructed oscillating fundamental 
string solution by solving the equations of motion (following from 
$S+S_{\sigma}$) of the target space background fields  
and imposing solution generating transformations.  
Note, all of such solutions do not correspond to the 
underlying string states.  To ensure that these solutions match onto 
the perturbative string states at the core, one has to additionally 
solve the equations of motion for string coordinates $X^M$ 
and the Virasoro constraints.  From the Virasoro constraints (or the level 
matching conditions), one extracts the relations between the {\it macroscopic} 
quantities of the oscillating fundamental string (\ref{oscilflat}) and 
the {\it microscopic} quantities of the perturbative string states. 
This allows to interpret the entropy of the target space solution 
in terms of the perturbative string state degrees of freedom. 
 
We start by choosing the following static gauge for the string coordinates 
$X^M$ in the coordinate frame corresponding to the solution in 
(\ref{oscil}):
\begin{equation}
U=U(\sigma^+,\sigma^-),\ \ \ \   V=V(\sigma^+,\sigma^-),\ \ \ \  X^m =0, 
\label{strcoor}
\end{equation}
where $\sigma^{\pm}=\tau\pm\sigma$ are the light-cone worldsheet 
coordinates.  Also, we choose the conformal gauge for strings, i.e. 
$\gamma_{\alpha\beta}={\rm diag}(-1,1)$.   From the Virasoro 
constraints $T_{++}=0=T_{--}$, one has the following form of  
string coordinates (\ref{strcoor}):
\begin{equation}
U=(2Rn+a)\sigma^{-},  \ \ \ \ \   V=2Rn\sigma^{+}, 
\label{solstrcoord}
\end{equation}
where $a\equiv {1\over\pi}\int^{2\pi Rn}_0\dot{F}^2$ is the zero mode of 
$\dot{F}^2$.  And the constant $Q$ in (\ref{oscil}) is expressed 
in terms of the perturbative string state quantities as 
\begin{equation}
Q={{n\kappa^2_D}\over{\pi\alpha^{\prime}(D-4)\Omega_{D-3}}}. 
\label{qexp}
\end{equation}

More information on matching of the spacetime solution onto states 
of the core string source is extracted by taking the flat 
spacetime limit $\kappa_D\to 0$, in which for example the Virasoro 
relations (\ref{lrmom}), (\ref{loscill}) and (\ref{stmass}) 
are valid.  For this purpose, we go to the frame represented by 
(\ref{oscilflat}), where the metric is manifestly 
asymptotically flat, by applying the transformations 
(\ref{tranflat}).  In this new frame (denoted by primes), $X^M$ take 
the form:
\begin{equation}
V^{\prime}=2Rn\sigma^{+}, \ \ \ 
U^{\prime}=(2Rn+a)\sigma^{-} +\int_{V^{\prime}}\dot{F}^2,  \ \ \ 
\vec{X}^{\prime}=\vec{F}(V^{\prime}), 
\label{primstcoord}
\end{equation}
manifestly showing that the core string is oscillating with profile $\vec{F}
(V^{\prime})$.    In the flat spacetime limit, i.e. $\kappa_D\to 0$ 
or $\langle\hat{\Phi}\rangle\to 0$, 
a perturbative string state has the momentum $p^{M}$ (conjugate to 
$X^{M}$, obtained from $S_{\sigma}$) 
and the winding vector $n^{M}$ given in the coordinates $(X^{\prime\,0},
\vec{X}^{\prime}, X^{\prime\,D-1})$ by 
\begin{equation}
n^{M}=(0,\vec{0},n), \ \ \ \ 
p^{M}=(2\alpha^{\prime})^{-1}(2nR+a,\vec{0},-a).
\label{widmom}
\end{equation}  
This expression for $p^{M}$ agrees with the $\kappa_D\to 0$ limit of 
the ADM momentum $\pi^{0,M}_{ADM}$ (\ref{admfmom}) of the target space 
solution (\ref{oscilflat}) with $Q$ given by (\ref{qexp}).  
This confirms that oscillating fundamental strings are  
matched onto perturbative states of the core macroscopic string.  

Since the momentum zero mode of a perturbative string state along the 
(compactified) $X^{D-1}$-direction is $m/R$ and $N_L=-nm$ ((\ref{loscill}) 
in the large $N_L$ limit), one can read off the expression for $p^{\mu}$ 
in (\ref{widmom}) to express $m$ and $N_{L}$ in terms of the {\it macroscopic} 
quantities of the fundamental string solution:
\begin{equation}
m=-{{Ra}\over{2\alpha^{\prime}}}, \ \ \ \ \ 
N_{L}={{nRa}\over{2\alpha^{\prime}}}. 
\label{levmatmo}
\end{equation}
Thus, we see that the oscillating fundamental string solution 
(\ref{oscilflat}) with $p(v)=0$ is matched onto the perturbative 
string state with $n$, $m$ and $N_L$ given in (\ref{levmatmo}).  
These are a subclass of solutions (\ref{oscilflat}) that can be matched 
onto perturbative states of the core source string.

\subsubsection{Black Holes as String States}\label{entfunbh}

When the longitudinal direction of the fundamental string 
(\ref{oscilflat}) is compactified, one has point-like object in 
$D-1$ dimensions.  Such a solution approaches Sen's BPS electric 
black hole \cite{SEN440} as $r\to\infty$.  This allows one 
to relate the {\it macroscopic} quantities,  
which are defined at spatial infinity, of black holes to the 
{\it microscopic} quantities 
of perturbative string states. 
Note, such point-like solutions in $D-1$ dimensions 
asymptotically approach only a subset of Sen's black holes that can 
be matched onto perturbative string states.   So, for example, 
the angular momenta of such solutions follow the Regge bound 
of perturbative string states, whereas Sen's rotating black holes 
\cite{SEN440}, 
in general, take arbitrary values of angular momenta, which do not 
satisfy the Regge bound.  
In the following, we discuss the $D=5$ case 
for the purpose of illustrating basic ideas.  
The generalization to an arbitrary $D$ 
is straight forward; one starts from $D$-dimensional solution 
(\ref{oscilflat}) with more general profile function $\vec{F}(v)$.  

One compactifies the longitudinal direction of the $D=5$ 
fundamental string (\ref{oscilflat}) with $p(v)=0$ on $S^1$ 
of radius $R$
to obtain a point-like solution in $D=4$.  
To make the resulting $D=4$ point-like solution 
approach a ``rotating'' black hole asymptotically, one  
chooses the following form of $\vec{F}$ that describes  
rotation in the $(x^1,x^2)$-plane with amplitude $A$ and angular frequency 
$\omega$
\footnote{The generalization to a $(D-1)$-dimensional rotational black hole
with $[{{D-2}\over 2}]$ angular momenta involves $\vec{F}$ 
representing independent rotations in $[{{D-2}\over 2}]$ mutually orthogonal 
planes.}:
\begin{equation}
\vec{F}=A(\hat{e}_1\,{\rm cos}\,\omega t +\hat{e}_2\,{\rm sin}\,\omega t),
\label{oscill}
\end{equation}
where $\hat{e}_i$ is a unit vector in the $x^i$-direction.  

Since the $D=4$ point-like solution depends on the compactified 
coordinate $x^4$ and the time coordinate $t$ through $v=x^4+t$, the 
compactification on $x^4$ (i.e. taking average over $x^4$ so that only 
the zero modes of fields are kept) is equivalent to taking the 
time-average.   
By taking the time-average of the leading order terms of the fields 
at large $r$, one can read off the following 
ADM mass $M_{BH}$, angular momentum $J$, the right- and left-handed electric 
charges $Q_{R,L}\equiv {{\pm Q^{(1)}+Q^{(2)}}\over \sqrt{2}}$, and 
the right- and left-handed magnetic moments $\mu_{R,L}$ of the rotating 
black hole that the point-like solution approaches 
asymptotically:
\begin{eqnarray}
M_{BH}&=&{{Q(1+A^2\omega^2)}\over 4}, \ \ \ \ \ \ \ 
J={{QA^2\omega}\over 4}, 
\nonumber \\
Q_L &=& -{Q\over{2\sqrt{2}}}(A^2\omega^2-1), \ \ \ \ 
Q_R=-{Q\over{2\sqrt{2}}}(1+A^2\omega^2), 
\nonumber \\
\mu_L&=& 0, \ \ \ \ \ \ \ \ \ \ \ 
\mu_R = -\sqrt{2}J,  
\label{bhpar}
\end{eqnarray}
with choice of unit in which $G_D=1$.  
With this choice of normalization, $Q$ in (\ref{qexp}) becomes 
$Q=4nR/\alpha^{\prime}$, and $p_{L,R}$ and $Q_{L,R}$ are related as 
$Q_{L,R}=2\sqrt{2}p_{L,R}$.  Furthermore, the periodicity of $x^4$ 
requires $\omega =\ell/(nR)$ for some integer $\ell$.  

With this identification, one has $M_{BH}=2p_R$, as one would expect 
from the fact that the target space solution with 
the ADM mass $M_{BH}$ is matched onto the string source with the right 
moving momentum $p_R$.  This proves the assumption that the 
perturbative string states are black holes.  Furthermore, $J$  
in (\ref{bhpar}), which reduced to the form $J={{A^2\ell}\over 
{\alpha^{\prime}}}$, satisfies the Regge bound $J\leq |2+\alpha^{\prime}
(p^2_R-p^2_L)|={{A^2\ell^2}\over{\alpha^{\prime}}}$, with $J$ forming 
the Regge trajectory when $\ell=1$.  
Finally, the gyromagnetic ratios $g_{L,R}$, defined by $\mu_{L,R}
={{g_{L,R}Q_{L,R}J}\over{2M_{BH}}}$, are
\begin{equation}  
g_{L}=0, \ \ \ \ \ \   g_{R}=2.
\label{gyro}
\end{equation} 
This result is consistent with the fact that the BPS black holes  
correspond to the perturbative string states with only left-mover excited, 
since the gyromagnetic ratios are related to the left- and right-moving 
angular momenta $J_{L,R}$ as $g_{L,R}=2{{J_{R,L}}\over{J_R +J_L}}$.  
Moreover, the right-moving gyromagnetic ratio is 2 as expected from the 
fact that the underlying states are fundamental.

\subsection{Dyonic Black Holes and Chiral Null Model}\label{entch}

So far, we discussed statistical interpretation for entropy of 
purely electric black holes in terms of microscopic degrees of 
freedom of perturbative string states.   
The BPS electric black holes have a nice virtue of being free of 
quantum corrections, thereby the ADM mass can be trusted.  
However, due to singularity at the horizon, the horizon gets shifted 
by $\sim\sqrt{\alpha^{\prime}}$ through the $\alpha^{\prime}$-corrections.  
Thus, entropy is known only up to the order 
of $\alpha^{\prime}$. 
Furthermore, such solutions are not black holes in the conventional 
sense, since the event horizon coincides with the singularity and has zero 
area.  

It is the construction of dyonic solutions \cite{CVEy672} in the heterotic 
string on $T^6$ that triggered renewed interests in black hole entropy 
and made the precise calculation of the statistical entropy possible.  
Such dyonic solutions not only do not receive quantum corrections, but 
also are free of classical $\alpha^{\prime}$-corrections\cite{CVEt366} 
since they are described by exact conformal $\sigma$-model and the event 
horizon is free of singularity.  
Such dyonic solutions contain as a subset the Reissner-Nordstr\"om solution 
and have non-zero event horizon area.  Since the event horizon is 
regular, the $\alpha^{\prime}$-corrections are under control at 
the event horizon. And as in the pure electric case, the dilaton is 
finite at the event horizon, implying that the string loop 
corrections are under control.  Being free of plagues 
(i.e. $\alpha^{\prime}$-corrections of purely electric solutions 
and the string loop corrections of the purely magnetic 
solutions) suffered by the previously known solutions in string theories, 
the dyonic solution \cite{CVEy672} is suitable for studying statistical 
origin of black hole entropy.  Such observation was first made in 
\cite{LARw375}, where it is proposed that the microscopic degrees of 
freedom of the dyonic black holes are due to the hair associated 
with the oscillations in the {\it internal} dimensions. 

The first attempt to explain the statistical origin of the BPS black holes 
with non-zero event horizon area is based upon a special class of string 
worldsheet $\sigma$-model called ``chiral null model''.  In this approach, 
the BPS black holes are embedded as background fields of the chiral null 
model and the throat region conformal model is studied for 
understanding microscopic degeneracy of string states.  
Remarkably, the throat region conformal theory approximates to 
the WZNW model of {\it perturbative} string theory with string tension 
rescaled
\footnote{In different approach \cite{HALkrs392,HALkrs075,HAL175,HAL068} 
based upon the Rindler geometry in throat region, black hole in the weak 
coupling limit is described by closed strings in the background of 
black hole carrying non-perturbative charges.  The effect of these 
non-perturbative charges on closed strings is to rescale the string 
tension, as in the case of chiral null model approach.}
by magnetic charges of the black holes.  So,  
the degeneracy of string states carrying magnetic charges is 
obtained by applying level density formula of {\it perturbative} string 
states.  In this section, we summarize the results of 
\cite{CVEt53,TSE11,TSE477}, which study chiral null model interpretation 
of black hole entropy.  

\subsubsection{Chiral Null Model}\label{entchmodel}

String theory is a promising candidate for consistent quantum gravity 
theory, being free from ultra-violet divergences, which plagued 
quantum gravity of point-like particles.  So, it is useful to study 
classical string solutions to address problems in quantum gravity.  
However, it is almost hopeless to obtain exact classical solutions 
to the equations of motion following from effective field theory of 
string massless states, since the effective action consists of infinite 
series of terms of all derivatives multiplied by powers of $\alpha^{\prime}$, 
which are also ambiguous due to the freedom of choosing different 
renormalization schemes (or field redefinitions).   So, the only {\it exact} 
classical string solutions that one can study are those that do not have  
$\alpha^{\prime}$-corrections.  In fact, there exist classes of string 
$\sigma$-models whose background fields do not receive 
$\alpha^{\prime}$-corrections in a special renormalization scheme.   
One starts from a string $\sigma$-model which is shown to 
be conformal to all orders in $\alpha^{\prime}$ and looks for classical 
solutions to the leading order (in $\alpha^{\prime}$) effective field 
theory which can be embedded as target space background field configurations 
of the $\sigma$-model, or vice versa.  

This approach of studying classical solutions of string theory is based 
upon a remarkable relationship between the conformaly invariant string 
$\sigma$-model and the extremum of the effective action.  
To the leading order in string coupling, the string field 
equations are obtained by the conformal (Weyl) invariance condition 
of $\sigma$-model, which are equivalent to the stationary conditions of 
the effective action due to the proportionality between the Weyl anomality 
coefficients and derivatives (with respect to fields) of the effective 
action.   Given a conformal $\sigma$-model, one obtains a string solution 
not modified by $\alpha^{\prime}$-corrections.  

The general bosonic $\sigma$-model describing string propagation in  
background of massless fields $G_{MN}$, $B_{MN}$ and $\Phi$ is
\begin{equation}
I={1\over{\pi\alpha^{\prime}}}\int dz^2\left[(G_{MN}+B_{MN})(X)
\partial X^M\bar{\partial}X^N+\alpha^{\prime}{\cal R}\Phi(X)\right].
\label{bosonicsigma}
\end{equation}
The chiral null model is a special case of  
(\ref{bosonicsigma}) with the Lagrangian:
\begin{eqnarray}
L&=&F(x)\partial u\left[\bar{\partial}v+K(x,u)\bar{\partial}u
+2{\cal A}_i(x,u)\bar{\partial}x^i\right]
\cr
& &\ \ \ \ \ \ \ \ \ \ \ \ \ \ \ \,
+(G_{ij}+B_{ij})(x)\partial x^i\bar{\partial}x^j+{\cal R}\Phi(x),
\label{chiralnullag}
\end{eqnarray}
where $X^M$ are splitted into `light-cone' coordinates 
$u,v$ and `transverse' coordinates $x^i$, i.e. $X^M=(u,v,x^i)$.  
Note, $F$ does not depend on $u$.  

There exists a special renormalization scheme in which the $\sigma$-model 
(\ref{chiralnullag}) is conformal to all orders in $\alpha^{\prime}$, 
provided $(i)$ the transverse $\sigma$-model $L_{\perp}=(G_{ij}+B_{ij})
(x)\partial x^i\bar{\partial}x^j+{\cal R}\phi(x)$, where $\phi\equiv 
\Phi-{1\over 2}\ln\,F$, is conformal and $(ii)$ $F$, $K$, ${\cal A}_i$ 
and $\Phi$ satisfy conformal invariance conditions, which for 
the case $L_{\perp}$ has at least $(4,0)$ supersymmetry  
take the form
\footnote{This follows \cite{HORt51} from the standard leading order 
conformal invariance conditions $\hat{R}_{-MN}+2\hat{D}_{-M}\hat{D}_{-N}
\Phi=0$ of the general $\sigma$-model (\ref{bosonicsigma}).  
Here, the Ricci tensor $\hat{R}_{-MN}$ and covariant derivative 
$\hat{D}_{-M}$ are defined in terms of the generalized connection 
$\hat{\Gamma}^{P}_{\pm MN}=\Gamma^{P}_{MN}\pm{1\over 2}H^P_{MN}$ with 
torsion $H_{MNP}$.} 
\cite{CVEt53}:
\begin{eqnarray}
& &-\textstyle{1\over 2}\nabla^2F^{-1}+\partial^i\phi\partial_iF^{-1}=0,
\ \ \ 
-\textstyle{1\over 2}\nabla^2K+\partial^i\phi\partial_iK
+\partial_u\nabla_i{\cal A}^i=0,
\cr
& &-\textstyle{1\over 2}\hat{\nabla}_i{\cal F}^{ij}+\partial_i\phi
{\cal F}^{ij}=0,\ \ i.e.\ \ 
\nabla_i(e^{-2\phi}{\cal F}^{ij})-\textstyle{1\over 2}e^{-2\phi}
{\cal F}_{ik}H^{ikj}=0,
\label{confcondofchnul}
\end{eqnarray} 
where ${\cal F}_{ij}=\partial_i{\cal A}_j-\partial_j{\cal A}_i$, 
$H_{ijk}=3\partial_{[i}B_{jk]}$ and the covariant derivative 
$\hat{\nabla}\equiv\nabla(\hat{\Gamma})$ is defined in terms of 
the generalized connection $\hat{\Gamma}^i_{jk}\equiv 
\Gamma^i_{jk}+{1\over 2}H^i_{jk}$ with torsion.  

The chiral null model has one null Killing vector which generates shifts 
of $v$: the action is invariant under an affine symmetry $v\to 
v^{\prime}=v+h(\tau+\sigma)$.  The associated null Killing vector 
$\partial/\partial v$ gives rise to the conserved current 
$J_v=F(x)\partial u$ on the string worldsheet.  
A balance between the metric and the antisymmetric tensor ($G_{ui}=B_{ui}$) 
implies that the conserved current $J_v$ is chiral, which is a crucial 
condition for the conformal invariance \cite{HORt50,HORt51}.   
The action (\ref{confcondofchnul}) is also invariant under the following 
subgroup of coordinate transformations on $v$ combined with a gauge 
transformation of $K$ and ${\cal A}_i$:
\begin{equation}
v\to v-2\eta(x,u),\ \ \ K\to K+2\partial_u\eta,\ \ \ 
{\cal A}_i\to {\cal A}_i+\partial_i\eta.
\label{coordgaugcombtrn}
\end{equation}
Unless $K$, ${\cal A}_i$ and $\Phi$ do not depend on $u$, one can choose 
a gauge in which $K=0$ by applying (\ref{coordgaugcombtrn}).   When the 
fields are independent of $u$,  
the chiral null model turns out to be self-dual.  Namely, a 
leading-order duality transformation along any non-null direction in 
the $(u,v)$-plane (say along the $u$-direction, where $v=\hat{v}+au$ with 
$a$ constant) leads to a $\sigma$-model of the same form with duality 
transformed background fields given by
\begin{equation}
F^{\prime}=(K+a)^{-1},\ \  K^{\prime}=F^{-1},\ \ 
{\cal A}^{\prime}_i={\cal A}_i,\ \ 
\Phi^{\prime}=\Phi-\textstyle{1\over 2}\ln[F(K+a)]. 
\label{dualtranchnul}
\end{equation}  

When background fields are independent of $u$, the conformal invariance 
conditions (\ref{confcondofchnul}) take the form of  the Laplace equations 
in the transverse space.  Background field solutions to the conformal 
invariance conditions are then parameterized by harmonic functions in the 
transverse space.  Since the equations are linear, one can superpose 
harmonic functions to generate multi-center solutions.  
One can further generalize background fields to depend 
on $u$ in such a way that the conformal invariance conditions 
(\ref{confcondofchnul}) are still satisfied.  Such changing of background 
fields is viewed as `marginal deformations' \cite{HORt51} of the 
conformal field theory.   In particular, adding zero modes 
of ${\cal A}_i$ has the effect of adding a Taub-NUT charge, angular 
momenta or extra electric/magnetic charges to the original solutions.  

The chiral null model (\ref{chiralnullag}) generalizes $K$-model 
(plane fronted wave solution) and $F$-model (a generalization of the 
fundamental string solution).   First, in the limit $F=1$, 
(\ref{chiralnullag}) describes a class of plane fronted 
wave backgrounds which have a covariantly constant null vector 
$\partial/\partial v$ ($K$-model).  For this case, one can add another 
vector coupling $2\bar{\cal A}_i(x,u)\partial x^i\bar{\partial}u$ to the 
Lagrangian while still preserving conformal invariance but breaking 
chiral structure.   Second, when $K=0$ with background fields independent 
of $u$, (\ref{chiralnullag}) reduces to the $F$-model, which has two 
null Killing vectors $\partial/\partial u$ and $\partial/\partial v$   
associated with affine symmetries $u^{\prime}=u+f(\tau-\sigma)$ and 
$v^{\prime}=v+h(\tau+\sigma)$.  
Since the coupling to $u$ and $v$ is chiral ($G_{uv}=B_{uv}$), the 
associated 2 conserved currents $\bar{J}_u=F\bar{\partial}v$ and 
$J_v=F\partial u$ are chiral.  
The $F$ model and the $K$ model are related by a duality transformation 
(\ref{dualtranchnul}) along $u$ with $a=0$.

The dimensional reduction of chiral null model leads to various 
charged (under the KK or 2-form gauge field) black hole 
and string solutions, which can also carry angular momenta or the 
Taub-NUT charge.  The chiral coupling leads to a no-force condition 
(a characteristic of BPS solutions) on the solutions, which allows 
the construction of multi-centered solutions.  
The balance between $G_{MN}$ and $B_{MN}$ in chiral theories manifests 
in lower dimensional solutions as the fixed ratio of mass and charge, i.e. 
the BPS condition.  

In the following, we discuss black hole solutions that satisfy the conformal 
invariance conditions (\ref{confcondofchnul}) and therefore are exact to 
all orders in $\alpha^{\prime}$.  
For the purpose of obtaining general $D=4,5$ black holes, we split the 
transverse coordinates $x^i$ into non-compact ones $x^s$ and 
compact ones $y^n$, i.e. $x^i=(x^s,y^n)$, where $s=1,...,D-1$ 
and $n=1,...,9-D$.  We decompose the 8-dimensional transverse space 
into the direct product $M^4\times T^4$ of some 4-space $M^4$ and $T^4$.  
The chiral null model Lagrangian (\ref{chiralnullag}) is accordingly 
expressed as the sum of two terms associated with each 
space.  Here, $M^4$ has $SO(3)$ [$SO(4)$] symmetry for the 
$D=4$ [$D=5$] black holes and, therefore, $M^4$ is parameterized by 
$(x^s,y^1)$ [$x^m$] with background fields depending on the  
coordinates only through $r=\sqrt{x^sx^s}$ [$r=\sqrt{x^mx^m}$].   
We consider the case where $M^4$ has the torsion related 
to dilaton $\phi$ in the specific way $H^{mnk}=-{2\over\sqrt{G}}
\epsilon^{mnkl}\partial_l\phi$, so that the last conformal invariance 
condition in (\ref{confcondofchnul}) simplifies to a 
Laplace-form $\nabla_i(e^{-2\phi}{\cal F}^{ij}_+)=0$, where 
${\cal F}^{ij}_+\equiv{\cal F}^{ij}+\star{\cal F}^{ij}$ with 
$\star{\cal F}^{ij}={1\over{2\sqrt{G}}}\epsilon^{ijkl}{\cal F}_{kl}$.  

\paragraph{General Four-dimensional, Static, BPS Black Hole}

We consider the case where 4-dimensional transverse part of 
the metric has the form $G_{ij}=f(x)g_{ij}$ where $\nabla^2f=0$, 
$\nabla\equiv\nabla(g)$ and $g_{ij}$ is a hyper-K\"ahler metric 
with a translational isometry in the $x^4$-direction.   
The $D=6$ part $(u,v,x^1,...,x^4)$ of (\ref{chiralnullag}) 
then takes the special form:  
\begin{eqnarray}
L&=&F(x)\partial u\left[\bar{\partial}v+K(x)\bar{\partial}u
+2A(x)(\bar{\partial}x^4+a_s(x)\bar{\partial}x^s)\right]
+\textstyle{1\over 2}{\cal R}\ln F(x)+L_{\perp},
\cr
L_{\perp}&=&f(x)k(x)(\partial x^4+a_s(x)\partial x^s)
(\bar{\partial}x^4+a_s(x)\bar{\partial}x^s)
+f(x)k^{-1}(x)\partial x^s\bar{\partial}x^s
\cr
& &\ \ \ \ \ \ \ 
+b_s(x)(\partial x^4\bar{\partial}x^s-\bar{\partial}x^4
\partial x^s)+{\cal R}\phi(x),
\label{4dbhchnullag}
\end{eqnarray}
where $x^s=(x^2,x^2,x^3)$ are non-compact coordinates and compact 
coordinates are $x^4=y_1$ and $u=y_2$.  Here, we chose  
${\cal A}_i$ in (\ref{chiralnullag}) to take the form 
${\cal A}_4=A$ and ${\cal A}_s=Aa_s$ so that the 
$D=4$ metric has no Taub-NUT term.  

The Lagrangian (\ref{4dbhchnullag}) is invariant under $T$-duality 
transformations in the $x^4$-direction ($P_1\leftrightarrow P_2$ and 
$q\to-q$): 
\begin{equation}
f\to k^{-1},\ \ \ k\to f^{-1},\ \ \ a_s\leftrightarrow b_s,\ \ \ 
A\to (fk)^{-1}A,
\label{4dchnulxftdual}
\end{equation}
and in the $u$-direction ($Q_1\leftrightarrow Q_2$):
\begin{equation}
F\to K^{-1},\ \ \ K\to F^{-1},\ \ \ \Phi(F)\to \Phi(K^{-1}).
\label{4dchnulutdual}
\end{equation}
When $A=0$, the Lagrangian has remarkable manifest invariance 
under $D=4$ $S$-duality, under which (\ref{4dchnulutdual}) 
transforms as $u\leftrightarrow x^4$ and 
\begin{equation}
F\to f^{-1},\ \ \ K\to k^{-1},\ \ \ f\to F^{-1},\ \ \ k\to K^{-1},
\label{4dchnulsdual}
\end{equation}
and under the $D=6$ string-string duality ($G\to e^{-2\Phi}G$, 
$dB\to e^{-2\Phi}\star dB$, $\Phi\to -\Phi$) between the heterotic 
string on $T^4$ and the type-II string on $K3$: 
\begin{equation}
F\to f^{-1},\ \ \ K\to K,\ \ \ f\to F^{-1},\ \ \ k\to k.
\label{4dchnulstrstr}
\end{equation}
Note, the invariance of (\ref{4dbhchnullag}) 
under the $T$-duality is manifest only when all the 
charges associated with 4 harmonic functions $F$, $K$, $f$ and 
$k$ are non-zero.  The self-dual case $F=K^{-1}=f^{-1}=k$ and 
$a_s=b_s$ corresponds to the $D=4$ Reissner-Nordstr\"om 
solution.  As expected, the combined transformation of 
$T$-duality and the string-string duality yields the $D=4$  
$S$-duality (\ref{4dchnulsdual}).  

One can obtain $D=4$ black hole solution which is exact to all orders 
in $\alpha^{\prime}$ by solving (\ref{confcondofchnul}) with 
(\ref{4dbhchnullag}) and all the background fields 
depending on non-compact transverse coordinates $x^s$, only.  
Solutions for background fields are expressed in 
terms of harmonic functions $f$, $k$, $F$ and $K$, which satisfy 
(linear) Laplace equations.  
Particularly, $A$ is given in terms of harmonic functions by 
$A=q_1k^{-1}+q_2f^2k$ ($q_{1,2}=$ const).  If one further 
assumes the asymptotic flatness condition 
(i.e. $k\to 1$, $f\to 1$, $A\to 0$ as $r=\sqrt{x^sx_s}\to\infty$), 
coefficients in $A$ are restricted such that $q_0:=q_1=-q_2$.  
The solutions for background fields are:
\begin{eqnarray}
F^{-1}&=&1+{{Q_2}\over r},\ \ \ K=1+{{Q_1}\over r},\ \ \ 
f=1+{{P_2}\over r},\ \ \  k^{-1}=1+{{P_1}\over r},
\cr
a_sdx^s&=&P_1(1-\cos\theta)d\varphi,\ \ \ \ \ \ \ \ \ \ \ 
b_sdx^s=P_2(1-\cos\theta)d\varphi,
\cr
A&=&{q\over r}\cdot{{r+\textstyle{1\over 2}(P_1+P_2)}\over 
{r+P_1}},\ \ \ \ \ \ \ \ \ \ 
e^{2\Phi}=Fe^{2\phi}={{r+P_2}\over{r+Q_2}},
\label{4dbckgrdsolchnul}
\end{eqnarray}
where $q\equiv 2q_0(P_1-P_2)$.  Since the resulting (conformal invariance 
condition) equations are of Laplace-type, one can superpose harmonic 
functions to obtain multi-center generalization of the above. 

The $D=4$ spherically symmetric solution in (\ref{chnulsol})
is obtained by applying the standard 
KK procedure with all the background field in 
(\ref{chiralnullag}) properly identified with those in 
(\ref{bosonicsigma}) and setting $u=y_2$, $v=2t$, $x_4=y_1$.  

\paragraph{General Five-Dimensional, Rotating, BPS Black Hole}

We consider the case where $M^4$-part is (locally) 
$SO(4)$-invariant.  The $D=6$ part of  
chiral null model Lagrangian is again given by 
(\ref{4dbhchnullag}) with ${\cal A}_i=({\cal A}_m,0)$ and the 
transverse part $L_{\perp}$ replaced by
\begin{equation}
L_{\perp}=f(x)\partial x^m\bar{\partial}x^m+B_{mn}(x)\partial x^m
\bar{\partial}x^m+{\cal R}\phi(x),
\label{5dtranchnullag}
\end{equation}
where $m,n,...=1,...,4$, and the fields are given in terms of a harmonic 
function $f(x)$ ($\partial^2f=0$) by $\phi={1\over 2}\ln f$, $G_{mn}=
f\delta_{mn}$ and $H_{mnk}=-2\sqrt{G}G^{pl}\epsilon_{mnkp}\partial_l\phi=
-\epsilon_{mnkl}\partial_lf$.  By solving the conformal invariance 
conditions (\ref{confcondofchnul}) with above Ans\"atze, one obtains
\begin{eqnarray}
f&=&1+{P\over{r^2}},\ \ \  F^{-1}=1+{{Q_2}\over{r^2}},\ \ \ 
K=1+{{Q_1}\over{r^2}},
\cr 
e^{2\Phi}&=&Ff={{r^2+P}\over{r^2+Q_2}},\ \ 
{\cal A}_{\varphi_1}={\gamma\over{r^2}}\sin^2\theta,\ \ 
{\cal A}_{\varphi_2}={\gamma\over{r^2}}\cos^2\theta,
\label{5dbgndfucchnul}
\end{eqnarray}
where $r\equiv\sqrt{x^mx^m}$ and $\gamma$ is related to 
the angular momenta as 
$J_1=J_2={\pi\over{4G_N}}\gamma$.  Note, the 
conformal invariance condition $\nabla_m(e^{-2\phi}{\cal F}^{mn}_+)=0$ 
is solved by imposing the flat-space anti-self-duality condition 
${\cal F}_{+mn}=0$ on the field strength of the potential ${\cal A}_m$, 
and as a result the 2 angular momenta are the same.  
By superposing harmonic functions, one obtains the multi-center 
generalization of the above.  

The dimensional reduction along $u$ leads to 
$D=5$, rotating, BPS black hole with charge configuration 
$(Q_1=Q^{(1)}_1,Q_2=Q^{(2)}_2,P)$, where $P$ is a magnetic charge 
of the NS-NS 3-form field strength (or an electric charge of its 
Hodge-dual).  The Einstein-frame metric is
\begin{eqnarray}
ds^2_E&=&-\lambda^2(dt+{\cal A}_mdx^m)^2+\lambda^{-1}dx^mdx_m
\cr
&=&-\lambda^2[dt+{\gamma\over{r^2}}(\sin^2\theta d\varphi_1+
\cos^2\theta d\varphi_2)]^2
\cr
& &+\lambda^{-1}[dr^2+r^2(d\theta^2
+\sin^2\theta d\varphi^2_1+\cos^2\theta d\varphi^2_2)]
\cr
\lambda&=&(F^{-1}Kf)^{-1/3}={{r^2}\over{[(r^2+Q_1)(r^2+Q_2)(r^2+P)]^{1/3}}}.
\label{5dchnulmet}
\end{eqnarray}

\subsubsection{Level Matching Condition}\label{entchmat}

To calculate statistical entropy of black holes, 
one has to relate macroscopic quantities 
of black holes to microscopic quantities 
of perturbative string states through ``level matching condition'' 
\cite{DABghw}.  Strictly speaking, level matching process is possible 
for electric solutions, only, since 
perturbative string states do not carry magnetic charges and 
string momentum [winding] modes are matched onto ``electric'' charges of 
KK [2-form field] $U(1)$ fields.  Furthermore, 
magnetic solutions can be supported without source at the core (Cf. 
magnetic solutions are regular everywhere including the core),  
since they are topological in character.  However, it turns out \cite{CVEt53} 
that the dyonic solution found in \cite{CVEy672}, which is a `bound state' 
of fundamental string and solitonic 5-brane, still needs a source for 
its support and satisfies the same form of level matching condition as 
the fundamental string. 

The crucial point in the level matching of such dyonic solutions onto 
the perturbative string spectrum is that as in the purely magnetic 
case  fields are perfectly regular near the horizon (or the throat 
region), making it possible to describe the solutions at the throat 
region in terms of WZNW conformal model \cite{NOV37,WIT92,KNIz247,GEPw278}.  
Since the solution is regular and the dilaton is finite near the event 
horizon (implying that the classical $\alpha^{\prime}$  
and the string loop corrections are under control), such 
effective WZNW model near the horizon can be trusted.
For large magnetic charges (or large level), the theory effectively 
looks like a free $\sigma$-model for perturbative string theory with the 
string tension rescaled by magnetic charges.  Namely, for large 
magnetic charges, the dyonic solutions are matched onto perturbative string 
states with string tension rescaled by magnetic charges.  

To match background field solutions onto the macroscopic 
string source at the core, one considers the combined action of 
string $\sigma$-model and the effective field theory.  Among the 
equations of motion of the combined action, the relevant parts 
are the Einstein equations for target space metric, 
equations of motion for $X^{\mu}$ and the Virasoro conditions.   
Requiring that all the solutions are supported by sources, one obtains 
the level matching condition: 
\begin{equation}
\overline{E(u)}\equiv{1\over{2\pi R}}\int^{2\pi R}_0\,du\,E(u)=0,\ \ \ 
(E(u)\equiv[F(x)K(u,x)]_{x\to 0}=0). 
\label{levmatch}
\end{equation}
This condition is satisfied without modification even when solutions 
carry magnetic charges.

\subsubsection{Throat Region Conformal Model and Magnetic Renormalization 
of String Tension}\label{entchthr}

\paragraph{Four-Dimensional Dyonic Solutions}\label{entchthrfou}

The $\sigma$-model (\ref{4dbhchnullag}) of dyonic black hole 
(\ref{gensol}) with charge configuration 
$(P^{(1)}_1,Q^{(1)}_2,P^{(2)}_1,Q^{(2)}_2)\equiv(P_1,Q_1,P_2,Q_2)$ 
takes the following form near the horizon ($r\to 0$) 
\cite{CVEt53,TSE11,TSE477}:
\begin{eqnarray}
I&=&{1\over{\pi\alpha^{\prime}}}\int d^2\sigma L_{r\rightarrow 0}=
{1\over{\pi\alpha^{\prime}}}\int d^2\sigma\left(e^{-z}\partial u
\bar{\partial}v +Q_1Q^{-1}_2\partial u\bar{\partial}u\right)
\cr
& &+{{P_1P_2}\over{\pi\alpha^{\prime}}}\int d^2\sigma\left(\partial z
\bar{\partial}z +\partial\tilde{y_1}\bar{\partial}\tilde{y}_1+\partial
\varphi\bar{\partial}\varphi+\partial\theta\bar{\partial}\theta -2{\rm cos}
\theta\partial\tilde{y}_1\bar{\partial}\varphi\right).
\label{4dwznw}
\end{eqnarray}
This is the $SL(2,R)\times SU(2)$ WZNW model with the 
level $\kappa={4\over{\alpha^{\prime}}}P_1P_2$.  (Since the 
level has to be an integer, one has the quantization condition 
$4P_1P_2/\alpha^{\prime}\in {\bf Z}$.)  
Here, the coordinates are defined as $z\equiv {\rm ln}{{Q_2}\over
r}\to \infty$, $\tilde{u}\equiv(Q^{-1}_1Q_2P_1P_2)^{-1/2}u$, 
$\tilde{v}\equiv(Q_1Q^{-1}_2P_1P_2)^{-1/2}v$, and $\tilde{y}_1\equiv 
P^{-1}_1y_1+\varphi$.  

For large $P_{1,2}$ (or $\kappa$), the transverse 
$(\rho,\tilde{y}_1,\varphi,\theta)$ part of (\ref{4dwznw}) looks like a 
free theory of perturbative string with the string tension $T=1/(2\pi
\alpha^{\prime})$ renormalized by $P_{1,2}$:
\begin{equation}
{1\over{\alpha^{\prime}}} \to {1\over{\alpha^{\prime}_{\bot}}}=
{{P_1P_2}\over{\alpha^{\prime}R^2_1}}={{P_1P_2}\over{\alpha^{\prime\,2}}}, 
\label{4dtension}
\end{equation}
where $R_1=\sqrt{\alpha^{\prime}}$ is the radius of the internal coordinate 
associated with $P_{1,2}$. 

\paragraph{Five-Dimensional Dyonic Solution}\label{entchthrfiv}

The $\sigma$-model with the target space configuration given by 
the 3-charged, BPS, non-rotating black hole, i.e. (\ref{5dhetBPS}) 
with $J=0$, takes the following form in the limit $r\to 0$ 
\cite{TSE11,TSE477}:
\begin{eqnarray}
I&=&{1\over{\pi\alpha^{\prime}}}\int d^2\sigma L_{r\to 0}= 
{1\over{\pi\alpha^{\prime}}}\int d^2\sigma\left(e^{-z}\partial 
u\bar{\partial}v +Q_1 Q^{-1}_2\partial u\bar{\partial}u\right)
\cr
&+&{P\over{4\pi\alpha^{\prime}}}\int d^2\sigma\left(\partial z
\bar{\partial}z
+\partial\varphi_2\bar{\partial}\varphi_2+\partial\varphi_1
\bar{\partial}\varphi_1
+\partial\theta\bar{\partial}\theta -2{\rm cos}\theta\partial\varphi_2 
\bar{\partial}\varphi_1\right), 
\label{5dwznw}
\end{eqnarray}
where $z\equiv{\rm ln}\,{{Q_2}\over {r^2}}\to\infty$. 
This is the $SL(2,R)\times SU(2)$ WZNW model
\footnote{The chiral null model corresponding to rotating 
black hole (\ref{5dchnulmet}) also approaches the $SL(2,{\bf R})
\times SU(2)$ WZW model in the throat limit.  The requirements 
that the level $\kappa=P/\alpha^{\prime}$ is an integer and 
$\varphi^{\prime\prime}_2=\varphi^{\prime}_2+2\gamma P^{-1}Q^{-1}_2u$ 
($\varphi^{\prime}_2\equiv\varphi_1+\varphi_2$) has the period $4\pi$ 
lead to the quantization of $P$ and $J$: $P=\alpha^{\prime}\kappa$ and 
$J=\kappa wl$ with $k,l\in{\bf Z}$ ($w=$ string winding number).  
The regularity of the underlying conformal model requires 
that $J^2<\kappa mw$, i.e. the `Regge bound'.} 
with the level $\kappa\equiv {1\over{\alpha^{\prime}}}P$.
In the limit of large $P$ or large $\kappa$, 
the transverse part ($\rho,\varphi_1,\varphi_2,\theta$) of  
(\ref{5dwznw}) reduces to free perturbative string theory 
with the renormalized string tension:
\begin{equation}
{1\over{\alpha^{\prime}}}\to {1\over{\alpha^{\prime}_{\bot}}}= 
{P\over{4\alpha^{\prime\,2}}}. 
\label{5dtension}
\end{equation}

\subsubsection{Marginal Deformation}\label{entchmar}

The degeneracy of micro-states responsible for statistical entropy is 
traced to the degrees of freedom associated with oscillations or 
marginal deformations around the classical solutions.  
The marginal deformations lead to a family of all the possible 
solutions (obeying conformal and BPS conditions)
with the same values of electric/magnetic charges but different short 
distance structures that depend on a choice of oscillation profile function.  
Thus, one has to consider the region near the horizon (at $r=0$) 
to determine the microscopic degrees of freedom, 
since in this region degeneracy of solutions is lifted. 

General chiral null model action which represents deformation from 
the classical BPS solutions in section \ref{entchmodel} 
and preserves the BPS and conformal invariance properties is given by 
(\ref{chiralnullag}) with $K$ and ${\cal A}_i=({\cal A}_m,{\cal A}_a)$ 
having an additional dependence on $u$, where $u\equiv z-t$ with the 
longitudinal direction $z$ satisfying the periodicity condition 
$z\equiv z+2\pi R$.  Here, ${\cal A}_m\sim q_m(u)/r^2$ and ${\cal A}_a=
q_a(u)/r^2$ ($r^2\equiv x_mx_m$) are respectively `deformations' in 
the non-compact $x^m$ and the compact $y^a$ directions.  On the other hand, 
the perturbation $K(u,x)=h_m(u)x^m+k(u)/r^2$ does not contribute to the 
degeneracy, since $h_m(u)x^m$ drops out and $k(u)$ has zero mean value.

The perturbations $K$, ${\cal A}_a$ and ${\cal A}_m$ represent various 
`left-moving' waves propagating along the string and are invisible far 
away from the core.  
The mean values $\overline{q^2_m(u)}$ and $\overline{q^2_a(u)}$ are 
related to the oscillation numbers of the macroscopic string at the core as
\begin{equation}
N^{(m)}_L={{\pi^2}\over{16G^2_N}}\overline{q^2_m(u)}, \ \ \ \ \ 
N^{(a)}_L={{\pi^2\alpha^{\prime}}\over{16G^2_N}}\overline{q^2_a(u)},
\label{margoscil}
\end{equation}
thereby contributing to the microscopic degrees of freedom. 

These marginal deformations do not contribute to the microscopic degeneracy 
of black holes with the same order of magnitude \cite{TSE477}.  
This can be inferred from the fact that the classical BPS black holes are 
solutions of both heterotic and type-II string.  
Namely, although the thermal entropy is the same whether one 
embeds the solutions within heterotic or type-II string, one faces the 
discrepancy in factor of 2 in the statistical entropies within the two 
theories, if one takes the degeneracy contributions of all the oscillators 
to be of the same order of magnitude.  In fact, as can be 
seen from the conformal invariance condition $\nabla_i(e^{-2\phi}
{\cal F}^{ij}_+)=0$, the perturbations ${\cal A}_a$ in the compact 
directions $y_a$ are decoupled from the non-trivial non-compact parts of 
the solution.  
On the other hand, the perturbations ${\cal A}_m$ in the non-compact 
directions  $x_m$ are non-trivially coupled to the magnetic harmonic 
functions, with the net effect being the rescaling of $q^2_m(u)$ terms 
by the magnetic charges. 
(This is related to the scaling of the string tension in 
the transverse directions by the magnetic charge(s).) 
So, the marginal deformation contributions 
from the compact directions, which are different for the two theories, 
are suppressed relative to those of non-compact directions 
by the factor of the inverse of magnetic charge(s), thereby 
negligible for a large magnetic charge(s) or the large level $\kappa$. 
Only the marginal deformations from the non-compact directions and the 
compact direction associated with non-zero magnetic charge(s), which are 
common for both theories, have the leading contribution 
to the degeneracy.  Only these 4 string coordinates get their tension 
effectively rescaled by the magnetic charge(s).  
Furthermore, the marginal deformations on the 
original black hole solutions have to be only left-moving (i.e. depend 
only on $u$, not on $v$) so that marginally deformed $\sigma$-models 
are conformaly invariant.  This is related to the `chiral' condition on 
the $\sigma$-model; only left-moving deformations lead to supersymmetric 
action.  Thus, for large magnetic charge(s), the statistical 
entropy calculations within heterotic and type-II strings agree.  

As pointed out, the marginal deformations ${\cal A}_a(u,x)=q_a(u)/r^2$ 
in the compact directions $y_a$ contribute to the microscopic 
degeneracy to sub-leading order (suppressed by the inverse of 
magnetic charge(s)), which can be neglected for large value of magnetic 
charge(s).  But the zero modes $\bar{q}_a$ of the Fourier expansions 
of $q_a(u)=\bar{q}_a+\tilde{q}_a(u)$ ($\tilde{q}_a(u)$ denoting the 
oscillating parts) produce additional left-handed electric charges 
\cite{CVEt53} of $D=6$ strings.  Namely, the internal marginal 
deformation ${\cal A}_a(u)={{q_a(u)}\over{r^2}}$ on the $\sigma$-model 
associated with $D=4$ 4-charged BPS black hole (\ref{gensol}) leads to 
5-charged BPS black hole solution (\ref{4dbckgrdsolchnul}) with the 
zero mode $\bar{q}_a$ corresponding to an additional charge parameter $q$.  

The mean oscillation values $\overline{\tilde{q}^2_m(u)}$ of the marginal 
deformations ${\cal A}_m(u,x)\sim q_{m}(u)/r^2$ ($q_m(u)=\bar{q}_m+
\tilde{q}_m(u)$ with $\bar{q}_m$ and $\tilde{q}_m(u)$ respectively denoting 
the zero modes and oscillating parts) in the non-compact directions $x_m$ 
contribute to $N_L$ to the leading order.
Meanwhile, the zero modes $\bar{q}_m$ have an interpretation 
as angular momenta \cite{TSE381} of black holes.  
Namely, the rotational marginal deformation 
corresponding to $SU(2)$ Cartan current deformation  
${\cal A}_m(u,x)dx^m={{\gamma(u)}\over{r^2}}({\rm sin}^2\theta\,d\varphi_1 
+{\rm cos}^2\theta\,d\varphi_2)$ on the $\sigma$-model action of the $D=5$ 
3-charged non-rotating solution leads to 
the rotating solution (\ref{5dchnulmet}) \cite{TSE381,TSE11}. 

To calculate the statistical entropy associated with degeneracy of 
solutions (i.e. all the possible marginal deformations of the 
original classical solution), one has to determine $N_L$ 
of the macroscopic string at the core.  
For this purpose, we write the marginally deformed $\sigma$-model action 
(\ref{chiralnullag}) in the throat region in the form:
\begin{eqnarray}
I^{\prime}&=&I+{1\over{\pi\alpha^{\prime}}}\int\,d^2\sigma
({\rm marginal\ deformation\ terms})_{r\to 0}
\cr
&=&{1\over{\pi\alpha^{\prime}}}\int\,d^2\sigma\,[e^{-z}\partial u
\bar{\partial}v+E(u)\partial u\bar{\partial}u]+{\rm other\ terms}, 
\label{magidegen}
\end{eqnarray}
where $I$ stands for the throat limit WZNW models (\ref{4dwznw}) and 
(\ref{5dwznw}) of the (undeformed) classical solutions.  

First, for the $D=4$ dyon, the marginal deformation (\ref{chiralnullag}) 
gives rise to the following expression for $E(u)$: 
\begin{eqnarray}
E(u)&=&Q_1Q^{-1}_2-(P_1P_2)^{-1}Q^{-2}_2q^2_n(u)-Q^{-2}_2q^2_a(u)
\cr
&=&(P_1P_2)^{-1}Q^{-2}_2[Q_1Q_2P_1P_2-q^2_n(u)-P_1P_2q^2_a(u)].
\label{4dmagi}
\end{eqnarray}
Note, since the string tension $\alpha^{\prime}_{\perp}$ 
(\ref{4dtension}) of the transverse parts is rescaled by the magnetic 
charges, the coefficient in front of the term $\overline{q^2_n(u)}$ is 
rescaled by $(P_1P_2)^{-1}$.      
Applying the level matching condition $\overline{E(u)}=0$ (\ref{levmatch}), 
one finds that $\overline{q^2_n(u)}=P_1P_2(Q_1Q_2-\bar{q}^2_a)$, where 
$\bar{q}_a$ denote the zero modes of the oscillations $q_a(u)$ in the 
compact directions.  
Thus, the statistical entropy is \cite{CVEt53}
\begin{equation}
S_{stat}\approx 2\pi\sqrt{N_L}={{\pi^2}\over{2G_N}}\sqrt{P_1P_2(Q_1Q_2-
\bar{q}^2_a)}, 
\label{4dentchnul}
\end{equation}
in agreement with the thermal entropy. 

Second, we consider $D=5$ solutions.  For the non-rotating 
solutions, the marginal deformation (\ref{chiralnullag}) leads to
\begin{equation}
E(u)=Q_1Q^{-1}_2+Q^{-1}_2k(u)-P^{-1}Q^{-2}_2q^2_m(u)+{\cal O}(P^{-2}). 
\label{5dstatmarg}
\end{equation}
Applying the level matching condition $\overline{E(u)}=0$, one finds 
that $\overline{q^2_m(u)}=Q_1Q_2P$ for large $P$, which reproduces the 
thermal entropy \cite{TSE11,TSE477}:
\begin{equation}
S_{stat}\approx 2\pi\sqrt{N_L}={{\pi^2}\over{2G_N}}\sqrt{Q_1Q_2P}.
\label{5dstatent}
\end{equation}
For the rotating solution, one introduces the marginal 
deformation ${\cal A}_m$ in a non-compact direction.  Then, one has 
\begin{equation}
E(u)=Q_1Q^{-1}_2-P^{-1}Q^{-2}_2\gamma^2(u)=
P^{-1}Q^{-2}_2[Q_1Q_2P-\gamma^2(u)],
\label{5drotmarg}
\end{equation}
where $\overline{\gamma^2(u)}=\bar{\gamma}^2+\overline{\tilde{\gamma}^2(u)}$ 
with $\bar{\gamma}$ and $\tilde{\gamma}(u)$ respectively denoting 
zero and oscillating modes of $\gamma(u)$.  From the level matching 
condition $\overline{E(u)}=0$, one has $\overline{\tilde{\gamma}(u)}=
Q_1Q_2P-\bar{\gamma}^2$, reproducing entropy of the rotating 
black hole \cite{TSE11,TSE381} in the limit of $P_1\approx P_2$ and 
$Q_2$ large:
\begin{equation}
S_{stat}\approx 2\pi\sqrt{N_L}={{\pi^2}\over{2G_N}}\sqrt{Q_1Q_2P-
\bar{\gamma}^2}. 
\label{5drotent}
\end{equation}

\section{$D$-Branes and Entropy of Black Holes}\label{dbr}
 
Past year or so has been an active period for investigation on  
microscopic origin of black hole entropy.   The construction of 
general class of BPS black hole solution in heterotic string 
on $T^6$ \cite{CVEy672} motivated renewed interest 
\cite{LARw375} in the study of black hole entropy within perturbative 
string theory.  The explicit calculation of statistical 
entropy of BPS solution in \cite{CVEy672} by the method of WZNW model 
in the throat region of black hole reproduced the Bekenstein-Hawking 
entropy.  Realization \cite{POL75} that $D$-branes in open string 
theory can carry R-R charges motivated the explicit 
$D$-brane calculation of statistical entropy of non-rotating BPS solution 
in $D=5$ with 3 charges \cite{STRv}.  
This is generalized to rotating black hole \cite{BREmpv}
in $D=5$, near extreme black hole \cite{CALm472,HORs77} 
in $D=5$ and near extreme rotating black hole \cite{BRElmpsv} 
in $D=5$.   Meanwhile, the WZNW model approach 
was generalized to the case of non-rotating BPS $D=5$ black hole 
\cite{TSE11} and rotating BPS $D=5$ black hole \cite{TSE477}.  
$D$-brane approach was soon extended to $D=4$ cases: non-rotating BPS 
case in \cite{MALs77,JOHkm378}, near extreme case in \cite{HORlm77} 
and extreme rotating case in \cite{HORlm77}.  Later, it is shown 
\cite{HORp146} that the microscopic counting argument in string theory 
can be extended even to non-extreme black holes as well, 
provided the entropy is evaluated at the proper transition point of 
black hole and $D$-brane (or perturbative string) descriptions.    

In this chapter, we review the recent works on $D$-brane interpretation 
of black hole entropy.  With  realization \cite{POL75} that R-R 
charges, which were previously known to be decoupled from string states, 
can couple to $D$-branes \cite{DAIlp}, it became possible to do 
conformal field theory of extended objects ($p$-branes) within string 
theories and to perform counting 
\cite{SEN54,SEN53,VAF088,VAF463,BERsv398,BERsv463}
of string states that carries R-R charges as well as NS-NS charges. 

To apply $D$-brane techniques to the calculation of microscopic 
degeneracy of black holes, one has to map non-perturbative NS-NS charges 
of the generating black hole solutions of heterotic string on tori 
to R-R charges by applying subsets of $U$-duality transformations.  
In $D$-brane picture of black holes \cite{CALm472}, the microscopic 
degrees of freedom are carried by oscillating open strings which are 
attached to $D$-branes.  Whereas the effect of magnetic charges 
in the chiral null model and Rindler geometry approaches is to rescale the 
string tension, the effect of R-R charges on open strings in the 
$D$-brane description is to alter the central charge (i.e. the bosonic and 
fermionic degrees of freedom) of open strings from the free open string 
theory value.  In the $D$-brane picture of \cite{CALm472},  
the number of degrees of freedom of open strings is increased relative 
to the free open string value because of an additional factor 
(proportional to the product of $D$-brane charges) related to all the 
possible ways of attaching the ends of open strings to different 
$D$-branes.  So, the net calculation results of statistical entropy in 
both descriptions are the same.  

This chapter is organized as follows.  In section \ref{dbrint}, we 
summarize the basic facts on $D$-branes necessary in understanding 
$D$-brane description of black holes.  In section \ref{dbrbh}, we discuss 
the $D$-brane embeddings of black holes.  The $D$-brane counting 
arguments for the statistical entropy of black holes are discussed  
in section \ref{dbrent}.

\subsection{Introduction to $D$-Branes}\label{dbrint}

We discuss basic facts on $D$-branes necessary in understanding 
$D$-brane description of black holes.  Comprehensive 
account of the subject is found, for example, in 
\cite{DAIlp,LEI,POL75,POLcj,POL050,JOH52}, which we follow 
closely.  The basic knowledge on string theories is referred to 
\cite{SCH89,GREsw}.  

Each end of open strings can satisfy two types of 
boundary conditions.  
Namely, from the boundary term ${1\over{2\pi\alpha^{\prime}}}
\int_{\partial{\cal M}}d\sigma\delta X^{\mu}\partial_n X_{\mu}$, where 
$\partial{\cal M}$ is the boundary of the worldsheet ${\cal M}$ swept 
by an open string and $\delta X^{\mu}$ [$\partial_nX_{\mu}$] is the 
variation [the derivative] of bosonic coordinates $X^{\mu}$ parallel 
to [normal to] $\partial{\cal M}$, in the variation 
(with respect to $X^{\mu}$) of the worldsheet action, one sees that 
the ends of the string either can have zero normal derivatives
\begin{equation}
\partial_nX^{\mu}=0,
\label{neumann} 
\end{equation}
called {\it Neumann boundary condition}, or have fixed position 
in target spacetime
\begin{equation}
X^{\mu}={\rm constant},
\label{dirichlet}
\end{equation}
called {\it Dirichlet boundary condition}.  

In order for the $T$-duality to be an exact symmetry of the string theory, 
open string has to satisfy both the Neumann and Dirichlet boundary 
conditions \cite{DAIlp}.  Under the $T$-duality of open string theory with 
the coordinate $X^i=X^i_R(z)+X^i_L(\bar{z})$ compactified on $S^1$ of 
radius $R_i$, $R_i\to R^{\prime}_i=1/R_i$ and $X^i\to Y^i=
X^i_R(z)-X^i_L(\bar{z})$.  
So, the Neumann and the Dirichlet boundary conditions get interchanged 
under $T$-duality:
\begin{equation}
\partial_nX^i=\left({{\partial z}\over{\partial\tau}}\right)
\partial X^i-\left({{\partial \bar{z}}\over{\partial\tau}}\right)
\bar{\partial} X^i=
\left({{\partial z}\over{\partial\tau}}\right)
\partial Y^i+\left({{\partial \bar{z}}\over{\partial\tau}}\right)
\bar{\partial} Y^i=\partial_{\tau}Y^i,
\label{neudirex}
\end{equation}
where $\tau$ is the worldsheet time coordinate, which is tangent to 
$\partial{\cal M}$.  
Starting from the $D=10$ open string with the Neumann boundary conditions 
and with the coordinates $X^i$ ($i=p+1,...,9$) compactified on circles of 
radii $R^i$, one obtains open string theory with the ends of the dual 
coordinates $Y^i$ confined to the $p$-dimensional hyperplane in the 
$R_i\to 0$ limit (or the decompactification limit, i.e. $1/R_i\to\infty$, 
of the dual theory).  Such $p$-dimensional hyperplane is called 
$D$-{\it brane} \cite{POL75}.  A further $T$-duality in the direction 
tangent [orthogonal] to a $D\,p$-brane results in a $D\,(p-1)$-brane 
[$D\,(p+1)$-brane].  

$D$-brane is a dynamical surface \cite{POL75} with the states of 
open strings (attached to the $D$-brane) interpreted as excitations of 
fluctuating $D$-brane.   The massless
\footnote{The tachyon is removed by the GSO projection of supersymmetric 
theory.} 
bosonic excitation mode in the open string spectrum is 
the photon with the vertex operator $V_A=A_M\partial_tX^M$, 
where $\partial_t$ is the derivative tangent to $\partial{\cal M}$.  
So, the bosonic low energy effective action is that of $N=1$, 
$D=10$ Yang-Mills theory \cite{WIT335}:
\begin{equation}
{1\over {2g}}\int d^{10}x\,{\rm Tr}[F_{\mu\nu}F^{\mu\nu}].
\label{ymlag}
\end{equation}
In the $T$-dual theory, the vertex operator $V_A$ of the photon 
is decomposed into $V_A=\sum^p_{\mu=0}A_{\mu}(X^{\mu})\partial_tX^{\mu}$ 
and $V_{\varphi}=\sum_{i>p}\varphi_i(X^{\mu})\partial_{\sigma}X^i$, 
corresponding respectively to $U(1)$ gauge boson and scalars on the 
$p$-brane worldvolume.  The scalars $\varphi_i$ are regarded as 
the collective coordinates for transverse motions of the $p$-brane.  
The bosonic low energy effective action of the $T$-dual theory is, 
therefore, obtained by compactifying (\ref{ymlag}) down to $p+1$ 
dimensions \cite{WIT335}.  In this action, the worldvolume scalars 
$\varphi_i$ have potential term $V=\sum_{i,j}\,{\rm 
Tr}[\varphi_i,\varphi_j]^2$.  

Note, open strings can have non-dynamic degrees of freedom 
called {\it Chan-Paton factors} $(i,j)$ at both ends of strings 
\cite{CHAp10,GREsw,POLcj,POL050,WIT335}.  
The indices $(i,j)$, which label the state at each end of the string, run 
over the representation of the symmetry group $G$.  When the state 
$|\Lambda,ij\rangle$ describes a massless vector, $(i,j)$ 
run over the adjoint representation of $G$.  Each vertex 
operator of an open string state carries antihermitian 
matrices $\lambda^a_{ij}$ ($a=1,...,{\rm dim}G$) representing the 
algebra of $G$ and $\lambda^a_{ij}$ describe the Chan-Paton degrees of 
freedom of the open string states.  The global symmetry $G$ of the 
worldsheet amplitude manifests as a gauge symmetry in target space. 

For the oriented open strings, $G$ is $U(N)$ and each end of the open 
string is respectively in complex and complex conjugate representations 
of $U(N)$.  When open string states are invariant under the 
{\it worldsheet parity} transformation $\Omega$ ($\sigma\to\pi-\sigma$ 
or $z\to -\bar{z}$), i.e. the exchange of two ends plus reversal of the 
orientation of an open string, the open string is called {\it unoriented}.  
For this case, the representations $R$ and $\bar{R}$ on both ends of the 
string are equivalent.  
For unoriented open strings, the transformation property of the Chan-Paton 
matrix $\lambda_{ij}$ under the worldsheet parity $\Omega$ 
determines $G$ \cite{SCH89,MARs119}.  Under the worldsheet parity 
symmetry, the open string state $\lambda_{ij}|\Lambda,ij\rangle$ 
transforms to
\begin{equation}
\Omega\lambda_{ij}|\Lambda,ij\rangle = 
\lambda^{\prime}_{ij}|\Lambda,ij\rangle, \ \ \ 
\lambda^{\prime}=M\lambda^TM^{-1}.
\label{wshtprt}
\end{equation}
If $M$ is symmetric, i.e. $M=M^T=I_N$, then 
the photon $\lambda_{ij}\alpha^{\mu}_{-1}|k\rangle$ 
survives the projection under the gauged worldsheet 
parity and the Chan-Paton factor is antisymmetric 
($\lambda^T=-\lambda$), giving rise to $SO(N)$ gauge group. 
If $M$ is antisymmetric, i.e. $M=-M^T=
i\left(\matrix{0&I_{N/2}\cr -I_{N/2}&0}\right)$, then the 
gauge group is $USp(N)$, i.e. $\lambda=-M\lambda^TM$.  

When the Chan-Paton factors are present at the ends of open 
string, a Wilson line, say, for the $U(N)$ oriented open string theory 
with the coordinate $X^9$ compactified on $S^1$ of radius $R$ 
given by
\begin{equation}
A_{9}={\rm diag}(\theta_1,...,\theta_N)/(2\pi R)
=-i\Lambda^{-1}\partial_9\Lambda,
\label{wilson}
\end{equation}
where $\Lambda={\rm diag}(e^{iX^9\theta_1/(2\pi R)},...,
e^{iX^9\theta_N/(2\pi R)})$, receives non-trivial 
phase factor ${\rm diag}(e^{-i\theta_1},...,e^{-i\theta_N})$ 
under the transformation $X^9\to X^9+2\pi R$.  As a result,  
the momentum number of an open string along $S^1$ can have a 
fractional value.  So, in the $T$-dual theory the winding number takes 
on fractional values \cite{POL50}, meaning that two ends of open 
strings can live on $N$ different hyperplanes ($D$-branes) located at 
$Y^9=\theta_kR^{\prime}=2\pi\alpha^{\prime}A_{9,kk}$ ($k=1,...,N$):
\begin{equation}
Y^9(\pi)-Y^9(0)=\int^{\pi}_{0}d\sigma\partial_{\sigma}Y^9=
(2\pi+\theta_j-\theta_i)R^{\prime},
\label{chpaends}
\end{equation}
where $R^{\prime}=\alpha^{\prime}/R$ is the radius of the circle 
in the dual theory.  Note, without the Chan-Paton factors taken 
into account, both ends of open strings of the dual theory 
are confined to the same hyperplane up to the integer multiple of 
periodicity $2\pi R^{\prime}$ of the dual coordinate $Y^9$.  

The similar argument can be made for unoriented open string theories.  
With $SO(N)$ Chan-Paton symmetry, the Wilson line can be brought to the 
form:
\begin{equation}
{\rm diag}(\theta_1,-\theta_1,\cdots,\theta_{N/2},-\theta_{N/2}). 
\label{sonwilson}
\end{equation}
Note, in the dual coordinate $Y^m$, worldsheet 
parity reversal symmetry ($z\to-\bar{z}$) of the original theory is 
translated into the product of worldsheet and spacetime parity 
operations.  Since unoriented strings are invariant under the 
worldsheet parity, the $T$-dual spacetime is a torus modded by 
spacetime parity symmetry ${\bf Z}_2$.  The fixed planes 
$Y^m=0,\pi R^{\prime}$ under spacetime parity symmetry are called 
{\it orientifolds} \cite{DAIlp}.  Away from the orientifold plane, the 
physics is that of oriented open strings, with a string away from the 
orientifold fixed plane being related to the string at the image point.  
Open strings can be attached to the orientifolds, 
but the orientifolds do not correspond to the dynamic surface 
since the projection $\Omega=+1$ \cite{HOR327,HOR231,PRAs216}
removes the open string states corresponding to collective motion of 
$D$-branes away from the orientifold plane.   In this $T$-dual theory, 
there are ${N\over 2}$ $D$-branes on the segment $0\leq Y^m <\pi 
R^{\prime}$ and the remaining ${N\over 2}$ are at the image points under 
${\bf Z}_2$.  Open strings can stretch between pairs of ${\bf Z}_2$ 
reflection planes, as well as between different planes on one side.  

With a single coordinate $X^9$ compactified on $S^1$ of radius $R$, 
the mass spectrum of the dual open string is
\begin{equation}
M^2=\left({{[2\pi n+(\theta_i-\theta_j)]R^{\prime}}\over
{2\pi\alpha^{\prime}}}\right)+{1\over{\alpha^{\prime}}}(N-1).
\label{openmss}
\end{equation}
Thus, the massless states arise in the ground state 
($N=1$) with no winding mode ($n=0$) and both ends of the string 
attached to the same hyperplane ($\theta_i=\theta_j$):
\begin{eqnarray}
\alpha^{\mu}_{-1}|k,ii\rangle,\ \ \ \  V&=&A_{\mu}\partial_tX^{\mu},
\cr
\alpha^9_{-1}|k,ii\rangle,\ \ \ \  V&=&\varphi\partial_tX^9=
\varphi\partial_nY^9,
\label{msslessopen}
\end{eqnarray}
respectively corresponding to $D$-brane worldvolume photon and 
scalar.    
For the case of oriented open string theories, when all the $N$ 
hyperplanes (located at $\theta_kR^{\prime}$) do not coincide 
(i.e. $\theta_i\neq\theta_j$, $\forall i,j$), $U(N)$ 
is broken down to $U(1)^N$, corresponding to $N$ massless 
$U(1)$ gauge fields at each hyperplane located at $\theta_kR^{\prime}$ 
($k=1,...,N$).   When $m$ hyperplanes ($m\leq N$) coincide, 
say $\theta_1=\cdots=\theta_m$, the additional massless $U(1)$ gauge 
fields (associated with open strings originally stretched 
between these $m$ hyperplanes) contribute to the restoration of the 
symmetry to $U(m)\times U(1)^{N-m}$ \cite{POL50,WIT335}. 
Furthermore, $m$ massless scalars (interpreted as positions 
\cite{DAIlp,WIT335} of $m$ distinct hyperplanes) are promoted to 
$m\times m$ matrix of $m^2$ massless scalars when these $m$ hyperplanes 
coincide \cite{WIT335}.  
For unoriented open strings with $SO(N)$ symmetry
\footnote{The same argument can be applied for the case of 
$USp(N)$ symmetry.}, 
the generic gauge group with all the ${N\over 2}$ $D$-branes 
distinct is $U(1)^{N/2}$.  When $m$ $D$-branes coincide, 
the symmetry is enhanced to $U(m)\times U(1)^{N/2-m}$ as in the 
oriented case.  But when $m$ $D$-branes are located at 
an orientifold plane, the symmetry is enhanced to $SO(2m)\times 
U(1)^{N/2-m}$, due to additional massless $U(1)$ gauge bosons 
arising from open strings that originally stretched between 
pairs of ${\bf Z}_2$ image branes.   In the language of effective 
field theory, this symmetry enhancement or reversely symmetry 
breaking to Abelian group is interpreted as Higgs mechanism 
with scalars associated with location and separation of $D$-branes 
interpreted as Higgs fields.  

The fermionic spectrum of open strings is divided into subsectors 
according to the boundary conditions that fermions $\psi^{\mu}$ 
($0\leq\sigma\leq \pi$, $-\infty <\tau <\infty$) satisfy 
at one end of string.  There are two types of boundary 
conditions on the fermion:
\begin{eqnarray}
{\rm R}&:&\ \ \ \ \  \psi^{\mu}(0,\tau)=\tilde{\psi}^{\mu}(0,\tau)\ \ \ \ 
\psi^{\mu}(\pi,\tau)=\tilde{\psi}^{\mu}(\pi,\tau)
\cr
{\rm NS}&:&\ \ \ \ \  \psi^{\mu}(0,\tau)=-\tilde{\psi}^{\mu}(0,\tau)\ \ \ 
\ 
\psi^{\mu}(\pi,\tau)=\tilde{\psi}^{\mu}(\pi,\tau).
\label{rnsbdy}
\end{eqnarray}
Defining $\psi^{\mu}(2\pi-\sigma,\tau)\equiv\tilde{\psi}^{\mu}
(\sigma,\tau)$, one sees that the Ramond (R) [Neveu-Schwarz (NS)] 
boundary condition becomes the periodic [anti-periodic] boundary 
condition on the redefined fermion $\psi(\sigma,\tau)$ 
($0\leq\sigma\leq 2\pi$, $-\infty <\tau <\infty$),  
leading to integer [half-integer] modded Fourier series decomposition.  

In the NS sector, the ground state consists of 
8 transverse polarizations $\psi^{\mu}_{-1/2}|k\rangle$ of
massless open string photon $A_{\mu}$.  In the R sector, the ground state 
is degenerate, transforming as $\bf 32$ spinor representation 
of $SO(8)$.  The Virasoro conditions pick out 2 irreducible 
representations ${\bf 8}_s$ and ${\bf 8}_c$.  Here, the subscript 
$s$ [$c$] means eigenstates $|s_0,s_1,s_2,s_3,s_4\rangle$ ($s_0,s_i=
\pm{1\over 2}$) with even [odd] number of eigenvalues $-{1\over 2}$ of 
$S_0=iS^{01}$ and $S_i=S^{2i,2i+1}$, where $S^{\mu\nu}=-{1\over 2}\sum_r
\psi^{[\mu}_{-r}\psi^{\nu]}_r$ are the fermionic part of the 
$D=10$ Lorentz generators.  These 2 representations are physically 
equivalent for open strings.  The GSO projection picks out 
${\bf 8}_s$ and, therefore, the ground state of the open 
string theory is ${\bf 8}_v\oplus{\bf 8}_s$, forming a vector multiplet of 
$D=10$, $N=1$ theory.   Including the Chan-Paton 
factors, the gauge group $G$ of the $N=1$, $D=10$ super-Yang-Mills 
theory is $U(N)$ [$SO(N)$ or $USp(N)$] for an oriented [an unoriented] 
open string theory.  For an open string theory with $9-p$ coordinates 
compactified, the massless spectrum of $D\, p$-brane worldvolume theory 
of dualized open string is described by the $D=10$, $N=1$ supersymmetric 
gauge theory compactified to $D=p+1$.

We briefly discuss some aspects of type-II closed string relevant for 
understanding $D\, p$-branes of open string theory.  
For type-II string, i.e. the closed string theory with supersymmetry on 
both left- and right-moving modes, the two choices of the GSO projections 
in the R sector are not equivalent.  So, there are two types of 
type-II theories defined according to the possible inequivalent choices 
of the GSO projections on the left- and the right-moving modes.  
The massless sectors of these two type-II theories are
\begin{eqnarray}
{\rm Type\ IIA}&:&\ \ \ \ \ \ \ 
({\bf 8}_v\oplus{\bf 8}_s)\otimes({\bf 8}_v\oplus{\bf 8}_c)
\cr
{\rm Type\ IIB}&:&\ \ \ \ \ \ \ 
({\bf 8}_v\oplus{\bf 8}_s)\otimes({\bf 8}_v\oplus{\bf 8}_s).
\label{iiab}
\end{eqnarray}
The massless modes ${\bf 8}_v\otimes{\bf 8}_v$ in the NS-NS sector of 
the both theories are the same: dilaton, gravitino and the 2-form field.  
In the R-R sector, the massless modes ${\bf 8}_s\times{\bf 8}_c$ 
[${\bf 8}_s\times{\bf 8}_s$] of type-IIA [type-IIB] theory 
are 1- and 3-form potentials [0-, 2- and self-dual 4-form potentials] 
\cite{FRIms271}.  The massless modes in NS-R and 
R-NS sectors contain 2 spinors and 2 gravitinos of the same 
[opposite] chirality for the type-IIB [type-IIA] theory.  

When an oriented type-II theory with a coordinate 
compactified on $S^1$ is $T$-dualized \cite{DINhs322,DAIlp}, 
the chirality of the right-movers gets reversed.  
So, when odd [even] number of coordinates in type-IIA/B theory 
are $T$-dualized, one ends up with type-IIB/A [type-IIA/B] theory.  
The effect of ``odd'' $T$-duality, which exchanges type-IIA and 
type-IIB theories, on massless R-R fields is to add [remove] 
the indices (of $(p+1)$-form potential) corresponding to the $T$-dualized 
coordinates, if those indices are absent [present] in the $(p+1)$-form 
potential.  For example, the $T$-duality on $B_{\mu\nu}$ [$B_{\mu\rho}$] 
along $x^{\rho}$ ($\mu\neq\rho\neq\nu$) produces $B_{\mu\nu\rho}$ 
[$B_{\mu}$].  

A worldsheet parity symmetry $\Omega$ in a closed string, defined as 
$\sigma\to-\sigma$ or $z\to\bar{z}$, interchanges left- and right-moving 
oscillators.  The unoriented closed string is defined by projecting 
only even parity states, i.e. $\Omega|\psi\rangle=+|\psi\rangle$, 
as in the open string case.  
When type-II string is coupled to open superstring (type-I string), 
the orientation projection of type-I string picks up only one 
linear combination of 2 gravitinos in type-II theory, resulting in 
an $N=1$ theory.  The only possible consistent coupling of  
type-I and closed superstring theories is between (unoriented) 
$SO(32)$ type-I theory
\footnote{$T$-duality on the type-II theory leads to 
16 $D$-branes on a $T_{9-p}/{\bf Z}_2$ orbifold.  (The restriction 
to 16, i.e. $SO(32)$ gauge symmetry, comes from the conservation 
of R-R charges.)  However, in non-compact space, one can have a consistent 
theory with an arbitrary number of $D$-branes.} 
\cite{POLc296,CALlny308} and unoriented $N=1$ type-II theory. 
But in the $T$-dual theory, type-II theories without $D$-branes 
are invariant under $N=2$ supersymmetry, with orientation projection 
relating a gravitino state to the state of the image gravitino.  
The chiralities of these 2 gravitinos are the same [opposite] if 
even [odd] number of directions are $T$-dualized.  In the presence 
of $D$-branes, only one linear combination of supercharges in the 
$T$-dual type-II theory is conserved, resulting in a theory with 
$1/2$ of supersymmetry broken ($N=1$ theory), i.e. the BPS state 
\cite{POL75}.   For this case, the left- and right-moving supersymmetry 
parameters (of the $T$-dual type-II theory) are constrained by the 
relation \cite{GUBhkm472,POLcj,POL050}
\begin{equation}
\varepsilon_R=\Gamma^0\cdots\Gamma^p\varepsilon_L.
\label{bpsspin}
\end{equation} 

The conserved charges carried by $D$-branes are charges of the 
antisymmetric tensors in the R-R sector.  The worldvolume of a $D\,p$-brane 
naturally couples to a $(p+1)$-form potential in the R-R sector, 
with the relevant space-time and $D\, p$-brane actions given by
\begin{equation}
{1\over 2}\int G_{(p+2)}\star G_{(p+2)}+i\mu_p\int_{p-brane}C_{(p+1)},
\label{dbractn}
\end{equation}
where $\mu^2_p=2\pi(4\pi^2\alpha^{\prime})^{3-p}$ is the $D\,p$-brane 
charge \cite{POL75}.  
The worldvolume action is given by the following 
Dirac-Born-Infeld type action \cite{FRAt163,LEI} describing interaction 
of the world-brane $U(1)$ vector field and scalar fields with the 
background fields \cite{LEI}:
\begin{equation}
-T_p\int d^{p+1}\xi e^{-\varphi}{\rm det}^{1/2}(G_{ab}+B_{ab}
+2\pi\alpha^{\prime}F_{ab}),
\label{dbiact}
\end{equation}
where $T_p$ is the $D\,p$-brane tension \cite{GRElpt384,DEA388}, and 
$G_{ab}$ and $B_{ab}$ are the pull-back of the spacetime fields to the 
brane.  

In the amplitudes of parallel $D$-brane interactions, terms involving 
exchange of the closed string NS-NS states and the closed string R-R 
states cancel \cite{POL75}, a reminiscence of no-force condition 
of BPS states.  
Furthermore, the $D$-brane tension, which measures 
the coupling of the closed string states to $D$-branes, has the 
$g^{-1}_s$ behavior \cite{LEI}, a property of R-R $p$-branes 
\cite{HORs360,HULt438,TOW350}.  
The field strengths which couple to $D\,p$- and $D\,(6-p)$-branes are 
Hodge-dual to each other, and the corresponding conserved R-R charges 
are subject to the Dirac quantization condition \cite{TEI167,NEP31,POL75}
$\mu_{6-p}\mu_p=2\pi n$.

\subsection{$D$-Brane as Black Holes}\label{dbrbh}

In section \ref{dbrint}, we observed that $D\,p$-branes  
have all the right properties of the R-R $p$-branes in 
the effective field theories.   As solutions of the effective field 
theories, which are compactified from the $D=10$ string effective 
actions, black holes can be embedded in $D=10$ as {\it bound states} 
\cite{WIT335,LI460,DOU077}of  $D\,p$-branes.   Here, $p$ takes the even [odd] 
integer values for the type-IIA [type-IIB] theory.  
Such $p$-branes of the effective field theories correspond to 
the string background field configurations \cite{CALhs359,CALhs367}
(with the $p$-brane worldvolume action being the source of 
$(p+1)$-form charges) to the leading order in string length scale 
$l_s=\sqrt{\alpha^{\prime}}$ and describe the long range fields away 
from $D\,p$-branes.  As long as the spacetime curvature of 
the soliton solutions (at the event horizon in string frame) is small 
compared to the string scale $1/l_s^2$, the effective field theory 
solutions can be trusted, since the higher order corrections to the 
spacetime metric is negligible.  The effective field 
theory metric description of solitons in superstring theories 
is valid only for length scales larger than a string.
The $D$-brane picture of black holes, or more generally black $p$-branes, 
is as follows. 

The string-frame ADM mass $M$ of $p$-branes carrying 
NS-NS electric charge \cite{DAB357,HUL357}, NS-NS magnetic charge 
\cite{STR343} and R-R charge \cite{HULt438} behaves as $\sim 1$, 
$\sim 1/g^2_s$ and $\sim 1/g_s$ (in the unit where $l_s\sim 1$), 
respectively.  Since the gravitational constant
\footnote{The gravitational constant in $D=d$ is $G^d_N=G^{10}_N/V_{10-d}$, 
where $V_{10-d}$ is the volume of the $(10-d)$-dimensional internal space 
and $G^{10}_N=8\pi^6g^2_s\alpha^{\prime\,4}\sim g^2_sl^2_s$.} 
$G_N$ is proportional to $g^2_s$, the gravitational field strength  
($\propto G_NM$) of the NS-NS electric charged and 
the R-R charged $p$-branes vanishes as $g_s\to 0$.   Namely, 
in the limit $g_s\to 0$, strings live in the flat spacetime 
background.  
Note, in the limit $g_s\to 0$, the description of R-R charged 
configurations in terms of black $p$-branes is not valid, 
since the size of the $p$-brane horizon ($A\sim g^2_s$) is smaller than 
$D$-brane size; 
the black hole is surrounded by a halo which is large compared to its 
Schwarzschield radius.  In the limit $g_s\gg 1$, one can integrate 
out massive string states (with their masses increasing 
in Planck units, defined as $l^2_p=1$
\footnote{The Planck length is defined as $l_p=m^{-1}_p=
(\hbar G_N/c^3)^{1/2}$.},  
as $g_s$ increases) to obtain string effective field theories.  
(So, one can trust black hole solutions in the effective 
field theories in the strong string coupling limit.)   
In this limit, the massive string states form black holes 
and these degenerate massive states (whose mass is identified with 
black hole mass) are degenerate black hole
microscopic states, which are origin of statistical entropy.  
To summarize, the weak string coupling description of R-R charged 
configuration is the perturbative $D$-branes in flat background, and in the 
strong string coupling limit the horizon size ($\sim G_NM$) becomes 
larger than the string scale (with string states undergoing 
gravitational collapse inside the horizon), thereby, the black 
$p$-brane description emerges.   
The transition point of the two descriptions occurs at the point 
where the horizon size $r_0$ is of the order of the string length 
scale $l_s$ \cite{HORp146}.  This occurs when $g_sN^{1/4}\sim 1$ or 
$g_sQ\sim 1$, where $N$ is the string excitation level and $Q$ is 
an R-R charge.  This implies that $g_s$ is very 
small for large $N$ or $Q$.  (Note, however that the effective 
string coupling of these bound states is $g^{eff}_s\sim 
g_sN^{1/4}$ or $g_sQ$, which is of order 1.)  At this transition 
point, the mass of a string state $M^2_s\sim N/l^2_s$ becomes comparable 
to black hole mass: $M^2_{bh}\sim r^2_0/G^2_N$. 

The essence of the $D$-brane description of black hole entropy 
is that the number of degenerate BPS states is a topological 
invariant which is independent of (continuous) moduli fields, 
including $g_s$ \cite{SEIw426,SEIw431,SEI49}.  
Furthermore, mass of the BPS states is not renormalized \cite{OLIw78}. 
It is argued \cite{MAL125,DAS146} that even for near BPS $D$-brane 
states the $D$-brane counting results can be extrapolated invariantly to 
strong coupling limit.  It is also shown that $D$-brane approach 
reproduces entropy of non-supersymmetric extreme black holes 
\cite{HORlm77,DAB050} and even non-extreme case \cite{HORp146} as well.
So, the statistical entropy of $p$-branes can be 
calculated by counting the number of degenerate perturbative 
($g_s\to 0$) string states in $D\,p$-brane configuration.   

Each $D\,p$-brane carries one unit of R-R charge and the 
$g_s\to\infty$ limit of $Q_p$ $D\,p$-branes is black 
$p$-brane carrying R-R $(p+1)$-form charge $Q_p$.   
R-R $p$-branes have a $p$-volume tension behaving as $\sim g^{-1}_s$ 
\cite{TOW350}.  So, although these are non-perturbative, R-R 
$p$-branes have singularities, except for $p=3$.  The transverse 
[longitudinal] directions of R-R $p$-branes correspond to open string 
coordinates with Dirichlet [Neumann] boundary condition.   The same 
$T$-duality rules of $D$-branes hold for R-R $p$-branes:  
$T$-duality on the transverse [longitudinal] directions of  
$p$-branes produces $(p+1)$- [$(p-1)$-] branes.  

When longitudinal directions are compactified, single-charged $p$-branes 
become black holes having singular horizon with zero surface area and 
diverging dilaton at the horizon.  
This is due to the brane tension which makes the volume parallel 
[perpendicular] to the brane shrinks [expands] as one 
gets close to the brane \cite{CALm472}.  Black holes having 
regular horizon with non-zero area in the BPS limit are constructed 
from bound states of $p$-branes (with a momentum) to balance 
the tension to stabilize the volume internal to all the 
constituent $p$-branes.  Such regular black holes are obtained with 
the minimum of 4 [3] $p$-brane charges for the $D=4$ [$D=5$] black holes
\footnote{It is shown \cite{KLEt475} that the stringy BPS black holes 
with non-zero horizon area are not possible 
for $D\geq 6$.  This can be seen from the explicit solutions discussed 
in section \ref{n4bhhigh}.}.  

The basic constituent of black holes is the 
R-R $p$-brane in $D=10$ \cite{HORs360}:
\begin{eqnarray}
ds^2_{string}&=&g^{string}_{\mu\nu}dx^{\mu}dx^{\nu}=
f^{-1/2}_p(-dt^2+dx^2_1+\cdots+dx^2_p)
\cr
& &+f^{1/2}_p(dx^2_{p+1}+\cdots+dx^2_9), 
\cr
e^{-2\varphi}&=&f^{{p-3}\over 2}_p, \ \ \ \ \ 
A_{0\cdots p}=-{1\over 2}(f^{-1}_p-1),
\label{pbran}
\end{eqnarray}
where $f_p=1+Q_pc^{10}_p/r^{7-p}$ ($r\equiv(x_{p+1}+\cdots+x_9)^{1/2}$).  
Here, $c^{10}_p$ is related to the basic $(p+1)$-form potential charge 
and can be estimated by comparing the ADM mass of (\ref{pbran}) to 
the mass of $D$-brane state carrying one unit of the $(p+1)$-form 
potential charge.  The Killing spinors of this solution are constrained 
by \cite{HORs360}:
\begin{equation}
\varepsilon_L=\Gamma^1\cdots\Gamma^p\varepsilon_L, \ \ \ \ 
\varepsilon_R=-\Gamma^1\cdots\Gamma^p\varepsilon_R, 
\label{pbrspcst}
\end{equation}
where $\varepsilon_{L,R}$ denotes the left/right handed chiral spinor 
($\Gamma^{11}\varepsilon_{L,R}=\varepsilon_{L,R}$).   
One can construct solutions for bound states of  
$p$-branes by applying the intersection rules.  
(See section \ref{bprmltint} for details on intersection rules.)   
In particular, the dilaton is the product of individual factors 
associated with those of the constituent $p$-branes: 
$e^{-2\varphi}=f^{{p_{1}-3}\over 2}_{p_{1}}\cdots f^{{p_{k}-3}
\over 2}_{p_{k}}$.   

To add a momentum along an isometry direction $x^i$, one oscillates  
$p$-brane so that it carries traveling waves along the $x^i$-direction.  
(See section \ref{entfunsol} for the detailed discussion on the 
construction of such solutions.)  At a long distance region, the 
solutions approach the form where all the oscillation profile 
functions are (time or phase) averaged over.  So, the long-distance region 
$p$-brane solution carrying a momentum along the $x^i$-direction is 
obtained by just imposing $SO(1,1)$ boost among the coordinates 
$(x^0,x^i)$, with the net effect on the metric being the 
following substitution:
\begin{equation}
-dt^2+dx^2_i \to -dt^2+dx^2_i+k(dt-dx_i)^2,
\label{momisocrd}
\end{equation}
where $k=c^{10}_pN/r^{7-p}$ with $N$ interpreted as a momentum along the 
$x^i$-direction.   The momentum along the $x^i$-direction adds one 
more constraint on the Killing spinor:
\begin{equation}
\varepsilon_R=\Gamma^0\Gamma^i\varepsilon_R, \ \ \ \ \ 
\varepsilon_L=\Gamma^0\Gamma^i\varepsilon_L.
\label{spcnstmm}
\end{equation}

In general, the intersecting $n$ $p$-branes preserve 
at least $1/2^n$ of supersymmetry;  
since a single $p$-brane breaks 1/2 of supersymmetry 
with one spinor constraint (\ref{pbrspcst}), as one increases 
the number of constituents more supersymmetry get broken.  
One obtains BPS configurations if spinor 
constraints of constituent $p$-branes are compatible with non-zero spinor 
$\varepsilon_{R,L}$.  The intersecting $D\,p$- and $D\,p^{\prime}$-branes 
preserve $1/2^2$ of supersymmetry iff $p=p^{\prime}$ mod 4 \cite{DOU077}.  
All the supersymmetries are broken when the dimension of the 
relative transverse space is neither 4 nor 8.  
When there is a momentum in the $x^i$-direction, 
the additional Killing spinor constraint (\ref{spcnstmm}) breaks 1/2 of 
the remaining supersymmetry.  

Black holes in lower dimensions are obtained by compactifying 
(intersecting) $p$-branes in $D=10$.  In the language 
of $p$-branes, this compactification procedure corresponds to wrapping 
$p$-branes along the cycles of compact manifold.  Since the 
compactified space is very small, the configuration looks point-like 
(0-brane) in lower dimensions.  

In the following subsections, we discuss various $D\,p$-brane 
embeddings of  $D=4,5$ black holes having the regular BPS limit with  
non-zero horizon area.

\subsubsection{Five-Dimensional Black Hole}\label{dbrbhfiv}

We discuss $D=5$ type-IIB black hole originated from intersecting $Q_1$ 
$D\,1$-branes (along $x^9$) and $Q_5$ $D\,5$-branes (along $x^5,...,x^9$) 
with a momentum $P$ flowing in the common string direction \cite{CALm472}, 
i.e. the $x^9$-direction.  The 1- and 5-brane charges 
are electric and magnetic charges of the R-R 2-form field, 
and the momentum corresponds to the KK electric charge associated with 
the metric component $G^{(10)}_{09}$.   

To obtain a black hole in $D=5$, one wraps $Q_1$ $D\,1$-branes around 
$S^1$ (along $x^9$) of radius $R$ and wrap $Q_5$ $D\,5$-branes 
around $T^5=T^4\times S^1$.  Here, $T^4$ has coordinates ($x_5,...,x_8$) 
and volume $V$.  The momentum (of open string) $P=N/R$ flows around $S^1$.  

The resulting $D=5$ solution has the form \cite{CALm472,CVEy476}:
\begin{equation}
ds^2_5=-f^{-2/3}(r)\left(1-{{r^2_0}\over{r^2}}\right)dt^2
+f^{1/3}(r)\left[\left(1-{{r^2_0}\over{r^2}}\right)^{-1}dr^2
+r^2d\Omega^2_3\right],
\label{iibfivsph}
\end{equation}
where $f(r)$ is given by:
\begin{equation}
f(r)=\left(1+{{r^2_0\sinh^2\delta_1}\over{r^2}}\right)
\left(1+{{r^2_0\sinh^2\delta_5}\over{r^2}}\right)
\left(1+{{r^2_0\sinh^2\delta_p}\over{r^2}}\right).
\label{lamfiv}
\end{equation}
The three charges carried by the black hole are:
\begin{equation}
Q_1={{Vr^2_0}\over{2g_s}}\sinh 2\delta_1, \ \ \ \ 
Q_5={{r^2_0}\over{2g_s}}\sinh 2\delta_5, \ \ \ \ 
N={{R^2Vr^2_0}\over{2g^2_s}}\sinh 2\delta_p.
\label{fivch}
\end{equation}

The ADM energy $E$, entropy $S$, and the Hawking 
temperature $T_H$ of (\ref{iibfivsph}) are
\begin{eqnarray}
E&=&{{RVr^2_0}\over{2g^2_s}}(\cosh 2\delta_1 +\cosh 2\delta_5 +
\cosh 2\delta_p),
\cr
S&=&{A\over{4G^5_N}}={{2\pi RVr^3_0}\over{g^2_s}}\cosh\delta_1
\cosh\delta_5\cosh\delta_p,
\cr
T_H&=&{1\over{2\pi r_0\cosh\delta_1\cosh\delta_5\cosh\delta_p}}.
\label{fivphys}
\end{eqnarray}
From the asymptotic values of $G^{(10)}_{99}$ and $G^{(10)}_{ii}$ 
($i=5,...,8$), one obtains the following expressions for pressures in 
the $x^9$- and $x^i$-directions:
\begin{eqnarray}
P_9&=&{{RVr^2_0}\over{2g^2_s}}\left[
\cosh 2\delta_p-{1\over 2}(\cosh 2\delta_1+\cosh 2\delta_5)\right],
\cr
P_i&=&{{RVr^2_0}\over{2g^2_s}}(\cosh 2\delta_1 -\cosh 2\delta_5).
\label{fivpress}
\end{eqnarray}
So, in the $x^i$-directions, which are parallel to 
5-brane but perpendicular to 1-brane, shrinking effect of 5-brane 
and expanding effect of 1-brane compete and become balanced when 
$\delta_2=\delta_5$.  In the $x^5$-direction, which are parallel to 
both 5- and 1-branes, the shrinking effects of 5- and 1-branes are 
compensated by momentum in the $x^5$-direction.  

The 6 parameters $r_0$, $\delta_1$, $\delta_5$, $\delta_p$, $V$ and 
$R$ of the solution (\ref{iibfivsph}) can be traded with the numbers of 
1-branes, anti-1-branes, 5-branes, anti-5-branes, right-moving momentum, 
and left-moving momentum, respectively given by \cite{HORms383}
\begin{eqnarray}
N_1&=&{{Vr^2_0}\over{4g_s}}e^{2\delta_1}, \ \ \ \ \ 
N_{\bar{1}}={{Vr^2_0}\over{4g_s}}e^{-2\delta_1},
\cr 
N_5&=&{{r^2_0}\over{4g_s}}e^{2\delta_5}, \ \ \ \ \ 
N_{\bar{5}}={{r^2_0}\over{4g_s}}e^{-2\delta_5},
\cr
N_R&=&{{r^2_0R^2V}\over{4g^2_s}}e^{2\delta_p}, \ \ \ \ \ 
N_L={{r^2_0R^2V}\over{4g^2_s}}e^{-2\delta_p}.  
\label{brnnum}
\end{eqnarray}
These parameters are related to the charges in (\ref{fivch}) as 
$Q_1=N_1-N_{\bar{1}}$, $Q_5=N_5-N_{\bar{5}}$ and $N=N_R-N_L$.  
In terms of the new parameters, the ADM energy and the Bekenstein-Hawking 
entropy take simple and suggestive forms \cite{HORms383}:
\begin{eqnarray}
E&=&{R\over {g_s}}(N_1+N_{\bar{1}})+{{RV}\over {g_s}}(N_5+N_{\bar{5}})+
{1\over R}(N_R+N_L),
\cr
S&=&2\pi(\sqrt{N_1}+\sqrt{N_{\bar{1}}})(\sqrt{N_5}+\sqrt{N_{\bar{5}}})
(\sqrt{N_L}+\sqrt{N_R}), 
\label{nadment}
\end{eqnarray}
where 
\begin{equation}
V=\left({{N_1N_{\bar{1}}}\over{N_5N_{\bar{5}}}}\right)^{1/2}, \ \ \ \ \ 
R=\left({{g^2_sN_RN_L}\over{N_1N_{\bar{1}}}}\right)^{1/4}.
\label{nvr}
\end{equation}

\subsubsection{Four-Dimensional Black Hole}\label{dbrbhfou}

There are various ways in which one can construct $D=4$  
black holes having regular BPS-limit with non-zero 
horizon area.  The criteria for such construction are that 
supersymmetries are preserved and all the contributions of $D$-brane 
tensions from constituents compensate for each other so that  
internal space is stable against the shrinking effect near the horizon.

The first type of configuration is obtained by first $T$-dualizing the 
type-IIB $D$-brane bound state discussed in section \ref{dbrbhfiv} 
along $x^4$, resulting in bound state of $Q_2$ $D\,2$-branes 
(along $x^4,x^9$) and $Q_6$ $D\,6$-branes 
(along $x^4,...,x^9$) with the momentum $N$ 
flowing along the  $x^9$-direction.  This configuration has a zero horizon 
area in the BPS limit, since the radii along the directions $x^4,x^9$ 
shrink as one approaches the horizon, due to the tensions of $D\,2$- and 
$D\,6$-branes and the momentum along the $x^9$ direction.  
This is compensated by adding solitonic 5-brane along $x^5,...,x^9$ and 
with magnetic charge of the NS-NS 2-form field $B_{\mu 4}$, where 
$\mu=0,...,3$.   
Since the spinor constraint of the additional solitonic 
5-brane does not further reduce the Killing spinor degrees of freedom, 
the solution still preserves $1/8$ of supersymmetry.  
$D=4$ black hole is obtained by compactifying this $p$-brane 
bound state on $T^6=T^4\times S^{1\,\prime}_1\times S^1$ 
\cite{HORlm77,MALs77}.   
Here, $T^4$ with the coordinates $(x^5,...,x^8)$ has the volume 
$V$ and $S^{1\,\prime}$ [$S^1$] with the coordinate $x^4$ 
[$x^9$] has the radius $R_4$ [$R_9$].  Namely, the $Q_6$ $D\,6$-branes 
wrap around $T^6$, $Q_2$ $D\,2$-branes wrap around $S^{1\,\prime}\times S^1$, 
solitonic 5-branes wrap around $T^4\times S^1$ and the momentum 
flows along $S^1$.   

The resulting $D=4$ solution has the form 
\cite{CVEy672,CVEy127,CVEy675,HORlm77}:
\begin{eqnarray}
ds^2_4&=&-f^{-1/2}(r)\left(1-{{r_0}\over r}\right)dt^2
\cr
& &\ \ \ +f^{1/2}(r)\left[\left(1-{{r_0}\over r}\right)^{-1}dr^2
+r^2(d\theta^2+\sin^2\theta d\phi^2)\right],
\cr
f(r)&=&\left(1+{{r_0\sinh^2\delta_2}\over r}\right)
\left(1+{{r_0\sinh^2\delta_5}\over r}\right)
\left(1+{{r_0\sinh^2\delta_6}\over r}\right)
\cr
& &\times\left(1+{{r_0\sinh^2\delta_p}\over r}\right).
\label{iiafobh}
\end{eqnarray}
The electric/magnetic charges carried by the black hole are
\begin{eqnarray}
Q_2&=&{{r_0V}\over {g_s}}\sinh 2\delta_2, \ \ \ \ \ 
Q_5=r_0R_4\sinh 2\delta_5,
\cr
Q_6&=&{{r_0}\over {g_s}}\sinh 2\delta_6, \ \ \ \ \ 
N={{r_0VR^2_9R_4}\over{g^2_s}}\sinh 2\delta_p. 
\label{iiafoch}
\end{eqnarray}

The pressures along the directions $x^4$, $x^i$ ($i=5,...,8$) 
and $x^9$ are
\begin{eqnarray}
P_4&=&{{r_0VR_4R_9}\over{g^2_s}}(\cosh 2\delta_5-\cosh 2\delta_p),
\cr
P_i&=&{{r_0VR_4R_9}\over{g^2_s}}(\cosh 2\delta_2-\cosh 2\delta_6),
\cr
P_9&=&{{r_0VR_4R_9}\over{g^2_s}}(\cosh 2\delta_2+\cosh 2\delta_6
-\cosh 2\delta_5-\cosh 2\delta_p).
\label{iiafopres}
\end{eqnarray}
When $\delta_2=\delta_5=\delta_6=\delta_p$, 
all the contributions to pressures get balanced ($P_4=P_i=P_5=0$) and, 
as a result, all the scalars become constant everywhere.  

One can trade the 8 parameters $r_0$, $\delta_2$, $\delta_5$, 
$\delta_6$, $\delta_p$, $V$, $R_4$, and $R_9$ with the 
numbers of right- (left-) moving momentum modes, (anti-) 2-branes, 
(anti-) 5-branes, and (anti-) 6-branes:
\begin{eqnarray}
N_R&=&{{r_0VR_4R^2_9}\over{2g^2_s}}e^{2\delta_p}, \ \ \ \ \ \ 
N_L={{r_0VR_4R^2_9}\over{2g^2_s}}e^{-2\delta_p}, 
\cr
N_2&=&{{r_0V}\over{2g_s}}e^{2\delta_2}, \ \ \ \ \ \ \ \ \ \ \ 
N_{\bar{2}}={{r_0V}\over{2g_s}}e^{-2\delta_2},
\cr
N_5&=&{{r_0R_4}\over 2}e^{2\delta_5},  \ \ \ \ \ \ \ \ \ \ 
N_{\bar{5}}={{r_0R_4}\over 2}e^{-2\delta_5},
\cr
N_6&=&{{r_0}\over{2g_s}}e^{2\delta_6}, \ \ \ \ \ \ \ \ \ \ \ 
N_{\bar{6}}={{r_0}\over{2g_s}}e^{-2\delta_6}.
\label{iiafonum}
\end{eqnarray}
In terms of these parameters, the ADM mass and entropy take 
forms \cite{HORlm77}:
\begin{eqnarray}
M&=&{1\over{R_9}}(N_R+N_L)+{{R_4R_9}\over {g_s}}(N_2+N_{\bar{2}})
\cr
& &+{{VR_9}\over{g^2_s}}(N_5+N_{\bar{5}})+{{VR_4R_9}\over {g_s}}
(N_6+N_{\bar{6}}),
\cr
S&=&2\pi(\sqrt{N_R}+\sqrt{N_L})(\sqrt{N_2}+\sqrt{N_{\bar{2}}})
(\sqrt{N_5}+\sqrt{N_{\bar{5}}})(\sqrt{N_6}+\sqrt{N_{\bar{6}}}),
\label{iiafonph}
\end{eqnarray}
where 
\begin{equation}
V=\sqrt{{N_2N_{\bar{2}}}\over{N_6N_{\bar{6}}}}, \ \ \ \ 
R_4=\sqrt{{N_5N_{\bar{5}}}\over{g^2_sN_6N_{\bar{6}}}}, \ \ \ \ 
R_4R^2_9=\sqrt{{g^2_sN_RN_L}\over{N_2N_{\bar{2}}}}.
\label{iiafovol}
\end{equation}

We discuss couple of other ways \cite{JOHkm378} to construct 
$D=4$ black holes.  The first configuration is a type-IIB 
configuration where $Q_1$ parallel $D\,1$-branes (along $x^5$) and 
$Q_5$ parallel $D\,5$-branes (along $x^5,...,x^9$) intersect in 
the $x^5$-direction, along which momentum 
$P^5=N/L$ flows, and magnetic monopole in the subspace 
$(t,r,\theta,\phi,x^4)$.  The $D=4$ black hole is obtained by first 
wrapping $Q_5$ $D\,5$-branes around 4-cycles of $K3$ surface 
(with the coordinates $x^6,...,x^9$), resulting in  
a $D$-string bound state along with $Q_1$ $D\,1$-branes 
in $D=6$.  This bound state in $D=6$ has momentum 
$P^5$ along $x^5$ and the KK magnetic charge associated with the 
metric component $g_{4\phi}$.  One further compactifies 
the coordinates $(x^4,x^5)$ on $T^2$ to obtain $D=4$ black hole.  
Entropy of this solution is $S={A\over{4G_N}}=2\pi\sqrt{Q_1Q_5N}$.  

The second configuration is a type-IIA solution obtained by 
$T$-dualizing the above configuration along $x^5$.  
Namely, this is a bound state of $Q_0$ $D\,0$-branes and $Q_4$ 
$D\,4$-branes (along $x^6,...,x^9$) with open strings wound around 
the $x^5$-direction (with the winding number $W$) and the KK monopole 
in the subspace $(t,r,\theta,\phi,x^4)$.  Similarly as in the first 
case, to have a $D=4$ black hole, one first wrappes $Q_4$ $D\,4$-branes 
around 4-cycles of $K3$ surface (with the coordinates $x^6,...,x^9$), 
resulting in a $D$ particle bound state in $D=6$, and then 
compactifies the coordinates $(x^4,x^5)$ on $T^2$.  
Such a solution has entropy $S={A\over{4G_N}}=2\pi\sqrt{Q_0Q_4W}$.

\subsection{$D$-Brane Counting Argument}\label{dbrent}

In this section, we discuss $D$-brane interpretation of black hole 
entropy.  The number of degenerate BPS states 
is calculated in the weak string coupling regime ($g_s\approx 0$), 
in which the $D$-brane picture, rather than black $p$-brane description, 
of charge configurations is applicable.  
Since mass of R-R charge carrier behaves as $\sim 1/g_s$, $D$-branes 
become infinitely massive in the perturbative region.  So, the leading 
contribution to the degeneracy of states is from perturbative states of 
open strings, which are attached to $D$-branes. 
So, perturbative $D$-brane configurations 
are described by conformal field theory of open strings in the target 
space manifold determined by $D$-branes and the internal space.  
In the following, we discuss conformal field theories 
of $D$-branes which correspond to specific $D=4,5$ black 
holes and calculate the number of degenerate perturbative open string 
states.  
More intuitive picture of $D$-brane argument is discussed in the 
subsequent subsection.  

\subsubsection{Conformal Model for $D$-Brane Configurations}\label{dbrconf}

We saw that the worldvolume theory of massless modes in 
a bound state of $N$ $D\,p$-branes is described by $D=10$ 
$U(N)$ super-Yang-Mills theory compactified to $D=p+1$ 
\cite{WIT335}.   Namely, dynamics of collective coordinates 
\cite{CALk465} of $N$ parallel $D\,p$-branes is described by a 
supersymmetric $U(N)$ gauge theory.  
The BPS states of $D$-brane configuration correspond to the 
supersymmetric vacuum of the corresponding super-Yang-Mills theory.  

The $U(N)$ group is decomposed into a $U(1)$ group, describing the center 
of mass motion of $D$-branes, and an $SU(N)$ group \cite{WIT335}.  It 
is the supersymmetric $SU(N)$ vacuum that contains states with a mass gap, 
which are relevant for degeneracy of $D$-brane bound states.  
When there is a mass gap, the number of ground states in the $SU(N)$ 
super-Yang-Mills theory is the same as the degeneracy of the $D$-brane 
bound states.  
Here, the ground states are identified with the cohomology elements of 
the $SU(N)$ instanton moduli space ${\cal M}^N$, implying that  
degeneracy of $D$-brane bound states is given by the number 
of cohomology elements of ${\cal M}^N$ \cite{VAF088,VAF463}.     
$D$-brane bound states are effectively described by the $\sigma$-model 
on the instanton moduli space
\footnote{In particular, the singular points in the moduli space 
\cite{WIT443,ASPm151,ASP357}, at which gauge symmetry is enhanced, 
correspond to the case where $D$-branes coincide.} 
${\cal M}^N$.  Generally, the $\sigma$-model with target space manifold 
given by hyperK\"ahler manifold of dimension $4k$ has the central charge 
$c=6k$.  For example, the moduli space ${\cal M}^N_k$ of $SU(N)$ 
instantons on $K3$ with the instanton number $k$ has the dimension 
$4[N(k-N)+1]$.  It has been checked 
\cite{SEN54,SEN53,VAF088,VAF463,BERsv398,BERsv463} that  
calculation of the degeneracy of $D$-brane bound states based on this  
idea yields the results which are consistent with conjectured string 
dualities that relate perturbative string states to $D$-brane 
bound states.   In particular, $P^2$ of BPS perturbative string states 
at the level $N_L=1+{1\over 2}P^2$
is dualized to the intersection number of $D$-brane bound states  
\cite{VAF088}.  

One of setbacks in study of $D$-brane bound states is that  
a bound state of $m$ $D\,p$-branes is {\it marginally stable}, i.e.  
there is no energy barrier against decay into $m$  
$D\,p$-branes (each carrying the unit R-R $(p+1)$-form charge).  
Such a problem is circumvented \cite{SEN54} by compactifying an extra 
one direction, say $y^1$, on $S^1$ so that states in the 
multiplet carry momentum or winding number $n$ along $S^1$.  If the pair 
$(m,n)$ is relatively prime, the corresponding states in the $D=9-p$ 
theory, i.e. $D=10$ string bound states compactified 
on $T^p\times S^1$, become {\it absolutely stable} against decay.  
Such worldvolume theory is described by the $N=1$ $U(m)$ gauge theory 
compactified to $D=p+1$ with states characterized by $n$ units of 
$U(1)\subset U(m)$ electric flux along $y^1$.  The non-trivial information 
on degeneracy of the BPS $D$-brane bound states is contained in the 
supersymmetric ground states of the $SU(m)$ part of theory on 
the base manifold $T^p\times {\bf R}$, where $\bf R$ is labeled by the 
time coordinate.

It is shown \cite{VAF088} that for the type-IIA bound state of $m$ 
0-branes and 1 4-brane (or the bound state of $m$ 1-branes and 1 5-brane 
in the $T$-dual theory) the instanton moduli space ${\cal M}$ of the 
corresponding $U(m)\times U(1)_c$ super-Yang-Mills theory is $(T^4)^m/S_m$, 
where $S_m$ is the permutation group on $m$ objects.  The degeneracy of the 
cohomology in this space is in one-to-one correspondence with the 
partition function of left-movers of the superstring.  To put it another 
way, the ground states of this gauge theory are related to the degenerate 
string states at level $m$.  The quantum numbers ${1\over 2}(F_L+F_R,F_L-F_R)$ 
of the cohomology of $(T^4)^m/S_m$, where $F_L$ [$F_R$] is the [anti-] 
holomorphic degrees of the homology shifted by half the complex dimension, 
is mapped to the light-cone helicities of the left oscillator states of 
type-II superstring \cite{VAFw431}.  

The generalization to $D$-brane bound states wrapped 
around $K3$ surface is carried out in \cite{BERsv463,VAF463}.  
The bound state of $N$ $D\,4$-branes wrapped around $K3$ is described 
by a $U(N)$ gauge theory on $K3\times{\bf R}$ with ${\bf R}$ parameterized 
by the time coordinate.  For this $D$-brane configuration, it is shown 
\cite{BERsv463} that the quantum corrections induce the effective $D\,0$-brane 
charge $M=k-N$.  So, the momentum square $P^2=P^2_R-P^2_L$  
is $P^2/2=NM=N(k-N)$.  If $N$ and $k$ are relatively prime, there is a 
mass gap and the moduli space ${\cal M}^N_k$ is smooth with a discrete 
spectrum, with cohomology of ${\cal M}^N_k$ being that of $N(k-N)+1$ 
unordered points on $K3$.  
When $N=2$ and $k$ is odd, the moduli space is 
the symmetric product of $K3$'s:  ${\cal M}^2_k=(K3)^{2k-3}/S_{2k-3}$.  
The Euler characteristic $d(2k-3)$ of ${\cal M}^2_k$ is 
the same as the degeneracy $d(N_L)=d({1\over 2}P^2+1)$ of string states 
at level $N_L$, or the degeneracy of $D$-brane bound states with charges 
$(N,M)$.  Here, $d(n)$ is defined in terms of the Dedekind eta function 
$\eta(q)$ as $\eta(q)^{-24}=q^{-1}\sum_n d(n)q^n$.

\paragraph{Applications to Statistical Entropy of Black Holes}
\label{dbrentconfbh}

We consider a type-IIB $D=5$ black hole carrying electric 
charges $Q_F$ and $Q_H$ of $U(1)$ fields, 
respectively, originated from R-R 2-form potential 
$\hat{A}^{RR}_{M_1M_2}$ and NS-NS 2-form potential 
$\hat{B}^{NS}_{M_1M_2}$ \cite{STRv,BRElmpsv,BREmpv}.   
Here, the $D=5$ $U(1)$ field strength which is associated 
with R-R 2-form potential in $D=10$ is Hodge-dual to 
the field strength of R-R 2-form field $\hat{A}^{RR}_{\mu_1\mu_2}$ 
in $D=5$.  From the $D=10$ perspective, $Q_F$ [$Q_H$] 
is magnetic [electric] charge of $\hat{A}^{RR}_{M_1M_2}$ 
[$\hat{B}^{NS}_{M_1M_2}$], which couples to the worldvolume of 
R-R 5-brane [NS-NS string].   The $D=5$ $U(1)$ field 
originated from $\hat{A}^{RR}_{M_1M_2}$, therefore, carries 
electric charge.  
The internal index $m$ in the $D=5$ NS-NS $U(1)$ field 
$\hat{B}^{NS}_{\mu m}$ is along $S^1$, upon which an extra 
coordinate of the $D=6$ type-IIB theory is compactified.  
So, this black hole is regarded as a bound state 
of 5-brane with R-R charge $Q_F$ and string with N-N charge $Q_H$ 
with the 5-brane wrapped around a holomorphic 4-cycle of $K3$ and 
partially around $S^1$, and the string wound round $S^1$.  
$T$-duality along $S^1$ leads to a type-IIA solution corresponding to 
4-brane with 5-form charge $Q_F$ and momentum flowing along 
$S^1$.  

The non-rotating BPS black hole with the above charge configuration has 
spacetime of the Reissner-Nordstrom black hole \cite{TSE11,CVEy476,STRv}:
\begin{equation}
ds^2=-\left(1-\left({{r_0}\over r}\right)^2\right)^2dt^2+
\left(1-\left({{r_0}\over r}\right)^2\right)^{-2}dr^2+
r^2d\Omega^2_3,
\label{stdfiv}
\end{equation}
where $r_0=(8Q_HQ^2_F/\pi^2)^{1/6}$.  The scalars are constant 
everywhere and expressed in terms of the (integer-valued) electric 
charges $Q_{H,F}$ \cite{FERks52}.   The Bekenstein-Hawking entropy is 
\begin{equation}
S_{BH}={A\over 4}=2\pi\sqrt{{Q_HQ^2_F}\over 2}.
\label{stfivent}
\end{equation}

For $D=5$ rotating solutions, the regular BPS limit is possible 
only for the case where 2 angular momenta have the same magnitude, 
i.e. $|J_1|=|J_2|=J$.  The solution is 
\cite{TSE381,CVEy476,BREmpv}
\begin{eqnarray}
ds^2&=&-\left(1-{{\mu}\over{r^2}}\right)
\left[dt-{{\mu\omega\sin^2\theta}\over{(r^2-\mu)}}d\phi_1
+{{\mu\omega\cos^2\theta}\over{(r^2-\mu)}}d\phi_2\right]^2
\cr
& &+\left(1-{{\mu}\over{r^2}}\right)^{-2}dr^2
+r^2(d\theta^2+\sin^2\theta d\phi^2_1+\cos^2\theta d\phi^2_2),
\label{rotdfiv}
\end{eqnarray}
with $Q_H=\mu/\lambda^2$, $Q_F=-{{\pi}\over{2\sqrt{2}}}\mu\lambda$ and 
$J_1=-J_2={{\pi}\over 4}\mu\omega$.  The Bekenstein-Hawking 
entropy is 
\begin{equation}
S_{BH}={A\over 4}=2\pi\sqrt{{{Q_HQ^2_F}\over 2}-J^2}.
\label{rotfivent}
\end{equation}

We discuss the $D$-brane interpretation of 
the entropies (\ref{stfivent}) and (\ref{rotfivent}).  
In terms of $D$-brane language, the above black holes 
are the bound state of $Q_F$ $D\,5$-branes wrapped around 
$K3\times S^1$ and a fundamental string wound around 
$S^1$ with the winding mode $Q_H$.  The product $Q_F\cdot Q_F$ 
is the intersection number of $D\,5$-branes in the $K3$ homology.  
We consider the case where $K3$ is small compared to the 
size of $S^1$.  For this case, the solution looks like a fundamental 
string in the 5-brane background.  The theory is effectively described 
by the conformal field theory on $S^1\times {\bf R}$ with the target 
space manifold given by the symmetric product of $K3$ surfaces 
\cite{STRv}:
\begin{equation}
{\cal M}={{(K3)^{\otimes[{1\over 2}Q^2_F+1]}}\over 
{S_{[{1\over 2}Q^2_F+1]}}}.
\label{fivdtgt}
\end{equation}
This conformal field theory has central charge $c=6({1\over 2}Q^2_F+1)$, 
since the real dimension of this manifold is $2(Q^2_F+2)$.  

First, we consider the non-rotating black hole (\ref{stdfiv}) with the 
thermal entropy (\ref{stfivent}) \cite{STRv}.   The statistical entropy of 
the BPS solutions is given in terms of degeneracy $d(n,c)$ of conformal 
states with the left-moving oscillators at level $L_0=n$ ($n\gg 1$) and the 
right-moving oscillators in ground state:
\begin{equation}
S_{stat}=\ln\,d(n,c)\sim 2\pi\sqrt{{nc}\over 6}.
\label{dbrstent}
\end{equation}
For the solution (\ref{stdfiv}), $n=Q_H$ and, therefore, the statistical 
entropy (\ref{dbrstent}) reproduces the thermal entropy (\ref{stfivent}): 
\begin{equation}
S_{stat}\sim 2\pi\sqrt{Q_H({1\over 2}Q^2_F+1)}.
\label{5ddstent}
\end{equation}

Next, we consider the thermal entropy (\ref{rotfivent}) of the rotating 
black hole (\ref{rotdfiv}) \cite{BREmpv}.  The rotation group 
$SO(4)$ of the $D=6$ lightcone-frame theory is identified with the 
$SU(2)_L\times SU(2)_R$ symmetry of the $N=4$ superconformal algebra.  
The charges $(F_L,F_R)$ of $U(1)_L \times U(1)_R\subset SU(2)_L\times 
SU(2)_R$ are interpreted as spins of string states \cite{VAFw431,BREmpv}, 
and are related to angular momenta of the black hole (\ref{rotdfiv}) as
\begin{equation}
J_1={1\over 2}(F_L+F_R), \ \ \ \ \ 
J_2={1\over 2}(F_L-F_R).
\label{angconf5}
\end{equation}
Note, the $U(1)_L$ current $J_L$ can be bosonized as 
$J_L=\sqrt{c\over 3}\partial\phi$ and a conformal state 
$\Phi_{F_L}$ carrying $U(1)_L$ charge $F_L$ is obtained 
by applying an operator $\exp({{iF_L\phi}\over{\sqrt{c/3}}})$ 
to the state $\Phi_0$ without $U(1)_L$ charge.  The net effect 
is to shift
\footnote{For detailed discussions on this point, see section 
\ref{entnearstat}.}  
the left-moving oscillator level $n$ of string 
states carrying the $U(1)_L$ charge $F_L$ with respect to 
the level $n_0$ for string states without the $U(1)_L$ charge 
$F_L$: $n_0=n-{{3F^2_L}\over{2c}}$.  For a large value 
of $n_0$, one can safely use the string level density formula 
$d_{n,c}$, which includes contributions of string states with 
all the possible spin values, in calculating the statistical entropy.  
Here, $c=6({1\over 2}Q^2_F+1)$ and $n=Q_H$, as in the non-rotating case.  
Substituting all of these values into (\ref{dbrstent}), one obtains the 
following statistical entropy in agreement with the thermal entropy 
(\ref{rotfivent}):
\begin{equation}
S_{stat}\sim 2\pi\sqrt{{n_0c}\over 6} \sim 
2\pi\sqrt{Q_H({1\over 2}Q^2_F+1)-
{1\over 4}(|J_1|+|J_2|)^2},
\label{5ddrotent}
\end{equation}
where $J_1=-J_2=J$.

\subsubsection{Another Description}\label{dbrentint}

We discuss more intuitive $D$-brane picture  
of black holes based upon the arguments in \cite{CALm472}.  
In this picture, one counts all the possible oscillator contributions 
of open strings attached to $D$-branes.  $D$-brane 
configurations, which are the weak string coupling limit of 
black holes, are described by the $\sigma$-model of open strings 
with the target space background determined by $D$-branes, upon which 
the ends of the open strings live.   

The statistical entropy of black holes is due to degenerate states in 
fixed oscillator levels $N_{L,R}$ (or a fixed mass) of open strings 
attached to $D$-branes:
\begin{equation}
S_{stat}=\ln\,d(N_L,N_R;c)\sim 
2\pi\sqrt{c\over 6}(\sqrt{N_L}+\sqrt{N_R}), 
\label{dbrstat}
\end{equation}
in the limit $N_{L,R}\gg 1$.  The central charge $c$ has an additive 
contribution of 1 [$1\over 2$] for each bosonic [fermionic] coordinate.   

$N_{L,R}$ are determined by the NS-NS electric charges (i.e. 
momentum and winding modes in the compactified space) from the mass 
formula of open string states, which is derived from the Virasoro 
condition $L_0-a=0$.  Here, $a$ is the zero point energy (or the normal 
ordering constant of oscillator modes).  
The contribution to the zero point energy $a$ is additive with 
the contribution of $-{1\over{24}}$ [$1\over{48}$] for each bosonic 
coordinate with the periodic [anti-periodic] boundary condition, 
i.e. the integer [half-integer] moding of oscillator, 
and for a fermionic coordinate there is an extra minus sign.  

Note, the Virasoro operator $L_0$, say of the bosonic coordinates, is 
defined in terms of the oscillator modes as
\footnote{The fermionic coordinates have the similar expressions.  
In the presence of both bosonic and fermionic coordinates, the 
Virasoro operator is sum of the 2 contributions, i.e.  
$L_0=L^{boson}_0+L^{fermion}_0$.} 
$L_0={1\over 2}\alpha^{\mu}_0\alpha_{0\,\mu}+\sum_n\alpha^{\mu}_{-n}
\alpha_{n\,\mu}$, where $\mu$ is a $D=10$ vector index.   Here, 
$\alpha^{\mu}_0=p^{\mu}$ is the center-of-mass frame momentum of an 
open string and $N\equiv{{\alpha^{\prime}}\over 4}\sum_n\alpha^{\mu}_{-n}
\alpha_{n\,\mu}$ is the oscillator number operator.  
For string theory compactified to $D=d$ ($d<10$), we 
divide $D=10$ momentum $p^{\mu}$ into the $D=d$ part 
$p^{\bar{\mu}}$ and the internal part $p^m$, i.e. $(p^{\mu})=
(p^{\bar{\mu}},p^m)$.   Since mass in $D=d$ is defined as 
$M^2=-p_{\bar{\mu}}p^{\bar{\mu}}$, one obtains the following mass 
formula from the Virasoro condition $L_0-a=0$:
\begin{equation}
M^2=(\vec{p}_L)^2+{4\over{\alpha^{\prime}}}(N_L+a_L)=
(\vec{p}_R)^2+{4\over{\alpha^{\prime}}}(N_R+a_R),
\label{ddimmass}
\end{equation}
where the subscripts $L$ and $R$ denote left- and right-moving sectors, 
and $\vec{p}_{L,R}$ are momenta in the internal directions. 
Note, this expression for mass has all the contributions from 
bosonic and fermionic coordinates.  In particular, for the compactification 
on $S^1$ of radius $R$, the left- and right-moving momenta 
in the $S^1$-direction is $p_{L,R}={n\over R}\mp{{mR}\over{\alpha^{\prime}}}$, 
where $n$ and $m$ are respectively momentum and winding modes 
around $S^1$.  

Mass of the BPS states is $M^2_{BPS}=(\vec{p}_R)^2$. 
So, for the BPS states, the right-movers are in ground 
state and only left-movers are excited with their 
total oscillation numbers determined (from the mass formula 
(\ref{ddimmass})) by the left- and right-moving momenta 
in the internal space:
\begin{equation}
N_R=-a_R, \ \ \ \ \ 
N_L={{\alpha^{\prime}}\over 4}\left[(\vec{p}_R)^2-
(\vec{p}_L)^2\right]-a_L.  
\label{bpsoscil}
\end{equation}
For non-BPS states ($M>M_{BPS}$), the right movers, as well as the 
left movers, are excited:
\begin{equation}
N_R=-a_R+k, \ \ \ \ \ 
N_L={{\alpha^{\prime}}\over 4}\left[(\vec{p}_R)^2-
(\vec{p}_L)^2\right]-a_L+k\ \ \ \ \ \ (k\in{\bf Z}).  
\label{nonbpsoscil}
\end{equation}
Since $N_{L,R}$ are fixed by NS-NS electric charges,  
the main problem of calculating the statistical entropy in 
$D$-brane picture is to determine the value of $c$, i.e. the total 
degrees of freedom of bosonic and fermionic coordinates of open strings. 

There are two types of open strings. The first type is 
open strings that stretch between the same type of $D$-brane, 
called the $(p,p)$ type.  The second type is open 
strings that stretch between different types of $D$-branes, called the 
$(p,p^{\prime})$ type, $p\neq p^{\prime}$.  For the second type, 
$(p,p^{\prime})$ and $(p^{\prime},p)$ are not equivalent, since 
open strings are oriented.  Open strings of the $(p,p)$ type can be 
ignored in the calculation of the open string state degeneracy, since 
the excited states of open strings which stretch between $D$-branes 
of the same type become very heavy as the relative separation between 
a pair of $D$-branes gets large.  

We briefly discuss some aspects of open string states of the 
$(p,p^{\prime})$ type.  The $D=10$ spacetime coordinates 
$X^{\mu}$ for this type of open strings are divided into 3 classes.  
The first [second] type, called the DD [NN] type, corresponds to 
coordinates for which both ends of strings satisfy the Dirichlet 
[Neumann] boundary conditions (i.e. coordinates which are 
transverse [longitudinal] to both branes).  For these cases, the bosonic 
string coordinates are integer modded and, thereby, have the zero point 
energy contribution of $-{1\over{24}}$ for each bosonic coordinate.  
The third type, called DN or ND type
\footnote{These two types are different since open strings are oriented.}, 
corresponds to coordinates for which each end of open strings satisfies 
different boundary conditions (i.e. coordinates which are transverse to 
one brane and longitudinal to the other brane).  
For this type, the bosonic string coordinates are half-integer modded and, 
therefore, have the zero point energy contribution of $1\over{48}$ for 
each bosonic coordinate.   The fermionic coordinates have the same moding 
(as the bosonic coordinates) in the R sector and the opposite in the 
NS sector.  So, the total zero point energy is always 0 for the R  
sector.  For the NS sector, the zero point energy depends on the total 
number $\nu$ of the ND and DN coordinates, which is always even:
\begin{equation}
(8-\nu)\left(-{1\over{24}}-{1\over{48}}\right)+
\nu\left({1\over{24}}+{1\over{48}}\right)=
-{1\over 2}+{{\nu}\over 8}.
\label{ndzero}
\end{equation}
Only when $\nu$ is a multiple of 4, $D$-brane 
configurations are supersymmetric and degeneracy between the NS and R 
sectors is possible.  

When $\nu=4$, $N=2$ supersymmetry is preserved in $D=4$ and 
the zero point energy is zero in both the R and NS sectors.  
An example is the intersecting $D$ string and $D\,5$-brane, 
corresponding to the $D=5$ black hole that we consider in the following.  
For this type of open strings, degrees of freedom of the fermionic 
and bosonic coordinates are determined as follows.  In the NS sector, 
among 4 fermionic coordinates $\chi$ in the ND directions 
only 2 of them survive the GSO projection $\Gamma^{i_1}\Gamma^{i_2}
\Gamma^{i_3}\Gamma^{i_4}\chi=\chi$, where $i_1,...,i_4$ correspond to 
the ND directions.  In the R sector, also only 2 of 4 periodic 
transverse fermions (in the NN and DD directions) survive the GSO 
projection.  Furthermore, there are $Q_pQ_{p^{\prime}}$ ways to attach 
open strings between $D\,p$- and $D\,p^{\prime}$-branes. 
Since open strings are oriented, we have the same numbers of 
fermionic and bosonic degrees of freedom satisfying the above 
constraints in the $(p,p^{\prime})$ and the $(p^{\prime},p)$ sectors.  
So, there are $4Q_pQ_{p^{\prime}}$ bosonic and $4Q_pQ_{p^{\prime}}$ 
fermionic degrees freedom for each $(p,p^{\prime})$ and 
$(p^{\prime},p)$ types.   Thus, the central charge of open strings of 
$(p,p^{\prime})$ and $(p^{\prime},p)$ types is
\begin{equation}
c=4Q_pQ_{p^{\prime}}(1+{1\over 2})=6Q_pQ_{p^{\prime}}.
\label{centpppr}
\end{equation}

We comment on different interpretation of the entropy expression 
(\ref{dbrstat}).  Let us consider a BPS configuration ($N_R=0$ case) 
with the total momentum number $N$ along $S^1$ of radius 
$R$.  When the number of the bosonic [fermionic] degrees of freedom are  
$N_B$ [$N_F$], the central charge is $c=N_B+{1\over 2}N_F$.  
Then, the statistical entropy (\ref{dbrstat}) becomes the 
following entropy formula describing the $N_B$ bosonic and $N_F$ fermionic 
species with energy $E=N/R$ in a 1-dimensional space of length $L=2\pi 
R$ 
\cite{MALs77,DEAs189}:
\begin{equation}
S=\sqrt{\pi(2N_B+N_F)EL/6}.
\label{entint1}
\end{equation}

Note, the entropy formula (\ref{dbrstat}) is derived assuming that 
$N_{L,R}\gg 1$, i.e. the NS electric charges 
are much bigger than the number $Q_p$ of $D\,p$-branes.  
When $Q_p$ and the NS electric charges are of the same 
order in magnitude, the string level density formula fails to reproduce 
the Bekenstein-Hawking entropy \cite{MALs475}.   
This stems from the fact that (with $Q_p$ of the order of $N$) 
the notion of extensivity (of entropy, energy, etc.) fails 
when radius $R$ of $S^1$, around which $D$-branes 
are wrapped, is smaller than the size of black holes ({\it fat} black 
hole limit), since the wavelength ($\sim{1\over T}$, where $T$ is the 
effective temperature of the left movers) of a typical quantum exceeds 
the size $L=2\pi R$ of the system \cite{MALs475}.

The remedy to this problem proposed in \cite{MALs475} is as follows.  
One considers a bound state of $Q_5$ $D\,5$-branes and $Q_1$ $D\,1$-branes 
which, respectively, wrap around $M_4\times S^1$ and $S^1$.  Here, $M_4$ 
is a 4-dimensional compact manifold and $S^1$ has 
radius $R$.  One neglects the directions associated with $M_4$ and 
therefore the $D\,5$-branes are regarded as strings wrapped around $S^1$.  
Instead of taking $Q_p$ $D\,p$-branes as $Q_p$ numbers of 
strings wrapped around $S^1$ once, 
one regards it as a single string wrapped $Q_p$ times around $S^1$, 
thereby having length $2\pi Q_pR$.  
So, a bound state of $D\,1$- and $D\,5$-branes is described 
by a single string of the length $2\pi Q_1Q_5R$ wrapped $Q_1Q_5$ times 
around $S^1$.  This is interpreted as a single specie with the energy 
$E=N/R=N^{\prime}/R^{\prime}$ in a 1-dimensional space of length 
$L^{\prime}=2\pi R^{\prime}$ (thereby simulating a spectrum of fractional 
charges), where $N^{\prime}\equiv Q_1Q_2N$ and $R^{\prime}\equiv Q_1Q_2R$.  
Note, both the oscillator level and the radius of $S^1$ are multiplied 
by $Q_1Q_5$.  In this description of $D$-brain bound states, the size 
$L^{\prime}$ of systems is always bigger than the wavelength of a 
typical quantum, thereby the extensivity condition is always satisfied.  
Furthermore, since the (effective) oscillator level $N^{\prime}=Q_1Q_5N$ 
is always much bigger than the number of $D$-branes (which is 1) even when 
$Q_1\approx N\approx Q_5$, one can still use the statistical entropy 
expression of the type (\ref{dbrstat}).  The statistical 
entropy is then $S=\sqrt{2\pi EL^{\prime}}=2\pi\sqrt{N^{\prime}}$, 
where $L^{\prime}=Q_1Q_5L={1\over 6}(N_B+{1\over 2}N_F)L$ and 
$N^{\prime}=Q_1Q_5N$.  Note, this is the same as (\ref{entint1}) 
when expressed in terms of the original variables, but with this 
new description the approximation (of string level density) leading 
to the statistical entropy is valid.  

Near-extreme black holes correspond to $D$-brane configurations  
with a small amount of right moving oscillations excited, 
i.e. $N_{L,R}$ given by (\ref{nonbpsoscil}) with a small integer $n$.  
As long as the string coupling is very small and the density of strings 
($\propto 1/R$) is low, the interaction between the left- and 
the right-movers is negligible \cite{HORs77}.  In this limit, entropy 
contributions from the left- and the right-movers are additive, leading to 
\cite{HORs77,HORms383}
\begin{equation}
S_{stat}=\sqrt{\pi(2N_B+N_F)E_RL/6}+\sqrt{\pi(2N_B+N_F)E_LL/6}, 
\label{nearstatgas}
\end{equation}
where $L=2\pi R$ and $E_{L,R}=N_{L,R}/R$ with $N_R$ very small.  
This expression correctly reproduces the Bekenstein-Hawking entropy 
of near extreme black holes.  We comment that the above $D$-brane 
interpretation of near extreme black holes is valid when only one 
of the constituents of the $D$-brane bound state 
has energy contribution much smaller than the others.  For example, 
the above calculation is done in the limit where $R$ is larger 
than the size $V$ of the other internal manifold and, thereby, 
black holes are effectively described by (oscillating) strings 
in spacetime 1 dimension higher.  In this limit, the momentum modes 
(of open strings) are much lighter than the branes 
(Cf. (\ref{nadment})); the leading order contribution to the 
black hole entropy is from open string modes.  
For other limits where one of branes has much smaller energy than the 
other constituents, one can apply the similar argument in  
calculating statistical entropy of near extreme black holes 
\cite{HORms383} and the result is consistent with the conjectured 
$U$-duality.  (In these cases, the leading contribution to the 
degeneracy is from $D$-branes rather than from open strings.) 
For the case where more than one constituents are light or all 
the constituents have energies of the same order of magnitude, 
the statistical interpretation of near extreme black holes 
is not known yet.  

Generalization to the rotating black hole case 
\cite{BREmpv,BRElmpsv,HORlm77} is along the same line 
described in section \ref{entnearstat}.  Note, only the fermionic 
coordinates of the $(p,p^{\prime})$ type can carry angular 
momentum under the spatial rotational group.  These are 4 periodic 
transverse fermions (from the NN and DD directions) in the R sector, 
which are spinors under the Lorentz group of the external spacetime.  
When the GSO projection is imposed, only the positive 
chirality representation of the external Lorentz group survives.  

We concentrate on the $(1,5)$ type, 
which is related to the $D=5$ ($D=4$) black holes under consideration 
in this section (through $T$-duality).   For this case, the $D=10$ 
Lorentz group $SO(1,9)$ is decomposed as:
\begin{equation}
SO(1,9)\supset SO(1,1)\otimes SO(4)_E\otimes SO(4)_I,
\label{lorentz}
\end{equation}
where the first [third] factor acts on the $D\,1$-brane worldsheet 
[the space internal to $D\,5$-brane] and the second factor 
on the space external to the configuration.  
Worldsheet spinors are decomposed into the 2-dimensional  
positive chirality representation ${\bf 2}^+$ (with R-type quantization in ND 
directions of the NS sector) under the internal rotational group $SO(4)_I$ 
(thereby a boson under the spacetime $SO(1,5)$ Lorentz group) and the 
2-dimensional representation ${\bf 2}^+_+$ (with NS-type quantization 
in ND directions of the R sector) which is positive-chiral under the 
external group $SO(1,1)\times SO(4)_E\subset SO(1,5)$ (thereby a spacetime 
fermion).  

Here, the spatial rotation group $SO(4)_E$ (external to the $D$-brane 
configuration) is isomorphic to $SU(2)_R\times SU(2)_L$, which is 
identified with the symmetry of $(4,4)$ superconformal theories.  
This is related to the fact that worldsheet fermions, which carry 
angular momenta, manifest themselves as spacetime fermions.   
The $U(1)_{L,R}\subset SU(2)_{L,R}$ charges $F_{R,L}$ are related 
to angular momenta $J_{1,2}$ of the spatial rotational group 
$SO(4)_E$ as in (\ref{angconf5}).  

The effect of angular momenta is to reduce the total oscillation 
numbers \cite{BREmpv}:
\begin{equation}
N_{L\,0}=N_{L}-{{3F^2_L}\over{2c}}, \ \ \ \ \ 
N_{R\,0}=N_{R}-{{3F^2_R}\over{2c}},
\label{rotoscil}
\end{equation}
where $N_{L,R\,0}$ [$N_{L,R}$] are the total oscillation numbers of 
states that do not carry the $U(1)_{L,R}$ charges [have the $U(1)_{L,R}$ 
charges $F_{L,R}$].  To obtain the statistical entropy of black holes, one 
plug this expression for $N_{L,R\,0}$ into (\ref{dbrstat}).  

For non-extreme or near-extreme black holes, which 
do not saturate the BPS bound, the right moving as well as the left 
moving supersymmetries are preserved, and thereby $F_{R,L}$ are both 
non-zero.  The statistical entropy of near-extreme black holes is:
\begin{equation}
S_{stat}\sim 2\pi\sqrt{c\over 6}(\sqrt{N_{L\,0}}+\sqrt{N_{R\,0}}),
\label{nearstat}
\end{equation}
where $N_{L,R\,0}$ are given in (\ref{rotoscil}).  

However, for $D=5$ BPS black holes, $F_R=0$ (leading to 
$J_1=J_2=:J$) since only the left-moving supersymmetry 
survives (i.e. $(0,4)$ superconformal theory).  
So, for BPS black holes, $N_{R\,0}=N_{R}=0$ and $N_{L\,0}=
N_{L}-{6\over c}J^2$, leading to the statistical entropy \cite{HORlm77}:
\begin{equation}
S_{stat}\sim 2\pi\sqrt{c\over 6}\sqrt{N_{L\,0}}=
2\pi\sqrt{{c\over 6}N_L-J^2}.
\label{fivbpsstat}
\end{equation}
For the $D=5$ rotating black hole with charge configuration 
described in the above, $c=6Q_1Q_5$ and $N_L=N$.  
So, the statistical entropy (\ref{fivbpsstat}) correctly reproduces the 
Bekenstein-Hawking entropy: $S_{stat}=2\pi\sqrt{NQ_1Q_5-J^2}$.  

To obtain $D=4$ rotating black hole with regular extreme limit, 
one first applies $T$-duality along the common string direction of 
the intersecting $D\,1$- and $D\,5$-branes with open strings wound 
around the common string direction.  The resulting configuration is 
bound state of $D\,2$- and $D\,6$-branes with momentum flowing 
along one of the common intersection directions.  
To have non-zero horizon area in the BPS limit, one adds 
solitonic 5-brane with charge $Q_5$.  Here, $D\,6$-branes, $D\,2$-branes and  
solitonic 5-brane, respectively, wrap around $T^6=T^4\times 
S^{1\,\prime}\times S^1$, $S^{1\,\prime}\times S^1$ and $T^4\times S^1$, 
and momentum flows along $S^1$.  Since the right moving supersymmetry 
breaks in the presence of solitonic 5-branes, the $U(1)$ charge $F_R$ 
in the right moving sector vanishes (i.e. $J_1=J_2=J$).  
Particularly, $D=4$ extreme rotating black hole corresponds to the 
minimum energy configuration which is regular with non-zero angular 
momentum, which happens when $N_{L\,0}=0$ \cite{HORlm77}.   
So, the left movers are constrained to carry angular momentum, only.  
In this case, $N_{R\,0}=N_R$ and $N_{L}=6J^2/c$.  By plugging these into 
(\ref{nonbpsoscil}), one obtains $N_{R\,0}={6\over c}J^2-
{{\alpha^{\prime}}\over 4}[p^2_R-p^2_L]$, leading to the statistical entropy:
\begin{equation}
S_{stat}\sim 2\pi\sqrt{c\over 6}\sqrt{N_{R\,0}}=
2\pi\sqrt{J^2-{{\alpha^{\prime}c}\over {24}}[p^2_R-p^2_L]}.  
\label{4rotexstat}
\end{equation}
For the $D$-brane bound state corresponding to the 
$D=4$ black hole described above, $c=6Q_2Q_5Q_6$ (since there are 
$4Q_2Q_5Q_6$ species
\footnote{The extra factor of $Q_5$ comes from the fact that whenever 
$D\,2$-branes (which intersect the solitonic 5-brane along $S^1$) 
cross the solitonic 5-brane they can break up and ends separate in 
different positions in $T^4$.  Thus, $Q_2$ $D\,2$-branes break up 
into $Q_2Q_5$ $D\,2$-branes.} 
of bosons and fermions) and ${{\alpha^{\prime}}\over 4}[p^2_L-p^2_R]=N$ 
is the total momentum mode of open strings around $S^1$.  
So, the statistical entropy $S_{stat}=2\pi\sqrt{J^2+NQ_2Q_5Q_6}$ 
agrees with the Bekenstein-Hawking entropy.  

\par

\section*{Acknowledgments}{I would like to thank M. Cveti\v c for 
suggesting me to write this review paper, for discussion during the 
initial stage of the work and for proof-reading part of the review.  
Part of the work was done while the author was at University of 
Pennsylvania.  The work is supported by DOE grant DE-FG02-90ER40542.}

\clearpage

\end{sloppypar}
\end{document}